\pdfoutput=1
\documentclass{cernyrep} 
\usepackage{texnames}
\usepackage[T1]{fontenc}
\usepackage{xcolor}

\usepackage[bookmarks, colorlinks=true, linktoc=page, pdftex, linkcolor=blue, citecolor=blue, urlcolor=blue]{hyperref}

\begin{document}
\begin{flushright}
\footnotesize
CERN-TH-2018-142\\
CP3-Origins-2018-023 DNRF90\\
FERMILAB-PUB-18-264-T\\
IFT-UAM-CSIC-18-060\\
IPMU18-0109\\
KIAS-P18052\\
LCTP-18-17\\
TTP18-022\\
ULB-TH/18-09\\
UMD-PP-018-04\\
YITP-SB-18-16
\end{flushright}
\title{Long-Lived Particles at the Energy Frontier: \\ The MATHUSLA Physics Case}
 
\IfFileExists{../authors.tex}{



\author{
Editors:\\ 
David Curtin$^{1}$, \,
Marco Drewes$^{2}$, \,
Matthew McCullough$^{3}$, \,
Patrick Meade$^{4}$, \,
Rabindra N. Mohapatra$^{5}$, \,
Jessie Shelton$^{6}$, \,
Brian Shuve$^{7, 8}$.
\\[0.4cm]
Contributors: \\Elena Accomando$^{9}$, \,
Cristiano Alpigiani$^{10}$, \,
Stefan Antusch$^{11}$, \,
Juan Carlos Arteaga-Vel\'azquez$^{12}$, \,
Brian Batell$^{13}$, \,
Martin Bauer$^{14}$, \,
Nikita Blinov$^{8}$, \,
Karen Salom\'e Caballero-Mora$^{15, 16}$, \,
Jae Hyeok Chang$^{4}$, \,
Eung Jin Chun$^{17}$, \,
Raymond T. Co$^{18}$, \,
Timothy Cohen$^{19}$, \,
Peter Cox$^{20}$, \,
Nathaniel Craig$^{21}$, \,
Csaba Cs\'aki$^{22}$, \,
Yanou Cui$^{23}$, \,
Francesco D'Eramo$^{24}$, \,
Luigi Delle Rose$^{25}$, \,
P. S. Bhupal Dev$^{26}$, \,
Keith R. Dienes$^{27, 5}$, \,
Jeff A. Dror$^{28, 29}$, \,
Rouven Essig$^{4}$, \,
Jared A. Evans$^{30, 6}$, \,
Jason L. Evans$^{17}$, \,
Arturo Fern\'andez Tellez$^{31}$, \,
Oliver Fischer$^{32}$, \,
Thomas Flacke$^{33}$, \,
Anthony Fradette$^{34}$, \,
Claudia Frugiuele$^{35}$, \,
Elina Fuchs$^{35}$, \,
Tony Gherghetta$^{36}$, \,
Gian F. Giudice$^{3}$, \,
Dmitry Gorbunov$^{37, 38}$, \,
Rick S. Gupta$^{39}$, \,
Claudia Hagedorn$^{40}$, \,
Lawrence J. Hall$^{28, 29}$, \,
Philip Harris$^{41}$, \,
Juan Carlos Helo$^{42, 43}$, \,
Martin Hirsch$^{44}$, \,
Yonit Hochberg$^{45}$, \,
Anson Hook$^{5}$, \,
Alejandro Ibarra$^{46, 17}$, \,
Seyda Ipek$^{47}$, \,
Sunghoon Jung$^{48}$, \,
Simon Knapen$^{29, 28}$, \,
Eric Kuflik$^{45}$, \,
Zhen Liu$^{49}$, \,
Salvator Lombardo$^{22}$, \,
H.~J.~Lubatti$^{10}$, \,
David McKeen$^{50}$, \,
Emiliano Molinaro$^{51}$, \,
Stefano Moretti$^{9, 52}$, \,
Natsumi Nagata$^{53}$, \,
Matthias Neubert$^{54, 22}$, \,
Jose Miguel No$^{55, 56}$, \,
Emmanuel Olaiya$^{52}$, \,
Gilad Perez$^{35}$, \,
Michael E. Peskin$^{8}$, \,
David Pinner$^{57, 58}$, \,
Maxim Pospelov$^{59, 34}$, \,
Matthew Reece$^{57}$, \,
Dean J. Robinson$^{30}$, \,
Mario Rodr\'iguez Cahuantzi$^{31}$, \,
Rinaldo Santonico$^{60}$, \,
Matthias Schlaffer$^{35}$, \,
Claire H. Shepherd-Themistocleous$^{52}$, \,
Andrew Spray$^{33}$, \,
Daniel Stolarski$^{61}$, \,
Martin A. Subieta Vasquez$^{62, 63}$, \,
Raman Sundrum$^{5}$, \,
Andrea Thamm$^{3}$, \,
Brooks Thomas$^{64}$, \,
Yuhsin Tsai$^{5}$, \,
Brock Tweedie$^{13}$, \,
Stephen M. West$^{65}$, \,
Charles Young$^{8}$, \,
Felix Yu$^{54}$, \,
Bryan Zaldivar$^{55, 66}$, \,
Yongchao Zhang$^{26, 67}$, \,
Kathryn Zurek$^{29, 28, 3}$, \,
Jos\'e Zurita$^{32, 68}$.
\vspace*{2cm}}

\institute{
$^{1}$ Department of Physics, University of Toronto, Toronto, ON M5S 1A7, Canada \\ 
$^{2}$ Centre for Cosmology, Particle Physics and Phenomenology, Universit\'e Catholique de Louvain, Louvain-la-Neuve, B-1348, Belgium \\ 
$^{3}$ CERN, TH Department, CH-1211 Geneva, Switzerland \\ 
$^{4}$ C.N. Yang Institute for Theoretical Physics, Stony Brook University, Stony Brook, NY 11794, USA \\ 
$^{5}$ Maryland Center for Fundamental Physics, Department of Physics, University of Maryland, College Park, MD 20742-4111 USA \\ 
$^{6}$ Department of Physics, University of Illinois at Urbana-Champaign, Urbana, IL, 61801, USA \\ 
$^{7}$ Harvey Mudd College, 301 Platt Blvd., Claremont, CA 91711, USA \\ 
$^{8}$ SLAC National Accelerator Laboratory, Menlo Park, California 94025 USA \\ 
$^{9}$ School of Physics and Astronomy, University of Southampton, Highfield, Southampton SO17 1BJ, United Kingdom \\ 
$^{10}$ Department of Physics, University of Washington, Seattle, WA 98195, USA \\ 
$^{11}$ Department of Physics, University of Basel, Klingelbergstrasse 82, CH-4056 Basel, Switzerland \\ 
$^{12}$ Instituto de F\'\i sica y Matem\'aticas, Universidad Michoacana de San Nicol\'as de Hidalgo, Edificio C3, Ciudad Universitaria 58030, Morelia, Michoac\'an, M\'exico \\ 
$^{13}$ Pittsburgh Particle Physics, Astrophysics, and Cosmology Center, Department of Physics and Astronomy, University of Pittsburgh, PA 15260, USA \\ 
$^{14}$ Institut f\"ur Theoretische Physik, Universit\"at Heidelberg, Philosophenweg 16, 69120 Heidelberg, Germany \\ 
$^{15}$ Facultad de Ciencias en F\'isica y Matem\'aticas, Universidad Aut\'onoma de Chiapas (FCFM-UNACH), Ciudad Universitaria UNACH, Carretera Emiliano Zapata Km. 8 Rancho San Francisco, Ciudad Universitaria Ter\'an, Tuxtla Guti\'errez, Chiapas C.P. 29050 Chiapas, Mexico \\ 
$^{16}$ Mesoamerican Centre for Theoretical Physics, Universidad Autonoma de Chiapas (MCTP-UNACH), Ciudad Universitaria UNACH, Carretera Emiliano Zapata Km. 4, Real del Bosque (Ter/'an), Tuxtla Guti\'errez, Chiapas C.P. 29050 Chiapas, Mexico \\ 
$^{17}$ Korea Institute for Advanced Study, Seoul 02455, Korea \\ 
$^{18}$ Leinweber Center for Theoretical Physics, University of Michigan, Ann Arbor, MI 48109, USA \\ 
$^{19}$ Institute of Theoretical Science, University of Oregon, Eugene, OR 97403, USA \\ 
$^{20}$ Kavli IPMU (WPI), UTIAS, University of Tokyo, Kashiwa, Chiba 277-8583, Japan \\ 
$^{21}$ Department of Physics, University of California, Santa Barbara, CA 93106, USA \\ 
$^{22}$ Department of Physics, LEPP, Cornell University, Ithaca, NY 14853, USA \\ 
$^{23}$ Department of Physics and Astronomy, University of California, Riverside, California 92521, USA \\ 
$^{24}$ Dipartimento di Fisica ed Astronomia, Universit\'a di Padova, Via Marzolo 8, 35131 Padova, Italy \\ 
$^{25}$ INFN, Sezione di Firenze and Department of Physics and Astronomy, University of Florence, Via G. Sansone 1, 50019 Sesto Fiorentino, Italy \\ 
$^{26}$ Department of Physics and McDonnell Center for the Space Sciences,  Washington University, St. Louis, MO 63130, USA \\ 
$^{27}$ Department of Physics, University of Arizona, Tucson, AZ  85721  USA \\ 
$^{28}$ Department of Physics, University of California, Berkeley, CA 94720, USA \\ 
$^{29}$ Theoretical Physics Group, Lawrence Berkeley National Laboratory, Berkeley, California 94720, USA \\ 
$^{30}$ Department of Physics, University of Cincinnati, Cincinnati, Ohio 45221, USA \\ 
$^{31}$ Facultad de Ciencias F\'isico Matem\'aticas, Benem\'erita Universidad Aut\'onoma de Puebla, Av. San Claudio y 18 Sur, Edif. EMA3-231, Ciudad Universitaria 72570, Puebla, M\'exico \\ 
$^{32}$ Institute for Nuclear Physics (IKP), Karlsruhe Institute of Technology, Hermann-von-Helmholtz-Platz 1, D-76344 Eggenstein-Leopoldshafen, Germany \\ 
$^{33}$ Center for Theoretical Physics of the Universe, Institute for Basic Science (IBS), Daejeon, 34126, Korea \\ 
$^{34}$ Department of Physics and Astronomy, University of Victoria, Victoria, BC, V8P 5C2, Canada \\ 
$^{35}$ Department of Particle Physics and Astrophysics, Weizmann Institute of Science, Rehovot 7610001, Israel \\ 
$^{36}$ School of Physics and Astronomy, University of Minnesota, Minneapolis, MN, 55455, USA \\ 
$^{37}$ Institute for Nuclear Research of the Russian Academy of Sciences, 60th October Anniversary prospect 7a, Moscow 117312, Russia \\ 
$^{38}$ Moscow Institute of Physics and Technology, Institutsky per. 9, Dolgoprudny 141700, Russia \\ 
$^{39}$ Institute of Particle Physics Phenomenology, Durham University, Durham, UK DH1 3LE \\ 
$^{40}$ CP$^{3}$-Origins, University of Southern Denmark, Campusvej 55, DK-5230 Odense M, Denmark \\ 
$^{41}$ MIT, 77 Massachusetts Ave, Cambridge, Ma 02139, USA \\ 
$^{42}$ Departamento de F\'{i}sica,Facultad de Ciencias, Universidad de La Serena,Avenida Cisternas 1200, La Serena, Chile \\ 
$^{43}$ Centro-Cient\'{i}fico-Tecnol\'{o}gico de Valpara\'{i}so, Casilla 110-V, Valpara\'{i}so, Chile \\ 
$^{44}$ AHEP Group, Instituto de F\'{\i}sica Corpuscular, C.S.I.C./Universitat de Val{\`e}ncia, Edificio de Institutos de Paterna, Apartado 22085, E--46071 Val{\`e}ncia, Spain \\ 
$^{45}$ Racah Institute of Physics, Hebrew University of Jerusalem, Jerusalem 91904, Israel \\ 
$^{46}$ Physik-Department T30d, Technische Universit\"at M\"unchen, James-Franck-Stra\ss{}e, 85748 Garching, Germany \\ 
$^{47}$ Department of Physics and Astronomy, University of California Irvine, 4129 Frederick Reines Hall, Irvine, CA 92617, U.S.A. \\ 
$^{48}$ Department of Physics and Astronomy, Center for Theoretical Physics, Seoul National University, Seoul 08826, South Korea \\ 
$^{49}$ Theoretical Physics Department, Fermi National Accelerator Laboratory, Batavia, Illinois 60510, USA \\ 
$^{50}$ TRIUMF, 4004 Wesbrook Mall, Vancouver, BC V6T 2A3, Canada \\ 
$^{51}$ Department of Physics and Astronomy, University of Aarhus, Ny Munkegade 120, DK-8000 Aarhus C, Denmark \\ 
$^{52}$ Particle Physics Department, STFC Rutherford Appleton Laboratory, Chilton, Didcot, Oxon OX11 0QX, UK \\ 
$^{53}$ Department of Physics, University of Tokyo, Tokyo 113-0033, Japan \\ 
$^{54}$ PRISMA Cluster of Excellence \& Mainz Institute for Theoretical Physics, Johannes Gutenberg University, 55099 Mainz, Germany \\ 
$^{55}$ Departamento de Fisica Teorica and Instituto de Fisica Teorica, IFT-UAM/CSIC, Universidad Autonoma de Madrid, Cantoblanco, 28049, Madrid, Spain \\ 
$^{56}$ Department of Physics, King's College London, Strand, WC2R 2LS London, UK \\ 
$^{57}$ Department of Physics, Harvard University, Cambridge, MA 02138, USA \\ 
$^{58}$ Department of Physics, Brown University, Providence, RI 02912, USA \\ 
$^{59}$ Perimeter Institute for Theoretical Physics, 31 Caroline St. N., Waterloo, Ontario N2L 2Y5, Canada \\ 
$^{60}$ INFN Sezione di Roma Tor Vergata, Dipartimento di Fisica, Universit\`a di Roma Tor Vergata,Roma, Italy \\ 
$^{61}$ Ottawa-Carleton Institute for Physics, Carleton University, 1125 Colonel By Drive, Ottawa, Ontario K1S 5B6, Canada \\ 
$^{62}$ Instituto de Investigaciones F\'isicas (IIF) Laboratorio de F\'isica C\'osmica de ``Chacaltaya'' Universidad Mayor de San Andr\'es (UMSA), Campus universitario de Cota Cota, C/27 Av. Mu\~noz Reyes, La Paz, Bolivia \\ 
$^{63}$ Planetario ``Max Schreier'' Universidad Mayor de San Andr\'es (UMSA), C. Federico Zuazo N$^\circ$ 1976, La Paz, Bolivia \\ 
$^{64}$ Department of Physics, Lafayette College, Easton, PA 18042, USA \\ 
$^{65}$ Department of Physics, Royal Holloway University of London, Egham, Surrey, TW20 0EX, UK. \\ 
$^{66}$ Univ. Grenoble Alpes, CNRS, USMB, LAPTh, 74000, Annecy, France \\ 
$^{67}$ Service de Physique Th\'{e}orique, Universit\'{e} Libre de Bruxelles, Boulevard du Triomphe, CP225, 1050 Brussels, Belgium \\ 
$^{68}$ Institute for Theoretical Particle Physics (TTP), Karlsruhe Institute of Technology, Egesserstra{\ss}e 7, D-76128 Karlsruhe, Germany 
}}{}

\maketitle 

\newpage 

\begin{abstract}


We examine the theoretical motivations for long-lived particle (LLP) signals at the LHC in a comprehensive survey of Standard Model (SM) extensions. LLPs are a common prediction of a wide range of theories that address unsolved fundamental mysteries such as naturalness, dark matter, baryogenesis and neutrino masses, and represent a natural and generic possibility for physics beyond the SM (BSM). In most cases the LLP lifetime can be treated as a free parameter from the $\mu$m scale up to the Big Bang Nucleosynthesis limit of $\sim 10^7$m. Neutral LLPs with lifetimes above $\sim$ 100m are particularly difficult to probe, as the sensitivity of the LHC main detectors is limited by challenging backgrounds, triggers, and small acceptances. MATHUSLA is a proposal for a minimally instrumented, large-volume surface detector near ATLAS or CMS. It would search for neutral LLPs produced in HL-LHC collisions by reconstructing displaced vertices (DVs) in a low-background environment, extending the sensitivity of the main detectors  by orders of magnitude in the long-lifetime regime.  In this white paper we study the LLP physics opportunities afforded by a MATHUSLA-like detector at the HL-LHC.  We develop a model-independent approach to describe the sensitivity of MATHUSLA to BSM LLP signals, and compare it to DV and missing energy searches at ATLAS or CMS. We then explore the BSM motivations for LLPs in considerable detail, presenting a large number of new sensitivity studies.  While our discussion is especially oriented towards the long-lifetime regime at MATHUSLA, this survey underlines the importance of a varied LLP search program at the LHC in general. By synthesizing these results into a general discussion of the top-down and bottom-up motivations for LLP searches, it is our aim to demonstrate the exceptional strength and breadth of the physics case for the construction of the MATHUSLA detector.

\end{abstract}

 \addtolength{\voffset}{-1.3cm}
\addtolength{\textheight}{1.1cm}

\tableofcontents
\clearpage

\addtolength{\voffset}{1.3cm}
\addtolength{\textheight}{-1.1cm}

\clearpage
\newcommand{\nc}{\newcommand}

\newcommand{\met}{E\!\!\!\!/_T} 
\def\iab{ab$^{-1}$}
\def\ifb{fb$^{-1}$}
\def\ipb{pb$^{-1}$}
\def \gsim{\mathrel{\vcenter
     {\hbox{$>$}\nointerlineskip\hbox{$\sim$}}}}
\def \lsim{\mathrel{\vcenter
     {\hbox{$<$}\nointerlineskip\hbox{$\sim$}}}}

\newcommand{\be}{\begin{equation}}
\newcommand{\ee}{\end{equation}}
\def\bea{\begin{eqnarray}}
\def\eea{\end{eqnarray}}
\nc{\beq}{\begin{equation}}
\nc{\eeq}{\end{equation}}
\nc{\barray}{\begin{eqnarray}}
\nc{\earray}{\end{eqnarray}}
\nc{\barrayn}{\begin{eqnarray*}}
\nc{\earrayn}{\end{eqnarray*}}
\nc{\bcenter}{\begin{center}}
\nc{\ecenter}{\end{center}}
\nc{\mc}{\mathcal}

\newcommand{\MeV}{\textrm{ MeV}}
\newcommand{\GeV}{\textrm{ GeV}}
\newcommand{\TeV}{\textrm{ TeV}}

\newcommand{\Sec}[1]{Sec.~\ref{#1}}
\newcommand{\Secs}[2]{Secs.~\ref{#1} and \ref{#2}}
\newcommand{\Fig}[1]{Fig.~(\ref{#1})}
\newcommand{\Figs}[2]{Figs.~\ref{#1} and \ref{#2}}
\newcommand{\App}[1]{App.~\ref{#1}}
\newcommand{\vev}[1]{\langle #1 \rangle}

\nc{\onehalf}{\frac{1}{2}} 
\nc{\partialbar}{\bar{\partial}}
\nc{\psit}{\widetilde{\psi}}
\nc{\Tr}{\mbox{Tr}}
\nc{\hc}{\mbox{H.c.}}
\nc{\ev}{\;\mathrm{eV}}
\nc{\mev}{\;\mathrm{MeV}}
\nc{\gev}{\;\mathrm{GeV}}
\nc{\tev}{\;\mathrm{TeV}}
\def\stau{\widetilde{\tau}}
\def\squark{\widetilde{q}}
\def\sneutrino{\widetilde{\nu}}
\def\sb{\tilde{b}}
\def\su{\tilde{u}}
\def\sd{\tilde{d}}
\def\st{\tilde{t}}
\def\l{\ell}
\def\slepton{\tilde{\l}}
\def\chii0{\chi_i^0}
\def\chij0{\chi_j^0}
\def\chiipm{\chi_i^{\pm}}
\def\chijpm{\chi_j^{\pm}}
\def\chiimp{\chi_i^{\mp}}
\def\chijmp{\chi_j^{\mp}}

\newcommand{\figref}[1]{Fig.~\ref{fig:#1}}
\renewcommand{\eqref}[1]{Eq.~(\ref{eq:#1})}
\newcommand{\eqrefs}[2]{Eqs.\ (\ref{eq:#1}) - (\ref{eq:#2}) }
\newcommand{\aref}[1]{Appendix~\ref{sec:#1}}
\newcommand{\secref}[1]{Section~\ref{sec:#1}}
\newcommand{\tabref}[1]{Table~\ref{tab:#1}}

\hyphenation{MCFODO}

\def\PM#1{{\bf  \textcolor{violet}{[PM: {#1}]}}}
\def\MM#1{{\bf  \textcolor{blue}{[MM: {#1}]}}}
\def\JS#1{{\bf  \textcolor{red}{[JS: {#1}]}}}

\def\AQ#1{{\bf  \textcolor{orange}{\emph{[Question for Authors: {#1}]}}}}

\def\PMc#1{{\textcolor{violet}{{#1}}}}
\def\MMc#1{{\textcolor{blue}{{#1}}}}
\def\JSc#1{{\textcolor{red}{{#1}}}}
\def\MPc#1{{\textcolor{orange}{{#1}}}}

\def\DC#1{{\bf  \color{magenta}{[DC: {#1}]}}}
\def\DCc#1{ {\color{magenta} #1} }
\def\magentacolor{\color{magenta}} 


%
\section*{Foreword}

The MATHUSLA detector concept was proposed in~\cite{Chou:2016lxi} to detect neutral long-lived particles (LLPs) produced at the HL-LHC that could be missed by the main detectors. By providing a very low background environment without trigger limitations for the detection of ultra-long-lived particle decays, the MATHUSLA detector could greatly extend the LLP search capabilities of the LHC, at a relatively incremental cost.
This idea led to the rapid formation of an experimental collaboration and the recent deployment of the small-scale MATHUSLA Test Stand detector at CERN to study backgrounds and help develop a  full-scale detector proposal.

The present document is the result of an extensive study carried out by the high energy theory community to demonstrate the broad nature of the physics motivation for LLP searches at the LHC.\footnote{
Since the appearance of this manuscript as a preprint,  a narrow subset of the MATHUSLA physics case was further explored in~\cite{Beacham:2019nyx}, comparing reach of experiments like SHiP, MATHUSLA, CODEX and FASER to several low-energy simplified LLP models.
}
We survey motivations for new physics at the LHC, including (for instance) many approaches to the hierarchy problem, dark matter, baryogenesis, and neutrino masses, and demonstrate how LLPs emerge as natural and generic predictions of these theories. While our treatment here focuses on the long-lifetime regime that especially motivates the construction of the MATHUSLA detector, our general discussion applies to  LLP searches at the LHC more broadly.   
We additionally discuss how predictions from this wide range of theories map into the space of LLP signatures at both MATHUSLA and the LHC main detectors, and establish the regions in LLP parameter space where MATHUSLA offers unique sensitivity. 

Many of the sections represent a collaboration of many authors and editors. Other sections contain primarily the work of a few authors, who are then indicated at the beginning of such sections.

This document focuses on the theoretical motivation for LLP signals at MATHUSLA without going into great detail regarding the detector design, since the theoretical arguments and signal estimates are relatively independent of the precise instrumentation. 
The MATHUSLA experimental collaboration aims to present a Letter of Intent in the second half of 2018, and develop detailed proposals for a full-scale MATHUSLA detector. It is the intention of the authors  that the arguments presented in this document aid the preparation and theoretical justification of these proposals.

MATHUSLA also has significant capabilities to act as a cosmic ray telescope. Some of the potential physics studies in this field are briefly pointed out here. A companion document exploring this secondary physics case, which would represent a guaranteed return on the investment of building the detector, is currently in preparation.

\newpage
\section{Introduction}
\label{sec:intro}

The quest for physics beyond the Standard Model (SM) encompasses many frontiers and employs a multitude of methods.
While any genuine deviation from SM predictions is a sign of new physics, by far the most illuminating of these methods would be to directly produce new particle that can arise in theories Beyond the SM (BSM), and study their properties. 
The most basic properties of such new particles are their mass, charge, spin, and lifetime. 
While most direct searches for new physics often focus on short lifetimes, it is important to note that long-lived particles (LLPs), defined to have macroscopically detectable decay lengths, are ubiquitous in BSM physics.  
This can be trivially demonstrated by reference to known SM phenomena, which include a myriad of lifetimes ranging from $< 10^{-24}$ seconds for the top quark to at least $>10^{41}$ seconds for the proton.  Given this 
gargantuan
range of lifetimes in the SM and potentially in BSM physics, it is important to understand the simple origin of LLPs.  

The proper lifetime $\tau$ of any particle is given by the inverse of its decay width $\Gamma$, which can be calculated straightforwardly in Quantum Field Theory (QFT) as
\begin{equation}\label{eqn:lifetime}
d\Gamma \sim \frac{1}{M} \vert \mathcal{M}\vert^2 d\Pi.
\end{equation}
$M$ is the mass of the particle, $\mathcal{M}$ is the matrix element governing the decay, and $d\Pi$ is the phase space for the decay.  For a particle to be long-lived, the matrix element  and/or the available phase space must be small. 
It is straightforward to characterize scenarios that lead to either case, and to find examples realizing these possibilities in the SM.

The matrix element for decay could be suppressed due to an approximate symmetry which would forbid the decay if it was precise, or simply a small effective coupling constant.\footnote{While small effective couplings and approximate symmetries can coincide, this is not always the case and we distinguish them as logically distinct possibilities.}  A small coupling in the matrix element can be further distinguished by whether it originates from a dimensionless coupling constant, or a  dimensionful scale, larger than $M$, from a higher-dimension operator that mediates the decay. 
Phase space can also be suppressed due to the small breaking of an approximate symmetry that splits otherwise degenerate states, or it can arise due to accidental degeneracies in the spectrum.

Examples of these suppression mechanisms in the SM are plentiful.
The proton is perhaps the most extreme example of an approximate symmetry giving rise to a very long lifetime, since proton decay is forbidden by baryon number, which is an accidental symmetry of the SM. 
 The long lifetime of the $\mu$ arises from a small coupling corresponding to a large dimensionful scale, the Fermi constant $G_F$, arising due to the high mass of the $W$ boson.  
 The Higgs boson, while not macroscopically long-lived, has a lifetime significantly greater than the similarly massive top quarks or $W/Z$ bosons. It propagates for about a proton diameter before decaying, due to the small dimensionless bottom Yukawa coupling $y_b\sim 0.02$ that dominates Higgs decay in the SM.
The SM neutron is very long-lived, with a lifetime of about 15 minutes outside the atomic nucleus, in large part due to the  phase space suppression of its weak decay to a proton and leptons.  
The $b$-quark's relatively long lifetime is due to a combination of effects: phase space suppression, approximate flavor symmetry and large dimensionful scale in the decay. It is not sufficiently long-lived to pass through the detector, but has a macroscopic decay length $\sim$ mm at the LHC.  The macroscopic decays of  $B$-hadrons are utilized for particle ID, which is essential for studying properties of the Higgs and searches for BSM physics.

Given the abundance of LLPs in the SM, and that all classes of mechanisms for creating a long-lifetime occur within the SM, it is unsurprising that LLPs are commonplace in BSM theories as well.   However, the discovery of LLPs presents experimental challenges stemming from both their production and detection.  For instance, the nature of of modern general-purpose particle physics detectors at the energy frontier is biased towards ``prompt'' particle production, assuming relatively short lifetimes $\lesssim \mu$m and maintaining good geometrical acceptance for decays $\mathcal{O}(1\, \mathrm{m})$ away from the IP.
In part this is due to practical considerations, since requiring approximate $4 \pi$ coverage makes detector sizes above $\mathcal{O}(10\, \mathrm{m})$ logistically and financially unfeasible.

If the LLP is charged or colored, it can be detected as it passes through the detector. 
Neutral LLPs could be detected via scattering off a shielded detector target if they are light enough to be produced in very large numbers at fixed-target experiments and have significant SM interactions, as might be the case with new weakly interacting states or DM candidates coupling to dark photon mediators~\cite{Shrock:1978ft,  Gallas:1994xp, Izaguirre:2013uxa, Battaglieri:2016ggd, deNiverville:2011it, Batell:2014mga, Aguilar-Arevalo:2017mqx, Aguilar-Arevalo:2018wea, deNiverville:2016rqh}. 
Unfortunately, most neutral LLP species, especially those that can only be produced at the LHC, are either too heavy or too feebly interacting to be observed in this manner. Therefore, the only way of directly detecting these neutral LLPs is to observe their decay.
%
%
An LLP with decay length $\lambda \gg 10$ m only has small probability $\sim L/\lambda$ of decaying inside a detector of size $L$.
\emph{This small  probability makes ultra-long-lived neutral particles inherently rare signals.} Background suppression and good trigger efficiency are therefore vital to their discovery.

Intensity frontier experiments are natural settings for neutral LLP searches, since their smaller scale means they can be customized to search for low-mass hidden sectors with very low backgrounds.  
Such experiments have much lower center-of-mass energies than the LHC, instead aiming for large rates of low-energy processes due to the intensity of their beams. 
If the LLP is light enough to be produced, one can try and exploit the inverse decay process of the LLP for its production.
However, this precludes probing LLPs with masses above a few GeV, not to mention the weak or TeV scale.
Furthermore, if the LLP is long-lived due to a higher-dimensional operator suppressed by a high scale, then processes at energies above that scale could lead to much larger LLP production rates than are possible at intensity frontier experiments. 
Finally, in many theories the LLP couples not just to its decay products but also other heavy SM or BSM particles. Only by accessing higher energies do these new production processes become available.

It is therefore clear that searching for ultra-long-lived neutral particles at energy frontier experiments like the LHC, HL-LHC or HE-LHC has many advantages. 
One can exploit the irreducible production mechanism which corresponds to the LLP's inverse decay for both dimensionless couplings and dimensionful scale suppressions.  For dimensionful couplings it is trivial to understand why the energy frontier is more powerful than the intensity frontier, as the cross-section for production scales with energy.  Typically the energy frontier is also more powerful for dimensionless couplings as well.  For instance, although the well-motivated Higgs portal coupling to a hidden sector is fundamentally through a dimensionless coupling, at low-energy intensity frontier experiments this is effectively a dimensionful scale suppression set by the Higgs mass.  Therefore any energy frontier experiment that can produce the Higgs will be a more powerful search tool for LLPs coupling to the Higgs. %
Even in the case of a dimensionless coupling and a low mass scale, for example hidden photon models, the 
energy frontier can in principle be just as powerful as intensity frontier experiments, since energy frontier experiments like the HL-LHC also utilize very high intensity beams.
Energy frontier experiments are also well suited for more general models of LLPs, because the most efficient production mechanism for LLPs {\em is often completely unrelated} to the LLP decay mechanism.   For instance, a large swath of BSM physics models described by supersymmetry~\cite{Martin:1997ns} have LLPs.  The R-parity symmetry structure inherently favors energy frontier production.  Even if R-parity is slightly broken, allowing for macroscopic decays, the production of the heavy states is better suited for the energy frontier rather than using the inverse decay production mechanism typical of the intensity frontier. 

The HL-LHC and HE-LHC would therefore be ideal tools for studying the  \emph{Lifetime Frontier}, supplying both the energy reach and the luminosity needed to study possibly rare LLP signals.   However, as mentioned, the general purpose LHC detectors have intrinsic limitations that restrict their reach for very long-lived neutral particles. 
While LLP decays can be spectacular signals, the high-rate HL-LHC environment is unforgiving, and large QCD backgrounds as well as triggering limitations are major bottlenecks for many LLP searches. 
Furthermore, while missing energy (MET) searches have great utility in probing new physics giving rise to either more than several 100 GeV of MET or QCD-like production rates, sensitivity drops dramatically for rarer or even slightly softer signals. 
Finally, even if a very long-lived state were discovered via MET searches at ATLAS or CMS, a critical cosmological question remains: is the newly discovered state a dark matter candidate, or a meta-stable state?

To probe all accessible possibilities for physics beyond the SM, it would therefore be very useful to {\em combine} the reach of an energy frontier experiment with the shielding and lever arm of a long-lived particle detector.  
This has been explored before in a number of cases for the LHC, but there is a tradeoff between the volume of the new detector and the distance from an LHC IP~\cite{Meade:2009mu} that precludes the use of existing detectors and calls for a new design. 
The ideal LLP detector would be shielded from QCD backgrounds produced in LHC collisions, while being large and close enough to have sufficient acceptance for LLP decays.

Remarkably, a connection between collider physics and cosmology reveals a lifetime of interest that corresponds to a dedicated LLP detector of achievable size. 
If the LHC is capable of producing a certain LLP, the same LLP is likely to have been thermally produced in the early universe. As a result, its lifetime is bounded from above by the strict constraints on Big Bang Nucleosynthesis from primordial elemental abundances~\cite{Jedamzik:2006xz}.  In most cases, this upper bound is about $\tau \lesssim 0.1$s. 
Fortunately, a detector at the surface above an LHC experiment, with a realistically sized, instrumentable volume, is precisely in the range to probe particles with such lifetimes if they are produced with plausible LHC cross-sections, such as in sizable exotic Higgs decays~\cite{Chou:2016lxi}.  
This connection led to the genesis of the proposed MATHUSLA experiment: a general-purpose LLP detector that exploits energy frontier particle production, coupled with shielding and size that is powerful enough to probe an enormous array of LLP scenarios. 

The detector design for MATHUSLA is discussed in Section~\ref{sec:mathuslasummary}. In the simplest terms, MATHUSLA is a large tracker that can reconstruct displaced  vertices on the surface near ATLAS or CMS. 
Its surface location provides shielding from QCD backgrounds at the interaction point. The remaining backgrounds can be rejected, allowing MATHUSLA to operate in the near-background-free regime without trigger limitations.
The design is  scalable, making it highly flexible from the budgetary standpoint as well as allowing for upgradability depending on the physics benchmarks of interest.  
The MATHUSLA program also provides a number of exciting possibilities and benefits beyond the HL-LHC LLP search program.
As discussed in Section~\ref{s.mathuslamodularity}, the detector will also be able to act as a powerful and unique cosmic ray telescope, independent of the LHC and without interfering with the primary LLP search objective. 
Additionally, since MATHUSLA is  proposed to be above an LHC IP, it will be useful for the entire lifetime of the HL-LHC program {\em and} a possible HE-LHC successor. 
MATHUSLA not only can extend the reach of an LHC experiment, it can also complement other discovery channels.  If for instance a MET signature was discovered by the LHC, MATHUSLA could provide valuable additional information on the spectrum and properties of the hidden sector.

After the MATHUSLA detector was proposed in~\cite{Chou:2016lxi}, its reach has been the subject of several studies~\cite{Co:2016fln, Dev:2016vle, Dev:2017dui, Caputo:2017pit, Curtin:2017izq, Evans:2017kti, Helo:2018qej, DAgnolo:2018wcn, Berlin:2018pwi, Deppisch:2018eth, Jana:2018rdf}, but this was usually done in the context of specific models.
Recent complementary proposals for external LLP detectors at the LHC~\cite{Feng:2017vli, Feng:2017uoz, Evans:2017lvd, Gligorov:2017nwh}, as well as recent communal efforts to help guide the LLP search program at the LHC~\cite{LHCLLPwhitepaper}, underscore the need for a comprehensive, general examination of the physics motivation for LLP searches.  The purpose of this document is therefore to explore, in detail, the physics case for neutral LLP searches in general and for construction of MATHUSLA in particular. 

To this end, we develop in Section~\ref{sec:LLPSmathuslahllhc} a model-independent understanding of the MATHUSLA signal yield, the resulting mass and lifetime reach, and how to compare that reach to the ATLAS or CMS main detectors. Many of our methods can be applied, with minor modifications, to help understand other LLP detector proposals as well.

In Sections~\ref{sec:naturalness}, \ref{sec:dmtheory}, \ref{sec:baryogenesis} and \ref{sec:neutrinos} we examine the top-down motivations for LLPs with long lifetimes in theories that address the fundamental mysteries of Naturalness, Dark Matter, Baryogenesis and Neutrino Masses respectively. 
Section~\ref{sec:bottomup} examines generic bottom-up scenarios, including hidden valleys~\cite{Strassler:2006im,Strassler:2006ri} and minimal extensions of the SM with additional scalars or vectors. 
Hidden valleys in particular deserve mention as one of the most generic bottom-up BSM possibilities that gives rise to LLPs. The possibility of a separate sector with its own particles and forces, only connected to the SM by a small portal coupling or a heavy mediator, is a straightforward consequence of the structure of gauge theories. Any massive states in the hidden sector that are not absolutely stable are natural LLP candidates.
These and other bottom-up possibilities are not only plausible on general grounds, they also arise as components of more complete theories, including those discussed in earlier sections.

Our investigation demonstrates the extremely broad motivation of LLPs and the foundational importance of searches for their signatures. 
LLPs not only arise ubiquitously in BSM theories; in many cases, they are intrinsic to the underlying theory mechanism as well.
The lesson for the entire LHC search program is obvious: LLP searches need to be a priority, and they should be explored with the main detectors as well as dedicated experiments like MATHUSLA that take advantage of existing LHC facilities.
For many broad classes of BSM scenarios with LLPs, MATHUSLA is the first or only discovery opportunity, being able to detect new physics with TeV scale masses and very long lifetimes with sensitivities that can exceed the cross-section reach of main detector searches by orders of magnitude. 
Clearly, the discovery potential of such a general-purpose LLP detector is enormous.

We prepare an Executive Summary of our findings in Section~\ref{sec:signatures}, which can be read as a stand-alone document and serves as a big-picture guide to the studies and important lessons of this white paper. MATHUSLA represents a unique opportunity for CERN. The collider to produce LLPs is already in place. A relatively incremental upgrade to maximize our chances of actually detecting these possible harbingers of new physics is not only feasible, but highly motivated from a vast and comprehensive range of bottom-up and top-down theoretical perspectives.

\clearpage 

\nc{\eev}{\;\mathrm{EeV}}
\nc{\pev}{\;\mathrm{PeV}}

\section[The MATHUSLA Detector Proposal]{The MATHUSLA Detector Proposal\footnote{Martin Alfonso, Cristiano Alpigiani, Juan Carlos Arteaga-Velazquez, Mario Rodriguez Cahuantzi, 
David Curtin, Henry Lubatti, Caballero Mora, Karen Salome, Rinaldo Santonico, Arturo Fernandez Tellez, Subieta Vasquez, Charlie Young}}
\label{sec:mathuslasummary}

Here we briefly summarize the design of the MATHUSLA (MAssive Timing Hodoscope for Ultra-Stable neutraL pArticles)  LLP detector for the HL-LHC as first proposed in~\cite{Chou:2016lxi}
(Section~\ref{s.mathuslabasic}).
We then review how such a design can discover and analyze LLP decays (Section \ref{s.mathuslaLLPsearch}),
and why this design is expected to allow the search for these signatures to be conducted with  zero or very low backgrounds (Section \ref{s.mathuslabackgrounds}).

While the original proposal (and this work) analyzes signal sensitivities in the context of the HL-LHC, it is important to note that MATHUSLA would perform its function as a dedicated LLP detector equally well for any future collider built in the same tunnel as the LHC, i.e. the HE-LHC, where higher LLP production rates lead to a correspondingly improved reach. 
We emphasize the modularity and scalability of the MATHUSLA idea and discuss developments towards a more realistic detector design in Section \ref{s.mathuslamodularity}. We also justify the use of the original simplified 200m MATHUSLA benchmark geometry as a physics benchmark for the possible reach of a more realistic design. The recent deployment of a 5-meter scale MATHUSLA ``test stand'' detector at CERN is reviewed in Section \ref{s.mathuslateststand}.

Finally, it was
realized that MATHUSLA has impressive capabilities as a cosmic ray (CR) telescope. This secondary physics mission represents a guaranteed return on the investment of building the detector and will be the subject of its own dedicated studies~\cite{MathuslaCRWP}. For completeness, we qualitatively discuss in Section \ref{s.mathuslaCR} why MATHUSLA can make unique CR measurements.

\subsection{Basic Principles and Simplified Detector Design}
\label{s.mathuslabasic}

The basic motivation for the MATHUSLA detector is the search for LLPs with lifetimes much greater than the size of the LHC main detectors, $c \tau \gg 100$ m.
Any detector that can be reasonably constructed could only catch a small fraction of such LLPs decaying inside of its volume.
Even with potentially large LLP production rates in LHC collisions, suppression of backgrounds is then crucial for discovery. 

The primary signals of neutral LLPs in ATLAS or CMS are displaced vertices (DVs). A DV corresponds to two or more  charged tracks that are reconstructed to originate from the same point in space (and in principle time as well, but this is highly detector dependent), a macroscopic distance displaced from the beam collision point where the LLP originated.
Especially for LLP searches with high energy or leptonic final states, the spectacular geometrical nature of DVs generally leads to very low backgrounds. 
Any other class of LLP signature, such as DVs without high energy or leptonic final states, or the anomalous energy deposits produced when a LLP decays within the calorimeters, suffers from backgrounds and triggering limitations that can be very significant.  As we discuss further in Section~\ref{sec:LLPSmathuslahllhc}, this greatly curtails the main detectors' ability to discover LLPs with very long lifetimes.

To address this broad blind spot of the LHC, 
MATHUSLA is envisioned to be a (1) large, (2) relatively simple (3) surface detector that (4) can robustly reconstruct DVs with good timing resolution. This is to ensure that: (1) the detector has a similar geometric acceptance for LLP decays as the ATLAS or CMS main detectors, which makes it possible to detect LLPs with lifetimes near the generic BBN uppper bound of $\sim 10^7$m if there are no backgrounds; (2) it can be  constructed in time for the HL-LHC upgrade with a realistic budget; (3) it is shielded from QCD backgrounds of the main collision by $\sim 100$m of rock; (4) CR backgrounds to DV searches can be rejected with near-perfect efficiency.

\begin{figure}
\begin{center}
\includegraphics[width=0.6\textwidth]{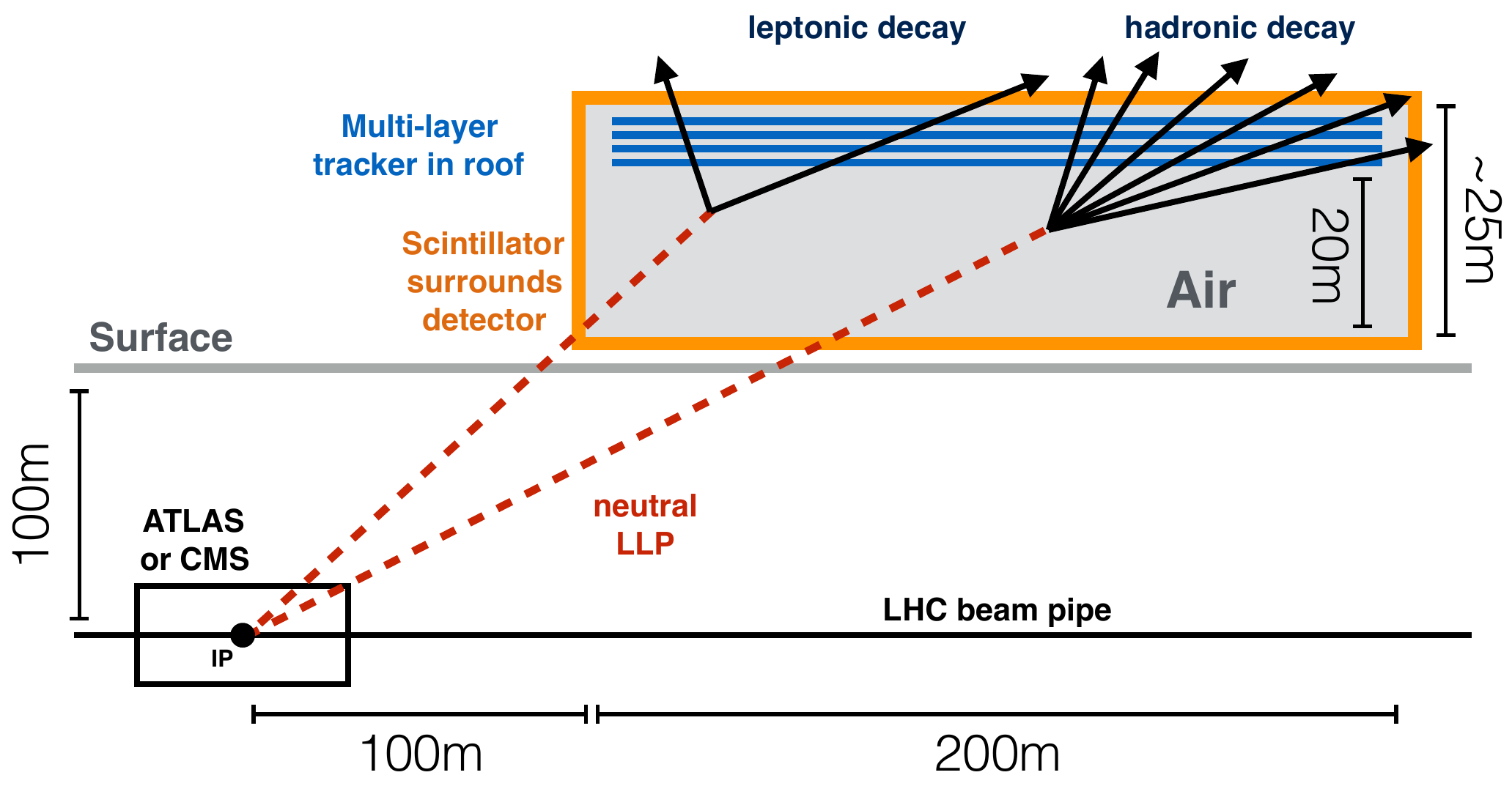}
\end{center}
\caption{
Simplified detector layout showing the position of the $200\,\mathrm{m} \times 200\,\mathrm{m}  \times 20\,\mathrm{m} $ LLP decay volume used for physics studies. The tracking planes in the roof detect charged particles, allowing for the reconstruction of displaced vertices in the air-filled decay volume. The scintillator surrounding the volume provides vetoing capability against charged particles entering the detector.
}
\label{f.mathuslalayout}
\end{figure}

A simplified detector design for MATHUSLA, showing its position on the surface near ATLAS or CMS, is shown in Fig.~\ref{f.mathuslalayout}.
(This is the geometry assumed for physics studies in subsequent Sections.)
The main component of the detector is an approximately 5-m thick tracker array
situated above an air-filled decay volume that is 20 m tall and $200 \,\mathrm{m}  \times 200 \,\mathrm{m} $ in area.
The tracker is envisioned to be composed of  five planes to provide highly robust tracking with a timing resolution of $\sim$1 ns. This allows the dominant background of downward going cosmic ray particles to be reliably separated from upward going LLP decay tracks. 
Each plane has a spatial resolution of $\sim$1 cm in each transverse direction, providing the vertexing capability necessary to confirm the DV signal topology. 
The entire bottom and sides of the decay volume\footnote{The diagram shows the top being covered in scintillator as well, but this might not be required depending on the triggering strategy.} are covered with scintillator to veto incoming charged particles such as high-energy muons coming from the primary pp interaction.

The sensor technology should be proven and cheap in order to achieve the requisite fiducial volume at a reasonable cost.
Resistive Plate Chambers (RPC) is the current default detector technology, though other options are not excluded at this early stage of the design process. 
Its tracking performance has
been proven in many earlier experiments. Indeed, the performance requirements for
MATHUSLA are less stringent than what has already been achieved in large-scale deployments.

For example, ATLAS has achieved a timing resolution of 1 ns and a spatial resolution of 1 cm, 
while CMS has achieved a timing resolution of 1 ns~\cite{Ying:2000pi} and a spatial resolution of 0.81 cm~\cite{Thyssen:2012zz}. RPCs operating in streamer mode at the YangBaJing laboratory for cosmic ray studies have demonstrated the required rate capability~\cite{Bacci:2003bg}. Higher rates can be achieved by operating in avalanche mode.  
RPCs have also been deployed in detectors with similar geometry and areas greater than $\sim 7000 m^2~$\cite{Aielli:2006cj, Iuppa:2015hna}. It is also worth noting that ARGO YBJ operated for 5 years almost unattended, testifying to the reliability of the technology. 
The construction procedure is straight-forward and has been industrialized, making its unit cost superior to the most obvious alternatives.
There are no fundamental obstacles to achieve the production rate
needed to match the HL-LHC time scale. Nevertheless, MATHUSLA will require a larger area of RPC than has been used in any single experiment before. 
Since the basic technology of RPC is
well understood, the ongoing effort in exploring this detector option is focused on cost performance optimization.

As we discuss in the next subsections, this minimal detector design is sufficient for LLP discovery and background rejection via geometrical DV reconstruction. It also allows for event-by-event measurement of the LLP boost~\cite{Curtin:2017izq}, which can reveal important information about the LLP mass and production mode. 

MATHUSLA is a unique detector with unusual requirements, and its detailed design will require further study. However, its reliance on proven and cost-effective technology means there is no fundamental obstacle for its deployment in time for the HL-LHC upgrade.

\subsection{Discovering and Analyzing LLP Decays with MATHUSLA}
\label{s.mathuslaLLPsearch}

\begin{figure}
\begin{center}
\includegraphics[width=0.8\textwidth]{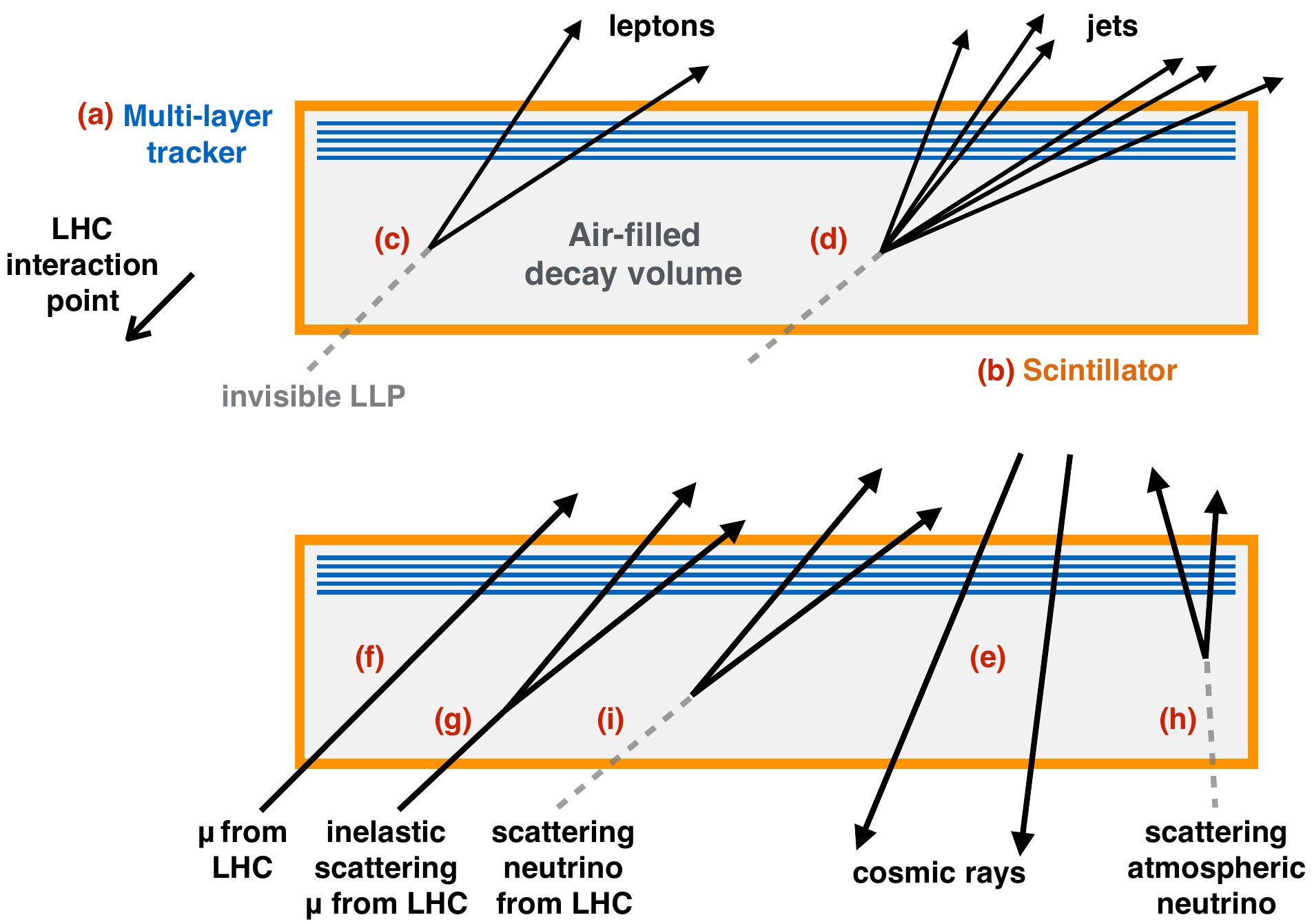}
\end{center}
\caption{
Schematic comparison of LLP Signal (top) and backgrounds (bottom) in MATHUSLA. Figure from~\cite{Chou:2016lxi}. 
The most stringent signal requirements are full 4-dimensional reconstruction of the DV from upwards traveling charged particle tracks measured with full spatial and timing information, as well as a veto on DVs involving charged tracks that originate outside the detector (provided by the scintillator).
Note that the multiplicity of the LLP final states alone provides important information on the decay mode. The signal requirements are very difficult to fake by the dominant cosmic ray background (d), especially (but not exclusively) for hadronic decays. Muons (f) either do not satisfy the signal requirement or give rise to displaced vertices that are easily vetoed (g). Neutrinos from atmospheric comic rays (h) and the LHC (i) can be vetoed due to the presence of non-relativistic protons in the final state, as well as geometrical cuts on the final state cone.}
\label{f.mathuslasigbg}
\end{figure}

Fig.~\ref{f.mathuslasigbg} (top) schematically shows the two main signals for MATHUSLA, LLPs decaying into at least two charged leptons (c), or into jets (d) that contain $\mathcal{O}(10)$ charged hadrons for LLP masses above a few GeV~\cite{Curtin:2017izq}. Hadronically decaying LLPs with mass below a few GeV would have lower final state multiplicity, but by charge conservation there would have to be at least two charged final states, making them similar to low-mass leptonically decaying LLPs. For simplicity we therefore focus our discussion on the two extremal scenarios in Fig.~\ref{f.mathuslasigbg} (top) as they roughly bracket the range of expected LLP signals.

For leptonic decay, both charged particles hit the tracker in the ceiling in $50-90\%$ of cases depending on the LLP boost, while for LLPs decaying to hadrons, almost all decays have 5 or more  charged particles hitting the ceiling~\cite{Curtin:2017izq}.
Since the tracker planes have $\sim$ cm spatial and $\sim$ ns timing resolution, the charged particle trajectories can be fitted to reconstruct a DV. Unlike traditional DV analyses in the main detectors, these DVs must satisfy the additional stringent requirement that all trajectories coincide \emph{in time} at the DV. 
The scintillator is used as a veto to ensure that the charged particles \emph{originated} at the DV. There should be no hits along the line between the vertex and the LHC IP, nor along the lines obtained by extrapolating the individual charged particle trajectories backwards.
Taken together, these exhaustive geometric and timing requirements make it very difficult for backgrounds to fake the LLP signal.

\begin{figure}
\begin{center}
\includegraphics[width=0.6\textwidth]{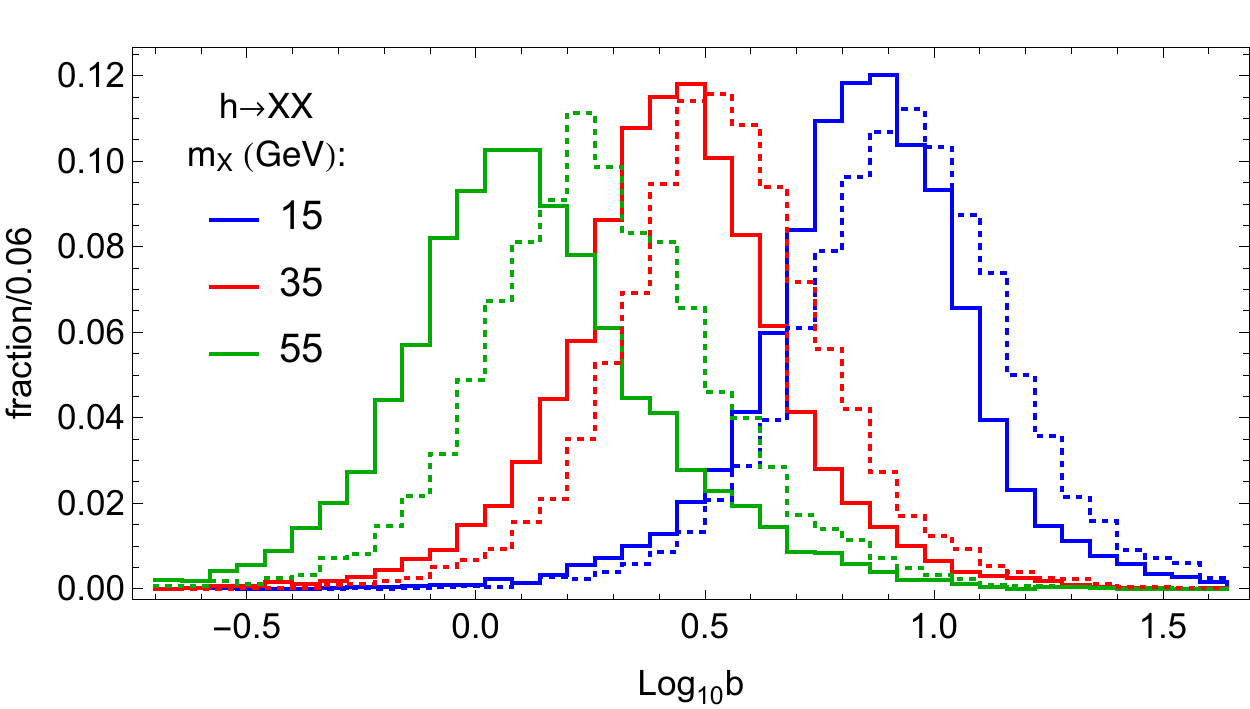}
\end{center}
\caption{
Distribution of LLP boost $b = |\vec p/m|$ for different masses of LLPs produced in exotic Higgs decays.  The solid histograms show the truth-level value of $b$, which is also close to the distribution of boosts reconstructed using MATHUSLA tracker information for LLP decay to 2 charged particles. The dotted histograms show the
distribution of reconstructed boosts for hadronic LLP decays using a sphericity-based
boost reconstruction method. For more details, see~~\cite{Curtin:2017izq}.
}
\label{f.mathuslaboostmeasurement}
\end{figure}

In addition to LLP discovery, MATHUSLA has significant capabilities to diagnose any discovered LLP decays. 
Even in the  absence of calorimetry or momentum measurement, the information supplied by MATHUSLA's tracker is sufficient to measure the LLP boost event-by-event using only the geometrical distribution of the final state trajectories~\cite{Curtin:2017izq}.
The basic principle is very simple: under the assumption that the LLP mass is significantly larger than the final state mass, the final state 4-momenta are ultra-relativistic and can be regarded as light-like, meaning they are fully determined up to an overall normalization by their direction as measured by the tracker. 
This allows the final state trajectories to be boosted back to the LLP rest frame, either exactly for two final states (back-to-back in rest frame) or approximately for many final states (assuming the LLP decay is on average forward-backward symmetric in its rest frame). 
The reconstructed boost distribution for LLPs originating in exotic Higgs decays is illustrated in Fig.~\ref{f.mathuslaboostmeasurement}. 
This analysis can be generalized to partially invisible LLP decays with some loss of event-by-event precision, but more work is needed to understand the fidelity of this method for extremely light LLPs where the mass of detected decay products cannot be neglected.

\begin{figure}
\begin{center}
\includegraphics[width=0.6\textwidth]{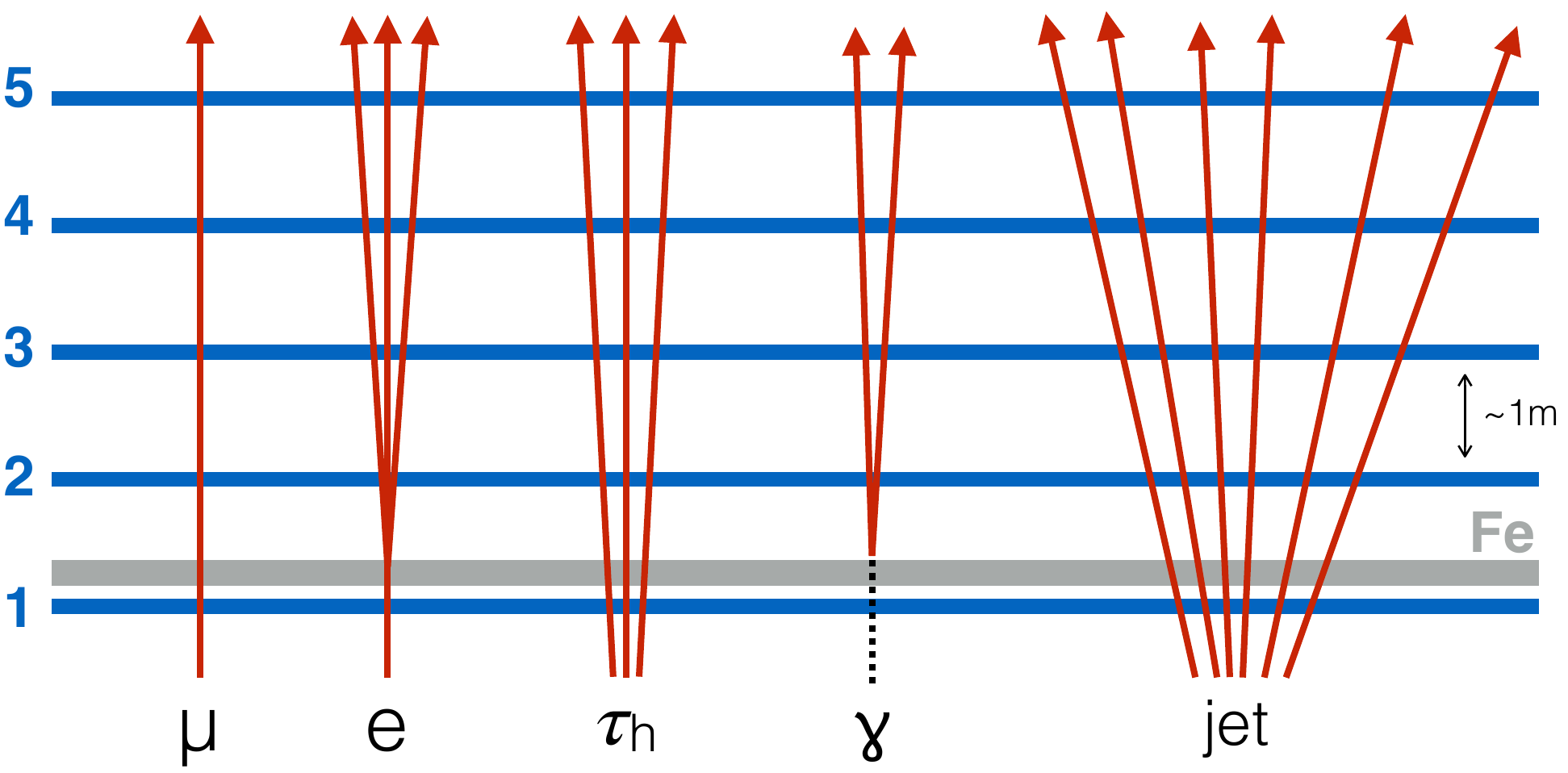}
\end{center}
\caption{
Possible particle ID in MATHUSLA with an extra layer of material between tracking layers~\cite{Curtin:2017izq}.
}
\label{f.mathuslaparticleID}
\end{figure}

Measuring LLP boost is important for several reasons.
The event-by-event boost determination allows production of the LLP to be associated with just a few candidate LHC bunch crossings. By correlating MATHUSLA and main detector data (especially but not exclusively if MATHUSLA could trigger the main detector), the production mode can be determined or at least constrained. 
As demonstrated by Fig.~\ref{f.mathuslaboostmeasurement}, the LLP boost distribution is tightly correlated with LLP mass once a production process is assumed. 
Correlating information from MATHUSLA with the main detectors therefore has the potential to determine or constrain the LLP mass and production mode.

The decay mode of the LLP can also be determined or constrained using MATHUSLA measurements: leptonic and hadronic LLP decays are straightforwardly distinguished  (for LLP masses significantly above a GeV)  by final state multiplicity.
However, it was also noted in~\cite{Curtin:2017izq} that MATHUSLA's capabilities could be extended by placing several cm of converter material like Iron between two of the RPC planes. 
This would allow photons to be detected by conversion and electrons to be distinguished from muons by the induced electromagnetic shower. 
As shown schematically in Fig.~\ref{f.mathuslaparticleID}, this permits event-by-event particle identification of the LLP final states, supplying important information on the nature of a discovered LLP.
Correlating LLP mass with final state multiplicity and possibly measured speed of the nonrelativistic hadrons would even supply information on the dominant \emph{flavor} of jets produced in LLP decay (e.g. $b$, light-quark, or gluon dominated).
Studies are underway on the amount and location of converter material to optimize particle
identification performance while taking into account practical issues such as weight and cost. It should be noted that the outcome of this study has no impact on the design of tracker planes, and is unlikely to affect discovery prospects for LLPs decaying to charged particles.

\subsection{Backgrounds to LLP Searches}
\label{s.mathuslabackgrounds}

We now briefly summarize the arguments and calculations put forth in \cite{Chou:2016lxi} that MATHUSLA could search for LLPs decaying into charged particles with little or no backgrounds.

The main backgrounds to LLP searches in MATHUSLA are represented in Fig.~\ref{f.mathuslasigbg} (bottom). Each of them can be rejected using a variety of strategies.
\begin{itemize}
\renewcommand\labelitemi{$\bullet$}
\itemsep=4mm
\parskip=2mm

\item  \textbf{Cosmic rays} (e) are by far the most dominant background and have a rate of $\sim \mathcal{O}(10 \mathrm{MHz})$ on the whole detector, resulting in $\sim 10^{15}$ charged particle trajectories over the whole HL-LHC run.

The overwhelming majority of CRs travel downwards, allowing them to be rejected based only on their direction of travel compared to the upwards-traveling LLP decay products. This can be demonstrated with a simple estimate: assuming gaussian spatial and timing resolutions of 1cm and 1ns for the tracking planes, and assuming that only four of the five tracking planes fire, the chance for a single downward traveling track to fake an upward traveling track is less than $10^{-15}$. Two such fakes are required to fulfill the basic multiplicity requirement of the LLP signal, and we have not yet made use of the DV requirement (tracks must coincide at a single point in time and space) nor the veto in the floor of the detector. 
%
Even accounting for non-gaussianities in tracking resolutions and other details, it is unlikely that the CR background is dominated by downward tracks if the above resolution requirements are met. 

The most likely source of CR background is 
CR albedo, or `splash-back' of CRs hitting the detector floor and ejecting unstable SM particles into the decay volume. This may give rise to signals that naively resemble LLP decays, but would also be correlated with downwards moving tracks in the tracker and signals in the floor detector.

This makes clear that the most plausible source of CRs faking LLP decays are Extended Air Showers (see Section~\ref{s.mathuslaCR}) with many charged particles coincident on MATHUSLA in a correlated manner. 
As illustrated in Fig.~\ref{f.mathusla3D}, this leads to a large number of charged particles occupying MATHUSLA near-simultaneously and is easily rejected with very little ``blind time'' for the LLP search. 

The impact of albedo from isolated single charged CR particles is expected to be small, but more work is required and underway to carefully quantify this background.

Detailed studies involving extensive cosmic ray and detector simulations are needed to verify the near-perfect rejection of CR background, and these studies are a high priority for the MATHUSLA experimental collaboration~\cite{Alpigiani:2018fgd}. However, the above arguments make it highly plausible that this rejection can be realized with a careful detector design. Furthermore, the CR background to LLP decays is intrinsically \emph{reducible} and falls rapidly if further tracking planes in the floor or ceiling or even walls of the detector volume are added. While at this point we do not expect such extensive modifications of the basic detector design to be necessary to reject CR backgrounds, the fact that this option exists makes the zero background regime a safe assumption for initial physics studies to assess the MATHUSLA physics case.

\begin{figure}
\begin{center}
\includegraphics[width=0.8\textwidth]{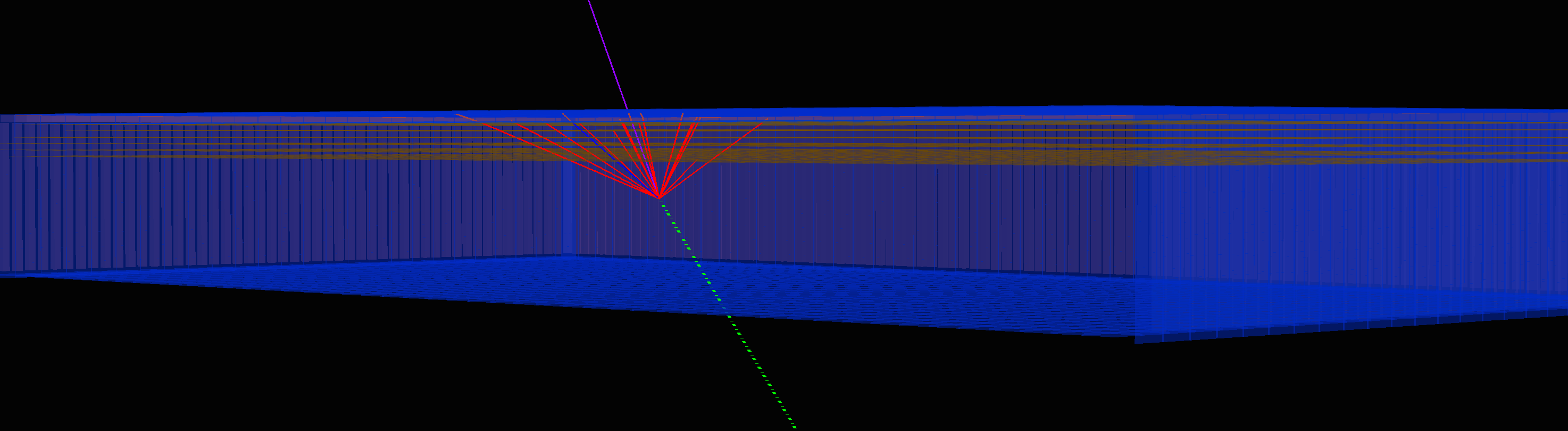}
\\
\includegraphics[width=0.8\textwidth]{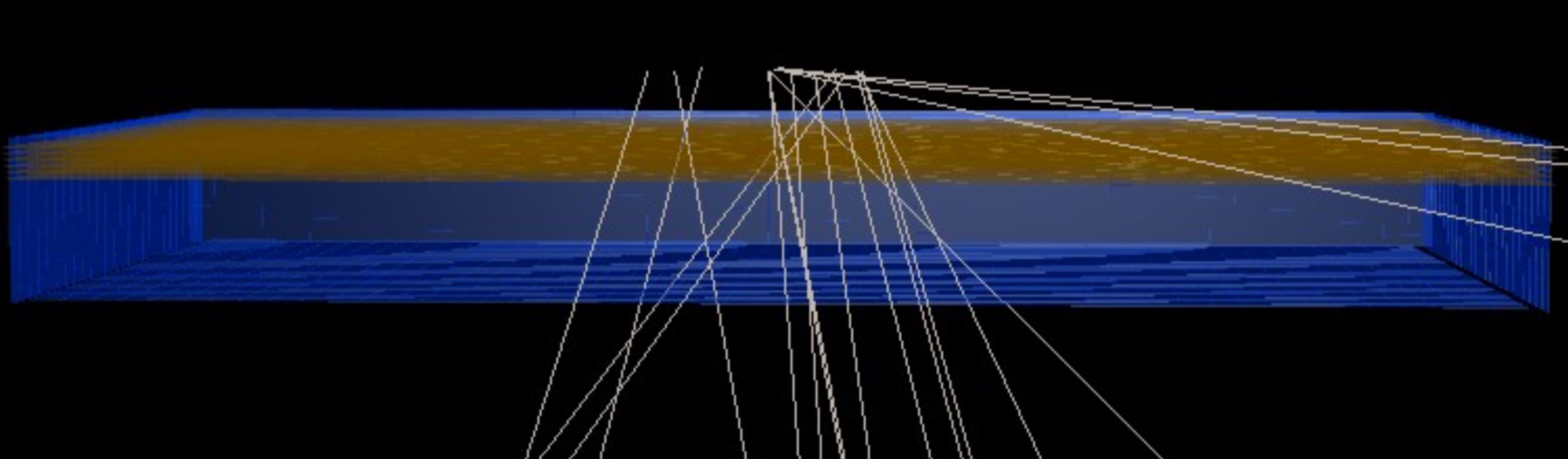}
\end{center}
\caption{
\emph{Top:} GEANT-VMC simulation of a 30 GeV LLP (green dashed line) decaying hadronically inside of the simplified MATHUSLA detector layout from Fig.~\ref{f.mathuslalayout}. Only charged hadrons (red lines) are shown. 
\emph{Bottom:} CORSIKA + GEANT-VMC simulation of atmospheric muon bundle event from Air Shower due to a iron cosmic ray primary with energy about $3\times10^{16}$ $\mathrm{eV}$ incident on the MATHUSLA decay volume (see sec. \ref{s.mathuslaCR}). White lines are atmospheric muons with energy threshold above 1 $\mathrm{GeV}$.
The total number of charged particles in the CR event is much larger than the number shown in the image, illustrating the obvious differences between CR and LLP events in MATHUSLA.
}
\label{f.mathusla3D}
\end{figure}

\item 
\textbf{Muons from the IP}: Muons that are produced at the LHC and have energy greater than $\sim 60 \gev$ could traverse the rock and reach the MATHUSLA decay volume. Their rate was initially analytically estimated in~\cite{Chou:2016lxi}, with an updated calculation utilizing GEANT4~\cite{Agostinelli:2002hh} for muon propagation through rock presented in~\cite{Alpigiani:2018fgd}. The total number of upwards traveling muons traversing the decay volume is $\mathcal{O}(10^7)$ over the run-time of the HL-LHC. 

A muon that simply passes through the decay volume (f) does not satisfy any of the signal requirements (no DV) and does not constitute a genuine background to the LLP search. The same is true for muons undergoing their most common decay $\mu \to e \nu \nu$. The rare decay $\mu \to eee \nu \nu$ or inelastic scatters off atomic nuclei in the air-filled decay volumne occur  $\lesssim 1$ times over the entire HL-LHC run and would easily be vetoed with a floor detector.

Depending on the assumptions made on geology and precise position and structural design of the detector, $\sim 10^2 - 10^3$ muons will liberate electrons from atoms in the air, or scatter inelastically in the support structure (g). The former can be vetoed with a floor detector. The latter can also be rejected with a material veto, which will not greatly reduce signal efficiency given the excellent tracking resolution required to deal with cosmic ray backgrounds. 

These estimates are currently being refined  by the experimental collaboration, but the main conclusions are unchanged. Rejecting muons from the LHC is clearly possible with a carefully designed MATHUSLA detector.

\item 
\textbf{Neutrinos} 
from atmospheric cosmic ray interactions (h)  and LHC collisions (i) could be traveling upwards and scatter with a nucleus in the decay volume, giving rise to a genuine DV of two or more charged particles originating at a single point in space and time, with no charged particle trajectories leading to the DV from the floor.  Even so, it can be vetoed with the capabilities of the MATHUSLA benchmark detector design. 

This background was studied analytically in~\cite{Chou:2016lxi} using the measured atmospheric neutrino flux and simulated hard and soft neutrino production at the HL-LHC. The cross section for neutrinos to scatter of nuclei is known theoretically and experimentally at the 30\% level or better~\cite{Formaggio:2013kya}. Using this information, the number of neutrinos scattering off air and producing a genuine DV with at least two charged particles in the final state can be analytically calculated. The multi-particle final state kinematics can be constrained using energy and momentum conservation, which is sufficient to formulate a simple rejection strategy. This conservative approach also means that the below background rates after cuts  are likely to be overestimates, since the detailed features of the background are not fully exploited.

It is helpful to divide the events into those which are exclusively defined to contain a proton in the final state, including many quasi-elastic scattering (QES) processes, and those which are not, like deep-inelastic scattering (DIS) events.
We also discuss atmospheric and LHC neutrino backgrounds separately.

In the 200m MATHUSLA benchmark geometry, the number of atmospheric neutrino scatters with a proton in the final state is about $\sim 60$ per year. Since atmospheric neutrinos are dominantly produced in secondary production (hadron decays) during CR showers, the distribution is dominated by neutrinos with energies below a GeV or so. If such a neutrino scatters off nuclei and releases a proton as well as other charged particles, the proton will be non-relativistic. Requiring a low-multiplicity DV to not contain a slow track ($v < 0.6 c$) was found to veto the large majority of these events. Reconstructing such tracks and measuring their speed is well within the capabilities of the detector if the time resolution of the tracking planes is $\mathcal{O}(1\ \mathrm{ns})$. It is also possible to veto DVs with a very narrow opening angle that point away from the LHC IP. This brings the number of these scatters to less than one per year. These cuts would not significantly reduce signal efficiency for the LLP signals we consider in this whitepaper.

Atmospheric neutrino scatters without a definite proton in the final state include higher-energy DIS events and occur about $\sim$10 times per year. Their rejection requires more detailed study, but owing to their higher energy they give rise to an even narrower DV opening angle than exclusive processes with final state protons, making the geometric cut on DV orientation relative to the LHC IP highly effective. Careful study is currently underway and likely to reveal additional features of this background that can be used for rejection. This  makes it highly likely that these events can be rejected down to levels of less than one per year. 

The discussion for neutrinos produced at the LHC follows similar lines, since that neutrino flux is also dominated by secondary production. The geometric veto on DV orientation is not available, but even so the estimated background rate of all neutrinos from the LHC after applying the cut on non-relativistic protons is less than about one per year. 

The MATHUSLA collaboration is currently refining these estimates with full GENIE~\cite{Andreopoulos:2015wxa} Monte Carlo simulation of neutrino interactions in the decay volume to  confirm the conclusion of this analytical calculation.

\end{itemize}

Crucial to the background rejection strategies discussed above is the assumption that the LLP decays into at least two charged particles that can be well-separated by the MATHUSLA tracking system to form a DV. 
Other LLP signatures are possible. For example, if the LLP boost is higher than $\sim \mathcal{O}(1000)$, the charged particle tracks may not be well separated and the LLP would show up as a ``one-pronged DV'', see Section~\ref{s.energythresholds}. A search for this signal is still possible, but may suffer from higher backgrounds. 
If MATHUSLA can detect photons, LLP decays to two or one photon could be reconstructed, but the indirect nature of photon detection via conversion in material may lower the spatial and temporal resolution of the DV reconstruction, with resulting higher levels of background. 
These alternative LLP signatures will be the subject of future study.

While the above arguments and estimates make the near-perfect rejection of CR and other backgrounds is plausible, detailed studies with full background Monte Carlo and detector simulation are clearly needed to prove that the zero background assumption can be achieved with a concrete detector design. These studies are beyond the scope of this theoretical whitepaper, and are currently being conducted by the MATHUSLA experimental collaboration. 
The aim of this white paper is to demonstrate the extensive reach of MATHUSLA for new physics if the zero background regime for LLP searches can be reached. 
This can be seen as providing the motivation for conducting these detailed background rejection studies. 
Reaching this zero-background regime will enable MATHUSLA to reach up to $\sim 10^3\times$ better sensitivity to LLP production cross-sections than ATLAS or CMS in the long-lifetime regime.\footnote{If backgrounds are ultimately nonzero, then LLP cross section sensitivity would be reduced by roughly a factor of few/$\sqrt{N_\mathrm{BG}}$. The sensitivity estimates in this paper are therefore easily rescaled to a given background assumption.}
The  model-independent LLP sensitivity of MATHUSLA will be discussed in more detail and compared to the capabilities of the LHC main detectors in Section~\ref{sec:LLPSmathuslahllhc}.

\subsection{Scalability and Modularity of a Realistic Detector Design}
\label{s.mathuslamodularity}

The MATHUSLA detector idea is highly flexible. This allows for a large variety of possible implementations, depending on detector technology, available space, and budget.

The number of LLPs decaying in MATHUSLA is a function of solid angle coverage, depth of the detector along the LLP trajectory, and distance from IP. Therefore, it only depends modestly on the precise geometry and location of the surface detector, as long as the decay volume is horizontally displaced from the IP by $\lesssim \mathcal{O}(100\mathrm{m})$. 
This modest dependence still motivates careful optimization of the precise detector geometry to maximize sensitivity for a given detector area.  For example, we have shown that a slightly more realistic non-square detector area on the potential MATHUSLA experimental site near CMS~\cite{Alpigiani:2018fgd} would achieve the same LLP sensitivity as the 200m benchmark geometry (Fig.~\ref{f.mathuslalayout}) used in this paper, while having only $\sim 1/3$ the area. This smaller size will be vital to achieve the MATHUSLA collaboration's goal of constructing the detector for a cost below 100 MCHF.

Importantly, up to a possible $\mathcal{O}(1)$ factor that depends on the final  detector geometry but does not affect the MATHUSLA physics case, the sensitivity projections we present in this work will be valid for any detector geometry which places $\sim 10^6\mathrm{m}$ of fiducial volume near ATLAS and/or CMS.
This leads us to three important conclusions:
\begin{itemize}
\item MATHUSLA lends itself to a \emph{modular} implementation, for example by arranging many smaller detector modules at ATLAS, CMS, or both to reach the required decay volume. 
This greatly simplifies construction of the full detector and allows for iterative deployment. The layout of the full detector complex can then be adapted to the chosen experimental site.

\item MATHUSLA is not an experiment with a fixed price tag, but rather a general detector concept which can be \emph{rescaled} to whatever funding level is available. For example, one could imagine as a first stage of deployment a ``mini-MATHUSLA'' of 1/10 the full volume (assembled of one or several sub-modules) which would have $\sim 1/10$ the sensitivity of the full detector, at approximately $1/10$ the cost, see Fig.~\ref{f.MATHUSLAsensitivitycartoon}. This would still improve LHC sensitivity to weak-scale, hadronically-decaying LLPs by orders of magnitude (even if the BBN lifetime limit cannot be reached with a smaller detector).

\item A modular construction also makes it natural to equip certain modules with special capabilities at a much lower cost than upgrading the whole detector. For example, some modules could be equipped with additional material between the tracking layers to add particle ID for a subset of observed LLP decays (or CRs, see Section~\ref{s.mathuslaCR}). One could also equip one or more of the modules with much higher resolution trackers than the rest, to allow very low-mass LLPs to be searched for without background (see Section~\ref{s.energythresholds}.)
\end{itemize}

While alternative technologies for MATHUSLA are not excluded, work is underway to finalize a realistic detector design using RPCs and possibly plastic scintillators.
Taking advantage of the possible modularity, coverage of the full 200 m $\times$ 200 m footprint would be achieved with a number of smaller identical modules, which will be entirely self-contained except for service connections. This facilitates construction, and the adoption of industrial practices for mass production is expected
to reduce costs. 
The modularity allows easy adaption to a different-sized or different-shaped
footprint so detector design and construction can proceed before a final decision on the
experimental site. 
Furthermore, each module will be made weather-tight so there is no need for an experimental hall to house the MATHUSLA detector.
Trigger information will be provided by the RPC tracking chambers, similar to what has been
done in experiments such as ATLAS and CMS. 
It is anticipated that each module will contribute a local trigger signal to the overall event trigger. Timing stability over the large area of MATHUSLA requires care; however, it should be noted that tighter timing requirements and greater distances than MATHUSLA have been dealt with in accelerator facilities.

\subsection{MATHUSLA Test Stand}
\label{s.mathuslateststand}

A test stand has been assembled at CERN and it was installed in the surface area above the ATLAS detector in November 2017. Figure \ref{f.mathuslateststand} shows the basic design of the test module and a picture of the final assembled structure in ATLAS SX1 building at CERN. The overall structure is $\sim 6.5$ m tall, with an active area of $\sim 2.5 \times 2.5$ m$^2$.

\begin{figure}
\begin{center}
\begin{tabular}{cc}
\includegraphics[height=7cm]{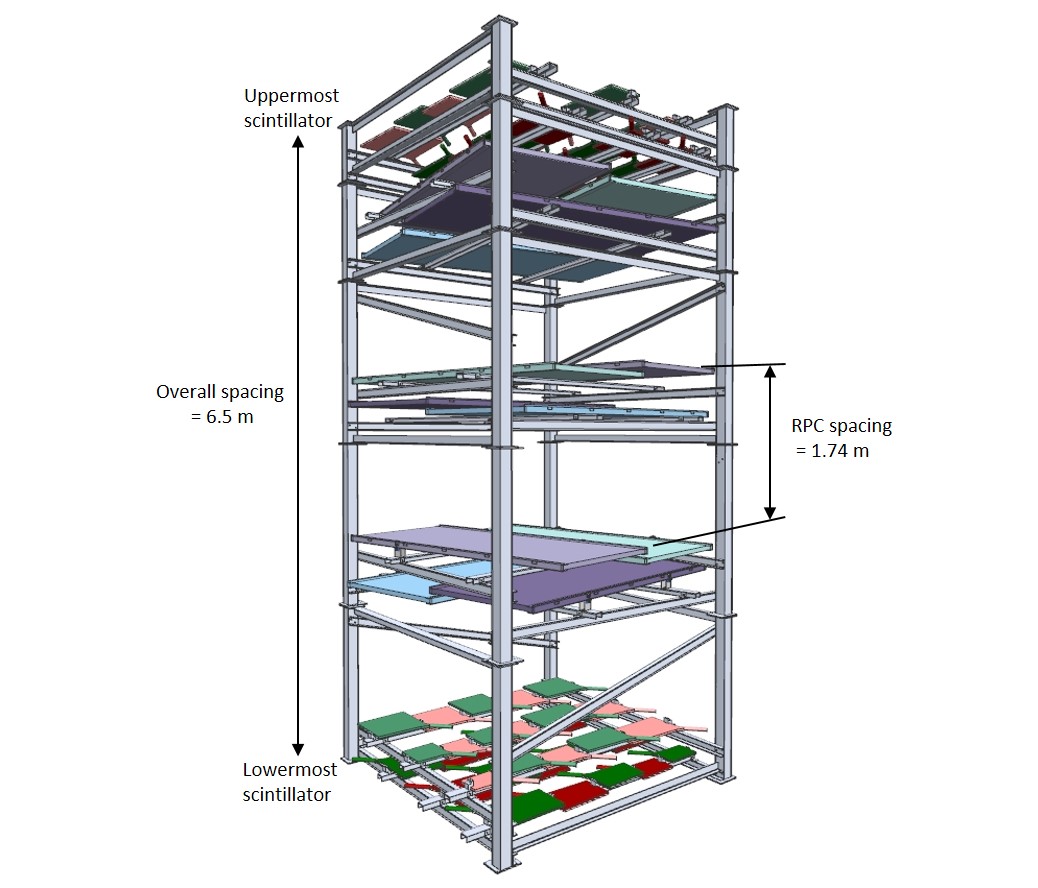}
&
\includegraphics[height=7cm]{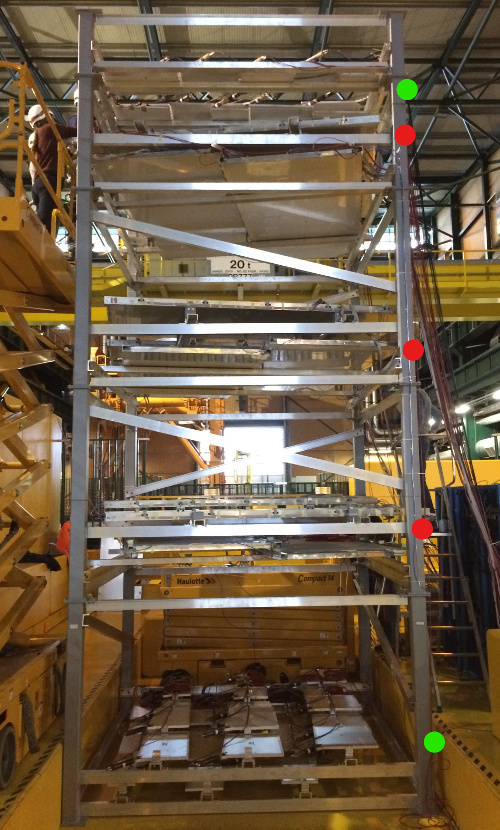}
\\
(a) & (b)
\end{tabular}
\vspace*{-3mm}
\end{center}
\caption{
(a): schematic view of the MATHUSLA test stand. (b): picture of the final assembled structure in his test area in the ATLAS SX1 building at CERN. The green dots identify the two scintillator layers used for triggering, while the red dots the three RPC layers used for tracking.
}
\label{f.mathuslateststand}
\end{figure}

Following the concept of the main detector, the test stand is made of three layers of RPCs between two layers of scintillator.  Scintillator detectors are used to trigger upward and downward charged tracks. 
The top and bottom scintillator layers are comprised of 28 and 31 scintillators, respectively, recycled from the Tevatron D0 experiment.
The RPCs are used for tracking and they were provided by Universit\`a di Tor Vergata, Rome. They are the same type of chambers used in the Argo-YBJ experiment at the YangBaJing Laboratory in Tibet (4300 m a.s.l.).

Several efforts are underway to develop simulations of the backgrounds expected in \mbox{MATHUSLA}. For muons and neutrinos traveling upwards, the idea is to create a ``MC particle gun'' that shoots particles into \mbox{MATHUSLA}, while for cosmics the plan is to use the standard cosmic ray simulations (e.g. CORSIKA). Nevertheless, the simulations need to be validated and tuned using real data, and for this reason the test stand is crucial.  

Since the main goal is to have a background-free MATHUSLA detector, the central purpose of the test stand is to measure the background from CRs and the LHC collisions in order to test the hypothesis that MATHUSLA could reject these most numerous of expected backgrounds. 
Nevertheless, the test stand should not be considered a prototype of the main detector:
the layout could be further optimized, especially with custom-built rather than repurposed components, and detector technologies could be considered.
Even so, it will provide very useful information for the design of the future MATHUSLA detector. All the tests that will be performed until the end of LHC Run 2 will be fundamental to understanding the cosmic ray rate in the test stand and to extrapolate the LHC-correlated background rate from the test stand to MATHUSLA. A precise measure of the charged particle flux in the test stand will provide the veto efficiency requirement for the main detector.  The goal is to achieve a sufficient timing resolution to guarantee that no cosmic particles can fake a charged particle coming from LHC.

The on-going analysis of the data collected during 2017 and the beginning of 2018, along with all the experience gained from the construction, assembling and commissioning of the test stand, will  be crucial for the preparation of the Letter of Intent that the MATHUSLA Collaboration plans to submit to the CERN Committee in late 2018.

\subsection{Cosmic Ray Physics with MATHUSLA}
\label{s.mathuslaCR}

The design of MATHUSLA is driven by the requirements of reconstructing upward-traveling displaced vertices and distinguishing them from downward-traveling cosmic rays. It therefore comes as no surprise that MATHUSLA has all the qualities needed to act as an excellent cosmic ray telescope. In fact, MATHUSLA's particular combination of robust tracking and large area allow it to make many unique measurements that could address important and long-standing questions in astroparticle physics. The study of cosmic rays is therefore an important secondary physics goal of MATHUSLA. These measurements, which in no way interfere with the primary goal of LLP discovery,  represent a ``guaranteed physics return'' on the investment of the detector, as well as an opportunity for CERN to establish a world-leading cosmic ray physics program.

The cosmic ray physics program at MATHUSLA warrants in-depth examination beyond the scope of this work.  
Some initial studies will be presented elsewhere \cite{MathuslaCRWP}. 
Here we only briefly comment on the  qualities that make MATHUSLA a uniquely interesting cosmic ray experiment, and outline some possible measurements that are of particular interest to the astroparticle physics community. 
To this end, we first review some basic facts about cosmic rays and how they are detected.

Cosmic rays, dominantly protons and heavier atomic nuclei, arrive at earth with an energy that spans some 12 orders of magnitude, from a few hundred MeV ($10^8$ eV) to 100 EeV ($10^{20}$ eV) \cite{Engel2009, Greiner15, Roulet17}, see for example, Fig.~\ref{f.CRspectrum}. 
They are produced in violent astrophysical scenarios  within our own galaxy (for energies $E \lesssim  10^{18} \, \mathrm{eV}$) and beyond (for $E > 10^{18} \, \mathrm{eV}$). At the highest energies, however, the origin of CRs is still mysterious, since propagation in galactic magnetic fields means CRs do not point back to far away sources~\cite{Roulet17, AugerScience2017}. 
In general, details of CR acceleration mechanisms,  composition, propagation through  space, and  features in their spectrum are not completely understood \cite{Roulet17, Giacinti15, Gaisser13, Deligny16}. The study of CRs offers a unique window on the most energetic natural phenomena of the cosmos \cite{Ackermann13, Abramowski16, Fang2018}, and their collisions with the atmosphere probe energies far in excess of the TeV scale \cite{Aab2015a,Aab2016a,Aab2017a}.

\begin{figure}
\begin{center}
\includegraphics[width=0.7\textwidth]{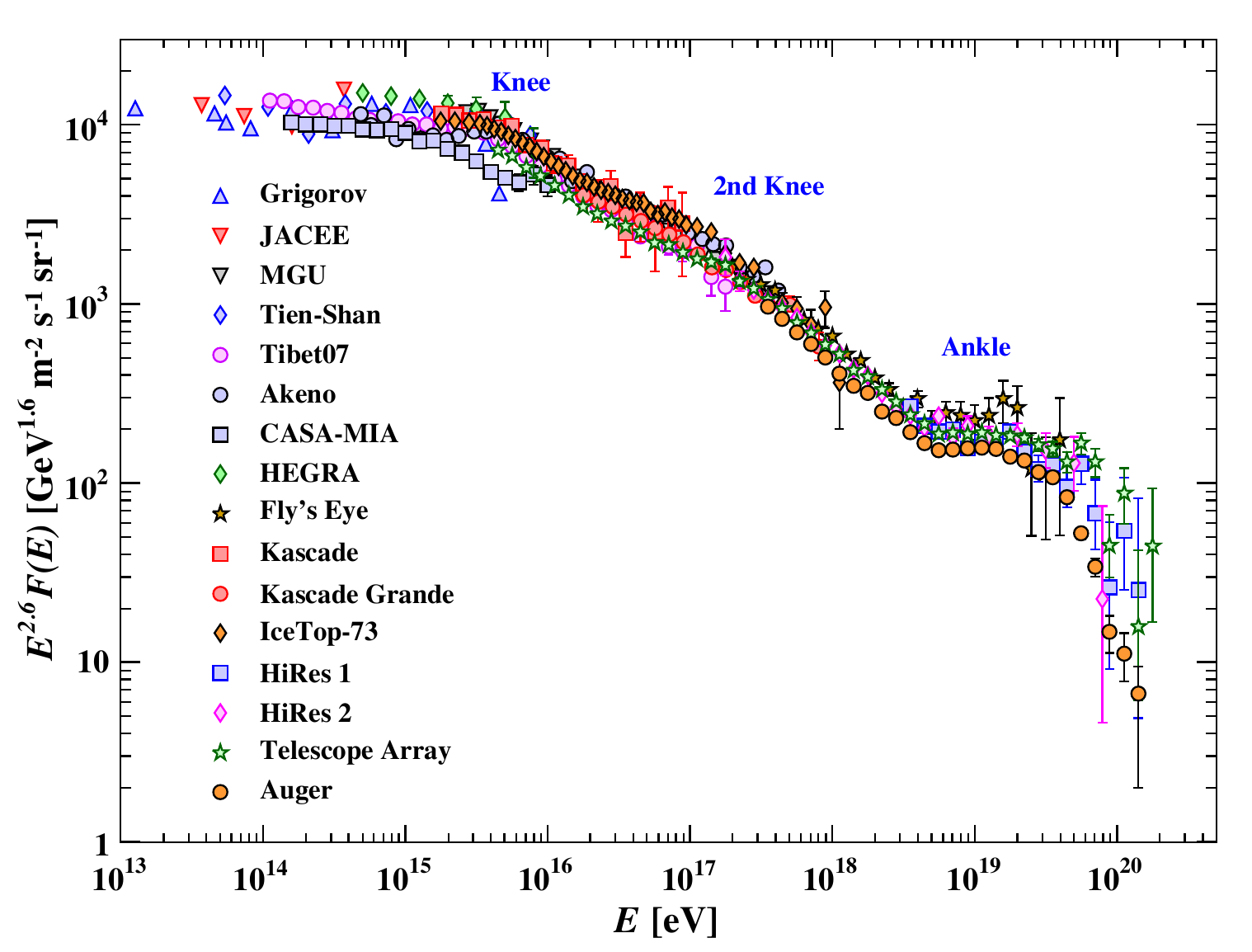}
\end{center}
\caption{
Global view of the all-particle cosmic ray energy spectrum (figure taken from \cite{Patrignani:2016xqp}). The total spectrum decreases quickly according to a power-law formula $E^{-\gamma}$, where $\gamma$ (the spectral index) varies from roughly $2.6$ to $3.3$ \cite{Roulet17}. Between $10^{15} \, \ev$ and $10^{19} \, \ev$, the spectrum exhibit three distinctive features created by a change in the value of $\gamma$: the knee, which is located at $\sim 4 \, \pev$ \cite{Patrignani:2016xqp, Kulikov1959}, the second knee, close to $100 \, \pev$ \cite{Apel2012, Aartsen2013, Prosin14} and the ankle, around $4 \, \eev$ \cite{Settimo2012, Abu2013}. There exists also a weaker structure called the low energy ankle \cite{Roulet17} at  $\sim 10 \, \pev$ \cite{Apel2012, Aartsen2013}. MATHUSLA is expected to be sensitive to hadronic EAS with primary energies around the knee, i.e. in the interval $E \sim 10^{14} - 10^{17} \, \ev$ according to its size $200 \times 200 \,\mathrm{m}^2$ and the atmospheric depth at which it will be located ($\sim 1000 \,\mathrm{g}/\mathrm{cm}^2$).
}
\label{f.CRspectrum}
\end{figure}

Primary CRs with relatively low energy can be directly characterized by balloon- or space-born particle physics experiments equipped with trackers and calorimeters, like AMS-02 \cite{Kounine2012,MAguilarAMS} and CREAM \cite{Ahn2007,CREAM2010} among
others (see, for example, \cite{Seo2012, Mocchiut2014, Maestro2015}). The small size of these detectors restricts this approach to energies below $\sim 100 \tev - 1 \pev$, both because the CR flux drops dramatically above this threshold and because the detector's magnetic field and radiation depth limits the maximum energy which can be reconstructed. 

To study higher energy CRs above $\sim 10^{14} \ev$, the atmosphere is used as a calorimeter  \cite{Engel2009, Roulet17, Kampert12,Engel2011}. The detection technique consists in observing the Extensive Air Showers (EAS) of SM particles that CRs induce in collisions with the atmosphere. The EAS typically originates  from $15 \,\mathrm{km}$ to $35 \,\mathrm{km}$ above sea level \cite{Engel2011, Grupen2012} and spreads out as the particle front travels towards the ground. 
Near sea level, the EAS consists mostly of muons, but also large fractions of eletrons and muons, as well as much
smaller fraction of hadrons (for vertical incidence) \cite{Grieder2009}. Depending almost linearly on the energy in logarithmic scale, the total number of particles in the EAS is in the range $\sim 10^4 - 10^{10}$, for primary energies $E = 10^{14} \, \,\mathrm{eV} - 10^{20} \,\mathrm{eV}$, and it is mostly contained within a cone of $\mathcal{O}(0.1 - 1 \,\mathrm{km})$ in diameter at ground level~\cite{Kampert12, AugerScience2017}.
Measurements of the EAS allow the primary CR's direction, composition and energy to be reconstructed~\cite{Grieder2009}. 
Earth-bound CR experiments employ two classes of techniques to probe the EAS: (1) particle detectors on the surface, like particle counters, trackers, and calorimeters to sample the air shower front and (2) various telescopes to observe electromagnetic emissions from the shower or from the interaction of the EAS with the atmosphere (Cherenkov radiation, radio, fluorescence light). 
Many experiments use a combination of both techniques like the Pierre Auger Observatory \cite{Abraham2004}, 
TUNKA \cite{Prosin14} and TALE \cite{Abbasi2018}. 
EAS observatories monitor in general large areas to compensate the low CR flux at high energies. For example, at $E \sim 10^{15} \, \ev$, CRs are received at a rate of $\sim 1 \,\mathrm{particle}/\mathrm{m}^2\cdot\mathrm{year}$ \cite{Engel2009} and shower arrays with areas of order 
$\gtrsim \mathcal{O}(10^4 \,\mathrm{m}^2)$ are required, like the KASCADE air shower detector ($200 \times 200 \,\mathrm{m}^2$) \cite{Antoni2003}. On the other hand, at extremely high energies $E \sim 10^{20} \, \ev$, the CR flux is so low ($\sim 1 \,\mathrm{particle}/\mathrm{km}^2\cdot\mathrm{century}$ \cite{Engel2009}) that sufficient exposure requires installations with very large areas of the order of 
$\mathcal{O}(10^3 \,\mathrm{km}^2)$, like the $3000 \,\mathrm{km}^2$ Pierre Auger Observatory \cite{Abraham2004, Abraham2010}.

EAS telescopes/antennas are able to observe the longitudinal development of the air shower, while surface detectors
measure the lateral structure of the EAS at the atmospheric depth of the site \cite{Grieder2009}. In the latter, 
particle detectors are arranged in arrays and are spaced at regular intervals to optimize the measurements for 
the energy range of interest. Hence, in most cases, they sample only a small fraction of the shower at 
the observation level. For instance, in case of KASCADE, the main array of 252 e/$\gamma$ detectors covered only $1.22\%$
of the total surface, the muon array of 192 detectors, only $1.55\%$,
while the muon tracking detector and the hadron calorimeter covered just $0.64\%$ and
$0.76\%$, respectively \cite{Antoni2003}. Just in a few cases, full coverage was achieved as in the case of the
ARGO-YBJ detector, which consisted of a $74  \times 78 \,\mathrm{m}^2$ carpet of RPC's with an active area of almost $93\%$ \cite{Aielli2006}.
In general, most surface detectors are insensitive to the energy of a single charged particle (although a counter can be equipped with shielding or buried underground to implement a desired minimum energy threshold). Rather, the focus is on collective shower properties, like the spatial and temporal distribution of particles, as well as some basic particle ID to separate the e/$\gamma$,
muon and/or hadron components in the EAS. This data is used to characterize the primary CR, assuming a certain hadron interaction model which governs the evolution of the EAS in the atmosphere \cite{Engel2011, Knapp2003}.

Hadron interaction models are a crucial part of air shower Monte Carlo simulations. They are tuned to available high energy physics data but rely on extrapolation in certain regions of phase space, in particular, the forward region at very high energies. This introduces unavoidable uncertainties in the determination of the properties of the primary cosmic rays \cite{Ostapchenko2006, Pierog2006, Antoni2005}. Verifying and tuning these hadron interaction models is therefore of fundamental importance to CR physics \cite{Engel2011}. Tests of hadron interaction models
can be performed with the same data from EAS observatories. That requires, however, the simultaneous measurement of different observables of the air showers. 
KASCADE made important contributions to this topic \cite{Antoni1999,Antoni2001,Antoni2003a, Antoni2005a, Antoni2006, Apel2007, Apel2009}, because of both the quality of its measurements and its different detection
systems, like the full-coverage central tracker and calorimeter \cite{Antoni2003}. This central detector only had less than $1\%$ the 
area of the full experiment but was crucial in allowing for more detailed analysis of the shower \cite{Antoni1999,Antoni2001,Antoni2005a, Apel2007, Apel2009}.


We can now understand why MATHUSLA would make important contributions to cosmic ray physics. 
MATHUSLA is the size of KASCADE and will be also located at a comparable atmospheric depth ($\sim 1000 \,\mathrm{g}/\mathrm{cm}^2$). 
Therefore, it is expected to be sensitive to a similar CR energy range of $10^{14} - 10^{17}$ eV.
After a few years of exposure ($\sim 3 \, \mathrm{yr}$), roughly $10^6$ air showers with $E > 10^{15} \ev$ would be recorded with 
their cores traveling through the MATHUSLA detector. 
Unique for a CR experiment of this size, MATHUSLA has full-coverage robust tracking with excellent position and timing resolution. 
The scintillator planes which enclose the LLP decay volume would supply additional information, especially for highly inclined showers. 
Even without track-by-track $e/\mu$ discrimination or calorimetry, this would allow for very detailed EAS measurements, including highly granular analysis of the shower's temporal and spatial structure, which has never been undertaken at this size scale at PeV energies.
Furthermore, for roughly half of those CR events, the so-called ``golden events'',  part of the shower's high-energy muon component ($E^{th}_\mu \gtrsim 50 \, \gev - 70 \, \gev$ for vertical incidence) also passes through the underground ATLAS or CMS detector and could be simultaneously registered with special CR triggers during the LHC runtime.
Therefore, whether working in standalone mode, or in tandem with the main underground detector, MATHUSLA constitutes a powerful cosmic ray experiment with unique capabilities that will open an era of precision EAS measurements in the PeV energy region.

The particular CR measurements which MATHUSLA can perform will be studied in more detail in a future document \cite{MathuslaCRWP}. Some of the most compelling targets for CR measurements include:
\begin{itemize}
\item \emph{Primary CR spectra and composition:}
The spatial and temporal distribution of charged particles in the shower would probe the energy and composition of the primary CR, potentially addressing open issues like the exact position of the ``light knee'' (a change in spectral index  in the spectrum of protons and helium,  
which ARGO-YBJ found at $\sim 700 \, \tev$ \cite{Bartoli2015} and KASCADE, around $3 \, \pev - 4 \pev$  \cite{Antoni2005, Apel2013}) and the shape 
of the spectra of the heavy elemental components of primary CRs at PeV energies \cite{Antoni2005, Apel2013}.

\item \emph{Cosmic Ray Anisotropies:}
MATHUSLA's enhanced resolution, compared to previous CR experiments, could allow to improve the measurements on the dipole component of the anisotropy in the diffuse CR flux at PeV energies, which has been poorly investigated (see \cite{Roulet17, Deligny16} and references therein).
Another important aspect is the search for local point sources in the northern celestial hemisphere, which would provide vital clues about the presence of nearby galactic accelerators of very high energy CRs \cite{Antoni2004,Over2008,Kang2015}.

\item \emph{Highly inclined showers:}
The vertical scintillator planes enclosing MATHUSLA's decay volume, together with its precise  full-coverage tracking, allow for the study of highly inclined air showers at large zenith angles $\theta > 60^\circ$. These showers are interesting for a variety of reasons. If they originate from charged primary CRs, they are dominated by muons since their electromagnetic component is attenuated after traversing large distances in the atmosphere. Observations of such events could help to study the high-energy muon content of EAS and to test hadron interaction models by looking for 
anomalies in this sector at PeV energies, such as those which have been observed at higher energies by observatories like the KASCADE-Grande
detector \cite{Apel2011, Apel2017}, the Pierre Auger observatory \cite{Aab2015a, Aab2014} or the Yakutsk experiment \cite{Abu2000}. 
The former has observed, for example, that the actual attenuation length of shower muons at $10^{17} \, \ev$ is bigger than the predictions of hadron interaction models \cite{Apel2017}, while the latter ones have measured an excess of muons in ultra-high-energy EAS in comparison with the models, a problem which is known as the \emph{muon puzzle} \cite{Petrukhin2014}.

Following \cite{Aab2015b}, atmospheric and/or astrophysical neutrinos with energies above $10^{15} \ev$ could also be detected in this way if they scatter deep in the atmosphere or interact with the rock of the nearby Jura mountains.  Such neutrinos could induce very inclined young EAS,
which could be distinguished from regular old showers produced by CRs due to MATHUSLA's superior tracking resolution. Young EAS are characterized by a richer electromagnetic component, a broader time signal and a larger EAS front curvature. Thus measurements of the particle content and the spatial/temporal structure of inclined EAS in MATHUSLA could allow the search for neutrino signals. MATHUSLA might also offer
a tool to look for upward-going EAS from Earth skimming $\nu_\tau$'s \cite{Aab2015b, Feng2002} and upward-going muons from $\nu_\mu$'s interacting with the rock below the detector or inside the MATHUSLA's instrumented volume \cite{Gaisser1995, Ambrosio2003}.  If MATHUSLA is able to detect also upward-going muons resulting from muon-flavored neutrinos interacting in rock, MATHUSLA could also provide complementary measurements for $\nu-$oscillations (see, for example, \cite{Fogli1998, Jung2001, Giacomelli2013}).

\item \emph{Study of EAS and tests of hadronic interaction models:}
Detailed measurements of the spatial and temporal structure of EAS, as well as data on the charged particle attenuation length and the muon components of highly inclined showers, may provide a number of clues to understand  several outstanding ambiguities in hadron interaction models \cite{Apel2011, Apel2017, Aab2014, Abu2000, Aab2015b}. Additional constraints would be supplied by correlating MATHUSLA's measurements with detection of the high-energy muon component in the underground detector.

\item \emph{High-multiplicity Muon Bundles:}
Muons with an energy greater than $\sim 50 \gev$ will penetrate down into the rock and reach the main detectors at CERN. ALEPH/DELPHI at LEP \cite{Grupen2003,Ridky2005} and ALICE at the LHC \cite{Adam2016} have studied this high-energy muon component of EAS's, observing events with more than 100 muons in the underground detector. During the LEP era, these high-multiplicity muon bundles could not originally be explained by hadron interaction models, and several BSM explanations were proposed \cite{Kankiewicz2017, Ridky2017}. Today, the data by ALICE points towards iron-rich CR primary composition at energies above $10^{16}$ eV, but the data has very low statistics and further measurements are needed to understand the origin of these muon bundles and their impact on primary CR studies. 

Muon bundles could be detected by the LHC main detector and correlated with data from MATHUSLA. This would give a much more complete picture of these special CR events and could allow for their origin to be unambiguously determined. These events are also valuable for constraining hadron interaction models.
\end{itemize}
It is worth noting that these measurements could be significantly improved if $e/\mu$ discrimination capability were added to MATHUSLA. Apart from making measurements of CR primary composition and spectra less dependent on hadron interaction models, the separate muon data would allow for many additional detailed probes of hadron interaction models, with great benefit to all future CR measurements at other experiments, e.g. \cite{Abbasi2018, Joshi2017, Budnev2017, Aartsen2017}.
This non-exhaustive lists of physics targets demonstrates that MATHUSLA could supply data which will be of unique value to the astroparticle and CR physics communities.

\clearpage 
\section{Model-Independent Considerations}
\label{sec:LLPSmathuslahllhc}

In this section, we provide some model-independent information that allows us to understand MATHUSLA's sensitivity to LLPs produced at the HL-LHC main interaction point, what mass scales and lifetimes it can hence probe, and how its resulting capabilities for discovering new physics compare to those offered by both MET and LLP searches at ATLAS and CMS. 
This will provide important context  for the signal estimates in the subsequent sections.
We also comment on the importance of energy thresholds in the low-mass regime, which has implications for the final detector design.

\subsection{LLPs at MATHUSLA}

\subsubsection{Signal Estimate}
\label{s.MATHUSLAsignalestimate}

The probability for each LLP in a  signal event sample to decay within MATHUSLA's assumed $(200m)\times(200m)\times(20m)$ decay volume of Fig.~\ref{f.mathuslalayout} is easily computed for a given proper lifetime $c\tau$. For an LLP that traverses the detector volume, this probability is given by
\begin{eqnarray}
\label{e.Pdecay}
P_\mathrm{decay}(b c \tau, L_1, L_2) &=& e^{- \frac{L_1}{b c \tau}} - e^{- \frac{L_2}{b c \tau}} 
\\
\nonumber &\approx& \frac{L_2 - L_1}{b c \tau} \ \ \  \ \ \ \ \ \ \ \ \ \ \mbox{for $(L_2-L_1) \ll b c \tau$} \ ,
\end{eqnarray}
where $L_1, L_2$ are the distances from the IP where the LLP enters and exits the decay volume, and 
\begin{equation}
b = \frac{|\vec p|}{m}
\end{equation}
is the boost of the LLP. 

The per-decay-detection-efficiency $\epsilon_\mathrm{LLP}^\mathrm{MATH}$ of the LLP within the detector volume will depend on the specific decay mode, as well the precise location of the decay within the detector. Since tracking has to be highly redundant to reject cosmic ray backgrounds, the dominant factor in determining signal efficiency for LLPs decaying into charged particles is simply geometric, i.e. whether the LLP decay products hit the tracker panes near the roof of MATHUSLA. 

We concentrate on LLPs decaying into at least two charged particles, assuming at least two charged tracks have to be associated with a displaced  vertex for signal reconstruction and background rejection. This was studied in \cite{Curtin:2017izq}, for LLPs with mass $\sim 10 $-$ 50 \gev$ having boosts of order $b \sim 1-10$, decaying to quarks, gluons or charged leptons. These hadronic decays produce $\sim 10$-$20$ charged hadrons, and the efficiency for more than 5 charged particles to hit the trackers is close to 100\% inside the decay volume. For leptonic decays, the efficiency for both leptons to hit the tracker is better than 90\% for lighter and more boosted LLPs, and about $50\%$ for relatively heavy and slow-moving LLPs. 
LLPs lighter than a few GeV decaying to a few hadrons would have similar efficiencies to light LLPs decaying leptonically.
These numbers help establish the expected range of the reconstruction efficiency $\epsilon_\mathrm{LLP}^\mathrm{MATH}$: approaching 1 for hadronic decays, and $\sim$ 0.5-1 for leptonic 2-body decays.  This efficiency may be somewhat reduced for LLP decays to  soft ($|\vec p| < \gev$) or highly collimated ($\Delta \theta < 0.01$) final states (as we discuss in Section~\ref{s.energythresholds}), or for decays that have a sizeable invisible component; however, this should not greatly affect the comparisons laid out below. In our signal estimates, we therefore either assume perfect efficiency, or quote all results normalized to an unknown $\epsilon_\mathrm{LLP}^\mathrm{MATH}$. 

It is not yet established whether MATHUSLA will be able to detect photons, see Section~\ref{s.mathuslabasic}. This would likely rely on a layer of material inserted between tracking panes to allow for conversion and subsequent detection of the electron-positron pair \cite{Curtin:2017izq}, see Fig.~\ref{f.mathuslaparticleID}. While this layer would add cost and complication, it would also allow for some particle identification, which would greatly aid diagnosis of the LLP decay mode and is highly motivated from the cosmic ray physics point of view (in particular the electron vs muon discrimination, see Section~\ref{s.mathuslaCR}). 
In the signal estimates for a few theories we therefore also examine LLPs decaying to one or two photons, to more closely examine the motivation for including this capability.

\subsubsubsection*{Analytical Approximation}

It is very helpful to have an analytical approximation of the LLP signal yield at MATHUSLA. This is often sufficient for simple signal estimates, especially in the important limit of long lifetime, and gives a very general understanding of the cross-sections and mass scales MATHUSLA can probe. For a given LLP production process with cross-section $\sigma_\mathrm{sig}^\mathrm{LHC}$  in the 14 TeV $pp$ collisions of the HL-LHC, the number of observed LLP decays over the HL-LHC run with $\mathcal{L} = 3000$ \ifb can be estimated as\begin{equation}
\label{e.NobsMATHUSLA}
N_\mathrm{obs}^\mathrm{MATHUSLA} \ \  \approx  \ \ 
(\sigma_\mathrm{sig}^\mathrm{LHC} \mathcal{L}) \ \
\epsilon_\mathrm{LLP}^\mathrm{MATH}
\ \  n_\mathrm{LLP} \ \ 
P_\mathrm{decay}^\mathrm{MATH}(c \tau)
\end{equation}
where $n_\mathrm{LLP}$ is the number of LLPs produced per event, and $P_\mathrm{decay}^\mathrm{MATH}$ is the chance that an LLP decays in the MATHUSLA detector volume. It is given approximately by 
\begin{equation}
P_\mathrm{decay}^\mathrm{MATH}(c \tau) \approx 
\epsilon_\mathrm{geometric} \ \ 
 P_\mathrm{decay}(\bar b c \tau, L_1, L_2)
 \end{equation}
where 
$\epsilon_\mathrm{geometric} \approx 0.05$ is the fraction of LLPs that fly through the MATHUSLA detector, and $\bar b$ is the average boost of that fraction.  The lengths $(L_1, L_2)$ are taken to be (200m, 230m). 

We have verified the above approximation for $c \tau \gg 200 m$ by explicitly computing the signal acceptance for a variety of LLP production modes and masses in the range of several hundred to $\ll 1$ GeV, simulated to lowest order in MadGraph~5 \cite{Alwall:2014hca} and showered in Pythia 6~\cite{Sjostrand:2006za}. Eqn.~\ref{e.NobsMATHUSLA} is very robust and agrees with the full simulation to within a factor of 2, usually underestimating the real signal yield. The average boost of LLP $X$ can be estimated from an effective parent mass scale $m_\mathrm{eff}$, 
\begin{equation}
\label{e.bbarmparent}
\bar b = \frac{m_\mathrm{eff}}{2 m_\mathrm{LLP}}
\end{equation}
which depends on the production processes in a physically intuitive way, up to a numerical prefactor which can be determined from simulation.\footnote{This is obviously reminiscent of the boost of a particle with mass $m$ that is pair produced in the decay of a stationary parent particle with mass $m_\mathrm{parent}$, $b =\frac{m_{parent}}{2m} \sqrt{1- 4 m^2/m_\mathrm{parent}^2}$, when the parent mass is large.} For the production processes we examined, $m_\mathrm{eff}$ is given in Table.~\ref{t.averageboost}. 

\begin{table}
\begin{center}
\begin{tabular}{|l|l|}
\hline
LLP production mode & effective parent mass scale $m_\mathrm{eff}$
\\
\hline \hline
s-channel scalar $g g \to \Phi, \Phi \to X X$ &  $\sim 1.5 m_\phi$
\\
s-channel vector $\bar q q \to Z'$, $Z' \to X X$ & $\sim m_{Z'}$
\\
s-channel vector  $\bar q q \to Z'$, $Z' \to X X$ with $m_{Z'} \gtrsim \tev$ & $\sim 1.5 \tev$ for $m_X \ll 700 \gev$
\\
$pp \to X X j j$ production via $W^\pm Y^\mp X$ coupling, $m_{Y} \lesssim \tev$ & $\sim 2 m_Y$
\\
$pp \to X X j j$ production via $W^\pm W^\mp X X$ effective coupling & $\sim 2.5 \tev$ for $m_X \ll 1.2 \tev$
\\
Heavy parent pair production $p p \to Y Y, Y \to X + \ldots$ & $\sim m_Y$
\\
Exotic decays of $B$-mesons & $\sim 14 \gev$
\\
\hline
\end{tabular}
\end{center}
\caption{
The average boost $\bar b$ of an LLP $X$ which flies through MATHUSLA, produced at $\sqrt{s} =14 \tev$ in the above production processes, can be estimated using an effective parent mass scale $m_\mathrm{eff}$ (second column) using $\bar b = m_\mathrm{eff}/2 m_\mathrm{LLP}$. The above table was empirically derived from simulation.
}
\label{t.averageboost}
\end{table}

For shorter decay lengths $c \tau \ll 200m$ the MATHUSLA signal is dominated by the tails of the LLP boost distribution.  Due to the exponential dependence of  Eqn.~\ref{e.Pdecay} on the LLP boost, the signal yield is not well captured by using an average boost. Therefore, Eqn.~\ref{e.NobsMATHUSLA} will significantly under-estimate the signal in the short lifetime limit. 

Note that in the above, we do not differentiate between two "simultaneous" displaced decays in MATHUSLA from the same LHC event, or two decays from different LHC events. For discovery, each DV is conspicuous enough at MATHUSLA that it can be treated as an independent signal.

\subsubsection{Sensitivity Estimate}

As argued in \cite{Chou:2016lxi} and reviewed in Section~\ref{sec:mathuslasummary}, MATHUSLA can operate in the background-free regime for LLPs decaying to two or more well-separated charged particles. We therefore obtain projected exclusion limits on the LHC production cross-section of the LLPs by setting $N_\mathrm{obs}^\mathrm{MATHUSLA} = 4$ in Eqn.~\ref{e.NobsMATHUSLA}.
\begin{equation}
\label{e.MATHUSLAxseclimit}
(\epsilon_\mathrm{LLP}^\mathrm{MATH} \cdot \sigma_\mathrm{sig}^\mathrm{MATHUSLA\ limit}) 
\approx
\frac{4}{\mathcal{L} n_\mathrm{LLP}  P_\mathrm{decay}^\mathrm{MATH}(c \tau)} \ \ , 
\end{equation}
and analogously for estimates using full simulation of LLP acceptance. As we show below this gives sensitivity to cross-sections above a fb. For discovery, we assume $N_\mathrm{obs}^\mathrm{MATHUSLA} = 10$ is required. 

Detailed study of backgrounds and signal reconstruction at MATHUSLA may slightly increase the required number of events for exclusion and discovery, but the above criterion is  expected to be a good approximation for the majority of LLP decay modes and sufficient for studies to motivate the detector. Detailed studies may also reveal that the zero-background assumption does not hold for some final states, e.g. decays to photons (which may not be detectable, see Section \ref{s.MATHUSLAsignalestimate}), one-pronged decays of an LLP (e.g. to two collinear jets + an invisible particle, or photon + invisible particle),  decays to electrons only (which may shower in the detector material, making exact DV reconstruction more challenging), or decays of LLPs resulting in very tightly collimated final states (which could mimic neutrinos scattering off air, see Section~\ref{s.energythresholds}). In that case, sensitivity estimates for those final states may have to be adjusted accordingly.

\subsubsection{Benchmark Signal Cross-Sections}
\label{s.benchmarksignalxsecs}

Limits on LLP production cross-section $\sigma$ as a function of lifetime $c \tau$ will have a minimum (best  limit) at some $c \tau_\mathrm{best}$. For $c \tau \gtrsim c \tau_\mathrm{best}$ the limit depends linearly on $c \tau$ as long as the search requires only a single LLP decay (as is the case at MATHUSLA), while the short-distance behavior is slightly more complicated. 
In Eqn.~(\ref{e.NobsMATHUSLA}) and in simulations, $P_\mathrm{decay}^\mathrm{MATH}(c \tau)$ is maximized for 
\begin{equation}
[\bar b c \tau]_\mathrm{best} \approx 200 \ \mathrm{m}
\end{equation}
giving
\begin{equation}
P_\mathrm{decay}^\mathrm{MATH}([\bar b c \tau]_\mathrm{best}) \approx 2 \times 10^{-3}
\end{equation}
Assuming for simplicity that $\epsilon_\mathrm{LLP}^\mathrm{MATH} = 1$ (which is accurate enough for this estimate), this means that for \emph{some range of lifetimes}, a model with LLPs will produce an observable MATHUSLA signal over the HL-LHC run if the LLP production cross-section is larger than 
\begin{equation}
\label{e.MATHUSLAminxsec}
\sigma_\mathrm{sig}^\mathrm{LHC} \gtrsim \mathrm{fb}
\end{equation}
(This lower bound will be reduced if the number of produced LLPs per event is very large, as in some dark shower models.)
In deriving this lower bound we required $N_\mathrm{obs}^\mathrm{MATHUSLA} = 4$, but at this level of precision, the distinction between exclusion and discovery is not important. If the LLP production cross-section is larger than $\sim$ fb, then the maximum lifetime that can be probed is roughly
\begin{equation}
\label{e.bctaumax}
\bar b c \tau_\mathrm{max} \sim (10^3 \ \mathrm{m}) \left( \frac{\sigma_\mathrm{sig}^\mathrm{LHC}}{\mathrm{fb}}\right)
\end{equation}
(assuming $\mathcal{O}(1)$ LLPs per production event) 
since the linear long-lifetime regime starts at a lifetime a factor of a few larger than $[b c \tau]_\mathrm{best}$. This model-independent schematic sensitivity of MATHUSLA is shown in Fig.~\ref{f.MATHUSLAsensitivitycartoon}.

To emphasize the scalability of the MATHUSLA design (see Section~\ref{s.mathuslamodularity}) we also show the sensitivity of a detector with only $1/10$ the volume of the $200m \times 200m \times 20m$ benchmark geometry, which is assumed throughout the rest of this paper. Since the LLP cross-section that can be discovered scales inversely with detector volume, all the expressions in this Sections and indeed the results of this entire whitepaper are easily rescaled for a smaller version of MATHUSLA. While such a mini-MATHUSLA may not probe BBN lifetimes, it would still extend the LLP sensitivity of the LHC main detectors by orders of magnitude.

\begin{figure}
\begin{center}
\includegraphics[width=11cm]{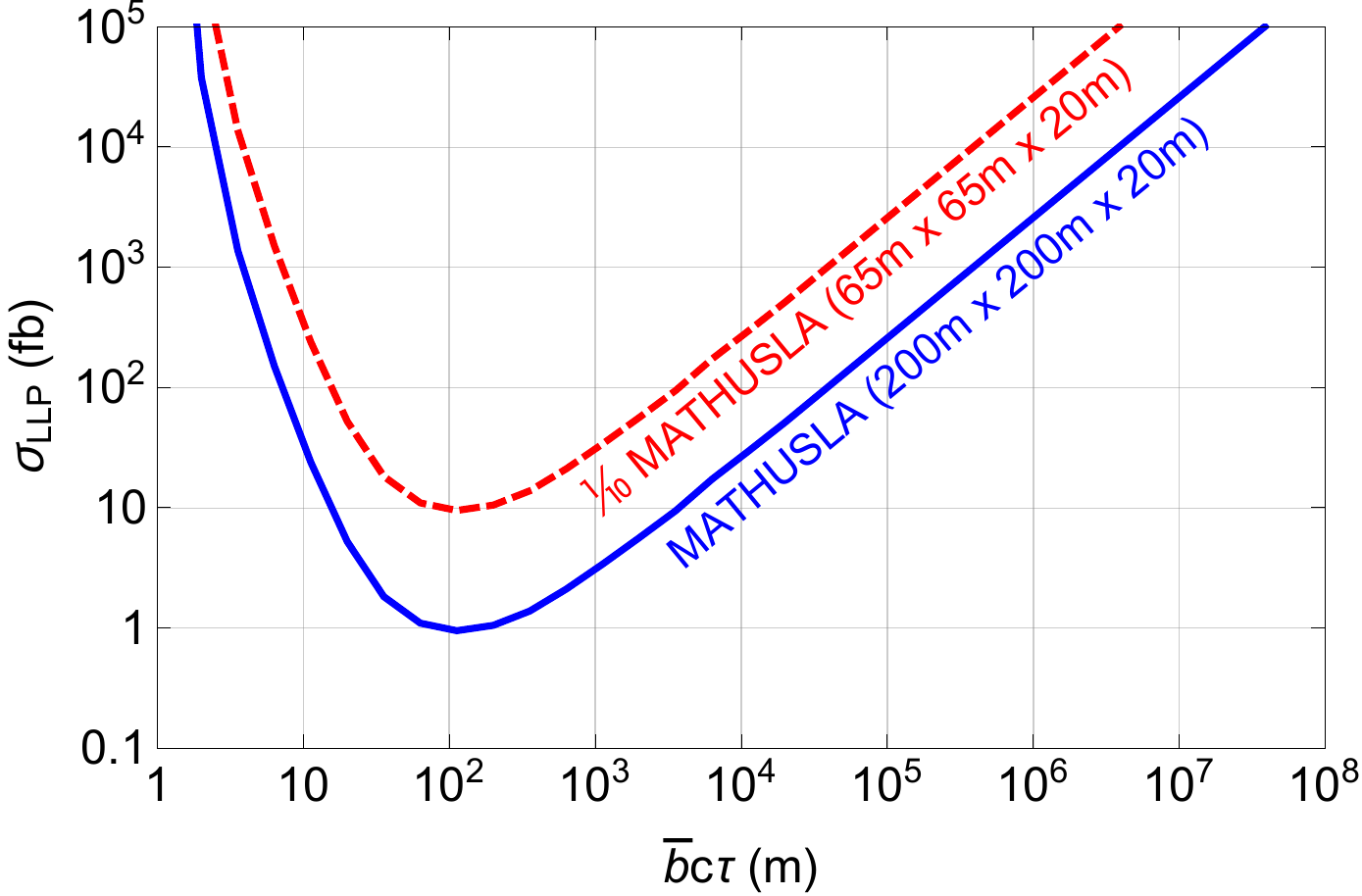}
\end{center}
\caption{
Schematic order-of-magnitude sensitivity of MATHUSLA, assuming $\mathcal{O}(1)$ produced LLPs per production event at the HL-LHC. $\bar b$ is the mean boost of the produced LLPs. The shape of the exclusion/discovery region at short lifetimes depends on the detailed boost distribution, but for long lifetimes $\bar b c \tau \gg 200 m$ depends only on the mean boost and is very model-independent up to an $\mathcal{O}(1)$ factor. Note that LLPs near the BBN lifetime limit of $c \tau \sim 10^7$m can be probed if they are produced with cross-sections in the pb range at the HL-LHC.
To emphasize the scalability of the MATHUSLA design, we also show the reach achievable with a version of MATHUSLA with only $1/10$ the detector volume of the $200m \times 200 m \times 20 m$ benchmark geometry. 
}
\label{f.MATHUSLAsensitivitycartoon}
\end{figure}

\begin{figure}
\begin{center}
\includegraphics[width=13cm]{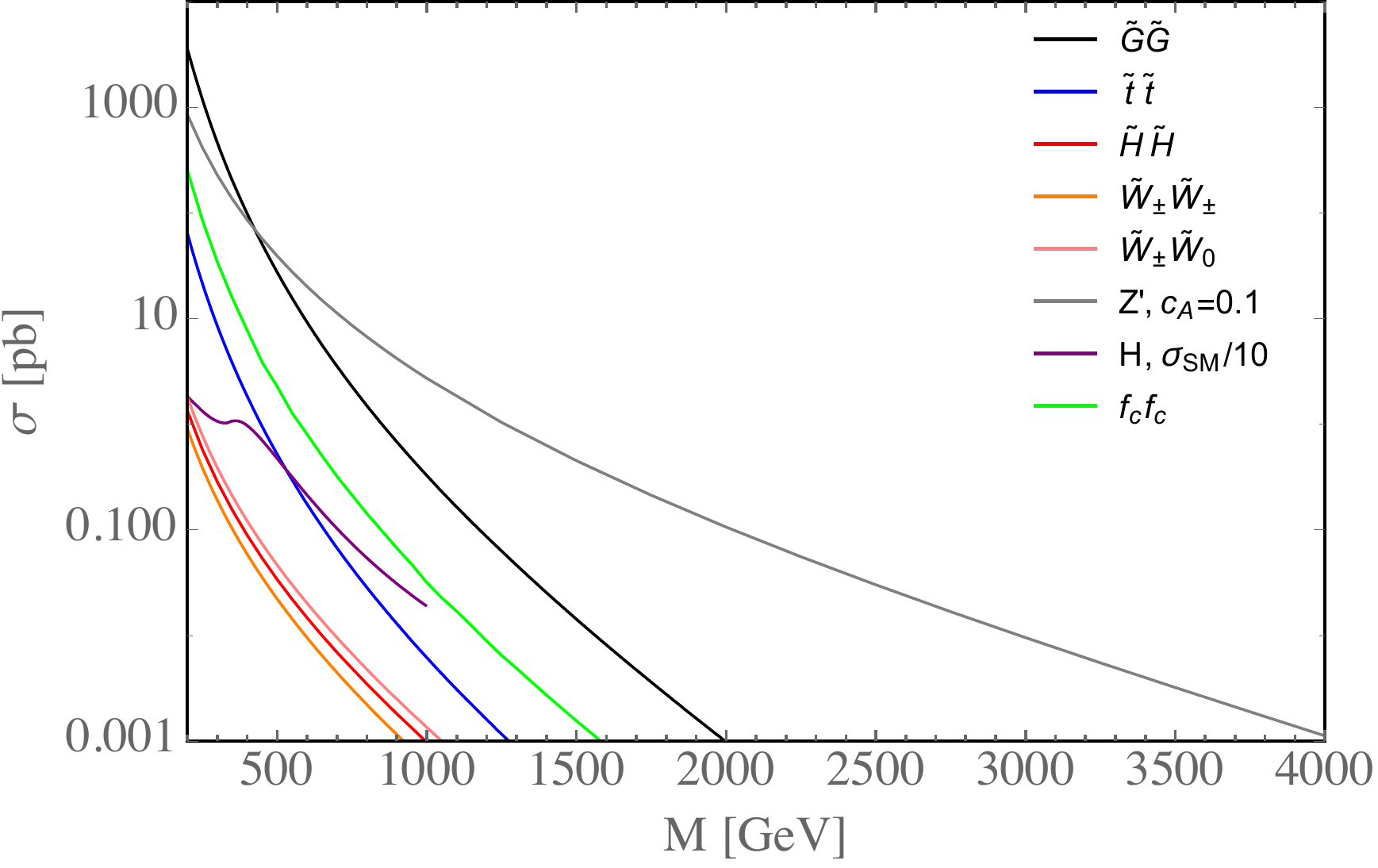}
\end{center}
\caption{
Benchmark LLP production cross-sections at the 14 TeV HL-LHC, as a function of either parent particle or LLP mass.
}
\label{f.benchmarkxsecs}
\end{figure}

This understanding of the model-independent LLP reach allows us to understand which BSM mass scales MATHUSLA can probe. In Fig.~\ref{f.benchmarkxsecs}, we show benchmark LLP signal cross-sections at the 14 TeV HL-LHC, either as a function of parent particle mass $m_\mathrm{parent}$ or as a function of the LLP mass $m_\mathrm{LLP}$, for the most important benchmark processes: 
\begin{itemize}

\item Pair production of color octet fermions (gluinos $\tilde G \tilde G$), color fundamental scalars (stops $\tilde t \tilde t$) or fermions ($f_c f_c$) which can either be LLPs or decay to LLPs. See Sections~\ref{sec:rpv},
\ref{sec:minisplit},
\ref{sec:stealthsusy},
\ref{sec:compositehiggs},
\ref{sec:adm},
\ref{sec:simps},
\ref{sec:wimpbg},
\ref{sec:hiddenvalleys}.

\item Pair production of EW charged states like SUSY higgsinos $(\tilde H \tilde H)$, and winos ($\tilde W_\pm \tilde W_\pm, \tilde W_\pm \tilde W_0$) which can either be LLPs or decay to LLPs. See Sections~\ref{sec:rpv},
\ref{sec:gaugemediation},
\ref{sec:axino-ewino},
\ref{sec:coannihilation},
\ref{sec:freezein},
\ref{sec:wimpbg},
\ref{sec:inertdoubletportal}.

\item S-channel resonance production of an $Z'$-type vector boson which can decay to LLPs. The plot assumes coupling $g = c_A g_W$, with $c_A = 0.1$ but the cross-section $\sigma \propto c_A^2$ can be easily rescaled for lower couplings. This also stands in for a kinetically mixed dark photon with $c_A \sim \epsilon$, and behaves similarly to the production cross-section for a charged $W'$. See Sections~\ref{sec:simps},
\ref{sec:bminusl},
\ref{s.BLnuRLLPscalars},
\ref{sec:LRSYMMnuR},
\ref{sec:LLPscalar_LR},
\ref{sec:darkphotons}, and also Sections~\ref{sec:darkphotons},
\ref{sec:alp}. for exotic decays of the SM $Z$ boson.

\item S-channel resonance production of a SM-like Higgs produced $1/10$ the SM cross-section, which can decay to LLPs. This can be the case for e.g. singlet scalars which mix with the SM Higgs. Again, the cross-section can be rescaled to account for lower mixings. 
See Sections~\ref{sec:wimpbg},
\ref{sec:hiddenvalleys},
\ref{sec:exhdecays}, as well as Section~\ref{sec:exhdecays} and \ref{sec:neutralnaturalness},
\ref{sec:Higgportalneut},
\ref{sec:singlets},
\ref{sec:darkphotons} for exotic decays of the 125 GeV Higgs boson.

\end{itemize}
The SUSY cross-sections are taken from \cite{Borschensky:2014cia}, with the results available in tabulated form in \cite{SUSYworkinggroup}.  The cross-sections for the $Z'$ boson and coloured triplet fermion are calculated using MadGraph \cite{Alwall:2014hca}, with $Z'$ coupling conventions as in \cite{Boveia:2016mrp}.  The heavy Higgs cross-section is taken from \cite{Higgsworkinggroup}, with the 14 TeV results rescaled by the luminosity ratio.

These processes are important for many theoretically motivated scenarios, as we discuss in more detail in the following sections. 
Some universally applicable statements can be made based on the mass for which the above cross-sections are $\sim$ fb: for strong production processes, MATHUSLA can probe LLP or parent particle masses in the 1.3 - 2 TeV range. Depending on the coupling, vector mediators can be produced with masses of several TeV. Electroweak LLPs or parents, like higgsinos or winos, can be probed with masses up to $\sim \tev$.

Of course, exotic decays of the 125 GeV Higgs Boson are some of the most well-motivated and discoverable LLP production modes. At the HL-LHC, the Higgs production cross-section is about 50 pb, meaning branching ratios to LLPs of $\sim 10^{-5}$ can be probed for lifetimes near $c \tau_\mathrm{best} \sim 200$m. Importantly, for branching ratios of $\sim 10\%$, which are not excluded by current measurements, LLP lifetimes near the BBN limit of $c \tau < 0.1$ seconds can be probed, see Section~\ref{sec:exhdecays} and also \ref{sec:neutralnaturalness},
\ref{sec:Higgportalneut},
\ref{sec:singlets},
\ref{sec:darkphotons}. 

Finally, an important possible source of LLPs with masses in the GeV range or below are the decay of SM hadrons, especially $B$-mesons, which are produced at the LHC with $\sim 0.6$~mb cross-section~\cite{Aaij:2016avz}. The resulting $\sim 10^{15}$ produced $B$-mesons produced at the HL-LHC can give rise to displaced signals even for extremely tiny exotic branching fractions to LLPs, see Section~\ref{sec:sgoldstinos},
\ref{sec:exoticbaryonoscillations},
\ref{sec:minimalRHN},
\ref{sec:singlets}.

Note that for the case of LLP production in parent particle decays, these cross-sections have to be multiplied by the appropriate parent particle branching fraction. Similarly, in some cases production rates may be suppressed by small couplings or unknown mixing factors. We do, however, assume that the LLP decays to final states involving SM particles 100\% of the time (with possibly different rates to different SM final states), since a displaced decay with a partial branching fraction into a hidden sector would require a seemingly unnatural coincidence of unrelated scales or couplings.

All the model-dependent sensitivity estimates computed in Section~\ref{sec:naturalness} - \ref{sec:bottomup}  are consistent with the discussion presented here.
The signal estimates involving LLPs from exotic Higgs decays 
(Sections~\ref{sec:exhdecays}, \ref{sec:neutralnaturalness},
\ref{sec:Higgportalneut},
\ref{sec:singlets}, 
\ref{sec:darkphotons}) 
are generated at hadron-level within the Higgs Effective Theory framework in Madgraph5~\cite{Alwall:2014hca}, CalcHEP~\cite{Belyaev:2012qa} and/or Pythia~\cite{Sjostrand:2006za, Sjostrand:2014zea} to account for the dominant gluon-fusion production process, and normalized to the results of the Higgs Cross Section Working Group~\cite{Higgsworkinggroup}. Different choices of generators do not give significant differences in the corresponding sensitivity estimates for this standard process.
The signal estimates involving LLPs from exotic $B$-decays
(Sections~\ref{sec:minimalRHN}, \ref{sec:singlets})
were obtained from $B$-meson distributions generated in Pythia 8, which yields compatible cross sections to LHC experimental measurements~\cite{Aaij:2016avz} and FONLL~\cite{Cacciari:2012ny,Cacciari:2015fta}.
Other processes, like SUSY pair production, are produced in Madgraph5 at lowest order and normalized by $K$-factors if available (or appropriate to the precision of the study). In some cases, like the high-multiplicity hidden valley (Section~\ref{sec:hiddenvalleys}), only kinematic distributions are generated using Monte Carlo, with the LLP reach expressed as an upper bound on some unknown BSM production cross section. 
In all cases, the MATHUSLA geometry was either fully accounted for using three-dimensional ray-tracing and weighing events by their decay probability within the detector volume, or (where indicated) an approximate signal estimate was obtained within a factor of $\sim 2$ by using the analytical expression in Eqn.~\ref{e.NobsMATHUSLA}.

\subsubsection{Impact of Detector Resolution and Thresholds}
\label{s.energythresholds}

The possibility of background-free LLP detection in MATHUSLA relies on being able to assign at least two separate, upwards-going particle tracks to a DV in the detector's decay volume.
If the LLP daughter particles are too soft, the DV may not be reliably detected. If the LLP daughter particles are too collimated, it may only give rise to a `merged' one-pronged DV. Detailed study of this signal is needed, but it would likely suffer significantly higher backgrounds than well-separated multi-pronged DVs. 
Physics reach is therefore maximized if well-motivated LLP scenarios can be detected as multi-pronged DVs. Here we discuss the impact of detector spatial resolution and energy thresholds to determine the regions of LLP parameter space that fall into the multi-pronged DV regime, with important implications for the final design of MATHUSLA.

Since LLPs produced in LHC collisions are generally very energetic compared to minimum ionization energies, energy thresholds of the MATHUSLA detectors are expected to play a less crucial role in determining sensitivity than spatial resolution. However, in cases where LLPs decay intrinsically to soft final states (e.g. dark shower models as in Section.~\ref{sec:hiddenvalleys}), we assume the following minimum thresholds on charged particle three-momenta $| \vec p |$ for detection: pions, 200 MeV; charged Kaons, 600 MeV; muons, 200 MeV; electrons: 1 GeV; protons: 600 MeV; photons: 200 MeV. 

To discuss the impact of spatial resolution, assume for simplicity that an LLP decays into two massless charged SM particles. (This discussion can be easily extended to higher-multiplicity final states or decays closer to kinematic threshold, but this does not qualitatively change the conclusions.) The characteristic opening angle of the decay products is then
\begin{equation}
\theta \sim \frac{1}{\bar b}
\end{equation}
where $\bar b$ is the average boost of the LLP. The spatial resolution of the tracker panes required to separate the decay products is therefore
\begin{equation}
\Delta x \sim (10 m) \ \theta \sim \frac{10 m}{b} \ ,
\end{equation}
corresponding to a maximum LLP boost for detection of a multi-pronged DV, 
\begin{equation}
\label{e.bLLPmax}
b_{LLP}^{max} \sim 1000 \ \left( \frac{1 cm}{\Delta x} \right)
\end{equation}
If the LLP boost can be expressed in terms of an effective parent mass as in Eqn.~(\ref{e.bbarmparent}), this can be translated into a \emph{minimum LLP mass} for multi-pronged DV detection:
\begin{equation}
m_{LLP}^{min} \sim \frac{m_{parent}}{2 b_{LLP}^{max}} \sim \left( \frac{m_{parent}}{2000} \right) \left( \frac{\Delta x}{1 cm} \right)
\end{equation}
For LLPs produced in the decay of a  0.1 - 1 TeV parent, this corresponds to a minimum LLP mass of $\sim 0.1 - 1 \gev$ for cm spatial resolution. For light LLPs produced in $B$-hadron decays, the minimum mass is about 10 MeV, see Table.~\ref{t.averageboost}.\footnote{Note that if photon detection is possible in MATHUSLA, the minimum mass for multi-pronged DV detection of an LLP decaying to $2 \gamma$ would be higher, since photons detection by conversion in material  degrades angular resolution.}

The benchmark detector described in Section~\ref{sec:mathuslasummary} assumes a $\sim cm$ spatial resolution. Clearly, one could lower the minimum discoverable LLP mass by improving this resolution, but this has to be balanced against cost. Options include having different resolutions in the horizontal $x$ and $y$ direction, or using finer segmentation in only a part of the MATHUSLA detector. 
However, even the baseline resolution allows MATHUSLA to discovery LLPs with very low masses below a GeV and perhaps even close to the MeV-scale.

\subsection{Comparing LLP reach at MATHUSLA to the HL-LHC Main Detectors}
\label{s.MATHUSLAvsMAINDETECTORS}

To understand the physics case for MATHUSLA, it is important to compare its capabilities for discovering  very long-lived neutral BSM particles with those of the HL-LHC main detectors. 
ATLAS or CMS could discover such particles in two ways:
\begin{enumerate}
\item  as missing energy in MET searches; or
\item through dedicated LLP searches.
\end{enumerate}

We compare the projected reach of HL-LHC MET searches to the MATHUSLA reach in Section~\ref{s.METcomparison} for  several important simplified models, and  demonstrate that MATHUSLA can probe large regions of parameter space inaccessible to the main detectors. 
 Further, even if a new particle is detected first as MET at the HL-LHC,
 MATHUSLA will still have an important role to play in  characterizing its lifetime. This is obviously a question of great cosmological significance. 

A quantitative comparison of MATHUSLA to direct HL-LHC LLP searches is much more challenging, as ultimate trigger capabilities and background rates for HL-LHC LLP searches are less well-established.   The main detectors can search for neutral LLP decays as (i) displaced tracks or vertices in the tracker, (ii)  isolated energy deposits in the calorimeters, and (iii) displaced vertices in the ATLAS Muon System. Since the detectors were not designed for LLP searches, reconstruction and triggering require dedicated algorithms and can be challenging. While LLP signals can be spectacular and are inherently low-background compared to prompt searches, the backgrounds that do exist are frequently non-collisional and hence difficult to characterize from first principles.   These backgrounds will typically become increasingly important as the LHC luminosity increases, and must be taken into account in establishing the ultimate reach of LLP searches at the HL-LHC.

As we show in Section~\ref{s.LHCLLPsignal}, MATHUSLA and the main detectors have very similar geometric acceptances for LLP decays in the long lifetime regime. 
The crucial advantage of MATHUSLA is thus not its enormous volume (which merely compensates for its distance from the IP), but that it operates almost entirely without backgrounds or triggering issues. 
Thus, for any LLP search which suffers backgrounds or is challenging to trigger on at the main detectors, MATHUSLA will beat the HL-LHC in cross-section sensitivity by up to several orders of magnitude. 

There are very general and well-motivated classes of neutral LLP scenarios for which triggering and backgrounds are both major obstacles at the HL-LHC.  For instance, many models where LLPs are produced in exotic Higgs decays \cite{Chou:2016lxi} typically yield low-mass ($m\lesssim \mathcal{O}(100 \gev)$), hadronically-decaying LLPs without accompanying  hard or leptonic prompt objects in the final state.   The characteristic low-mass, hadronic final states in these models present challenges for both triggering and background rejection. 
Conversely, there are some LLP scenarios where the relative advantage enjoyed by MATHUSLA is much smaller: for example, an LLP with a TeV-scale mass decaying leptonically will likely be easy to trigger on and is unlikely to have much background. 
For scenarios in between, quantitative statements about the sensitivity gain offered by MATHUSLA are more difficult to make. They have to rely on results from the relatively small number of published LLP searches and studies, which can often be difficult to extrapolate to the different running conditions, detector capabilities, and search strategies available at the HL-LHC.
Nevertheless, we can make very universal qualitative statements about how important MATHUSLA will be to cover the LLP parameter space, and parameterize our ignorance of LLP backgrounds at the main detectors in such a way that the results of future studies, or rough experimental intuition, can be utilized to understand the sensitivity gain from MATHUSLA in more detail. This is discussed  in Section~\ref{s.LHCLLPcomparison}.

Our primary focus is neutral LLPs, but charged or colored LLPs can also be considered, since for masses above a few hundred GeV and sufficiently long lifetime, only a small fraction of such LLPs will be stopped in the rock before they reach the surface~\cite{Mackeprang:2006gx,Mackeprang:2009ad}. Their decay can be reconstructed at MATHUSLA, possibly with different background considerations since the passage of such LLPs will register in the scintillator veto surrounding the decay volume.
In general, the HL-LHC coverage for such LLPs is quite good if they have long lifetime~ \cite{Khachatryan:2016sfv, Aaboud:2016uth, ATLAS:2014fka}, since they are not invisible and leave signals in most detector subsystems. In that case, MATHUSLA will offer complementary information.


\subsubsection[Comparing MATHUSLA reach to MET searches with the Main Detectors]{Comparing MATHUSLA reach to MET searches with the Main Detectors\footnote{Philip Harris}}
\label{s.METcomparison}

It is natural to ask whether missing energy searches could effectively probe LLPs with very long lifetimes.  
In this section, we present updated projections of the monojet + MET search reach at the HL-LHC  for three canonical scenarios: exotic Higgs decays to invisible particles ($h \to \mathrm{invis}$), DM simplified models, and supersymmetry. 
One can then compare the MET reach to the reach of MATHUSLA by assuming that the invisible neutral particle is instead unstable. 
We also compare the MATHUSLA reach to the reach of a simple $\mathrm{MET}_\mathrm{PV}$ + LLP search at the main detectors, which adds a DV requirement to the MET search and computes MET using only primary vertex information.

The $h \to \mathrm{invis}$ projections are computed following current LHC practices for dark matter searches. 
We assume a generic LHC detector that stands in for either ATLAS or CMS, using a simple in-house detector simulation that models current running conditions and gives equivalent results to Delphes~\cite{deFavereau:2013fsa}.
The MET trigger is assumed to be efficient above 200 GeV, and assumed to work in HL-LHC conditions. This is consistent with studies shown in~\cite{Collaboration:2272264}, where a L1 Trigger with particle flow and PUPPI (for pile-up mitigation) is presented, and shown to give a consistent MET trigger across the full intensity range of the upgraded LHC.

The monojet search is very inclusive, requiring at least one jet and missing energy above the 200 GeV trigger threshold. Leptons are also vetoed with rapidity up to $|\eta| < 4$ assuming realistic inefficiencies that contribute a residual background, dominantly from $W \to \tau \nu$. 
The dominant Higgs signal comes from VBF, but gluon fusion (ggF) also contributes. VBF and ggF contributions are separately constrained using a two category fit of MET and $m_{jj}$.  
A series of five separate control regions consisting of a single muon/electron/photon and double muon/electron are used in a simultaneous fit \emph{in situ} with the signal regions to constrain both the $Z\rightarrow\nu\nu$ and $W\rightarrow\ell\nu$ backgrounds. 
This method can be extended to constrain top background, but we do not make use of this method here: instead, we use standard MC predictions and apply the same extrapolation uncertainties as for the $W$ background.
To extrapolate from the control regions to the signal region we apply the extrapolation uncertainty scheme following the NNLO QCD+NLO EW predictions~\cite{Lindert:2017olm}. %
As a check, a more conservative uncertainty scheme is applied, which consists of the predicted uncertainties scaled up by an order of magnitude. This uncertainty scheme corresponds to the NLO QCD+NLO EW predictions where the full EW scale corrections are taken as uncertainty.

\begin{figure}
\begin{center}
\includegraphics[width=0.5\textwidth]{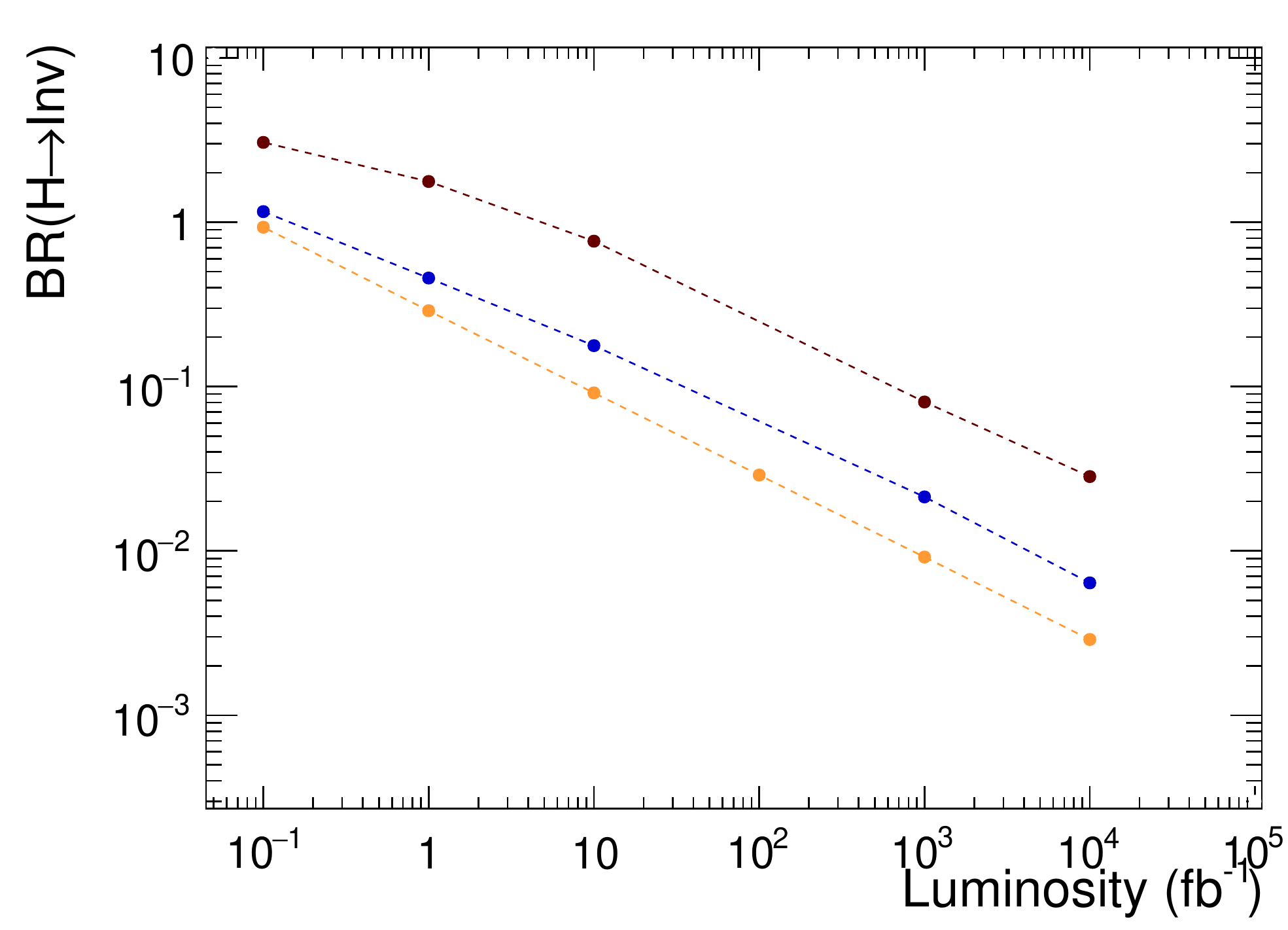}
\end{center}
\caption{Projected $\mathrm{Br}(H \to \mathrm{invis})$ monojet bounds as a function of HL-LHC luminosity. 
The blue curve is uses the systematic uncertainty extrapolated from the NNLO QCD+NLO EW predictions following~\cite{Lindert:2017olm}. 
The purple curve inflates that uncertainty by a factor of 10, while the orange curve assumes no systematic uncertainty. 
}
\label{fig:Hinv}
\end{figure}

\begin{figure}
\begin{center}
\includegraphics[width=0.45\textwidth]{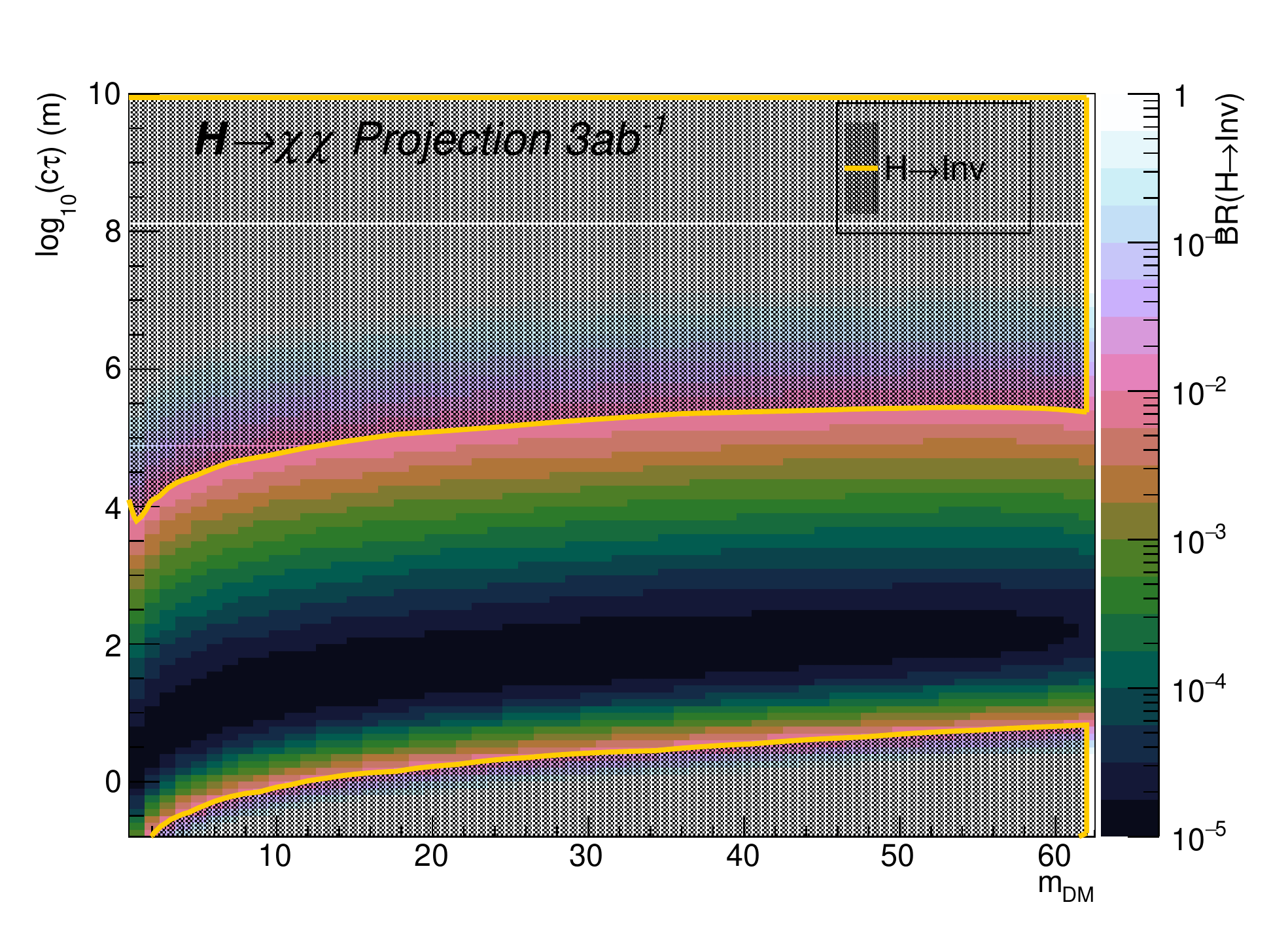}
\includegraphics[width=0.45\textwidth]{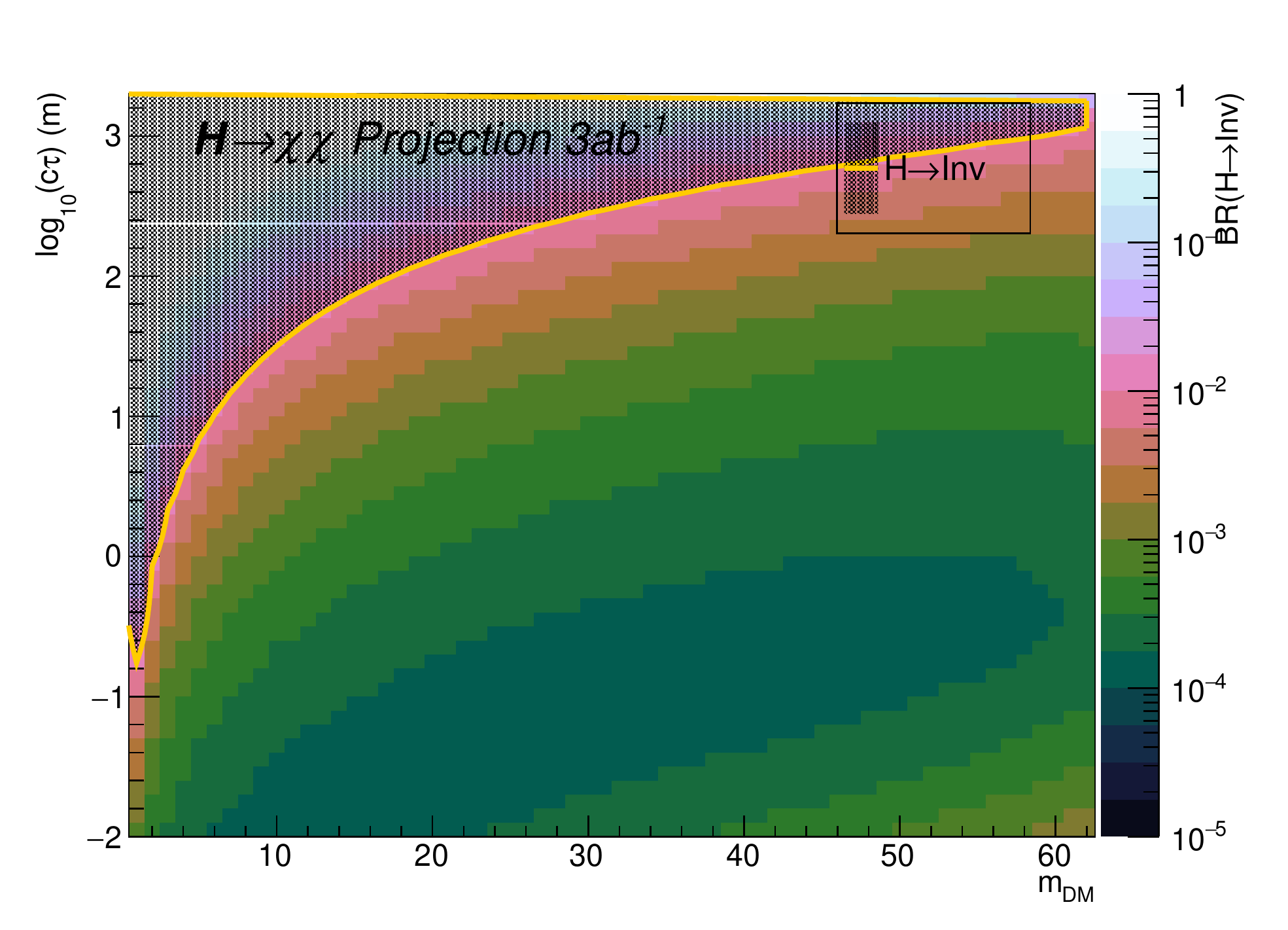}
\end{center}
\caption{$\mathrm{Br}(h \to XX)$ projected  bounds at the HL-LHC as a function of LLP mass, from a background-free DV search at the MATHUSLA detector (left) or from the $\mathrm{MET}_\mathrm{PV}$ + DV search using the main detectors (right). 
In the shaded regions, the updated $\mathrm{Br}(h \to \mathrm{invis})$ bounds from Fig.~\ref{fig:Hinv} are stronger than the direct LLP bounds, but it is important to keep in mind that detection of an invisible Higgs decay would only add motivation to an LLP search at that signal rate.}
\label{fig:Hinv_LL}
\end{figure}

\begin{figure}
\begin{center}
\begin{tabular}{cc}
\includegraphics[width=0.45\textwidth]{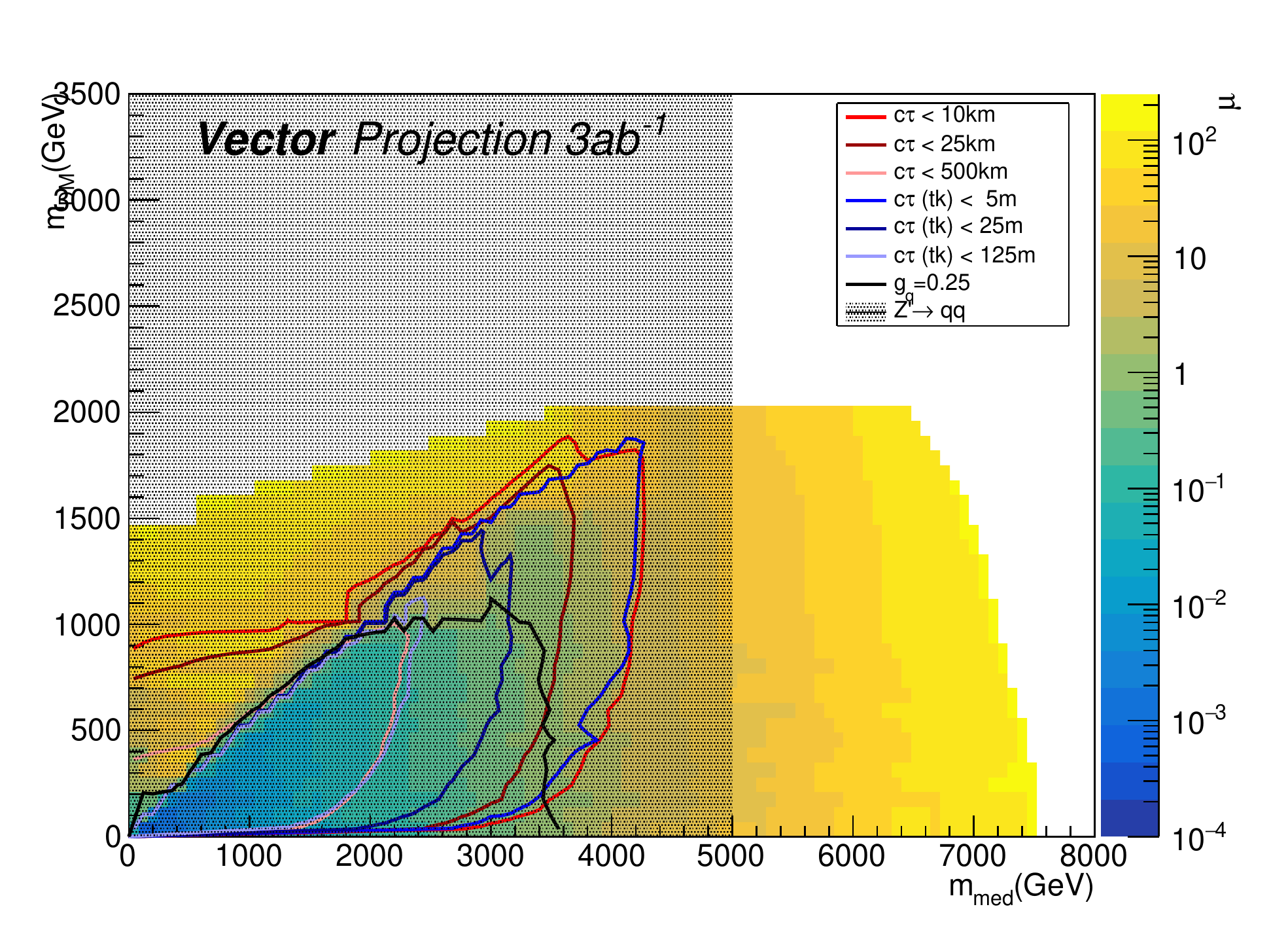}
&
\includegraphics[width=0.45\textwidth]{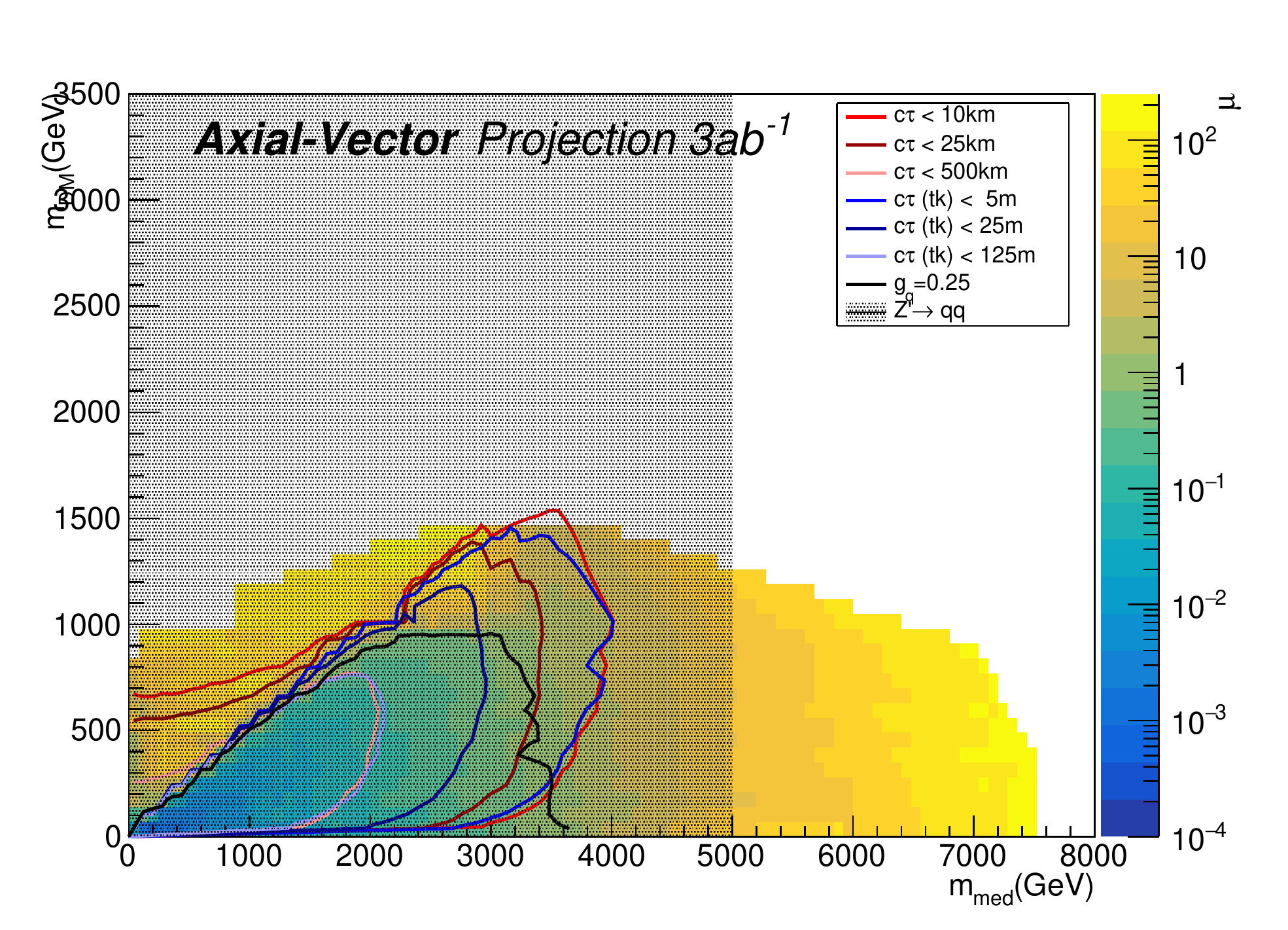}
\\
\includegraphics[width=0.45\textwidth]{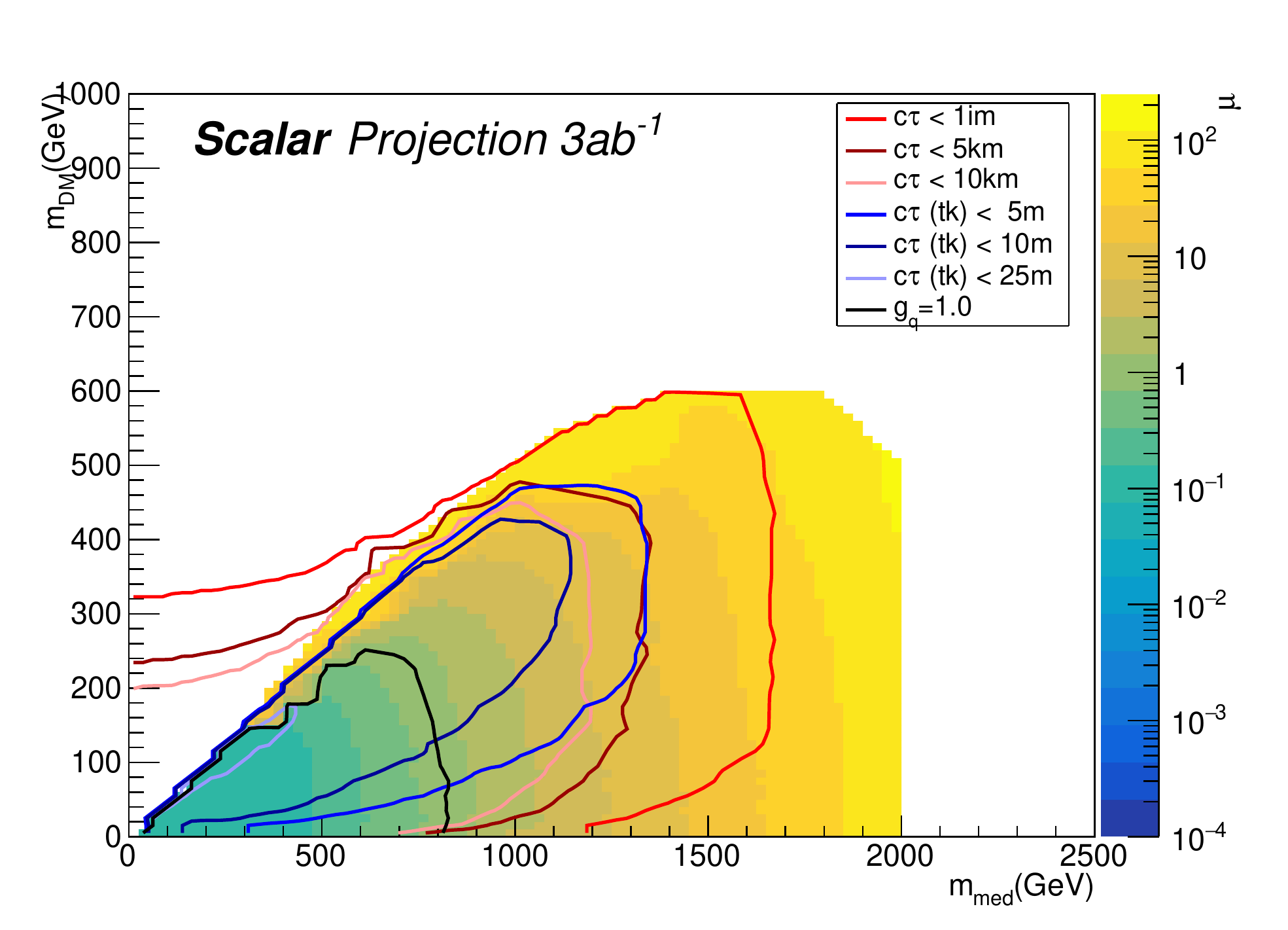}
&
\includegraphics[width=0.45\textwidth]{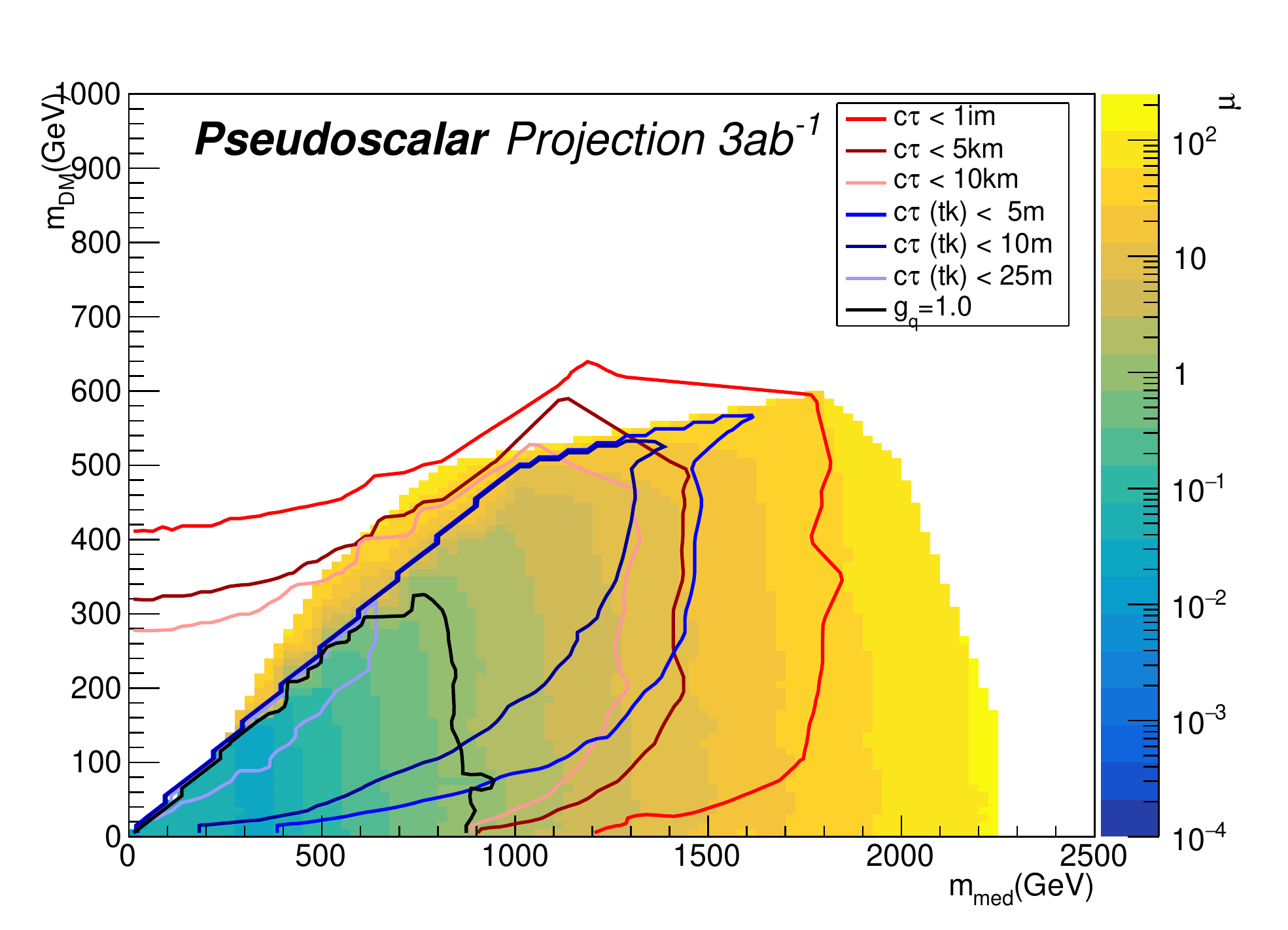}
\end{tabular}
\end{center}
\caption{Projected HL-LHC limits on simplified dark matter models from the MET-only monojet dark matter search (black), a $\mathrm{MET}_\mathrm{PV}$+DV search assuming a range of lifetimes for the invisible particle (blue), and the MATHUSLA detector (red) for a different set of lifetimes. 
\emph{Top:} The bounds are shown for a spin-1 mediator ($g_{q}=0.25$ and $g_{DM}=1.0$ with no assumed mediator to lepton couplings) for vector couplings (left) and for axial-vector couplings (right).
The shaded black area corresponds to the expected bound from dijet searches projected out to 3000$fb^{-1}$ (these searches lose sensitivity for $g_q \lesssim 0.1$), see text.
\emph{Bottom:} The bounds are shown for a spin-0 mediator ($g_{q}=1.0$ and $g_{DM}=1.0$ with SM Higgs-like Yukawa couplings rescaled by $g_q$ to the SM fermions for scalar couplings (left) and for pseudoscalar couplings (right).
}
\label{fig:DMmodelsLLPbounds}
\end{figure}

\begin{figure}
\begin{center}
\begin{tabular}{cc}
\includegraphics[width=0.45\textwidth]{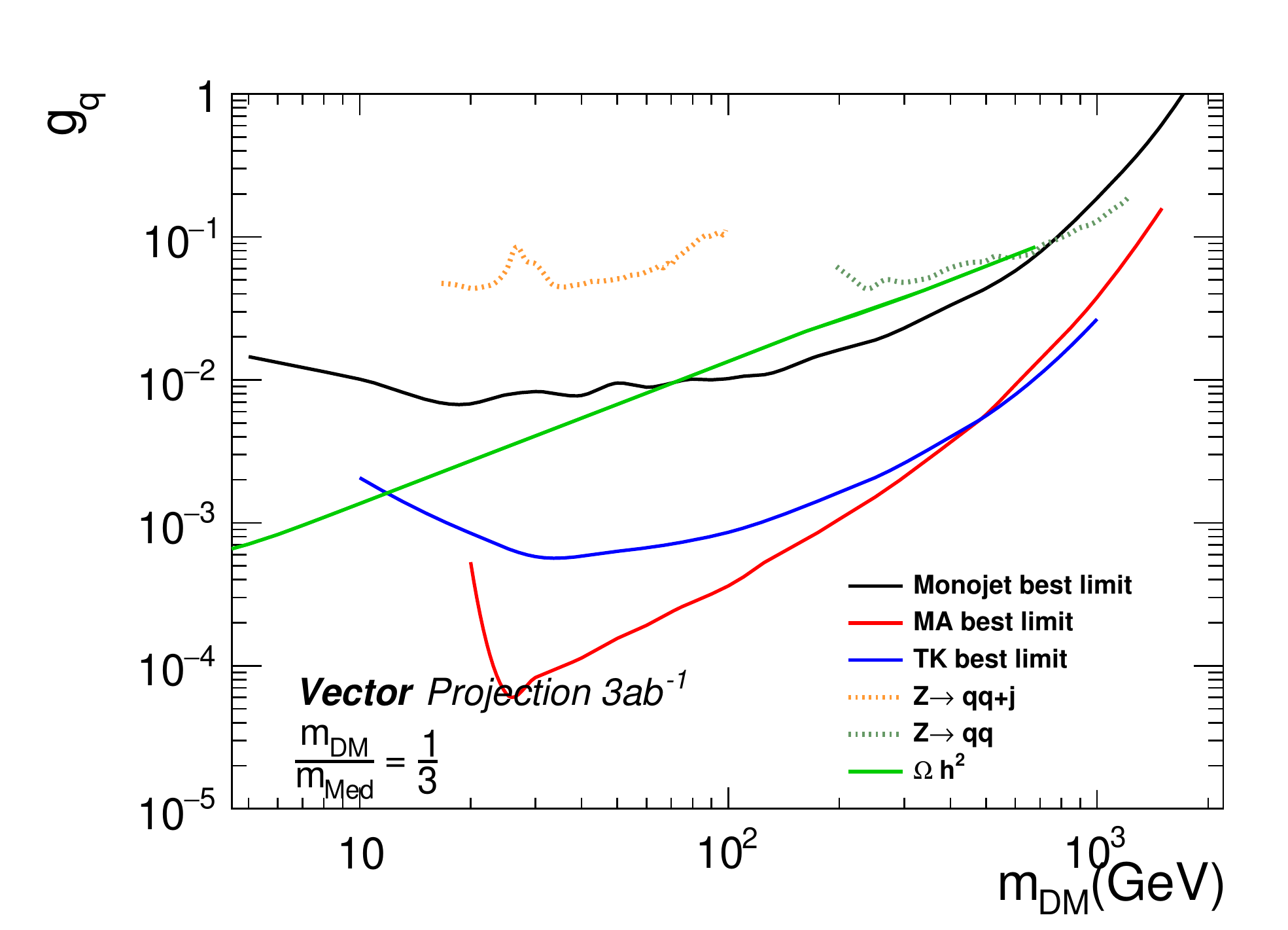}
&
\includegraphics[width=0.45\textwidth]{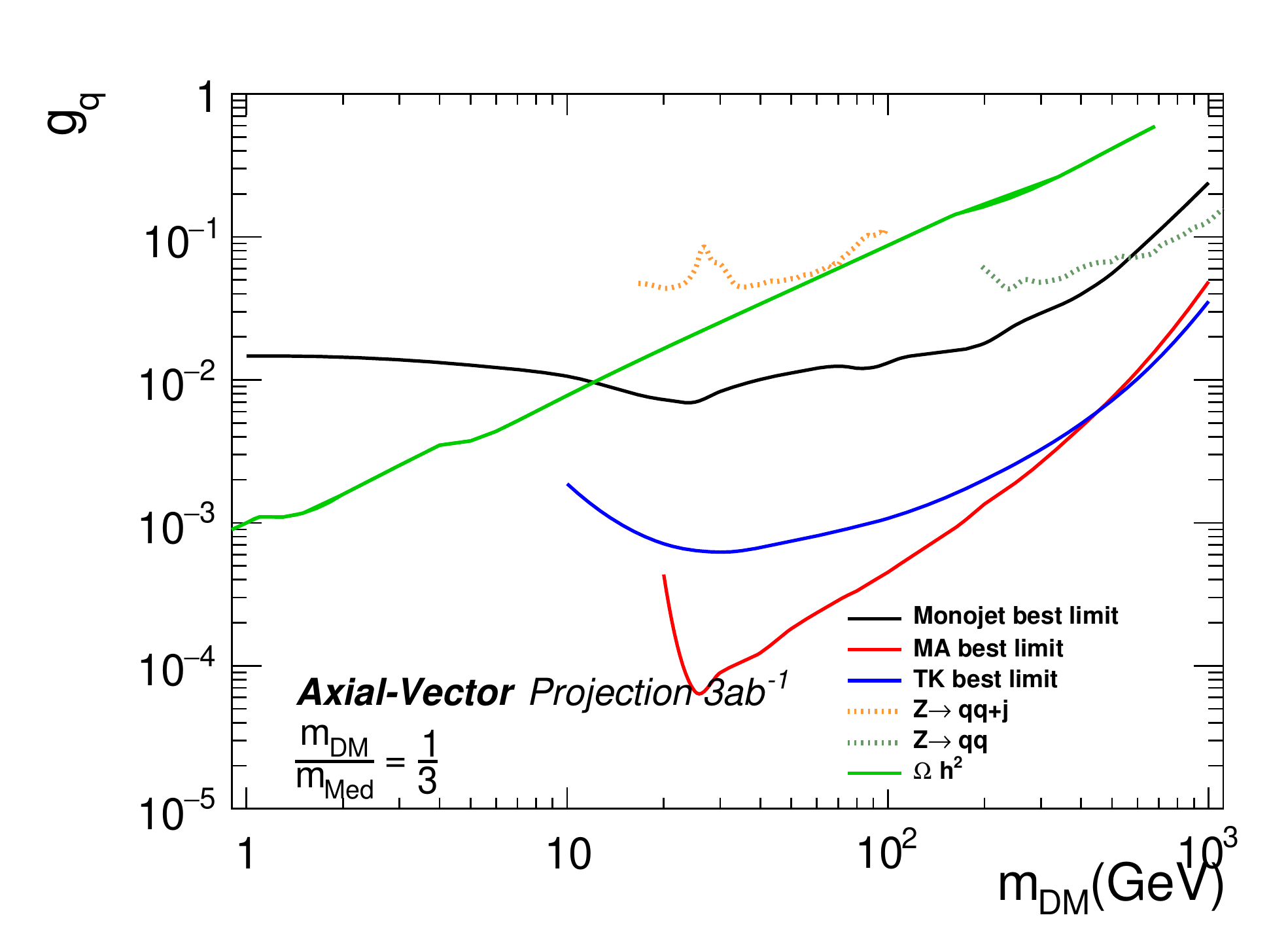}
\\
\includegraphics[width=0.45\textwidth]{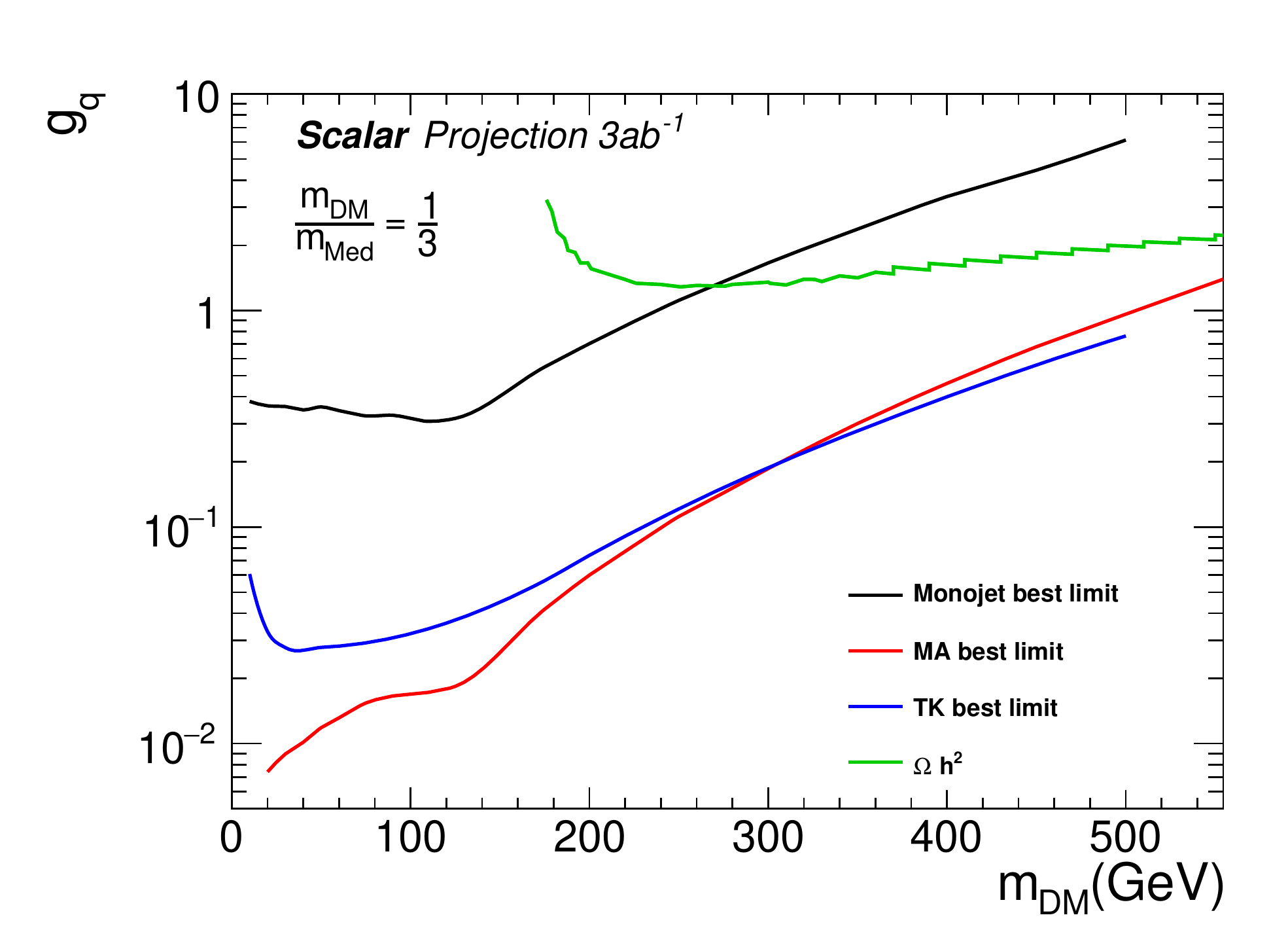}
&
\includegraphics[width=0.45\textwidth]{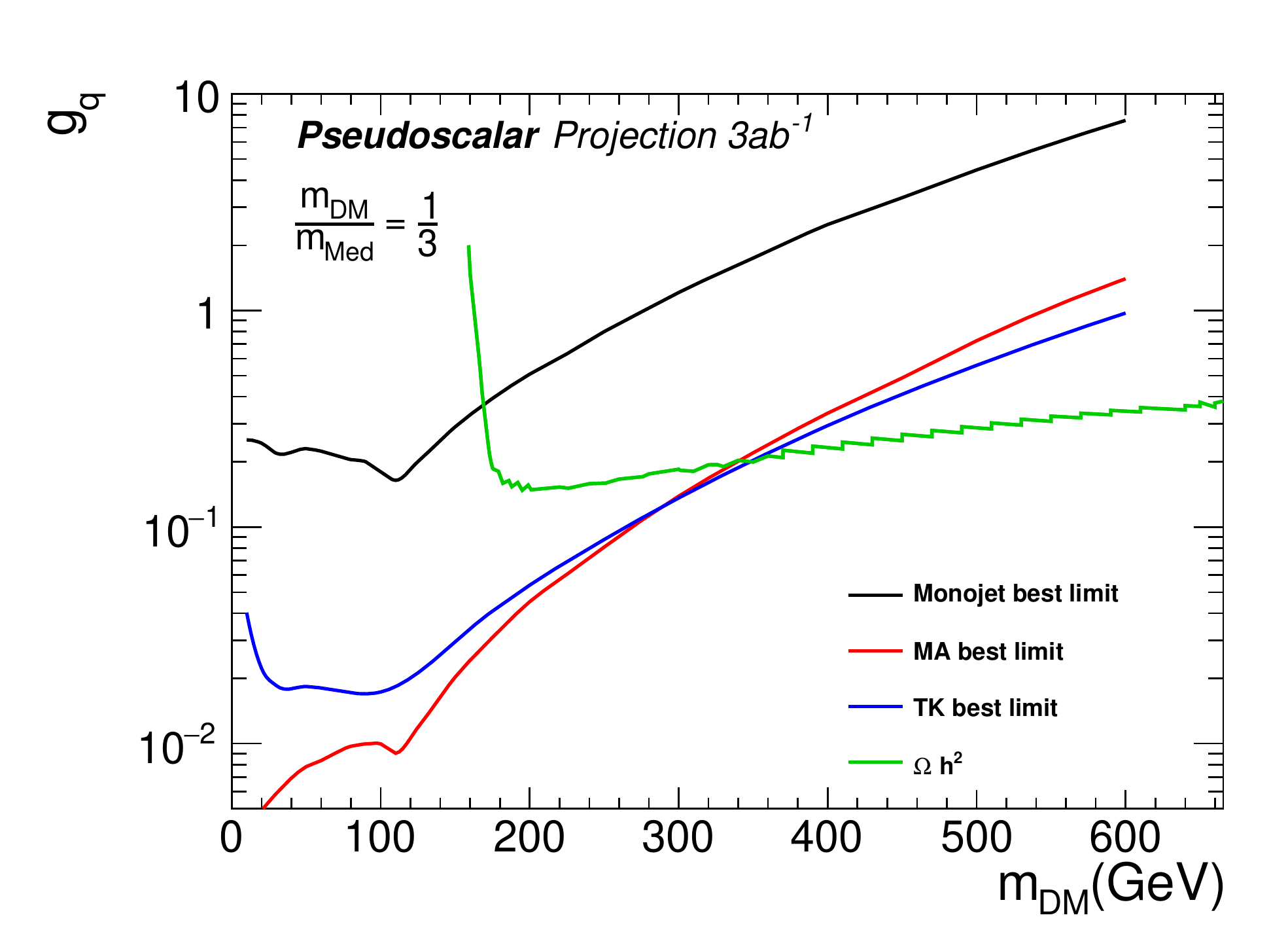}
\end{tabular}
\end{center}
\caption{ Expected minimum coupling $g_{q}$ probed with 3000$fb^{-1}$ of data for the optimal lifetime value (scanning all lifetimes) for spin-1 vector (top left) and axial vector (top right) mediators, and spin-0 scalar (bottom left) and pseudosclar (bottom right) mediators, as a function of dark matter mass $m_{DM}$. 
In this plot $g_{DM} = 1$, and the mediator mass $m_{med}$ is fixed to be exactly three times the mass of the dark matter.
The black line corresponds to the minimum coupling probed in a MET only search, the blue line corresponds to the case where a secondary vertex is identified in the missing energy search, and the red line corresponds to the minimum coupling with the MATHUSLA detector. 
The dashed lines correspond to visible search in either a boosted jet (dashed-orange) or a di-jet resonance (dashed-green), for both searches the dark matter branching ratio is accounted for. 
Lastly the solid green line $\Omega h^{2}$ corresponds to the minimum allowed coupling that will not overproduce dark matter (assuming a single dark matter candidate)
for the shown spin-1 models where no lepton couplings are added, and the shown spin-0 models with a Yukawa coupling. 
While this coupling has no physical significance if the invisible particle is an LLP, it allows the shown coupling reaches to be compared to the sensitivity goal of the monojet searches.}
\label{fig:DMLLPminimumcoupling}
\end{figure}

The resulting limit projection on $\mathrm{Br}(h \to \mathrm{invis})$ is shown in Fig.~\ref{fig:Hinv}. With our assumptions about systematic uncertainty, the invisible branching ratio limit with $3000$ \ifb of luminosity is $\sim 1-2\%$, a significant improvement on the earlier projection of $\sim 7\%$~\cite{CMS:2013xfa}.

The corresponding reach at MATHUSLA on $\mathrm{Br}(h \to X X)$ for LLPs X is readily computed by requiring 4 LLP decays within the detector volume, reproducing the analysis in~\cite{Chou:2016lxi}. MATHUSLA's branching ratio reach as a function of LLP mass and lifetime is shown in Fig.~\ref{fig:Hinv_LL} (left). MATHUSLA is  orders of magnitude more sensitive than the MET search for a large range of lifetimes from meters to 100 km, probing branching ratios as small as $\sim 10^{-5}$. 

A very powerful extension of the MET searches at the HL-LHC makes use of main detector upgrades that will allow some tracking information to be used at L1~\cite{Gershtein:2017tsv}. 
This would allow tracks from the LLP decay to be removed from the MET at L1, either by explicit DV reconstruction if possible at L1, or simply because a L1 track trigger is likely to remove tracks not originating from the IP.\footnote{Current upgrade trigger designs in CMS are considering a vertex constraint on the MET algorithm using Puppi, a PUPPI MET trigger. Through PUPPI this vertex constraint is capable of associating both neutrals and charged particles to the PV and neglecting any unassociated tracks or neutrals near these tracks. This would preserve the MET trigger even in the instance of displaced tracks, provided they are either not reconstructed or pointing sufficiently far away from the PV.}
The LLP would therefore not contribute to MET, allowing the MET trigger to be used for DV searches, where the LLP decay in the tracker is reconstructed in the off-line analysis. 
To distinguish this cleaned-up MET variable from the conventional MET discussed in this section, we refer to MET computed using only primary vertex information as $\mathrm{MET}_\mathrm{PV}$.

For the secondary (displaced) vertex identification, multiple schemes were considered consisting of progressively tighter secondary vertex identifications. The final scheme adopted follows the most recent displaced vertex search performed by ATLAS~\cite{Aaboud:2017iio}. 
The final selection consists of a fit of the MET (without the LLP) and using the same control region constrained fit as used for the dark matter searches.
The second vertex identification efficiency and background rates are taken from the efficiency maps and fake rate estimates shown in~\cite{Aaboud:2017iio}. 
These results were cross checked on a related SUSY model and found to give consistent results, including for compressed spectra which somewhat mimic the kinematics of the exotic Higgs decay final state.
The  $\mathrm{Br}(h \to X X)$ reach of this $\mathrm{MET}_\mathrm{PV}$ + DV search at the HL-LHC main detectors is shown in  Fig.~\ref{fig:Hinv_LL} (right). It is clearly highly complementary to MATHUSLA, with great sensitivity for much shorter lifetimes. 
The trigger upgrades are crucial for this projected sensitivity, since it allows the MET trigger to be used even if the LLP decays in the detector.

Next we examine some canonical DM simplified models \cite{Abercrombie:2015wmb, Buchmueller:2014yoa, Malik:2014ggr, Harris:2014hga, Haisch:2015ioa, Abdallah:2015ter}  where a fermionic dark matter candidate   couples to a vector, axial vector, scalar or pseudoscalar mediator with coupling $g_{DM}$ while the mediator couples to quarks with coupling $g_{q}$ (spin 1) or Higgs-like Yukawa couplings (spin-0) scaled by a flavor-universal prefactor $g_q$. The analysis proceeds exactly like in the invisible Higgs decay case, as does the $\mathrm{MET}_\mathrm{PV}$+DV search and the MATHUSLA DV search, assuming the invisible particle to be unstable instead of a DM candidate. 
Fig.~\ref{fig:DMmodelsLLPbounds} compares bounds in the mediator-DM mass plane, from the MET-only monojet search (black contours), the $\mathrm{MET}_\mathrm{PV}$+DV search for a range of fixed lifetimes (blue) and from the MATHUSLA DV search for a different range of fixed lifetimes (red). 
In all cases, MATHUSLA significantly extends the mass reach for large ranges of lifetimes compared to the MET and $\mathrm{MET}_\mathrm{PV}$+DV searches.
While the $\mathrm{MET}_\mathrm{PV}$+DV search is again complementary to MATHUSLA at shorter lifetimes, only MATHUSLA has sensitivity in the regime where the mediator is off-shell, since in that case the $p_T$ spectrum of signal events is very difficult to distinguish from the $Z \to \bar \nu \nu$ background.
The sensitivity gain is especially pronounced for spin-0 mediators, due to their lower mono-jet efficiency compared to vector mediators that are produced dominantly in valence quark collisions with more additional radiation.

For vector mediators, much of the mass range that is accessible by either MET, $\mathrm{MET}_\mathrm{PV}$ + DV or MATHUSLA searches will also covered by $Z'$ dijet resonance searches. 
We obtain an HL-LHC sensitivity projection of dijet resonance search by rescaling the current limits \cite{Sirunyan:2017nvi, Sirunyan:2016iap, Aaboud:2018fzt, ATLAS:2016bvn, CMS-PAS-EXO-16-056} and show the resulting reach as the black shaded regions in Fig.~\ref{fig:DMmodelsLLPbounds}. 
Note however that dijet searches loose sensitivity for $g_q \lesssim 0.1$, while LLP searches would continue to see a signal (albeit with reduced mass range compared to the examples shown here), see Fig.~\ref{fig:DMLLPminimumcoupling}. In that case, MATHUSLA may be the only way to see these models. Furthermore, a dijet resonance signal will not reveal that the produced resonance has a non-SM decay mode, let alone into unstable particles.

The sensitivity of dedicated LLP searches, either in the main detector or at MATHUSLA, is also illustrated by Fig.~\ref{fig:DMLLPminimumcoupling}, where we show the minimum invisible particle coupling to the mediator (for the optimal range of lifetimes) that can be probed by the MET, $\mathrm{MET}_\mathrm{PV}$ + DV and MATHUSLA searches. The spectacular nature of the LLP signal, with zero or greatly reduced background, means that the smallest couplings that can in principle be probed are one or two orders of magnitude smaller than the reach of the corresponding MET search.

Finally, we perform exactly the same analysis and comparison for a SUSY simplified model with a single light squark being pair-produced and decaying to two jets and two neutralinos. 
The mass reach of the MET-only monojet search, the $\mathrm{MET}_\mathrm{PV}$ + DV search and the MATHUSLA LLP search is shown in Fig.~\ref{fig:SUSY_2D}. 
This simple example of a SUSY decay chain is different from the scenarios discussed above, since the invisible particle is always produced in association with hard jets. As a result, the MET trigger is much more efficient than in the Higgs- and DM-related searches. For a heavy squark and a very light neutralino, there is little background and the reach of the MET search is signal-limited. 
Therefore, the $\mathrm{MET}_\mathrm{PV}$ + DV and MATHUSLA LLP searches do not significantly extend mass reach in the light neutralino case (though they would again be required to correctly diagnose the invisible particle as being an LLP). 
However, if the neutralino mass is even an $\mathcal{O}(1)$ fraction of the squark mass, the MET search becomes much less efficient and squark mass reach decreases drastically. On the other hand, the $\mathrm{MET}_\mathrm{PV}$+DV and MATHUSLA LLP searches are unaffected, and greatly extend mass reach in regions of parameter space where the neutralino is of comparable mass to the squark.

While the SUSY scenario we studied was that of a single light squark species, broadly similar conclusions can be drawn for other searches where the LLP is produced in a decay chain, like gluinos, additional light squarks, EW SUSY partners, and non-SUSY scenarios with similar topologies. 
The MET search will be very efficient and signal limited if the invisible particle is very light, and at high parent particle masses.
For moderately heavy LLPs compared to the parent mass, not to mention highly compressed regions, the MET search looses sensitivity. While strategies exist to probe these challenging spectra at the main detectors~\cite{Schwaller:2013baa} it is likely for many scenarios that $\mathrm{MET}_\mathrm{PV}$ + DV and MATHUSLA LLP searches are the only discovery channels.

\begin{figure}
\begin{center}
\includegraphics[width=0.5\textwidth]{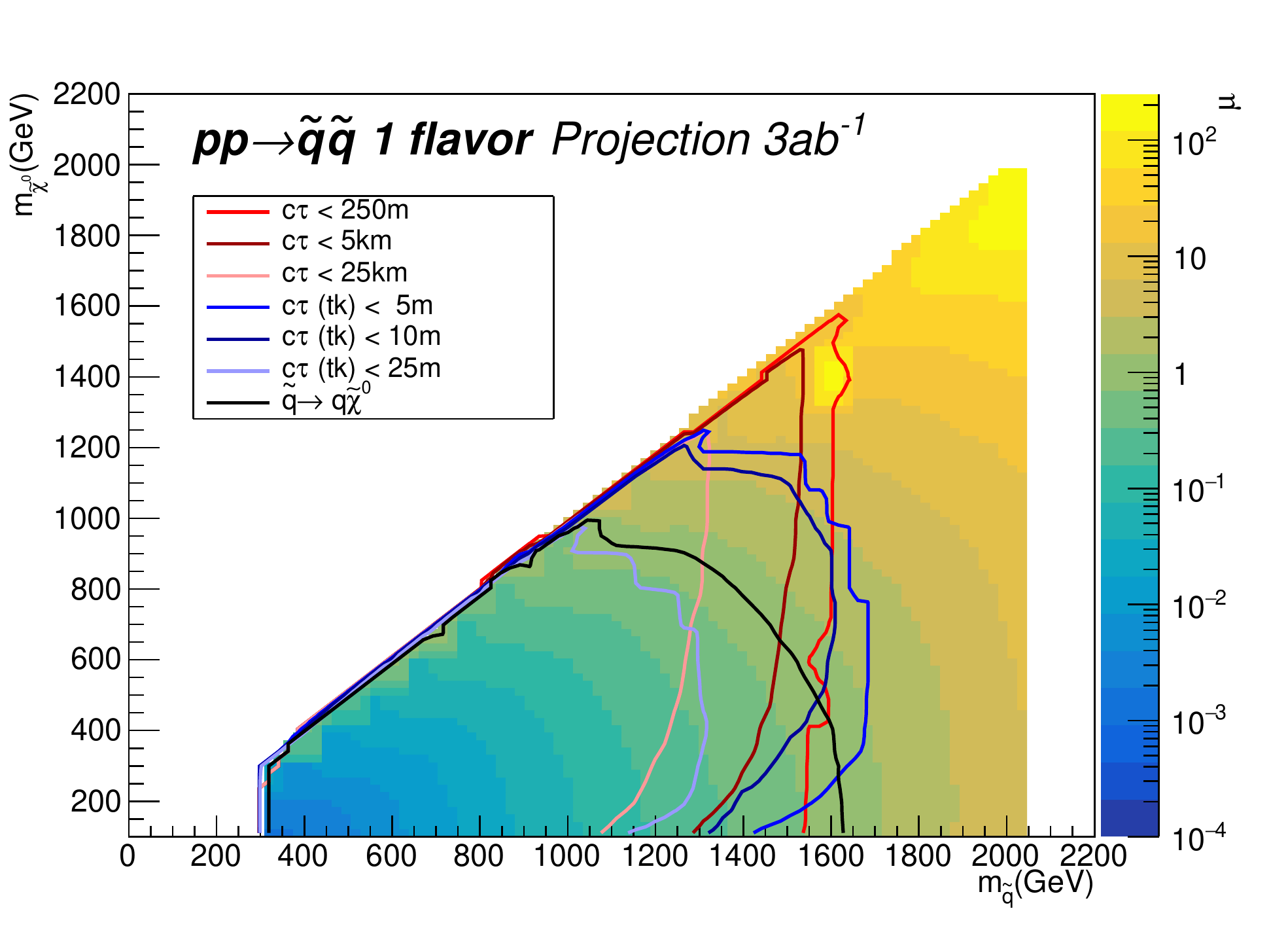}
\end{center}
\caption{Projected HL-LHC limits on single squark pair production and decay into a neutralino, with   bounds for projected limits to a total for 3000$fb^{-1}$ of the monojet dark matter search (black), a search for a displaced vertex for fixed lifetimes (blue), and the MATHUSLA detector (red) for additional fixed lifetimes.The bounds are shown for squark pair production using a cross-section corresponding to a single light squark pair, typical bounds use four light squark flavors in the final state. 
  }
\label{fig:SUSY_2D}
\end{figure}

The benchmark scenarios studied here allow for some universal conclusions to be drawn. For invisible particles produced directly in Higgs decays or via mediators, MATHUSLA and other LLP searches significantly extend the mass range in both mediator/parent particle mass and invisible particle mass, since MET searches rely on additional radiation to trigger.
For invisible particles produced in decay chains, LLP searches extend mass reach into (even very slightly) compressed regions. 
We also reiterate the fact that even if an LLP is first discovered in MET searches, detection of its decay is the only way to ascertain whether the newly produced particle can have a cosmological role as a DM candidate, or whether it is part of a BSM spectrum of unstable states. 
It is clear that MATHUSLA offers great discovery potential that is inaccessible to HL-LHC MET searches.


\subsubsection{LLP Signal Estimate for the Main Detectors}
\label{s.LHCLLPsignal}

To understand the general HL-LHC sensitivity to LLP decays compared to MATHUSLA, it is helpful to begin with a few general comments about the size of a potentially observed LLP signal.
The following discussion applies regardless of how a particular search is constructed (i.e., trigger requirements, prompt cuts, etc.) or how the LLP is reconstructed in detail (i.e., as a displaced vertex or displaced track, in the tracker or muon system, calorimeter deposition, etc.). 

For a given LLP search, the number of observed signal events with a reconstructed LLP decay at the HL-LHC can be estimated as follows:
\begin{equation}
\label{e.NobsLHC}
N_\mathrm{sig}^\mathrm{LHC} \ \approx \ (\sigma^\mathrm{LHC}_\mathrm{sig} \ \mathcal{L}) \ \ 
n_\mathrm{LLP}
\ \ 
\epsilon_\mathrm{LLP}^\mathrm{LHC}  \ 
P_\mathrm{decay}^\mathrm{LHC}(\bar b c \tau) \ 
\epsilon_\mathrm{cuts}^\mathrm{LHC} \ ,
\end{equation}
where $\epsilon_\mathrm{LLP}^\mathrm{LHC}$ is the efficiency for reconstructing an LLP decay that occurred in the specified main detector subsystem (tracker, muon system, calorimeter, depending on the analysis), $n_{LLP}$ is the number of LLPs produced in a single event, $P_\mathrm{decay}^\mathrm{LHC}(\bar b c \tau)$ is the chance that an LLP decays in that detector subsystem, and $\epsilon_\mathrm{cuts}^\mathrm{LHC}$ is the chance that \emph{those events with reconstructed LLP decays} also pass the \emph{trigger} and \emph{off-line} analysis cuts which do \emph{not} pertain to the displaced nature of the LLP decay. (Note the order in which these efficiencies are defined.) 

Arriving at a realistic signal estimate through Eq.~\ref{e.NobsLHC} for a specific signal requires a dedicated collider study.  In the long-lifetime regime, however, several major simplifications occur, which
helps make the comparison with MATHUSLA's capabilities transparent and robust.
\begin{itemize}

\item The chance $P_\mathrm{decay}^\mathrm{LHC}(\bar b c \tau)$ that an LLP with a lifetime $\bar b c \tau \gg 200 m$ decays in a given main detector subsystem can be expressed relative to the corresponding MATHUSLA LLP acceptance in a nearly process-independent way. For the LLP production processes listed in Table~\ref{t.averageboost}, we observe:
\begin{equation}
\color{black}
\label{e.Pdecayratio}
\frac{P_\mathrm{decay}^\mathrm{LHC}}{P_\mathrm{decay}^\mathrm{MATH}}\
\approx
\left\{
\begin{array}{ll}
2.2 & \mbox{ATLAS Muon System}
\\
0.8 & \mbox{ATLAS HCAL}
\\
1.0 & \mbox{ATLAS or CMS tracker (full volume)}
\\
0.25 & \mbox{ATLAS tracker (DV reconstruction volume)}
\end{array}
\right. 
\end{equation}
The DV reconstruction volume of the ATLAS tracker refers to the analysis in \cite{Aad:2015uaa}. For this purpose, the ATLAS muon system barrel and endcap can be combined and their detection efficiencies $\epsilon_\mathrm{LLP}^\mathrm{LHC}$ averaged. The HCAL endcap has very low acceptance relative to the barrel and is neglected at this level of precision. The above relation holds for all examined LLP production modes at the $\sim 10\%$ level.

\item The LHC LLP detection efficiency, relative to the corresponding efficiency in MATHUSLA, will be in the range 
\begin{equation}
\frac{\epsilon_\mathrm{LLP}^\mathrm{LHC}}{\epsilon_\mathrm{LLP}^\mathrm{MATH}}
\sim 0.1 - 1
\end{equation}
since $\epsilon_\mathrm{LLP}^\mathrm{MATH} \sim 1$ for hadronic decays  given the redundant tracking required for cosmic ray background rejection, and $\epsilon_\mathrm{LLP}^\mathrm{MATH} \gtrsim 0.5$ for a LLP decaying to two widely separated leptons. On the other hand, characteristic reconstruction efficiencies for LLPs in the main detectors are $\epsilon_\mathrm{LLP}^\mathrm{LHC} \sim 0.3$ (ATLAS MS, \cite{Aad:2015uaa}), 0.1 (DV in ATLAS tracker, \cite{Aad:2015uaa, Aad:2015rba}) and 0.5 (CMS displaced jet in tracker, \cite{CMS:2014wda}).

\item The final unknown factor, $\epsilon_\mathrm{cuts}^\mathrm{LHC}$, can be estimated using relatively simple simulations, provided the signal requirements are known. 
This requires an understanding of how various trigger and analysis thresholds would change at the HL-LHC compared to run 2. 
In many cases, especially if $\epsilon_\mathrm{cuts}^\mathrm{LHC}$ is dominated by a trigger requirement, it can also be analytically estimated from known kinematic distributions and branching ratios (which are important, for example, if the analysis relies on leptons which are only present in a fraction of signal events). 

\end{itemize}
As we now discuss, this schematic understanding will be very helpful to understand MATHUSLA's advantage over the HL-LHC alone.

\subsubsection{Comparing MATHUSLA reach to LLP searches with the Main Detectors}
\label{s.LHCLLPcomparison}

For purely geometric reasons, the HL-LHC main detectors will have superior sensitivity to MATHUSLA if the LLP has a relatively short lifetime $\bar b c \tau \ll 200$m.   We are interested in understanding the relative sensitivity of the two experiments in the long-lifetime regime $\bar b c \tau \gtrsim 200$m. Below, we therefore discuss both hypothetical and performed searches that require a \emph{single} observed LLP decay in the main detector, which will give the best limit in this regime. 

If a given HL-LHC LLP search has $N_{BG}^{LHC}$ background events, let us parameterize this background simply by an effective background cross-section after analysis cuts: 
\begin{equation}
N_\mathrm{BG}^\mathrm{LHC} \ \equiv \ \sigma_\mathrm{BG\ after\ cuts} \ \mathcal{L}.
\end{equation}
The exclusion limit of the LLP search at the HL-LHC can then be estimated by solving $N_\mathrm{sig}^\mathrm{LHC}/\sqrt{N_\mathrm{BG}^\mathrm{LHC} }= 2$, giving
\begin{eqnarray}
\label{e.LHCxseclimitBG}
\sigma_\mathrm{sig}^\mathrm{LHC\ limit} \ \ \approx \ \ 
 \frac{2}{\epsilon_\mathrm{LLP}^\mathrm{LHC}  \ 
P_\mathrm{decay}^\mathrm{LHC}(c \tau) \ 
\epsilon_\mathrm{cuts}^\mathrm{LHC}
n_\mathrm{LLP}
}
\sqrt{\frac{\sigma_\mathrm{BG\ after\ cuts}}{\mathcal{L}}}.
\end{eqnarray}
In the absence of background, the HL-LHC sets an exclusion limit corresponding to $N_\mathrm{sig} = 4$ in Eqn.~(\ref{e.NobsLHC}):
\begin{equation}
\label{e.LHCxseclimitnoBG}
\sigma_\mathrm{sig}^\mathrm{LHC\ limit}  \ \ \approx \ \ 
\frac{4}{
 \mathcal{L} \ \ 
\epsilon_\mathrm{LLP}^\mathrm{LHC}  \ 
P_\mathrm{decay}^\mathrm{LHC}(c \tau) \ 
\epsilon_\mathrm{cuts}^\mathrm{LHC}
n_\mathrm{LLP}
}
\end{equation}
We now define an important figure of merit, $R_s$, the \emph{long-lifetime sensitivity gain of MATHUSLA}:
\begin{equation}
R_s \equiv \left. \frac{\sigma_\mathrm{sig}^\mathrm{LHC\ limit} }{\sigma_\mathrm{sig}^\mathrm{MATHUSLA\ limit}} \right|_\mathrm{b c \tau \gg 200 \mathrm{m}}
\end{equation}
which is independent of lifetime in this regime. 
Making use of Eqns.~(\ref{e.MATHUSLAxseclimit}), (\ref{e.LHCxseclimitBG}) and (\ref{e.LHCxseclimitnoBG}), we can express $R_s$ in the following way:
\begin{eqnarray}
\label{e.RS}
R_s &\approx& 
\left( \frac{P_\mathrm{decay}^\mathrm{MATH}}{P_\mathrm{decay}^\mathrm{LHC}}\right)
\ 
\left(
\frac{\epsilon_\mathrm{LLP}^\mathrm{MATH}}{\epsilon_\mathrm{LLP}^\mathrm{LHC}}
\right)
\
\frac{1}{\epsilon_\mathrm{cuts}^\mathrm{LHC}}
\
\times
\
\mathrm{Max}\left[1,  \left(\frac{\sigma_\mathrm{BG \ after \ cuts}}{\mathrm{10^{-3}\  \mathrm{fb}}}\right)^{1/2}\right]
\end{eqnarray}
Eqn.~(\ref{e.RS}) summarizes important information about the relative capabilities of MATHUSLA and the HL-LHC main detectors in the long lifetime regime:
\begin{itemize}
\item When the HL-LHC search is background-free, the arguments of Section~\ref{s.LHCLLPsignal} show that $R_s \gtrsim 1$, and possibly $R_s\gg 1$ if the trigger or reconstruction efficiency is low in the main detector, or if the analysis relies on a subdominant production or decay mode of the LLP (e.g. leptons). MATHUSLA will therefore never do much worse than the HL-LHC, and will do better in many cases.
\item When there is \emph{any} background above an ab level, $R_s > 1$ automatically. The relative sensitivity gain of MATHUSLA can then be very large, and can be estimated by inserting the effective background cross-section into Eqn.~(\ref{e.RS}).
\end{itemize}
For many possible searches for a single LLP decay at the HL-LHC, the size of the backgrounds is still unknown, but once the corresponding experimental studies are completed, the relative sensitivity gain can be estimated using Eqn.~(\ref{e.RS}).   However we can already make some general statements based on information from experimental analyses at LHC Runs 1 and 2.

 First, the LHC main detectors have excellent capabilities for LLPs that decay to well-separated pairs of leptons ($e^+e^-$, $\mu^+\mu^-$) in the tracker.  Here lepton triggers can be used to record the event without any requirements on associated prompt objects \cite{CMS:2014hka}.  This provides excellent and inclusive sensitivity to LLPs with masses as low as $\mathcal{O}(10\mathrm{\, GeV})$.   Backgrounds are negligible in the Run I searches even in the absence of additional cuts on the momenta of the displaced lepton vertex or on prompt objects \cite{CMS:2014hka}, suggesting that the background cross-section for well-separated displaced lepton pairs will likely be very small even at the HL-LHC, in the ab-range or below. 
 
If the LLP has a mass in the few GeV range or below, its decay to leptons gives rise to the displaced lepton jet (LJ) final state. To date, no search for a single displaced LJ has been performed, but ATLAS conducted a search \cite{ATLAS:2016jza} at 13 TeV for at least \emph{two} displaced LJs decaying in the tracker, calorimeters, or lower regions of the muon system. The search is very inclusive, with no additional prompt signal requirements, and acceptance-times-efficiency (per event) is in the 0.1-0.3 range. However, even with the requirement of two reconstructed LLP decays instead of one, the background in the signal region corresponds to $\sigma_\mathrm{BG\ after\ cuts} \sim 10$fb. 
This demonstrates that backgrounds for a single displaced LJ in the ultra-long lifetime limit would be much higher than for well-separated leptons, representing an opportunity for MATHUSLA to make very large sensitivity gains compared to the HL-LHC main detectors.

 LLPs decaying hadronically in the tracker are more challenging for the main detectors, especially at low mass.  This difficulty begins in the trigger.   Triggering options for displaced decays are limited by the need to pass the Level 1 (L1) hardware triggers.  While it may be possible to implement  L1 triggers based on properties of the LLP decay itself, many analyses  rely on L1 triggers optimized for prompt physics even if higher-level triggers are designed for LLP decays \cite{Aad:2013txa, CMS:2014wda}. Often in the long lifetime regime of interest, analyses using the MET and $H_T$ triggers will be most important.  This reliance on prompt triggers means that events must typically be relatively energetic to be recorded to tape, thus limiting LHC sensitivity to low-mass or low-energy ($\lesssim 100$ GeV) final states.  Trigger thresholds will generally rise at the HL-LHC to cope with the increased luminosity, exacerbating the issue.  %
Even the dedicated displaced triggers at ATLAS targeting decays in the outer tracker and HCAL \cite{Aad:2013txa} are based on trigger objects with $E_T$ thresholds of $\sim 50 \gev$ of GeV that will also be expected to increase.

For this reason, the searches for single LLPs decaying hadronically in the tracker which have been carried out by ATLAS and CMS typically rely on high-$p_T$ objects to clear Level 1 triggers \cite{CMS:2014wda,Aad:2015rba,ATLAS-CONF-2017-026}.  These searches have typical signal reconstruction efficiencies in the $10-60\%$ range, with acceptance and efficiencies increasing with increasing signal mass.  In Run 1 these searches were effectively or nearly background free. The CMS displaced dijet search  \cite{CMS:2014wda}, which requires $H_T > 325 \gev$ for triggering but does not require full displaced vertex reconstruction, sees $\sigma_\mathrm{BG\ after\ cuts} \sim 60$ab; the ATLAS DV+MET search is less inclusive \cite{Aad:2015rba}, imposing tighter vertex identification and a MET requirement of $\met >180$ GeV, resulting in $\sigma_\mathrm{BG\ after\ cuts} \sim 0.6$ab.  

The dominant background for the ATLAS tracker searches for multi-track DVs  occurs when a low-mass vertex is crossed by an unrelated high-$p_T$ track, and will thus increase with luminosity \cite{Aad:2015rba,ATLAS-CONF-2017-026}. Indeed, the analogous ATLAS search at Run 2 maintained $\sigma_\mathrm{BG\ after\ cuts} \sim 0.6$ab, at the cost of increasing the requirement on MET, $\met >250$ GeV.   This makes clear that backgrounds to LLP searches in the inner tracker will depend sensitively on the details of trigger and reconstruction, but will generically increase with luminosity. The already sizeable cuts imposed on event $E_T$ scales in these analysis may also be expected to increase at the HL-LHC, both to pass triggers and to control the increasing backgrounds.  Thus we generically expect searches for hadronically decaying LLPs in the trackers with masses below a few 100 GeV to become increasingly challenging at the HL-LHC, while the prospects for high-mass LLPs will benefit from increasing luminosity.

On the other hand, LLPs decaying hadronically in the  ATLAS Muon System can be triggered on directly using a L1 muon trigger seed and a dedicated higher-level LLP trigger \cite{Aad:2013txa, Aad:2015uaa}, with only weak dependence on the energy scale of the event for LLP masses $\gtrsim 5-10 \gev$.   In the ultralong lifetime regime, a search for a single LLP decay in the Atlas Muon System is likely the best LHC search for low-mass, hadronically-decaying LLPs produced without additional prompt objects \cite{Coccaro:2016lnz}. The background cross-section found by \cite{Coccaro:2016lnz}, derived using public data from \cite{Aad:2015uaa}, is of order $\sigma_\mathrm{BG\ after\ cuts} \sim 100$fb.
This is the leading way to probe hadronically decaying very long-lived particles produced in exotic decays of the 125 GeV Higgs. 

Given this background cross-section and the reconstruction efficiencies $\epsilon_\mathrm{LLP}^\mathrm{LHC}\sim 0.5, \epsilon^\mathrm{MATH}_\mathrm{LLP}\sim 1$, along with $\epsilon_\mathrm{cuts}^\mathrm{LHC} \sim 1$, MATHUSLA's sensitivity gain is substantial, corresponding to improving the cross-section reach by a factor of $R_s \sim 1000$. This agrees  with the findings of the full study in \cite{Chou:2016lxi}. 
Since the muon system is physically separated from the IP, and since the main background for displaced vertices in the muon system can be traced back to unusual high-energy QCD events, this analysis is somewhat unique in that its conclusions can probably be applied to the HL-LHC with a reasonable degree of confidence.

ATLAS and CMS also search for LLPs decaying in the calorimeters.  Here
a useful general lesson for very long-lived particles can be extracted by comparing the ATLAS search for two LLP decays in the HCAL \cite{Aad:2015asa} to the search for two LLP decays in the muon system (or one in the MS and one in the inner tracker) \cite{Coccaro:2016lnz}. Both searches feature an inclusive search with no MET requirements that is sensitive to  LLPs produced in exotic Higgs decays and in turn decaying hadronically. With 20.3 \ifb of 8TeV data, the HCAL search has about 24 background events, while the MS search has about 2.  This already indicates that for such LLPs, the ATLAS Muon System search described above provides the better inclusive sensitivity at the main detector. This is not surprising, since the ATLAS MS search has full displaced vertex reconstruction, while the information supplied by the HCAL is much less detailed.  In general we expect that LHC LLP searches in the inner tracker or muon system will be the most powerful at the HL-LHC, as tracking capabilities will help control pileup events that spoil HCAL signal isolation and contribute non-collision backgrounds.

Depending on the final design chosen, it is possible that MATHUSLA may be able to detect photons from LLP decays by inducing conversion in material \cite{Curtin:2017izq}. To date, CMS performed the only LHC search for a single LLP decaying to a single photon + MET \cite{CMS:2015sjc} in the context of Gauge Mediation. Under the assumption that a neutralino NLSP is produced in the decay of heavy colored supersymmetric particles, and decays as $\tilde \chi_0^1 \to \gamma + \tilde G$, the search reconstructed a single LLP decay using timing measurements and required more than 60 GeV of MET, 2 jets with $p_T > 35 \gev$ and the leading photon to have $p_T > 80 \gev$. Even with these additional kinematic cuts, the resulting background cross-section was non-negligible, $\sigma_\mathrm{BG\ after\ cuts} \sim 0.5 $~fb. The importance of these non-LLP cuts to reduce backgrounds is illustrated by the ATLAS search \cite{Aad:2014gfa}, which looked for \emph{two} LLPs decaying to $\gamma$ + invisible, but imposed no additional cuts beyond MET $> 75 \gev$. That search had to contend with an effective $\sigma_\mathrm{BG\ after\ cuts} \sim 20 $~fb. Given these examples, and the sensitivity of photon reconstruction to pile-up considerations, it is clear that searches for a single LLP decaying to a single photon will have orders of magnitude more background at the HL-LHC than searches for leptonic LLP decays.

Despite the difficulty of quantitatively extrapolating some of the above cases to the HL-LHC, these examples provide a useful point of reference for  understanding MATHUSLA's advantages compared to the main detectors. The greatest sensitivity gains, possibly by several orders of magnitude, apply for hadronically decaying LLPs with less than a few 100 GeV of visible energy (prompt or displaced), leptonically decaying LLPs with masses below 10 GeV, and (if detectable) LLPs decaying to photons.

\subsection{Summary}
The capabilities of MATHUSLA can be summed up in a few simple lessons:
\begin{enumerate}

\item If the LLP signal cross-section is greater than a fb, then MATHUSLA can see a signal for some range of lifetimes, see Section~\ref{s.benchmarksignalxsecs}. In the long-lifetime limit, MATHUSLA's sensitivity to LLP production is readily estimated using Eqn.~(\ref{e.NobsMATHUSLA}) and Table~\ref{t.averageboost}. The model-independent LLP cross-section sensitivity is shown in  Fig.~\ref{f.MATHUSLAsensitivitycartoon}.

\item LLPs with average boosts lower than $\mathcal{O}(1000)$ can be reconstructed as DVs with two or more prongs, see Eqn.~(\ref{e.bLLPmax}). This results in excellent rejection of cosmic rays and other backgrounds, justifying the zero-background assumption for LLP searches.  Higher boosts (and hence lower LLPs masses) would require additional analysis and likely suffer higher backgrounds, unless the detector resolution is increased.

\item Geometrically, MATHUSLA has very similar acceptance for LLP decays with $c \tau \gg 200\mathrm{m}$ as the HL-LHC main detectors, see Eqn.~(\ref{e.Pdecayratio}), though it may have significantly higher reconstruction efficiency. Therefore, if the corresponding LLP search at the main detectors has any appreciable background ($\sigma_\mathrm{BG\ after\ cuts} > \mathrm{ab}$) or low signal efficiency (small $\epsilon_\mathrm{cuts}^\mathrm{LHC}$, e.g. due to trigger issues, or requirements on the LLP production or decay mode), MATHUSLA will have better sensitivity.

\item The greatest sensitivity gains, possibly by several orders of magnitude, apply for dominantly hadronically decaying LLPs produced with less than a few 100 GeV of visible energy (prompt or displaced), leptonically decaying LLPs with masses below 10 GeV, and (if MATHUSLA can detect them) LLPs decaying to photons. See Section ~\ref{s.LHCLLPcomparison}. 

\item MATHUSLA is sensitive to mass regions that MET searches cannot reach, both in parent and LLP mass, and provides a means of diagnosing any MET signal that is found. See Section~\ref{s.METcomparison}.

\end{enumerate}
This will provide important intuition in assessing the physics case for the theoretically motivated LLP scenarios described in the following sections.
Qualitatively, the above also applies to a mini-MATHUSLA with $1/10$ the decay volume, see Fig.~\ref{f.MATHUSLAsensitivitycartoon}, though of course with the sensitivity gain reduced by one order of magnitude.

\clearpage 


\section{Theory Motivation for  LLPs: Naturalness}
\label{sec:naturalness}

With the discovery of the Higgs at the LHC, or something very similar to it, a resolution to the Hierarchy problem or Naturalness of the EW scale is an ever more pressing concern. 
From the modern Wilsonian understanding of renormalization, since the Higgs is a scalar field,  the EW scale in the SM is in conflict with treating the SM as an EFT valid to energies greater than the EW scale.  To avoid this inherent tension in any model where a scalar field is lighter than the cutoff of the theory requires some mechanism which makes the theory ``natural".    This mechanism could be a symmetry; the Higgs might not be an elementary field; or there could be some sort of selection mechanism for the EW scale.  There have been a proliferation of new ideas and more complicated models of naturalness in recent years given the dearth of BSM signals at the LHC typically associated with canonical models of naturalness.  
A common thread amongst all the classes of naturalness motivated models surveyed here is the presence of neutral long-lived states.

The nature of the Hierarchy Problem singles out the EW scale for new physics searches. 
In the theory frameworks of Supersymmetry (Section~\ref{sec:susy}) and Neutral Naturalness (Section~\ref{sec:neutralnaturalness}), this leads to predictions of LLP signatures that are intimately connected with the mechanism that stabilizes the weak scale. These LLPs would be produced with appreciable cross-sections at the LHC, and neutral long-lived states can naturally have lifetimes that are in the relevant range for MATHUSLA. 
In other models, like the Composite Higgs (Section~\ref{sec:compositehiggs}) or Relaxions (Section~\ref{sec:relaxion}), the existence of LLPs occurs as part of various possible complete models with lifetime as more of a free parameter. MATHUSLA then provides a new window beyond ATLAS and CMS into this theory space.
In the rest of this section we discuss the motivation and predictions for LLPs in these theory frameworks.

\subsection{Supersymmetry}
\label{sec:susy}

Supersymmetry is the most well known and theoretically under control solution of the Hierarchy problem.  It also has a wide array of naturally long-lived particles which are highly motivated.  In particular, a mechanism which causes a long-lifetime, potentially measurable at MATHUSLA, exists solely within the MSSM alone without any additional model building or fine-tuning.   This occurs when the gravitino is the LSP and as such the particles of the MSSM can decay to it.  What is particularly compelling is that this mechanism only depends on the scale of SUSY breaking, $\sqrt{F}$.  The gravitino mass in SUSY scales as
\begin{equation}
m_{3/2}\sim \frac{F}{M_{pl}},
\end{equation}
while the coupling of the NLSP to the gravitino scales as $1/F$.  It turns out there is a range of $F$ where $m_{3/2}$ is small enough and the coupling is sufficiently suppressed that the NLSP decays to the gravitino with a macroscopic lifetime.  This region is commonly understood as Gauge Mediated SUSY breaking (GMSB), which is discussed in Section~\ref{sec:gaugemediation}.    More generally the concept can be extended to Sgoldstinos, the SUSY partner of the goldstino which is eaten by the gravitino in the super-Higgs mechanism (see Section~\ref{sec:sgoldstinos}).  While the identity of the sGoldstino can be more general than the GMSB version of the MSSM, the ultimate cause of the long-lifetimes and scalings are the same.

Within the MSSM alone, it is also possible that there are long-lived particles coming from a hierarchy amongst SUSY particles.  In particular if Gauginos are light and Sparticles are sufficiently heavy, then Gaugino decays are highly suppressed and have macroscopic lifetimes.   This idea was originally put forth in the context of Split-Supersymmetry at the expense of tuning.  However, in light of the Higgs discovery and other considerations the idea of Mini-Split supersymmetry has the same structure {\em and} Gaugino lifetimes in the MATHUSLA range (see Section~\ref{sec:minisplit}).

Given that as of yet there are no excesses attributable to SUSY at the LHC, there has been increased interest in extensions of the MSSM.  Nevertheless, these extensions have ubiquitous long-lifetime possibilities as well.  The most commonly known possibility is R-parity violation (RPV), discussed in Section~\ref{sec:rpv}.  To avoid the pitfalls of not-having a preserved R-parity, the dimensionless couplings of these operators are typically very small.  Therefore they naturally give long-lifetimes when SUSY particles decay through an RPV operator. 

Stealth SUSY (Section~\ref{sec:stealthsusy}) is a more recent extension of the MSSM designed to avoid LHC bounds. It introduces another sector that the MSSM superpartners can decay to and  which is approximately supersymmetric thus avoiding typical large MET signatures.  However, the decay in this sector can also use the same mechanism as gravitino decays in GMSB and long lifetimes are a part of the experimentally preferred region.

Finally, there are natural extensions of the MSSM built to address the strong CP problem using an axion (Section~\ref{sec:axino-ewino}).   These models will naturally have SUSY partners of the axion, in particular the Axino.  The Axino is a natural DM candidate to be the LSP.  However, the decays of other SUSY particles are suppressed through the PQ breaking scale.   Within the range of well-motivated PQ scales, the decay of SUSY particles to the Axino is also in the natural range of MATHUSLA.

\subsubsection[RPV Supersymmetry]{RPV Supersymmetry\footnote{Csaba Csaki, Eric Kuflik, Salvator Lombardo, Jared Evans, Brock Tweedie, Tim Cohen, Zhen Liu, Patrick Meade}}
\label{sec:rpv}

\begin{table}[t!]
 \begin{center}
\begin{tabular}{| c | c | c | }
\hline
\multicolumn{3}{ |c| }{Topologies} \\
\hline
LSP &  Decay & Operator \\ \hline
\multirow{3}{*}{$\tilde{t}$}  
 & $\bar{d}\, \bar{d}^\prime$ & $\lambda^{\prime\prime}$, $\eta^{\prime\prime}$   \\
 & $u\, \bar{\nu}$ & $\eta^\prime$   \\
 & $d\, \ell^+ $ & $\lambda^\prime$, $\eta$   \\ \hline
\multirow{3}{*}{$\tilde{g}$} 
 & $t \, d \, d^\prime+c.c$  &$\lambda^{\prime\prime}$, $\eta^{\prime\prime}$  \\
 & $t\,\bar{u}\, \bar{\nu}+c.c$ & $\eta^\prime$   \\
 & $t \bar{d}\, \ell^- +c.c$ & $\lambda^\prime$, $\eta$ \\ \hline 
\multirow{3}{*}{$\tilde{H}^0/\tilde{H}^\mp$} 
& $(t/b) \, d \, d^\prime+c.c$  & $\lambda^{\prime\prime}$, $\eta^{\prime\prime}$ \\
 & $(t/b)\,\bar{u}\, \bar{\nu}+c.c$ & $\eta^\prime$  \\
 & $(t/b)\, \bar{d}\, \ell^- +c.c$ &$\lambda^\prime$, $\eta$ \\ \hline
\end{tabular}
 \end{center}
\caption{Summary of various LSPs and their 
decay channels. The third column denotes the RPV operators from (\ref{eq:WRPV}) or (\ref{eq:KRPV}).\label{tab:summarychannels}}
\end{table}

\begin{figure}
\begin{center}
\begin{tabular}{cc}
\includegraphics[width=0.4\textwidth]{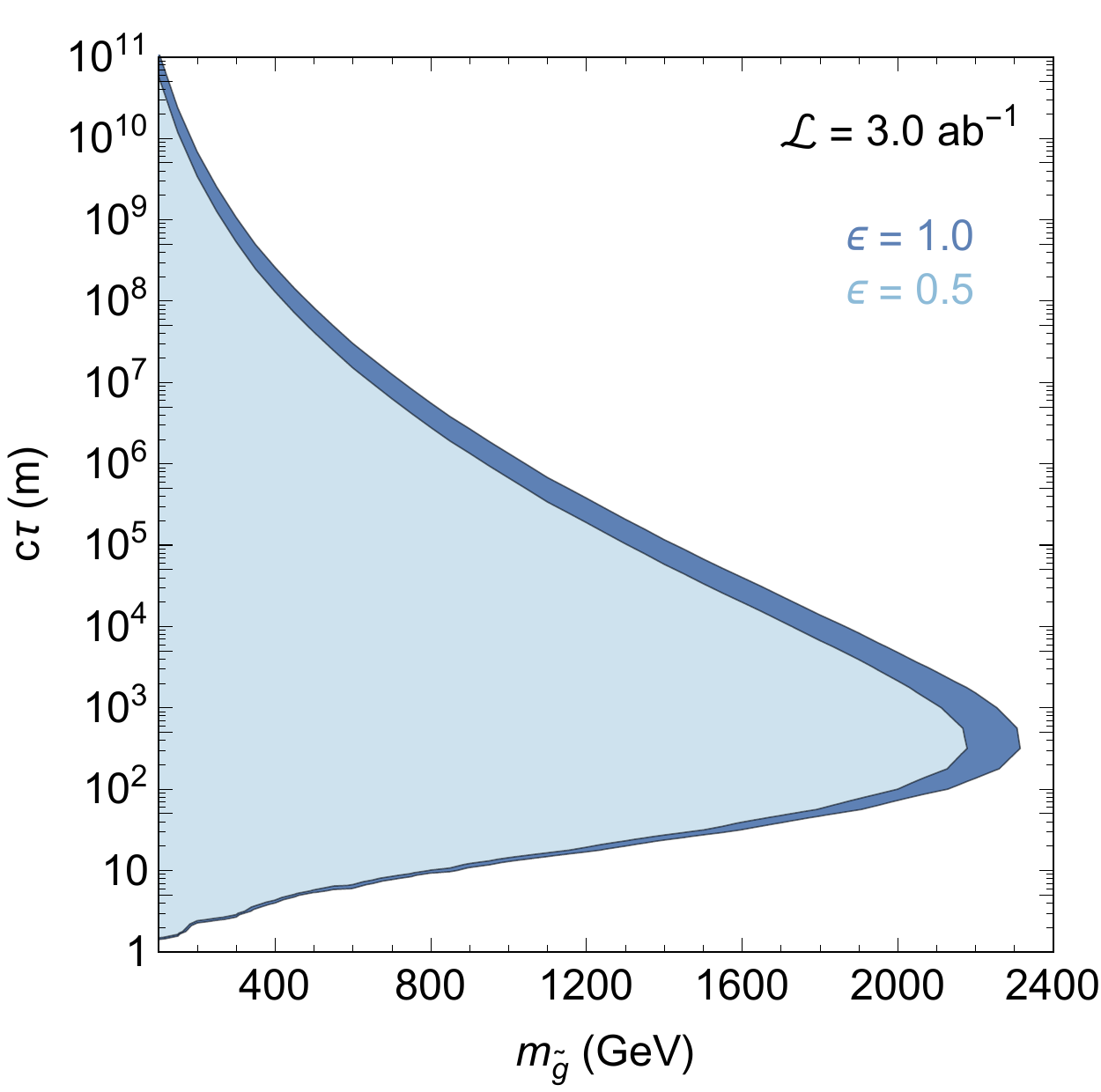}
&
\includegraphics[width=0.4\textwidth]{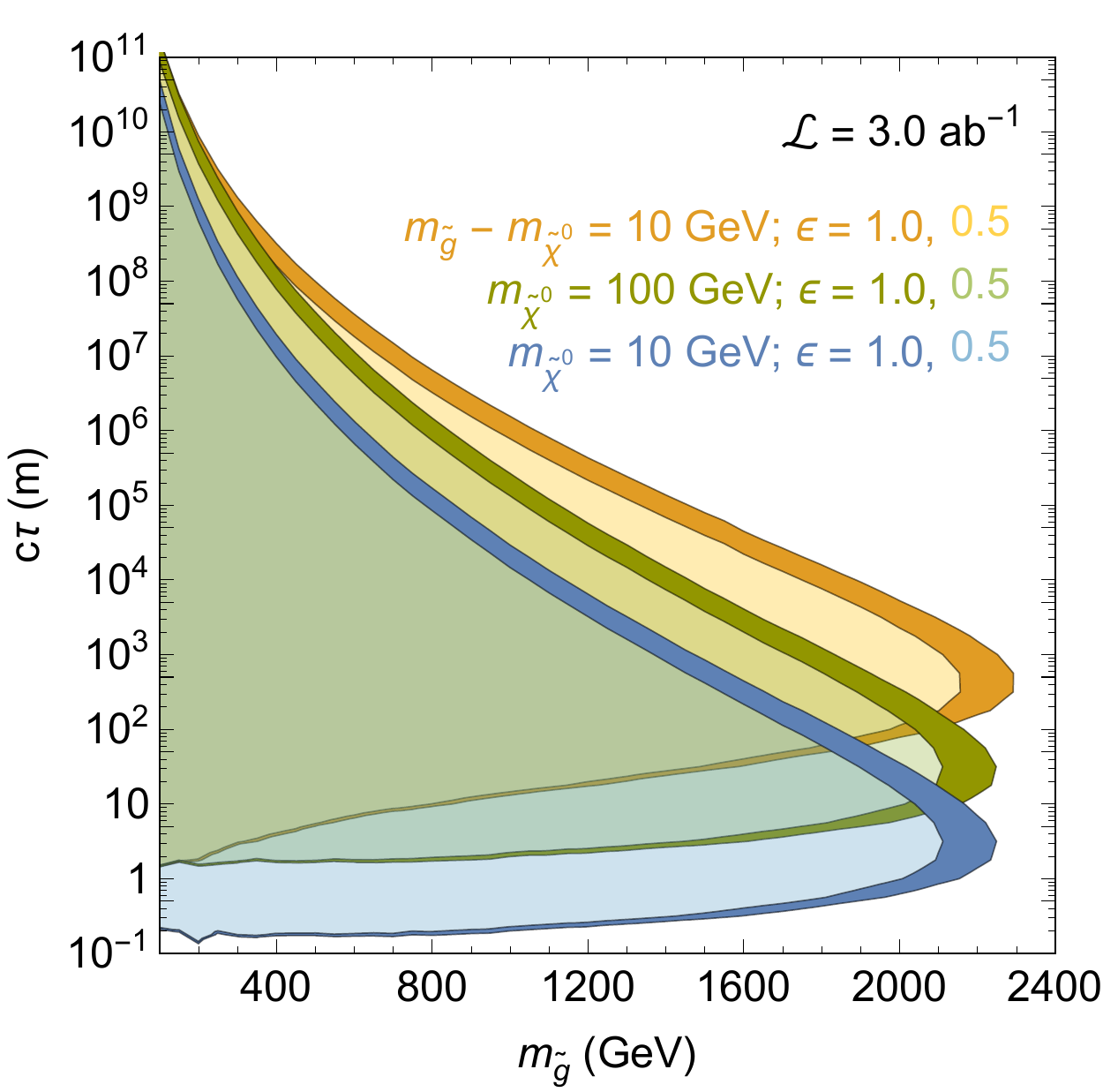}
\\
\includegraphics[width=0.4\textwidth]{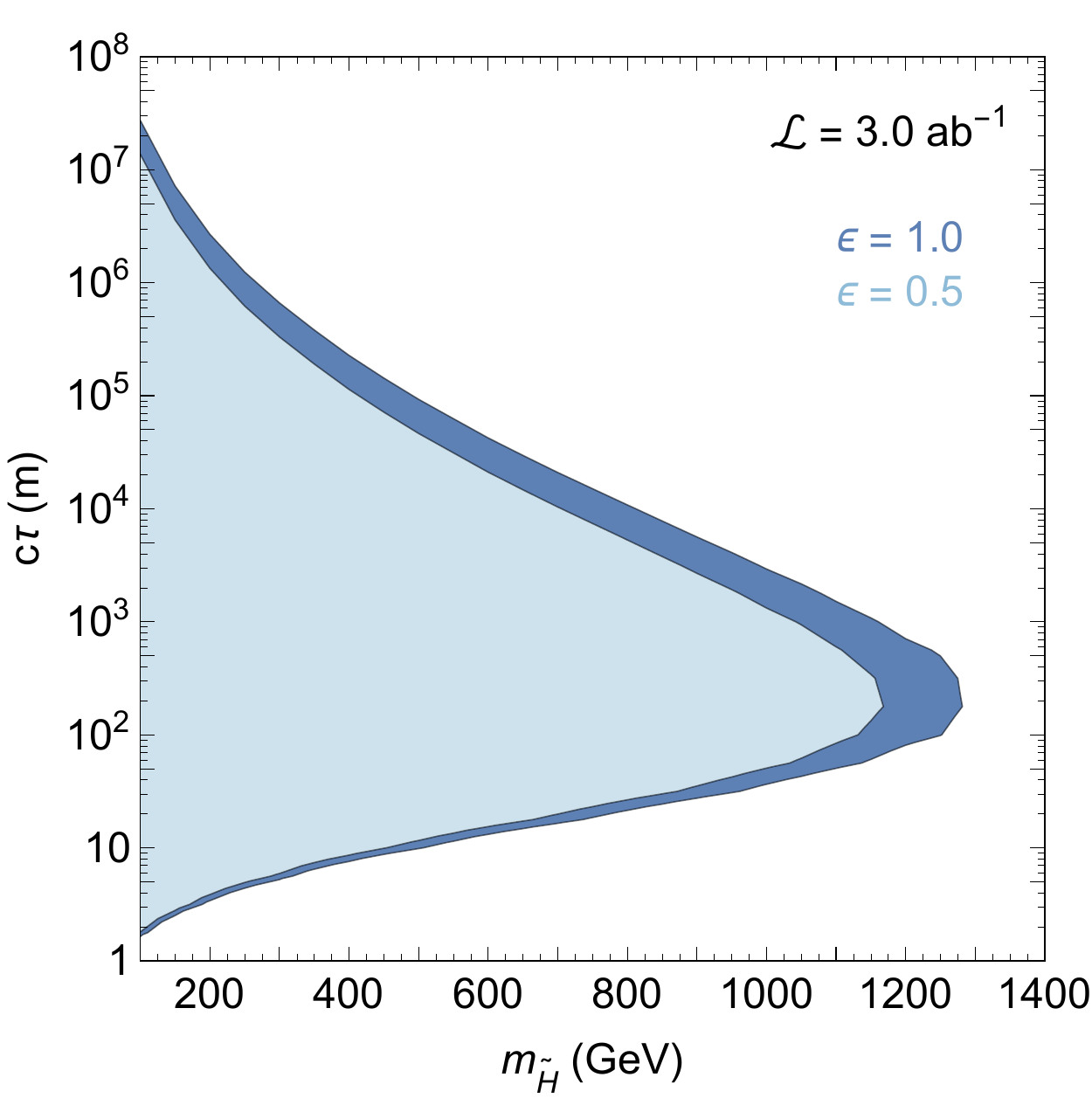}
&
\includegraphics[width=0.4\textwidth]{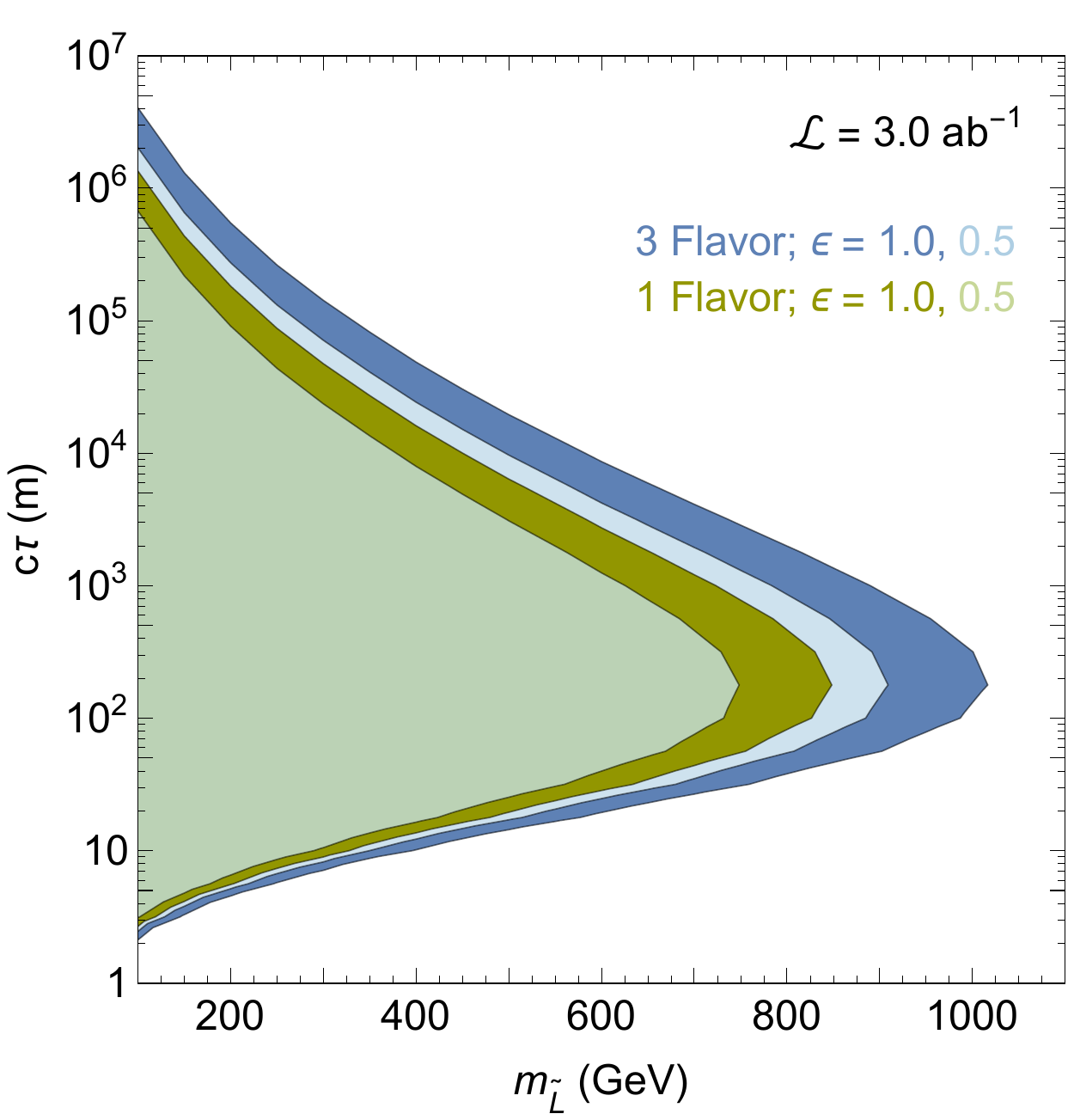}
\\
\includegraphics[width=0.4\textwidth]{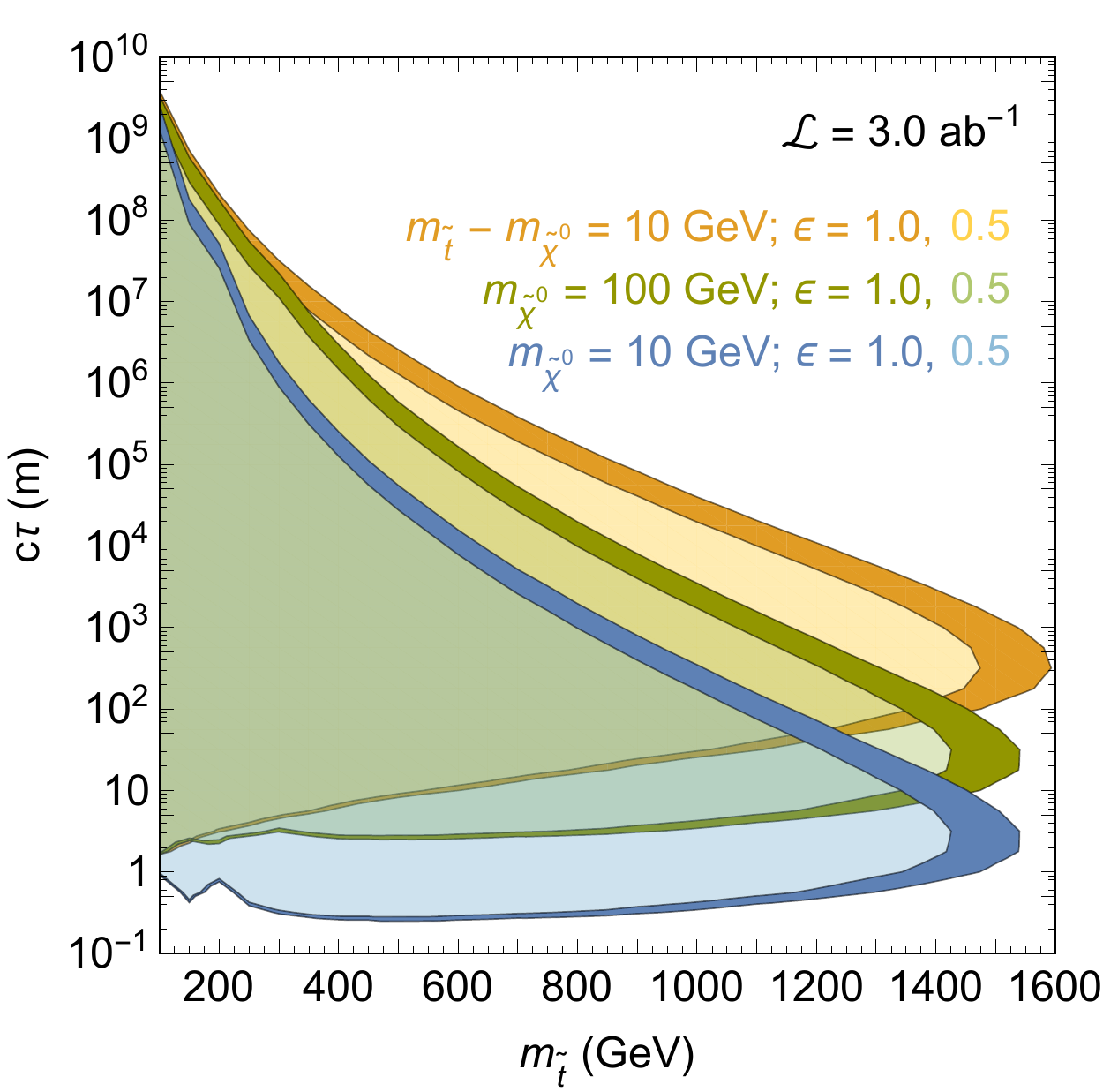}
&
\\
\end{tabular}
\end{center}
\caption{Projected MATHUSLA reach (assuming the $200m \times 200m \times 20m$ benchmark geometry of Fig.~\ref{f.mathuslalayout}) for several LSP and production scenarios with 3 ab$^{-1}$. We assume that the decay of the LSP leads to visible particles in the MATHUSLA detector. In order: direct gluino pair-production, prompt gluino pair decay to long-lived binos (for discrete bino mass choices), direct Higgsino pair-production, direct sneutrino pair-production, and prompt stop pair decays to long-lived binos (for discrete bino mass choices). The efficiency is varied from 0.5-1 (darker bands) in order to take into account the difference in efficiencies for each particular LSP decay mode.
}
\label{fig:RPVplots}
\end{figure}

Perhaps the most commonly studied BSM framework to stabilize the Higgs mass is the Minimal Supersymmetric Standard Model (MSSM). The MSSM predicts new scalar fields charged under SM gauge interactions, allowing one to write tree-level, renormalizable operators which violate baryon (B) and lepton (L) number. Since such flavor violation is highly constrained by low-energy experiments~\cite{Chemtob:2004xr,Barbier:2004ez}, the most minimal solution to the B and L constraints is to forbid such operators by imposing a global $Z_2$ symmetry known as R-parity, under which all SM particles carry charge +1 and all supersymmetric partners carry charge -1. In this case, the lightest supersymmetric particle (LSP) is pair-produced, stable, and escapes the detector leading to MET signatures at the LHC.

However, given the null results for LHC SUSY searches thus far, perhaps SUSY manifests itself non-minimally, and one should consider R-parity violating (RPV) interactions. If R-parity is violated, the source must be small since these operators are highly constrained. If this is the case, the LSP may decay to SM particles via a hierarchically small RPV interaction, motivating one to search for for LSPs with macroscopic lifetimes. For a review of RPV phenomenology, see~\cite{PhysRevD.40.2987, Dreiner:1997uz, Barbier:2004ez}.

The most general renormalizable Lagrangian allowing R-parity breaking is parametrized by the following superpotential written in terms of left-handed chiral superfields.

\begin{equation}
{\mathcal W}_{\rm RPV} =  \mu_i \ell_i h_u + \lambda_{ijk}  \ell_i\ell_j\bar{e}_k +\lambda^\prime_{ijk}  \ell_i q_j \bar{d}_k+\lambda^{\prime\prime}_{ijk} \bar{u}_i\bar{d}_j\bar{d}_k
\label{eq:WRPV}
\end{equation}
RPV interactions are usually parametrized by $\mu_i$, $\lambda_{i,j,k}$, $\lambda_{i,j,k}'$, and $\lambda_{i,j,k}''$. The $\lambda''$ operator together with $\lambda'$ lead to proton decay, while the $\mu$ term mixes neutrinos with the higgsinos and gauginos. 

Depending on the UV completion, the usual RPV operators in Eq.~\ref{eq:WRPV} may not be the most important ones. For example R-parity may be a good symmetry in the UV of the visible sector but may be broken spontaneously by a SUSY-breaking sector and mediated to the visible sector dynamically~\cite{Csaki:2013jza, Csaki:2015fea}. In this case, the dominant RPV operators could arise from the Kahler potential, allowing for the following non-holomorphic operators. 
\begin{equation}
{\cal K}_{\rm nhRPV} = \frac{1}{M} \left( \kappa_i^\prime \ell_i^* h_D +
 \eta_{ijk} \bar{u}_i\bar{e}_j\bar{d}_k^* + \eta^\prime_{ijk} q_i\bar{u}_j \ell_k^* +\eta^{\prime\prime}_{ijk} q_i q_j \bar{d}_k^* \right)
\label{eq:KRPV}
\end{equation}
Here the RPV interactions are suppressed by the messenger scale $M$, explaining the small size of the interactions. 

There are also cosmological considerations depending on the type of RPV operators that are active for the LLP decays.  To avoid most searches at the LHC it is easier to have B violating operators which are potential hidden within QCD backgrounds.  However, there is an interplay between introducing B violation and the baryon asymmetry that we observe in the universe.  In particular, if the B violating RPV operator has a large coefficient, then its interactions could destroy any initial baryon asymmetry created through a standard baryogenesis mechanism.  Whether the baryon asymmetry is washed out depends upon the reheat temperature which governs whether the interactions of the SM with SUSY particles are in equilibrium.  Nevertheless, it is quite natural to expect long-lived LSPs when there is RPV from cosmological considerations~\cite{Barry:2013nva}.

The long-lived LSP has rich collider phenomenology since any of the superpartners could in principle be the LSP, and the largest RPV coupling determines the dominant final state of the LSP decay. Recently, the phenomenology and existing constraints on long-lived LSP scenarios was studied in detail~\cite{Csaki:2015uza, Liu:2015bma}. Motivated by naturalness, a summary of possible decay channels for stop, gluino, or higgsino LSP is provided in  Table~\ref{tab:summarychannels}. Assuming the long-lived LSP decays to visible particles, we project the MATHUSLA reach for 3 ab$^{-1}$ in Fig.~\ref{fig:RPVplots} assuming a particular LSP and production channel (without assuming a particular LSP decay mode) for several scenarios including direct gluino pair-production, gluino pair decays to long-lived binos, direct Higgsino pair-production, direct sneutrino pair production, and pair-produced stops decaying to long-lived binos. The efficiency is varied from 0.5-1 in order to take into account the varying reconstruction efficiency for different LSP decay modes.

MATHUSLA can place strong constraints neutral LSPs which are competitive with MET searches performed by the LHC in the long lifetime limit. 
Even if a MET signal is observed at the LHC, the lifetime of the LSP can be probed by late decays in the MATHUSLA detector. 
Furthermore, there are regions of parameter space in which the MET in the event is suppressed, $\textit{e.g.}$ LSPs with mass below a few hundred GeV or scenarios with a compressed spectrum, where MATHUSLA can have more discovery potential than LHC MET searches. 
As argued in Section.~\ref{sec:LLPSmathuslahllhc}, MATHUSLA would also have a significant advantage over LLP searches at the main detectors, especially for LLPs decaying hadronically with masses below a few 100 GeV.
On the other hand, if the LSP is colored ($\textit{e.g.}$ a squark or gluino) and has a lifetime longer than $\Lambda_{QCD}^{-1}$, it hadronizes to form an R-hadron. LHC searches for heavy stable charged particles ($\textit{e.g.}$~\cite{Aaboud:2016dgf,Khachatryan:2016sfv}) are sensitive to these scenarios. In this case, MATHUSLA can be complementary to these searches and can provide lifetime information of the R-hadron.

\subsubsection[Gauge Mediation]{Gauge Mediation\footnote{Matthew Reece}}
\label{sec:gaugemediation}

In gauge mediated supersymmetry breaking (GMSB), Standard Model superpartners decay with a potentially long lifetime to a much lighter gravitino $\widetilde G$. A classic review of this scenario is \cite{Giudice:1998bp}; a recent update on the experimental status is \cite{Knapen:2016exe}. Here we will not assume gauge mediation in the strict sense \cite{Meade:2008wd}, but any theory of low-scale SUSY breaking with a light gravitino, which could (for example) include Yukawa mediation effects as well \cite{Chacko:2001km,Evans:2013kxa}. An important parameter dictating the long-lifetime phenomenology is the order parameter $F_0$ for supersymmetry breaking, which is related to the gravitino mass via
\begin{equation}
m_{3/2} = \frac{F_0}{\sqrt{3} M_{\rm Pl}} \approx 1~{\rm keV}\, \left(\frac{\sqrt{F_0}}{2000~{\rm TeV}}\right)^2,
\end{equation}
where $M_{\rm Pl} \approx 2.4 \times 10^{18}~{\rm GeV}$ is the reduced Planck mass and $F_0  \equiv \sqrt{3} \langle e^{K/(2 M_{\rm Pl}^2)} W\rangle/M_{\rm Pl}$ is determined by the full set of $F$-term VEVs of the theory. Standard Model superpartners acquire masses set by loop factors multiplying $F/M_{\rm mess} = k F_0/M_{\rm mess}$ where $F$ is the dominant source of SUSY breaking mediated to the Standard Model, $k$ is a model-dependent parameter satisfying $k \lesssim 1$, and $M_{\rm mess} \ll M_{\rm Pl}$ is the mass scale of ``messengers'' of SUSY breaking. The lightest MSSM superpartner, often called the LOSP (Lightest Ordinary SuperPartner), decays to Standard Model matter and the gravitino via higher-dimension operators that are suppressed by $1/F_0$. For example, for a neutralino LOSP, the lifetime is \cite{Ambrosanio:1996jn}
\begin{align}
c\tau({\widetilde \chi}^0_1 \to {\widetilde G} + {\rm SM}) &= \frac{16 \pi F_0^2}{\xi m_\chi^5} \approx 100~{\rm m}\, \frac{1}{\xi} \left(\frac{\sqrt{F_0}}{10^7~{\rm GeV}}\right)^4 \left(\frac{250~{\rm GeV}}{m_\chi}\right)^5, 
\end{align}
with $\xi$ a constant depending on the neutralino mixing matrix and $m_\chi$ denoting the mass of the LOSP ${\widetilde \chi}^0_1$. For example, for a pure bino, $\xi_{\rm bino} = c_W^2 + s_W^2 \left(1 - \frac{m_Z^2}{m_\chi^2}\right)^4$; for a pure higgsino, 
\begin{equation}
\xi_{\rm higgsino} = \frac{1}{4} \left[\left(s_\beta + {\rm sgn}(\mu) c_\beta\right)^2 \left(1 - \frac{m_Z^2}{m_\chi^2}\right)^4 + \left(s_\beta - {\rm sgn}(\mu) c_\beta\right)^2 \left(1 - \frac{m_h^2}{m_\chi^2}\right)^4\right].
\end{equation}
The large phase space suppression factors when $m_\chi \approx m_Z, m_h$ arise because the goldstino is derivatively coupled. The range of $F_0$ to consider ranges from low-scale SUSY breaking with $F_0 \sim 10^5~{\rm GeV}$ up to the intermediate scale $F_0 \sim 10^{10}~{\rm GeV}$, where gravity-mediated effects become dominant. Across this range, the LOSP lifetime varies from prompt to macroscopic. 

\begin{figure}
\begin{center}
\includegraphics[width=1.0\textwidth]{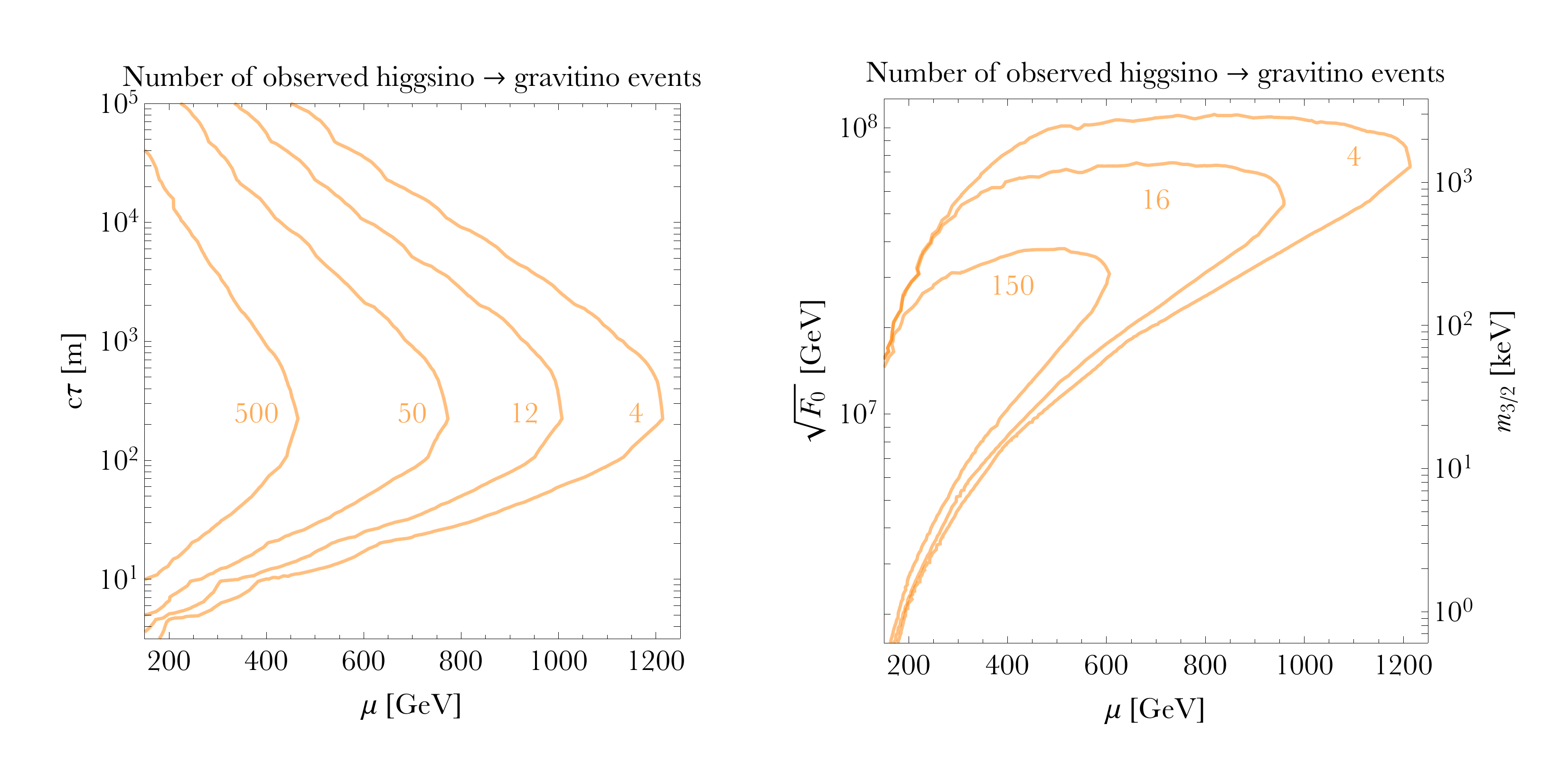}
\end{center}
\caption{Number of ${\widetilde H} \to {\widetilde G} + (Z,h)$ events that MATHUSLA would detect from electroweak production of higgsinos at the LHC operating at $\sqrt{s} = 14~{\rm TeV}$ with an integrated luminosity of 3 ab$^{-1}$. Left: higgsino mass $\mu$ versus lifetime $c\tau$ in meters. Right: higgsino mass $\mu$ versus the SUSY breaking scale as parametrized by $\sqrt{F}$ in GeV (label on left axis) or $m_{3/2}$ in keV (label on right axis). In a wide swath of parameter space with LOSP lifetimes ranging from $10$ to $10^5~{\rm m}$, MATHUSLA could provide a discovery of new physics with electroweak cross-sections for which the LHC would fail to discover new physics. This estimate assumes the $200m \times 200m \times 20m$ benchmark geometry of Fig.~\ref{f.mathuslalayout}.}
\label{fig:GMSBhiggsinoreach}
\end{figure}

In \figref{GMSBhiggsinoreach}, we show the number of events that MATHUSLA would detect for electroweak higgsino production at the LHC with a subsequent ${\widetilde H}^0_1 \to (h, Z) + {\widetilde G}$ decay in MATHUSLA. Kinematics of the higgsino events and leading order cross-sections were calculated using Pythia 8 \cite{Sjostrand:2014zea}. Several points are noteworthy. First, the number of events observed at MATHUSLA could range up to several hundred, even for an electroweak production process with a relatively small cross-section. For $c\tau \sim 100~{\rm m}$, MATHUSLA could exclude higgsino masses up to above 1 TeV. This may be compared to the LHC reach for higgsino production with a detector-stable higgsino LSP, which would not reach beyond masses of about 200 GeV even with optimistic assumptions about systematics \cite{Low:2014cba, Ismail:2016zby}. Furthermore, because the $Z$ and $h$ dominantly decay hadronically, HL-LHC searches for displaced higgsino decays using the main detectors will be less sensitive than MATHUSLA, especially for less than several hundred GeV of jet energy per decay (see Section \ref{s.LHCLLPcomparison} and also Section~\ref{sec:axino-ewino} for a discussion of this signal). It is well-established that measuring energies and angular distributions for macroscopically displaced decay events {\em inside} the LHC could provide a powerful handle on the underlying nature of GMSB physics, such as the mass and identity (e.g.~wino or higgsino) of the LOSP \cite{Kawagoe:2003jv,Meade:2010ji,Park:2011vw}. A similar conclusion should be true of MATHUSLA as well: assuming a particular production mode, the mass and decay mode of the higgsino could be determined using the track geometry and multiplicity of the daughter products \cite{Curtin:2017izq}. Complementary information from the HL-LHC main detectors could also help to unambiguously determine the properties of the higgsino. Production of strongly interacting particles (squarks or gluinos) that cascade down to LOSPs could lead to larger cross-sections and a greater discovery potential (cf.~Section \ref{s.benchmarksignalxsecs}), in parameter space where new physics might first be spotted using the LHC main detectors (see the comparison of MATHUSLA to MET searches in Section \ref{s.METcomparison}). 

The right-hand panel of \figref{GMSBhiggsinoreach} shows the number of observed events at MATHUSLA plotted against $\sqrt{F_0}$ and $m_{3/2}$ on the left and right axes respectively. We see that MATHUSLA is sensitive to $\sqrt{F_0} \sim 10^6~{\rm to}~10^8~{\rm GeV}$ and $m_{3/2} \sim 1~{\rm keV}~{\rm to}~1~{\rm MeV}$. This significantly extends the LHC's GMSB discovery potential toward larger SUSY-breaking scales.

\subsubsection[Mini-Split Supersymmetry]{Mini-Split Supersymmetry\footnote{Tim Cohen}}
\label{sec:minisplit}

\label{sec:minisplit}
It is possible that nature could be fundamentally supersymmetric, but the weak scale could still be highly fine tuned.\footnote{Perhaps this tuning can be alleviated by additional physics, \emph{e.g.}~\cite{Batell:2015fma} in the case of a relaxion extension.} This paradigm is  known as Split SUSY~\cite{Wells:2004di, ArkaniHamed:2004fb, Giudice:2004tc, ArkaniHamed:2004yi}.  Phenomenologically, this is motivated by the notion that SUSY would still explain dark matter via relic neutralinos while additionally accommodating gauge coupling unification.  Furthermore, the $R$-symmetry implicit to the MSSM provides a rationale for superpartner fermion masses to be parametrically lighter than the scalar masses -- this spectrum can be additionally motivated by models where the SUSY breaking is communicated to the scalar sector via gravity mediation, and the gauginos via anomaly mediation~\cite{Randall:1998uk, Giudice:1998xp} such that there is a loop factor difference between the scalar masses and gaugino mass~\cite{Randall:1998uk, Giudice:1998xp, Ibe:2012hu, Acharya:2012tw} (this hierarchy can also occur in models of gauge mediation, \emph{e.g.}~\cite{Cohen:2015lyp}).  

One of the interesting consequences of Split SUSY (as applied to the MSSM) is that the Higgs mass is predicted to lie within a finite range, with an upper bound given by $\sim 140$ GeV~\cite{Arvanitaki:2004eu}.  The discovery of the Higgs at 126 GeV, along with the lack of any BSM discoveries thus far, has reinvigorated interest in this paradigm, which has been re-coined Mini-split SUSY~\cite{Arvanitaki:2012ps} due to the fact that the scalar superpartners cannot have their masses $M_\text{SC}$ arbitrarily close to the Planck scale~\cite{Giudice:2011cg, Arvanitaki:2012ps, ArkaniHamed:2012gw, Bagnaschi:2014rsa, Vega:2015fna}, see  Fig.~\ref{fig:MiniSplitHiggsMass}.

\begin{figure}[h!]
\begin{center}
\includegraphics[width=0.5\linewidth]{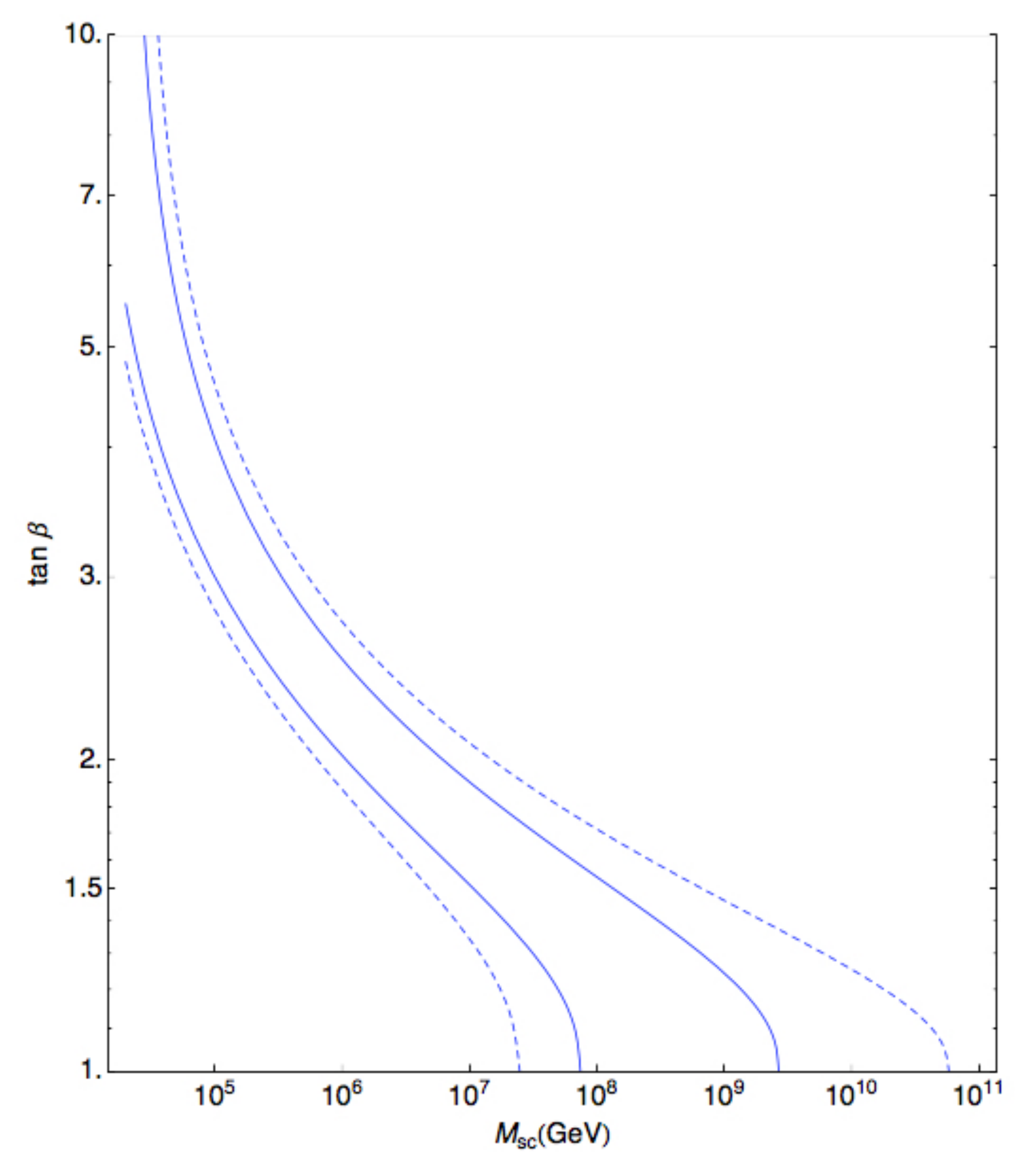}
\caption{The SM-like Higgs boson mass contours in the $\tan\beta$ versus $M_\text{SC}$ plane.  The parameter $M_\text{SC}$ is the common mass scale of the scalar superpartners.  This figure is taken from~\cite{ArkaniHamed:2012gw}.}
\label{fig:MiniSplitHiggsMass}
\end{center}
\end{figure}

The only particles that are expected to be collider accessible for Mini-Split models are the gauginos.\footnote{In principle the higgsinos could also be light enough to be collider accessible.  For our purposes here, we will assume that the Higgsino mass scale is set by SUSY breaking in the context of the Giudice-Masiero mechanism~\cite{Giudice:1988yz}.  This is in keeping with the desire to motivate Mini-split SUSY using minimality~\cite{ArkaniHamed:2012gw}.}  As will be emphasized in the next paragraph, it is straightforward to find parameter space where the gluino lifetime is long, making it a natural target for MATHUSLA.  For contrast, the electroweak gauginos all decay via the weak interactions, which do not lead to long enough lifetimes to be of interest without extremely small splittings.  This case will be discussed briefly towards the end of this section.

The gluino decays via a higher dimension operator suppressed by inverse powers of the squark masses~\cite{Gambino:2005eh}:
\begin{align}
\mathcal{O}_{\text{decay}}^{(6)} \sim \frac{g_s^2}{m_{\tilde{q}}^2} \bar{q} \tilde{g} \bar{\tilde{\chi}} q\qquad \mathcal{O}_{\text{decay}}^{(5)} \sim \frac{g_s^2}{16\pi^2 m_{\tilde{q}}} \tilde{g} \sigma_{\mu\nu} \tilde{\chi} G^{\mu\nu}\,,
\end{align}
where the superscript on the operator corresponds to the mass dimension.  The dimension-6 operator comes from the tree-level exchange of an off-shell squark, while the dimension-5 operator comes from a one-loop diagram.  
\begin{figure}[h!]
\begin{center}
\includegraphics[width=0.5\linewidth]{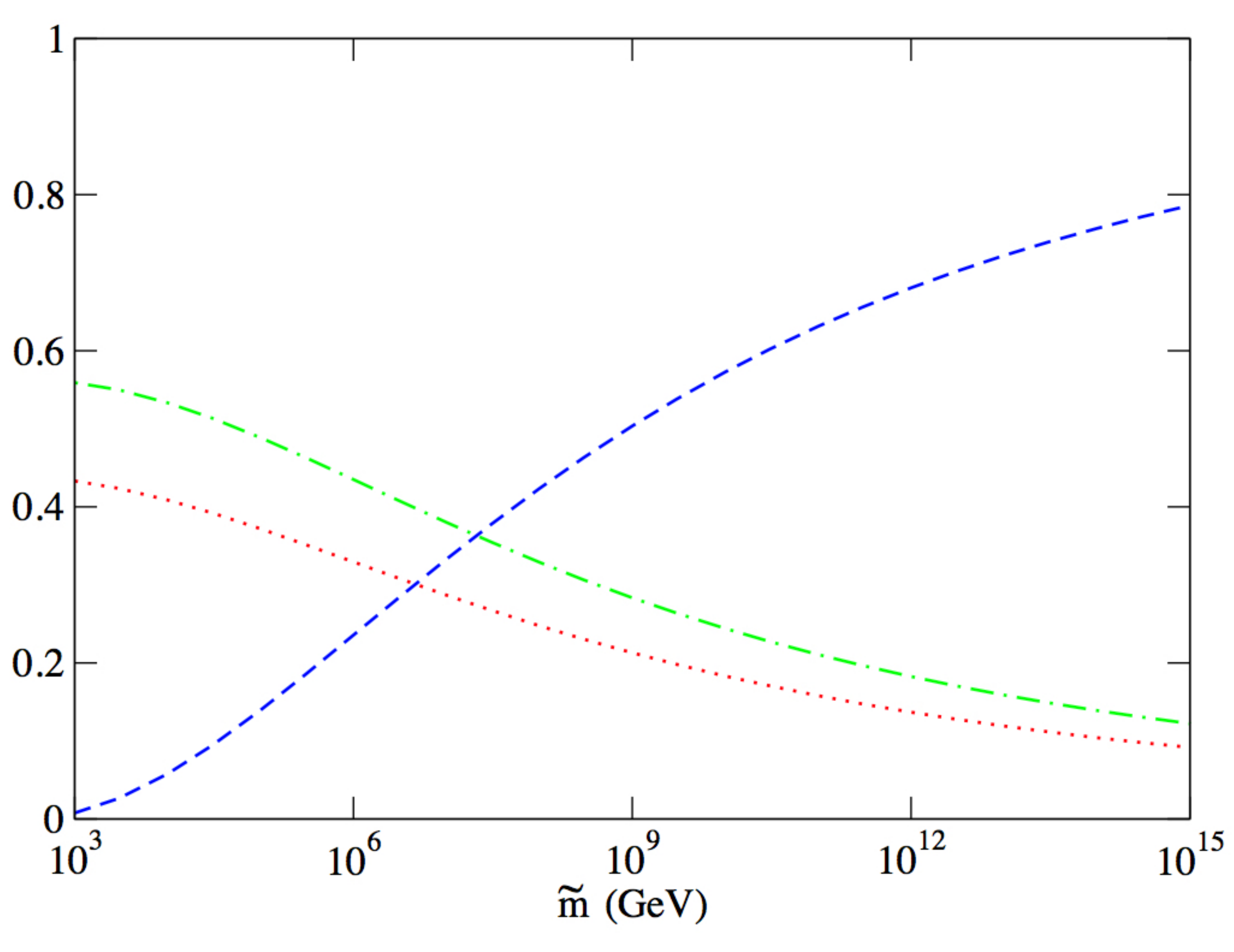}
\caption{
Branching ratios for the gluino decay channels $\tilde{g} \rightarrow \tilde{\chi} g$ (dashed lines), $\tilde{g} \rightarrow \tilde{\chi} q\,\bar{q}$ (dotted) and $\tilde{g} \rightarrow \tilde{\chi}^\pm q\,\bar{q}'$(dot-dashed), summed over all possible neutralino or chargino states, as a function of $m_{\tilde{q}}$ and with $m_{\tilde{g}} = 1 \text{ TeV}$.  This figure is taken from~\cite{Gambino:2005eh}.
}
\label{fig:MiniSplitGluinoDecays}
\end{center}
\end{figure}

The gluino pair production cross-section is provided in Sec.~\ref{sec:LLPSmathuslahllhc}.  For reference, we provide an estimate of the lifetime using just the dimension-6 operator:
\begin{align}
c\tau \sim 100 \text{ m } \left(\frac{m_{\tilde{q}}}{50\text{ PeV}}\right)^4 \left(\frac{\text{TeV}}{m_{\tilde{g}}}\right)
\end{align}
Comparing this estimate to Fig.~\ref{fig:MiniSplitHiggsMass}, we conclude that there is ample room for this model to be consistent and observable by MATHUSLA.  There is clearly a competition between the two decay modes, as illustrated in Fig.~\ref{fig:MiniSplitGluinoDecays}.  Furthermore, there are variations due to both the mass spectrum of the squarks (which effects which flavor dominates in the gluino decays) and the mass spectrum of the gauginos.  For example, motivated by the renormalization group evolution of the squark masses, and taking the gaugino mass ordering to be $m_{\tilde{g}} > m_{\tilde{W}} > m_{\tilde{B}}$, the decay $\tilde{g} \rightarrow t\,\bar{t}\,t\,\bar{t} \,h\,h\,\tilde{\chi}\,\tilde{\chi}$ could dominate~\cite{ArkaniHamed:2012gw}.

Clearly the different channels involve search strategies for prompt searches at the LHC, which is not the focus of the study here.  In order to estimate the reach for MATHUSLA, one must account for the interactions of the long-lived gluinos with matter, both to estimate the fraction of $R$-hadrons that are neutral, along with the energy one would expect them to have when they reach the MATHUSLA detector -- for a discussion of these issues see Sec.~\ref{sec:LLPSmathuslahllhc}.  Then as long as the gluino decay products are energetic enough to light up the MATHUSLA scintillators, then there should be a detectible signal regardless of final state.  The reach for a long-lived gluino is given in Fig.~\ref{fig:RPVplots} and is in excess of 2 TeV, providing a complementary discovery channel to the HL-LHC main detectors.
It is also worth noting that if we are in the exciting situation where a long-lived gluino has been discovered at ATLAS and CMS, MATHUSLA would also provide valuable information about the properties of $R$-hadron interactions with matter.

Finally, we note that there is a region of parameter space where the electroweak gauginos could be long-lived enough to be detectable at MATHUSLA~\cite{Han:2014xoa, Nagata:2015pra}.  This can be additionally motivated by an interest in testing the bino-wino coannihilation region~\cite{Han:2014xoa, Nagata:2015pra}.  The candidate long-lived state is a nearly pure neutral wino, which dominantly decays to the bino and either a photon or an off-shell Higgs, see Fig.~\ref{fig:MiniSplitEWInoDecays} for a schematic of the relevant decay modes.  For the version of Mini-split SUSY of interest here, where the Higgsino mass and scalar masses are all set by gravity mediated physics, a lifetime relevant for MATHUSLA can be achieved for $\mu \sim 100 \text{ TeV}$~\cite{Nagata:2015pra}.  Consistency with the measured Higgs boson mass then requires $\tan \beta \sim 1$.  Exploring the detailed implications for the parameter space that could be probed at MATHUSLA, along with correlating these projections with the Higgs boson mass and relic density is an interesting subject that deserves further study. However, we expect MATHUSLA to significantly extend the sensitivity of the main detectors for these scenarios, for identical reasons as for the Higgsino LLPs discussed in Sections.~\ref{sec:gaugemediation} and \ref{sec:axino-ewino}.

\begin{figure}[h!]
\begin{center}
\includegraphics[width=0.7\linewidth]{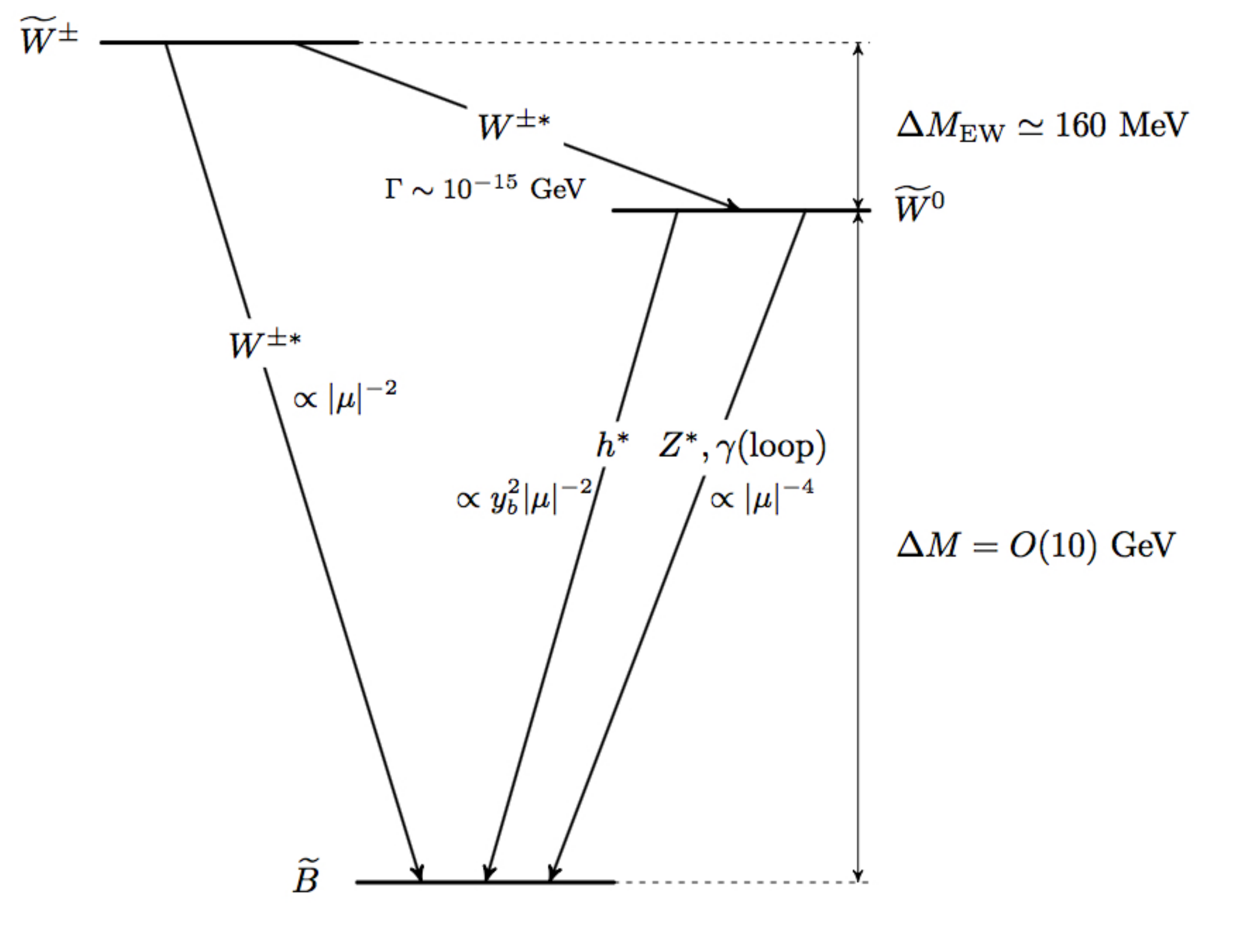}
\caption{An illustration of the possible decay patterns for long-lived electroweak gauginos in mini-Split SUSY scenarios.  This figure is taken from~\cite{Nagata:2015pra}. The very long-lived particle visible at MATHUSLA would be $\tilde W_0$ for $\mu \gtrsim 100 \tev$.}
\label{fig:MiniSplitEWInoDecays}
\end{center}
\end{figure}

\subsubsection[Stealth Supersymmetry]{Stealth Supersymmetry\footnote{Matthew Reece and David Pinner}}
\label{sec:stealthsusy}

Stealth Supersymmetry is a scenario in which collider signals of supersymmetry involve very little missing momentum. It is thus less constrained than more typical models of supersymmetry, given stringent LHC bounds on the rate of events with large missing momentum. Stealth SUSY phenomenology relies on the existence of particles that are nearly degenerate with their superpartners, naturally squeezing the phase space available for invisible particles \cite{Fan:2011yu}. The key decay step in any Stealth Supersymmetry model has the form
\begin{equation}
\widetilde{X} \to X + {\widetilde N}, \quad X \to {\rm SM~particles}.
\end{equation}
Here $\widetilde X$ and $\widetilde N$ are $R$-parity odd, and $m_{\widetilde N} \ll m_{\widetilde X} \approx m_X$. The particle $\widetilde N$ is neutral and escapes the detector, carrying away missing momentum, but due to the kinematics of the decay it is necessarily soft (even in the presence of hard initial-state radiation). In order for the stealth mechanism to be operative, this decay must occur inside the detector; otherwise, $\widetilde{X}$ contributes a large missing momentum. In the simplest realization of Stealth SUSY, the neutral particle $\widetilde N$ is the gravitino $\widetilde G$. However, this scenario is under some strain from searches for displaced decays at colliders: if $m_{\widetilde X} \approx 100~{\rm GeV}$, $\delta m = m_{\widetilde X} - m_X \approx 10~{\rm GeV}$, and $\sqrt{F_0} \approx 100~{\rm TeV}$, the decay length of ${\widetilde X} \to X {\widetilde G}$ is about 8 cm \cite{Fan:2011yu}. Such macroscopic lifetimes will be in tension with a range of inclusive searches for long-lived particles that have already been carried out at ATLAS and CMS \cite{CMS:2014wda, CMS:2014hka, Aad:2015asa, Aad:2015uaa, Aad:2015rba}. Although these searches have not been specifically recast as constraints on Stealth SUSY, they are known to strongly constrain a wide variety of R-parity violating SUSY scenarios with broadly similar kinematics \cite{Liu:2015bma, Csaki:2015uza,Zwane:2015bra}. This disfavors the Stealth SUSY scenario with $\widetilde{X} \to X \widetilde{G}$ decay. Although we could consider smaller $\sqrt{F_0} \approx 10~{\rm TeV}$, it is challenging to build a model of SUSY breaking that operates at such a low scale. For instance, two recent attempts to build models of very low-scale SUSY breaking achieve $m_{3/2} \sim 1~{\rm eV}$, with $\sqrt{F_0} \gtrsim 65~{\rm TeV}$ \cite{Hook:2015tra,Ibe:2016kyg}. On the other hand, values of $\sqrt{F_0}$ below $\sim 140~{\rm TeV}$ are favored by cosmological constraints from CMB lensing and cosmic shear \cite{Osato:2016ixc}. Thus the preferred range of $\sqrt{F_0}$ is around, but perhaps modestly smaller than, 100 TeV.

The theoretically simplest alternative decay is the case where $\widetilde N$ is an {\em axino} $\widetilde a$ \cite{Fan:2012jf}. This is also known as a Goldstone fermion (not to be confused with a goldstino): that is, $\widetilde a$ is the supersymmetric partner of a (pseudo-)Goldstone boson from a symmetry that is broken in an approximately supersymmetric manner. In this case, $\widetilde a$ can enjoy derivative couplings to ${\widetilde X}, X$, allowing for the desired decay ${\widetilde X} \to X {\widetilde a}$; for some range of symmetry-breaking scales, this decay is prompt enough on collider timescales to evade the ATLAS and CMS searches for displaced vertices. In that case, the bounds on natural realizations of Stealth SUSY can still be significantly weaker than bounds on standard SUSY scenarios \cite{Fan:2015mxp}.

\begin{figure}[!h]
\begin{center}
\includegraphics[width=0.65\textwidth]{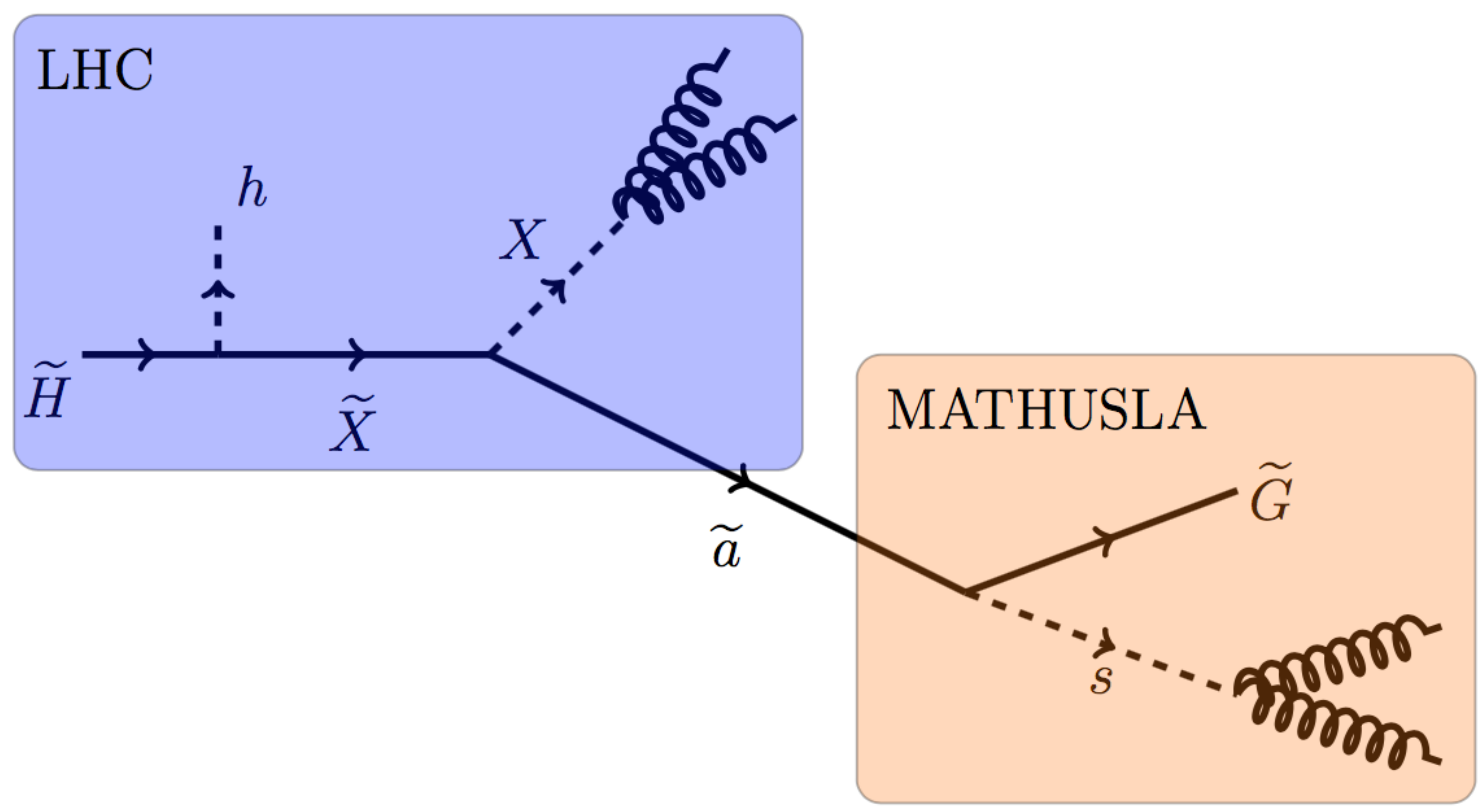}
\end{center}
\caption{A full Stealth SUSY decay chain. It is crucial that the ${\widetilde X} \to X {\widetilde a}$ decay step happen promptly inside the LHC, to evade missing momentum and displaced vertex searches. Significantly later, the axino decay can produce a saxion which in turn can produce visible Standard Model particles inside MATHUSLA.}
\label{fig:stealthdecaychain}
\end{figure}

\begin{figure}[!h]
\begin{center}
\includegraphics[width=0.65\textwidth]{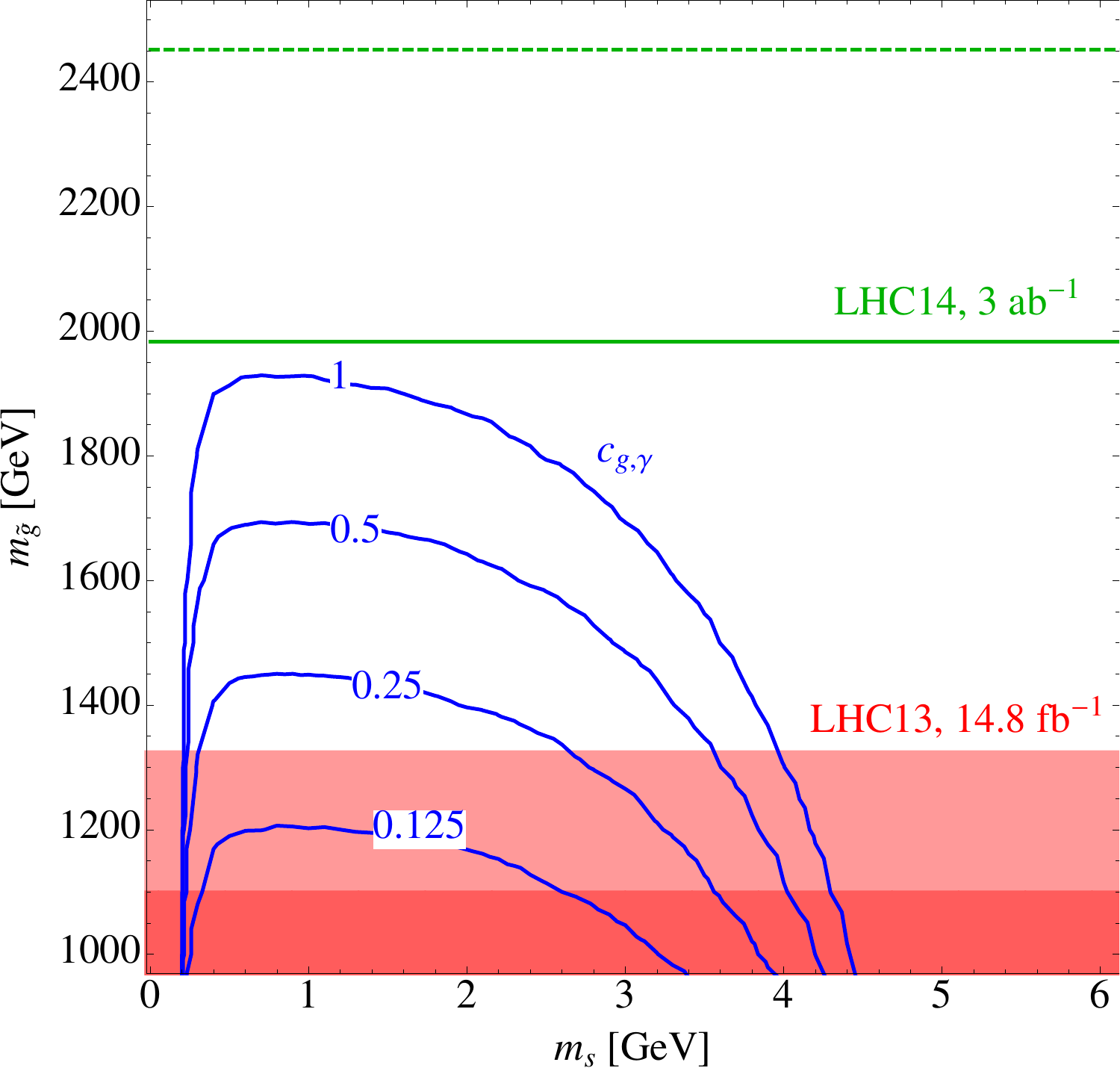}
\end{center}
\caption{MATHUSLA discovery potential for Stealth SUSY. The signal arises in gluino pair production events (gluino mass is on the vertical axis), and the observed decay is a light saxion (mass on the horizontal axis) decaying to two gluons or photons. Blue curves are the reach (corresponding to 4 detected events) for various values of $c_g$ and $c_\gamma$ defined in equation (\ref{eq:stealthcouplingdefinition}).  As the saxion mass approaches the axino mass (fixed here to 5~GeV), the axino begins to decay invisibly to an axion and a gravitino.  Conversely, as the saxion mass falls below the two-pion threshold, the visible branching fraction of the saxion quickly becomes negligible, with the decay proceeding predominantly to a pair of axions.  The red shaded regions show the current LHC exclusions for gluinos decaying to singlinos from the ATLAS RPV search \cite{ATLAS:2016nij}. The projected discovery reach at HL-LHC with 3 \iab is shown in green. The dashed green line and lighter red region are estimated discovery and exclusion curves, respectively, with the solid green line and darker red region corresponding to an efficiency reduced by 50\%, shown here as a rough guide to possible systematic error in recasting. Note that even in the event of a positive signal at the HL-LHC, MATHUSLA would be required to discover the long-lived axino produced in the event.
This estimate assumes the $200 \,\mathrm{m} \times 200 \,\mathrm{m} \times 20 \,\mathrm{m}$ benchmark geometry of Fig.~\ref{f.mathuslalayout}.
}
\label{fig:stealthreach}
\end{figure}

Interestingly, the axino scenario naturally predicts a {\em secondary} displaced decay, as depicted in \figref{stealthdecaychain}, which may be visible to MATHUSLA. The axino is contained in a supermultiplet with two real scalar fields, $s$ (the saxion) and $a$ (the axion). It is possible to parametrize the interactions of this supermultiplet with a simple effective field theory \cite{Higaki:2011bz, Cheung:2011mg}, from which one learns that $s$ and $\widetilde{a}$ will generically obtain a mass on the order of the gravitino mass or larger, although 
$a$ may remain lighter. The axino mass allows for a later decay
\begin{equation}
{\widetilde a} \to a + {\widetilde G}~{\rm or}~s + {\widetilde G}.
\end{equation}
In turn, $a$ or $s$ may decay to Standard Model particles.

Let us fill in some estimates of the relevant regimes of parameter space. The decay to the axino must be prompt. We can consider an effective theory in which the stealth field $X$ couples to a chiral Goldstone superfield $A$ with a shift symmetry,
\begin{equation}
K \supset \frac{1}{f} (A + A^\dagger) X^\dagger X.
\end{equation}
In the approximation of negligible axino mass and a small stealth mass splitting $\delta m_X \equiv m_{\widetilde X} - m_X$, this leads to a decay width (correcting a formula in  \cite{Fan:2012jf} by a factor of 4):
\begin{equation}
\Gamma({\widetilde X} \to X {\widetilde a}) \approx \frac{m_{\widetilde X} \delta m_X^2}{\pi f^2} \approx \frac{1}{6~\mu {\rm m}} \left(\frac{m_{\widetilde X}}{100~{\rm GeV}}\right) \left(\frac{\delta m_X}{10~{\rm GeV}}\right)^2 \left(\frac{10^7~{\rm GeV}}{f}\right)^2.
\end{equation}
Collider searches for displaced vertices will not be sensitive to this decay provided the decay length $bc\tau \lesssim 100~{\rm \mu}m$. We require $\delta m_X \lesssim 10~{\rm GeV}$ for stealth phenomenology; furthermore, much larger splittings or much smaller values of $f$ are potentially associated with sizable tadpole effects for the singlet $X$ that could induce direct (non-stealthy) decays to axinos. Because of this target for $\delta m_X$, we require that the axino, saxion, and axion masses are all below $10~{\rm GeV}$.

The axino decay to gravitino has width
\begin{equation}
\Gamma(\widetilde{a} \to (s,a) + \widetilde{G}) = \frac{m_{\widetilde a}^5}{32\pi F_0^2} \left(1 - \frac{m_{s,a}^2}{m_{\widetilde a}^2}\right)^4 \approx \frac{1}{1~{\rm km}} \left(\frac{m_{\widetilde a}}{4~{\rm GeV}}\right)^5 \left(\frac{100~{\rm TeV}}{\sqrt{F_0}}\right)^4,
\end{equation}
where in the last step we have dropped the phase space factor and summed the two partial widths. We see that the requirement for the stealth decay to be kinematically accessible, $m_{\widetilde a} < \delta m_X$, already pushes the lifetime of the axino decay to gravitino to order kilometers even when the SUSY breaking scale is low. This is in the right range for sensitivity at MATHUSLA. Much larger values of $\sqrt{F_0}$ will have decay lengths that are substantially longer, but the lowest possible values of $\sqrt{F_0}$ are favored for naturalness.

The final decay is that of the saxion or axion itself. In general, the axion may be much lighter than the saxion. The saxion always has a decay width to axions arising from the kinetic term, 
\begin{equation}
\Gamma(s \to aa) = \frac{m_s^3}{32 \pi f_a^2} \approx \frac{1}{3~{\rm cm}} \left(\frac{m_s}{4~{\rm GeV}}\right)^3 \left(\frac{10^7~{\rm GeV}}{f_a}\right)^2,
\end{equation}
so once the axino has decayed to a saxion and gravitino, the saxion will then decay relatively promptly. Of course, the $s \to aa$ decay will not be observable. But the saxion can additionally couple to Standard Model fields through couplings like
\begin{equation}
\frac{c_g \alpha_s}{\pi f}  s G^a_{\mu \nu}G^{a\mu\nu} + \frac{c_\gamma \alpha}{\pi f} s F_{\mu \nu}F^{\mu \nu}.
\label{eq:stealthcouplingdefinition}
\end{equation}
Provided that the effective coupling scale $f/c_g, f/c_\gamma$ appearing in these couplings is not much larger than $f_a$, a fraction of the decays will be to gluons (producing hadrons in the MATHUSLA detector) or to photons (detectable through conversions). 

The MATHUSLA reach for saxion decays in Stealth SUSY is depicted in Figure \ref{fig:stealthreach}. These events originate with gluino pair production in the LHC, with the decay chain:
\begin{equation}
pp \to {\widetilde g} {\widetilde g},\quad {\widetilde g} \to g {\widetilde X}, \quad {\widetilde X} \to X {\widetilde a}, \quad X \to gg, \quad {\widetilde a} \to s {\widetilde G}, \quad s \to gg.
\end{equation}
In this figure we assume low-scale SUSY breaking with $\sqrt{F_0} = 50$ TeV, along with $m_{\widetilde X} = 100~{\rm GeV}$, $m_X = 90~{\rm GeV}$, an axino mass of 5~GeV and an approximately massless axion, corresponding to a rest-frame axino lifetime of approximately 20~m.  Combined with an average boost factor of $\mathcal{O}(10)$ from the gluino production and decay, these choices approximately maximize the MATHUSLA sensitivity to Stealth SUSY.  Nevertheless, these parameters fall within reasonable ranges for Stealth SUSY, given the requirements discussed above for prompt singlet decays in the collider, minimal MET, the naturalness pressure towards a low SUSY breaking scale, and cosmological constraints favoring a very light gravitino. On the other hand, because the axino lifetime is highly sensitive to $\sqrt{F_0}$ and $m_{\widetilde a}$, much of the parameter space predicts a longer lifetime and a low acceptance for MATHUSLA.  As an illustration, raising $\sqrt{F_0}$ to 100~TeV reduces the maximum MATHUSLA reach by approximately 250~GeV in the gluino mass, taking $c_g = c_{\gamma} = 1$.

Figure \ref{fig:stealthreach} also shows the current LHC exclusion and the HL-LHC discovery reach for this scenario. The gluino pair production events contain six energetic gluons (plus ISR or FSR) and very little missing energy. They thus have some similarity to R-parity violating decays in the MSSM, which can produce six quark jets from gluino decays. Thus we have relied on an ATLAS RPV search \cite{ATLAS:2016nij} to estimate the LHC's capabilities. Validating our recasting against the reported ATLAS RPV signal efficiency suggests that the solid red region, in which we have reduced our calculated Monte Carlo efficiency by a factor of 2, is approximately accurate. The lighter shaded red region is the unscaled result, which we expect overstates the current exclusion. To estimate the HL-LHC discovery reach, we have first computed the mass for which $\epsilon \times \sigma$ is $5/2$ the expected 95\% CL$_s$ exclusion at the current 13 TeV search, as a rough estimate for the gluino mass which could already have been discovered at 5$\sigma$ confidence. We have then rescaled to the 14 TeV LHC with 3 ab$^{-1}$ using the Collider Reach estimation procedure of Salam and Weiler \cite{SalamWeiler}, which has been argued to produce approximately accurate results based on simple considerations of parton luminosities.

From the figure, it is apparent that the HL-LHC has good discovery potential over a wide range of gluino masses. MATHUSLA would discover an additional displaced decay over a smaller region of parameter space. On the other hand, the LHC signal would be in a background-dominated region and could prove hard to interpret. For example, both R-parity violating SUSY and Stealth SUSY could produce the same several-jet final state. MATHUSLA could be crucial to avoid misdiagnosing the nature of the new physics, demonstrating that R-parity violation is not the origin of the physics and that a model containing a low mass, long-lived particle is necessary. Furthermore, even a handful of events at MATHUSLA would be sufficient to provide this information. The axino lifetime is long enough, and its mass low enough, that MATHUSLA is much better suited than LLP searches at the HL-LHC to detecting it (see \S\ref{s.LHCLLPcomparison}). 

Notice that a decay chain along the lines we have discussed,
\begin{equation}
{\widetilde \chi} \to {\rm SM} + {\widetilde a}, \quad {\widetilde a} \to s + {\widetilde G}, \quad s \to {\rm SM},
\end{equation}
can happen in low-scale SUSY breaking scenarios completely independently of Stealth SUSY, and a MATHUSLA signal from a late-decaying saxion-like particle is even more generic. What Stealth SUSY adds is a strong motivation for having an axino $\widetilde a$ available for this decay chain to proceed. Precisely the challenge of obtaining a stealthy decay that is prompt on collider timescales leads to the added ingredient that gives us an auxiliary signal of a potentially very long lifetime.

\subsubsection[Axinos]{Axinos\footnote{Eung Jin Chun, Sunghoon Jung}}
\label{sec:axino-ewino}

Another concern for naturalness resides in the QCD sector, which is called the strong CP problem \cite{Kim:2008hd}.
The gauge invariance does not forbid a CP-odd term:
\begin{equation}
     {\cal L}_{\theta} =  \theta {g_s^2 \over 32\pi^2}  G^a_{\mu\nu} \tilde G_a^{\mu\nu}
\end{equation}
where $\theta$ is a dimensionless parameter which is generically of order one. However, the actual value of $\theta$ is strongly constrained by measurements of nucleon electric dipole moments: $|\theta| < 10^{-10}$.  The problem of  such a (vanishingly) small $\theta$ is elegantly resolved by introducing an axion which is a Goldstone boson of a QCD-anomalous Peccei-Quinn (PQ) symmetry\cite{Peccei:1977ur, Wilczek:1977pj, Weinberg:1977ma}.
The PQ symmetry can be realized typically by introducing heavy quarks (KSVZ) \cite{Kim:1979if,Shifman:1979if} or
by extending the Higgs sector (DFSZ) \cite{Dine:1981rt,Zhitnitsky:1980tq}.
The PQ symmetry is supposed to be broken spontaneously by PQ-charged (dominantly SM-singlet) fields $\phi=v_\phi e^{i a_\phi/v_\phi}/\sqrt{2}$ carrying the PQ charge $x_\phi$ and thus the axion is a linear combination of the phase degrees of freedom $a_\phi$: $a\equiv \sum x_\phi v_\phi a_\phi/v_{PQ}$ where the overall breaking scale is given by $v_{PQ}\equiv \sqrt{\sum_\phi x_\phi^2 v_\phi^2}$.  Due to the QCD anomaly of the PQ symmetry, there arises an axion-gluon-gluon couling 
\begin{equation}
 {\cal L}_{agg} = {g_s^2 \over 32\pi^2} {a\over f_a} G^a_{\mu\nu} \tilde G_a^{\mu\nu}
\end{equation}
where $f_a \equiv v_{PQ}/N_{\rm DW}$ and $N_{\rm DW}$ is the domain wall number counting the QCD anomaly.  Note that the $\theta$ term can be absorbed into the dynamic degree $a$ whose potential is generated after the QCD phase transition:
\begin{equation}
 V_{QCD}[a]  \approx {m_\pi^2 f_\pi^2} \cos\left( a\over f_a\right).
\end{equation} 
This settles  the effective $\theta$ term to zero: $\theta_{eff}\equiv  \langle a\rangle/f_a =0$, and induces a non-vanishing axion mass $m_a \approx m_\pi f_\pi/f_a$. 

 The conventionally allowed window of the axion scale is $10^{9} \lesssim f_a / {\rm GeV} \lesssim 10^{12}$. The lower limit comes from star cooling processes \cite{Raffelt:2006cw, Chang:2018rso} and the upper limit from the axion cold dark matter contribution taking the initial mis-alignment angle $\theta_i$ of order one~(see e.g.~\cite{Fox:2004kb}).  
One may allow higher $f_a$ if  initial $\theta_i \ll 1$ is taken depending on cosmological scenarios.

The existence of such a high scale causes quadratic divergences to the Higgs boson mass and thus requires a huge fine-tuning
to keep stable two scales, the electroweak scale and the axion scale (or a generic UV scale).
Supersymmetry (SUSY) would be the best-known framework to avoid such a hierarchy problem.
However, the electroweak symmetry breaking in SUSY suffers from a certain degree of fine-tuning
to maintain a desirable potential minimization condition:
\begin{equation} \label{mZmin}
{m_Z^2 \over 2} = {m^2_{H_d} - m^2_{H_u} \tan^2\beta \over \tan^2\beta -1 } - \mu^2
\end{equation}
where $m_{H_{u,d}}$ are the soft masses of the two Higgs doublets, $\tan\beta \equiv v_u/v_d$ is the ratio of their vacuum
expectation values, and $\mu$ is the Higgs bilinear parameter in the superpotential.
As LHC finds no hint of SUSY, it pushes up
the soft mass scale above TeV range, the minimization condition (\ref{mZmin}) requires a fine cancellation among different terms.
Barring too huge a cancellation, one may arrange $m_{H_{u,d}}$ and  $\mu$ not too larger than $m_Z$. This has been advocated as ``natural SUSY'' \cite{Kitano:2006gv,Brust:2011tb,Papucci:2011wy,Baer:2012up} 
 implying stops/sbottoms at sub-TeV and light higgsinos with $\mu\sim {\cal O}(100)$ GeV.

\medskip

The origin of $\mu$ at the electroweak scale may be related to the PQ symmetry in the manner of DFSZ  which introduces a non-renormalizable superpotential in the Higgs sector \cite{Kim:1983dt,Chun:1991xm}:
\begin{equation} \label{Wdfsz}
 W= \lambda_\mu {P^2 \over M_P} H_u H_d
\end{equation}
where $P$ and thus $H_u H_d$ carries a non-trivial PQ charge and $M_P$ is the reduced Planck mass.
Upon the PQ-symmetry-breaking $v_{PQ} \sim \langle P \rangle$,
a $\mu$ term is generated by $\mu = \lambda_\mu \langle P \rangle^2/M_P$.
{\color{black} Once PQ symmetry is broken, there appear the axion $a$,  its scalar partner, the saxion $s$, and
the fermion super-partner, the axino $\tilde a$.}
Forming an axion superfield $A= (s+ ia, \tilde a)$, one can schematically write down
the effective $\mu$-term superpotential;
\begin{equation} \label{Weff}
 W = \mu H_u H_d + c_H {\mu \over v_{PQ} } A H_u H_d
\end{equation} 
where $c_H$ is a parameter depending on the PQ symmetry breaking sector; we take $c_H=2$ in this work. 
The axino mass is expected to be of order of the soft SUSY breaking scale, but it
is in general model-dependent  \cite{Goto:1991gq,Chun:1992zk,Chun:1995hc}.

Although axino interactions are suppressed by the axion scale, cosmic axinos can be abundantly produced through thermal particle interactions (see \secref{freezein} on Freeze-In Scenarios) like decays, inverse-decays and scattering of higgsinos and Higgs bosons into the axino:
\begin{equation}
\label{eq:rhoatildeaxinosection}
\frac{\rho_{\tilde a}}{\rho_{DM}} \approx \xi_H \left(\frac{m_{\tilde a}}{\mathrm{MeV}} \right)\left( \frac{\mu}{300 \gev}\right)^r \left(10^{12} \mbox{GeV} \over v_{PQ}\right)^2 
\end{equation}
$\xi_H$ is a model-dependent parameter involving the soft SUSY breaking parameters.
$r = 1$ if Higgsino decay to Higgs and axino is allowed only after the electroweak phase transition~\cite{Chun:2011zd,Bae:2011jb,Bae:2011iw},  and $r = 2$ if it is allowed before EWSB as well, see Section~\ref{subsec:AxinoDM}.
Thus a stable axino has to be lighter than about MeV for $v_{PQ} \lesssim 10^{12}$ GeV, unless mechanisms to dilute its relic abundance are present. Such dilution could be provided, for example, by a decaying saxion condensate, see Section~\ref{sec:DFSZ}. 
We therefore consider axino masses above an MeV as well, though the precise axino mass will not greatly affect the LLP phenomenology we study.

\begin{figure} \centering
\includegraphics[width=0.49\textwidth]{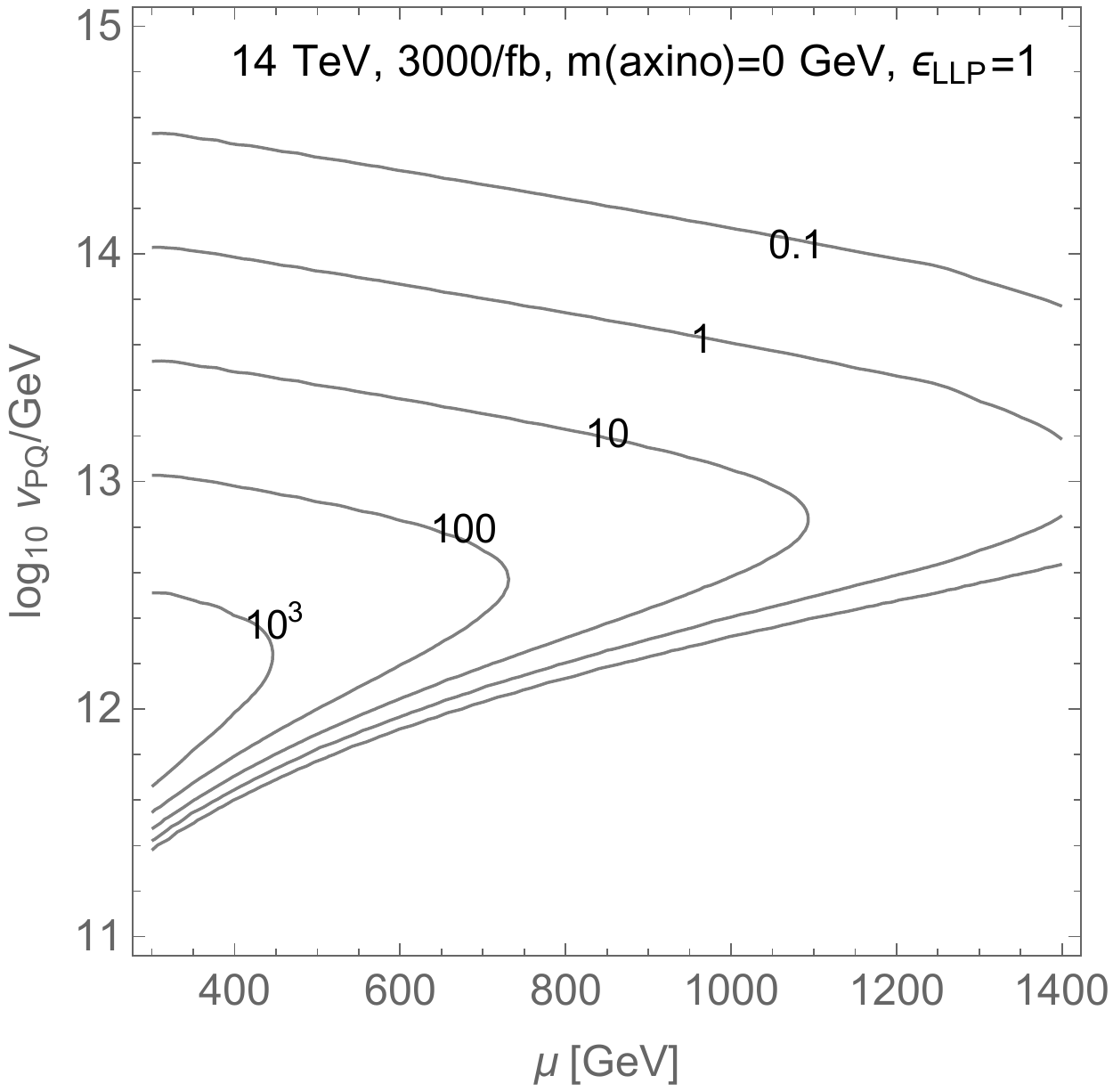}
\caption{
Number of long-lived Higgsino decays $\widetilde{H} \to \widetilde{a} Z, \widetilde{a} h$  observed in MATHUSLA for the DFSZ axino model, assuming only Drell-Yan like Higgsino production at the HL-LHC. A perfect LLP detection efficiency $\epsilon_{LLP} = 1$ is assumed. Dependence $\tan \beta$ and axino mass (except near kinematic threshold for the decay) is weak. 
This estimate assumes the $200m \times 200m \times 20m$ benchmark geometry of Fig.~\ref{f.mathuslalayout}.
}
\label{fig:axinos}
\end{figure}

Motivated by natural SUSY realizing the DFSZ axion, one could therefore have a Higgsino  NLSP and an axino LSP \cite{Barenboim:2014kka}.
Light higgsinos can be copiously produced at the LHC via Drell-Yan production (and even greater rates are possible if heavy colored particles decay to higgsinos, see Fig.~\ref{f.benchmarkxsecs}).
They decay to axinos plus the EW bosons $h$ or $Z$  through the coupling in Eq.~(\ref{Weff}):
\beq
\widetilde{H}^0 \to \widetilde{a}  Z, \, \widetilde{a} h \to \textrm{displaced dilepton/dijet + MET}
\eeq
The typical decay rate of the Higgsino NLSP is estimated as
\begin{equation}
 \Gamma_{\tilde H^0} \approx {c_H^2\over 32\pi} {\mu^3 \over  v_{PQ}^2}
\end{equation}
corresponding to 
\begin{equation}
c \tau_{\tilde H^0} \approx {620 \mbox{m}  \over (c_H/2)^2}  
\left( 200 \mbox{GeV} \over \mu\right)^3
\left( v_{PQ} \over 10^{12} \mbox{GeV}\right)^2,
\end{equation}
where we assumed a massless axino (but the precise mass does not change this drastically, as long as the final-state EW boson is on-shell).
Especially for PQ scales near the upper end of the motivated range, it is clear that higgsinos have the right lifetime to be detected by MATHUSLA.

We show the expected number of observed Higgsino decays in MATHUSLA at the HL-LHC in Fig.~\ref{fig:axinos}. MATHUSLA can probe a wide range of PQ scales and higgsino masses, $\mu \sim 100 - 1000$ GeV and $v_{PQ} \sim 10^{11} - 10^{14}$ GeV. 

How does the MATHUSLA reach compare to the achievable reach of the HL-LHC main detectors in the long lifetime limit?
For Higgsino masses below a few hundred GeV, or heavier higgsinos with axinos that are comparable in mass, the amount of visible hadronic energy per LLP decay is likely too low for background-free searches using the LLP decay along, see Section~\ref{s.LHCLLPsignal}.
Therefore, while MATHUSLA is sensitive to any visible final state of the long-lived higgsino decay, 
the most sensitive HL-LHC search would make use of the leptonic $Z$ decay to suppress hadronic backgrounds and obtain a clean sample for offline LLP reconstruction. 
The signal for the main detector search is therefore suppressed, relative to MATHUSLA, by the branching ratio
\begin{equation}
\mathrm{Br}_\mathrm{LHC} = \mathrm{Br}(\tilde H^0 \to \tilde a Z) \ \mathrm{Br}(Z \to \ell^+ \ell^-)
\end{equation}
For  high $\tan \beta = 50$, $\mathrm{Br}_\mathrm{LHC} \sim 0.05$ for either sign of $\mu$, while for low $\tan \beta = 2$, $\mathrm{Br}_\mathrm{LHC} \sim 0.08$ ($\mu > 0)$ or $0.02$ $(\mu < 0)$. 
As a result, the factor $\epsilon_{cut}$ in Sections~\ref{s.LHCLLPsignal} and \ref{s.LHCLLPcomparison} is $\mathrm{Br}_\mathrm{LHC}$ $\times$ (various trigger and cut efficiencies) even without main detector backgrounds.
For heavier higgsinos with more than a few hundred GeV of hadronic energy in their decay, displaced jet analyses at the main detectors may be able to achieve sensitivities closer to those of MATHUSLA. 

MATHUSLA significantly extends our axino sensitivity to very high $v_{PQ}$ scales. Given that the main detectors would have good sensitivity at shorter lifetimes, MATHUSLA would enable us to probe the entire  $v_{PQ}$ range motivated by axion DM (produced via mis-alignment) or axino DM (produced via freeze-in, see Sec.~\ref{sec:DFSZ}). 
Furthermore, as Higgsino NLSPs are produced in almost any cascade decays of heavier supersymmetric particles, the signal can be enhanced and probe heavier supersymmetric particles too.

\subsubsection[Sgoldstinos]{Sgoldstinos
\footnote{Dmitry Gorbunov}}
\label{sec:sgoldstinos}

There is a lot of supersymmetric extensions of the Standard Model
differing from each other in many respects, but exhibiting a common
feature: in each model the supersymmetry must be spontaneously
broken. The majority of models exploit for this purpose a chiral
superfield, 
\[
\Phi=\varphi+\sqrt{2}\theta{\psi}+F_\varphi\theta\theta\,,
\]
whose auxiliary component acquires non-zero vacuum expectation value (v.e.v.)
\[
\langle F_\varphi \rangle \equiv F  \neq 0 \,,
\]
that breaks the supersymmetry. In accordance with the Goldstone
theorem, there is a massless particle in the spectrum, which is
fermion in case of supersymmetry, {\it goldstino} $\psi$. Its
superpartners, $\varphi$ and $\varphi^*$ form scalar and pseudoscalar {\it
sgoldstinos},
\[
\frac{1}{\sqrt{2}}\left(\varphi+\varphi^*\right)\equiv{
S}\,,\;\;\;\;\;\frac{1}{i\sqrt{2}}\left(\varphi-\varphi^*\right)
\equiv{P}\,, 
\]
respectively. As we discuss below, these sgoldstinos can be long-lived and are therefore a natural target for searches at MATHUSLA. They are an very challenging signal, but production in $B$-meson decays can give rise to LLP decays that MATHUSLA can detect, in a complementary manner to searches at SHiP~\cite{Alekhin:2015byh}.

\subsubsubsection{Sgoldstino couplings and lifetime}

The auxiliary field $F_\varphi$ has a dimension of mass squared, and its
v.e.v. is of the order of the squared energy scale of the
supersymmetry breaking,
\[
F \sim \left(E_{\,\text{SUSY-breaking-scale}}\right)^2\,.
\]
Goldstino couples to the non-conserved (super)current \cite{Cremmer:1978iv}
as follows from the
general Goldberger--Treiman formula,
\begin{equation}
\label{goldstino-TG} 
{\cal L}_{
\psi}\propto\frac{1}{{F}}J^{\mu}_{{SUSY}}
\partial_\mu{\psi}\,. 
\end{equation}
When supersymmetry is promoted to the local symmetry, it becomes
supergravity and goldstino gets eaten in the super-Higgs mechanism,
giving mass to gravitino, 
\[
m_{\tilde G}= \sqrt{\frac{8\,\pi}{3}}\frac{F}{M_{\text{Planck}}}\,,
\]
and forming its longitudinal component.  

Sgoldstinos remain massless at tree level as well, but gain masses due
to higher order corrections. Their scale is very model-dependent and
for phenomenological purposes sgoldstino masses $m_S$ and $m_P$ can be
considered as free parameters. In particular, sgoldstinos are
naturally expected to be light in models with no-scale
supergravity\,\cite{Ellis:1983sf} 
and models with gauge mediated supersymmetry
breaking\,\cite{Giudice:1998bp,Dubovsky:1999xc}. Meanwhile,
sgoldstino couplings to the Standard Model (SM) fields are fixed by
supersymmetry and in most cases the coupling constants are
proportional to the ratio of some supersymmetry breaking
parameters (soft masses, trilinear couplings) and supersymmetry
breaking parameter $F$ \cite{Cremmer:1978iv}.
The explicit expression may be obtained by either
performing the supersymmetry transformation of goldstino
interaction (\ref{goldstino-TG}) or exploiting the spurion
technique\,\cite{Brignole:1996fn}. Then, to the leading order in $1/F$
sgoldstino couplings to gravitino $\tilde G$,
leptons $f_L$, up- and down-quarks $f_U$,
$f_D$, photons $F_{\mu\nu}$ and gluons $G^a_{\mu\nu}$
read\,\cite{Gorbunov:2000th,Gorbunov:2001pd}
\begin{align}
{\cal L}_{eff}=-\frac{1}{2\sqrt{2}F}\left( m_S^2S\bar{\tilde{G}}\tilde{G}+
im_P^2P\bar{\tilde{G}}\gamma_5\tilde{G}\right)\label{eef}
-\frac{1}{2\sqrt{2}}\frac{M_{\gamma\gamma}}{F}SF^{\mu\nu}F_{\mu\nu}+
\frac{1}{4\sqrt{2}}\frac{M_{\gamma\gamma}}{F}
P\epsilon^{\mu\nu\rho\sigma}F_{\mu\nu}F_{\rho\sigma}\\
-\frac{1}{2\sqrt{2}}\frac{M_3}{F}SG^{\mu\nu~a}G_{\mu\nu}^a+
\frac{1}{4\sqrt{2}}\frac{M_3}{F}
P\epsilon^{\mu\nu\rho\sigma}G_{\mu\nu}^a G_{\rho\sigma}^a\\
-\frac{\tilde{m}_{D_{ij}}^{LR~2}}{\sqrt{2}F}S\bar{f}_{D_i}f_{D_j}-
i\frac{\tilde{m}_{D_{ij}}^{LR~2}}{\sqrt{2}F}P\bar{f}_{D_i}\gamma_5f_{D_j} 
-\frac{\tilde{m}_{U_{ij}}^{LR~2}}{\sqrt{2}F}S\bar{f}_{U_i}f_{U_j}-
i\frac{\tilde{m}_{U_{ij}}^{LR~2}}{\sqrt{2}F}P\bar{f}_{U_i}
\gamma_5f_{U_j}\\\label{eefd}
-\frac{\tilde{m}_{L_{ij}}^{LR~2}}{\sqrt{2}F}S\bar{f}_{L_i}f_{L_j}-
i\frac{\tilde{m}_{L_{ij}}^{LR~2}}{\sqrt{2}F}P\bar{f}_{L_i}\gamma_5f_{L_j}\;. 
\end{align}
Hereafter $\theta_W$ is the weak mixing angle,
$M_{\gamma\gamma}=M_1\sin^2\theta_W+M_2\cos^2\theta_W$ and $M_i$,
$i=1,2,3$ are gaugino masses with index corresponding to SM gauge groups,
$U(1)_Y$, $SU(2)_W$ and $SU(3)_c$, and  $\tilde{m}_{D_{ij}}^{LR~2}$
and $\tilde{m}_{U_{ij}}^{LR~2}$ are left-right down- and up-squark
mass terms. Above we omit two-sgoldstino coupling terms,
see Refs.\,\cite{Perazzi:2000id,Perazzi:2000ty,Gorbunov:2000th,Demidov:2011rd} 
which are strongly suppressed by $1/F^2$. 

These interactions determine sgoldstino production and decay
rates. Their palletes depend on the patterns of MSSM soft terms. Without
any specific hierarchy there, most naturally coupling to gluons
dominate. Generically, sgoldstino couplings become weaker with
increase of supersymmetry breaking scale, $\sim \sqrt{F}$, and ratio of
superpartner mass and supersymmetry breaking scales,
$M_{soft}/\sqrt{F}$, which must not exceed unity while the unitarity
is conserved. The most attractive feature of sgoldstino phenomenology is that
the measurement of sgoldstino couplings gives an opportunity to probe
the scale of supersymmetry breaking in the whole theory. Sgoldstinos
are $R$-even, contrary to $R$-odd gravitino, and so if SM
superpartners are heavy (an assumption consistent with LHC results),
their production is less suppressed as compared to (light) gravitinos,
which couplings start at ${\cal O}(1/F^2)$ level only. Sgoldstino production
at LHC with couplings (\ref{eef})-(\ref{eefd}) has been studied in
Refs.\,\cite{Gorbunov:2000ht,Gorbunov:2002er,Demidov:2004qt,Asano:2017fcp}. 

For interesting sgoldstino masses above 1\,GeV, sgoldstino decays into
photons and into gluons are always open.
An order-of-magnitude estimate of sgoldstino life-time,
\begin{equation}
\label{sgoldstino-lifetime}
\tau_{S(P)}\sim 4\pi \frac{F^2}{M_{soft}^2\,m_{S(P)}^3}\sim
10^{-8}\times 
\left(\frac{\sqrt{F}/1000\,\text{TeV}}{M_{soft}/1\,\text{TeV}}\right)^2
\left(\frac{\sqrt{F}}{1000\,\text{TeV}}\right)^2 \left(\frac{10\,\text{GeV}}{m_{S(P)}}\right)^3\,\text{s}\,,  
\end{equation}
shows that in order to allow sgoldstino 
to cover the distance of a hundred meters and  reach the
MATHUSLA detector, supersymmetry breaking scale must be high and
well-separated from the scale of MSSM soft terms. However, in this
limit the sgoldstino coupling constants to SM fields become tiny. As we show below, its direct production rate is much below the critical 1\,fb scale. However, sizable production rates are possible if its mass lies below the $B$-threshold.

\subsubsubsection{Sgoldstino production mechanisms}

Sgoldstinos can be produced in gluon fusion,  the same way as
the SM Higgs boson $h$. Hence, the sgoldstino production cross-section
at LHC can be estimated as 
\begin{equation}
\label{sgoldstino-cross-section}
\sigma_{gg\to S(P)}(m_{S(P)})=\sigma_{gg\to h}(m_{S(P)})\,\frac{\Gamma_{S(P)\to gg}(m_{S(P)})}
{\Gamma_{h\to gg}(m_{S(P)})}=\sigma_{gg\to
h}(m_{S(P)})\,\left(\frac{3\,\pi}{\alpha_s(m_{S(P)})}\right)^2 \frac{\sqrt{2}\,M_3^2}{F^2G_F}\,
\end{equation}
where $\sigma_{gg\to h}(M)$ and $\Gamma_{h\to gg}(M)$ are production and
decay width of the SM Higgs boson into gluons, if its mass would be
$M$. Putting numbers into (\ref{sgoldstino-lifetime}) one finds
\begin{equation}
\label{sgoldstino-production}
\sigma_{gg\to S(P)}(m_{S(P)})\simeq0.1\,\text{fb}\times\left(\frac{\sigma_{gg\to
h}(m_{S(P)})}{10\,\text{pb}}\right)\left(\frac{M_3}{1\,\text{TeV}}\right)^2
\left( \frac{100\,\text{TeV}}{\sqrt{F}}\right)^4\,.
\end{equation}
It is clear from eqs.\,(\ref{sgoldstino-lifetime})
and (\ref{sgoldstino-production}) that it is impossible to have both
reasonably high production rate and long-lived sgoldstino. The product
of eqs.\,(\ref{sgoldstino-lifetime})
and (\ref{sgoldstino-production}) is
\[
\sigma_{gg\to S(P)}(m_{S(P)})\times \tau_{S(P)}\simeq
10^{-13}\,\text{fb}\,\text{s}
\times\left(\frac{10\,\text{GeV}}{m_{S(P)}}\right)^3 \left(\frac{\sigma_{gg\to
h}(m_{S(P)})}{10\,\text{pb}}\right)\,.
\]

Sgoldstino can be also produced in decays of heavy SM particles
emerged in proton-proton collisions at LHC: top-quark, $Z$-boson and
Higgs boson are potential sources. The sgoldstino interaction with
$t$-quark is governed by unknown flavor-violating structures of the
squark squared mass matrix, see Ref.\,\cite{Gorbunov:2000ht} for
details, and top-quark can decay into sgoldstino and light quark with
branching ratios as high as if $\sqrt{F}\lesssim 10$\,TeV, which
implies short-lived sgoldstinos untestable at MATHUSLA, see
eq.\,(\ref{sgoldstino-lifetime}). $Z$-bosons are much more abundant at
LHC than $t$-quarks, and can decay into sgoldstino and photon due to
coupling\,\cite{Perazzi:2000id}
\begin{equation}
\label{Z-gamma-sgoldstino}
{\cal L}_{Z\gamma}=-\frac{\left( M_2-
M_1\right) \cos\theta_W\sin\theta_W}{\sqrt{2}\,F}\,F_{\mu\nu}Z^{\mu\nu}S+
\frac{\left( M_2-
M_1\right) \cos\theta_W\sin\theta_W}{2\sqrt{2}\,F}\,F_{\mu\nu}Z_{\lambda\rho}\epsilon^{\mu\nu\lambda\rho}P
\end{equation}
with branching
\begin{equation}
\label{branching-Z-gamma-sgoldstino}
\text{Br}\left( Z\to\gamma
S(P)\right) \sim 10^{-7}\times\left(\frac{M_{soft}/1\,\text{TeV}}{\sqrt{F}/10\,\text{TeV}}\right)^2
\left(\frac{10\,\text{TeV}}{\sqrt{F}}\right)^2\,. 
\end{equation}
The SM Higgs boson can decay into a couple of sgoldstinos, if
kinematically allowed. Sgoldstino interaction with the Higgs sector
is considered in Refs.\,\cite{Astapov:2014mea,Asano:2017fcp}, the
relevant couplings are suppressed by $1/F^2$ (each sgoldstino leg brings factor
$1/F$). Assuming all the MSSM massive parameters are of the same order
$M_{soft}$ we can estimate the branching as
\begin{equation}
\label{branching-H-to-sgoldstinos}
\text{Br}\left( h\to SS\right) \sim
10^{-5}\times\left(\frac{M_{soft}/1\,\text{TeV}}{\sqrt{F}/10\,\text{TeV}}\right)^8\,. 
\end{equation}
In all the cases above with superpartner scale of order 1\,TeV the
sgoldstino is either too short-lived or too rare produced.

The best target for an ultra-long-lived particle search is the sgoldstino mass regime below $B$-threshold, $m_{S(P)}<5$\,GeV, with sgoldstino production in
beauty meson decays. In this case sgoldstino is much lighter, and so
long-lived, and, for the production, the feeble sgoldstino couplings must
compete not with strong but only with weak interactions.
Thus, beauty mesons can decay into sgoldstino with branchings as
large as $10^{-4}$  \cite{Astapov:2015otc}. There may be contributions
from quark flavor-conserving and flavor-violating sgoldstino
couplings. In this case MATHUSLA will compete with SHiP
experiment \cite{Alekhin:2015byh} operating on SPS 400\,GeV proton beam in a beam-dump mode. 
The number of beauty-quarks at ATLAS/CMS is expected to exceed
significantly that at the SHiP. 
Even though the SHiP geometry is optimized for the flux of outgoing particles, it has been shown that MATHUSLA has significantly better acceptance for LLP decays than SHiP, provided the lifetime is above $\sim$ 100 m \cite{Evans:2017lvd} (see also Section~\ref{sec:singlets}). 
The reach of SHiP has been investigated in detail in
Ref.\,\cite{Astapov:2015otc}. It was found,
that depending on the MSSM soft term pattern
and the scale of SM superpartners, SHiP can probe the scale of
supersymmetry breaking as high as $\sqrt{F} \sim 10^2\!-\!10^4$\,TeV.     
Since the sgoldstino decay length is in the long-lifetime regime at the upper limit of this $\sqrt{F}$ sensitivity range (and much shorter at the lower limit), we expect that MATHUSLA will be able to significantly extend the reach of SHiP and allow access to  higher SUSY breaking scales.

\subsection[Neutral Naturalness]{Neutral Naturalness\footnote{David Curtin, Nathaniel Craig, Yuhsin Tsai}}
\label{sec:neutralnaturalness}

The discovery of a light, apparently elementary Higgs boson at the LHC has heightened the severity of the electroweak hierarchy problem, while increasingly severe bounds on new colored particles have begun to disfavor conventional solutions such as supersymmetry or compositeness. Models of neutral naturalness provide a compelling alternative, in which the lightest states protecting the weak scale are not colored (and, in some cases, entirely neutral under the Standard Model). Such protection of the weak scale is achieved primarily through discrete symmetries, rather than continuous symmetries. Realizations of neutral naturalness include the Twin Higgs \cite{Chacko:2005pe}, Orbifold Higgs \cite{Craig:2014aea}, Quirky Little Higgs \cite{Cai:2008au}, and Folded Supersymmetry \cite{Burdman:2006tz}. 

Addressing the hierarchy problem via discrete symmetries naturally leads to specific hidden valleys (Section~\ref{sec:hiddenvalleys}) without Standard Model quantum numbers. Rather, the new states in these sectors primarily couple to the Standard Model through various portal-type interactions, including the Higgs portal $\lambda |H|^2 \mathcal{O}$ and the neutrino portal $y (LH) \mathcal{O}$. 
If the discrete symmetry between the two sectors extends to hypercharge, the photon portal  $\epsilon \tilde F_{\mu \nu} \hat F^{\mu \nu}$ may also mix the hidden and visible abelian gauge bosons.
In contrast with generic hidden sectors, in models of neutral naturalness both the size of these portal interactions and the mass scale of hidden particles are typically dictated by naturalness considerations, providing a motivated range of rates and lifetimes that can be effectively probed by MATHUSLA.  

\subsubsection{LLP Signatures of the Hidden Sector}

The discrete symmetries in successful models of neutral naturalness must be nearly exact in the top sector, given its relevance to the hierarchy problem, but may be approximate for states more remote from the Higgs sector \cite{Craig:2015pha}. 
In particular, since the QCD gauge coupling gives the dominant contribution to the renormalization of the top yukawa, the preservation of a near-exact discrete symmetry in the top sector requires the existence of one or more new QCD-like hidden gauge groups whose couplings are comparable to their Standard Model counterpart. 
Solution of the Hierarchy Problem via Neutral Naturalness therefore leads to the existence of specific confining Hidden Sectors. 
The confinement scales of these gauge groups are then typically of the same order as $\Lambda_{QCD}$, giving rise to hidden sector bound states whose masses range from $\mathcal{O}(1-100)$ GeV, allowing for their production at the LHC. 

\subsubsubsection{Production of Hidden Glueballs in Exotic Higgs Decays}

The coupling between the Standard Model-like Higgs and the top partners required to address the hierarchy problem induces loop-level couplings to hidden gluons, much as the top quark generates the leading coupling of the Higgs to Standard Model gluons. The effective coupling between the SM-like Higgs $h$ and the hidden gluon field strengths $\widehat G^a_{\mu \nu}$ takes the general form
\begin{equation} \label{eq:hgg}
\mathcal{L} \supset \theta^2\frac{\widehat \alpha_{3}}{12 \pi}  \frac{h}{v} \widehat G^a_{\mu \nu} \widehat G_a^{\mu \nu} 
\end{equation}
where $\theta$ is a mixing angle that varies between different realizations of neutral naturalness. In Twin Higgs or Quirky Little Higgs models, $\theta^2 \simeq -v^2/f^2$, where $f$ is a scale of spontaneous global symmetry breaking, while for Folded Supersymmetry $\theta^2 \simeq m_t^2/2\widehat m_{\tilde t}^2$, where $\widehat m_{\tilde t}$ is the mass of the QCD-neutral scalar top partners. Naturalness considerations bound the scales $f, \widehat m_{\tilde t}$ to lie at or below the TeV scale, so that the relevant mixing angles are naturally $\mathcal{O}(0.1-1)$. 

The coupling in \eqref{hgg} provides a predictive portal for the production of states in the hidden QCD sector, as well as an avenue for them to decay back to the Standard Model when kinematically allowed \cite{Craig:2015pha, Curtin:2015fna}. In particular, for generic hidden sector masses and mixing angles, \eqref{hgg} predicts exotic Higgs decays into the hidden sector with branching ratios of order $0.01-1\%$, corresponding to rates on the order of 5-500 fb at the 14 TeV LHC.

Once produced, states in the hidden sector  cascade down to the lightest accessible hidden sector state -- typically a bound state of hidden QCD -- which then can decay back to the Standard Model. The lifetime for these decays varies depending on the nature of the lightest bound states of hidden QCD. 
If the lightest hidden QCD states are mesons, or if even lighter hidden photons or neutrinos are part of the low-energy spectrum (as in the original Mirror Twin Higgs~\cite{Chacko:2005pe}), the hidden sector relics produced during the Big Bang are usually stable and abundant enough to be in conflict with $\Delta N_{eff}$ bounds from CMB measurements. 
We discuss this scenario in more detail below, but a straightforward way to avoid cosmological bounds is a hidden sector without such light states. 
This is always the case in Folded Supersymmetry, where the folded superpartners carry electroweak charge and have to be heavier than $\sim 100 \gev$ to respect LEP limits.
It may also be the case in Twin-Higgs-like models such as the Fraternal Twin Higgs \cite{Craig:2015pha}) where the discrete symmetry only applies to the third fermion generation, and there are no light mirror photons or neutrinos. 
In either scenario, the hidden QCD confines with zero light quark flavors, and the lightest accessible hidden sector states are hidden glueballs whose decay back to the Standard Model is governed by the dimension-5 operator in \eqref{hgg}. Decays proceeding through this operator generically lead to lifetimes within the BBN limit. 

In this case, the relevant process is the decay of a hidden glueball, typically the $J^{PC} = 0^{++} = G_0$ glueball, expected to lie at the bottom of the glueball spectrum with $m_{G_0} \approx 7 \hat \Lambda_{QCD}$, though decays of higher glueball excitations are also possible. 
$G_0$ decays to light Standard Model states via an off-shell Higgs: $0^{++} \to h^* \to YY$, where $Y$ are SM fields. The amplitude for this process in terms of the glueball mass $m_0$ is \cite{Juknevich:2009gg}
\begin{equation}
\frac{\widehat \alpha_3 \theta^2}{6 \pi v^2} \langle YY| m_f \bar f f + m_Z^2 Z_\mu Z^\mu + 2m_W^2 W_\mu^+ W^{\mu -} |0 \rangle \frac{1}{m_h^2 - m_0^2} \langle 0 | S |0^{++} \rangle
\end{equation}
which results in a width
\begin{equation}
\Gamma_{0^{++} \to YY} = \left( \frac{\widehat \alpha_3  \theta^2}{6 \pi(m_h^2 - m_0^2)} \frac{ {\mathfrak{f}}_0}{v} \right)^2 \Gamma_{h \to YY}^{SM}(m_{0^+}^2)
\end{equation}
where $\mathfrak{f}_0$ is the hidden $0^{++}$ decay constant. The corresponding mean decay length is
\begin{equation}
c \tau_0 \sim 20 \; {\rm m} \; \times \left( \frac{10 \; {\rm GeV}}{m_0} \right)^7 \left( \frac{f}{750 \; {\rm GeV}} \right)^4 \, .
\end{equation}
(A similar expression applies in FSUSY.)
The decay width of these glueballs is a steep function of their mass and mixings. For typical values the mean decay length ranges from $10^{-6}-10^7$ m, giving rise to the distinctive signal of exotic displaced decays where lifetime should be regarded as an almost free parameter within the BBN limit.

\begin{figure}
\begin{center}
\includegraphics[width=0.7\textwidth]{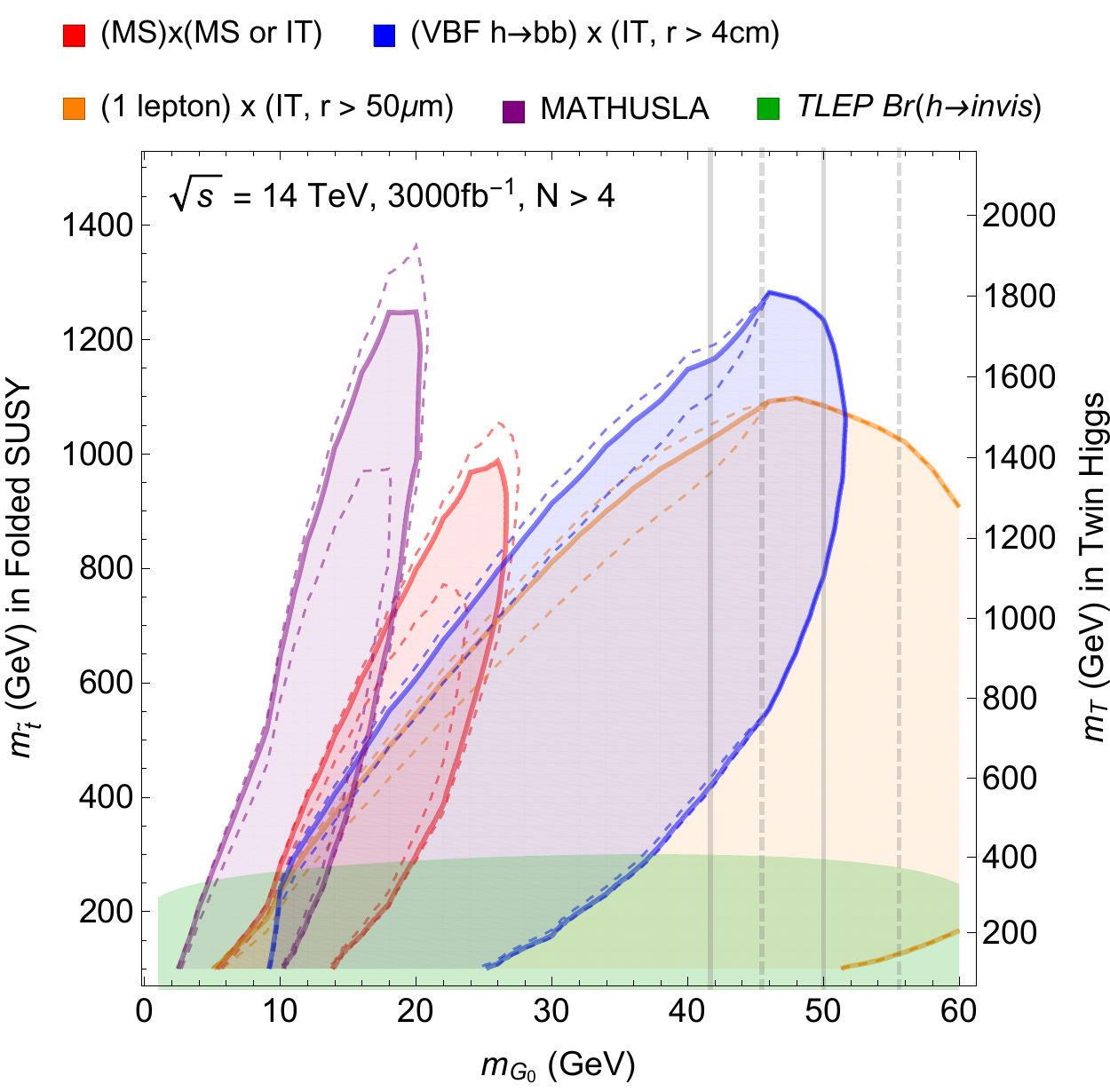}
\end{center}
\caption{
Simplified Neutral Naturalness parameter space of lightest glueball mass $m_{G_0}$ and top partner mass in Folded SUSY or the Fraternal Twin higgs.
Shown is projected reach of HL-LHC LLP searches for glueballs produced in exotic Higgs decays~\cite{Curtin:2015fna} in the ATLAS Muon System (red) or in the tracker in association with VBF jets or leptons from Higgs production (blue, orange).
The reach of MATHUSLA (assuming the $200m \times 200m \times 20m$ benchmark geometry of Fig.~\ref{f.mathuslalayout}) is shown in purple, and covers the regime of long-lived glueballs with masses $\lesssim 15 \gev$.
Sensitivity in all searches is conservatively estimated by assuming two glueballs produced per Higgs decay, and dashed contours indicate uncertainties due to details of hidden sector hadronization. See text for additional details.
}
\label{f.NNLLPbounds}
\end{figure}

These displaced decays are sufficiently distinctive and occur with sufficient rate at the LHC that they may be distinguished from Standard Model backgrounds provided appropriate triggering strategies, which were explored in~\cite{Curtin:2015fna}.
The projected reach is summarized in  Fig.~\ref{f.NNLLPbounds}. The plot shows a simplified parameter space of lightest glueball mass $m_{G_0}$ and top partner mass, either in Folded SUSY assuming no stop mixing (left vertical axis) or for a Fraternal Twin Higgs like model (right vertical axis). Glueball lifetime depends most dramatically on $m_{G_0}$, with proper decay lengths at the cm to sub-mm level for high glueball masses near $m_H/2$, and long lifetimes approaching the BBN limit for glueball masses below 10-20 GeV.

Uncertainties of hidden sector glueball hadronization make a precise prediction of glueball multiplicity in exotic Higgs decays difficult~\cite{Chacko:2015fbc}. For the purpose of these estimates, we conservatively assume that only two glueballs are produced per Higgs decay. This is likely to be reasonably accurate at high glueball masses. At lower glueball masses, there many more glueballs, each of which has an accordingly lower boost than under our simple assumption. Since the long glueball lifetime in that regime is the main bottleneck for searches with any detector, a more realistic treatment of hidden sector hadronization would only improve all reach projections. This makes our assumption suitable for a pessimistic estimate of the LHC's ability to probe Neutral Naturalness. 
Since the $0^{++}$ glueball is the state with the shortest lifetime, the fraction of glueballs that end up in the $0^{++}$ state is another important factor in estimating the LLP signal rate. However, the large ratio between the lightest glueball mass and the hidden QCD string tension suggests that $0^{++}$ states form a majority or at least a significant fraction of the produced states~\cite{Juknevich:2010rhj}, based on modeling of hadronization processes as thermal emissions~\cite{VanApeldoorn:1981gx}.
In Fig.~\ref{f.NNLLPbounds} we therefore assume that the $0^{++}$ and other glueball fractions are given by spin-weighted Boltzmann factors for all kinematically available states. The dashed contours indicate the variation of reach estimates from varying that $0^{++}$ up or down by a factor of 2. 
Finally, vertical solid (dashed) lines show where the production rate of $0^{++}$ glueballs may be additionally enhanced or suppressed due to non-perturbative mixing effects~\cite{Craig:2015pha}.

This simplified model of Neutral Naturalness then produces the signal of LLP pair production in exotic Higgs decays, with subsequent LLP decay through the Higgs portal. (See also Sec.~\ref{sec:exhdecays}.) 
As explained in \cite{Curtin:2015fna}, there are three particularly promising search strategies using the HL-LHC main detectors: 
(1) Search for two LLPs using the dedicated displaced decay trigger in the ATLAS Muon System, similar to \cite{Aad:2015uaa} (red in Fig.~\ref{f.NNLLPbounds}). (2) Search for a single displaced vertex in the tracker, in association with VBF jets from Higgs production (blue). (3) Search for a single displaced vertex in the tracker, in association with a lepton from associated Higgs production (orange). In this case we optimistically assume a displaced vertex can be reconstructed less than a mm from the primary  vertex to demonstrate how the short glueball lifetime regime may be probed. (Such a search would require significant further experimental study.)
Each search requires either two observed LLP decays, or one LLP decay in association with a conspicuous prompt object like VBF jets or leptons. Search strategies along these lines are likely to have low or zero background.
The sensitivity estimates for these three searches in Fig.~\ref{f.NNLLPbounds} therefore show regions with at least 4 observed LLP decays passing the requirements of each analysis. 
Also shown for comparison is the $\mathrm{Br}(h \to \mathrm{invisible})$ bound achievable with a TLEP-like lepton collider~\cite{Dawson:2013bba} (green), which may provide clues for very production of very long-lived glueballs.\footnote{In the short glueball lifetime regime (masses close to 60 GeV), prompt or displaced $h \to 4b$ searches at lepton colliders are likely to have better sensitivity~\cite{Liu:2016zki}.}

It is clear that the HL-LHC has the capability to probe Neutral Naturalness at the TeV scale for hidden glueball masses above $\sim 15 \gev$ using displaced vertex searches. Given the absence of conspicuous colored top partner signatures, this reach is impressive, but it misses the light glueball region of parameter space $m_{G_0} \lesssim 15 \gev$, corresponding to long glueball lifetimes $c \tau \gg 100 \mathrm{m}$.
This is a very important region to probe, not only because the lifetime of the glueballs should be treated as an essentially free parameter in theories of Neutral Naturalness, but also because RG-arguments~\cite{Curtin:2015fna} can favor relatively low glueball masses for theories like the Fraternal Twin Higgs.
This long-lifetime regime is an ideal target for the MATHUSLA detector, due to its low background and absence of trigger thresholds. 
We show the reach of MATHUSLA as the purple shaded region in Fig.~\ref{f.NNLLPbounds}. It provides the only direct probe of this important region of Neutral Naturalness parameter space.

\subsubsubsection{Other LLP states and production modes}


Additional processes may lead to the production of hidden sector bound states, though these channels are model-dependent. Here we briefly discuss LLP signatures due to hidden Bottomonia, Quirk production and production of heavy UV states. 

\emph{Hidden Bottomonium production:} In theories like the Fraternal Twin Higgs, where partner particles are entirely neutral under the Standard Model, hidden QCD bound states may also be produced in Higgs decays to bottom partners. Depending on $m_{\hat b}$ and $\hat \Lambda_{QCD}$, this may lead to the production of an excited quirky boundstate \cite{Kang:2008ea} of hidden $\hat {\bar b} \hat b$, or the production of hidden bottomonium. Since this exotic Higgs decay proceeds through the mirror bottom Yukawa coupling and is otherwise only suppressed by the Higgs mixing factor $v^2/f^2$, it can have $\sim 10\%$ branching ratio, leading to much larger hidden sector production rates than the hidden gluon coupling of Eqn.~(\ref{eq:hgg}).
If these mirror bottom states cascade down to lighter hidden glueballs, the LLP searches at MATHUSLA and the main detectors would cover even more of the parameter space shown in Fig.~\ref{f.NNLLPbounds}.

The lightest bottomonium states are the pseudoscalar $\hat{\eta}_{b}(0^{-+})$, the vector $\hat{\Upsilon}(1^{--})$, and the scalar $\hat{\chi}_{b0}(0^{++})$.
These states have masses $\sim 2 m_{\hat b} + \mathcal{O}(\hat \Lambda_{QCD})$. If the discrete symmetry is respected by the bottom Yukawa couplings, the Higgs coupling bound $f/v \gtrsim 3$~\cite{Burdman:2014zta} implies they are heavier than $\sim 35 \gev$. While this makes it natural for glueballs to be lighter, it is also possible for threshold corrections to hidden QCD RG running near the discrete symmetry breaking scale to lift the glueball mass above $m_H/2$~\cite{Craig:2015pha}. 
In that case, hidden bottomonia that are produced in exotic Higgs decays can only decay back to the SM via the Higgs portal and would be detectable as LLPs. 
The lightest pseudoscalar state is extremely long-lived, but the scalar decays back to SM states through the Higgs portal, with a decay length that can be estimated as~\cite{Cheng:2015buv}
\begin{eqnarray} \label{B0decay_quirk}
c\tau_{\hat{\chi}_{b0}}&\simeq& 8.3\,\text{cm}\left(\frac{m_b}{m_{\hat{b}}}\right)^5\left(\frac{f}{1\;\mathrm{TeV}}\right)^4\left(\frac{5\,\text{GeV}}{{\hat \Lambda}_{QCD}}\right)^2\left[\frac{9}{5}\left(\frac{\sqrt{s}}{3m_{\hat{b}}}\right)^2-\frac{4}{5} \right]^{-1}\quad (m_{\hat{b}}\gg {\hat \Lambda}_{QCD}),
\\ \label{B0decay_NDA}
c\tau_{\hat{\chi}_{b0}}&\simeq& 3.8\,\text{cm}\left(\frac{m_b}{m_{\hat{b}}}\right)^2\left(\frac{f}{1\;\mathrm{TeV}}\right)^4\left(\frac{5\,\text{GeV}}{{\hat \Lambda}_{QCD}}\right)^5 \left(\frac{\sqrt{s}}{3{\hat \Lambda}_{QCD}}\right)^{-2}\qquad\qquad\,\,\quad\,\, (m_{\hat{b}}\ll {\hat \Lambda}_{QCD}).
\end{eqnarray}
This estimate is valid in the range $2m_b < \sqrt{s}\ll m_h$, where $\sqrt{s}$ is the mass of the hidden bottomonium (which may be somewhat greater than $m_{\chi_{b0}}$ if the state is excited).

It is also possible for the discrete symmetry to be broken in the bottom sector of the theory. In that case, hidden bottomonia could be much lighter than 35 GeV, with correspondingly longer lifetimes. 
In the original Fraternal Twin Higgs model, this would cause cosmological problems. An extremely large abundance of hidden sector states is produced during the Big Bang~\cite{Garcia:2015toa,Farina:2015uea}, and if the scalar is much longer-lived than a meter, this abundance would no longer efficiently deplete itself via decays to the SM~\cite{Cheng:2015buv}. Late decays of the very long-lived pseudoscalar $\eta$ would then disrupt BBN. 
However, this regime would still be permitted if there was a kinetic mixing between a massive hidden hypercharge gauge boson and SM hypercharge. This arises naturally, for example, if the UV completion of the theory contains particles charged under both hypercharges. 
The hidden vector meson $\hat{\Upsilon}$ could then decay to SM particles, which sufficiently depletes the hidden hadron abundance at early times if it has a decay length below $\sim$ meter~\cite{Cheng:2015buv}. Again, light bottomonia and cosmological constraints lead to some LLPs with relatively short lifetimes. 
Such lifetimes are well-suited to main detector LLP searches, and may also be good targets for LHCb~\cite{Pierce:2017taw}. MATHUSLA can then play a vital role searching for longer-lived hidden hadron states and diagnosing the connection between the hidden valley and naturalness.

\emph{Quirk Production:} 
Models like Folded Supersymmetry and the Quirky Little Higgs feature top partners that are charged under the SM electroweak force. Drell-Yan production of top partners then provides an additional portal into the hidden sector. 
Once produced, these states remain connected by a hidden QCD flux tube and eventually undergo quirk-like annihilation decays \cite{Kang:2008ea} to hidden glueball states and possibly SM EW final states, depending on the details of the theory and spectrum.
These additional processes can enhance the production of hidden sector states by an order of magnitude or more, as has been studied in \cite{Chacko:2015fbc}. 
This would greatly enhance the reach of LLP searches at MATHUSLA and the main detectors in the parameter space of Fig.~\ref{f.NNLLPbounds}.

\emph{Production of heavy UV states:} The similarity between the SM and hidden gauge symmetries indicate the existence of a ``unification" between the two sectors, and most of its UV-completion scenarios contain particles carrying both the SM and hidden charges and provide extra portals between the two sectors~\cite{Chacko:2005pe, Cheng:2015buv, Chang:2006ra, Craig:2014fka, Craig:2013fga, Geller:2014kta, Low:2015nqa, Barbieri:2015lqa, Batra:2008jy, Craig:2014roa, Craig:2014aea}.  

For example, in many UV-completion of the twin Higgs models, there are exotic-fermions that carry either twin QCD $+$ SM weak charge, or SM QCD $+$ twin weak charge. These fermions are likely to carry masses close to the cutoff scale ($\gsim 5$ TeV) near the scale of the UV-completion, and an observation of them will provide valuable information about the SM-twin unification~\cite{Cheng:2015buv}. We can produce these bi-charged fermions at the LHC either through strong or electroweak production. For the exotic-quark that carries SM color, once being produced inside detector, they promptly decay into a pair of SM tops and two twin $Z$ bosons~\cite{Cheng:2016uqk}. Each of the twin $Z$ decays into twin quarks and form long-lived twin hadrons, and the whole event contains prompt high $p_T$ leptons from the top decay plus displaced lepton or jet signals. The hard leptons provide simple triggering and background rejection, and the search at the LHC will be able to probe the exotic-quark mass up to $\approx 2.5$ TeV that is getting close to the UV scale of the model. 

Besides the fermion that carries SM color, LHC can also produce the exotic-fermion that carry SM electroweak charge and twin QCD through Drell-Yan process~\cite{Li:2017xyf}. The fermions can be produced as a bound state bind by the mirror QCD force, which annihilates into SM fermions or gauge bosons for various resonance searches. 
Depending on details of the UV model and the SM charge of the produced exotic-fermion bound state, their decay can produce SM gauge bosons and/or hidden glueballs, giving different displaced signal topologies. 
If these bound states decay without leptons or hard SM jets in the final state, the main detector LLP searches will likely have at least some backgrounds or suffer from reduced trigger/cut efficiencies. In those scenarios, MATHUSLA will play an important role in discovering and diagnosing these LLP signals of the Neutral Naturalness UV completion. 


\subsubsection{LLP Signatures of Late-Time Reheating}

In addition to the states directly connected to the stabilization of the weak scale, models of neutral naturalness predict a variety of additional light degrees of freedom whose couplings and masses are comparable to their Standard Model counterparts and may give rise to additional distinctive signals at MATHUSLA.

For example, models of neutral naturalness based on global symmetries (such as the Twin Higgs and its relatives) generically predict additional neutrino species in the hidden sector. The cosmology of these neutrinos gives rise to both observational constraints and potential discovery channels. In the case of the simplest realization of neutral naturalness, the mirror Twin Higgs \cite{Chacko:2005pe}, the energy density stored in twin neutrinos is in tension with CMB and BBN constraints on dark radiation \cite{Barbieri:2016zxn, Chacko:2016hvu,  Craig:2016lyx}. 
One way to avoid these $\Delta N_{eff}$ constraints is to remove the light degrees of freedom in the hidden sector which are not instrumental for stabilizing the Higgs mass, leading to the Fraternal Higgs like models and their LLP signatures discussed in the above subsection.
Another way of mitigating this tension is by diluting the energy density of the hidden sector at the time of BBN via late decays that preferentially reheat the visible sector\cite{Chacko:2016hvu,  Craig:2016lyx}. In particular, the late decay of right-handed neutrinos, motivated to explain the active neutrino masses and mixings (see Section~\ref{sec:neutrinos}), can reconcile cosmological constraints with the existence of light relics from the hidden sector which stabilizes the Higgs mass.

For the decay of right-handed neutrinos to sufficiently dilute the energy density stored in twin active neutrinos, the right-handed neutrinos must decay preferentially to the Standard Model. 
This is possible in a restricted region of parameter space where the right-handed neutrino masses $m_N$ in the $\mathcal{O}(1 - 10 \gev)$ range (see \cite{Chacko:2016hvu} for details).
However, the allowed parameter space of the theory opens up dramatically if the right-handed neutrinos in both sectors acquire part of their mass through Higgs portal-type couplings of the form
\begin{equation} \label{eq:rhnportal}
\mathcal{L} \supset \frac{1}{2 \Lambda}\left( |H_A|^2 N_A^2  + |H_B|^2 N_B^2 \right)
\end{equation}
where $H_A, H_B$ are respectively Higgs doublets in the Standard Model and mirror twin sectors, and $N_A, N_B$ are corresponding right-handed neutrinos. The couplings in \eqref{rhnportal} ensure that the lightest right-handed neutrino mass eigenstates decay preferentially to the Standard Model, such that the energy density stored in the hidden sector is diluted consistent with observed limits. 

The right-handed neutrino couplings required for viable cosmology in the mirror Twin Higgs also predict displaced decays with mean decay lengths relevant for MATHUSLA. In particular, the couplings in \eqref{rhnportal} give rise to rare displaced decays of the Higgs into right-handed neutrinos,
\begin{equation}
\label{eq:MTHRHneutrinolifetime}
\Gamma(h \rightarrow NN) = \frac{m_h}{16 \pi} \frac{v^2}{\Lambda^2} \left(1 - \frac{4 M_N^2}{m_h^2}\right)^{3/2}
\end{equation}
In order for \eqref{rhnportal} to lead to an appropriate asymmetry in right-handed neutrino decays, the scale $\Lambda$ must be of order $10-10^4$ TeV, corresponding to an exotic Higgs branching ratio of $3 \times 10^{-5} - 3 \times 10^{-1}$ and therefore rates on order of femtobarn or larger at the 14 TeV LHC. 

\begin{figure}
\begin{center}
\includegraphics[width=0.5 \textwidth]{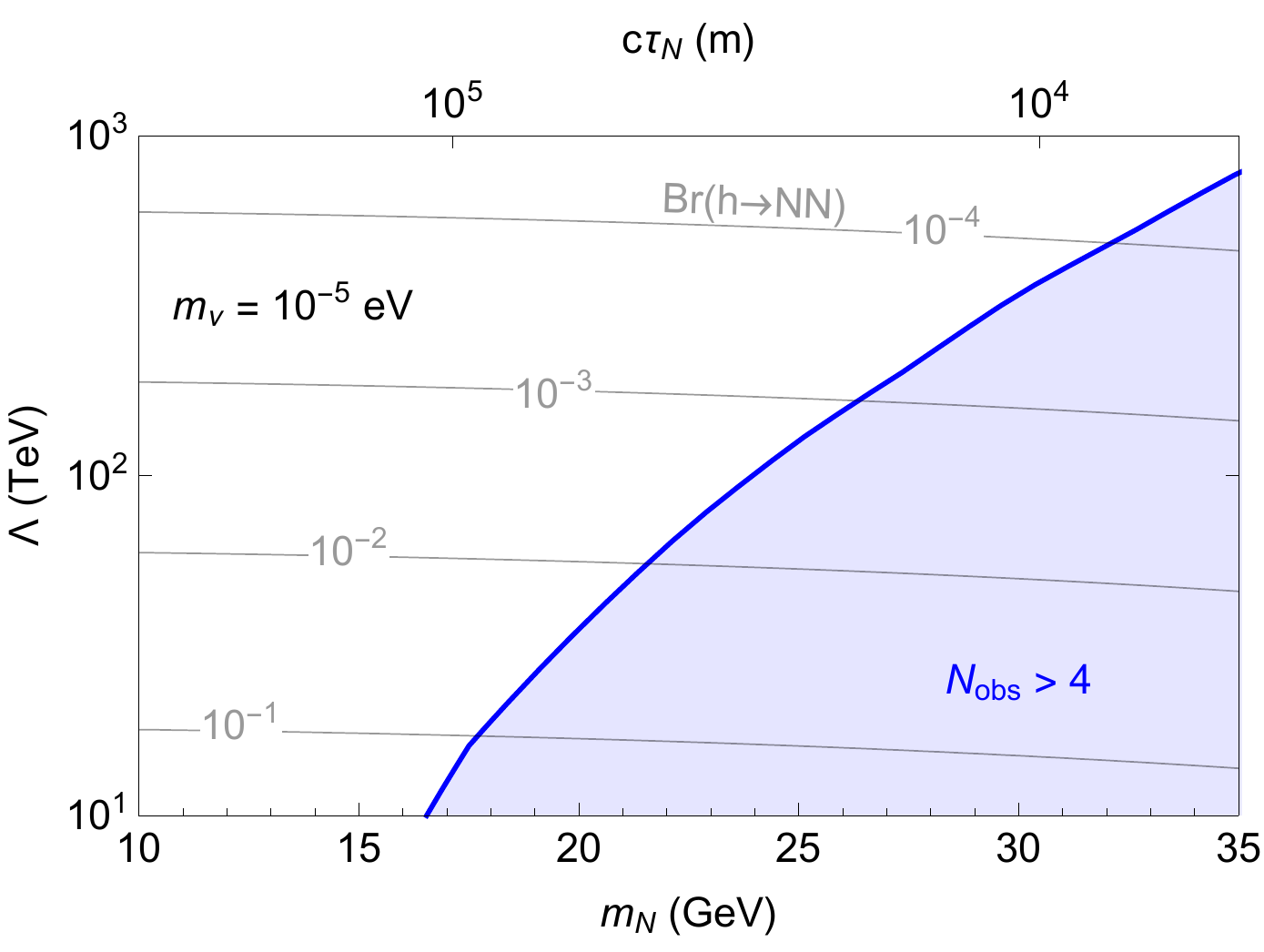}
\end{center}
\caption{
Blue shaded regions: MATHUSLA reach for RH neutrino LLPs in the asymmetrically reheated MTH model with soft $\mathbb{Z}_2$ breaking in the neutrino sector as per Eqn.~(\ref{eq:rhnportal}), setting $C = 1$ in Eqn.~(\ref{eq:MTHRHneutrinolifetime}) and assuming the corresponding active neutrino is very light, with mass $m_\nu = 10^{-5}$ eV. Gray contours show the exotic Higgs decay branching ratio into RH neutrinos.
In this scenario, cosmological constraints on the Mirror Twin Higgs model can be satisfied for $m_N \lesssim 30 \gev$~\cite{Chacko:2016hvu}
These constraints are obtained by remapping the general $\mathrm{Br}(h \to XX)$  MATHUSLA reach projections from Section~\ref{sec:exhdecays}.
This estimate assumes the $200m \times 200m \times 20m$ benchmark geometry of Fig.~\ref{f.mathuslalayout}.
}
\label{f.MTHRHneutrinoMATHUSLAreach}
\end{figure}

Once produced, decays of the right-handed neutrinos then proceed dominantly through the weak interactions back to light Standard Model fermions, with widths of order
\begin{equation}
\Gamma(N \rightarrow {\rm SM})\sim C \frac{G_F^2}{192 \pi^3} \left( \frac{m_{\nu}}{M_N} \right) M_N^5
\end{equation}
where $C$ is a $\mathcal{O}(1)$ number and $m_{\nu}$ is of order the masses of the appropriate active neutrino species. 
In the region of parameter space where Twin Higgs cosmology is consistent with CMB and BBN observables, the mean decay length of right-handed neutrinos ranges from $\sim 10^3$ m to the BBN limit at $\sim 10^7$ m. While this range of rates and lifetimes for exotic Higgs decays is challenging to probe at the LHC main detectors, it is ideally suited for MATHUSLA, as illustrated in Fig.~\ref{f.MTHRHneutrinoMATHUSLAreach}. 
Given the long-lifetime regime of this Higgs-portal LLP search, MATHUSLA will have several orders of magnitude better reach than the HL-LHC main detectors, see Section~\ref{sec:exhdecays}.

In this section, we only discussed a few possible scenarios in which the states and couplings required by neutral naturalness give rise to displaced decays at the LHC. In all cases, production rates at the 14 TeV LHC are considerable, but constraining the parameter space is challenging using the LHC main detectors alone. In this respect, MATHUSLA would contribute significantly to comprehensive coverage of scenarios of neutral naturalness.

\subsection[Composite Higgs]{Composite Higgs\footnote{Peter Cox, Tony Gherghetta, Andrew Spray}}
\label{sec:compositehiggs}

An intriguing possibility is that the Higgs boson could be a composite state of some new, underlying strong dynamics that confines above the TeV scale. Assuming that a global symmetry of the strong dynamics is spontaneously broken at a scale $f$, the Higgs boson can then be identified as a Nambu-Goldstone boson, whose mass is protected by a shift symmetry.  To actually generate a Higgs potential and mass, this symmetry must be explicitly broken. Motivated by partial compositeness, this is achieved via a linear mixing of elementary fields and composite operators with couplings that are related to the SM gauge and Yukawa couplings (see Ref.~\cite{Panico:2015jxa} for a review). Currently, direct searches at the LHC for new resonances and deviations in Higgs couplings, which would provide evidence of the strong dynamics, place an approximate lower limit of $f\gtrsim$\, TeV, suggesting that composite Higgs models are becoming less natural. Furthermore, there are indirect limits from flavor and precision electroweak observables. While the precision electroweak constraints from the T parameter are avoided with a custodial symmetry, and those from the S parameter are ameliorated with sufficiently heavy vector resonances, the most stringent constraints actually arise from flavor observables, which give rise to an approximate lower bound on the scale of spontaneous symmetry breaking $f \gtrsim 10$ TeV~\cite{Panico:2015jxa}. It is therefore clear that composite Higgs models require additional model building in order to maintain naturalness and satisfy these constraints.

Instead, a more minimal approach is to simply assume that $f \gtrsim 10$\,TeV. Of course this simplicity comes at the price of a tuning in the Higgs potential, of order $v^2/f^2 \simeq 10^{-4}$. This meso-tuning is still a many orders of magnitude improvement compared to that encountered in the Standard Model with a Planck scale cutoff, and leads to an unnatural (or split) version of composite Higgs models, 
akin to models of Split Supersymmetry which preserve SUSY's various attractive features at the cost of meso-tuning (see Section~\ref{sec:minisplit}).
  Interestingly, even though the resonances are now very heavy, these models can still give rise to distinctive experimental signals, such as LLPs. The crucial requirement involves improving gauge coupling unification due to a composite right-handed top quark~\cite{Agashe:2005vg}. The minimal coset preserving this one-loop result, together with a discrete symmetry needed for proton and dark matter stability, is $SU(7)/SU(6)\times U(1)$ with $f\lesssim 100-1000$~TeV~\cite{Barnard:2014tla}. This coset contains twelve Nambu-Goldstone bosons, forming a complex $\bf 5$, comprising the usual Higgs doublet, $H$, a color triplet partner, $T$, and a complex singlet, $S$ that can be a stable dark matter candidate. In addition, the composite right-handed top quark, needed for gauge coupling unification, is part of a complete $SU(6)$ multiplet containing extra exotic states, $\chi^c$, that will be degenerate with the top quark. These states can be made sufficiently heavy by pairing them with top companions, $\chi$, to form a Dirac mass of order $f$.

The particle spectrum of the unnatural (or split) composite Higgs model therefore consists of the pseudo Nambu-Goldstone bosons, $H, T$ and $S$ with masses $\ll f$, which are split from the resonances with masses $> f$, while the top companions have Dirac masses of order $f$. Thus for $f \gtrsim 10$\, TeV, the color triplet partner, $T$, of the Higgs doublet will generically be the lightest new colored state~\cite{Barnard:2014tla}. Its dominant decay mode is $T\rightarrow t^c b^c S S$ which arises from a dimension-six term, where $t^c(b^c)$ are the right-handed top (bottom) quarks and $S$ is the singlet scalar. The decay length is given by
\begin{equation}
       c\tau = 100~{\rm m}\, \biggl( \frac{1}{c_3^T} \biggr)^2 \biggl( \frac{8}{g_\rho} \biggr)^3 \biggl( \frac{3~{\rm TeV}}{m_T} \biggr)^5 \biggl( \frac{f}{200~{\rm TeV}} \biggr)^4 \frac{1}{J(m_t, m_S)} \,,
\end{equation}
where $c_3^T$ is an order one constant, $g_\rho$ a strong-sector coupling, $m_T$ $(m_S)$ is the color triplet (singlet) scalar mass and $J(m_t, m_S)$ is a phase space factor (see Ref.~\cite{Barnard:2015rba} for details). Thus, since the scale $f\geq 10$ TeV, the color triplet is long-lived and can decay via displaced vertices or outside the LHC main detectors with lifetimes in the most sensitive range for MATHUSLA.

The possible LLP signals at the LHC were analyzed in Ref.~\cite{Barnard:2015rba}.  The color triplet scalar is pair-produced via QCD and then hadronizes to form an R-hadron. Roughly 50\% of these are charged and can leave a track in the inner detector, and possibly the muon chambers.  If the triplet is collider-stable (i.e.\ decays outside the main detectors), R-hadron searches at the LHC can be used to place limits on its mass. 
Limits from Run-I results~\cite{ATLAS:2014fka} forbid a collider-stable color triplet with a mass below 845~GeV. At Run-II and beyond similar searches will be performed, and with 300~fb$^{-1}$ of integrated luminosity triplet masses up to about 1.4~(1.5)~TeV can be discovered (excluded) for lifetimes corresponding to $c\tau\gtrsim 10$~m. These results are depicted in Figure~\ref{fig:13TeVf10}.  As shown, the efficiency of these searches decreases for lighter DM masses; displaced vertex searches have greater sensitivity when the triplet predominantly decays inside the LHC detectors.  Larger values of $f$ increase the triplet lifetime, and correspondingly enlarge the region where the R-hadron searches dominate.  The mass reach remains unchanged at 1.5~TeV since it is set by the $f$-independent QCD production cross-section.

 \begin{figure}[h]
  \centering
  \includegraphics[width=0.6\textwidth]{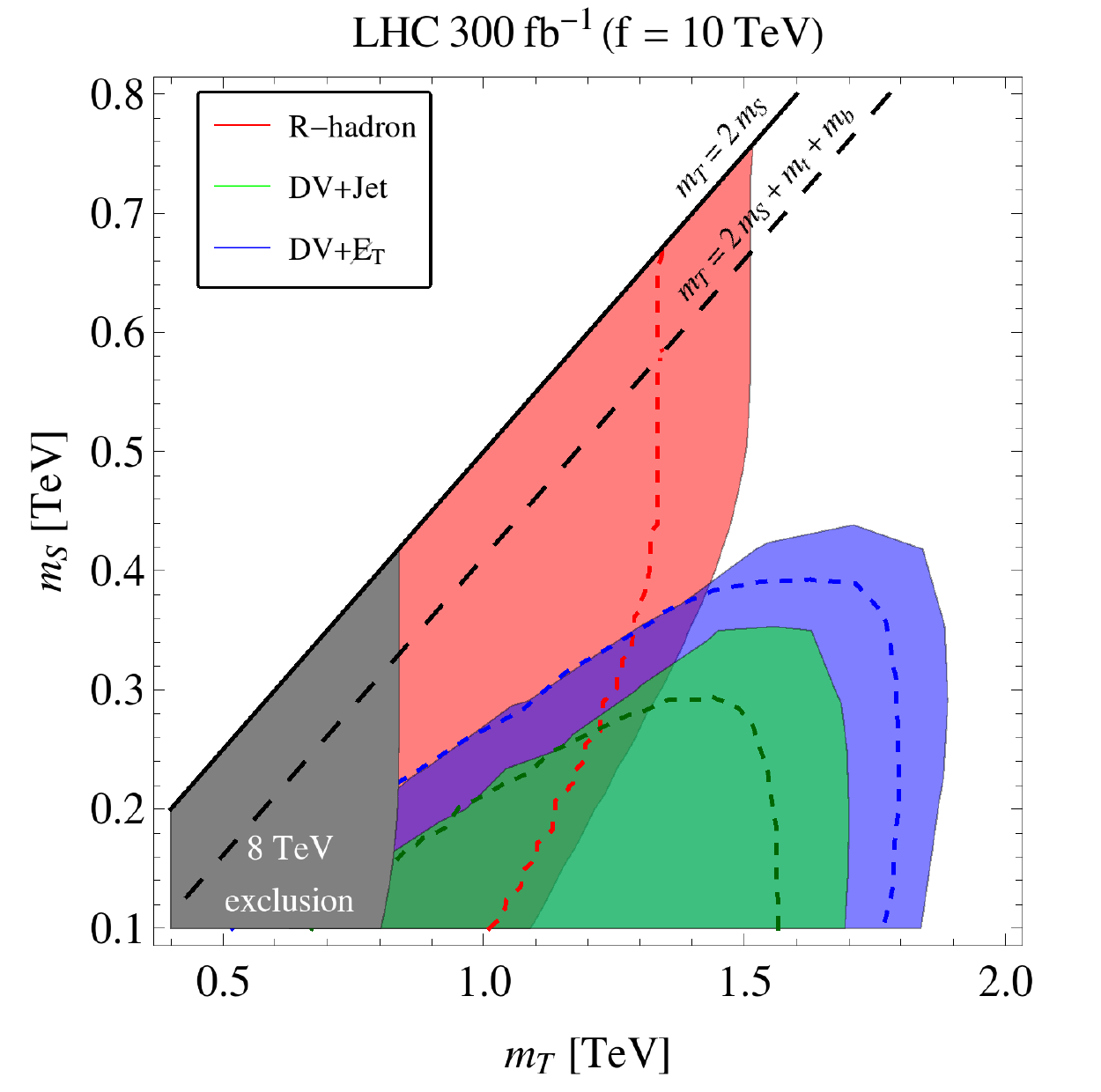}\ 
  \caption[]{Projections for the R-hadron (collider-stable) and displaced-vertex searches at the LHC with 300 fb$^{-1}$ of integrated luminosity at $\sqrt{s}=13$~TeV, as functions of the scalar mass $m_S$ and triplet mass $m_T$. The shaded regions can potentially be excluded at 95\% CL and the dashed lines denote the $5\sigma$ discovery reach. The grey shaded region is excluded by current R-hadron searches at $\sqrt{s}=8$~TeV. This figure is taken from Ref.~\cite{Barnard:2015rba}}
  \label{fig:13TeVf10}
\end{figure}

To be relevant for the MATHUSLA experiment the R-hadrons will need to pass through the substantial rock layer.  An estimate of the survival probability suggests a large fraction will indeed make it to the surface.  This is simply because the R-hadrons are heavy (much more than several hundred GeV), the energy loss per interaction is less than a GeV, and the interaction length is of order 0.5\,m in rock.  Furthermore, the charge of an R-hadron can change upon interactions with matter, such that the majority of R-hadrons formed by the color triplet are expected to be neutral when they reach the surface~\cite{Mackeprang:2009ad}.

We use the procedure outlined in section~\ref{sec:LLPSmathuslahllhc}, and in particular the estimate of Eq.~(\ref{e.NobsMATHUSLA}), to project the expected number of R-hadron decays in the MATHUSLA detector volume.  The color triplet $T$ is pair-produced ($n_{\text{LLP}} = 2$) via QCD, with a production cross-section $\sigma^{\text{LHC}}_{\text{sig}} \sim 0.4$~fb at $m_T = 1.5$~TeV~\cite{Borschensky:2014cia}.  Production is dominantly near threshold, \emph{i.e.} a boost factor $b \approx 1$.  We show in Figure~\ref{fig:ch_MTH_rch} the regions in the $(m_T, m_S)$ plane where we expect $N_{\text{obs}}^{\text{MATHUSLA}}$ greater than 4 and 1, for two different values of $f$.  The pattern of the exclusions is simple to understand.  The number of R-hadrons produced is set by QCD, and so is a function of $m_T$ (only).  This determines the maximum mass that can be excluded to be $\sim 1.55$~TeV.  MATHUSLA has optimal sensitivity for decay lengths $c\tau \approx$~200\,m, which for smaller values of $f$ requires an additional phase space suppression; the probed parameter space is then close to the kinematic boundary $m_T = 2m_S + m_t + m_b$.  As $f$ increases, less of a suppression is required and so the exclusions move to smaller values of the DM mass.

 \begin{figure}[h]
  \centering
  \includegraphics[width=0.6\textwidth]{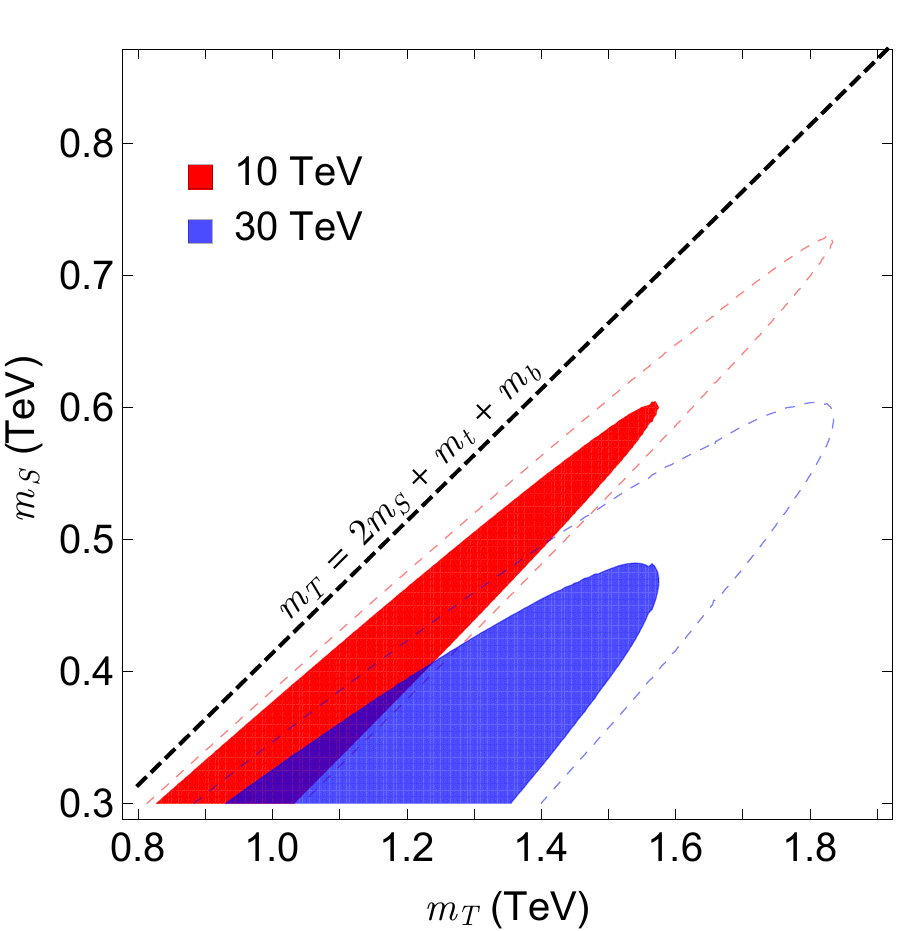}
  \caption[]{Projected sensitivity of MATHUSLA to the minimal unnatural composite Higgs, for compositeness scale $f = 10$ and $30 \tev$, and a total HL-LHC integrated luminosity $\mathcal{L} = 3000$\,fb$^{-1}$.  In the shaded regions, we expect 4 signal events, while the dashed contours bound the region where at least one event is predicted.  We also show the boundary where the four-body triplet decay is kinematically allowed.
  This estimate assumes the $200m \times 200m \times 20m$ benchmark geometry of Fig.~\ref{f.mathuslalayout}.}
  \label{fig:ch_MTH_rch}
\end{figure}

Comparing Figures~\ref{fig:13TeVf10} and~\ref{fig:ch_MTH_rch}, we see that MATHUSLA is unlikely to set stronger limits than the LHC in this minimal model.  
This is consistent with the arguments laid out in Section~\ref{s.LHCLLPcomparison}. The main detector searches for heavy hadronic particles that are either detector-stable or give rise to displaced vertices have very low background and suffer no particular trigger limitations. 
However, just as is the case for long-lived gluinos in split-SUSY (Section~\ref{sec:minisplit}), MATHUSLA will provide an important complementary discovery channel for these long-lived hadronic particles. Any LLP search, when correlated with the stable charged particle search that reveals mass and cross-section information, will reveal information on the lifetime of the LLP and hence the compositeness scale $f$. MATHUSLA will also provide  information about the propagation of heavy hadronic states in ordinary matter.

\subsubsection{Beyond the minimal model}

The unnatural composite Higgs should be understood as a framework, rather than purely the specific minimal model discussed above.   As such, while MATHUSLA has 
rather limited exclusion potential in the minimal scenario, it may be able to probe alternative implementations more effectively.  The limitations found above derive from the LLP forming charged R-hadrons significantly often,  
about one-half of the time, which in turn is due to the LLP having the gauge quantum numbers of a down quark.  
Models with an electrically neutral LLP may be more promising, and we discuss several possibilities below.

As noted previously, all composite Higgs models consistent with gauge coupling unification will have top companion fermions $\chi$, $\chi^c$.  In the above we took them to be heavy, $m_\chi \sim f \gg m_T$.  However, their masses are determined by the Yukawa couplings, $y_\chi$ of the elementary states $\chi$ to the composite sector, $m_\chi \sim y_\chi f$.  Taking all the $y_\chi \sim 1$ is the simplest possibility, but it is technically natural for one or more to be small.  With a modest $y_\chi \sim 0.01$\,--\,$0.1$ and $f = 10$\,--\,100~TeV, top companions in the TeV range are possible.  Their decays will proceed through the strong sector and be suppressed by $f$, so they will also be LLPs if $m_\chi < m_T$.  They then determine the phenomenology in appropriate regions of parameter space.

Applying this reasoning to the minimal model, we note that the top companions comprise a $\mathbf{5}$ and incomplete $\mathbf{10}$ of $SU(5)$.  In particular, the former contains a lepton-like doublet $\tilde{l}$, which has electroweak production cross-sections and the charged component will decay quickly to the neutral one.  SM gauge symmetries plus the requirement that the low-energy physics respects baryon and lepton number force the six-body decay $\tilde{l} \to S^\dagger S^\dagger S^\dagger qqq$ via an off-shell $T$, where $q$ is the third generation quark doublet.  In the strictly minimal model, a number of accidental symmetries suppress this decay further to the two-loop level, resulting in a lifetime too large for MATHUSLA to have any sensitivity.  However, since these additional symmetries are not phenomenologically required, we can imagine breaking them so that the decay length is
\begin{equation}
	c\tau_{\tilde{l}} \sim 100\, \text{m} \, \frac{1}{\epsilon^2} \, \biggl( \frac{8}{g_\rho} \biggr)^3 \biggl( \frac{\text{800 GeV}}{m_{\tilde{l}}} \biggr)^{12} \biggl( \frac{m_T}{\text{3 TeV}} \biggr)^4 \biggl( \frac{f}{\text{60 TeV}} \biggr)^8 \,,
\end{equation}
where $\epsilon \ll 1$ parameterizes the size of the breaking of the accidental symmetries.  We see that this is generically long-lived.  Based on the production cross-sections for an electroweak doublet~\cite{Fuks:2012qx,Fuks:2013vua} we expect MATHUSLA to have a potential sensitivity up to $m_{\tilde{l}} \lesssim 1.25$~TeV.  This is most relevant when the $S$ is on the Higgs resonance, $2m_S \sim m_h$, since in that case DM searches have little sensitivity.

It is also possible to consider different global symmetries of the composite sector.  This will lead to additional Nambu-Goldstone bosons that can potentially be long-lived for the same reasons that the triplet is in the minimal model.  If any of these are color singlets, they will then be more susceptible to MATHUSLA searches.  We outline two particularly promising possibilities.  First, the symmetry breaking pattern $SU(7)/SU(6) \times U(1)$ is minimal in the number of Goldstones, but is non-minimal in the symmetries preserved by the strong sector.  Gauge coupling unification requires only that we preserve an $SU(5)$ global symmetry; dark matter then requires an appropriate additional symmetry to distinguish the Goldstones.  In particular, noting that $SU(7)/SU(6) \times U(1)$ breaking can be achieved by a spurion adjoint, we consider an alternative adjoint-induced breaking $SU(7)/SU(5) \times U(1) \times U(1)$.  The Goldstones additionally include a second complex $\mathbf{5}$ of $SU(5)$, which is constrained to decay to DM plus visible sector particles. This will include an inert doublet that can potentially be a neutral LLP.\footnote{The third $SU(5)$-consistent possibility for an adjoint spurion, $SU(7)/SU(5)\times SU(2) \times U(1)$, has the second $\mathbf{5}$ but no longer has the complex singlet.  The inert doublet is then forced to be the dark matter.}  Of particular note is that the SM couplings to the composite sector in this model are unchanged from the minimal scenario, while there are fewer top companions.

Second, recent developments~\cite{Barnard:2013zea, Ferretti:2013kya,Cacciapaglia:2015eqa} in the UV completions of composite Higgs models have identified a number of symmetry breaking patterns as particularly promising.  The most relevant possibility for an $SU(5)$ GUT is $SU(N)\times SU(N)/SU(N)$, where symmetry breaking is induced by vector-like fermions in a complex representation of the confining gauge group.  The smallest such group consistent with gauge coupling unification and where the Goldstones include a Higgs is $SU(6) \times SU(6)/SU(6)$; dark matter stability requires that we extend this to $U(6) \times U(6)/U(6)$.  The pseudo Nambu-Goldstone bosons comprise a complex $\mathbf{5}$ of $SU(5)$ that contains the Higgs, as well as a real $\mathbf{24}$ and a real singlet.   The details of the Nambu-Goldstone spectrum, top companions and elementary-composite couplings remain to be explored.  However, we note that the $\mathbf{24}$ contains a neutral color octet charged under the DM symmetry.  In some regions of parameter space, it can be an LLP, while its quantum numbers prefer the production of neutral R-hadrons, weakening LHC bounds. Thus the MATHUSLA experiment could provide valuable information on these types of models.

\newcommand{\sint}{s_{\theta}}
\newcommand{\mphi}{m_{\phi}}

\subsection[Relaxion Models]{Relaxion Models\footnote{Jason L. Evans, Tony Gherghetta, Natsumi Nagata}}
\label{sec:relaxion}

A novel idea to address the hierarchy problem introduces a new axion-like field, the relaxion,
that couples to the Higgs field and dynamically relaxes to a field value that partially
cancels the quadratic divergence in the Higgs mass squared~\cite{Graham:2015cka}. The
relaxion coupling to the Higgs induces a Higgs-relaxion mixing which can lead to long-lived decays of the
relaxion. Furthermore, a two-field supersymmetric generalization of the relaxion mechanism~\cite{Espinosa:2015eda, Batell:2015fma, Evans:2016htp}, that addresses the little-hierarchy problem in supersymmetric theories,
can also incorporate inflation where the second field is identified as the inflaton~\cite{Evans:2017bjs}. This has several possible signals at MATHUSLA, where the Higgs mixing with the inflation sector leads to a long-lived relaxion and the Higgsino-relaxino mixing leads to long-lived gauginos.

\subsubsection{Higgs-Relaxion mixing}
\label{sec:Higgsrelaxionmixing}

The relaxion can be produced through Higgs-relaxion mixing as discussed in \cite{Graham:2015cka, Espinosa:2015eda, Batell:2015fma, Evans:2016htp,Evans:2017bjs}. The relevant interaction, which is present in all of these models, is
\begin{eqnarray}
{\cal L} \supset C\frac{h^2}{\Lambda} \Lambda_N^3 e^{i \frac{\phi}{f}}+h.c. =2|C|\frac{h^2}{\Lambda} \Lambda_N^3\cos\left(\frac{\phi}{f}+\delta\right)~,
\label{eq:RelaxStop}
\end{eqnarray}
where $h$ is the real component of the SM Higgs field which obtains a VEV, $\phi$  is the relaxion, $f$ is the decay constant associated with the breaking of a global U(1) symmetry, $\Lambda$ is a UV scale typically associated with integrating out a heavy particle, $\Lambda_N$ is the confinement scale of some strongly coupled gauge theory, and we have taken $C=|C|e^{i\delta}$ in the second equality. The Lagrangian term (\ref{eq:RelaxStop}) leads to mixing with the Higgs boson under two different conditions. For single field relaxion models, we can take $\delta=0$ and still have non-zero mixing because $\phi$ stops for $\sin\left(\frac{\phi}{f}\right)\sim 1$.  For the two-field relaxion model, the relaxion mass continues to evolve after EWSB, due to the dynamical evolution of the second field. This causes  $\sin\left(\frac{\phi}{f}\right)$, which determines the Higgs-relaxion mixing, to be quite small in our vacuum today. Therefore for some parameters of the two field relaxion model, the decays we discuss below may need $\delta \ne 0$, although for the two-field relaxion model in \cite{Evans:2016htp}, $\delta \sim 0$ is needed in order to stop the relaxion and this generically  leads to a long-lived relaxion. Constraints on this operator already exist and can be found in  \cite{Choi:2016luu,Flacke:2016szy}, although Ref.~\cite{Choi:2016luu} assumed a single field relaxion model.

Expanding the stopping potential (\ref{eq:RelaxStop}) around the local minimum
$\phi_0=\langle \phi \rangle$ leads to the interactions
\begin{eqnarray}
{\cal L} \supset 2 \lambda' \sin\left(\frac{\phi_0}{f}+\delta\right)\frac{v^2}{f} h^2\phi +  \lambda'\cos\left(\frac{\phi_0}{f}+\delta\right) \frac{v^2}{f^2}h^2\phi^2+..
\end{eqnarray}
where we have assumed
\begin{eqnarray}
|C|=\lambda'\frac{v^2\Lambda}{\Lambda_N^3}~,
\end{eqnarray}
with $\lambda'\lesssim 1$ in order that the relaxion VEV does not lead to a Higgs mass contribution larger than the weak scale. This gives a branching fraction
\begin{eqnarray}
{\rm BR}(h\to \phi \phi) = 4\times 10^{-4} \left(\frac{\lambda'}{1}\right)^2\left(\frac{c_\delta}{1}\right)^2 \left(\frac{10^4~{\rm GeV}}{f}\right)^4~,
\end{eqnarray}
where $c_\delta=\cos\left(\frac{\phi_0}{f}+\delta\right)$ and we have assumed $m_\phi \ll m_h$. This branching fraction falls right in the middle of the range considered in \cite{Chou:2016lxi}.

Next we look at the decays of the relaxion.  For the models we consider, the dominant decay modes of the relaxion arise via its mixing with the SM Higgs due to the term proportional to $\sin\left(\frac{\phi_0}{f}+\delta\right)$ in Eq.(\ref{eq:RelaxStop}).  Since this mixing is small compared to the Higgs mass, it has little effect on the relaxion and Higgs masses. In the small angle approximation we obtain\footnote{In \cite{Evans:2016htp} for $m_{SUSY}\sim 10$ TeV, this size of mixing is realized for $\delta \simeq 0$.}
\begin{eqnarray}
\theta_{\phi h} \simeq  10^{-3} \left(\frac{\lambda'}{1}\right)\left(\frac{s_\delta}{10^{-2}}\right) \left(\frac{10^4~{\rm GeV}}{f}\right)~,
\label{eq:HiggsRelaxMix}
\end{eqnarray}
where $s_\delta=\sin\left(\frac{\phi_0}{f}+\delta\right)$ and we have  assumed $m_\phi \ll m_h $. This scenario has several constraints coming from many different experiments as shown in Ref.~\cite{Alekhin:2015byh}, that constrains the scalar-Higgs mixing, $\theta_{\phi h} \gtrsim 10^{-3}$ for much of the parameter space.
However, in the relaxion model, since the mixing angle $\theta_{\phi h}$ can be controlled by adjusting $\delta$ without affecting the production, much of the unconstrained parameter space from previous experiments can be realized. This means that relaxion models have mixing angles and production rates which can be seen at the MATHUSLA experiment.

Now we examine the decay length of the relaxion to verify that it does indeed decay inside the detector.  If the mixing in Eq.(\ref{eq:HiggsRelaxMix}) produces the dominant decay mode(s) for the relaxion, the relevant perturbative interactions are of the form
\begin{eqnarray}
{\cal L} \supset \sin \theta_{\phi h} \frac{m_f}{v} \phi \bar f  f~,
\label{eq:Phifbarf}
\end{eqnarray}
where $m_f$ is the mass of the fermion $f$, and $v$ is the SM Higgs VEV.  A perturbative calculation leads to the decay width
\begin{eqnarray}
\Gamma_{\phi\to \bar f f} = \frac{1}{8\pi}\left(\frac{\sin\theta_{\phi h} m_f}{v}\right)^2\left[1-\frac{4m_f^2}{m_\phi^2}\right]^{3/2}m_\phi~,
\end{eqnarray}
which will be the dominant decay mode for certain masses of the relaxion.  However, for most masses the dominant decay mode will be to mesons, where non-perturbative effects dominate and the above perturbative approximation breaks down.  The lifetime and branching fractions of a particle with a dominant decay mode coming from the interactions in Eq.(\ref{eq:Phifbarf}) can be found  in Ref. \cite{Alekhin:2015byh, Evans:2017lvd}.
Since this signal of the relaxion is identical to the LLP signal in the SM+S simplified model discussed in Section~\ref{sec:singlets}, the region which can be detected for this model at the MATHUSLA experiment can be read off of Fig.~\ref{fig:HBRScalar} by identifying $\theta \equiv \theta_{\phi h}$ and $S \equiv \phi$. It is clear that MATHUSLA can probe deep into the model's parameter space, using both exotic Higgs decays and Meson decays as a source of LLPs.

\subsubsection{Mixing with the Inflation Sector}
\subsubsubsection{Higgs mixing}
An interesting feature of the supersymmetric two-field relaxion model discussed in \cite{Evans:2017bjs}
is the identification of the inflaton with the second field that controls the amplitude of the stopping potential. The SM Higgs can then mix with fields in the inflation sector. In this model inflation occurs along a $D$-term flat direction which is lifted by radiative corrections, as is typical of $D$-term inflation. The inflation scale is pushed down to values less than a few GeV by taking the $U(1)$ gauge coupling associated with the $D$ flat direction to be very small. One of the consequences of this very small gauge coupling is a very light waterfall field, that also has a very large VEV of order the GUT scale. This combination makes it hard to reheat since anything the waterfall field couples to strongly will be heavier than the waterfall field. This difficulty can be circumvented if the waterfall field is coupled to a supersymmetric $F$ flat direction which involves the Higgs field~\cite{Evans:2017bjs}
\begin{eqnarray}
W\supset \kappa_1 R \phi_+\langle M_-\rangle  +\kappa_2 R H_uH_d +m_R R\bar R~.\label{eq:ReheatDtermInf}
\end{eqnarray}
In this expression $\phi_+$ is the waterfall field, $\langle M_-\rangle$ is a residual VEV from some heavy field responsible for generating $\langle D \rangle$ during inflation, $R$ is a singlet whose $F$-term contribution to the potential leads to interactions between the waterfall field and the SM Higgs, and $\bar R$ has no other interactions and adjusts its VEV so that $F_R=0$ is preserved\footnote{This cancels large corrections to the Higgs $B$-term so the relaxion process still works.}.  This superpotential gives rise to
the Higgs interaction
\begin{eqnarray}
-{\cal L} \supset \kappa_2' m_{\phi_+}\phi_+ h^2 +h.c.
\end{eqnarray}
where
\begin{eqnarray}
\kappa_2=\kappa_2' \frac{2m_{\phi_+}}{\sin2\beta~\kappa_1\langle M_-\rangle }~.
\end{eqnarray}
In the small angle limit this leads to a mixing between the SM Higgs and $\phi_+$ given by
\begin{eqnarray}
\theta_{\phi h} \simeq 10^{-1} \left(\frac{\kappa_2'}{1}\right)\left(\frac{m_{\phi_+}}{5~{\rm GeV}}\right)~,\label{eq:HiggsRelaxMixInf}
\end{eqnarray}
and we have again assumed $ m_{\phi_+} \ll m_h$. This situation is very analogous to the Higgs-relaxion mixing in Section~\ref{sec:Higgsrelaxionmixing}.  By varying $\kappa_2$ or $\kappa_1$,  we can reduce the mixing angle
and extend the decay length.  Furthermore, the $\phi_+$ mass is independent of the mixing angle, so that our constraints are similar\footnote{There is another possible constraint on these models that we have not mentioned. If the coupling $\kappa_{1,2}$ causes the universe to reheat to too high of a temperature, the universe may become overclosed. This constraint can be avoided but is somewhat model-dependent.} to those found in Ref.~\cite{Alekhin:2015byh}.
Again, this relaxion implements the SM+S simplified model, and the region of parameter space that is discoverable at MATHUSLA is shown in  Fig.~\ref{fig:HBRScalar} (Section~\ref{sec:singlets}).

Note also that while the decay width is similar to the Higgs-relaxion case, the production mode is very different. The waterfall field $\phi_+$ is instead generated through $B$ meson decays, and since the LHC will produce many $B$-mesons the production should be sufficient.  More details can be found in Section~\ref{sec:singlets} and Ref.~\cite{Evans:2017lvd}.

\subsubsubsection{Higgsino Mixing}
\label{sec:higgsinomixing}
Given the superpotential interactions in Eq.(\ref{eq:ReheatDtermInf}), we next describe the
effect that these interactions have on the neutralino lifetimes. Since $\tilde R$ will generally be much lighter than the MSSM higgsinos, the higgsinos can decay to $\tilde R$. However, the higgsinos of this model are much too heavy to be produced at the LHC. Because the interaction in Eq.(\ref{eq:ReheatDtermInf}) also generates mixing of $\tilde R$ with the higgsinos,
\begin{eqnarray}
{\cal L}\supset \kappa_2\tilde R \left(v_u \tilde H_d +v_d \tilde H_u\right) +h.c.\label{eq:RmixNeut},
\end{eqnarray}
interactions between $\tilde R$ and the MSSM neutralinos will be generated. As we will see, for certain ranges of  $\kappa_{1,2}$ these interactions can lead to events in the MATHUSLA detector.

There is one complication: since $\tilde R$ is stable, decays of the neutralinos to $\tilde R$ can overclose the universe. The dominant production mode of $\tilde R$ is through Higgsino decays to $\tilde R$. Thus, the production of $\tilde R$ can be drastically reduced by reheating below the Higgsino mass\footnote{The production of $\tilde R$ can also be suppressed by taking $\kappa_2$ very small. For a reheat temperature above the weak scale, as was considered in \cite{Evans:2017bjs}, this method of suppression only works for some of the parameter space.}. For the model in \cite{Evans:2017bjs}, this means the reheat temperature can be as high as $10^6$ GeV. If the reheat temperature is larger than the Bino and/or Wino mass, the mixing in Eq.(\ref{eq:RmixNeut}) could still cause the Universe to overclose.  However, if the decay width is sufficiently small that the decay of the Bino/Wino occurs after freeze out, then as long as the Bino or Wino thermal density is small enough, the decay to $\tilde R$ will not overclose the Universe.  The Bino or Wino will freeze out at a temperature of order $T_f\sim \frac{M_{i}}{25}$, where $M_i$ for $i=1,2$ is the Bino or Wino mass respectively. Since the Wino or Bino will not decay until $\Gamma \sim H$, we can estimate the upper bound on the decay width of the Wino and Bino,
\begin{eqnarray}
\Gamma(\chi\to XX) \lesssim H(T_f) &=& 7.5\times 10^{-16}~{\rm GeV}\left(\frac{g_\rho}{106.75}\right)^{1/2}\left(\frac{M_i}{10^3~{\rm GeV}}\right)^2~,\\
&=& \left(0.26~{\rm m}\right)^{-1}\left(\frac{g_\rho}{106.75}\right)^{1/2}\left(\frac{M_i}{10^3~{\rm GeV}}
\right)^2~,
\end{eqnarray}
where $g_\rho$ is the number of relativistic degrees of freedom and $M_i$ is the LSP mass, and
$\chi$ denotes either a Wino or Bino. Interestingly, this decay length falls right in the range where MATHUSLA is sensitive.

Therefore to obtain a viable reheating that does not over produce $\tilde R$,  we need to reheat with a temperature less than the Higgsino mass while giving the LSP a decay length longer than about a meter. The reheat temperature which is generated through the interactions in Eq.(\ref{eq:ReheatDtermInf}) is
\begin{eqnarray}
T_R=153~ {\rm GeV} \times \left(\frac{106.75}{g_\rho}\right)^{1/4}\left(\frac{\langle M_-\rangle}{10^{16}~{\rm GeV}}\right)\left(\frac{10^3~{\rm GeV}}{m_{\phi_+}}\right)^{1/2}
\left(\frac{\kappa_1}{10^{-13}}\right)\left(\frac{\kappa_2}{10^{-7}}\right)~,
\end{eqnarray}
where the normalizations of the couplings are typical values for an inflation scale of $H_I=1$~GeV.

Next we determine the decay length of the Wino and Bino. The easiest way to find the Bino/Wino interactions with $\tilde R$ is to integrate out the higgsinos which gives
\begin{eqnarray}
{\cal L}\supset  \kappa_2 \frac{M_Z}{\mu} \left(s_\beta^2-c_\beta^2\right)\left(-c_W \tilde R \lambda_3+s_W\tilde R\lambda_0\right)\frac{h}{\sqrt{2}} +h.c.
\end{eqnarray}
where $h$ is the SM Higgs, $\lambda_3$ the neutral Wino and $\lambda_0$ the Bino. Using these interactions, the Wino decay width in the limit $M_2\gg m_h,m_R$ is given by
\begin{eqnarray}
\Gamma_{\lambda_3\to h \tilde R} &=& \frac{|\kappa_2|^2}{32\pi} \left(\frac{M_Z}{\mu}\right)^2 c_{2\beta}^2 c_W^2 M_2~,\\ \nonumber &\simeq & (150~{\rm m})^{-1} \left(\frac{|\kappa_2|}{10^{-7}}\right)^2\left(\frac{10~{\rm TeV}}{\mu}\right)^2 \left(\frac{c_{2\beta}}{0.5}\right)^2 \left(\frac{M_2}{1~{\rm TeV}}\right)~,
\end{eqnarray}
where $c_{2\beta}=\cos2\beta$. For a Bino LSP, the decay width is obtained by substituting $c_W\to s_W$ and $M_2\to M_1$.  As is clear from the above expressions, this is a prime candidate for detection at MATHUSLA.

Although the direct production rate of TeV neutralinos at the LHC is negligible, they can be produced through gluino decays.  Since the decay rate of the gluino to $\tilde R$ is quite small, all gluinos produced will eventually become the lightest neutralino.  Because of this, the lightest neutralino production rate will be equal to the gluino production rate, which is large enough to produce MATHUSLA signals in a range of LLP lifetimes for gluino masses up to a few TeV, see Fig.~\ref{f.benchmarkxsecs}. The reach projection in gluino mass will then be similar to the RPV search for gluinos decaying into neutralino LLP, shown in Fig.~\ref{fig:RPVplots}. 

\subsubsection{Decays to Relaxinos}
Another possible signal of relaxion models which could be seen at MATHUSLA are neutralino decays to the relaxino.  In supersymmetric relaxion models, the relaxion superfield couples to the field-strength superfield in a similar manner to the axion,
\begin{eqnarray}
{\cal L} \supset \int d^2 \theta \frac{C}{32\pi^2} \frac{S}{f} W^{a\alpha} W^a_\alpha + h.c.
\end{eqnarray}
where $C$ is an order one coefficient\footnote{Decays from this operator were also discussed in \cite{Batell:2015fma} with $S$ being the axion superfield.}.  This superpotential term leads to the following interactions
\begin{eqnarray}
{\cal L} \supset \frac{\sqrt{2}Cg_a^2}{32\pi^2}  \bar \lambda^a \Sigma^{\mu\nu} \gamma_5 \tilde S F_{\mu\nu}^a~,
\end{eqnarray}
where $\lambda^a$ is the gaugino Majorana fermion, $\tilde S$ is the relaxino Majorana fermion, and $F_{\mu\nu}^a$ is the field-strength tensor with gauge group index $a$. This leads to the gaugino decay width
\begin{eqnarray}
\Gamma_{\lambda^a\to \tilde S +\gamma / Z} &=& \frac{|C|^2}{128\pi}\left(\frac{g_a^2}{8\pi^2}\right)^2 \left(\frac{M_{\lambda^a}}{f}\right)^2 M_{\lambda^a}~,\\ &=& \nonumber  \left(171~{\rm m}\right)^{-1}
\left(\frac{|C|}{1}\right)^2\left(\frac{M_{\lambda^a}}{1~{\rm TeV}}\right)^{3} \left(\frac{3\times 10^8~{\rm GeV}}{f}\right)^2~,
\end{eqnarray}
where for the $\lambda^a\to \tilde S Z$ decay mode, we have absorbed some order one coefficients depending on the weak mixing angle into $C$. This decay length can be seen to be within the range of MATHUSLA, where the production mechanism occurs through gluinos, as discussed in Section~\ref{sec:higgsinomixing}.

\subsubsection[Relaxion models with additional ultra long-lived BSM particles (ULLPs)]{Relaxion models with additional ultra long-lived BSM particles (ULLPs)\footnote{Thomas Flacke, Claudia~Frugiuele, Elina~Fuchs, Rick~S.~Gupta, Gilad~Perez and Matthias~Schlaffer}}

A heavy relaxion could yield interesting signatures in MATHUSLA if the relaxion can also decay into ultra long-lived BSM particles (ULLPs) $X$. The presence of such particles in relaxion models has been proposed in \cite{Choi:2016kke} in order to solve a cosmological problem. In some cases, reheating can de-stabilize the electroweak vacuum established by the relaxion mechanism. This occurs because the relaxion field starts to re-roll after re-heating, and adding a coupling of the relaxion to an ULLP, which in  \cite{Choi:2016kke}  is chosen to be a dark photon, can provide an additional friction term in the evolution equation of the relaxion which can stop the relaxion from rolling. 

If such a ULLP is present and it is produced from relaxion decays, then its production cross-section is dictated by the relaxion production discussed in Ref.~\cite{Flacke:2016szy}, while its lifetime follows from the ULLP couplings to Standard Model particles. Realizations of scalar or dark photon ULLPs are possible to construct. In the following we give estimates of the number of events which could be measured in MATHULSA in a more general description, in terms of the branching ratios, life-times, and masses of the relaxion and the ULLP. This effectively realizes a well-motivated extension of the SM+S simplified model of Section~\ref{sec:singlets}, where $S$ (the relaxion) decays into additional hidden-sector LLP states.

\vspace{4mm} \noindent \emph{Production from $B$-meson decays} \vspace{2mm}
 \\GeV-scale relaxions can be produced in $B \to K \phi$ decays. The $b \bar{b}$ production cross-section at the LHC is around 500$\,\mu b$~\cite{Altarelli:2008xy}.  These $B$-mesons have a branching ratio into relaxions given by\footnote{See Sec.~\ref{sec:singlets}.}
\be
Br(B\to K\phi)\approx 6.2 \sin^2 \theta\,,
\ee
where $\sin \theta$ is the Higgs-relaxion mixing angle.  A current bound on $Br(B\to K\phi)\times Br(\phi \to XX)$ arrises from the bound on $Br(B \to K \nu\bar{\nu})\lesssim 1.4\times 10^{-5}$ (see Eq.(14) of Ref.~\cite{Clarke:2013aya} ). Thus, the largest potential LLP production cross-section from B-decays to a relaxion mixed with the Higgs is
\be
\sigma_{pp \to XX} \lesssim 7 \text{ nb} ~~,
\ee
corresponding to a relaxion-Higgs mixing angle $s_\theta \lesssim 10^{-3}$.  This cross-section is well above the sensitivity expected for MATHUSLA, corresponding to $\sigma \sim 1$ fb, demonstrating that MATHUSLA would be sensitive down to mixing angles $\sin \theta \sim 10^{-6}$.

\vspace{4mm} \noindent \emph{Production from Higgs decays}  \vspace{2mm} \\
The SM Higgs cross-section at 14 TeV is $\sigma_H \approx 50$ pb.  Furthermore, the anticipated limit on invisible Higgs decays from the main detectors at the end of HL-LHC running will be $\mathcal{O}(1-10\%)$.  Thus the largest potential LLP production cross-section from Higgs decays to relaxions mixed with the Higgs which then subsequently decay to LLPs is
\be
\sigma_{pp \to H \to  XX} \lesssim 5 \text{ pb} ~~.
\ee
The branching ratio into relaxions depends not only on the mixing angle, but also other scalar potential parameters, thus the cross-section cannot be simply related to the mixing angle as for B-meson decays.  Nonetheless, the branching ratio can easily surpass $BR(H\to\phi\phi) \gtrsim 2 \times 10^{-5}$,\footnote{See Sec.~\ref{sec:singlets}.} thus MATHUSLA has sensitivity to new LLPs through relaxion-Higgs mixing.

\clearpage


\section{Theory Motivation for  LLPs: Dark Matter}
\label{sec:dmtheory}

The existence of dark matter (DM), comprising some $26\%$ of the present-day energy budget of our universe  \cite{Ade:2015xua}, has been solidly established by several independent lines of gravitational evidence, and provides some of the sharpest evidence for new particle physics at potentially accessible mass scales.  The particle nature of DM remains a mystery.  Null results to date in indirect detection (ID), direct detection (DD), and missing energy searches at colliders have forced models of WIMP DM into severely constrained regions of parameter space, and have helped to stimulate a broader investigation into possible signals of particle dark matter.  There are a wide variety of possible DM candidates, whose experimental signals are intimately connected to the mechanism responsible for populating DM in the early universe.  These DM models often require new BSM states in addition to DM itself, or multi-component DM. In many cases, the mechanism that yields the correct relic density for DM then naturally and generically results in one or more of these BSM states having a proper decay length on collider scales.  In other cases, long lifetimes are not a direct consequence of the mechanism that determines the DM relic abundance, but are a generic feature of models that implement it.

There are several reasons why a particle may have a collider-scale lifetime: a renormalizeable coupling controlling its decay may simply be tiny; the phase space available for its decay may be  small; and/or its available decays may be suppressed by high mass scales.  As we demonstrate in this chapter, all of these mechanisms are naturally realized in well-motivated DM models.  For instance, small phase space is a generic prediction of models where WIMPs coannihilate with an additional particle in the early universe.  
In this case the cosmically-mandated small mass splitting $\Delta$ between DM and its coannihilating partner can frequently result in long partner lifetimes, as discussed in  Sec.~\ref{sec:coannihilation}.   Decays suppressed by high mass scales naturally arise in theories of asymmetric DM, which, motivated by the apparent coincidence $\Omega_{DM} \approx 5 \Omega_b$, relate SM baryon (or lepton) number to a conserved dark number $D$. Relating a baryon number asymmetry to a dark number asymmetry requires new interactions to transfer asymmetries between SM and dark sectors, which can be described through the introduction of transfer operators $\mathcal{O}_{ADM}$ that carry both $B-L$ and dark number $D$.   These operators are necessarily non-renormalizeable, and can lead to displaced decays of (e.g.) SM superpartners, as demonstrated in Sec.~\ref{sec:adm}.  

Models like SIMPs and ELDERs, discussed in Sec.~\ref{sec:simps}, require DM to have rapid number-changing self-interactions and to be (at least initially) in thermal contact with the SM.  These models are perhaps most naturally realized when DM lives in a confining hidden sector.  In this case many of the hidden sector hadrons, notably vector mesons, are natural and attractive targets for collider LLP searches; lifetimes are in this case rendered long thanks in large part to a small portal coupling between the SM and the hidden sector.  Along with ADM scenarios, SIMPS and its relatives represent hidden valleys (Section~\ref{sec:hiddenvalleys}) realizing both Dark Matter and collider signatures.

When DM lives in a sector that is not in thermal equilibrium with the SM in
the early universe, a variety of novel possibilities open up for the
thermal history of DM and thus for its signals today
\cite{Pappadopulo:2016pkp,Berlin:2016gtr,Bernal:2017kxu}.  Such thermally decoupled
hidden sectors require the leading coupling between the HS and the SM
to be very small, and thus generically these hidden sectors will contain
LLPs.  However, the leading coupling between the SM and the HS is then
generally too small to allow that same coupling to mediate production of HS
particles at rates large enough to be observable at the LHC and
MATHUSLA \cite{Kahlhoefer:2018xxo, marcoduccio}.  Probing non-equilibrated dark
states at both LHC and MATHUSLA thus requires the LLP to be produced
in cascade decays of a BSM parent particle.  This is the case in e.g. freeze-in scenarios (discussed in Sec.~\ref{sec:freezein}), as well as co-decaying models (Sec.~\ref{sec:simps} below).  Conversely, the same observation allows us to immediately
conclude that any LLPs observed at MATHUSLA whose production and decay
are governed by the same couplings would have been in thermal
equilibrium with the SM in the early universe.  
In that case, their lifetime is bounded from above so as not to disrupt Big Bang Nucleosynthesis, see Section~\ref{sec:bbn}. 
Hidden sector dark matter therefore provides strong motivations for LLPs: for instance, LLPs in the simple models
SM+S, SM+V (see Secs.~\ref{sec:singlets} and~\ref{sec:darkphotons})  are the leading collider signal of a
class of secluded DM models \cite{Evans:2017kti}, and a scenario where MATHUSLA can
probe unique territory.

An important exception to the BBN constraint is provided by the Dynamical Dark Matter framework, see Section~\ref{sec:dynamicaldm}, which generalizes the notion of a single or a few hyperstable DM states to an ensemble of states with varying lifetimes. Some of these constitute the DM abundance today, while heavier states in the ensemble may be produced at the LHC and decay with lifetimes observable by MATHUSLA.

\subsection[Coannihilation]{Coannihilation\footnote{Felix Yu}}
\label{sec:coannihilation}

Dark matter coannihilation~\cite{Griest:1990kh} offers an attractive
and useful twist to the standard dark matter relic abundance
calculation.  Typically, dark matter is assumed to be in thermal
equilibrium with Standard Model particles in the early universe, and
as the universe expands and cools, the comoving number density of the
dark matter falls exponentially.  As the temperature cools below about
$T \sim m_{\text{DM}} / 25$, however, the condition for chemical
equilibrium fails, and the dark matter relic abundance is set.
Moreover, in generic dark sectors, the additional particles in the
model can significantly affect the ultimate dark matter abundance.
Since the thermal freeze-out of dark matter occurs at finite
temperature, these particles can have significant number densities as
long as they are relatively close in mass to the dark matter, with
lifetimes that are cosmologically negligible but very attractive
targets for \textsc{Mathusla}.  These particles, which are not the
dark matter, will contribute new channels to the effective thermally
averaged cross-section for dark matter annihilations to Standard Model
particles.  From Ref.~\cite{Griest:1990kh}, the effective annihilation
cross-section is
\begin{align} 
\sigma_{\text{eff}} = \frac{g_{\rm DM}^2}{g_{\text{eff}}^2} \bigg\{ & \sigma_{\rm DM \, DM} + 2 \sigma_{\rm DM \, X} \frac{g_{\rm X}}{g_{\rm DM}} (1 + \Delta)^{3/2} \exp(-x \Delta) \nonumber \\
  & + \sigma_{\rm X \, X} \frac{g_{\rm X}^2}{g_{\rm DM}^2} (1 + \Delta)^3 \exp(-2 x \Delta) \bigg\} \ ,
\label{eq:sigmaeff}
\end{align}
where $\Delta = (m_{\text{X}} - m_{\text{DM}} )/m_{\text{DM}}$, $x =
m_{\text{DM}} / T$, $g_a$ counts the number of degrees of freedom
(spin, color, etc.)  for $a = \text{DM}$ or X, the coannihilation
partner, and $g_{\text{eff}} = \sum_{i = 1}^N g_i (1 + \Delta_i)^{3/2}
\exp (-x \Delta_i)$, $\Delta_i = \Delta$ for X and $\Delta = 0$ for DM,
is the effective number of degrees of freedom in the dark sector. 
We assume that DM and coannihilation partners are in thermal and chemical equilibrium. 
 In particular, we note that when the dark matter has a vanishingly small
self-annihilation cross-section $\sigma_{\text{DM}\,\text{DM}}$, the
dominant contributions to $\sigma_{\text{eff}}$ are then from the
coannihilation processes $\sigma_{\text{DM}\,\text{X}}$ and
$\sigma_{\text{X}\,\text{X}}$.

The coannihilation process in the early universe can also play a
dominant role in determining the phenomenology of dark matter
production at colliders~\cite{Izaguirre:2015zva, Baker:2015qna,
  Buschmann:2016hkc}.  In particular, since the mass splitting between
the dark matter and the coannihilation partner is typically small, the
coannihilation partner can be long-lived on collider timescales.  Even
if the mass splitting is relatively large, the dominant Standard Model
coannihilation products may be heavily kinematically suppressed, and
this suppression can also lead to long-lived signatures.  For
\textsc{Mathusla}, we illustrate these ideas with a concrete
simplified model.

\subsubsection{Coannihilation through Higgs mediator}
\label{subsec:Higgscoannihilation}

We study the following Lagrangian, in the broken phase of electroweak symmetry,
\begin{align}
\mathcal{L} \supset \bar{\chi} (i \slashed{\partial} - m_\chi) \chi +
\bar{\psi} (i \slashed{\partial} - m_\psi) \psi + (y h \bar{\chi} \psi
+ \text{ h.c.}) \ ,
\end{align}
where $m_\psi > m_\chi$ and hence $\psi$ is long-lived while $\chi$ is
the dark matter.  This structure can be realized in, for example, the
minimal supersymmetric model when the lightest and second-to-lightest
neutralinos are both dominantly bino-Higgsino admixtures\footnote{In
  this case, the third lightest neutralino is also a bino-Higgsino
  admixture, and the lack of a direct Higgs-mediated $\bar{\chi} \chi$
  annihilation channel can be motivated by taking a very small
  bino-Higgsino mixing angle for the lightest neutralino.}, and the
relevant mass range can be taken in $\mathcal{O}(100)$~GeV to several
hundred GeV range (cf.~Fig.~11 of Ref.~\cite{Cheung:2012qy}),
corresponding to pair production cross-sections of 1-100 fb, see also Fig.~\ref{f.benchmarkxsecs}.
Separately, these states can be produced in cascade decays of heavier
colored particles, as in supersymmetric quarks, extending the mass
reach to roughly 1.2~TeV, but additional coannihilation channels
should be included if the mass difference between the squarks and the
dark matter is too compressed.
In this simplified Higgs mediator scenario, as shown in Section~\ref{s.METcomparison}, direct searches for the unstable coannihilation partner $\psi$ at \textsc{Mathusla} can have much larger sensitivity than direct MET searches, especially if the squark-neutralino mass difference is even moderately compressed. 

We have three free parameters: $\Delta = (m_\psi - m_\chi) / m_\chi$,
$m_\chi$, and $y$.  The partial decay width of $\psi$ is
\begin{align}
\Gamma (\psi \to f \bar{f} \chi) = \frac{N_c y^2 y_{f}^2}{15 \pi^3}
\Delta^5 \frac{m_\chi^5}{m_H^4} \ ,
\end{align}
where $N_c = 3$ for quark final states and $1$ for leptons, and the SM
final state masses are neglected.  For simplicity, if we only consider
one three-body decay mode, and if the fractional mass splitting
$\Delta = 0.01$, the corresponding $\psi$ decay length is macroscopic,
\begin{align}
c \tau \sim 2.24 \times 10^{1} \text{ m} \frac{1}{N_c y^2} 
\left( \frac{10^{-3}}{y_f} \right)^2 
\left( \frac{10^{-2}}{\Delta} \right)^5 
\left( \frac{100 \text{ GeV}}{m_\chi} \right)^5 \ .
\end{align}
We remark that the decay length will be modified by $\mathcal{O}(1)$
factors depending on the interplay between the available phase space,
given by $m_\chi \Delta$, and the QCD phase transition, but this
estimate immediately points to the $\mathcal{O}(100)$m lifetimes in
the target zone for \textsc{Mathusla}.  Moreover, in this benchmark
model, the Higgs-mediated decays are dominantly hadronic and hence
very difficult to detect from comparable LLP searches at the HL-LHC for masses below a few hundred GeV. MATHUSLA is therefore expected to increase sensitivity by orders of magnitude due to its background-free environment, as
estimated in Section~\ref{s.LHCLLPcomparison}.

We also remark that when annihilating to heavy flavors, such as top
quarks, the multiplicities of on-shell final state particles will
counterbalance the large Yukawa enhancement and provide additional
corrections to the above lifetime estimate.  Nevertheless, this
estimate demonstrates that macroscopic decays are a characteristic
signature of the coannihilation partner with a small fractional mass
splitting from the dark matter.  If we generalize the Yukawa
interaction to a singlet scalar mediator whose couplings to pairs of
SM particles are free parameters, then the $y_f$ Yukawa parameters
above become model-dependent and new targets for \textsc{Mathusla}
open up.

\subsection[Asymmetric Dark Matter]{Asymmetric Dark Matter\footnote{Kathryn Zurek}}
\label{sec:adm}

Asymmetric Dark Matter (ADM) is a class of hidden sector or hidden valley DM models where the DM density is set by its coupling to SM baryon or lepton number.  Because $\rho_DM/\rho_b \sim 5$ observationally, the natural mass scale for DM in this model is $m_X \sim 5 m_p$, though depending on the details of the model, other masses are possible.  Higher dimension operators share a primordial asymmetry between the two sectors, and then decouple at low temperatures to separately freeze-in the asymmetry in the visible and dark (hidden valley) sector.  See Ref.~\cite{Zurek:2013wia} for a review of these models.   Asymmetric dark matter (ADM) arising from a hidden sector naturally gives rise to LLPs at the LHC \cite{Kaplan:2009ag}, as we discuss here,  following Ref.~\cite{Kim:2013ivd}.

To transfer the asymmetry between sectors, we need a higher dimension operator connecting the SM operator ${\cal O}_{B-L}$ which carries no Standard Model gauged quantum number, but carries $B-L$, to a hidden valley operator ${\cal O}_X$ carrying DM number:
\begin{equation}
\label{eq:admop}
{\cal O}_{ADM} = \frac{{\cal O}_{B-L} {\cal O}_X}{M^{n+m-4}},
\end{equation}
where ${\cal O}_{B-L}$ and ${\cal O}_X$ have dimension $m,~n$, respectively. 
These interactions may be embedded in either a supersymmetric or non-supersymmetric theory.  
In the supersymmetric case, the simplest operators are 
\begin{equation}
\label{eq:basicADM}
W_{\rm ADM} = X \ell H,~~\frac{X u_i^c d_j^c d_k^c}{M_{ijk}},~~\frac{X q_i \ell_j d_k^c}{M_{ijk}},~~~\frac{X \ell_i \ell_j e_k^c}{M_{ijk}}.
\end{equation}
These ADM interactions de-stabilize the lightest ordinary supersymmetric particle (LOSP) to decay into the $X$-sector plus additional SM particles, and can naturally lead to very long LOSP lifetimes.  In non-supersymmetric implementations of ADM, the interactions of Eq.~\ref{eq:admop} can also naturally lead to LLPs, but the precise details of the LLP signatures will generally depend on the specific UV completion.  For simplicity and generality, we will thus focus on the supersymmetric case, although qualitatively similar conclusions apply to non-supersymmetric models as well.

\begin{figure}[t]
\centering
\begin{tabular}{ccc}
\includegraphics[width=5cm]{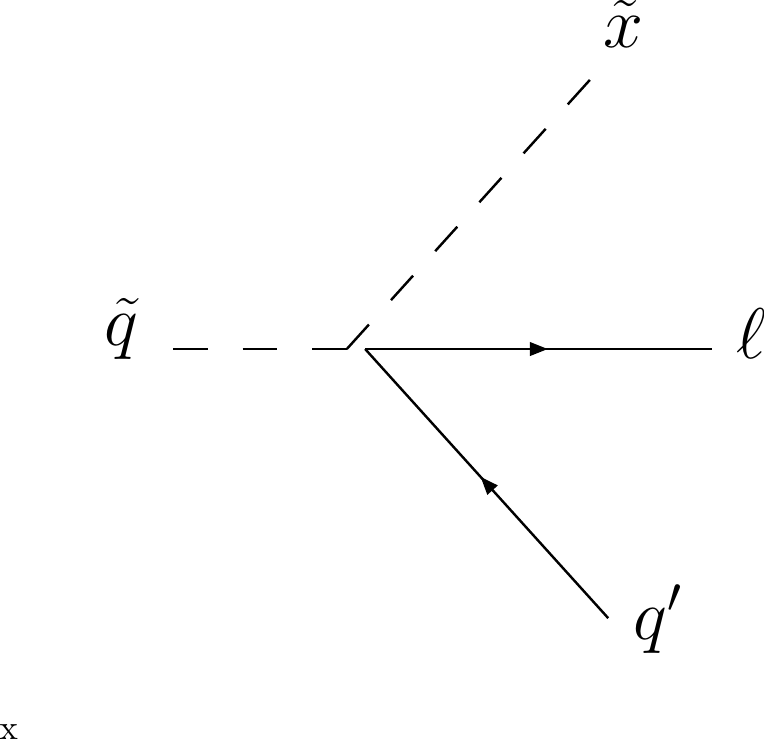}
&
\hspace{3mm}
&
\includegraphics[width=5cm]{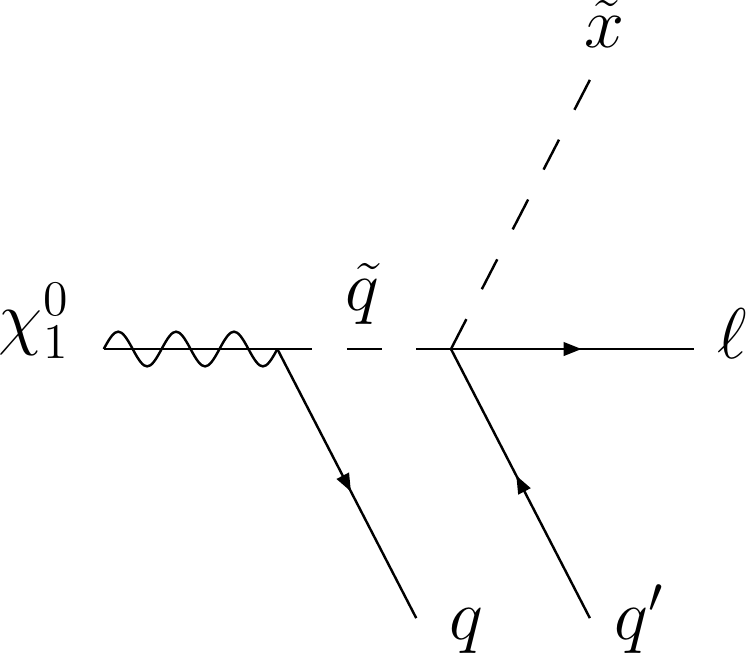}
\\
(a) & & (b)
\end{tabular}
\caption{\label{fig:feynman_xqld_decay}  {\em Left}: 3-body decay of a squark LOSP directly through the interaction of Eq.~(\ref{eq:basicADM}), and {\rm Right}: 4-body decay of a neutralino LOSP through an off-shell squark for $q \ell d^c$ models. Here, the quark flavors $q$ and $q'$ are generically different. $\tilde{x}$ denotes the scalar component of the ADM supermultiplet $X$. Decay of a slepton LOSP and a neutralino LOSP through an off-shell squark is also given by the same diagrams trading a squark and a lepton with a slepton and a quark, respectively. Figure from Ref.~\cite{Kim:2013ivd}. } 
\end{figure}

For example, as shown in Fig.~\ref{fig:feynman_xqld_decay}, in the model with ${\cal O}_{B-L} = q \ell d^c$, a squark LOSP decays to $\tilde X,~\ell,~q$, while a neutralino LOSP decays via an off-shell squark to $q,~q',~\ell,~\tilde X$, where the tilde denotes the scalar superpartner.  As detailed in Ref.~\cite{Kim:2013ivd}, the lifetime of the LOSP in the decay process depends on the scale $M$ of the operator.  This scale is in turn constrained by flavor physics.  

The lifetime for the squark or slepton LOSP decay process, shown in the left side of Fig.~\ref{fig:feynman_xqld_decay}, and ignoring masses of final state particles, is
\begin{eqnarray}
c \tau \sim 100 \mbox{ m} \times \left(\frac{10^{-5} \mbox{ mm}}{F^{\rm (3-body)}}\right)\left(\frac{1~{\rm TeV}}{m_{\rm LOSP}} \right)^3 \times 
\left(\frac{M_{ijk}}{3 \times 10^{11}~{\rm GeV}} \right)^2.  \label{eq:displaced_vertex_losp_3_body_generic}
\end{eqnarray}
The lifetime for the 4-body neutralino LOSP, as shown in the right side of Fig.~\ref{fig:feynman_xqld_decay} is
\begin{eqnarray}
 c \tau  &\sim &  100 \mbox{ m} \times \left(\frac{100 \mbox{ mm}}{F^{\rm (4-body)}}\right) \times \left(\frac{M_{ijk}}{10^8~{\rm GeV}} \right)^2 
\times \left( \frac{m_\phi}{1500~{\rm GeV}} \right)^4 
\times \left( \frac{500~{\rm GeV}}{m_{\rm LOSP}} \right)^7 \times 
\label{eq:displaced_vertex_losp_4_body_generic}  \\  
&& \qquad \times x^5\left[ 
 (10 x^3 - 120 x^2 + 120 x) + 60 (1-x) (2-x) \log (1-x) 
\right]^{-1} , \nonumber 
\end{eqnarray}
where $\phi$ is the intermediate squark or slepton and $x = (m_{\rm LOSP} / m_\phi )^2$.  The 3-body and 4-body coefficients, $F^{{\rm (3-body)}}$ and $F^{{\rm (4-body)}}$, are calculated in Ref.~\cite{Kim:2013ivd}.  Typical values are $(F^{{\rm (3-body)}})^{-1}\sim{\rm few} \times 10^{-5} \mbox{ mm}$ and $(F^{{\rm (4-body)}})^{-1}\sim{\rm few} \times 100 \mbox{ mm}$.  Both processes give rise to macroscopic proper lifetimes depending on the supersymmetric particle masses and the scale of the ADM operator $M_{ijk}$. 
Four-body neutralino decays in particular can easily give rise to proper decay lengths in the $\sim \mathrm{100 \,m}$ range well-suited to MATHUSLA.

The LLPs in supersymmetric ADM theories arise naturally from the need to transfer the matter asymmetry using higher-dimension operators, combined with small phase space and the hierarchy of scales imposed by flavor physics. Consequently, the existence of LLPs is largely independent of the detailed superpartner spectrum, though the precise value of the lifetime can depend sensitively on the masses of the LOSP and the lightest relevant sfermion.  Within the ADM scenario, the superpartner spectrum is mostly relevant for determining the LOSP lifetime and for setting the overall production of SUSY particles at the LHC. The collider signatures of ADM theories are broadly similar to those of RPV SUSY, for which MATHUSLA's sensitivity can be read off from Fig.~\ref{fig:RPVplots}. Heavier sfermions translate into longer lifetimes for the LOSP, so for a neutralino ULLP either direct electroweak pair production or production from parent gluino cascades are especially well-motivated.

In the cases when the operator mediating LLP decay involves a light lepton ($W_{\rm ADM} = X \ell H,~~\frac{X q_i \ell_j d_k^c}{M_{ijk}},~~~\frac{X \ell_i \ell_j e_k^c}{M_{ijk}}$ in Eq.~\ref{eq:basicADM}), the neutralino LOSP decay will be detectable at the HL-LHC with low background owing to the high-$p_T$ lepton produced in the decay. In this case MATHUSLA will have sensitivity comparable to or up to $\sim 10$ times better than the HL-LHC detectors, as discussed in Sec.~\ref{sec:LLPSmathuslahllhc}. Models that preferentially couple to taus through these leptonic operators will have higher backgrounds at the main detectors, and correspondingly greater relative advantage at MATHUSLA. For $W_{\rm ADM} = \frac{X u_i^c d_j^c d_k^c}{M_{ijk}}$, the LOSP has a purely hadronic decay and MATHUSLA can have much better sensitivity than the HL-LHC, by up to three orders of magnitude depending on the overall energy scale of the event.

\subsection[Freeze-In Scenarios]{Freeze-In Scenarios\footnote{Raymond T. Co, Francesco D'Eramo, Lawrence J. Hall, Jose Miguel No, Stephen M. West, Bryan Zaldivar}}
\label{sec:freezein}


Thermal freeze-out is one of the most popular mechanisms for dark matter (DM) production. DM particles have interactions with the thermal bath strong 
enough to be in thermal equilibrium at high temperatures. As the universe cools, and the radiation bath temperature drops below the DM mass, the expansion rate 
becomes larger than the annihilation rate and the DM particles go out of chemical equilibrium. This freeze-out happens at temperatures typically a factor 
of $20$ below the DM mass, and therefore the final DM abundance is insensitive to the history of the universe before freeze-out. 

Freeze-in is another motivated mechanism where DM production is dominated at IR temperatures of the order of the DM mass~\cite{Hall:2009bx}. In such 
scenarios, DM particles are extremely weakly coupled to the thermal bath and never achieve thermal equilibrium. Bath particles scatter and/or decay to 
final states containing the DM particle, and these reactions proceed only in one direction: the DM abundance increases towards equilibrium, but never 
reaches it.   Here, we discuss how DM is produced through freeze-in from the decay of parent particles in the thermal bath. If the interactions mediating this 
process are renormalizable, most of the DM is produced at temperatures of the order the bath particle mass. 
Interestingly, for a wide range of different cosmological evolutions, the decay length required for the observed dark matter abundance leads to displaced signals at colliders and in the MATHUSLA detector.

We review the set-up for freeze-in calculations in Sec.~\ref{sec:FIDM}, and we evaluate the DM relic density for two different cosmological histories. 
First, we consider the standard cosmology where the universe snapshot at the time of Big Bang Nucleosynthesis (BBN) is extrapolated to high temperatures. 
The freeze-in calculation for this radiation-dominated (RD) universe is presented in Sec.~\ref{subsec:FIRD}. 
A motivated modification of the standard history, 
naturally arising in extensions of the standard model of particle physics, is an early matter-dominated (MD) epoch.\footnote{Freeze-in taking place in a universe undergoing a \emph{faster} than standard expansion rate also leads to displaced signatures, as studied in \cite{DEramo:2017ecx}.} We compute DM freeze-in relic density for 
this case in Sec.~\ref{subsec:FIMD}. As a summary of this model-independent analysis, in Sec.~\ref{subsec:DisplacedFI} we present the prediction for the decay 
length of the parent particle under the condition of reproducing the observed DM abundance~\cite{Co:2015pka}. Remarkably, our predictions are in the ballpark 
suitable for MATHUSLA.

In freeze-in scenarios, the DM parent particle could have a variety of couplings to the SM model, but here we focus on neutral parents, since neutral LLPs are of greatest interest to MATHUSLA. 
In Sec.~\ref{subsec:freezeinHiggs}, 
we consider a simplified model of fermionic dark matter, where freeze-in production can proceed via the Higgs portal. 
This minimal benchmark scenario demonstrates the range of neutral LLP signatures predicted by a standard radiation-dominated cosmology, and demonstrates that MATHUSLA can probe a wide portion of the freeze-in DM parameter space.

In Sec.~\ref{sec:DFSZ} we discuss a more complete model which gives rise to freeze-in DM and the associated LLP signatures for a range of different RD and MD early universe cosmologies: the axino in supersymmetric DFSZ theories~\cite{Co:2016fln, Co:2017orl}. 
In this scenario, the Higgsino is the parent LLP which can be observed at colliders and at MATHUSLA, realizing the signatures described in Sec.~\ref{subsec:freezeinHiggs} as well as Sec.~\ref{sec:axino-ewino}.
Freeze-in on a RD background is discussed in Sec.~\ref{subsec:AxinoDM}, requiring very light axinos and hence a low inflationary reheating temperature to avoid the associated gravitino problem. 
This restriction can be avoided if there is a dilution mechanism to reduce the axino abundance to the observed value, realized via an early MD epoch. 
We show in Sec.~\ref{subsec:Saxion} how supersymmetric DFSZ theories naturally incorporate a dilution mechanism through the saxion condensate. 
The diluted axino abundance is computed in Sec.~\ref{subsec:SaxionDilution}, where we describe how this dilution effect leads to an axino abundance consistent with observations, and we generalize these conclusions to arbitrary dilution mechanisms in Sec.~\ref{subsec:generaldilution}. 
We summarize the predicted LLP signals of the DFSZ freeze-in axino DM scenario in Sec.~\ref{subsec:DisplacedAxinos} and show that MATHUSLA puts almost the entire motivated parameter space within our reach.

\subsubsection{Dark Matter Freeze-In}
\label{sec:FIDM}

We consider DM freeze-in production through decays of a parent particle in thermal equilibrium with the plasma. Consistently 
with the notation in Ref.~\cite{Co:2015pka}, we denote this process as follows:
\be
B \; \rightarrow \; A_{\rm SM} \, X \ ,
\label{eq:FIprocess}
\ee
where $B$ is the decaying bath particle, $X$ the DM and $A_{\rm SM}$ is one or more Standard Model particles. The DM abundance is initially negligible, and continuously increases as the bath particles decay. This production process is effective as long as 
the parent particle is relativistic. Once the temperature drops below $m_B$, the abundance of the decaying particle is exponentially suppressed 
and freeze-in is not effective anymore. Most $X$ particles are produced at temperatures $T_{\rm FI} \simeq m_B$.

The DM number density evolves according to the Boltzmann equation~\cite{Hall:2009bx}
\be
\frac{d n_X}{dt} + 3 H n_X =  \Gamma_B  \, n_B^{\rm eq} \,\frac{K_1[m_B/T]}{K_2[m_B/T]} \ ,
\label{eq:BoltzFI}
\ee
where $\Gamma_B$ is the decay rate for the process in Eqn.~\ref{eq:FIprocess} and $K_{1,2}[x]$ are the first and second modified Bessel functions 
of the 2nd kind. The bath particle equilibrium number density $n_B^{\rm eq}$ appearing on the right-hand side of the Boltzmann equation can be obtained 
using Maxwell-Boltzmann statistics 
\be
n_B^{\rm eq} = \frac{g_B}{2 \pi^2} \, m_B^2 T \, K_2[m_B/T] \ .
\ee
At high temperatures ($T \gg m_B$), we recover the $T^3$ dependence for a relativistic species, while at low temperatures ($T \ll m_B$) the number 
density has the Maxwell-Boltzmann exponential suppression. 

The Boltzmann equation Eqn.~\ref{eq:BoltzFI} is general and its validity extends beyond the standard RD cosmology. The details of the cosmological 
history enters through the Hubble parameter $H$ and the time vs temperature relation. In what follows, we present solutions to this Boltzmann equation 
for two different cosmological backgrounds.

\subsubsubsection{Freeze-In for Standard Cosmology}
\label{subsec:FIRD}

In a standard RD cosmological background, the expansion of the universe is driven by its radiation content and the Hubble parameter reads
\be
H = \frac{\sqrt{\rho}}{\sqrt{3} M_{\rm Pl}} = \frac{\pi \, g^{1/2}_*}{3 \sqrt{10}} \frac{T^2}{M_{\rm Pl}} \ .
\ee
Here, $g_*$ is the number of effective relativistic degrees of freedom, taken constant for this discussion. In such a background, the total entropy of 
the radiation bath is conserved, and therefore it is convenient to employ comoving variables. We define the comoving DM density $Y_X = n_X / s$, where $s$ 
is the entropy density. Furthermore, we describe the evolution in terms of the inverse temperature $x = m_B / T$. The Boltzmann equation in terms of these 
dimensionless variables reads
\be
\frac{d Y_X}{d x} = \frac{\Gamma_B}{H \, x} \, Y_B^{\rm eq}(x) \, \frac{K_1[x_B]}{K_2[x_B]} \ .
\ee
The equilibrium comoving number density for the bath particle is defined as $Y_B^{\rm eq} = n_B^{\rm eq} / s$. This differential equation can be integrated, 
with the initial condition that at very high temperatures the abundance of $X$ is vanishing. The final comoving DM number density results in
\be
Y^\infty_X = 4.4\times10^{-12} \, \left( \frac{g_B}{2} \right)  \left(\frac{106.75}{g_*}\right)^{3/2}
 \left(\frac{300 \, \GeV}{m_B}\right) \left(\frac{\Gamma_B / m_B}{1.8 \times 10^{-25}}\right) \ 
\label{eq:RDFIresult}
\ee
where $g_B$ is the internal degrees of freedom of $B$.

This result has to be compared with the observed DM density, which can be expressed in terms of a comoving energy density
\be
\frac{\rho^{\rm obs}_{DM}}{s} = 0.44 {\textrm{ eV}} \ ,
\ee 
close to the temperature of matter radiation equality, $T_{eq} \simeq 1{\textrm{ eV}}$. We can thus rewrite Eqn.~\ref{eq:RDFIresult} as follows
\be
\frac{\rho_{X}}{\rho^{\rm obs}_{DM}} = \frac{m_X Y_X^\infty}{\rho^{\rm obs}_{DM} / s } =
\left( \frac{g_B}{2} \right)  \left(\frac{106.75}{g_*}\right)^{3/2} \left(\frac{m_X}{100 \, \GeV}\right) \left(\frac{300 \, \GeV}{m_B}\right) \left(\frac{\Gamma_B / m_B}{1.8 \times 10^{-25}}\right) \ .
 \label{eq:xiX}
\ee
Upon requiring that this ratio is one, we obtain a prediction for the decay length of $B$ 
\be
c \tau_B \sim 4 \times 10^6 \text{ m } \left( \frac{g_B}{2} \right)  \left(\frac{106.75}{g_*}\right)^{3/2} \left(\frac{m_X}{100 \, \GeV}\right) \left(\frac{300 \, \GeV}{m_B}\right)^2.
\ee
The coupling $\lambda$, defined by 
\be
\Gamma_B = \frac{\lambda^2}{8 \pi}  m_B,
\label{eq:FIparameters}
\ee
must be very small to avoid overclosure. For the benchmark points chosen in Eqn.~\ref{eq:xiX}, the observed DM density results for 
$\lambda \simeq 2 \times 10^{-12}$. Well-motivated examples of such feeble couplings include the gravitino, through interactions suppressed 
by the Planck scale, and the axino, through interactions suppressed by the Peccei-Quinn scale \cite{Co:2016fln}.

As discussed in Sec.~\ref{s.MATHUSLAsignalestimate}, very long decay lengths near the BBN scale of $\sim 10^7$m may be observable at MATHUSLA if the parent particle has a large enough cross-section near the pb range. 
However, the greatest chance for discovery exists for lifetimes below $\sim$ km. 
In freeze-in scenarios, this can be realized in two ways: either either the DM candidate has a mass below the GeV scale (Sec.~\ref{subsec:freezeinHiggs}), or the relic density of the DM candidate is diluted by an earlier MD epoch (Sec.~\ref{sec:DFSZ}) as we describe below.

\subsubsubsection{Freeze-In for an Early Matter-Dominated Epoch}
\label{subsec:FIMD}

The standard cosmology in \Sec{subsec:FIRD} assumes that freeze-in occurs during the RD era and that total entropy is conserved after freeze-in, 
i.e.~$\rho_{DM}/s$ is constant between freeze-in and today. This conventional picture is drastically modified when there exists a late decaying 
matter field $M$. If $M$ decays after dominating the energy content of the Universe, this late decay injects a large amount of entropy and dilutes 
the dark matter abundance. Examples of such matter include inflatons and moduli, and we discuss the saxion as a 
well-motivated candidate in \Sec{subsec:Saxion}.

We now elaborate on the details of the cosmological evolution. 
We begin with the case when $M$ is not the inflaton so the matter energy density $\rho_M$ is initially subdominant to radiation. Due to the 
scaling of $\rho_M$ with the scale factor $a^{-3}$, $\rho_M$ will eventually dominate over that of radiation, which scales as $a^{-4}$. This onset of the matter-dominated epoch occurs at temperature $T_M$. This MD era ends when $M$ decays at the reheat 
temperature $T_R$, and is followed by a radiation-dominated era. 

The MD epoch itself consists of two phases-- adiabatic and non-adiabatic. When the Universe first enters the MD era, the decay of $M$ is 
still inefficient, i.e.~the Hubble rate is much larger than the decay rate of $M$ ($H \gg \Gamma_M$). This means that the radiation energy 
density originated from the existing red-shifted radiation. The total entropy is conserved during this phase, which is thus adiabatic and called MD$_{A}$. 
On the other hand, the relativistic decay products of $M$ will eventually outnumber the original radiation at the temperature we 
call $T_{NA} \sim (T_M T_R^4)^{1/5}$. Between $T_{NA}$ and $T_R$, as radiation is constantly produced by $M$ decay, the Universe 
is being reheated and therefore a large amount of entropy is actively injected. We call this a non-adiabatic phase, MD$_{NA}$. 

In the case of $M$ as the inflaton, inflation ends when $M$ starts to oscillate around the true minimum and at this time $M$ is the dominant 
contribution of the total energy density. Therefore, we enter a MD era immediately after inflation. The decay products of $M$ also quickly 
becomes the dominant source of radiation so the Universe enters a MD$_{NA}$ era without going through a MD$_A$ era. The MD era ends when the 
inflaton completely decays away at $T_R$.

If dark matter is produced before and during the MD$_A$ epoch, the abundance receives the full dilution 
factor $D \approx T_M/T_R \approx (T_{NA}/T_R)^5$. On the other hand, if the dark matter is dominantly produced 
during the MD$_{NA}$ epoch at temperature $T_X$, the partial dilution factor is $D(T_X)=(T_X/T_R)^5$.  In 
particular, for freeze-in $T_X = T_{FI}$. In addition to dilution, in calculating DM abundance, one needs to take into account the different Hubble rate during the MD epoch.

We now show the final results for freeze-in production in a matter-dominated background cosmology. The full derivations can be 
found in Ref.~\cite{Co:2015pka}. The dark matter particle $X$ is constantly produced from the decays of $B$ until $B$ becomes 
non-relativistic with a Boltzmann-suppressed number density. As a result, the freeze-in process is IR-dominated at $T \sim m_B$. 
Dropping numerical factors of $\mathcal{O}(1)$, the final yield today reads
\be
Y_{i} \, \sim \,  \frac{\lambda^2 M_{Pl}}{m_B}  
 \left( 1, \;\;\; 10^4 \, \frac{T_R^7}{m_B^7}, \;\;\; \frac{T_R m_B^{1/2}}{T_M^{3/2}}, \;\;\;\frac{T_R}{T_M} \right).
\label{eq:Y03}
\ee
where $i=1-4$ runs over the cases where freeze-in occurs in the (RD, MD$_{NA}$, MD$_A$, RD$'$) eras, with RD the usual RD 
era at $T<T_R$ and RD$'$ the early one at $T>T_M$. The first (RD) component reproduces the scaling behavior of Eqn.~\ref{eq:RDFIresult}. 
The last three terms in the parentheses of Eqn.~\ref{eq:Y03} are necessarily less than unity, and therefore the abundance is suppressed if 
freeze-in happens during MD$_{NA}$, MD$_A$, and RD$'$ eras. This depletion results from a larger Hubble rate and/or large dilution due to the entropy 
production of $M$ decays.

\begin{figure}
\begin{center}
\begin{tabular}{cc}
\includegraphics[scale=0.4]{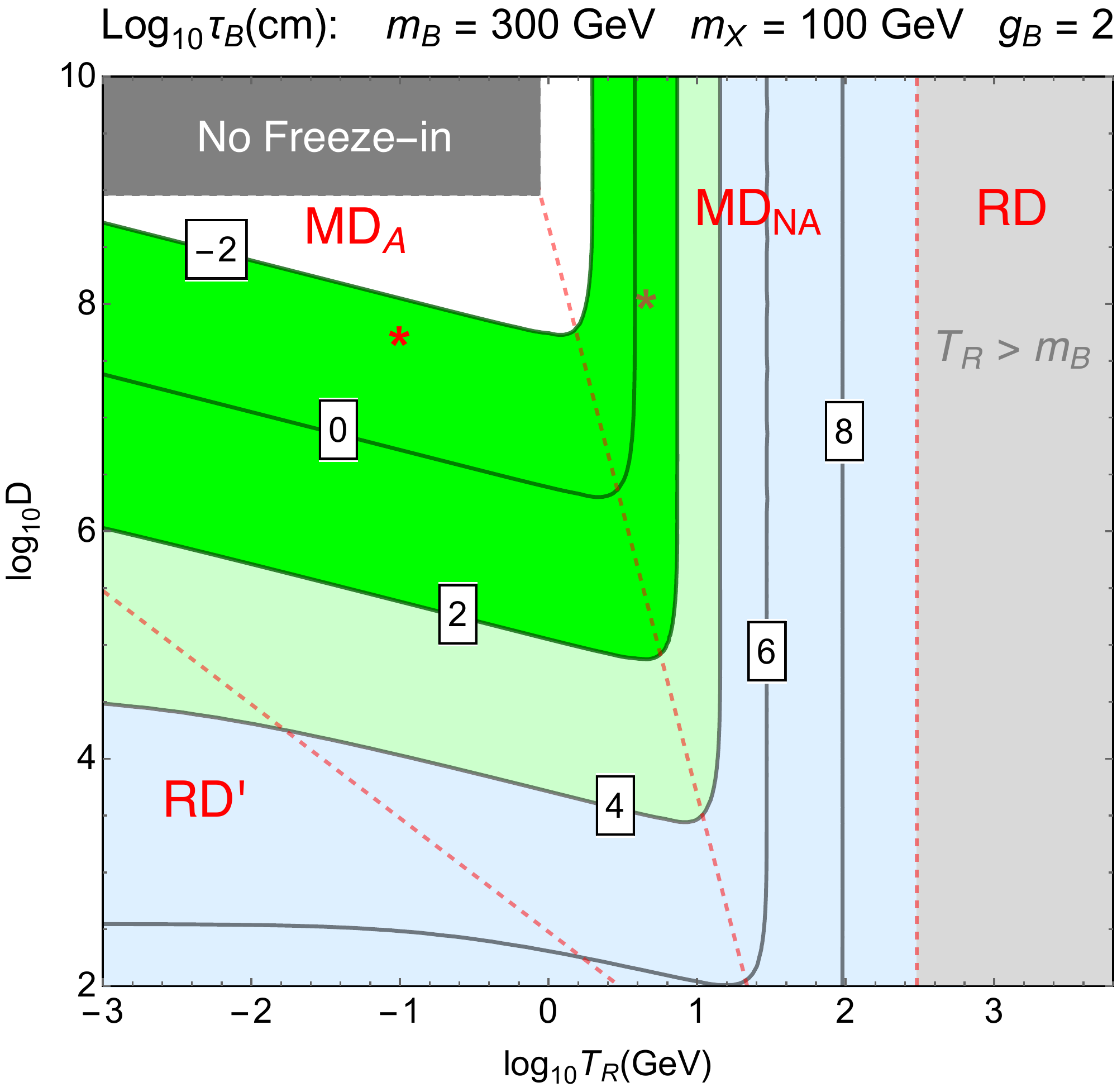} 
&
 \includegraphics[scale=0.4]{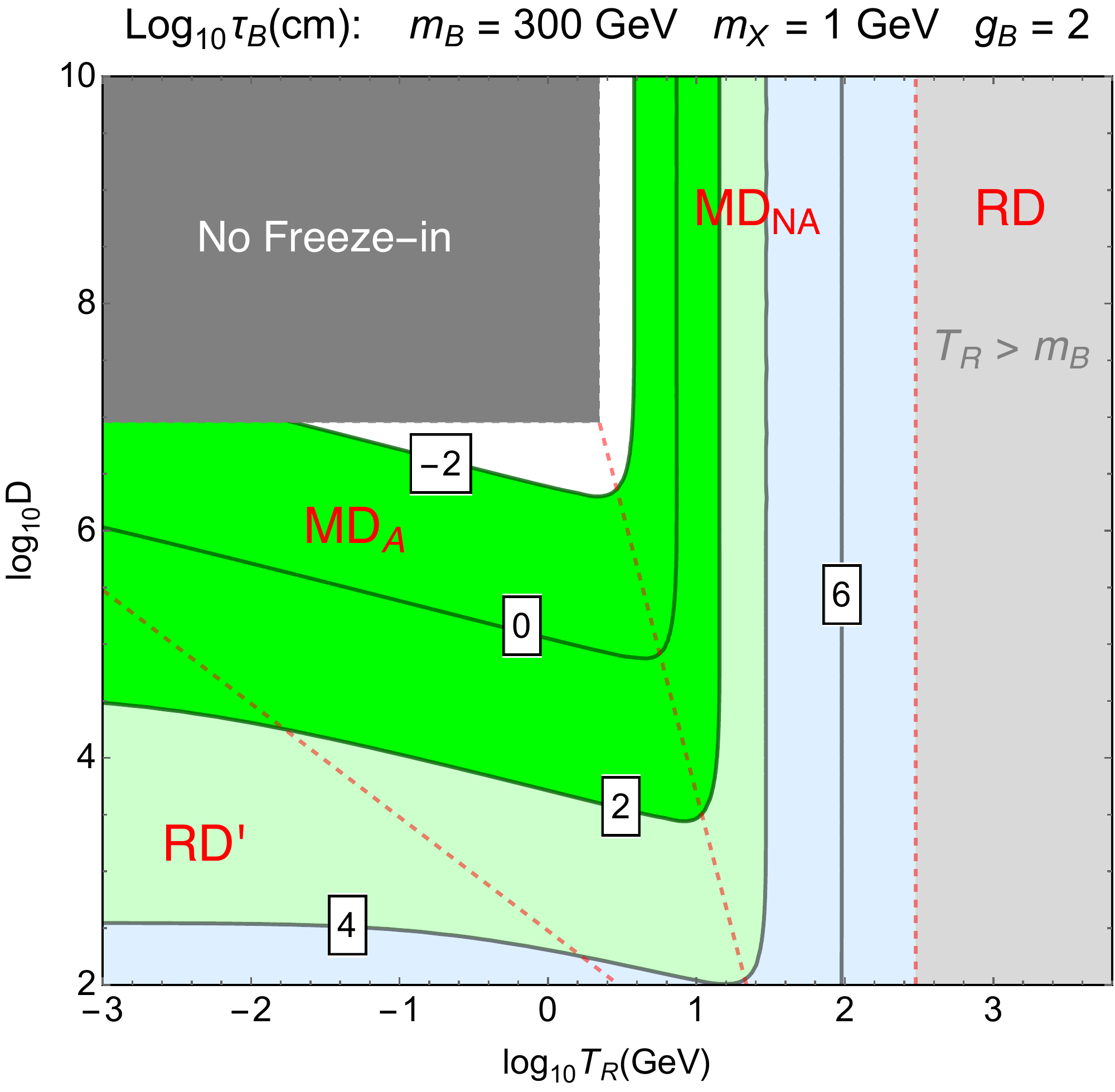} 
 \\
 \includegraphics[scale=0.4]{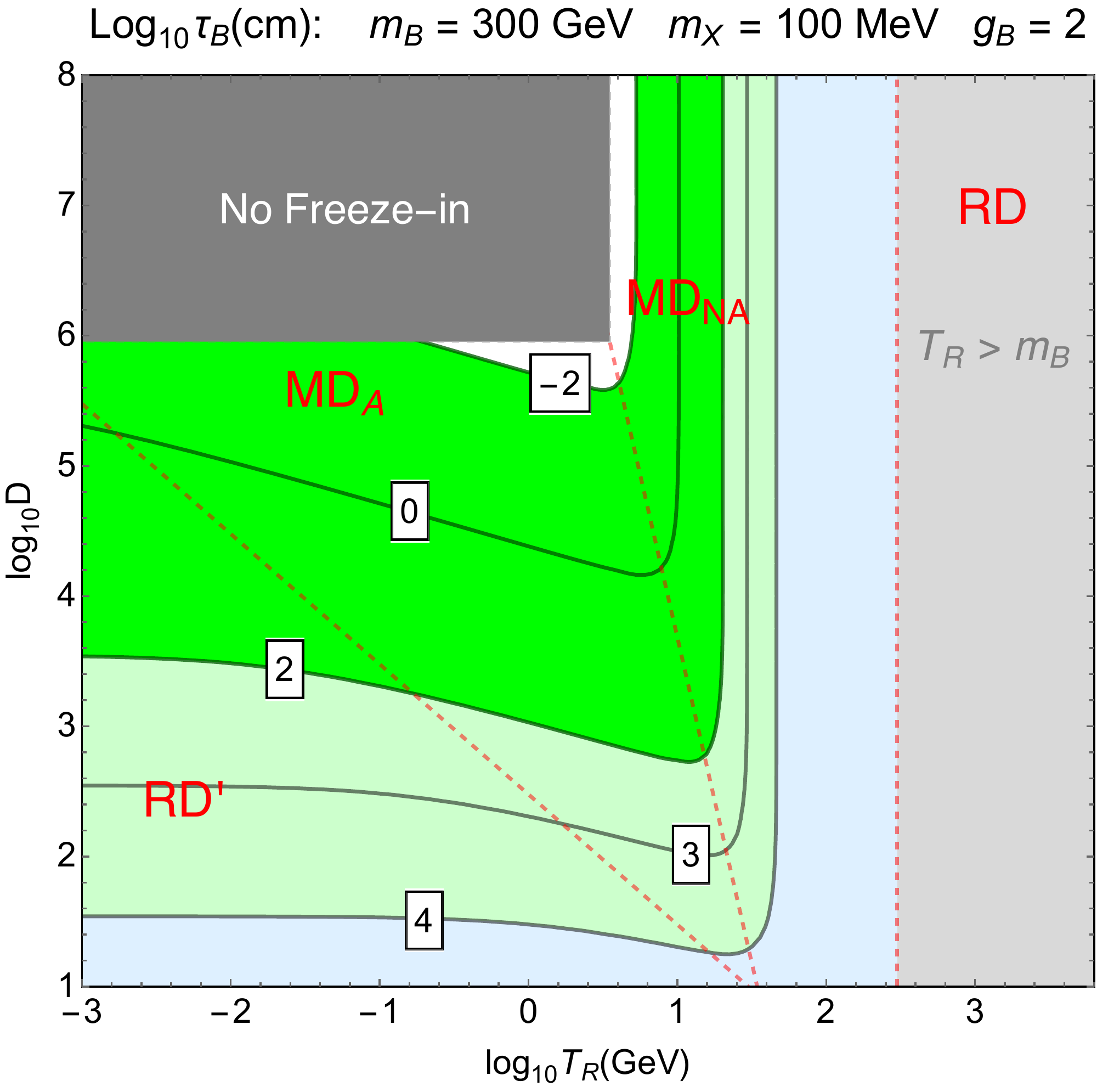} &
 \includegraphics[scale=0.4]{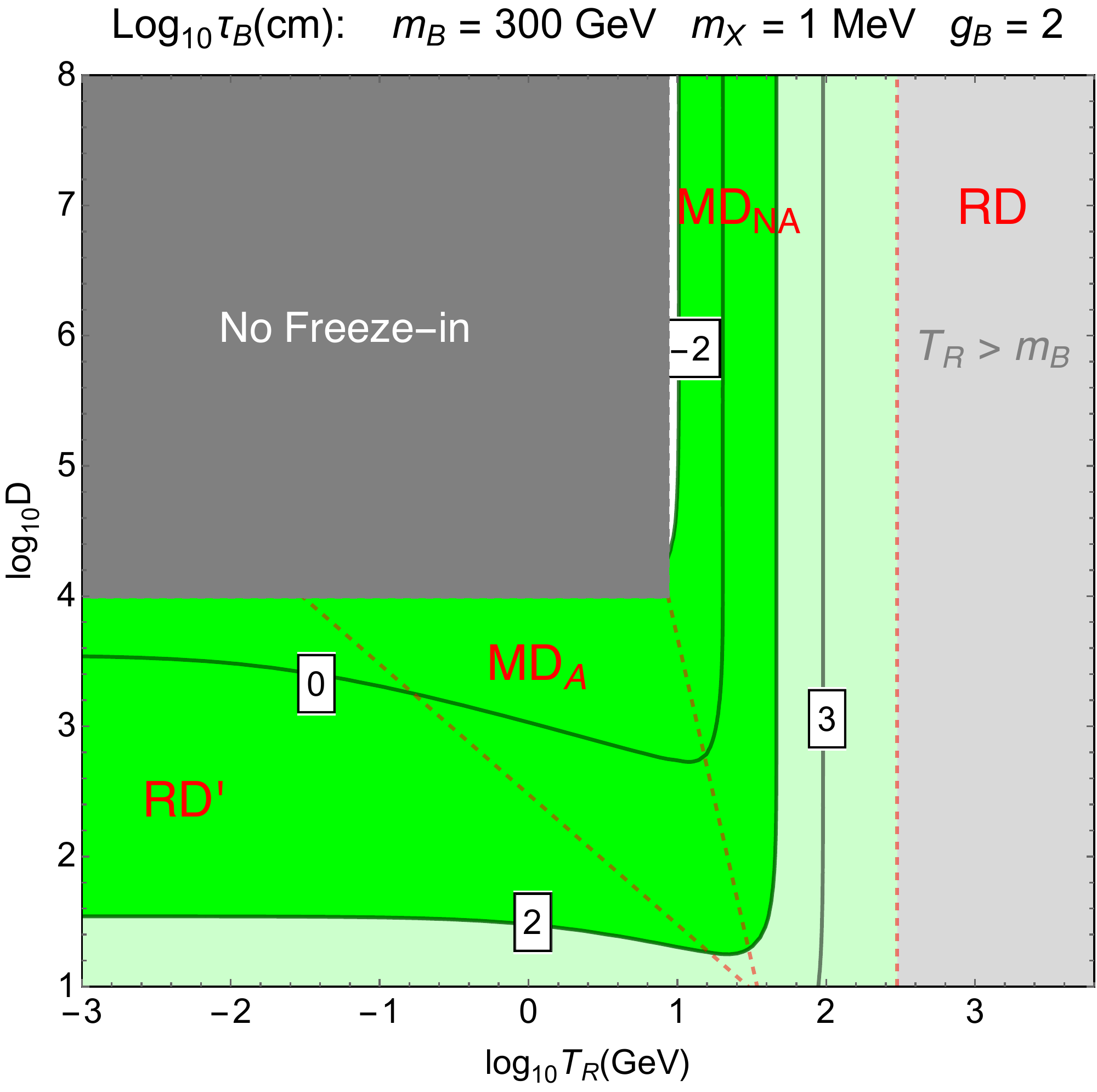}
 \end{tabular}
\end{center}
\caption{Contours of the parent particle lifetime $\tau_B$ (in cm) that give the observed DM abundance via FI 
,
assuming that some mechanism dumps entropy into the SM sector at temperature $T_R$ to reduce the DM abundance by dilution factor $D$.
We fix $m_B = 300 \, \GeV$ 
and in each panel we consider different values of $m_X$.  The upper-left gray region does not give FI, as $X$ thermalizes. The three dashed red lines 
separate four regions where FI occurs during (left to right) RD$'$, MD$_A$, MD$_{NA}$, and RD eras. 
{
Displaced collider signals occur in almost the entire parameter space.
Light green and blue shaded regions in particular are prime targets for MATHUSLA. 
}
(This figure is taken from~\cite{Co:2015pka}, which studied the benchmarks marked by the red and brown stars.)
} 
\label{fig:MasterPlots}
\end{figure}

\subsubsubsection{Displaced Signals at Colliders}
\label{subsec:DisplacedFI}

As shown in Eqn.~\ref{eq:Y03}, freeze-in production is altered when a long matter-dominated era is present. This implies that the observed DM abundance 
also requires a different decay rate of $B$. In particular, once $X$ is produced during the MD$_{NA}$, MD$_A$, or RD$'$ era, the decay rate of $B$ must 
be enhanced in order to compensate for the dilution effect, leading to a shorter decay length ideal for collider searches. Including all $\mathcal{O}(1)$ 
factors, the decay length of $B$ predicted by our numerical results for production in various eras can be approximated by
\begin{eqnarray}
c \, \tau_B &\simeq& 3\times10^6 \text{ m } \left( \frac{300 \GeV}{m_B} \right)^2 \, \left(\frac{m_X}{100 \GeV} \right)  \hspace{1.9in}  (\mbox{RD}) \\
c \, \tau_B &\simeq& 10 \text{ m } \left(\frac{T_R}{10 \GeV}\right)^7 
\left( \frac{300 \GeV}{m_B} \right)^9 \, \left(\frac{m_X}{100 \GeV} \right)  \hspace{1.3in}  (\mbox{MD}_{NA}) \\
c \, \tau_B &\simeq& 1 \text{ m } \left(\frac{T_R}{\GeV}\right)\left( \frac{10^5 \GeV}{T_M} \right)^{3/2}
\left( \frac{300 \GeV}{m_B} \right)^{3/2} \, \left(\frac{m_X}{100 \GeV} \right)  \hspace{0.5in}  (\mbox{MD}_A) \\
c \, \tau_B &\simeq& 1 \text{ m } \left(\frac{T_R}{10 \MeV}\right)\left( \frac{30 \GeV}{T_M} \right)
\left( \frac{3 \TeV}{m_B} \right)^2 \, \left(\frac{m_X}{10 \GeV} \right) .  \hspace{1in}  (\mbox{RD}')
\label{eq:tauB}
\end{eqnarray}
Figure \ref{fig:MasterPlots} from Ref.~\cite{Co:2015pka} shows the numerical calculations of $c\tau_B$ in the $(T_R, D)$ plane. For 
illustration, we fix the bath parent particle mass $m_B = 300$ GeV and vary the dark matter mass $m_X$ from 1 MeV to 100 GeV in the four panels. 
Reheating temperatures above the BBN bound of $\sim$ MeV are considered.
Remarkably, displaced collider signals can occur in almost the entire parameter space. 
We shade some regions  dark green, light green and light blue to indicate proper decay lengths greater than $0.1$ mm, 1 m and 100 m.
The precise phenomenology depends of course on the LLP production mode, but broadly speaking green shaded regions might be probed at the LHC main detectors, while light shaded regions are prime targets for MATHUSLA.

In the gray region of Fig.~\ref{fig:MasterPlots}, the dilution factor is so large that $\Gamma_B$ required for 
$\rho^{\rm obs}_{DM}/s$ is already sufficient to thermalize $X$ by scattering processes, and therefore freeze-in process does not occur. The red 
dashed lines separate the regions by the eras in which freeze-in occurs. In the light-gray RD region, freeze-in happens after the end of matter 
reheating and therefore the result reduces to the conventional case studied in Sec.~\ref{subsec:FIRD}. The prediction of $c\tau_B$ inside this RD 
region is the same as that on the left edge of the light-gray region. In the region labeled by MD$_{NA}$, freeze-in occurs during entropy production so the 
abundance receives only partial dilution and is insensitive to the total dilution $D$. As for MD$_{A}$, larger dilution $D$ allows for a larger 
decay rate and thus smaller $c\tau_B$. Besides, larger $T_R$ increases the Hubble rate at the freeze-in temperature, which also leads to smaller 
$c\tau_B$. Finally, when produced in RD$'$, DM abundance receives the full dilution factor with the usual Hubble scaling and is thus independent of $T_R$.

\subsubsection{A Simplified Model: Freeze-In through the SM Higgs}
\label{subsec:freezeinHiggs}

We now consider a simplified model of fermionic DM which is populated by the freeze-in mechanism assuming a standard RD cosmology before BBN. (For a projection of main detector reach for this scenario, see~\cite{Calibbi:2018fqf}.) This shows that even a minimal implementation of the freeze-in mechanism can give rise to neutral LLP signals at MATHUSLA across large regions of parameter space.

For the family of freeze-in models giving rise to processes like the one 
in Eq.~\ref{eq:FIprocess}, if $A_{\rm SM}$ is the SM Higgs, one of the simplest implementations is to 
add on top of the SM a Dirac fermion $\chi$, singlet under the SM gauge group, and an $SU(2)_L$ Dirac doublet  
$\psi$:
\begin{equation}
\psi = \left( 
\begin{array}{c}
\psi^{+}\\
\psi^0
\end{array}
\right)
\end{equation}
such that the Lagrangian reads:
\begin{equation}
\label{eq:DMSM_FI}
\mathcal{L} = \mathcal{L}_{\mathrm{SM}} +  
i \,\bar{\chi} \gamma^{\mu} \partial_{\mu} \chi + i \,\bar{\psi} \gamma^{\mu} D_{\mu} \psi - m_s \,\bar{\chi}\chi - m_D \bar{\psi}\psi
- y_{\chi} \, \bar{\psi} H \chi + h.c. 
\end{equation}
The system we consider could be regarded as a simplified model for a feebly interacting higgsino-bino or higgsino-singlino system 
(see also~\cite{Calibbi:2015nha}) as well as for the higgsino-axino system considered in Sec.~\ref{sec:DFSZ} for light axinos in an RD background,
bearing in mind that here $\chi$ and $\psi$ are Dirac fermions. The coupling $y_\chi$ is taken here to be very small (in the correct 
ballpark for the freeze-in regime, see below). Consequently, upon electroweak symmetry breaking, the neutral particles $\chi$ and $\psi^0$ acquire a tiny mixing, giving rise to mass eigenstates $\chi_{1}$ (mostly singlet) and $\chi_{2}$ (mostly doublet), with masses  $m_1$ and $m_2$, with 
$m_2 > m_1$ (assuming $m_{D}  > m_{s}$ in Eq.~\ref{eq:DMSM_FI}). The mixing is simply given by
\begin{equation}
\mathrm{sin} \theta \simeq \frac{y_{\chi} v}{\sqrt{2} (m_2 - m_1)}
\end{equation}
with $v = 246$ GeV the Higgs vev. The interactions of the DM candidate $\chi_{1}$ are given by $h\chi_2\chi_1$ (from the Yukawa term in Eq.~\ref{eq:DMSM_FI}) 
and $Z\chi_2\chi_1$, $W^{\pm}\psi^{\mp}\chi_1$ (from the singlet-doublet mixing), of comparable strength.
When kinematically possible, the decay widths for $\chi_2 \to h \chi_1$, $\chi_2 \to Z \chi_1$ and $\psi^{\pm} \to W^{\pm} \chi_1$ are given by
\begin{eqnarray}
\label{decayFIDM_1}
 \Gamma (\chi_2 \to h \chi_1) &=& \frac{y_{\chi}^2}{32\,\pi\,\, m_{2}^3} \left[(m_2+m_1)^2 - m_h^2\right] \lambda(m_2, m_1, m_h) \\
 \label{decayFIDM_2}
 \Gamma (\chi_2 \to Z \chi_1) &=&  \frac{y_{\chi}^2}{32\,\pi} 
 \frac{\left[(m_2 - m_1)^2 - m_Z^2\right]\left[(m_2+m_1)^2 + 2 m_Z^2\right]}{m_{2}^3\,\,(m_2-m_1)^2} \lambda(m_2, m_1, m_Z) \\
 \label{decayFIDM_3}
 \Gamma (\psi^{\pm} \to W^{\pm} \chi_1) &=&  \frac{y_{\chi}^2}{16\,\pi} 
 \frac{\left[(m_{\psi} - m_1)^2 - m_W^2\right]\left[(m_{\psi}+m_1)^2 + 2 m_W^2\right]}{m_{\psi}^3\,\,(m_{\psi}-m_1)^2} \lambda(m_{\psi}, m_1, m_W)
\end{eqnarray}
with 
\begin{equation}
\lambda(x, y, z) = \sqrt{x^4 + y^4 + z^4 - 2 x^2 y^2 - 2 x^2 z^2 - 2 y^2 z^2}
\end{equation}

As discussed in Sec.~\ref{sec:FIDM}, the DM relic abundance is obtained via slow $\chi_1$ production in the early Universe (during radiation domination, as discussed in Sec.~\ref{subsec:FIRD}) through the decays of 
the $\chi_2$ and $\psi^{+}$ states, which are in equilibrium with the thermal bath, and the subsequent DM 
freeze-in when the abundance of $\chi_2$ and $\psi^{+}$ becomes exponentially suppressed, 
around\footnote{For the obtention of the DM relic abundance we use for simplicity $m_2 = m_{\psi}$, as their mass difference 
$m_{\psi} - m_2 = \mathcal{O}(100 \,\mathrm{MeV})$
does not play a role in the freeze-in mechanism.} $T \sim m_2/3$. 
Assuming that the reheating temperature $T_R\gg m_2$, the DM relic abundance can be estimated as:
\begin{equation}
\label{DM_Relic_FIHiggs}
\Omega_{\mathrm{DM}}\,h^2 \simeq \frac{2\,m_1}{\rho_c/s_0} \frac{45\, M_{\mathrm{Pl}} \,\Gamma_{\mathrm{FI}}}{4\,\pi^4\,m_2^2} 
\int_0^{\infty} \dfrac{K_1(x) x^3}{[g_*(m_2/x)]^{3/2}} dx
\simeq  \frac{m_1}{\rho_c/s_0} \frac{135\, M_{\mathrm{Pl}} \,\Gamma_{\mathrm{FI}}}{2\,\pi^3\,m_2^2\,[g_*(m_2/3)]^{3/2}} 
\end{equation}
with $\Gamma_{\mathrm{FI}} = \Gamma (\chi_2 \to h \chi_1) + \Gamma (\chi_2 \to Z \chi_1) + \Gamma (\psi^{\pm} \to W^{\pm} \chi_1)$, 
$M_{\mathrm{Pl}} = 10^{19}$ GeV the Planck mass, $\rho_c/s_0 = 3.6 \times 10^{-9}$ GeV the critical energy density over the entropy density today, 
and $g_*(T)$ the number of relativistic degrees of freedom in the early Universe 
at temperature $T$. We note an extra factor 2 in Eq.~\ref{DM_Relic_FIHiggs} due to the fact that both 
$\chi_2$, $\psi^{+}$ and their antiparticles $\bar{\chi}_2$, $\psi^{-}$ 
are present in the early Universe plasma and contribute to the relic abundance through their decays. 
Demanding $\Omega_{\mathrm{DM}}\,h^2 = 0.12$ fixes $y_{\chi}$ in terms of $m_1$ and $m_2$.

\subsubsubsection{Production of Dark Matter $\chi_1$ at the LHC: MATHUSLA Sensitivity Estimate}

The states $\chi_2$ and $\psi^{\pm}$ can be produced at the LHC via the Drell-Yan processes $p p \to \chi_2 \chi_2$, 
$p p \to \chi_2 \psi^{\pm}$, $p p \to \psi^{+}\psi^{-}$. 
The state $\psi^{\pm}$ is short-lived and dominantly decays to $\chi_2\, \pi^{\pm}$, due to the electromagnetically induced radiative 
mass splitting $m_{\psi} - m_2 = 341\,\mathrm{MeV}$ (see e.g.~\cite{Cirelli:2005uq}).
The short lifetime and very soft pion in the final state make direct detection of $\psi^\pm$ very challenging at the LHC main detectors~\cite{Low:2014cba, Cirelli:2014dsa, Mahbubani:2017gjh}.
In contrast, the neutral state $\chi_2$ is very long-lived: 
combining the freeze-in DM relic abundance condition, Eq.~\ref{DM_Relic_FIHiggs}, with Eqs.~\ref{decayFIDM_1}-\ref{decayFIDM_2} in the limit 
$m_{2} \gg m_1,\,m_W,\,m_Z,\,m_h$ yields for the decay length $c\tau$ of $\chi_2$
\begin{equation}
\label{Lifetime_FIHiggs}
c \tau \simeq 3500 \,\,\text{m } \left(\frac{m_1}{100 \, \MeV}\right) \left(\frac{500 \, \GeV}{m_2}\right)^2 .
\end{equation}
Despite the very large decay length of $\chi_2$, we 
show in the following that MATHUSLA can be sensitive to a wide range of freeze-in DM masses. 
We implement our model in {\sc FeynRules}~\cite{Alloul:2013bka} and 
simulate the various Drell-Yan production processes for $\chi_2$ and $\psi^{\pm}$ (with subsequent decay into $\chi_2$) at LHC 13 TeV 
in {\sc Madgraph$\_$aMC@NLO}~\cite{Alwall:2014hca}, choosing to normalize the respective cross-sections to the corresponding NLO/NLL
charged/neutral higgsino production cross-sections at 13 TeV LHC given by the CERN LHC SUSY XS Working Group~\cite{SUSYworkinggroup}.
The probability for an LLP $\chi_2$ to decay inside MATHUSLA  is computed directly via a convolution of the MATHUSLA detector geometry with the distribution of produced LLPs.

\begin{figure}[h!]

\begin{center}
\includegraphics[width=0.49\textwidth]{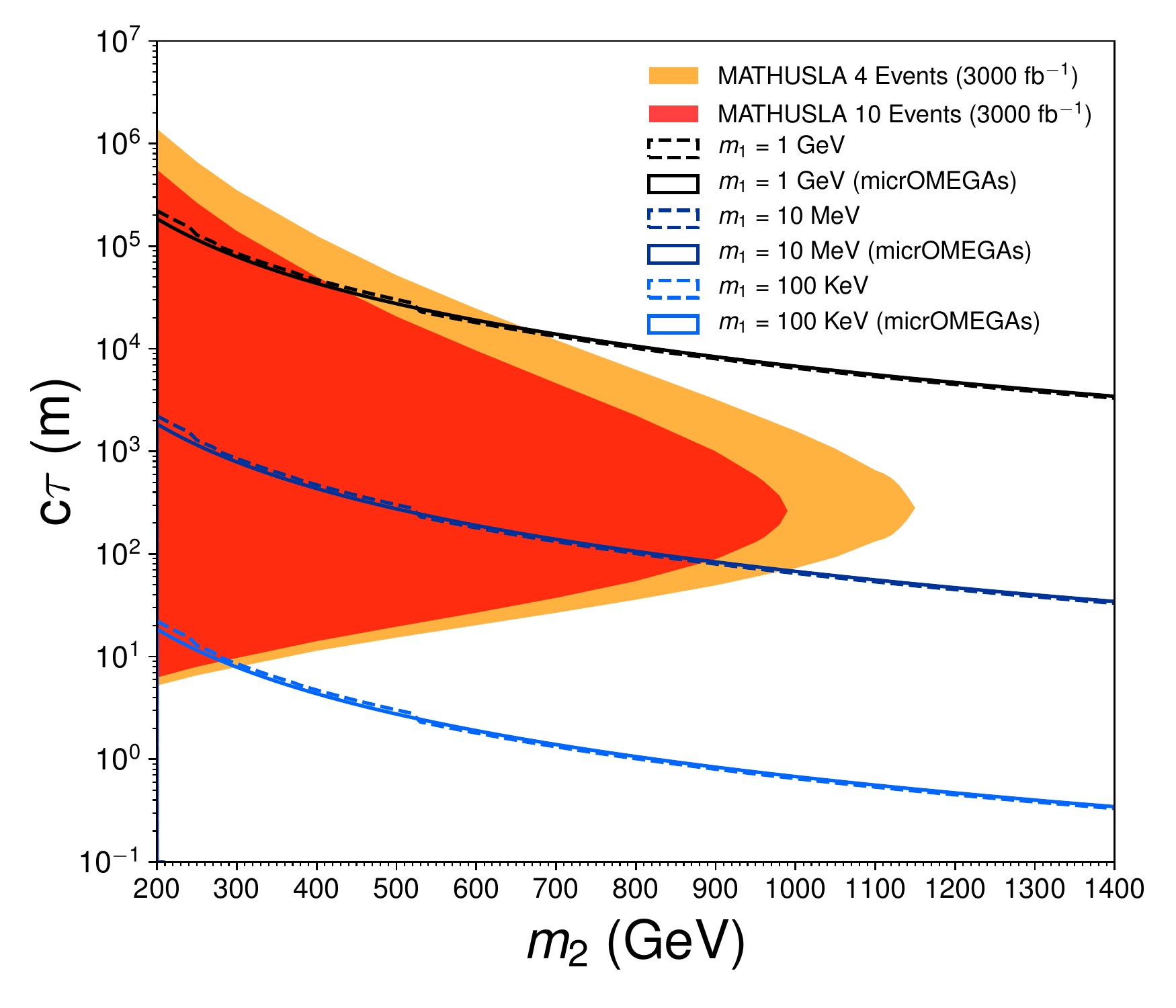}
\includegraphics[width=0.49\textwidth]{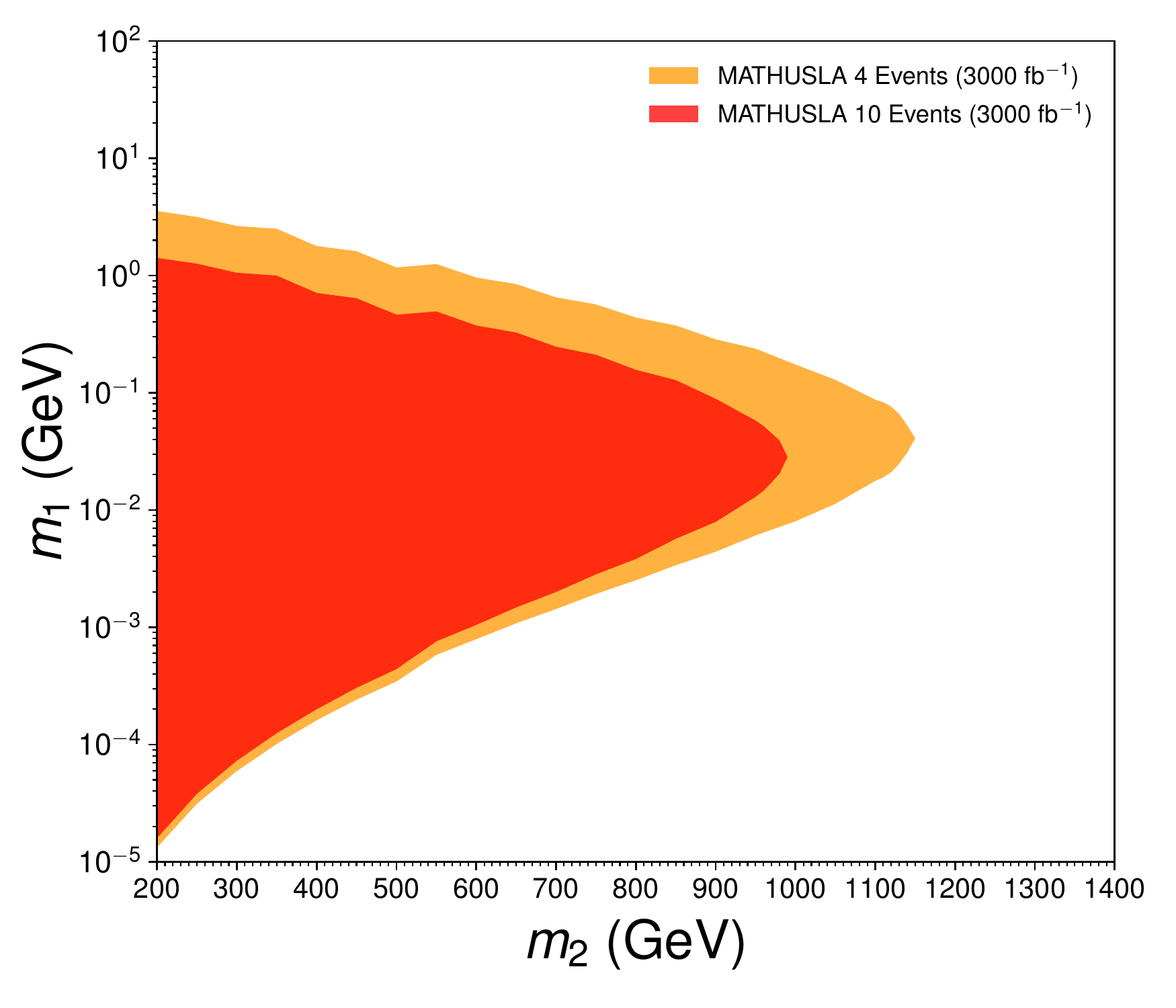}

\caption{\small Left: LLP MATHUSLA sensitivity in the ($m_2$, $c\tau$) plane. Lines yielding 
the observed DM relic density 
are shown for $m_1 = 1$ GeV (black), $10$ MeV (dark blue), $100$ keV (light blue),
with {\sc micrOMEGAs} (solid) and through the analytic approximation Eq.~\ref{DM_Relic_FIHiggs} 
(dashed).
Right: MATHUSLA sensitivity in the ($m_2$, $m_1$) plane.
This estimate assumes the $200m \times 200m \times 20m$ benchmark geometry of Fig.~\ref{f.mathuslalayout}.
}
\label{FI_Higgs_m1m2}
\end{center}

\end{figure}

In Fig~\ref{FI_Higgs_m1m2} (left) we show the MATHUSLA LLP ``exclusion'' (4 event reach) and ``discovery'' (10 event reach) sensitivity in the ($m_2$, $c\tau$) plane for 
$\mathcal{L} = 3000$ fb$^{-1}$ of integrated luminosity from HL-LHC, under the assumption of perfect MATHUSLA object reconstruction efficiency. 
We also show the curves in this plane yielding the observed DM relic abundance for $m_1 = 1$ GeV (black), $m_1 = 10$ MeV (dark blue) and $m_1 = 100$ keV (light blue), 
obtained both numerically using {\sc micrOMEGAs}~\cite{Belanger:2018ccd} (solid) and through the analytic approximation Eq.~\ref{DM_Relic_FIHiggs} 
(dashed), which are observed to agree to 
better than 20\%.
In Fig~\ref{FI_Higgs_m1m2} (right) we directly show the MATHUSLA 4 and 10 event reach in the ($m_2$, $m_1$) plane, highlighting 
that MATHUSLA is sensitive to a large range of freeze-in DM masses for LLP masses below the TeV scale.
As discussed in Section~\ref{sec:axino-ewino}, which examines the same signature as it arises from higgsinos decaying to axinos, this MATHUSLA sensitivity is at least 1-2 orders of magnitude better in cross-section than the corresponding HL-LHC LLP search, especially for parent particle masses below a few hundred GeV.

\subsubsection{A Complete Model: DFSZ axino}
\label{sec:DFSZ}

We now discuss a motivated freeze-in DM candidate that can arise as part of a more complete model with a variety of possible cosmological histories: the DFSZ axino. As we show below, this  generates displaced higgsino signals at colliders and at MATHUSLA. 

DFSZ axion 
models are theories based on a Peccei-Quinn (PQ) solution of the strong CP problem, where the Higgs doublets carry a PQ charge. If the theory is made 
supersymmetric, the axion is promoted to a supermultiplet and appears with its partners: the saxion (CP-even scalar) and axino (Weyl fermion). 

Stellar cooling bounds~\cite{Raffelt:2006cw, Chang:2018rso} severely constrain the PQ scale to be high, $V_{\rm PQ} \gtrsim 10^9 \, {\rm GeV}$. Thus we have a hierarchy 
between such a scale and the masses of the supersymmetric partners, which we consider around the scale $\tilde{m} \simeq {\rm TeV}$. We take 
advantage of this hierarchy between scales to write down an effective field theory (EFT) where the PQ symmetry is non-linearly realized. The axion 
supermultiplet in the language of superfields reads
 \be
 A=\frac{s+ia}{\sqrt 2}+\sqrt 2\theta \tilde a+\theta^2 F \ .
 \label{eq:axionsupermultiplet}
 \ee
If we perform a PQ rotation with angle $\alpha$, the axion superfield shifts
\be
A \; \rightarrow \; A + i  \, \alpha \, V_{\rm PQ} \ .
\label{eq:shiftofA}
\ee
Within this language, the QCD anomaly interaction is
\be
{\cal L}_{A W W} = - \frac{g_3^2 N_{\rm DW}}{32\pi^2 V_{\rm PQ}} 
\int d^2 \theta A \, W^\alpha W_\alpha + {\rm h.c.}  \ ,
\label{eq:LagAWW}
\ee
where $N_{\rm DW}$ is the color anomaly coefficient, also known as the domain wall number. (The suppression scale of this term is also often written as $f_a = V_{PQ}/N_{DW}$.)

We write down the most general set of interactions for the axion superfield, with all operators respecting the shift 
symmetry in Eqn.~\ref{eq:shiftofA}. Supersymmetric interactions include the K\"ahler potential and the superpotential 
\begin{align}
\label{eq:Kaler} K = & \, A^\dag A + \frac{\kappa}{2 \, V_{\rm PQ}} A^\dag A  \, (A + A^\dag) + \ldots \ , \\
\label{eq:Wdefinition} W = & \,  \mu H_u H_d  + q_\mu \frac{\mu}{V_{PQ}} \, A \, H_u H_d  + \ldots \ .
\end{align}
The dimensionless quantity $\kappa$ is a model-dependent coefficient, that depends on the charges and vacuum expectation values of the PQ 
breaking fields. For a single PQ-breaking field, we have $\kappa = 1$. The PQ charge of the $\mu$ term $q_\mu$ is also model-dependent, and in the 
minimal supersymmetric DFSZ theory it is $q_\mu = 2$. The renormalizable cubic coupling between the axion and the Higgs bosons is responsible for axino 
freeze-in production.

\subsubsubsection{Axino Dark Matter in a RD Background}
\label{subsec:AxinoDM}

We review the calculation of the axino freeze-in for a standard cosmology. For the purpose of this illustration, we consider higgsinos mass eigenstates, 
but they need not be the lightest observable supersymmetric particles (LOSP). The final axino DM density can be derived from 
Eqn.~\ref{eq:xiX}, after we replace $X \rightarrow \tilde{a}$ and $B \rightarrow \tilde{h}$. 

Accounting for both charged and neutral higgsinos leads to $g_{\tilde{h}} = 8$. The only missing information is the higgsino decay width. We can 
compute it from the cubic superpotential interaction in Eqn.~\ref{eq:Wdefinition}, and we find
\be
\Gamma_{\tilde{h}} =  \frac{q_\mu^2 \, \mu^3}{32 \pi V_{\rm PQ}^2} \ .
\label{eq:Gammahiggsino}
\ee
corresponding to a higgsino lifetime of
\be
\label{eq:higgsino_ctau}
c\tau_{\tilde{h}} = \Gamma_{\tilde{h}}^{-1} \simeq 180 \, {\rm m} \left(\frac{2}{q_\mu}\right)^{2} \left(\frac{300 \, \GeV}{\mu} \right)^3 
 \left( \frac{V_{\rm PQ}}{10^{12} \, \GeV}\right)^2 \ .
\ee
The resulting axino relic abundance in units of the observed density reads
\be
\frac{\rho_{\tilde{a}}}{\rho^{\rm obs}_{DM}} =   8 \, \times \, 10^{4}
\left(\frac{106.75}{g_*}\right)^{3/2}  \left(\frac{q_\mu}{2}\right)^{2}  \left(\frac{m_{\tilde{a}}}{100 \, \GeV}\right) 
  \left(\frac{\mu}{300 \, \GeV}\right)
 \left(\frac{{10^{12} \, \GeV}}{V_{\rm PQ}}\right)^2  \ .
  \label{eq:rhoaxinoRD}
\ee
Without additional dilution mechanisms or a small initial misalignment angle, coherent oscillations of the axion field at a temperature near $T = 160 \mev$ will overproduce axions unless  $V_{PQ} \lesssim 10^{12}$ GeV~\cite{Fox:2004kb}. 
Imposing this constraint on the PQ scale has two consequences. 
\begin{enumerate}
\item Eqn.~(\ref{eq:higgsino_ctau}) makes clear that the Higgsino lifetime lies in the range that is optimal for detection at MATHUSLA. 
The general collider phenomenology of Higgsino LLPs decaying into axions is studied in Sec.~\ref{sec:axino-ewino}. As shown in Fig.~\ref{fig:axinos}, MATHUSLA can probe $V_{PQ}$ up to $\sim 10^{13} \gev$. 
\item Reproducing the observed DM abundance in Eqn.~(\ref{eq:rhoaxinoRD}) requires very low axino masses at or below the MeV scale. This essentially realizes the simplified scenario of Section~\ref{sec:axino-ewino}.
\end{enumerate}
Light axino freeze-in DM therefore represents an excellent target for LLP detection at MATHUSLA. There is, however, significant motivation to also consider heavier axinos with masses near the weak scale. 

The axino mass is expected to be at least of order the gravitino mass~\cite{Cheung:2011mg}. This is due to non-renormalizable couplings between the PQ sector and the SUSY breaking sector that cannot be forbidden by symmetries (though certain extra-dimensional sequestering scenarios could change this argument). A light axino therefore implies a light gravitino, leading to the usual gravitino overclosure problem.
One way to address this problem is a low inflationary reheating temperature $\lesssim 10^5 \gev$, avoiding high-temperature overproduction of gravitinos. In that case, light axinos (or similar-mass gravitinos produced in axino decays) produced via freeze-in from higgsino decay would be a viable dark matter candidate.

While low reheating temperatures are a valid solution, they restrict the possible inflationary scenario and preclude any high-scale baryogenesis mechanism from accounting for the matter-antimatter asymmetry of the universe. 
Hence there is ample motivation for considering weak scale axinos to solve the gravitino overclosure problem. 
Eqn.~(\ref{eq:Gammahiggsino}) makes clear that this requires some dilution mechanism to reduce the axino relic density. 
As discussed in Section~\ref{subsec:FIMD}, this can be realized if axino freeze-in occurs during an early MD epoch.
Remarkably, supersymmetric PQ theories incorporate such a dilution mechanism  through the saxion condensate. 

\subsubsubsection{Saxion Cosmology}
\label{subsec:Saxion}

The CP-even scalar of the axion supermultiplet has no potential in the absence of SUSY breaking. 
In the vacuum today, PQ symmetry is broken, and the 
saxion takes a mass of the order of the superpartners scale. 
PQ may also have been broken during inflation and not restored afterwards, conveniently 
solving the domain wall problem for the minimal DFSZ model ($N_{\rm DW} = 6$). If this is the case, the SUSY breaking vacuum energy during inflation 
would also provide a potential for the saxion. Unless theories with specific symmetries are considered~\cite{Co:2015pka}, the minimum of the saxion 
potential today and during inflation are different. Inflation sets the initial condition for the evolution of the saxion field, displacing it 
from its current mininum by an amount $s_I \simeq V_{\rm PQ}$ or $s_I \simeq M_*$, where the latter is the cutoff of the theory.

The subsequent evolution of the saxion condensate can be tracked by solving the equation of motion. Right after inflation ends, the saxion field does 
not evolve due to the Hubble friction. Once the universe slows down enough, the saxion condensate starts harmonic damped oscillations. This happens at a 
temperature $T_{\rm osc}$ found by solving the condition $3 H(T_{\rm osc}) \simeq m_s$, and it approximately reads
\be
T_{\rm osc} \simeq \left( m_s M_{\rm Pl} \right)^{1/2} \simeq 10^{10} \, {\rm GeV} \left( \frac{m_s}{1 \, {\rm TeV}} \right)^{1/2} \ .
\ee
The damped saxion oscillations at lower temperature red-shift with the expansion as non-relativistic matter. As this red-shifting is milder than 
the one for radiation, at some temperature $T_M$ the saxion condensate energy dominates the universe:
\be
T_M \simeq 10 \, {\rm TeV} \, \left( \frac{m_s}{1 \, {\rm TeV}} \right)^{1/2} \, \left( \frac{s_I}{10^{15} \, {\rm GeV}}\right)^2 \ .
\ee 

At temperatures below $T_M$ the universe enters an early MD epoch where the saxion energy density controls the Hubble expansion. 
This epoch has to be terminated by the saxion condensate decay before the time of BBN. 
The possible saxion decay channels can be identified from the 
interactions in Eqs.~\ref{eq:LagAWW}-\ref{eq:Wdefinition}. 
The QCD anomaly term mediates interactions to gluon final states,
\begin{align}
\Gamma_{s\rightarrow gg} = N_{\rm DW}^2 \frac{\alpha_3^2}{64\pi^3} \frac{m_s^3}{V_{\rm PQ}^2} \ ,
\label{eq:GammaSaxGluons}
\end{align}
while the cubic term in the superpotential induces decays to Higgs bosons and longitudinal electroweak gauge bosons,
\be
\Gamma_{s \, \rightarrow \, hh, W_L W_L, Z_L Z_L} = \frac{q^2_\mu \mu^4}{4 \pi m_s V_{\rm PQ}^2} \ .
\label{eq:Gammastovisible}
\ee
Decay to EW bosons typically dominates over the loop-suppressed decay to two gluons. 

In order for the condensate to dilute the axino abundance, saxion decay to axions and axinos must be small.  
These dangerous decay channels are generated by the model-dependent cubic self-interaction in the K\"ahler potential:
\begin{align}
\label{eq:stoaa}
\Gamma_{s \, \rightarrow \, a a } = & \, \frac{\kappa^2 \, m_s^3}{64 \pi V_{\rm PQ}^2} \ , \\
\label{eq:stoaxinos}
\Gamma_{s \, \rightarrow \, \tilde{a} \tilde{a} } = & \,  \frac{\kappa^2 \, m^2_{\tilde{a}} m_s}{8 \pi V_{\rm PQ}^2} \ .
\end{align}
The second decay badly overproduces LSP dark matter, and we forbid it by assuming the saxion mass to be below twice the axino mass.
The first decay produces axion dark radiation, which is severely constrained by BBN and CMB bounds. The corresponding largest allowed value of $\kappa$ is~\cite{Co:2016vsi}
\begin{equation}
\label{eq:axinokappamax}
\kappa < \kappa_{max} \approx 2.1 \  \left(\frac{q_\mu}{2}\right) \ \left(\frac{\mu^2}{m_S^2}
\right) \  \left(\frac{\Delta N_{eff}}{0.45}\right)^{1/2} \ .
\end{equation}
Given the current bound of $\Delta N_{eff} \lesssim 0.45$~\cite{Ade:2015xua}, this constraint can be easily satisfied if $\kappa = 1$ for a single PQ-breaking field. $\kappa < \kappa_{max}$ then also guarantees that saxions decay dominantly to EW bosons.

Since the saxion decay width is dominated by Eqn.~\ref{eq:Gammastovisible}, we can use this expression to evaluate $T_{Rs}$, defined as the reheat temperature of the radiation bath after the saxion condensate has decayed:
\be
\label{eq:TRs}
T_{Rs} \simeq \left( \Gamma_s M_{\rm Pl} \right)^{1/2} \simeq 10 \, {\rm MeV} \, \left( \frac{\mu}{1 \, {\rm TeV}} \right)^{3/2} \,
 \left( \frac{\mu}{m_s} \right)^{1/2}  \left( \frac{10^{15} \, {\rm GeV}}{V_{\rm PQ}} \right) \ .
\ee
For TeV scale SUSY parameters, the reheat temperature is in the ${\rm MeV} - {\rm GeV}$ range. This range is interesting, since it is below the axino 
freeze-in production temperature while still being consistent with BBN bounds.
Furthermore, for reheating temperatures below 160 MeV, any initial axion abundance produced from misalignment is diluted away, allowing $V_{PQ}$ to be much larger than the usual limit of $10^{12} \gev$~\cite{Co:2016vsi}. 

As a result of the saxion early matter-dominated epoch, the axino freeze-in calculation presented in Sec.~\ref{subsec:AxinoDM} has to be revisited. 
There are two reasons why the final result will be different: the different time vs. temperature dependence for an early MD era and the dilution due 
to the entropy dumped into the radiation bath from saxion decays. We present the results for the axino relic density in Sec.~\ref{subsec:SaxionDilution}. 
Before we do that, we quantify the entropy due to saxion decays. As described in Sec.~\ref{subsec:FIMD}, the saxion MD epoch is made of two distinct phases. The 
temperature when we enter a non-adiabatic phase for the saxion condensate reads
\be
\label{eq:saxTNA}
T_{\rm NA} \simeq \left(T_M \, T_{Rs}^4 \right)^{1/5} \simeq 
0.2 \, {\rm GeV} \, \left( \frac{\mu}{1 \, {\rm TeV}} \right)^{13/10} \,
 \left( \frac{\mu}{m_s} \right)^{3/10} \left( \frac{s_I}{V_{\rm PQ}} \right)^{2/5} \left( \frac{10^{15} \, {\rm GeV}}{V_{\rm PQ}} \right) \ .
\ee
Axino produced through freeze-in at temperatures above $T_{\rm NA}$ get diluted by the full amount of entropy injected ($D \simeq T_M / T_{Rs}$). 
If the production happens at temperatures $T_{\rm FI}$ below $T_{\rm NA}$, the dilution factor is only $D(T_{\rm FI}) \simeq (T_{\rm FI} / T_{Rs})^5$.

\subsubsubsection{Freeze-In Axino Yield with Dilution}
\label{subsec:SaxionDilution}

\begin{figure}[t]
\begin{center}
\includegraphics[width=0.495\linewidth]{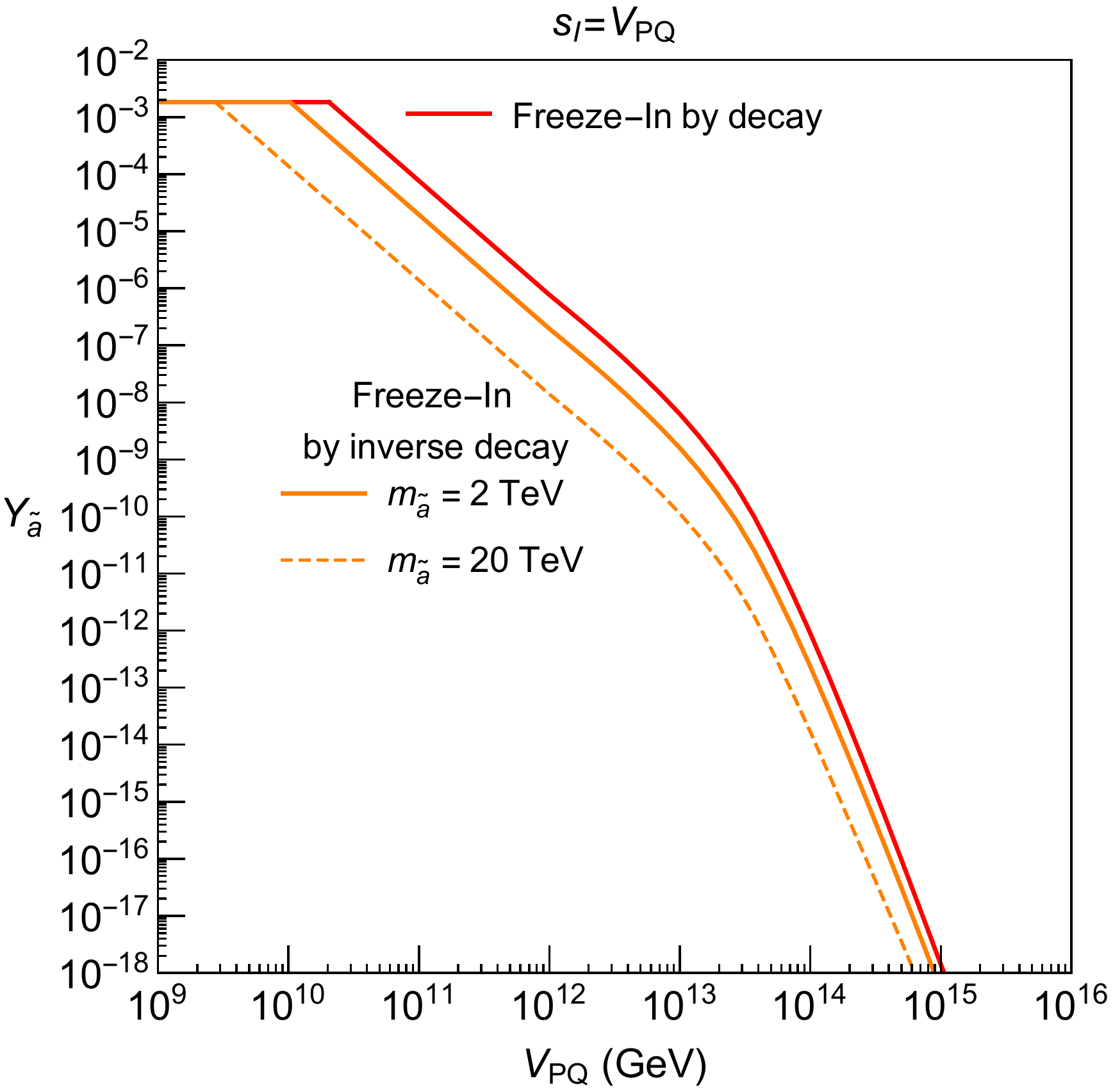} \includegraphics[width=0.495\linewidth]{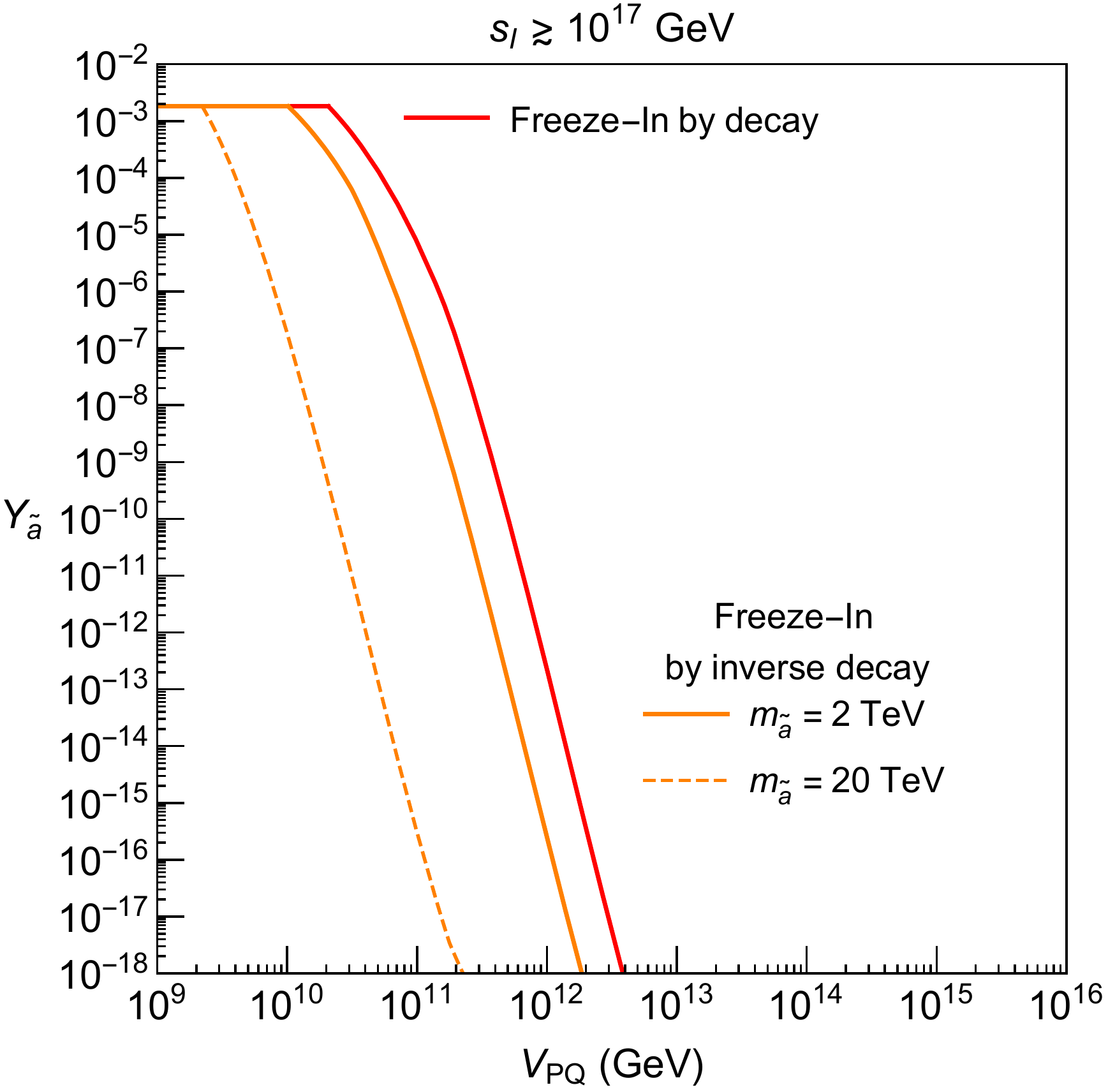}
\end{center}
\caption{The axino yield from neutralino decays to SM + $\tilde a$ (red) for $m_{\tilde{a}} \ll 1$ TeV, and neutralino inverse decays ($\tilde \chi^0$ + SM $\to \tilde a$, orange) for $m_{\tilde{a}} = 2$ 
and 20 TeV. In both panels, $2 M_1 = M_2= \mu =1$ TeV, $m_s=500$ GeV, $\tan \beta = 2$, $q_\mu = 2$, and 
$\mathcal{D}=4$; while $s_I = V_{PQ} \ (M_* \gsim 10^{17} \GeV)$ for the left (right) panel.
Figure taken from~\cite{Co:2016fln}.
}
\label{fig:AxinoYield}
\end{figure}

The final freeze-in axino abundance after saxion dilution can be obtained by using the general yield $Y_{i}$ in Eqn.~\ref{eq:Y03} with the 
saxion reheat temperature $T_{Rs}$ in Eqn.~\ref{eq:TRs} and $T_{NA}$ in Eqn.~\ref{eq:saxTNA}. The coupling constant $\lambda$ defined in 
Eqn.~\ref{eq:FIparameters}, in this case, is given by $q_\mu^2 \, \mu^2/32 \pi V_{\rm PQ}^2$ based on Eqn.~\ref{eq:Gammahiggsino}. 

The numerical result of the freeze-in axino yield is shown in Fig.~\ref{fig:AxinoYield}. Though our focus in the context of LLP signals is freeze-in by decay, this figure was taken from~\cite{Co:2016fln} which also considered freeze-in by inverse decay ($\tilde \chi^0$ + SM $\to \tilde a$). 
The general features of this plot can be understood as follows. 
Freeze-in by inverse decay is less efficient than by neutralino decay, because in the former case the axino mass is higher, and the inverse decay process stops at $T_{FI} \sim m_{\tilde{a}}$, which is earlier than $T_{FI} \sim \mu$ for decay.
 In the left panel with $s_I = V_{\rm PQ}$, the saxion condensate becomes insufficient to dominate the energy of the Universe when $V_{\rm PQ} \lsim 10^{13}$ GeV and therefore the result is identical to that of conventional RD cosmology. 
The effect of saxion dilution at higher $V_{\rm PQ}$ can be seen from the change of the slope at $V_{\rm PQ} \simeq 10^{13}$ GeV. In the right panel, where $s_I \gsim 10^{17}$ GeV, freeze-in occurs during the MD$_{NA}$ era so the abundance becomes insensitive to $s_I$, as can be seen from Eqn.~\ref{eq:Y03}. 
The decay of a larger saxion condensate results in a much more severe dilution of the final yield than in the left panel.

Given that $m_{\tilde{a}} Y_{\tilde{a}} \simeq 0.44$ eV for the observed DM abundance, 
the freeze-in by decay production of axino DM heavier than $\sim$ keV requires $V_{\rm PQ}$ in the range of $10^{10} - 10^{13}$ GeV. 
This corresponds to lifetimes in the ideal range for higgsino LLPs to be detected as displaced vertices at either the LHC main detectors or MATHUSLA. 

It is worth briefly commenting on the cosmological interplay between gravitinos and axinos, especially since the gravitino problem was a major motivation to consider weak-scale axino masses.
Since they are of similar mass, one can decay into the other via emission of an axion. 
Therefore, if the gravitino is lighter than the axino, the freeze-in mechanism effectively generates gravitino dark matter (with only $\mathcal{O}(1)$ modifications to the quantitative yields discussed here). 
Apart from allowing freeze-in weak-scale axino production, the saxion condensate has the additional feature that it allows for inflationary reheating temperatures in excess of $10^{12} \gev$, at which even TeV-scale gravitinos would be overproduced. 
In that scenario, $V_{PQ}$ would be fixed by the dilution of the thermal gravitino abundance required to reproduce the observed DM relic density, giving another motivation for higgsino LLPs. More details can be found in~\cite{Co:2016fln, Co:2017orl}.

\subsubsubsection{General Dilution Mechanisms}
\label{subsec:generaldilution}

\begin{figure}
\begin{center}
\begin{tabular}{cc}
\includegraphics[scale=0.4]{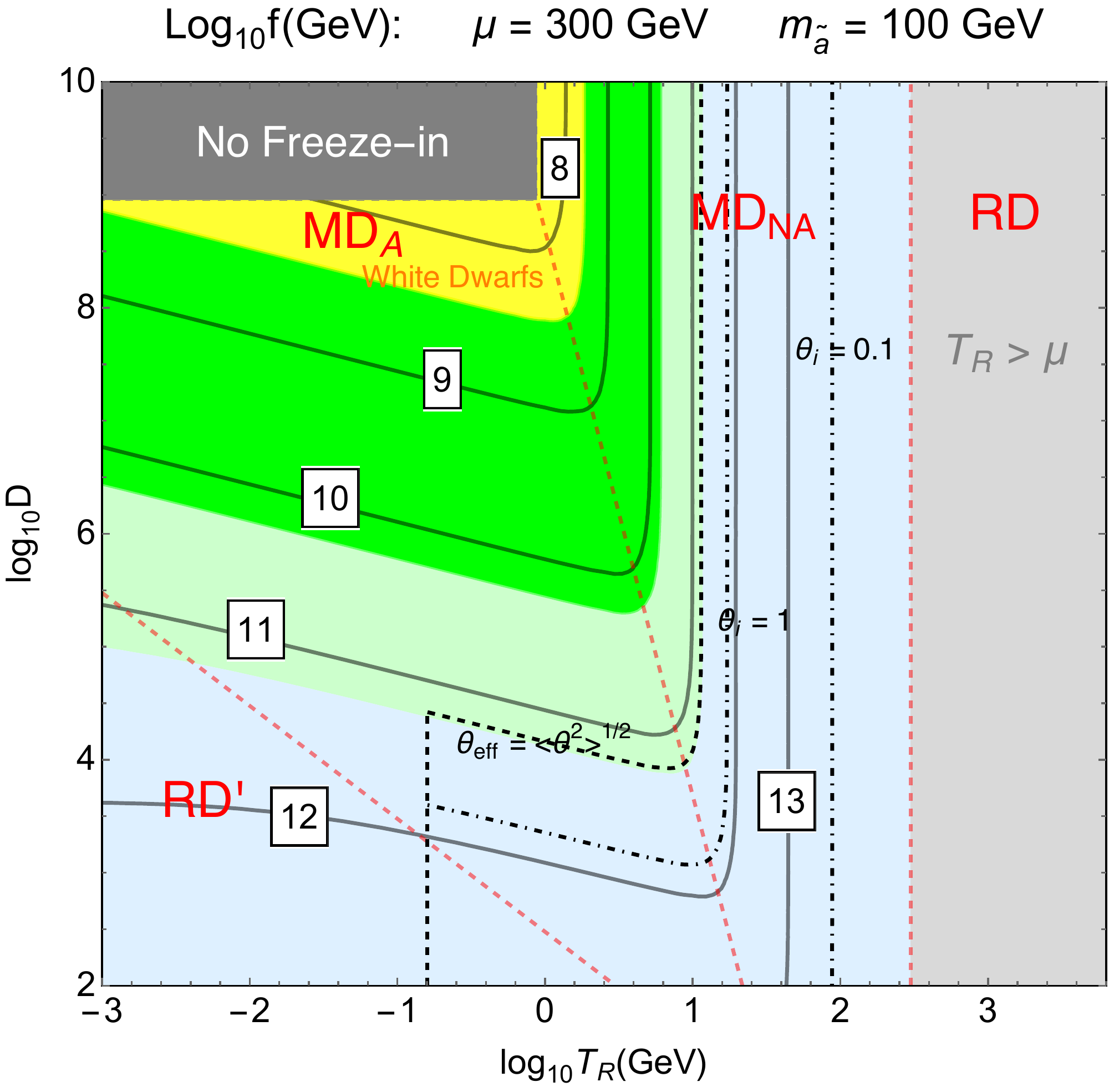} 
&
 \includegraphics[scale=0.4]{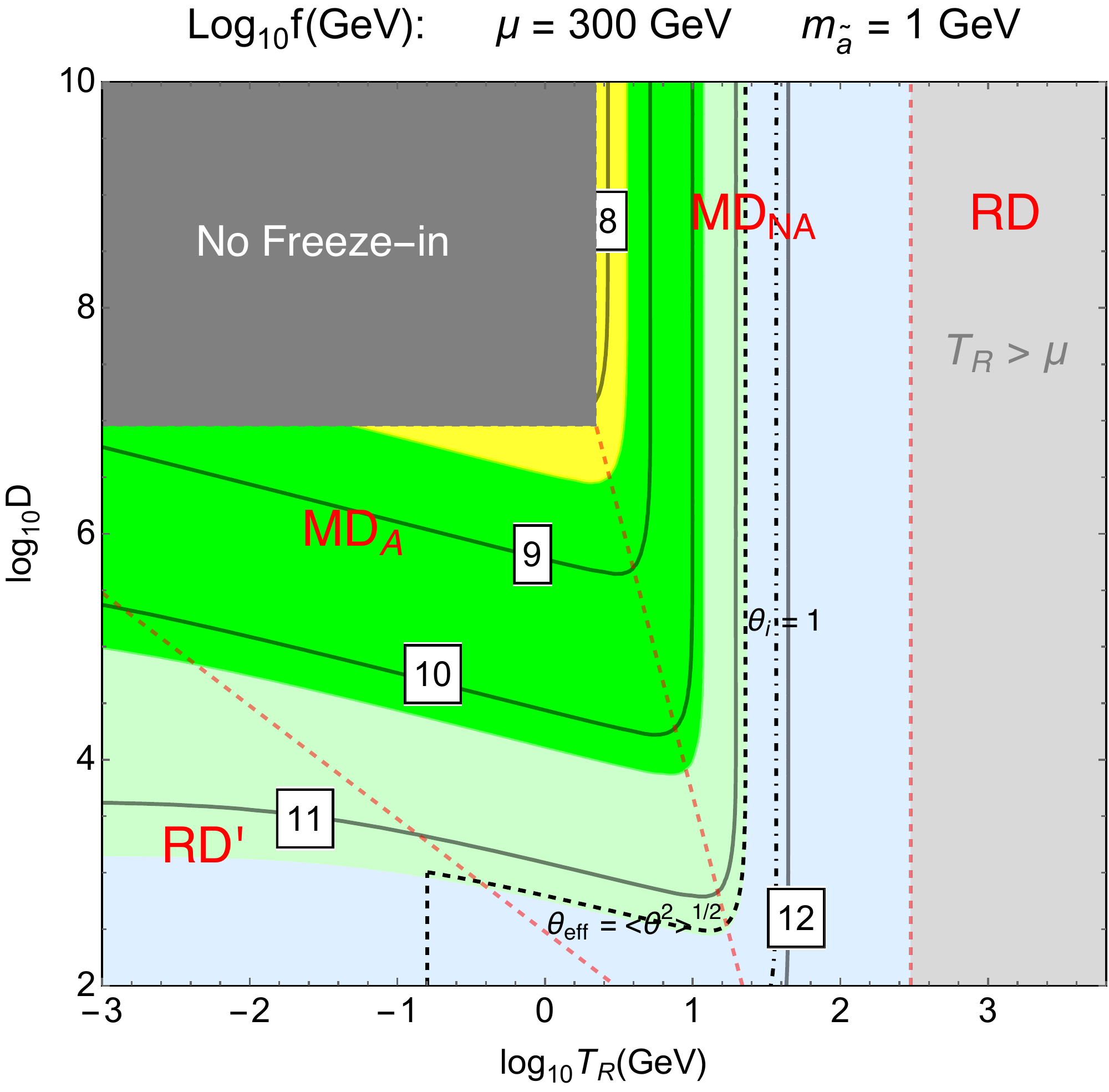} 
 \\
 \includegraphics[scale=0.4]{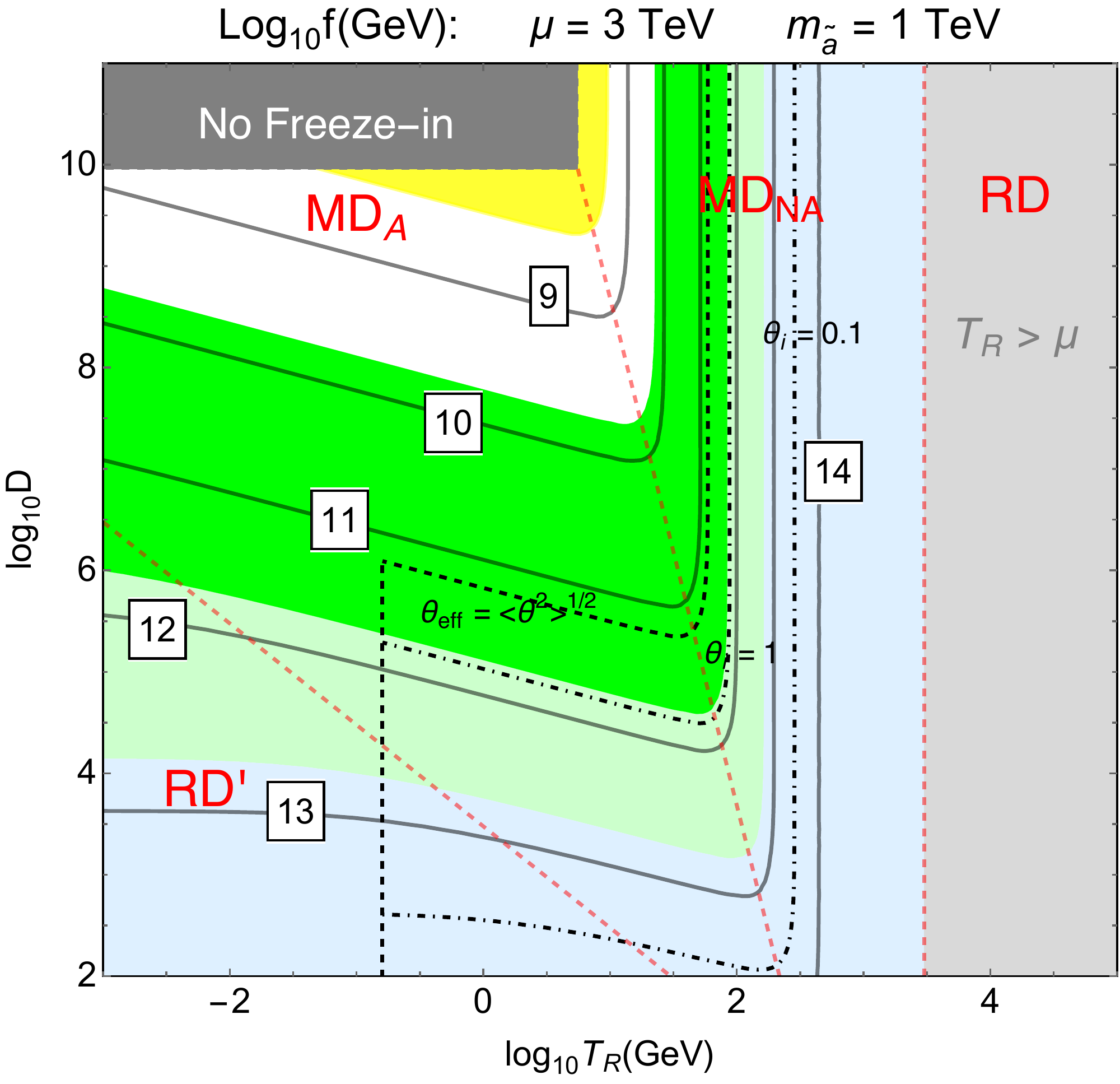} & \includegraphics[scale=0.4]{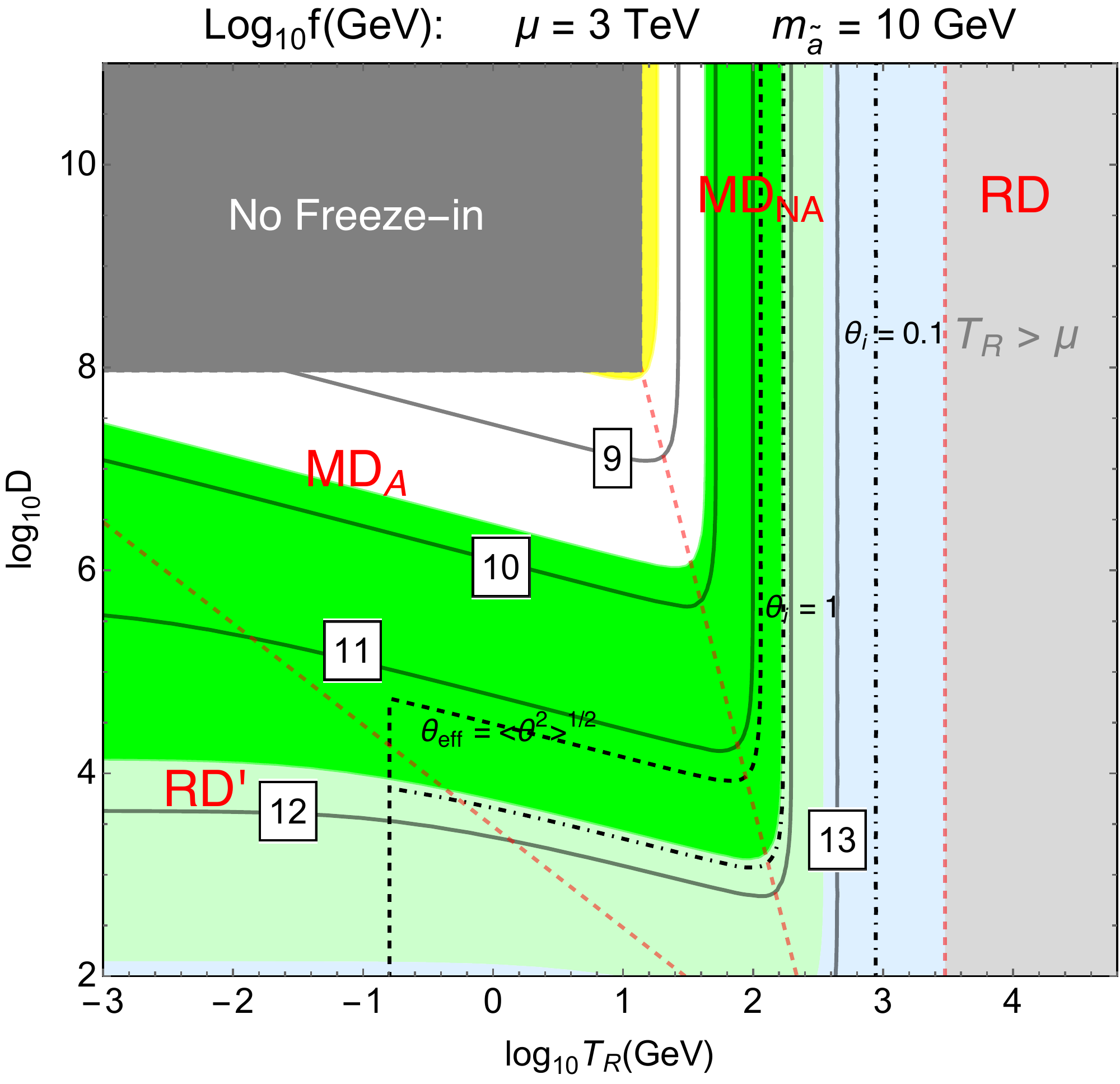}
\end{tabular}
\end{center}
\caption{
Black solid conours: $f = V_{PQ}/N_{DW}$ required for freeze-in axinos from Higgsino decay to have the observed DM relic density for different values of $\mu$ and $m_{\tilde a}$, assuming that some mechanism dumps entropy into the SM sector at temperature $T_R$ to reduce the axino abundance by dilution factor $D$. 
Axions could still be produced and constitute part of dark matter, but above and to the left of the dashed/dot-dashed lines, axions have less than the observed DM relic density for different effective misalignment angles, see text  for details.
If reheating takes place below $T = 160 \mev$, axions are diluted away. 
The yellow region is excluded by white dwarf constraints.  
The Higgsino lifetime in the light green and blue shaded regions is a prime targets for MATHUSLA, see Fig.~\ref{fig:MasterPlots}.
Figure taken from~\cite{Co:2015pka}.
}
\label{fig:MasterPlotsAxino}
\end{figure}

Above, we used the specific dilution mechanism of the decaying saxion condensate to argue that freeze-in weak-scale axino dark matter, produced in Higgsino decays, motivates PQ-breaking scales in the range of $V_{PQ} \sim 10^{10} - 10^{13} \gev$, resulting in observable displaced vertex signals.
It is important to emphasize that similar $V_{PQ}$ ranges, and hence higgsino LLP signals, are not limited to saxion dilution.
In fact, any scalar condensate can provide the necessary dilution for axino dark matter from freeze-in production.

In the generic dilution case, once the dilution temperature $T_R$ (corresponding to $T_{R_s}$ in the saxion condensate case) and dilution factor $D$ are computed for the scalar condensate, the PQ scale and hence the higgsino decay length are determined from the axino dark matter abundance. 
The resulting values of  $f_a = V_{PQ}/N_{DW}$ are shown as solid contours in \Fig{fig:MasterPlotsAxino} for four different choices of higgsino and axino masses.

In any axino scenario, the possible contribution to the DM energy density from axions must also be considered. 
In the pre-inflationary scenario, PQ symmetry is broken above the energy scale of inflation and axions are produced via coherent oscillation of the axion field around $T = 160 \mev$. The axion relic density therefore depends on the initial misalignment angle $\theta_i$, as well as $f_a$. 
The dash-dotted lines in \Fig{fig:MasterPlotsAxino} indicate where the axion relic density is equal to the observed DM abundance for a given $\theta_i$. Note that they trace the $f_a$ contours for $T_R > 160 \mev$, while for $T_R < 160 \mev$ the axion contribution to the energy density is practically eliminated by the dilution. 
Similarly, the dashed line corresponds do the post-inflationary scenario, where PQ symmetry is broken below the energy scale of inflation.
In most of the parameter space we show, the axion energy density can be negligible or subdominant to the axino freeze-in contribution. 
Observing the LLP decay into axinos can therefore provide direct information about the origin of dark matter.

The higgsino lifetime is indicated by the color shadings in~\Fig{fig:MasterPlotsAxino}, same as in Fig.~(\ref{fig:MasterPlots}): dark green, light green and light blue to indicate proper decay lengths greater than $0.1$mm, 1m and 100m.
The majority of the relevant parameter space corresponds to $V_{PQ} \lesssim 10^{13} \gev$ and lifetimes observable at MATHUSLA or the LHC main detectors.

\subsubsubsection{Upshot: Long-Lived Higgsinos at MATHUSLA}
\label{subsec:DisplacedAxinos}

Axinos can make up the observed DM abundance if they are produced via freeze-in from Higgsino decay. For a standard RD cosmology before BBN, the axino must have a mass around or below the MeV scale, generically calling for low inflationary reheating temperatures to avoid the gravitino problem. If some dilution mechanism is present, whether the saxion or a more general scalar condensate, weak-scale axino masses  and hence higher reheating temperatures are possible. 
As the arguments of the previous subsections show, all of these scenarios call for a PQ-breaking scale in the range $V_{PQ} \sim 10^9 - 10^{13} \gev$.
For Higgsino masses accessible at LHC energies, this leads to a wide range of possible lifetimes, from $\mu m$ to $10^5$ m.

The Higgsino decays to an invisible axino and a Higgs or EW gauge boson which in turn decays promptly to visible particles.
At the lower end of the motivated lifetime range, the LHC main detectors will therefore be able to constrain this scenario with LLP searches. 
For larger lifetimes above $\sim 10$m, MATHUSLA will be at least 1-2 orders of magnitude more sensitive than the main detectors due to small LLP branching ratios to leptons and non-negligible backgrounds for hadronically decaying LLPs with less than a few hundred GeV of visible energy, see Sec.~\ref{s.LHCLLPcomparison}.

Quantitative predictions for MATHUSLA's reach depend on the LLP production mode. higgsinos can be produced at the LHC either through Drell-Yan processes, or via cascade decay of heavy colored particles like gluinos. 
By far the most pessimistic assumption is that only direct Drell-Yan production is present. In that case, we can refer to Sec.~\ref{sec:axino-ewino} which studied the general collider phenomenology of Higgsino LLPs decaying to axions. As shown in Fig.~\ref{fig:axinos}, MATHUSLA can probe $V_{PQ}$ up to $\sim 10^{13} \gev$ even for TeV-scale higgsinos, putting almost the entire parameter space of freeze-in axino DM within our reach.


\subsection[SIMPs, ELDERs and Co-Decay]{SIMPs, ELDERs and Co-Decay\footnote{Jeff Asaf Dror, Yonit Hochberg, Eric Kuflik}}
\label{sec:simps}

In the WIMP paradigm, the dark matter relic abundance is entirely set by $2\to2$ annihilations of the dark matter, with no dependence on dark matter self-interactions or on its decay. 
Here we review three classes of dark matter candidates which differ from this vastly: Strongly Interacting Massive Particles (SIMPs), where the relic density is determined by the $3\to 2$ annihilation rate within the dark sector; ELastically DEcoupling Relics (ELDERs), where the relic density is determined by the elastic scattering cross-section with the SM; and Co-Decaying dark matter, where the relic density is determined by strong interactions between the dark matter and other dark sector particles.

\subsubsection{Strongly Interacting Massive Particles (SIMPs)}
\label{s.SIMP}

In the SIMP mechanism~\cite{Hochberg:2014dra} 
(see also~\cite{Hochberg:2014kqa, Choi:2017zww,Choi:2018iit}), dark matter self-interactions play a crucial role in setting its relic density. Here the dark matter abundance is set by the freeze-out of number-changing self-annihilations, typically of $3\to 2$ depletion processes where three dark matter particles collide and annihilate into two dark matter particles. In contrast to `the WIMP miracle', which predicts weak-scale masses for weak couplings, such $3\to2$ processes point to strong scale masses for strong couplings, hence dubbed `the SIMP miracle'.

The self-annihilation process pumps heat into the system, and so
the dark matter must be in thermal contact with the Standard Model bath in order to cool (or dump its entropy into other light degrees of freedom). Such interactions between the dark and visible sectors imply measurable signals that should be observable in a variety of upcoming experiments, including direct-detection, indirect-detection, and direct production at colliders. Moreover, the dark matter's strong $3\to2$ self-interactions typically predict sizable contributions to
$2\to2$ self-scattering processes which naturally address long-standing puzzles in structure formation.
The SIMP setup robustly predicts light dark matter (typically in the MeV to GeV mass range), with strong interactions with itself and weak interactions with ordinary matter.

The SIMP mechanism can be realized in various different ways.
In Ref.~\cite{Hochberg:2014kqa} it was found that SIMP dark matter emerges in generic classes of strongly coupled gauge theories. These are theories of dynamical symmetry breaking---resembling QCD---in which the pions play the role of dark matter, with the Wess-Zumino-Witten term generating the $3\to2$ interactions. $Sp(N)$, $SU(N)$ and $SO(N)$ gauge theories are all viable, provided the number of confining quarks is large enough ($N_f\geq 2, 3,3$ respectively). Dark matter then has the same type of mass, with the same type of interactions, in the same type of theory, as the strong force that binds nuclei together. 

There are many potential ways to mediate the requisite thermalizing interactions between the dark matter and the Standard Model. An attractive  possibility is the vector portal~\cite{Hochberg:2015vrg}, where a kinetically mixed hidden photon (see Sec.~\ref{sec:darkphotons}) communicates between the two sectors. The kinetically mixed mediator can be embedded in the symmetry structure of the theory by identifying the appropriate $U(1)$ subgroup of the residual global symmetries to be gauged, leading to a calculable and predictive framework of SIMP dark sectors.
 Detailed constraints and prospects on the relevant parameter space can be found in Ref.~\cite{Hochberg:2015vrg}. 

From a phenomenological point of view, SIMP DM can therefore be seen as a strong theoretical motivation for the existence of a hidden valley that can be produced at the LHC through various portals. 
For example, a the kinetically mixed dark photon can be produced and decay to hidden quarks, which will shower and fragment into dark mesons.
Depending on the masses and representations of the dark photon, the pions $\pi$ and the $\rho$'s, some dark mesons will decay entirely visibly into leptons and/or hadrons, entirely invisibly, or via off-shell hidden photons to $\pi+\ell^+\ell^-$ or $ \pi+jj$. A host of signatures is thus expected, including a mix of missing energy and $\rho$-decays into narrowly collimated small invariant mass lepton pairs (`lepton jets') or jets.  
It is generic for some or all of these dark sector particles to be long-lived. The dark hadron decay width scales as
\begin{equation}
\Gamma \sim \frac{\alpha_D \alpha \epsilon^2}{18 \pi} \frac{m_D^5}{m_V^4}
\end{equation}
where $\alpha_D$ is the coupling of dark fermions to the dark photon, $m_D$ is a dark hadron mass scale, $\epsilon$ and $m_V$ are the kinetic mixing and mass of the dark photon. As discussed in Sec.~\ref{sec:darkphotons}, the dark photon mass and kinetic mixing could easily allow for copious production of dark photons at the LHC, while the dark sector parameters lead to some or all of the dark hadrons having suppressed decay widths to the SM, giving rise to LLP signatures at the LHC and MATHUSLA. (See also Sec.~\ref{sec:hiddenvalleys} for related signatures)
Following the discussion in Sec.~\ref{s.LHCLLPcomparison}, MATHUSLA will have much better sensitivity than the main detectors if the dark hadrons decay hadronically or have GeV mass or below.

\subsubsection{ELasticaly DEcoupling Relics (ELDERs)}
\label{s.ELDER}

Until recently, in all known examples of proposed dark matter
frameworks, the relic abundance was determined by processes that change the dark matter number
density. A novel alternative is the ELastically DEcoupling Relic proposal~\cite{Kuflik:2015isi}, in which the dark matter relic density is determined
almost exclusively by the decoupling of the elastic scattering off Standard Model particles---a process
that does not change dark matter number density.
As was the case for WIMPs and SIMPs, couplings to
the Standard Model are a necessary of the mechanism, and lead to observable predictions in a host of different experimental frontiers.

Much as in the SIMP scenario, $3 \to 2$ self-interactions of dark matter are required in order for the ELDER mechanism to be viable. 
As the temperature of the universe drops below the DM mass, $3\to2$ annihilation depletes DM number density, while elastic scattering to the visible sector dumps entropy into the SM bath, reducing energy density in the dark sector. 
In the SIMP scenario, the $3\to2$ annihilation freezes out while the dark sector is still in thermal equilibrium with the SM. The annihilation rate therefore determines the DM relic density.
In the ELDER scenario, on the other hand, the elastic scattering between the two sectors freezes out before the $3 \to 2$ annihilation. 
The dark sector therefore enters a period of cannibalization after kinetic decoupling. 
Since it is in chemical equilibrium and its comoving entropy is conserved, the
comoving dark matter density redshifts logarithmically with the
expansion of the Universe. Therefore the DM density today is determined by the density at kinetic decoupling, which in turn depends on the strength of the elastic scattering interaction instead of the $3 \to 2$ annihilation rate. 
This opens up different regions of parameter space for strongly coupled hidden sectors to produce viable dark matter candidates. 
Like SIMPs, ELDERs therefore  represent another motivation for hidden valley type LLP signatures at both the LHC main detectors and MATHUSLA, as discussed above.

\subsubsection{Co-decaying Dark Matter}
\label{s.codecayingDM}

Dark sector self-interactions, {\it i.e}, strong interactions between the dark matter and other hidden sector particles, can also play a  role in determining the dark matter relic abundance.  Even though these dark sector interactions may not be directly observable in the lab, it can still be the case that the hidden sector interactions with the Standard Model play a critical role in determining the relic abundance, and are observable in current and future experiments.
One example of dark matter freezeout where this is true is Co-decaying dark matter~\cite{Dror:2016rxc}, which is generic in hidden sectors with  approximate degeneracies between one or more LLPs with a stable particle. (See also Ref.~\cite{Farina:2016llk,Okawa:2016wrr} for related ideas.) 
Crucial to the Co-decaying DM mechanism is that the dark sector includes an LLP. Either the dark sector LLP, or a visible sector LLP that is long-lived due to the same small portal, could then be produced at the LHC and observed in MATHUSLA.

Hidden sectors often have accidental symmetries as a residual of the symmetry used to keep the dark matter stable. This can lead to degenerate low-lying states with some of these particles remaining unstable. The decay of the unstable particles can efficiently deplete the dark sector, in analogy with co-annihilation (Section~\ref{sec:coannihilation}), leaving behind a dark matter candidate with the observed relic abundance. 

As a concrete example, consider a system of two  nearly-degenerate states $A, B$ of mass $m$. $A$ is the stable DM candidate, while $B$ is meta-stable and decays to the SM with lifetime $\tau_B$ at temperature $T_\gamma$. The process $AA \leftrightarrow BB$ is active in the plasma with some cross-section $\sigma$ that freezes out at temperature $T_\sigma$. Finally, there is a small interaction between the dark and visible sectors that keeps them in thermal equilibrium until it freezes out at temperature $T_d$. 
The Co-decaying DM mechanism corresponds to the regime  $T_d > m > T_\gamma > T_\sigma$. In other words, as the universe cools, the hidden sector thermally decouples from the SM at $T_d$. $A$ and $B$ stay in equilibrium with each other as they cool and become non-relativistic. $B$ starts decaying out-of-equilibrium into the SM at temperature $T_\gamma$, which depletes the dark sector energy density until $A$ and $B$ decouple at $T_\sigma$. The surviving relic density of $A$ makes up DM today.

\begin{figure}
  \begin{center}
\includegraphics[width=9cm]{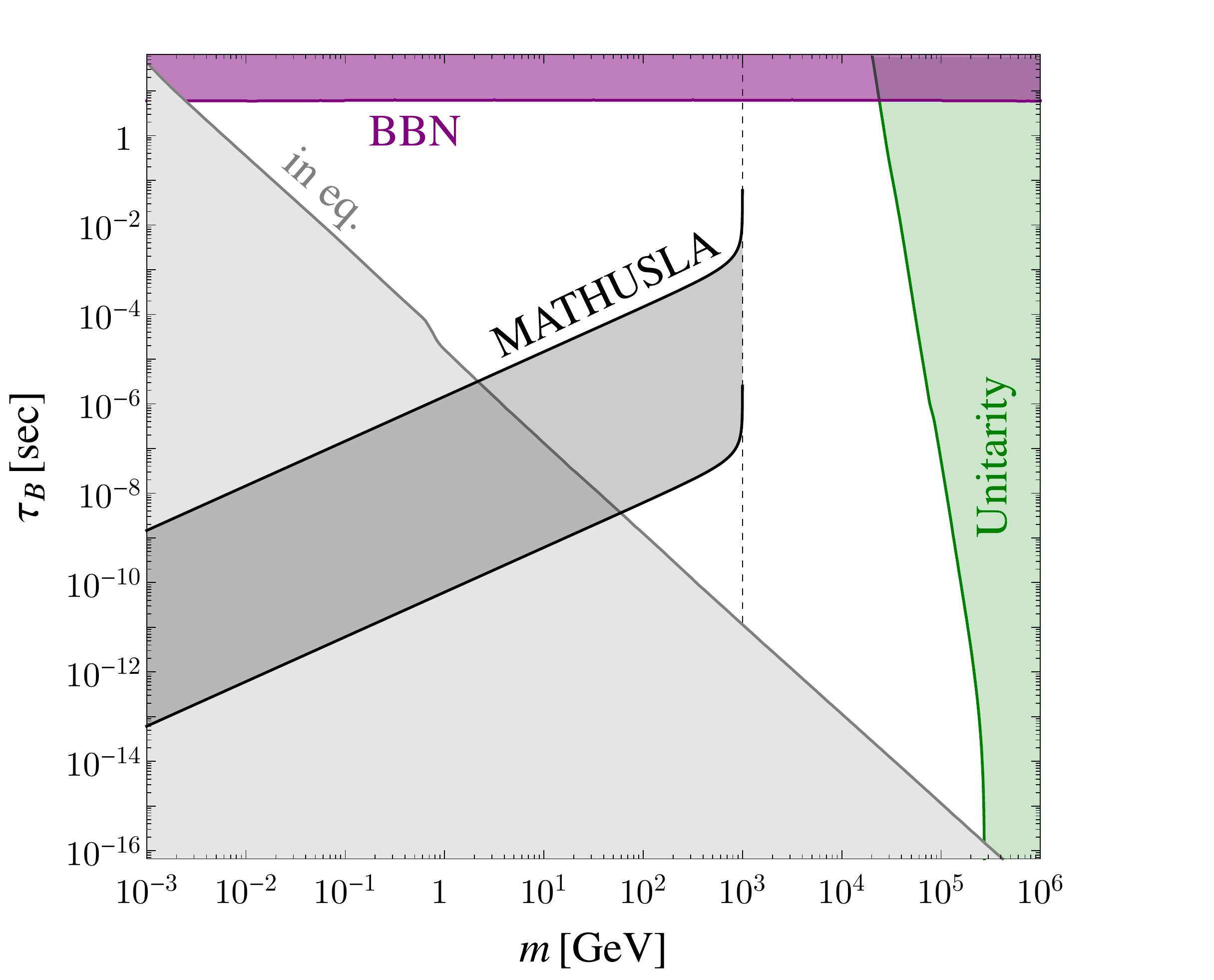}
\caption{
White region: the parameter space of co-decaying DM, assuming negligible cannibalization effects, that satisfies the out of thermal equilibrium ({\color[rgb]{0.3,0.3,0.3}{\bf gray}}), unitarity ({\color[rgb]{0,0.5,0} {\bf green}}) and Big Bang Nucleosynthesis ({\color[rgb]{.5,0,.5} {\bf purple}}) constraints.
Assuming LLP production in 1 TeV gluino decays, the most pessimistic MATHUSLA reach projection is shown as the dark gray shaded region. Details of the complete model implementation generically open up additional regions that MATHUSLA could probe,  see text for details.
This estimate assumes the $200m \times 200m \times 20m$ benchmark geometry of Fig.~\ref{f.mathuslalayout}.
}
\label{fig:codecaylimit}
\end{center}
\end{figure}

The decay of the LLPs  play the role of the Boltzmann suppression in depleting the dark sector. 
This framework neatly evades the stringent bounds from direct-detection due to the small couplings to the Standard Model particles. Furthermore, the temperature where freeze-out occurs can be delayed by many orders of magnitude compared to WIMP dark matter, which leads to a relatively smaller relic density for a given cross-section. 
This results in a robust prediction of an enhanced indirect-detection signal that can be relevant even for dark matter masses above the TeV scale.

The parameter space of Co-decaying dark matter spans many orders of magnitude in both mass and lifetimes of the unstable particle(s), as illustrated in Fig.~\ref{fig:codecaylimit}. At each point in the $(m, \tau_B)$ plane, $\sigma$ is chosen to obtain the correct relic density. For simplicity, we work here in the limit where DM cannibalization effects are negligible.\footnote{See Ref.~\cite{Dror:2016rxc} for a detailed discussion on the differences when cannibalization is important.} 
In the light gray region, the decay of $B$ does not occur out-of-equilibrium, i.e. $T_\gamma$ is too high. The purple region is conservatively excluded by BBN constraints on LLP lifetime (the exact constraint depends on the $B$ decay mode and may be more than an order of magnitude lower in $\tau_B$). The light green shaded region is excluded because the $\sigma$ required to avoid overclosure violates unitarity bounds.

Co-decaying DM is a very general framework for obtaining the observed DM relic density from dark sector dynamics. As such, a large variety of detailed phenomenologies are possible, depending on the model implementation. 
Interestingly, the co-decay mechanism can fit naturally in models which address the hierarchy problem, such as supersymmetry or composite Higgs, since these frameworks can yield degenerate particles as the lowest lying particles in a new sector with weak couplings to the Standard Model. 
Furthermore, the same small portal between the visible and dark sectors that makes $B$ long-lived can also result in a visible sector particle being long-lived  (like the lightest parity-odd particle in SUSY) before it decays to the hidden sector.
As a result, the co-decaying DM framework generically predicts LLPs production in the decay of heavier BSM particles. For states connected to the solution of the Hierarchy Problem, this leads to QCD-strength LLP production cross-sections at the LHC. 
In that case, the main detectors and MATHUSLA could probe large, complementary regions of Co-decaying DM parameter space.

As a relevant example, consider the case of LLPs produced from the production of 1 TeV gluinos. The most pessimistic reach estimate is obtained by assuming that the gluino decays directly to $B$ and that most of the gluino mass is converted to LLP boost. In that case, the region of co-decaying DM parameter space where MATHUSLA sees at least 4 LLP decays  is shown as the dark gray region in Fig.~\ref{fig:codecaylimit}. 
If less than all the gluino rest energy is converted to LLP boost due to other energetic decay products, the lab-frame LLP lifetime decreases, shifting sensitivity to larger values of $\tau_B$ and probing more regions which are not already excluded by the out-of-equilibrium bound. 

It could also be the case that the $B$ LLP is not produced directly in gluino decays, but rather as part of a visible-sector decay chain that terminates with particle $\chi$. If $\chi$ decays to the dark sector, the same small portal coupling which makes $B$ long-lived can also make $\chi$ long-lived, though its decay width is expected to scale as $\Gamma_\chi/\Gamma_B \sim m_\chi/m > 1$ (or some higher power). Given the that $\chi$ also has lower boost by a factor of $\sim m_\chi/m$, the lab-frame lifetime of $\chi$ is shorter than for $B$ by at least a factor of $(b_\chi c \tau_\chi)/(b_B c \tau_B) \sim (m/m_\chi)^2$.
MATHUSLA then has two chances of detecting an LLP: either $B$ itself (produced in $\chi$ decays) or $\chi$ (produced in gluino decays). This allows MATHUSLA to probe both the dark-gray region in Fig.~\ref{fig:codecaylimit} and regions at larger $\tau_B$ where $B$ escapes undetected but $\chi$ decays in MATHUSLA.

The comparison of MATHUSLA's reach main detector LLP searches depends sensitively on model details, i.e. the exact production and decay modes. However, as argued in Section~\ref{sec:LLPSmathuslahllhc}, the cross-section reach for LLP detection is likely to be better at MATHUSLA, possibly by orders of magnitude if the LLP is produced and decays without highly conspicuous jets, leptons or missing energy in the final state.
Of course, the main detectors would have better sensitivity at lower lifetimes. MATHUSLA and the main detectors together would therefore probe deep into the center of the co-decaying dark matter parameter space.

\subsection[Dynamical Dark Matter]{Dynamical Dark Matter\footnote{David Curtin, Keith R. Dienes, Brooks Thomas}}
\label{sec:dynamicaldm}

\def\tnow{t_{\mathrm{now}}}
\def\OmegaDM{\Omega_{\mathrm{CDM}}}

In this section we study how MATHUSLA could discover DM within the \emph{Dynamical Dark Matter} (DDM) framework~\cite{Dienes:2011ja,Dienes:2011sa,Dienes:2012jb,Dienes:2012iia,Dienes:2013qua,Dienes:2013rua}, which intrinsically gives rise to a large number of dark sector states with varying lifetimes from collider-scales to cosmological hyperstability. This is a particularly dramatic example of LLP signatures giving direct insight into the nature, as well as the cosmological and astrophysical role, of the dark sector. 

\subsubsection{Introduction\label{sec:DDMintro}}

In most models of DM, the dark sector is composed of one or several dark-matter particle(s) $\chi$ which 
carry the entire dark-matter cosmological abundance
$\Omega_{\rm CDM} \approx 0.26$~\cite{Hinshaw:2012aka}.
These particle(s) must be hyperstable, with lifetimes 
exceeding the age of the universe by many orders of magnitude:  $\tau_\chi \gsim 10^{26}$~s.
This stability is critical for traditional dark matter. 
Indeed, any particle which decays too rapidly into Standard-Model (SM) 
states is likely to upset BBN and light-element 
abundances,
and also leave undesirable imprints in the CMB and
diffuse X-ray/gamma-ray
backgrounds.
However, as a result of this stability,
the resulting dark sector is then essentially ``frozen'' in time, with
$\Omega_{\rm CDM}$ remaining constant in our late-time matter-dominated universe.
Moreover, as explained above, this stability also ensures that 
once such a dark-matter particle 
is produced in a collider,
it escapes without any subsequent observable decay.

Dynamical Dark Matter~\cite{Dienes:2011ja,Dienes:2011sa,Dienes:2012jb,Dienes:2012iia,Dienes:2013qua,Dienes:2013rua} generalizes this assumption by positing that the dark sector consists not merely of one or more hyperstable DM particles, but \emph{many} such particles which can have varying lifetimes. The number $N$ of dark-matter states can be order 10, 1000, or even much larger in some scenarios. 
Thus, instead of having a single dark-matter particle $\chi$,
the dark sector contains an entire {\it ensemble}\/ of dark-sector states $\chi_n$ ($n=1,...,N$).
In that case, no state individually needs to carry the full abundance $\Omega_{\rm CDM}$ so long as
the sum of their individual abundances  $\Omega_n$ matches $\Omega_{\rm CDM}$.
In particular, the individual dark components within the ensemble can carry a wide variety of
abundances $\Omega_n$, some relatively large but others relatively small.
This is a critical observation, because
{\it a given dark-matter
component $\chi_n$ need not be stable if its
abundance $\Omega_n$ at the time of its decay into SM states is sufficiently small}\/.
Indeed, a sufficiently small abundance assures that all of the disruptive effects
of the decay of $\chi_n$ into SM states will be minimal, and that all
constraints from BBN, CMB, {\it etc.}  will continue to be satisfied.

We are thus naturally led to an alternative concept~\cite{Dienes:2011ja}: balancing of 
SM-producing decay widths $\Gamma_n$ against cosmological 
abundances $\Omega_n$.
Dark-matter states with larger abundances must have smaller decay widths
and survive until (and potentially beyond) the present time,
but states with smaller abundances can have larger decay widths and decay at earlier times.
As long as decay widths are balanced against abundances in this way across our entire
dark-sector ensemble, all phenomenological constraints can potentially be satisfied.
Thus, dark-matter hyperstability is no longer required.

This is the basic principle of Dynamical Dark Matter: 
an alternative framework for dark-matter 
physics in which the notion of dark matter
stability is
replaced by a balancing of lifetimes against cosmological abundances
across an ensemble of $N$ individual dark-matter 
components $\chi_n$ with
different masses $m_n$, lifetimes $\tau_n\equiv \Gamma_n^{-1}$, and abundances $\Omega_n$.
DDM is in some sense a natural generalization of the standard scenario of a single hyperstable DM species, which is recovered in the $N \to 1$ limit. In general, DDM can give rise to far richer cosmology and phenomenology. As its name implies, the dark sector becomes truly {\it dynamical}\/,
with the different components
of the DDM ensemble decaying before,
during, and after the present epoch.
Indeed, some portions of the DDM ensemble
have already decayed prior to the present epoch,
and are thus no longer part of the dark sector.
However, other portions of the DDM ensemble have yet to decay.
It is these ensemble constituents whose abundances $\Omega_n$ together comprise
the specific dark-matter abundance $\Omega_{\rm CDM}\approx 0.26$ observed today.

Since the original DDM proposal~\cite{Dienes:2011ja,Dienes:2011sa,Dienes:2012jb},
there have been many explicit realizations of such DDM ensembles, i.e. different theoretical
scenarios for BSM physics which give rise to a large collection of dark states in which
the widths for decays into SM states are naturally inversely balanced against cosmological abundances.
These include theories involving large extra spacetime
dimensions~\cite{Dienes:2011ja,Dienes:2011sa,Dienes:2012jb}, theories involving strongly-coupled
hidden sectors~\cite{Dienes:2016vei,Dienes:2017ecw}, theories involving large spontaneously-broken
symmetry groups~\cite{Dienes:2016kgc},
and examples from string theory~\cite{Dienes:2016vei,Dienes:2017ecw,Chialva:2012rq}.
Indeed the dark states within these different realizations
can accrue suitable cosmological abundances in a variety of
ways, including not only through non-thermal generation 
mechanisms such as misalignment production~\cite{Dienes:2011ja,Dienes:2011sa,Dienes:2012jb}
but also through thermal mechanisms such as freeze-out~\cite{Dienes:2017zjq}. 
Mass-generating phase transitions
in the early universe can also endow collections of such states with non-trivial
cosmological abundances~\cite{Dienes:2015bka,Dienes:2016zfr,Dienes:2016mte}.

In these and other realistic DDM scenarios, the masses, lifetimes,
and abundances of these individual particles are not arbitrary.
Rather, these quantities are determined by the underlying physics model and take the form of {\it scaling relations} (either exact or approximate) which encode their dependence on each other and how they vary within the DDM ensemble.
These scaling relations completely specify the properties of the ensemble constituents through a relatively small number of free parameters. 
Thus, even though the number of particles which
contribute to the total dark-matter abundance is typically large,
DDM scenarios can nevertheless be very predictive.

The most fundamental of these scaling relations governs the spectrum
of masses for the DDM constituent particles $\chi_n$. 
In general, we assume a constituent mass spectrum of the form
\beq
         m_n ~=~ m_0 + (\Delta m) \, n^\delta
\label{ddmmasses}
\eeq
where $\lbrace m_0, \Delta m, \delta\rbrace$ are arbitrary parameters
and where $\Delta m, \delta > 0$ 
(so that $n$ labels the DDM constituents in order of increasing mass).
Most concrete realizations of DDM ensembles have mass spectra which take
this general form, either exactly or approximately. 
For example, if --- as in Refs.~\cite{Dienes:2011ja,Dienes:2011sa} --- 
the ensemble constituents consist of the Kaluza-Klein (KK) excitations of a scalar
field compactified on a circle of radius $R$ (or a $Z_2$ orbifold thereof), 
we have either $\lbrace m_0,\Delta m,\delta\rbrace = \lbrace m, 1/R, 1\rbrace$
or 
 $\lbrace m_0,\Delta m,\delta\rbrace =\lbrace m, 1/(2 m R^2), 2\rbrace$,
depending on whether $m R \ll1$ or $mR\gg 1$, 
respectively, 
where $m$ is the four-dimensional scalar mass.
In general, for arbitrary $m R$, we find that the latter behavior
holds for $n\ll mR$ and the former for $n\gg mR$.
Likewise, if the ensemble constituents consist of the bound states of a strongly-coupled
gauge theory, as in Refs.~\cite{Dienes:2016vei,Dienes:2017ecw}, we have $\delta = 1/2$, where $\Delta m$ and $m_0$ are related to
the Regge slope
and Regge intercept of the strongly-coupled theory, respectively.
Thus $\delta=1/2$, $\delta=1$, and $\delta=2$ may be considered as particularly compelling ``benchmark'' values.

Given a mass spectrum of this general form, we then typically take a scaling
relation for the decay widths $\Gamma_n$ of the form
\beq
            \Gamma_n ~=~ \Gamma_0 \left( {m_n\over m_0}\right)^y
\label{ddmdecaywidths}
\eeq
where $\Gamma_0$ is the decay width of the lightest DDM state and $y$ is an additional free parameter.   
Note that $\Gamma_n$ is assumed to be the decay width of the $n$th ensemble constituent $\chi_n$
into SM states, and we disregard the possibility of intra-ensemble decays
(or assume that the branching ratios for such decays are relatively small).
The corresponding $\chi_n$ lifetimes are then given by $\tau_n \equiv \Gamma_n^{-1}$.
In general, the scaling exponent $y$ can be arbitrary.
For example, if we assume that the dominant decay mode of $\chi_n$ is to a final state consisting of SM particles whose masses are all significantly less than $m_n$, and if this decay occurs through a dimension-$d$ contact operator of the form ${\cal O}_n \sim c_n \chi_n {\cal O}_{\rm SM}/\Lambda^{d-4}$ where $\Lambda$ is an appropriate mass scale and where ${\cal O}_{\rm SM}$ is an operator built from SM fields, we have 
\beq
                   y ~=~ 2d - 7~.
\label{ddmyd}
\eeq
In general one finds $y>0$ (such as $y = 5$ for hidden valley decays mediated by a dark photon, see Eqn.~(\ref{e.darkphotonHVdecaywidth})) but this need not be a strict requirement.
Indeed, since the fundamental couplings that underlie such decays can often themselves depend on $n$,
the scaling exponents $y$ could be large, depending on the scaling behavior of $c_n$ and the dimensionality of ${\cal O}_{\rm SM}$. 

Another important quantity is the spectrum of cosmological relic abundances  
$\Omega_n$ associated with each DDM constituent.  These are likewise assumed to satisfy an approximate
scaling relation of the form
\beq
                   \Omega_n ~=~ \Omega_0 \left( {m_n\over m_0}\right)^\gamma~.
\eeq
The precise value of the scaling exponent $\gamma$ generally depends on the particular 
dark-matter production mechanism
assumed.
One typically finds that $\gamma<0$ for misalignment production~\cite{Dienes:2011ja,Dienes:2011sa}, while
$\gamma$ can generally be of either sign for thermal freeze-out~\cite{Dienes:2017zjq}.

In a similar vein, for many investigations 
it is instructive to focus on the coupling coefficients $c_{m,n,...p}$ of Lagrangian operators
which involve multiple ensemble constituents $\lbrace \chi_{m},\chi_{n},...,\chi_p\rbrace$
together with some set of particles outside the ensemble.
Such couplings can ultimately be relevant for 
dark-matter production, scattering, annihilation, and decay.
In the analysis below, we are mostly interested in couplings that involve two dark-matter constituents $\chi_m$ and $\chi_n$ (or their antiparticles), and we further restrict our attention to the ``diagonal'' $m = n$ case. 
We then assume a scaling relation for such couplings of the form
\beq
                   c_{n,n} ~=~ c_0 \left( {m_n\over m_0}\right)^\xi
\label{ddmcouplings}
\eeq
where $c_0$ is an overall normalization and where $\xi$ is a corresponding scaling exponent.
For example, $\xi = 0$ corresponds to democratic decay into different final states that are much lighter than the parent particle, while $\xi = 1$ corresponds to a Yukawa-like coupling instead. 
Assuming a scaling relation of this form allows us to study a wide variety of underlying theoretical mechanisms that might generate such couplings.   
Once a particular scaling relation for the coupling is specified, the scaling behaviors of the  corresponding production, scattering, or annihilation cross-sections 
are also determined. 
Since these cross-sections also depend on kinematic factors, their behavior across the ensemble can deviate significantly from the simple power-law couplings we have assumed for the underlying couplings. 
For example, the results in Ref.~\cite{Dienes:2017zjq} can be interpreted as illustrating the large range of possible scaling behaviors that can be exhibited by an annihilation cross-section when the underlying couplings $c_{nn}$
are held fixed ($\xi=0$) across the entire DDM ensemble.

In general, the phenomenological viability of the DDM framework rests upon relations between these different scaling exponents. 
Two of the most important which underpin
the entire DDM framework are the relations~\cite{Dienes:2011ja,Dienes:2017zjq}
\beq
  \gamma y < 0
\eeq
and 
\beq
           -1 ~\lsim~ {1\over y} \left( \gamma + {1\over \delta}\right) ~< ~ 0~.
\eeq
The first of these relations simply ensures that states with larger abundances have smaller decay widths/longer lifetimes, as required within the DDM framework
The second relation
ensures a suitable effective equation of state for the  
collective DDM ensemble, with an effective equation-of-state parameter $w \approx 0$ that does not change appreciably over a significant portion of the recent cosmological past~\cite{Dienes:2017zjq}.
Moreover, this relation also ensures that the total energy density carried by the ensemble is finite in the $N \to \infty$ limit.

\subsubsection{Detecting DDM at MATHUSLA\label{sec:colliders}}

Scenarios within the DDM framework can give rise to distinctive signatures at
colliders~\cite{Dienes:2012yz,Dienes:2014bka,Dienes:2016udc}, at direct-detection
experiments~\cite{Dienes:2012cf}, and at indirect-detection experiments~\cite{Dienes:2013xff,Dienes:2014cca,Boddy:2016fds,Boddy:2016hbp}.
Such scenarios also give rise to enhanced complementarities~\cite{Dienes:2014via,Dienes:2017ylr} 
between different types of experimental probes.

If a production mode for DDM states is available at the LHC, such as a heavy BSM state with SM charge that decays into DDM states, then  the ensemble can give rise to a mixed variety of MET and DV signatures in the main detectors. Kinematic analysis of the visible final states could  provide evidence for the multi-component nature of DM and the existence of DDM scaling relations. 
On the other hand, if the decay length of accessible states is much larger than the main detector size then MATHUSLA could be the best discovery opportunity for DDM. This is the scenario we examine here.

As evidenced by Eqn.~(\ref{e.bctaumax}), MATHUSLA is capable of detecting LLPs with lifetimes at or even exceeding the BBN limit depending on the LHC production cross-section. Even so, it is clear that no possible LHC production rate would allow MATHUSLA to detect the decay of the most stable DDM states which constitute the DM abundance today. Fortunately, the very nature of the DDM ensemble gives rise to dark states with a large variety of possible lifetimes, with decay lengths generally decreasing with increasing mass. Production of heavier and detectably meta-stable DDM states is therefore possible at energy frontier machines like the HL- or HE-LHC. This could give rise to MATHUSLA signatures alongside MET signatures from both the MATHUSLA-detectable states and longer-lived states in the ensemble. 
If the production process of DDM states does not give rise to hard SM final states in decays, then MET searches have to rely on ISR and MATHUSLA could be our only probe of these DDM scenarios, see Section~\ref{s.METcomparison}. Careful analysis of the observed LLP decays within MATHUSLA could then reveal their varying masses and lifetimes~\cite{Curtin:2017izq} and provide evidence for the scaling relations of the DDM ensemble. 
On the other hand, if the DDM states are produced in the decays of heavy particles that also produce SM-charged final states, then correlating the MET and LLP signatures will clearly be an important tool to constrain the properties of the DDM ensemble. 
In either scenario, MATHUSLA would be invaluable to discover and diagnose DDM.

\subsubsection{MATHUSLA reach in a benchmark DDM parameter space\label{sec:results}}

In order to provide a quantitative assessment of the reach of MATHUSLA within the DDM parameter space,  
we conduct a toy study of the simplest scenario whereby DDM states $\chi_n$ are produced in the prompt two-body decay $\phi \to \chi_n \chi_n$ of a heavy state $\phi$ with mass $m_{\phi}$ and LHC production cross-section $\sigma_\phi$. At our level of analysis, the spin of $\phi$ and $\chi_n$ is irrelevant, and we fix kinematics of $\phi$ by assuming the $n^\mathrm{th}$ DDM state has an average boost factor $b_n = \frac{|\vec p_n |}{m_n} = \frac{m_{\phi}}{2 m_{n}}\sqrt{1  - 4 m_{n}^2/m_\phi^2}$.
The masses and decay widths of the $\chi_n$ are given by the scaling relations Eqns.~(\ref{ddmmasses}) and (\ref{ddmdecaywidths}), while their relative couplings to $\phi$ are determined by Eqn.~(\ref{ddmcouplings}). 
We therefore have a nine-dimensional parameter space
$\lbrace m_\phi, \sigma_\phi, m_0, \Delta m, \delta, \Gamma_0, y, c_0, \xi\rbrace$.
For each value of the chosen parameters, we can use the simple expressions in Section~\ref{s.MATHUSLAsignalestimate} to estimate the rate of observed decays within MATHUSLA for each state $\chi_n$.

For concreteness, we take $\Gamma_0$ to be determined by the traditional dark-matter hyperstability bound, i.e., $\Gamma_0 = (10^9 \, t_{\rm now})^{-1} = 10^{-26}~{\rm s}^{-1}$ where $t_{\rm now} = 10^{17}$~s is the current age of the universe.  (Larger values will simply linearly rescale the signal in the long-lifetime limit.) We also set $m_\phi = 2$~TeV as a concrete benchmark, to be discussed further below.  
If $\phi$ has couplings to SM or other non-DDM states, then the quantity $c_0$ determines the total branching fraction ${\rm BR}_{\chi\chi}\equiv\sum_{n=0}^\infty {\rm BR}(\phi\to{\chi_n}\chi_n)$ of $\phi$ into DDM states.
Since MATHUSLA is ultimately sensitive not to $\sigma_\phi$ alone but to the product $\sigma_\phi {\rm BR}_{\chi\chi}$, we therefore now have a seven-dimensional parameter space $\lbrace m_\phi, \sigma_\phi {\rm BR}_{\chi\chi}, m_0, \Delta m, \delta, y, \xi\rbrace$.  We shall therefore quantitatively assess the reach of the MATHUSLA detector in terms of the minimum value of $\sigma_\phi {\rm BR}_{\chi\chi}$  gives rise to four observed LLP decays within the MATHUSLA decay volume, given specific values of the remaining six parameters $\lbrace m_\phi, m_0, \Delta m, \delta, y, \xi\rbrace$.  In the zero-background regime, this can be interpreted as an exclusion limit on $\sigma_\phi {\rm BR}_{\chi\chi}$.  In the event that LLP decays are observed, this would correspond roughly to the minimum cross-section required for DDM discovery.

For any value of $m_\phi$, the decays of $\phi$ can potentially produce the ensemble 
constituents $\lbrace \chi_0,\chi_1,...\chi_{n_{\rm kin}}\rbrace$, where $n_{\rm kin}$ is the kinematic limit, defined as the maximum value of $n$ for which $m_n\leq m_\phi/2$.
We also define two further quantities $n_{\rm min}$ and $n_{\rm max}$ as those values of $n$ which delimit the range of ensemble constituents $\chi_n$ whose subsequent decays into SM states are responsible for approximately 90\% of the observed events within MATHUSLA.~
Thus $(n_{\rm min}, n_{\rm max})$ describes that subset of DDM ensemble states to which the MATHUSLA detector is 
most sensitive.  
Finally, we define $n_{\rm cs}$ as indicating the heaviest ensemble constituent $\chi_n$ which is cosmologically stable, with $\tau_n\equiv \Gamma_n^{-1} \geq t_{\rm now}$.
Thus only the ensemble constituents $\lbrace \chi_0,\chi_1,...,\chi_{n_{\rm cs}}\rbrace$ will have survived to the present time and have the potential to contribute to the total present-day dark-matter abundance 
$\Omega_{\rm CDM}\approx 0.26$.
We have already noted that the MATHUSLA detector, while capable of probing large portions of the DDM ensemble, 
cannot actually probe those elements of the ensemble which constitute dark matter today, and  $n_{\rm cs} < n_{\rm min}$.
Therefore, the $n_{\rm cs}<1$ contour in our plots demarcates the area of parameter space in which DDM realizes a more traditional DM model with only a single hyperstable DM particle $\chi_0$. However, even in those scenarios the DDM ensemble could contain long-lived states  that may have affected the early cosmological history of the universe.

\begin{figure*}[t]
\centering
\includegraphics[width=0.48\textwidth,keepaspectratio] {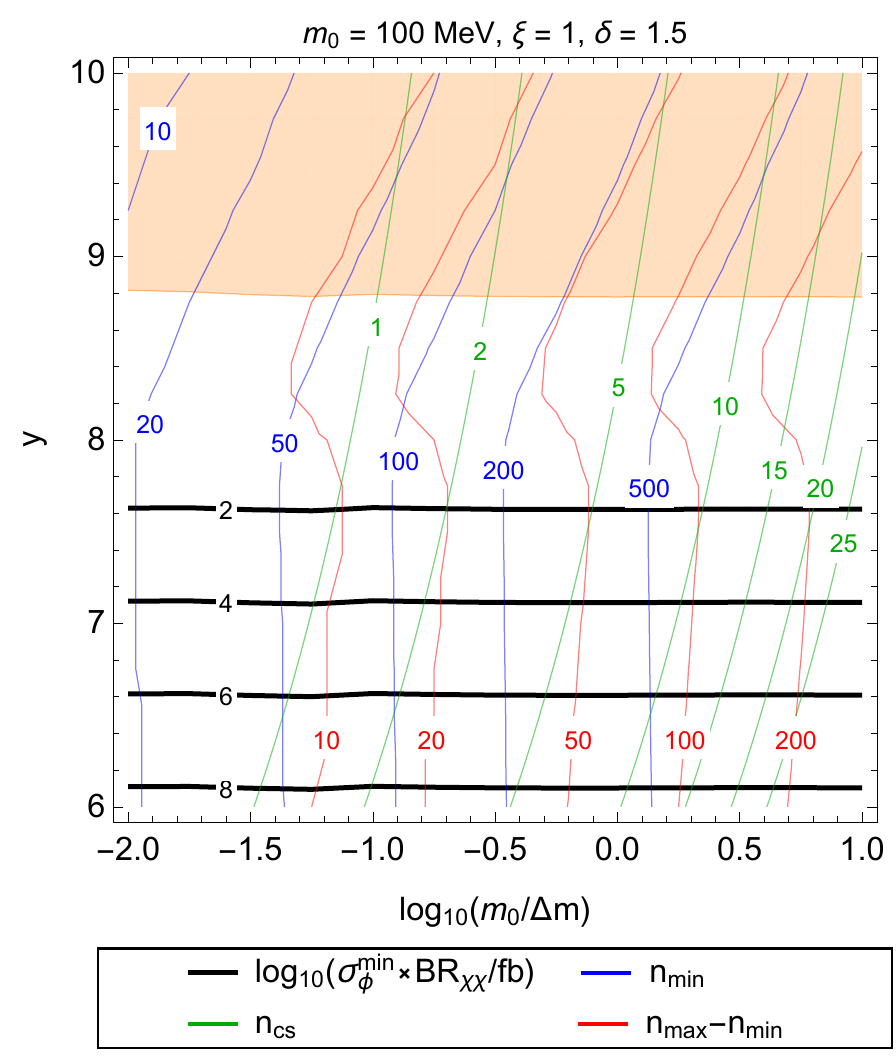}
\hfill
\includegraphics[width=0.49\textwidth,keepaspectratio] {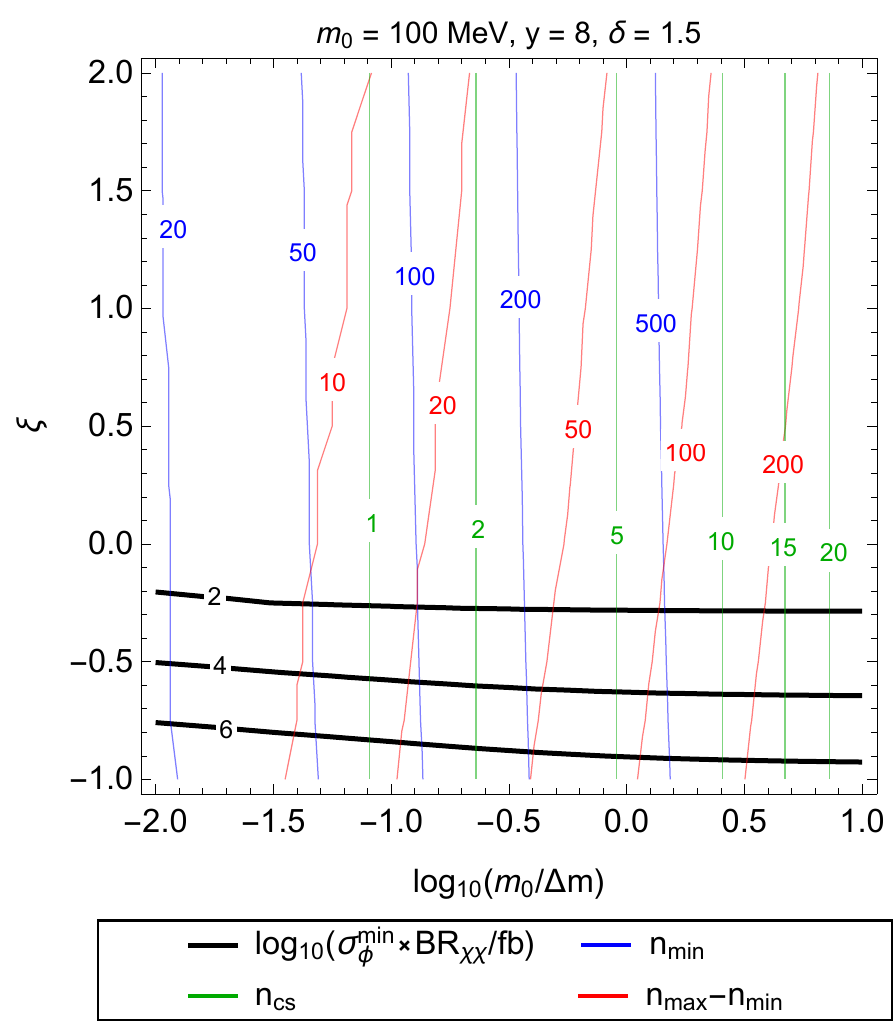}
\caption{The reach of MATHUSLA within the DDM parameter space for the benchmark values 
         $m_0=100$~MeV and $\delta=1.5$.
     Black curves indicate contours of  $\sigma_\phi^{\rm min} {\rm BR}_{\chi\chi}$, while
      blue, red, and green curves indicate contours of $n_{\rm min}$, 
     $n_{\rm max}-n_{\rm min}$, and $n_{\rm cs}$, respectively.
     Likewise, the orange shading indicates the region 
        of DDM parameter space in which at least one of the ensemble constituents $\chi_n$ has a  
  characteristic decay length $\beta\gamma c\tau_{\mathrm{min}} < 1$~m.
  As discussed in the text, the region with 
     $m_0 /\Delta m \gtrsim 0.1$, $7.5 \lesssim y \lesssim 8.8$, and  $\xi \gtrsim -0.3$ 
   is a particular ``sweet spot'' in which multiple light states within the DDM ensemble 
       comprise the present-day dark matter while numerous heavier states within the same ensemble
       can lead to an observable signal at MATHUSLA.
       This estimate assumes the $200m \times 200m \times 20m$ benchmark geometry of Fig.~\ref{f.mathuslalayout}.
       }
\label{fig:ddmfig1}
\end{figure*}

\begin{figure*}[t]
\centering
\includegraphics[width=0.48\textwidth,keepaspectratio] {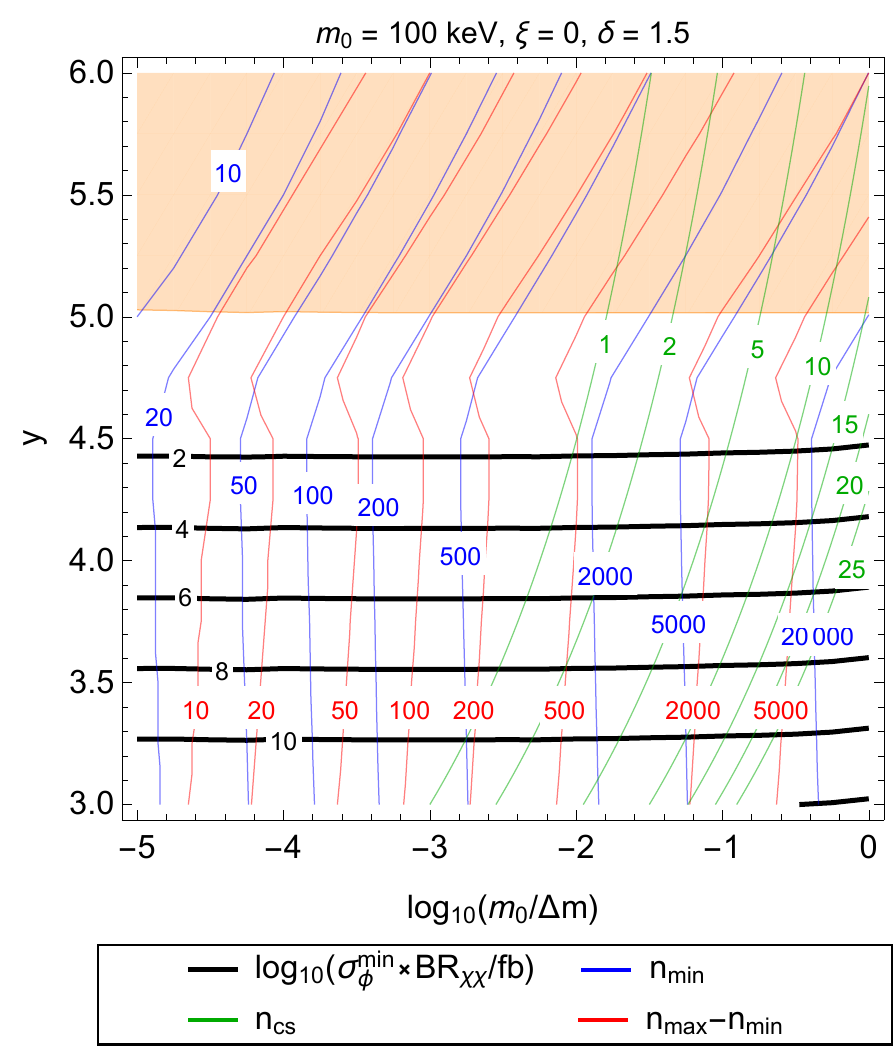}
\hfill
\includegraphics[width=0.49\textwidth,keepaspectratio] {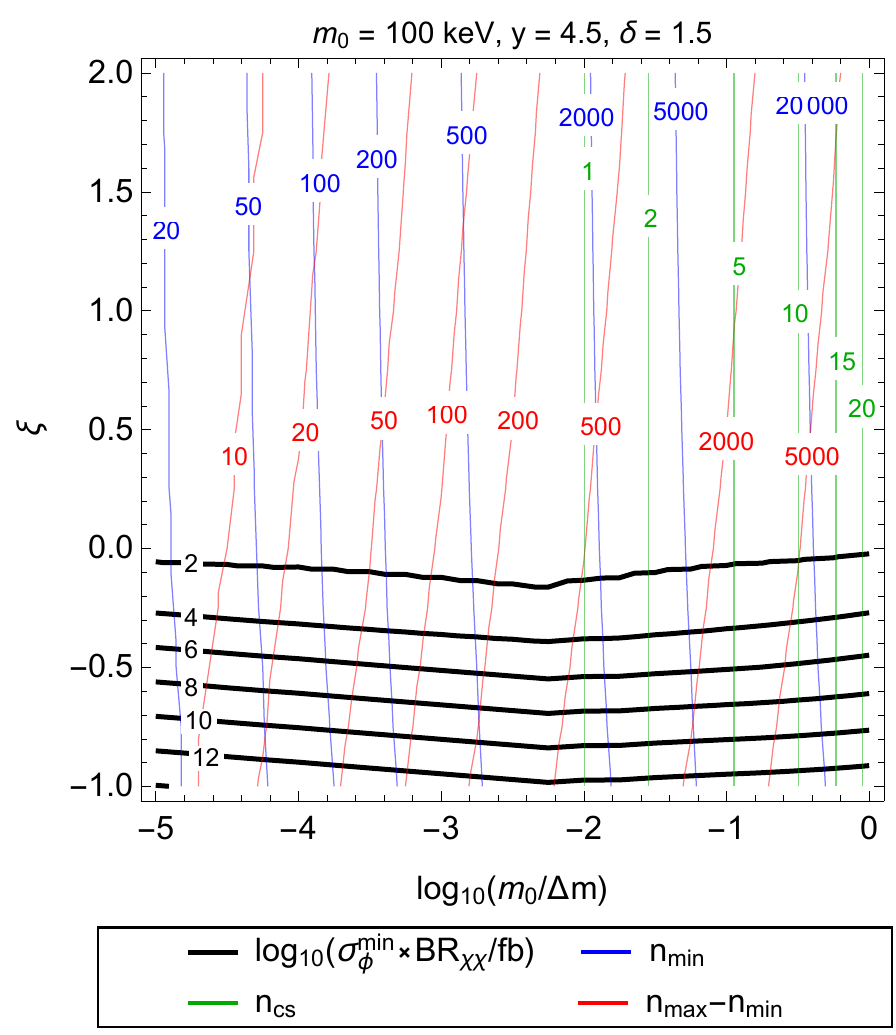}
\caption{Same as Fig.~\ref{fig:ddmfig1}, except that we have now shifted $m_0$ from $100$~MeV to $100$~keV.~
   This allows MATHUSLA to be sensitive to DDM ensembles with smaller values of $y$,
   leading to an even more compelling ``sweet spot'' with
      $m_0/\Delta m\gsim 0.01$,
     $4.3 \lsim y \lsim 5.0$, and $\xi \gsim -0.2$.}
\label{fig:ddmfig2}
\end{figure*}

Given these definitions, our results are as follows.
In Fig.~\ref{fig:ddmfig1}, we indicate the sensitivity of MATHUSLA by plotting contours of
$\sigma_\phi^{\rm min} {\rm BR}_{\chi\chi}$, where  $\sigma_\phi^{\rm min} $ is the minimum production cross-section for the parent particle $\phi$ 
which will produce at least four signal events within MATHUSLA.~
In the left and right panels of Fig.~\ref{fig:ddmfig1}, 
these contours (black curves) are plotted within the $( m_0/\Delta m, y)$ and
$(m_0/\Delta m, \xi)$ planes, respectively.
For each plot
we have chosen the benchmark values
$m_\phi = 2$~TeV, $m_0 = 100$~MeV, $\xi = 1$, and $\delta = 1.5$.
Within each panel we also show contours of $n_{\rm min}$ (blue curves), 
$n_{\rm max}-n_{\rm min}$ (red curves), and $n_{\rm cs}$ (green curves).  
The orange shaded region in the left panel is the region in which at least one of 
the $\chi_n$ has a characteristic decay length $\beta\gamma c\tau_{\mathrm{min}} < 1$~m.
The results in Fig.~\ref{fig:ddmfig1} correspond to the case in which the $\chi_n$ are real scalars,
but the results for spin-1/2 fermions are qualitatively similar.

In this connection, our benchmark value $m_\phi = 2$~TeV deserves further comment.  This benchmark is motivated in part by a self-consistency requirement:  in order for the ensemble to lead to a detectable signal at MATHUSLA during the HL-LHC run, the production cross-section $\sigma_\phi$ must exceed the sensitivity threshold $\sigma_\phi^{\mathrm{min}}$ at any point within the DDM parameter space.  For example, if $\phi$ is a real scalar that couples to quarks through a Yukawa-type interaction with a flavor-independent coupling constant $g_q$, the dominant production process for $\phi$ is resonant production of $\phi$ through quark fusion.  In this case, we find that the product of the production cross-section and this branching fraction is $\sigma_\phi \times \mathrm{BR}_{\chi\chi} \sim 100$~fb  for the choice $m_\phi = 2$~TeV (with $g_q = 0.15$ and $c_0$ chosen such that the total branching fraction $\mathrm{BR}_{\chi\chi}$ of $\phi$ to $\chi_n$ pairs is 0.5).  As $m_\phi$ increases beyond this benchmark value, $\sigma_\phi$ rapidly decreases, rendering nearly all of the DDM parameter space beyond the reach of MATHUSLA during the upcoming LHC run.  By contrast, while $\sigma_\phi$ can be significantly larger than $100$~fb for $m_\phi$ below our $2$~TeV benchmark, ATLAS and CMS searches for new physics in the monojet + MET~\cite{Aaboud:2017phn,Sirunyan:2017hci} and dijet~\cite{Aaboud:2017yvp,Khachatryan:2015dcf} channels impose stringent lower bounds on $m_\phi$.  Nevertheless, values of $m_\phi$ at or slightly below this benchmark are consistent with these constraints.  Thus, we see that the choice $m_\phi=2$~TeV corresponds to a MATHUSLA sensitivity in the range $\sigma_\phi^{\mathrm{min}} \times \mathrm{BR}_{\chi\chi} \sim 100$~fb and that values of $m_\phi$ near this benchmark are of particular phenomenological interest.

We see from the results shown in the left panel of Fig.~\ref{fig:ddmfig1} 
that there is indeed a substantial region of parameter space within which MATHUSLA 
is capable of detecting a DDM ensemble. 
As discussed in Section \ref{s.LHCLLPcomparison}, the main detector reach for our simple scenario depends strongly on the decay mode of the DDM states, which is not specified in our toy model. However, there are many general scenarios, like decay to hadrons or Yukawa- or gauge-ordered democratic decay to SM fermions, for which MATHUSLA is likely to exceed the main detector reach by orders of magnitude in cross-section.

The most obvious region for a MATHUSLA signal is 
$7.5 \lsim y \lsim 8.8$.  For $y \gtrsim 8.8$, the characteristic 
decay lengths of the heaviest states in the tower fall below 
$\beta \gamma c\tau_n \lesssim \mathcal{O}(1\,\mathrm{m})$.  Since a
significant number of particles with decay lengths in this regime would decay inside 
the main collider detector, ensembles with $y \gtrsim 8.8$ would either be detected
at the high-luminosity LHC without the help of MATHUSLA or would already have been detected 
during the current LHC run.  On the other hand, for $y \lesssim 7.5$, a parent-particle 
production cross-section $\sigma_\phi {\rm BR}_{\chi\chi} \gtrsim 10^3$~fb is required in order for the expected 
number of signal events in the MATHUSLA detector to exceed the detection threshold.
This is approaching the upper range of typical strong production rates for TeV-scale states, see Fig.~\ref{f.benchmarkxsecs}. Furthermore, given the sensitivity of monojet searches to invisible Higgs decays, see Section~\ref{s.METcomparison}, such large cross-sections are likely to be detectable  (and possibly excluded by) current or future (HL-)LHC monojet searches.

The right panel of Fig.~\ref{fig:ddmfig1} indicates how the sensitivity of 
MATHUSLA depends on $\xi$, the
scaling exponent for the couplings in Eqn.~(\ref{ddmcouplings}).
For this plot
where we have taken a fixed scaling exponent $y = 8$ for the decay widths of the $\chi_n$.  
We see from this figure that there is generally a loss of sensitivity for MATHUSLA
as $\xi$ decreases.  This behavior ultimately reflects the fact that
for $\xi < 0$, the width of $\phi$ is dominated by decays to the lightest states in
the DDM ensemble, which are also the states with the longest lifetimes.  

We see from Fig.~\ref{fig:ddmfig1} 
that the reach of the MATHUSLA detector is 
not particularly sensitive to the ratio $m_0/ \Delta m$ --- 
at least not within the region of parameter space shown.
However, we see that this ratio nevertheless plays a crucial role
in determining $n_{\rm cs}$, the number of $\chi_n$ states
which are cosmologically stable, with $\tau_n \gtrsim \tnow$.  
Indeed, given the contours of $n_{\rm cs}$ shown in Fig.~\ref{fig:ddmfig1},  
we see that a significant number of ensemble constituents $\chi_n$ are 
cosmologically stable for $m_0 /\Delta m \gtrsim 0.1$. 
By contrast, for $m_0/\Delta m \lesssim 0.1$, the only
contribution from the ensemble to $\OmegaDM$ is that associated with the single
lightest particle species $\chi_0$.  Thus, the region of parameter space
 in which $\xi \gsim -0.3$, $m_0/\Delta m\gsim 0.1$, and $7.5\lsim y \lsim 8.8$ is of particular interest from a DDM perspective,
with many individual dark-matter components $\chi_n$ potentially comprising
the total present-day dark-matter abundance $\Omega_{\rm CDM}$.

Taken together, the results in Fig.~\ref{fig:ddmfig1}
indicate that there 
exists a significant region of parameter space in which multiple light states in
the DDM ensemble can contribute non-negligibly to the present-day
dark-matter abundance --- all  while heavier states in
the same ensemble can lead to an observable signal at MATHUSLA.~ 
This alone demonstrates that MATHUSLA can be 
highly relevant for collider-based probes of the DDM framework. 
However, these results also 
indicate that the value of the scaling exponent $y$ in these
regions is relatively large.  Although there is no fundamental reason
why such values are problematic, 
it would be interesting 
from a theoretical  perspective
to know whether the same successes can be 
achieved with smaller values of $y$.

Such regions of parameter space also exist, and correspond to very light hyperstable DDM states.
In Fig.~\ref{fig:ddmfig2}, we plot essentially the same information as we plotted in Fig.~\ref{fig:ddmfig1},
the only change being that we have now taken $m_0=100$~keV rather than $m_0= 100$~MeV.~
We see that this shift in $m_0$ has not changed the gross features of these plots
relative to those in Fig.~\ref{fig:ddmfig1},
but has shifted the regions in which MATHUSLA is most sensitive down to smaller values of $y$ --- precisely 
as desired.
Indeed, we now see that MATHUSLA remains sensitive to the DDM ensemble even below $y\approx 5$ ---
a very natural value for $y$, given that this value corresponds to a dimension-six decay operator
according to Eqn.~(\ref{ddmyd}), see also Eqn.~(\ref{e.darkphotonHVdecaywidth}).
Once again, just as for greater $m_0$, we find that taking $m_0/\Delta m\lsim 0.01 $ leads to situations in
which only a single dark-matter component survives to the present day.   
However, for values of $m_0/\Delta m\gsim 0.01$,
we find that  multiple components of the ensemble survive to the present day and can potentially
contribute to $\Omega_{\rm CDM}$.
Thus, for $m_0\approx 100$~keV, MATHUSLA is sensitive to a theoretically particularly motivated region of DDM parameter space, 
      $m_0/\Delta m\gsim 0.01$,
     $4.3\lsim y \lsim 5.0$, and $\xi \gsim -0.2$.
 The low mass of the lightest DDM state(s) means that astrophysical and cosmological constraints may apply, but they are highly dependent on the mechanism which generates the DDM abundance. We leave analysis of these constraints for future work.

 In summary, DDM is a framework for DM which generalizes the scenario of a single or a few hyperstable DM constituents. 
 It arises naturally in a variety of top-down theoretical frameworks and gives rise to meta-stable states
that are related to, and/or are a part of, the states contributing to the DM abundance today. 
Given this continuum of realizable lifetimes, MATHUSLA will be an important discovery and diagnosis tool for DDM, along with cosmological, astrophysical, and direct detection searches. If all accessible ensemble states have decay lengths exceeding the main detector size, MATHUSLA could easily be the first or only discovery opportunity for DDM.

\clearpage


\section{Theory Motivation  for LLPs: Baryogenesis}
\label{sec:baryogenesis}

An overwhelming array of evidence indicates that the observable universe is expanding and originates from a very compressed dense state that was filled with a hot primordial plasma ``hot big bang''). If matter and antimatter had been present in exactly the same amount during that epoch, then they would have mutually annihilated, and no baryons would have been left to form galaxies, stars and eventually human beings.
Hence, the presence of baryons in the present day universe clearly indicates a matter-antimatter asymmetry in the early universe, see e.g. \cite{Canetti:2012zc}. The magnitude of this \emph{baryon asymmetry of the universe} (BAU) can be determined from the baryon to photon ratio, or equivalently the ratio $Y_B=(n_B-n_{\bar B})/s$ of the total comoving baryon density $n_B-n_{\bar B}$, and the entropy density $s$. The value of $Y_B$ can be consistently extracted from two very different measurements:~the Cosmic Microwave Background anisotropy power spectrum \cite{Ade:2015xua}, and the abundance of light elements in the intergalactic medium \cite{Fields:2014uja}. The current observed value for this ratio is $Y_B=(8.6\pm 0.1)\times 10^{-11}$ \cite{Ade:2015xua}. In inflationary cosmology, this number cannot be set as an initial condition for the universe because the rapid expansion would have diluted any pre-inflationary asymmetry. Hence, the BAU must be generated dynamically by \emph{baryogenesis} during or after inflation, but before the onset of Big Bang Nucleosynthesis when the Universe was at a temperature of $T \sim \mev$. 

Baryogenesis requires fulfillment of the Sakharov conditions~\cite{Sakharov:1967dj}:~i) violation of baryon number B, ii) violation of charge conjugation $C$ and charge+parity conjugation $CP$ and iii) a deviation from thermal equilibrium. While all these ingredients are in principle provided in the SM, the numerical value of $Y_B$ cannot be explained within the SM with a standard cosmological history because the amount of CP violation is too small \cite{Gavela:1994dt} and the deviation from equilibrium is insufficient \cite{Kajantie:1996mn, Bernreuther:2002uj}. 
The BAU is therefore a clear sign of BSM physics, and many possible extensions of the SM could generate the required baryon asymmetry. 

Many BSM models of baryogenesis have been studied, including electroweak baryogenesis with modifications to the SM Higgs potential that induce a first-order phase transition~\cite{Morrissey:2012db}, the Affleck-Dine mechanism \cite{Affleck:1984fy}, and Cogenesis mechanisms where the BAU is related to a dark matter asymmetry \cite{MarchRussell:2011fi}.
With the exception of a few cases (like electroweak baryogenesis, which a priori has to involve weak-scale degrees of freedom), baryogenesis models are generally very difficult to test, since the new physics can enter at many possible scales.

In this section, we concentrate on two baryogenesis scenarios that can give rise to LLP signals at the LHC and MATHUSLA: WIMP Baryogenesis, where the late-time decay of a weak-scale LLP is directly responsible for the production of the BAU, and Leptogenesis, which involves extensions of the SM neutrino sector (see Section~\ref{sec:neutrinos}) and can generate LLPs due to either the late-time decays required to generate lepton-number, or more generically due to the small couplings involved in the neutrino sector.

\subsection[WIMP Baryogenesis]{WIMP Baryogenesis\footnote{Yanou Cui, Seyda Ipek}}
\label{sec:wimpbg}

\subsubsection{Introduction}

In models of WIMP Baryogenesis~\cite{Cui:2012jh} (WIMP BG), the WIMP miracle is leveraged to explain the observed BAU. Since the observed baryon and DM abundances in our universe are within an order of magnitude of each other, and since WIMPs can give rise to the observed DM abundance, the late-time decay of a WIMP-like parent particle into more baryons (or leptons) than anti-baryons (or anti-leptons) could easily give rise to the observed BAU. 
This scenario is particularly exciting, since the meta-stable WIMP-like progenitor particles can be produced at colliders and have to be long-lived in order to decay out-of-equilibrium in the early universe and satisfy the Sakharov conditions. Not only does this give rise to LLP signatures, we might even be able to observe a $B$ or $L$-asymmetry in the decay of these LLPs, thereby allowing us to \emph{directly} study the same primordial process of creation that gave rise to all the matter we are made of today.

A particle decays out of equilibrium if its lifetime is longer than the Hubble time at a temperature comparable to its mass; $\tau_X > H^{-1}(T\sim M_X)$, where $M_X$ is the mass of the particle $X$ and $H(T)$ is the Hubble rate at temperature $T$. This gives a lower bound on the lifetime that depends on the particle mass. Since the baryon asymmetry needs to be produced before the BBN, there is also an upper bound on this lifetime: $ \tau_X\lesssim1~{\rm s}$, that is $c\tau_X \lesssim 10^8$~m in terms of proper decay length. Hence baryogenesis requirements push us to an interesting region: a particle that could be produced at the LHC with mass $M_X\lesssim {\rm TeV}$ could also have a lifetime long enough to travel $O(100 {\rm m})$ or longer.

The BBN limit on LLP lifetime is not the only cosmological reason why WIMP BG is a particularly attractive target for MATHUSLA. 
Relatively heavy LLPs at or above the TeV scale could decay out-of-equilibrium at lifetimes less than $\sim 10^{-11}$ seconds, before the time of the electroweak phase transition. These relatively short-lived LLPs could decay into either baryons or leptons, since a generated lepton asymmetry would be converted into the observed BAU via sphaleron processes that are still active in the plasma for $T \gtrsim 100 \gev$. 
On the other hand, progenitor LLPs with masses in the 100 GeV range have to have lifetimes longer than $\sim 10^{-10}$ seconds to decay out-of-equilibrium. It is also possible that heavier LLPs have a longer lifetime than the minimum required for WIMP BG. In either case, the LLP has to decay into baryons to generate a BAU, since the decay takes place after sphaleron processes have already switched off. 

In other words, WIMP BG strongly motivates LLPs with decay lengths above $\sim$ cm, and if they have such a long lifetime, they have to decay dominantly into hadrons. This is the perfect target for MATHUSLA, both because the lifetime above the $\sim$ cm minimum is arbitrary and can easily exceed 100m, and because LLPs decaying hadronically are more difficult to search for in the main detectors, especially if their masses are around or below the 100 GeV scale. MATHUSLA would be orders-of-magnitude more sensitive to production rates of such LLPs than main detector searches, see Section~\ref{s.LHCLLPcomparison}, and could be our only option for discovering for many WIMP BG scenarios. 

\subsubsection{Model-independent Features and Phenomenology with MATHUSLA}

Let us summarize at a more quantitative level the model-independent features behind the weak-scale baryogenesis scenario in which a weak scale particle X decays after its thermal freeze-out. The freeze-out temperature $T_{\rm fo}$ is given by
\begin{align}
n_X^{\rm eq} \langle \sigma_{\rm ann}v\rangle = H(T_{\rm fo}) \longrightarrow T_{\rm fo}\sim M_X \ln^{-1}\left[ 10^7 \left(\frac{M_X}{100~{\rm GeV}} \right)\left(\frac{\sigma_{\rm ann}}{{\rm fb}}\right) \right].
\end{align}
Here $H(T)=\sqrt{\frac{4\pi^3g_\ast(T)}{45}}\frac{T^2}{M_{\rm pl}}$ is the Hubble expansion rate with  $M_{\rm pl}\simeq 1.2\times 10^{19}~$GeV is the Planck mass and $g_\ast(T)$ is the effective number of relativistic degrees of freedom at temperature $T$ which we take to be $\sim 100$ for $T\sim 100~$GeV. For weak scale particles with annihilation cross-section $\sigma_{\rm ann}\sim$ fb, the freeze-out temperature is $T_{\rm fo}\sim \frac{M_X}{20}\ll M_X$.

Now let us assume $X$ decays to a final state $YZ$, $X\to YZ$. For example, if $X$ is a fermion, the final state $YZ$ can either contain 3 fermions or a fermion and a scalar. We also assume $X$ does not carry a lepton or baryon number while $YZ$ has $+1$ baryon number\footnote{Since we are interested in particles with long lifetimes which would decay after the EW sphalerons shut off, we do not consider lepton number violating decays.}. Furthermore, the decay needs to be $CP$ violating. The out-of-equilibrium condition for the decays requires that 
\begin{align}
\tau_X > H^{-1}(T\sim M_X)
\ \ \ \ \ \longrightarrow \ \ \ \ \ 
 c \tau_X \gtrsim 1~{\rm cm}\left(\frac{100~{\rm GeV}}{M_X}\right)^2.
\end{align}

If the decay temperature is less than the freeze-out temperature, $T_{\rm fo}>T_{\rm d}>T_{\rm BBN}$, and assuming that we can neglect washout processes, the baryon asymmetry is given by
\begin{align}
\Delta_B = \epsilon_{\rm CP} n_X(T_{\rm fo}),
\end{align}
where $\epsilon_{\rm CP}<1$ is a measure of $CP$ violation in the decays and is model-dependent. 

One important observation is that, in these models, the production mechanism for the new particles can be separate from the decay mechanism. Hence, even though their decays are suppressed, giving rise to long-lived particles, these particles could be copiously produced at the LHC. For example, if an approximately conserved $Z_2$ symmetry is responsible for the long lifetime, $X$ particles can still be produced in pairs via $Z_2$ conserving interactions. A detailed outline for an LHC study of simplified models for such baryogenesis mechanisms, including various production and decay channels, can be found in \cite{Cui:2014twa}, along with a recast of several dedicated displaced vertex searches by ATLAS and CMS.

\subsubsection{Motivated Model Examples} \label{sec:baryomodels}
Concrete, motivated models realizing the general idea outlined earlier have been proposed recently, with different mechanisms of generating the CP violation effect required by Sakharov conditions. Below we briefly describe three models, which correspond to the three benchmark cases illustrated in Fig.\ref{fig:WIMPBGexclusion}. Despite different sources of CP violation and model setups, these models utilize $B$-violating, out-of-equilibrium decay of a weak scale particle in order to satisfy the Sakharov conditions for baryogenesis, and thus generically predict displaced vertex signals at the LHC.

\begin{figure}
\begin{center}
\includegraphics[scale=.45]{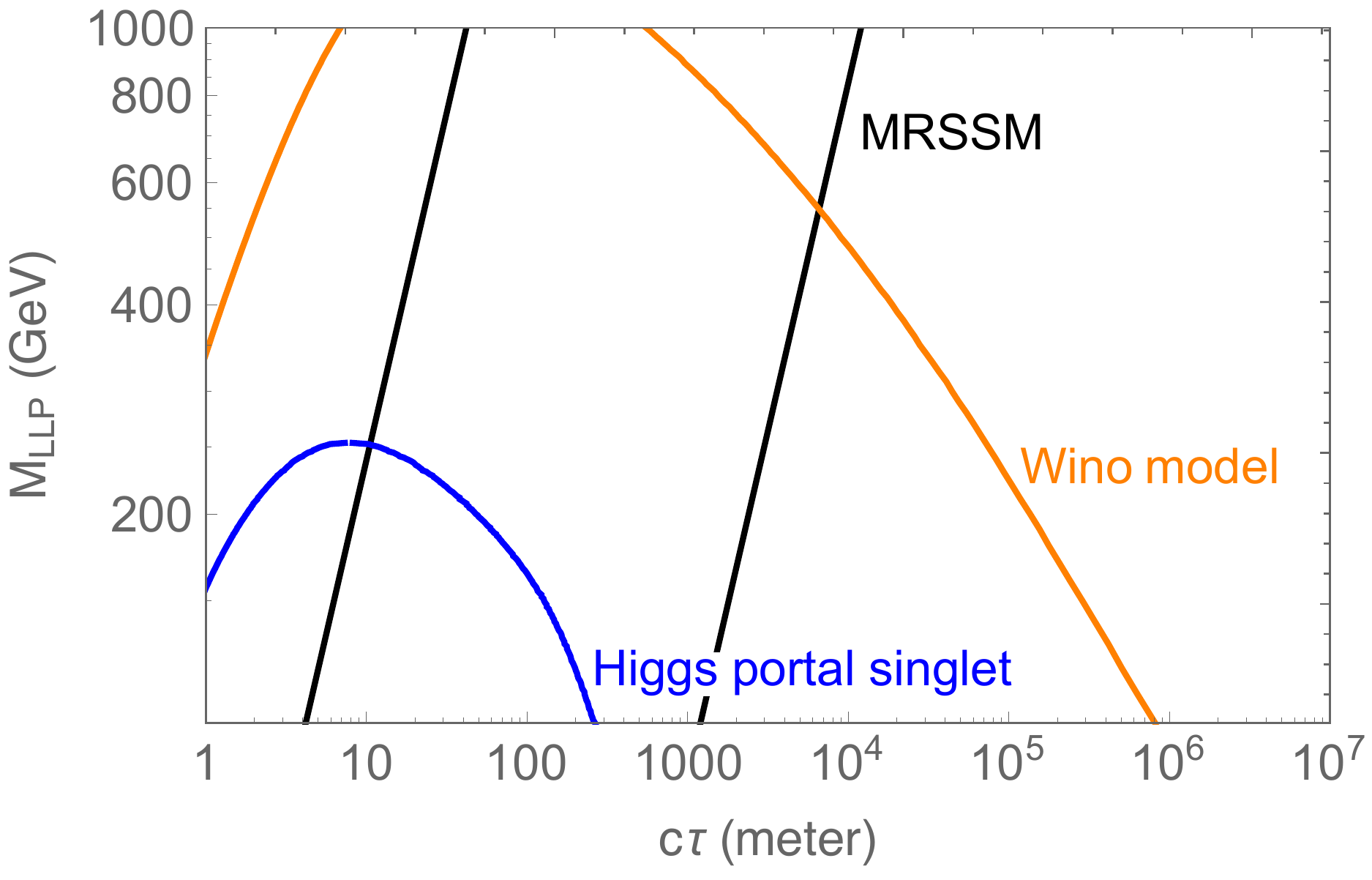}
\end{center}
\caption{MATHUSLA reach for the models described in Section~\ref{sec:baryomodels} for long-lived particle masses $M_{\rm LLP}=0.1-1$~TeV and decay lengths $c\tau>$~m for 14 TeV HL-LHC, computed using the analytical approximations outlined in Section~\ref{s.MATHUSLAsignalestimate}. (Regions between the solid lines give $N_{\rm obs}>4$ in MATHUSLA.) 
Note that these models give successful baryogenesis inside the full parameter space shown.  
The MRMSSM reach projection assumes Bino LLP production in the decay of a 1.8 TeV gluino. The Higgs portal singlet projection is shown for $c_H v/\Lambda_H=0.5$. 
This estimate assumes the $200m \times 200m \times 20m$ benchmark geometry of Fig.~\ref{f.mathuslalayout}.
} \label{fig:WIMPBGexclusion}
\end{figure}

\paragraph{WIMP baryogenesis-1: Higgs portal singlet}
The WIMP baryogenesis mechanism proposed in \cite{Cui:2012jh} is in part motivated by providing a novel way of addressing the ``coincidence'' between the dark matter and baryon abundances today, while retaining the merit of WIMP miracle. The first example model proposed in \cite{Cui:2012jh} includes the following beyond-the-SM Lagrangian terms:
 \bea
\Delta\mathcal{L}&=&\nonumber\lambda_{ij}\phi d_id_j + \varepsilon_i \chi\bar{u}_i\phi +M_{\chi}^2\chi^2 + y_i\psi\bar{u}_i\phi +M_{\psi}^2\psi^2\\ &+& \alpha\chi^2S + \beta|H|^2 S + M_S^2S^2 + \rm h.c.\label{minimalmodel}
\eea
$H$ is the SM Higgs, $d, u$ are right-handed SM quarks, with family indices $j=1,2,3$, $\phi$ is a di-quark scalar with same SM gauge charges as $u$. 
$\chi, \psi$ are SM singlet Majorana fermions, and $S$ is a singlet scalar. $\varepsilon_i\ll1$ are formal small parameters leading to long-lived $\chi$ which triggers baryogenesis upon its late decay to $udd$. All new particles involved are assumed to be of weak scale. Flavor structure of the model is assumed to be third generation dominated so as to be consistent with various constraints. A concrete realization of such a flavor structure based on minimal flavor violation (MFV) is presented in~\cite{Cui:2016rqt}. The CP violation in this model is realized through the interference between the tree level decay of the $\chi$ and loop processes involving intermediate $\psi$. This model example \cite{Cui:2012jh} can be embedded in Natural SUSY framework with R-parity violation augmented with a singlet field as baryon parent, where $\phi$ is identified as the right-handed top squark and CP violation arises from the interference with loop processes involving an intermediate neutralino. In this class of models, the baryon parent $\chi$ can be pair produced through the Higgs portal ($H$-$S$ mixing). The effective Lagrangian for LHC studies are as follows (this is just for illustration, the intermediate states may not be always heavier than $\chi$ and get integrated out):
\beq
\mathcal{L}_{\rm eff} \supset \frac{c_H}{\Lambda_H}\chi^2|H|^2+\frac{c_q}{\Lambda_{q}}g_{ijk}\chi u_id_jd_k.
\eeq
In Fig.\ref{fig:WIMPBGexclusion} we show the MATHUSLA reach for a benchmark case for such a Higgs portal model where $m_\chi=0.1 - 1$ TeV, $c_H v/\Lambda_H=0.5$ and all the other new particles decoupled.

\paragraph{WIMP baryogenesis-2: Split MSSM with RPV couplings}
A later related work \cite{Cui:2013bta} demonstrates an embedding of the general WIMP baryogenesis idea in mini-split SUSY without an additional singlet, where bino plays the role of the baryon parent. The CP violation comes from interference with loop processes involving wino. Due to the very high mass of sfermions in such models, the rate of direct production of binos is negligible at the LHC. Nevertheless, (nearly) pure wino, which plays an essential role for baryogenesis is also expected to be long-lived with RPV decay. The relevant effective interaction is the following:
\begin{align}
\mathcal{L}_{\rm eff} \supset \frac{\sqrt{2}g_2 \lambda_{ijk}''}{3m_{\rm sq}^2} T^a\tilde{W}^a\bar{u}_i\bar{d}_j \bar{d}_k + {\rm h.c.},
\end{align}
where heavy squarks with mass $m_{\rm sq}$ are integrated out. In Fig.\ref{fig:WIMPBGexclusion} we illustrate the MATHUSLA reach for a benchmark case for such a model where $m_{\tilde{W}}=0.1-1$ TeV. 

\paragraph{Baryogenesis via Pseudo-Dirac Bino Oscillations}
In supersymmetric models with a global $U(1)_R$ symmetry (MRSSM), the gauginos are pseudo-Dirac fermions. Similar to neutral mesons, pseudo-Dirac gauginos go under particle--antiparticle oscillations. CP violation can be enhanced in this quantum-mechanical phenomena and could be $O(1)$. This has been studied in \cite{Ipek:2016bpf} for a pseudo-Dirac bino with RPV couplings. The spectrum of the model is such that the lightest neutralino is a pure bino and is the NLSP. (A keV gravitino is the LSP.) The bino mass is $O(100~{\rm GeV})$ while other neutralinos and gluinos are $O({\rm TeV})$. The sfermions and other (new) scalars in the model are heavier than a few TeV. The baryon-number violating, effective Lagrangian is given by
\begin{align}
\mathcal{L}_{\rm eff} \supset -\frac{\sqrt{2}g_Y \lambda_{ijk}''}{3m_{\rm sq}^2} \tilde{B}\bar{u}_i\bar{d}_j \bar{d}_k - g''_{ijk}S\bar{u}_i\bar{d}_j \bar{d}_k + {\rm h.c.},
\end{align}
with heavy squarks with mass $m_{\rm sq}$ integrated out. $\lambda''$ is one of the usual RPV coefficients. $S$ is SM a singlet fermion that is the Dirac partner of the bino and the coefficient $g''$ depends on the parameters of the model. (See \cite{Ipek:2016bpf} for details.) The proper decay length of the bino is 
\begin{align}
c \tau \approx 30 \ \mathrm{m} \ 
\left(\frac{100 \gev}{M_{\tilde B}} \right)^5 
\left( \frac{m_{sq}}{10 \tev}\right)^4
\left(\frac{10^{-3}}{\lambda''}\right)^2
\end{align}

Due to the $U(1)_R$ symmetry, supersymmetric particles would be pair produced at the LHC. Furthermore, some of the production channels are not available, $e.g.$ s-channel $Z$ exchange for neutralinos, which reduces the electroweak production cross-section for the wino. For a spectrum with $M_{\tilde{B}}<1~$TeV, $M_{g}=1.8~$TeV and squarks heavier than 5~TeV, we find that the main bino pair-production occurs via gluino decays to neutralinos and jets. (Note that gluinos are not long-lived.) For 14~TeV LHC, this cross-section is 4.9~fb. MATHUSLA reach for this model is shown in Fig.\ref{fig:WIMPBGexclusion}.

\subsection[Baryogenesis via Exotic Baryon Oscillations]{Baryogenesis via Exotic Baryon Oscillations\footnote{David McKeen}}
\label{sec:exoticbaryonoscillations}

In Refs.~\cite{Aitken:2017wie,McKeen:2015cuz} a new scenario for baryogenesis through particle-antiparticle oscillations of heavy flavor baryons was proposed. Parts of the mechanism can be related to WIMP BG models described in Section~\ref{sec:wimpbg}, but since baryon number is generated via exotic baryon oscillations late in the hadronization era, this mechanism occupies a very different low-mass region of parameter space and is compatible with low inflationary reheating temperatures. It is therefore compatible with a wider variety of inflationary scenarios, in particular avoiding some of the problems associated with high reheating temperatures in axion, gravitino and relaxion models.

The mechanism of baryogenesis via exotic baryon oscillations relies on the existence of GeV-scale Majorana fermions, $\chi_{1,2}$, coupled to SM heavy quarks.  Stability of nuclear matter generically requires the $\chi_i$ to be long-lived, as discussed below. Focusing on a single Majorana fermion for simplicity, the relevant couplings are the four-fermion interactions with quarks,
\begin{equation}
\frac{g_{ijk}}{\Lambda}\chi u_i d_j d_k+{\rm h.c.},
\label{eq:chiudd}
\end{equation}
where $i,j,k$ label the up- or down-type quark flavors. This model is similar in field content and interactions to the baryogenesis scenario of Ref.~\cite{Ipek:2016bpf} (as described above) but populates a lower-mass parameter space. The heavy flavor baryons must be produced at late times out of equilibrium, and this can be done in a number of ways, most simply through the late decay of a weak scale (or below) particle.

The coupling of $\chi$ to heavy flavor quarks in Eq.~(\ref{eq:chiudd}) leads to the dimension-9 baryon-number--violating operator $(u_i d_j d_k)^2$ that sources baryon-antibaryon oscillations which lead to the BAU. For these oscillations to be efficient, $m_\chi$ has to be comparable to that of a heavy flavor baryon.  The decays of $\chi$ are mediated by the four-fermion interaction of Eq.~(\ref{eq:chiudd}) involving lighter quarks, with couplings suppressed relative to those of heavy flavors. For definiteness, focusing on decays to $uss$ quarks, the $\chi$ decay length is
\begin{equation}
\label{e.baryonoscillationminlifetime}
c \tau_\chi\gtrsim 100~{\rm m}\left(\frac{5~\rm GeV}{m_\chi}\right)^5\left(\frac{\Lambda/\sqrt{g_{uss}}}{20~\rm TeV}\right)^4.
\end{equation}
This lower bound on $\tau_\chi$, or equivalently on $\Lambda/\sqrt{g_{uss}}$, arises from avoiding dinucleon decay of $^{16}$O at a rate above the limit observed at Super-Kamiokande. Note that this baryogenesis scenario relies on new particles that must be long-lived for reasons that are distinct from the requirement of a departure from thermal equilibrium in the early Universe. (If the exotic baryons $\chi$ are produced in out-of-equilibrium decay as in WIMP BG, then the WIMP-like parent would constitute an additional LLP signature of the model.)

At the LHC, these long-lived Majorana fermions are produced through the decay of heavy flavor baryons with branchings on the order of $10^{-3}$.
A detailed study of the phenomenology of these long-lived particles has not yet been performed. However, results shown in Sections~\ref{sec:minimalRHN} and \ref{sec:singlets} demonstrate that MATHUSLA has excellent sensitivity to LLPs produced in $B$ decays, far exceeding the reach of the LHC main detectors and extending the reach of SHiP at long lifetimes. 
Beauty baryons $\Lambda_b$ are produced at the LHC with orders of magnitude smaller cross-section than $B$-mesons~\cite{Chatrchyan:2012xg}, but compared to e.g. the SM+S model studied in Section~\ref{sec:singlets}, this is compensated by the much larger exotic branching ratio to LLPs. 
Furthermore, the bound of Eqn.~\ref{e.baryonoscillationminlifetime} places the LLP lifetime exactly in the MATHUSLA  regime. We therefore expect that MATHUSLA would be sensitive to a large portion of the parameter space for this class of baryogenesis models.

\subsection[Leptogenesis]{Leptogenesis\footnote{Marco Drewes}}
\label{sec:bgleptogenesis}

\emph{Leptogenesis} is a particularly elegant mechanism that relates the BAU to the origin of neutrino masses~\cite{Fukugita:1986hr}.
The LLP phenomenology of neutrino extensions to the SM is discussed in detail in Section~\ref{sec:neutrinos}. Here, we briefly place leptogenesis scenarios in the context of these studied models.

If the SM is complemented by heavy right-handed Majorana neutrinos $\nu_R$ that give masses to the known light neutrinos via the seesaw mechanism ~\cite{Minkowski:1977sc,Mohapatra:1979ia,Yanagida:1979as,GellMann:1980vs,Glashow:1979nm}, then the CP-violating decay of the very same particles $\nu_R$ can generate a matter-antimatter asymmetry amongst the leptons in the primordial plasma, which is then partly transferred into a $B\neq 0$ by weak sphalerons \cite{Kuzmin:1985mm}. 
Meanwhile a plethora of leptogenesis scenarios has been proposed that adopt the idea that the $C/CP$ violation and out-of-equilibrium Sakharov conditions are fulfilled by colour-neutral particles that may or may not be related to the origin of neutrino masses. A recent review can be found in refs.~\cite{Dev:2017trv,Drewes:2017zyw,Dev:2017wwc,Biondini:2017rpb,Chun:2017spz,Hagedorn:2017wjy}.

One way to classify leptogenesis scenarios is through the manner in which the nonequilibrium Sakharov condition is realised. In most scenarios this occurs due to the freeze-out and out-of-equilibrium decay of some heavy particle in the early universe at temperatures above the temperature $T_{\rm sp}\sim 130$ GeV \cite{DOnofrio:2014rug} when sphalerons freeze out ("freeze-out scenario").\footnote{Contrast this to the WIMP Baryogenesis mechanism discussed in Section~\ref{sec:wimpbg}, where baryons can be created directly in out-of-equilibrium decay. This allows the mechanism to take place after the electroweak phase transition.}
The standard thermal leptogenesis proposed in ref.~\cite{Fukugita:1986hr} falls into this category.
A review of the most studied scenarios of this kind can e.g. be found in ref.~\cite{Davidson:2008bu}. Another alternative is that the asymmetry is generated by feebly coupled particles that do not reach thermal equilibrium before the temperature drops below $T_{\rm sp}$ ("freeze-in scenario", see also Section~\ref{sec:freezein}). Leptogenesis from neutrino oscillations \cite{Akhmedov:1998qx,Asaka:2005pn} in the \emph{Neutrino Minimal Standard Model}  ($\nu$MSM) \cite{Asaka:2005an,Asaka:2005pn} falls into this  latter category. This possibility is particularly interesting in the context of MATHUSLA because the feebly coupled particles are generally long-lived. Both scenarios can be realised within the type-I  seesaw model described by the Lagrangian Eq.~(\ref{lagrangian}) in Section~\ref{sec:minimalRHN} with experimentally accessible Majorana masses $M_i$, see e.g. \cite{Drewes:2013gca} for a review.
We shall use this well-known model as benchmark scenario in what follows.

In its minimal version, the type-I seesaw model only adds $2$ right-handed neutrinos $\nu_R$ to the SM.
In this case one can qualitatively distinguish between the cases with Majorana masses $M_i$ below vs.~above the electroweak scale.

For $M_i$ above the electroweak scale, the "freeze-out scenario" is realised because the BAU is generated in the decay of the $\nu_R$. This is  impossible for experimentally accessible $M_i$ \cite{Davidson:2002qv} unless the $M_i$ are quasi-degenerate ("resonant leptogenesis") \cite{Pilaftsis:2003gt}. Since the lifetime of $\nu_{R i}$ scales as $\propto U^{-2} M_i^{-5}$ \cite{Barbieri:1989ti,Gorbunov:2007ak,Atre:2009rg}, searches for $\nu_{R i}$ in this mass range usually focus on prompt decays  \cite{Aad:2015xaa,Khachatryan:2015gha,Khachatryan:2016olu,Sirunyan:2018mtv}. Here $U^2={\rm tr}\theta^\dagger\theta$ is the total heavy neutrino mixing.
Numerous authors have proposed strategies to refine such searches, cf. e.g. \cite{Atre:2009rg, Deppisch:2015qwa,Chun:2017spz, Cai:2017mow} for reviews, but it seems unlikely that MATHUSLA can access the viable leptogenesis parameter region of the minimal model in this mass range because the $\nu_R$ will either be too short lived or the number in which they are produced is too small. However, as we discuss below, details of the UV completion of such models may lead to LLP signatures.

For $M_i$ below the electroweak scale, the seesaw relation Eq.~(\ref{active-masses}) enforces comparably smaller Yukawa couplings, so that thermal $\nu_R$ production leading to equilibrium in the early universe is delayed and the freeze-in scenario can be realised. Two competing processes generate the asymmetry, the CP-violating oscillations of the $\nu_R$ \cite{Akhmedov:1998qx,Asaka:2005pn} and the decay of Higgs bosons into $\nu_R$ and SM leptons \cite{Hambye:2016sby,Hambye:2017elz}. The simplest model that realises this possibility is the $\nu$MSM, cf. \cite{Boyarsky:2009ix} for a review, in which two $\nu_{R_i}$ generate the neutrino masses and the BAU while the third one is a viable Dark Matter candidate (cf. \cite{Adhikari:2016bei} for a recent review). The leptogenesis parameter region in this scenario \cite{Shaposhnikov:2008pf,Canetti:2010aw,Asaka:2011wq,Canetti:2012vf,Canetti:2012kh,Shuve:2014zua,Hernandez:2015wna,Drewes:2016lqo,Drewes:2016jae,Hernandez:2016kel,Asaka:2016zib,Drewes:2016gmt,Asaka:2017rdj,Antusch:2017pkq,Ghiglieri:2017csp,Eijima:2017cxr,Ghiglieri:2017gjz,Asaka:2016rwd} 
and its realisation within inverse and linear seesaw models \cite{Abada:2015rta,Abada:2017ieq} has been studied by various authors, as well as
 the slightly more general case with $3$ heavy Majorana neutrinos \cite{Drewes:2012ma,Khoze:2013oga,Canetti:2014dka,Shuve:2014zua,Hernandez:2015wna,Drewes:2016lqo}. The $\nu_{R i}$ in this mass range tend to be long-lived and can be found in displaced vertex searches at ATLAS and CMS \cite{Helo:2013esa,Izaguirre:2015pga,Gago:2015vma,Cottin:2018kmq, Cottin:2018nms, Nemevsek:2016enw, Maiezza:2015lza} or LHCb \cite{Antusch:2017hhu}. For lower masses, fixed target experiments like NA62 \cite{Lazzeroni:2017fza,Talk:Lanfranchi,Drewes:2018gkc} can access the viable leptogenesis parameter range.
In the future the proposed SHiP experiment \cite{Anelli:2015pba,Graverini:2214085} or a similar detector at LBNE/DUNE \cite{Adams:2013qkq} or T2K \cite{Asaka:2012bb} can achieve a higher sensitivity for masses of a few GeV, while the future colliders FCC-ee \cite{Blondel:2014bra,Antusch:2016vyf,Caputo:2016ojx,Asaka:2016rwd,Antusch:2017pkq}, ILC \cite{Antusch:2016vyf,Antusch:2017pkq} and CEPC \cite{Antusch:2016vyf,Antusch:2017pkq} can probe heavier masses up to a few tens of GeV. 
Remarkably, MATHUSLA can access the viable leptogenesis parameter space, and has the potential to be the world's most sensitive experiment in part of this mass range, see.~Fig.~\ref{fig:Sterile2} in Section~\ref{sec:minimalRHN}.

Finally, we return to the freeze-out leptogenesis scenario. While the minimal version of these models does not predict LLPs, various UV completions which implement this scenario in the type-1 see-saw can give rise to LLPs due to the mechanism of discrete symmetry breaking at high scales. 

In Sec.~\ref{BhupalClaudiaEmiliano} we present a scenario
that features resonant leptogenesis with  $\nu_R$ of masses between 100 GeV to a few TeV \cite{Hagedorn:2016lva}, which  also give rise to long-lived particles that could be detectable at MATHUSLA~\cite{usinpreparation}.
This scenario belongs to a class of models in which flavour and CP symmetries and their residual symmetries ($G_\nu$ and $G_e$ in the neutrino and charged lepton sectors)
explain the measured values of the lepton mixing angles, make predictions for leptonic CP violation in neutrino oscillations and  neutrinoless double beta decay as well as connect low energy CP phases with those relevant for leptogenesis.
The desired breaking scheme of the flavour and CP symmetries to $G_\nu$ and $G_e$ can be realized in explicit models~\cite{King:2013eh} in which
 flavour symmetry breaking fields obtain peculiarly aligned vacuum expectation values, achieved with an appropriately
constructed potential.
One can observe that the symmetry preserved by the latter vacuum can be  larger than $G_\nu$ and $G_e$. In this case one encounters a point of {\it enhanced residual symmetry}.
   In Ref.~\cite{Feruglio:2013hia}, a model has been constructed in which the slight breaking of such larger symmetry
via higher-dimensional operators can be connected to the smallness of the reactor mixing angle $\theta_{13}$. In scenarios of resonant leptogenesis this type
of  breaking can not only
 be correlated with the possibility to maximize the CP asymmetries, but also with the longevity  of heavy neutrinos.
   Additional effects, arising from the mixing of the
different symmetry breaking sectors, disturb the mentioned symmetry breaking pattern and eventually lead to the breaking of $G_\nu$ and $G_e$.
The size of the parameters, controlling the different symmetry breakings, is given in powers of the symmetry breaking parameter, whose size is expected to be a few percent. 
The effectiveness of  symmetry breaking in the different sectors of the theory ultimately  depends on the explicit model. 
However, the crucial insight is that details of the UV completion of freeze-out leptogenesis models may violate the intuition that the associated LLPs always have relatively short lifetime, leading to the possibility of probing these models at MATHUSLA.

\clearpage


\section[Theory Motivation for LLPs: Neutrinos]{Theory Motivation for LLPs: Neutrinos\footnote{Section Editors: Marco Drewes, Rabindra Mohapatra, Brian Shuve}}
\label{sec:neutrinos}

Since the minimal standard model predicts that neutrino masses are zero, the observations of neutrino oscillations have provided the first definitive evidence for new particle physics beyond the standard model. However, neither the scale nor the detailed nature of the new states responsible for neutrino masses is known, and there are a variety of experiments under way to learn more about the physics of neutrino masses. Here we explore scenarios in which new states responsible for neutrino masses are long-lived. 

A wide class of theories that explain the smallness of neutrino  masses  predict that neutrinos are their own anti-particles, \emph{i.e.,}~ they are Majorana fermions. This implies that the interactions responsible for neutrino masses may break lepton number symmetry (or, more generally, $B-L$), a symmetry which is preserved in the SM. A widely discussed class of such models is based on the seesaw mechanism~\cite{Minkowski:1977sc,Mohapatra:1979ia,Yanagida:1979as,GellMann:1980vs,Glashow:1979nm, Appelquist:2002me},  in which a set of right-handed neutrinos with Majorana masses exist in addition to the SM leptons. In the simplest models,  the only new particles are right-handed neutrinos and the tiny observed left-handed neutrino masses suggest that the right-handed neutrinos have very small couplings relative to SM fermions. If the right-handed neutrinos are within kinematic reach of current experiments, these small couplings can generically predict that the right-handed neutrinos have a long lifetime. Unfortunately, the same small couplings also tend to predict very tiny production rates, making the RHNs challenging to produce at colliders such as the LHC.  We begin the study by considering the minimal scenario involving only the right-handed neutrinos in Section~\ref{sec:minimalRHN}.

UV complete versions of the seesaw mechanism tend to have degrees of freedom beyond the new right-handed neutrino particles, including 
gauge bosons associated with a broken local $B-L$ symmetry or left-right symmetry. While each model can lead to differing signals at high-energy colliders depending on the details of the model, they naturally provide new production mechanisms for right-handed neutrino LLPs with masses in the GeV-TeV range.  Enhanced right-handed neutrino production from $B-L$ gauge bosons and left-right symmetric models are discussed in Section~\ref{sec:bminusl_lowmass} and Section~\ref{sec:LRSYMMnuR} respectively.  In these models, new particles associated with the seesaw mechanism can, in largely unexplored parameter ranges, lead to displaced vertex signatures at the LHC observable at the MATHUSLA detector. Observation of any of these signals will provide crucial insight into the origin of neutrino masses and potentially other physics beyond the standard model.  Furthermore, the UV-completion of these scenarios can involve light scalar bosons which may also be produced via the gauge bosons and have their own displaced decays.  Although such states are not directly connected with the neutrino mass generation mechanism, they may provide an avenue for discovery of the underlying framework.  We study their phenomenology in Sections~\ref{s.BLnuRLLPscalars} and ~\ref{sec:LLPscalar_LR}.

Following this we consider two alternative portals into the neutrino sector.  In Section~\ref{sec:Higgportalneut} the Higgs portal, motivated by unification, and in Section~\ref{sec:inertdoubletportal} the inert doublet portal, which connects a FIMP DM candidate (see Section~\ref{sec:freezein}) with discrete symmetries which generate the active neutrino masses.  Finally in Section~\ref{BhupalClaudiaEmiliano} we turn to global symmetry structures and consider points of enhanced residual symmetry which may lead to small couplings, in turn greatly suppressing decays and enhancing the lifetime of the RHNs.

\subsection[Minimal ``Sterile'' Right-Handed Neutrinos]{Minimal ``Sterile'' Right-Handed Neutrinos\footnote{S.~Antusch,~B.~Batell, M.~Drewes,~O.~Fischer,~J.~C.~Helo, M.~Hirsch, A.~Ibarra, D.~Gorbunov, J.~M.~No}}
\label{sec:minimalRHN}

\paragraph*{\bf Type-I Seesaw mechanism and neutrino masses}
The simplest implementation of the seesaw mechanism is the addition of $n$ copies of right-handed neutrinos $\nu_R$ to the SM, which
permits a mass term for all SM neutrinos.
The renormalizable addition to the SM lagrangian reads
\begin{equation}
\label{lagrangian}
      {\cal L}_{N} = i\,\overline{\nu_{R i}}\lefteqn{/}{\partial}\nu_{R i}
      -{f_{\alpha i}\overline{L}_\alpha\tilde H \nu_{R i}}
      -{\frac{M_i}{2}\overline{\nu_{R i}}^c\nu_{R i}} + \text{h.c.}\,,
\end{equation}
with $\nu_{R i}$, $i=1,2,\ldots,n$ denoting $n$ sterile neutrinos with Majorana
masses $M_i$, $H$ being the SM Higgs boson doublet\footnote{The tilde denotes the usual contraction of $\mathrm{SU}(2)$ indices with the antisymmetric tensor.} and $L=(\nu_L, e_L)^T$ is the
SM lepton doublet; $f_{\alpha i}$ are the elements of a  $3\times n$ matrix of Yukawa coupling
constants (the $\alpha$ indices take on the values of $e$, $\mu$, and $\tau$).
When the Higgs field gains a non-zero vacuum expectation
value, the Yukawa terms in Eq.~\ref{lagrangian} give rise to mass mixing between active and sterile
neutrinos. In the field basis where  the mass matrix is diagonal,
the neutrino flavour states mix, which gives rise to
 neutrino oscillations among SM neutrinos. There is also mixing between active SM neutrinos and
sterile Majorana neutrinos.

In the neutrino sector with mass term $(\overline{\nu_{L^c}} \ , \  \overline{\nu_R}){\cal M}^{(\nu)}(\nu_L^c \ , \ \nu_R)^T$, the $(3+n) \times (3+n)$ mass matrix ${\cal M}^{(\nu)}$ can be diagonalised by
a unitary rotation
 \begin{eqnarray}\label{rot-1}
 && V^T {\cal M}_{(\nu)}V =
\textrm{Diag}\{m_1, m_2, m_3, M_1 \cdots ,M_n\}.
\label{mass-eigen}
\end{eqnarray}
The resulting spectrum contains the three very light
active (SM) neutrinos $\nu_{\alpha}$ ($\alpha = e, \mu, \tau$) and $n$
heavy states, denoted by $N_i$ with masses $M_i$.
Since the mixing between $\nu_L$ and $\nu_R$ is tiny, it is convenient to perform the diagonalisation in two steps. We first block-diagonalise $ {\cal M}_{(\nu)}$ into a light neutrino mass matrix $M_\nu$ and a heavy neutrino mass matrix $M_N$.
It is
convenient to introduce a so-called Dirac mass matrix
\[
\left( M_D\right)_{\alpha i}\equiv \frac{v}{\sqrt{2}}\,f_{\alpha i}
\]
where $v=246$\,GeV is the Higgs vacuum expectation value. Suppose, its
elements are small compared to the sterile neutrino masses. Then after
rotation the active neutrinos mix with one another and have a  non-diagonal  mass matrix
\begin{equation}
  \label{active-masses}
M_\nu = -  M_D\frac{1}{
  M_N}
M_D^T
\end{equation}
while the masses in the sterile neutrino sector remain almost intact. 
\begin{equation}
M_N\simeq {\rm diag}(M_1,\ldots,M_n).
\end{equation}
The $n$ heavy neutrino states therefore have mass
\begin{equation}
m_{Ni} = M_{N_i} \simeq M_i.
\end{equation}
The
rotation induces a mixing between active and sterile neutrinos parameterized by the mixing
matrix $\theta$,
\begin{equation}
  \label{active-sterile-mixing}
\theta =  M_D  M^{-1}_N\,,
\end{equation}
so that the mass eigenstates of the active neutrinos are related to the SM and Majorana neutrino flavour
states as
follows,
\begin{eqnarray}
\nu_i=(U_{PMNS}^\dagger)_{i \alpha }\nu_\alpha - (U_{PMNS}^\dagger \theta)_{i j}\nu_{R j}^c\,.
\end{eqnarray}
with $U_{PMNS}$ being the PMNS mixing matrix in the active neutrino
sector.
The formulae above show that the active-sterile mixing angles $\theta_{\alpha   j}$ control the sterile neutrino contribution to the active neutrino masses $
M_\nu$.
The $\nu_R$ are gauge singlets, but the heavy mass eigenstates 
\begin{equation}
N \simeq \nu_R + \theta^T \nu_L^c
\end{equation}
feel a $\theta$-suppressed weak interaction due to their mixing with the $\mathrm{SU}(2)$-doublet components $\nu_L$. They also directly couple to the Higgs field via the Yukawa interaction. In summary, the sterile neutrino mass eigenstates $N_i$ have the following couplings:
\begin{eqnarray}\label{WeakWW}
&&-\frac{g}{\sqrt{2}}\overline{N^c}_i \theta^\dagger_{i \alpha}\gamma^\mu e_{L \alpha} W^+_\mu
-\frac{g}{\sqrt{2}}\overline{e_{L \alpha}}\gamma^\mu \theta_{\alpha i} N_i^c W^-_\mu\nonumber\\
&&- \frac{g}{2\cos\theta_W}\overline{N_i^c} \theta_{i \alpha}^\dagger\gamma^\mu \nu_{L \alpha} Z_\mu
- \frac{g}{2\cos\theta_W}\overline{\nu_{L \alpha}}\gamma^\mu \theta_{\alpha i} N_i^c Z_\mu\nonumber\\
&&-\frac{g}{\sqrt{2}}\frac{M_i}{m_W}\theta_{\alpha i} h \overline{\nu_{L \alpha}}N_i
-\frac{g}{\sqrt{2}}\frac{M_i}{m_W}\theta^\dagger_{i \alpha} h \overline{N_i}\nu_{L \alpha}.
\end{eqnarray}
The last line is the Yukawa coupling to the physical Higgs field $h$ in unitary gauge re-expressed using the definition of $\theta$ and the relation $m_W=\frac{1}{2}v g$. 
For $M_i \lesssim 5 \gev$, RH neutrino decay modes are complicated by hadronic effects, but decays with at least one charged lepton in the final state have $\mathcal{O}(1)$ branching fraction. For $M_i \gtrsim 10 \gev$, the branching ratios follow the perturbative prediction, so (partially) leptonic decays constitute a $\mathcal{O}(1)$ majority  fraction.

There are generically $7n-3$ new physical parameters in the seesaw model.
These comprise the mixing angles and phase in the matrix $U_{PMNS}$, the $n$ heavy neutrino masses, and additional mixing angles and phases amongst the sterile neutrinos. The connection between these parameters and observables has been studied by various authors \cite{Shaposhnikov:2008pf,Gavela:2009cd,Ruchayskiy:2011aa,Asaka:2011pb,Hernandez:2016kel,Caputo:2016ojx,Drewes:2016jae,Drewes:2018gkc,Das:2014jxa,Das:2015toa,Das:2016hof,Das:2018hph}. 
The present knowledge of light neutrino oscillation parameters also allows us to obtain probability distributions for the patterns of mixing with heavy RH neutrinos
i.e., the relative size of the mixings $U_{ai}^2=|\theta_{ai}|^2$, see Fig.~\ref{fig:Chi2NO}.
    For the minimal model with $n=2$, all parameters in the Lagrangian can be constrained from measurements of the $U_{ai}^2$  if in addition the Dirac phase $\delta$ in $U_{PMNS}$ is measured in light neutrino oscillation experiments, making this a fully testable model of neutrino masses and (possibly, see Section~\ref{sec:bgleptogenesis}) leptogenesis \cite{Hernandez:2016kel,Drewes:2016jae}.

\begin{figure}
\begin{tabular}{c c}
\includegraphics[width = 0.5\textwidth]{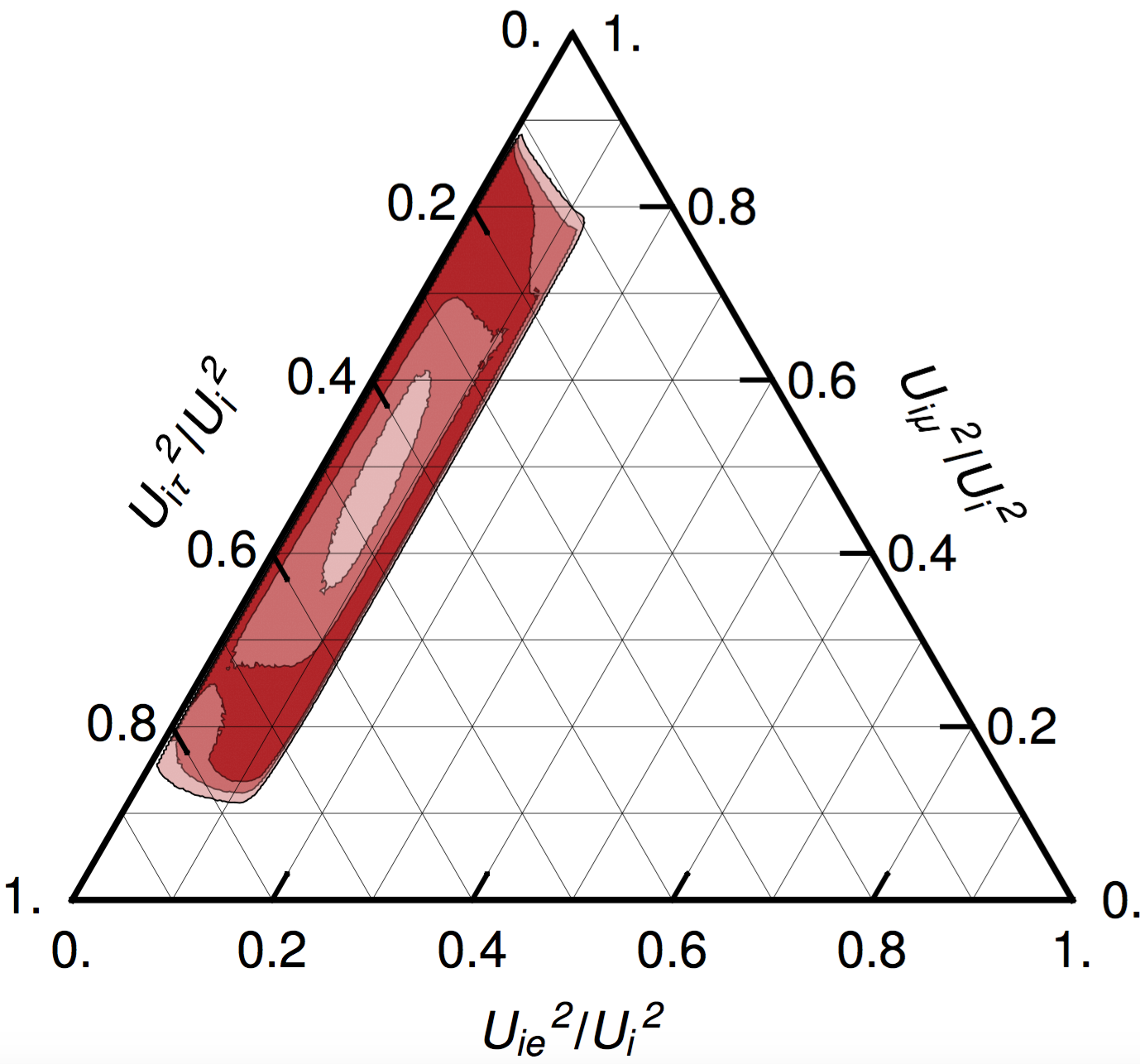} & \includegraphics[width = 0.5\textwidth]{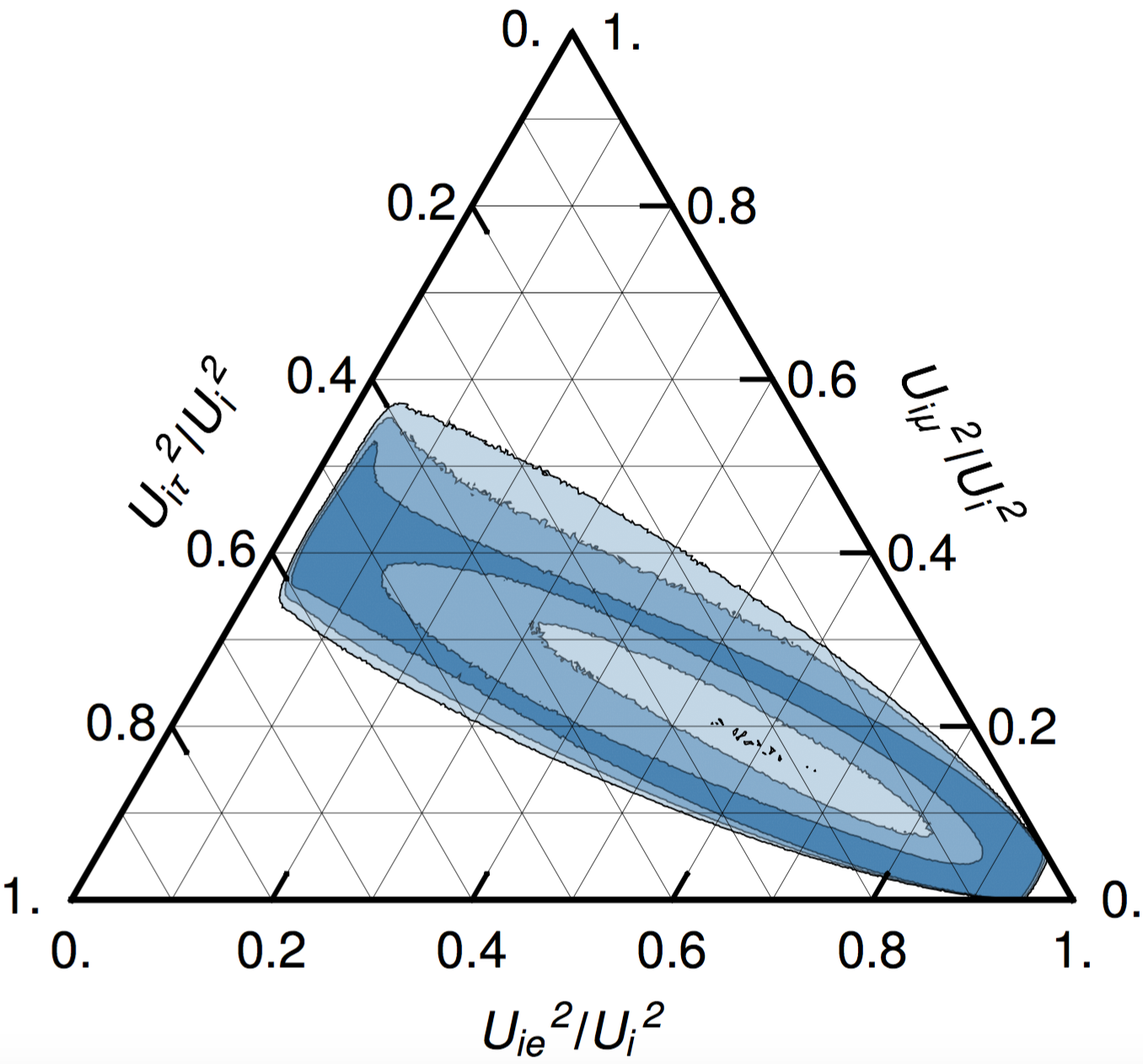}
\\
Normal Ordering & Inverted Ordering
\\
\end{tabular}
\caption{
In the $n=2$ model, the ratios $U_{ai}^2/U_i^2$  with $U_i^2=\sum_a U_{ai}^2$ that are large enough to be tested in experiments can in good approximation be expressed in terms of the parameters in the light neutrino mixing matrix $U_{PMNS}$ \cite{Hernandez:2016kel,Drewes:2016jae}. 
With the exception of the Majorana phase, all parameters in $U_{PMNS}$ are constrained by neutrino oscillation data. These constraints can be translated into probability contours for the heavy neutrino flavour mixing pattern \cite{Drewes:2018gkc}.
The different shades indicate the 1$\sigma$ (darkest), 2$\sigma$ and 3$\sigma$ (lightest) regions that can be obtained from the NuFIT~3.1 global fit to neutrino oscillation data ~\cite{Esteban:2016qun, nufit}, assuming a flat prior for the unconstrained Majorana phase. 
The results depend only mildly on the choice of this prior.
For $n=3$ the $U_{ai}^2/U_i^2$ in general depend on more unconstrained parameters.
However, neutrino oscillation data still allows us to constrain the heavy neutrino flavor mixing pattern. In particular, for a hierarchical spectrum of light neutrinos, values far outside the region displayed here can only be realized with considerable tuning.  
Figure taken from ref.~\cite{Drewes:2018gkc}. 
\label{fig:Chi2NO}
}
\end{figure}

\paragraph*{\bf Connection with cosmology}
The heavy right-handed neutrinos $\nu_R$ in the type-I seesaw model can, in addition to generating the light neutrino masses, also help to address important outstanding questinos in cosmology. In particular, they can explain the baryon asymmetry of the universe via \emph{low-scale leptogenesis} (see Section~\ref{sec:bgleptogenesis}) . This can occur either during the RH neutrinos' freeze out and decay ("freeze-out scenario")\cite{Fukugita:1986hr} or during their production ("freeze-in scenario") \cite{Akhmedov:1998qx,Asaka:2005pn}, cf. Section~\ref{sec:bgleptogenesis}. For masses in the GeV range, where MATHUSLA is expected to have the highest sensitivity, the freeze-in scenario is at work.
The minimal number $n$ of $\nu_{R i}$ for which this mechanism works is $n=2$ \cite{Asaka:2005pn}, which is also the minimal number that is required to explain the observed neutrino oscillation data within the seesaw mechanism.
This minimal scenario has been studied by a number of authors \cite{Asaka:2005pn,Shaposhnikov:2008pf,Canetti:2010aw,Asaka:2011wq,Canetti:2012vf,Canetti:2012kh,Shuve:2014zua,Abada:2015rta,Hernandez:2015wna,Drewes:2016lqo,Drewes:2016jae,Hernandez:2016kel,Asaka:2016zib,Drewes:2016gmt,Asaka:2017rdj,Abada:2017ieq,Antusch:2017pkq}.
For $n\ge2$, the additional states can either have similar properties and participate in  leptogenesis \cite{Drewes:2012ma,Khoze:2013oga,Canetti:2014dka,Shuve:2014zua,Hernandez:2015wna,Drewes:2016lqo}, or some subset of RH neutrinos can have very different properties from those responsible for leptogenesis.
An interesting possibility is that very "sterile" RH neutrinos with  tiny mixing angles $U_{\alpha i}^2 < 10^{-8}$ and masses in the keV range are candidates for Dark Matter (DM) \cite{Dodelson:1993je,Shi:1998km}, see \cite{Adhikari:2016bei} for a recent review.
This possibility is realised, for example, in the {\it Neutrino minimal extension of the SM} ($\nu$MSM) \cite{Asaka:2005pn,Asaka:2005an}, see \cite{Boyarsky:2009ix} for a review. From the viewpoint of MATHUSLA, the $\nu$MSM is equivalent to the minimal $n=2$ scenario because the couplings of the DM candidate must be so feeble that it is not produced efficiently at colliders. Moreover, its contribution to the seesaw mechanism and leptogenesis is negligible. However, the GeV-scale neutrinos that can be probed at MATHUSLA do play an important role in generating the low-scale lepton asymmetry \cite{Shaposhnikov:2008pf,Canetti:2012kh} necessary for production of  DM candidate \cite{Laine:2008pg}.
Since this scenario has been studied in much detail, we use it as a benchmark in what follows.

One  requirement for successful leptogenesis is that the oscillating sterile neutrinos must remain out of equilibrium down to temperatures near or below the electroweak phase transition; otherwise, the primordial lepton asymmetry is destroyed and no appreciable baryon number asymmetry is produced. This requirement places an upper limit on the sterile-active
mixing $U^2={\rm tr}(\theta^\dagger\theta)$, which for $n=2$ can be roughly approximated as \cite{Canetti:2010aw}
\begin{equation}
  \label{limit-on-mixing-from-BAU}
\left( \frac{m_N}{10\,\text{GeV}}\right) \left( \frac{U^2}{10^{-8}}\right)<1,
\end{equation}
where more recent studies suggest that an order of magnitude larger mixing angles are possible  \cite{Drewes:2016gmt,Drewes:2016jae}.
 For $n>2$ this upper limit is believed to be weaker \cite{Canetti:2014dka}. In particular, for $n = 3$, the leptogenesis region extends to relatively large mixing angles $U^2 \sim 10^{-5}$~\cite{lucente_michele_2018_1289773, Canetti:2014dka}, all the way up to the DELPHI bounds in Fig.~\ref{fig:Sterile2}.
A lower limit on $U^2$ comes
from the requirement for the two sterile neutrino
to give a contribution to active
neutrino masses large enough to explain the light neutrino oscillation data in the
active neutrino sector, namely the so-called atmospheric neutrino mass
$m_{atm}\simeq 0.05$\,eV. This limit can be approximated as
\begin{equation}
  \label{limit-on-mixing-from-seesaw}
\left( \frac{m_N}{10\,\text{GeV}}\right) \left( \frac{U^2}{0.5\times10^{-11}}\right)>1\,.
\end{equation}
Also this bound is significantly weakened for $n>2$ \cite{Gorbunov:2013dta,Drewes:2015iva}.

There is, however, a lower bound from cosmology that is independent of $n$, neutrino oscillation data and leptogenesis.
The $N_i$ are unstable particles, and their decay in the early universe can modify the formation of primordial elements or the Cosmic Microwave Background. The requirement to decay sufficiently long before primordial nucleosynthesis implies that they must be heavier than about 100 MeV \cite{Ruchayskiy:2012si} unless they are so feebly coupled that they are effectively stable in the early universe, in which case they cannot significantly contribute to the seesaw mechanism \cite{Hernandez:2014fha}.

\paragraph*{\bf Experimental Projections and MATHUSLA Sensitivity Estimate}

MATHUSLA can probe the sterile neutrinos $N$ in a region of parameter space similar to that accessible with present and proposed fixed-target experiments (see e.g.~\cite{Gorbunov:2007ak,Gninenko:2013tk} for details) like the SHiP project at CERN~\cite{Gninenko:2013tk,Bonivento:2013jag,Alekhin:2015byh} and similar facilities (e.g. based on DUNE or T2K). In such cases the sterile neutrinos are produced in leptonic meson  decays, which are kinematically limited to be sensitive to masses $m_N\lesssim5$ GeV. The mass region $m_N > 5$ GeV is inaccessible  from meson decays (and in some experiments, even the region $m_N>2$ GeV is inaccessible due to the low production rate of $B$-mesons).

\begin{figure}
\hspace*{-5mm}
\begin{tabular}{m{0.5\textwidth}m{0.5\textwidth}}
\includegraphics[width=0.498\textwidth]{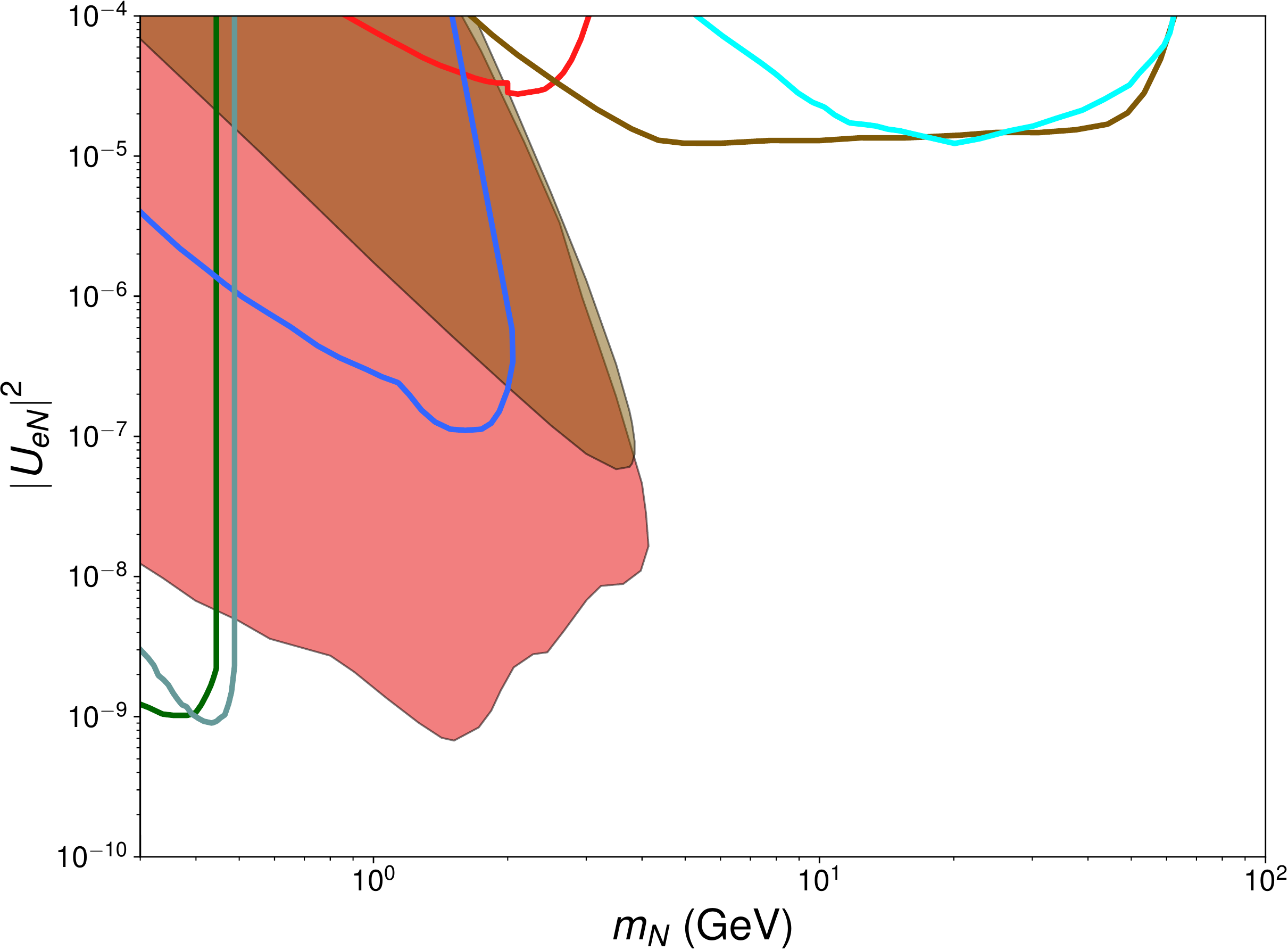}
&
\includegraphics[width=0.498\textwidth]{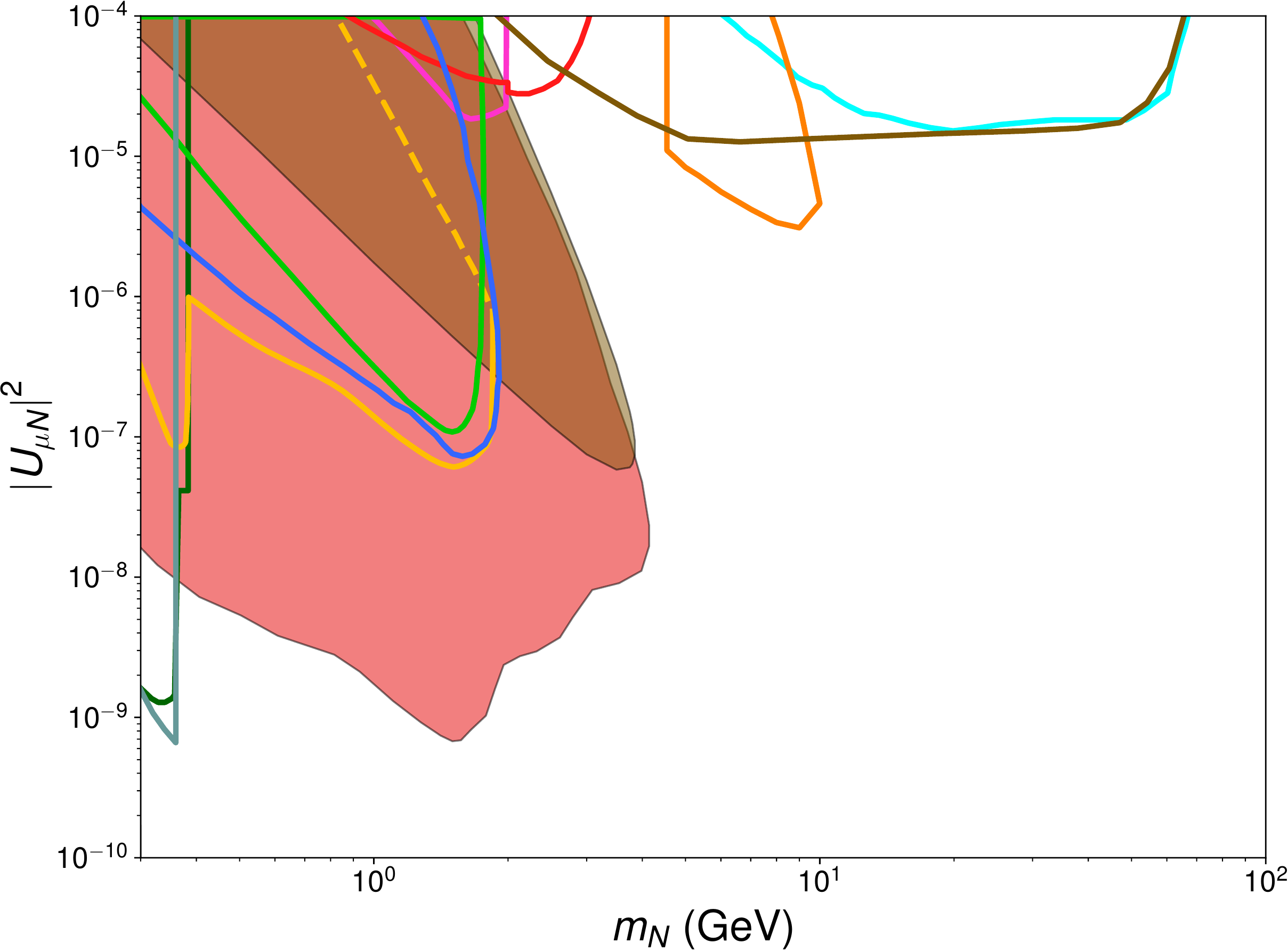}
\\
\includegraphics[width=0.498\textwidth]{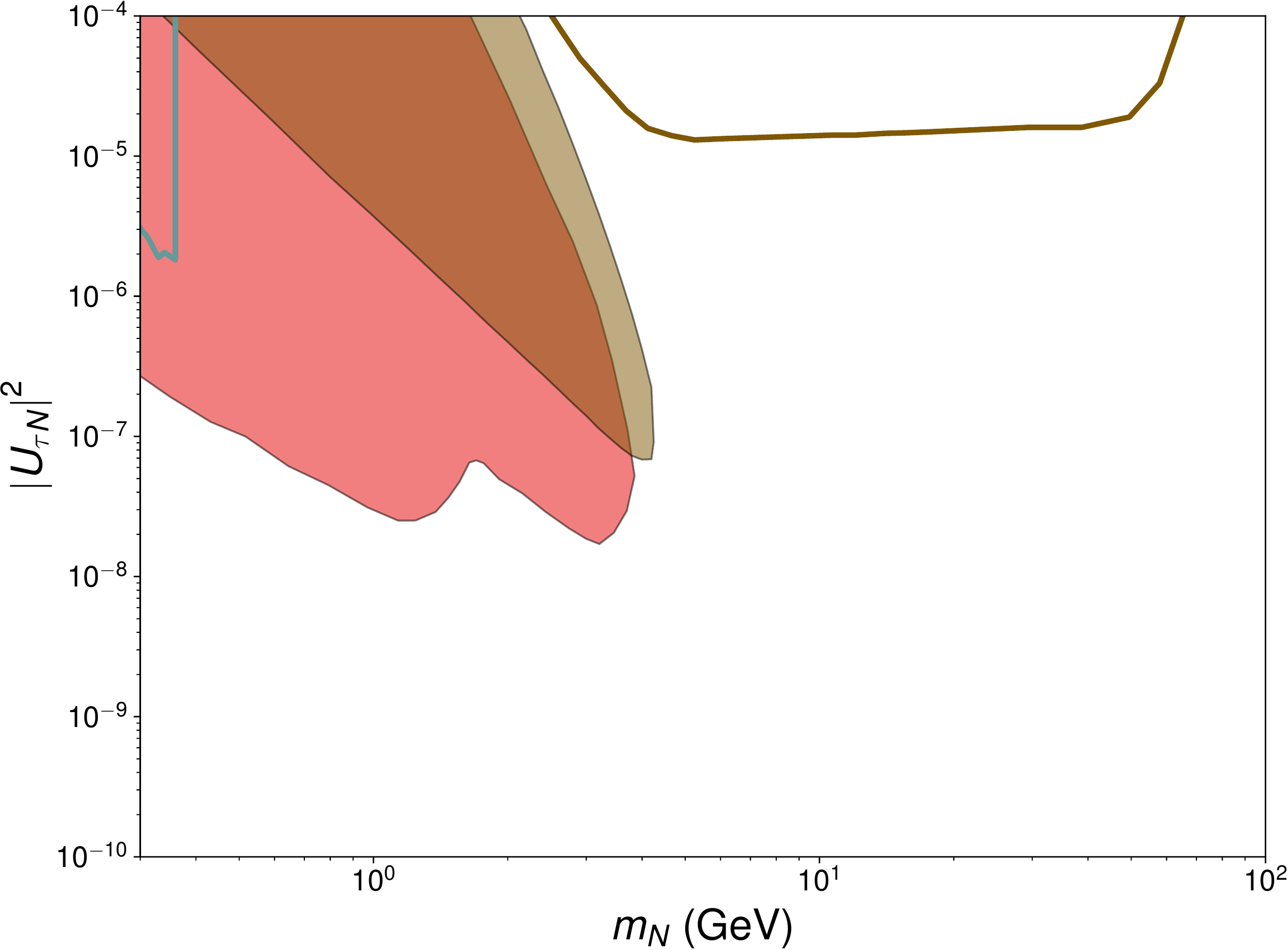}
&
\begin{tabular}{m{0.02\textwidth}m{0.25\textwidth}m{0.23\textwidth}}
&
\includegraphics[width=0.2\textwidth]{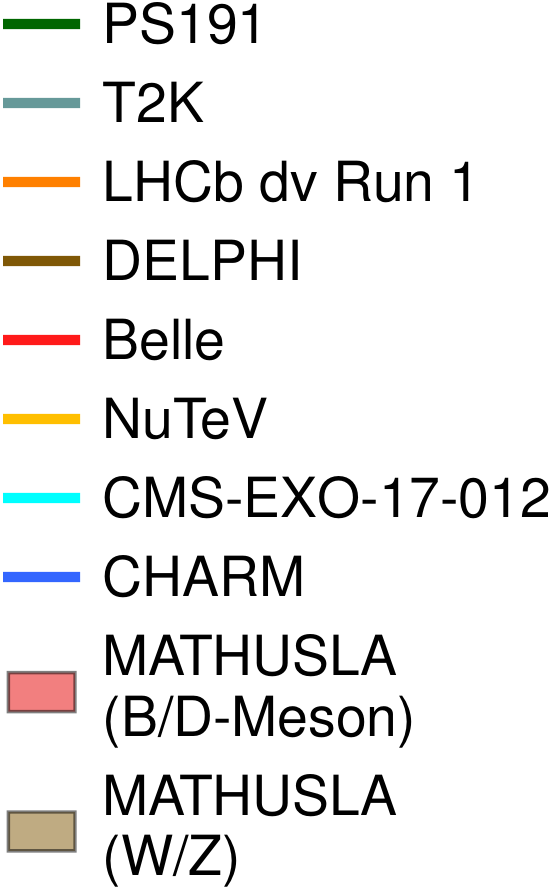}
&
\end{tabular}
\end{tabular}
\caption{Projected MATHUSLA sensitivity (assuming the $200m \times 200m \times 20m$ benchmark geometry of Fig.~\ref{f.mathuslalayout}) in the ($m_N,\, |U_{\alpha N}|^2$) plane to sterile neutrinos, $N$, produced
in $W/Z$ decays (brown regions) and in $B/D$-meson decays (light red region) for $\alpha = e$ (top-left), 
$\alpha = \mu$ (top-right) and $\alpha = \tau$ (bottom-left). 
Also shown are the present exclusion limits (solid lines) at 90\% C.L. from Belle (as given in the Erratum)~\cite{Liventsev:2013zz} (red),
PS191~\cite{Bernardi:1987ek} (using the reinterpretation from~\cite{Ruchayskiy:2011aa}) (dark-green), T2K~\cite{Abe:2019kgx} (gray-blue),
CHARM~\cite{Bergsma:1985is} (blue), NuTeV~\cite{Vaitaitis:1999wq,Vaitaitis:2000vc} (yellow; the dashed line shows the decrease in sensitivity due to the
short-lived nature of $N$, as taken from~\cite{Liventsev:2013zz}), NA3~\cite{Badier:1986xz} (purple), BEBC~\cite{CooperSarkar:1985nh} (light green), and
at 95\% C.L. from DELPHI~\cite{Abreu:1996pa} (brown),
LHCb (using the reinterpretation from Ref.~\cite{Antusch:2017hhu} of the displaced vertex search in Ref.~\cite{Aaij:2016xmb}) (orange) and 
$\sqrt{s} = 13$ TeV CMS~\cite{Sirunyan:2018mtv} (cyan).
}
\label{fig:Sterile2}
\end{figure}

At the LHC, the main production mechanism for the sterile neutrinos $N$ are rare
decays of heavy flavour mesons ($B$ and $D$ mesons for $M_N < 5$\,GeV and $M_N < 2$\,GeV, respectively)
and decays of $W,Z$ weak bosons into SM leptons and sterile neutrinos, $W^{\pm} \to \ell^{\pm} N$ and $Z \to \nu N$ (for $M_N < 80$\,GeV and
$M_N < 91$\,GeV, respectively). In the following we estimate the sensitivity of
MATHUSLA to sterile neutrinos in this mass range. (Note that the LHC can produce heavy neutrinos at higher masses via DY production, but in that case the heavy neutrino is typically not long-lived, with some exceptions, see Section~\ref{BhupalClaudiaEmiliano}.)

We adopt a phenomenologically driven approach to determining the MATHUSLA sensitivity, and we consider a simplified model of only one
sterile neutrino $N$ with mass $m_{N}$ and mixing $U_{\alpha N}$ with the active flavours $\alpha=e, \mu, \tau$.
The sterile neutrino decay rate  $\Gamma_{N}$ is given in Refs.~\cite{Gorbunov:2007ak,Helo:2010cw} (see also \cite{Bondarenko:2018ptm}) and
scales parametrically like (neglecting phase space, color factors, and overall constants)
\begin{eqnarray}
\label{eqn:RHNwidth}
\Gamma_N \sim G_{\rm F}^2\, m_N^5 \sum_\alpha |U_{\alpha N}|^2.
\end{eqnarray}
For $m_N\ll m_W$ and small mixing $|U_{\alpha N}|^2$, the decay length is macroscopic.
MATHUSLA is most sensitive to the parameters yielding a sterile neutrino decay length of $\sim 200$ m, which implies
\[
L\simeq 2 \gamma\left( \frac{3\,\text{GeV}}{m_N}\right)^5
\left( \frac{10^{-9}}{U^2}\right)\,200\,\text{m}\,,
\]
where $\gamma=E_N/m_N$ is the sterile neutrino $\gamma$-factor and $U^{2} = \sum_\alpha |U_{\alpha N}|^2$. See also Fig.~\ref{fig:decay.prod}~(a).

The strategy we follow in deriving the MATHUSLA sensitivity is similar to earlier proposals for sterile neutrino $N$ searches via displaced vertices
(see e.g.~\cite{Helo:2013esa}). For the case of weak boson decays, we study the sensitivity to
$\alpha = e, \mu, \tau$ via the processes
$p p \rightarrow W^{\pm} \rightarrow \ell^{\pm} N$ and $p p \rightarrow Z \rightarrow \nu N$ (for $\alpha = \tau$ and $m_N < m_{\tau}$ the process 
$p p \rightarrow W^{\pm} \rightarrow \tau^{\pm} \nu$ with a subsequent tau decay into the sterile neutrino $N$ also has to be taken into account) 
at LHC with $\sqrt{s} = 14$ TeV. We compute the cross-sections
using Madgraph5\_aMC@NLO \cite{Alwall:2014hca} and Pythia 8.2 \cite{Sjostrand:2014zea}, and assume an integrated luminosity of ${\cal L} = 3000$ fb$^{-1}$.
The probability of detecting sterile neutrino decays at MATHUSLA is calculated using the corresponding simulated kinematic distributions and
geometrical acceptance $\epsilon_{\mathrm{geometric}}$ of MATHUSLA
($\epsilon_{W^{+}} \simeq 0.026$, $\epsilon_{W^{-}} \simeq 0.038$, $\epsilon_Z \simeq 0.029$ on average for $m_N\ll m_W$).
The decay lengths are calculated from~\cite{Helo:2013esa}, and in determining the sensitivity to MATHUSLA
including all the $N$ decay modes that contain at least two charged particles.

We also estimate the sensitivity of MATHUSLA to sterile neutrinos produced in rare $D$ and $B$ meson decays,
specifically focussing on the following four channels: $D \rightarrow K \ell N$, $D_s \rightarrow \ell N$, $B \rightarrow D\ell N$, $B \rightarrow \ell N$
(again, for $\alpha = \tau$ and $m_N < m_{\tau}$ the production of tau leptons 
with a subsequent tau decay into $N$ also plays an important role).
Charm and bottom production at $\sqrt{s}$ = 14 TeV are simulated with Pythia 8.2 \cite{Sjostrand:2014zea}.
A dedicated simulation is then used to decay the mesons to sterile neutrinos and compute the probability for the sterile neutrinos
to decay visibly within the MATHUSLA detector.

The MATHUSLA 4 event (``exclusion") and 10 event (``discovery") sterile neutrino sensitivities (under the assumption of zero background)
in the ($m_N,\, |U_{\alpha N}|^2$) plane are shown in
Fig.~\ref{fig:Sterile2} for $\alpha = e$ (top-left), $\alpha = \mu$ (top-right) and $\alpha = \tau$ (bottom-left). 
In each case, it is assumed that the $N$ mixes only with a single flavour of SM neutrino and the other mixing angles are zero. 
The results are shown together with the present limits from
Belle~\cite{Liventsev:2013zz} (as given in its Erratum), DELPHI~\cite{Abreu:1996pa}, CHARM~\cite{Bergsma:1985is},
NuTeV~\cite{Vaitaitis:1999wq,Vaitaitis:2000vc}, PS191~\cite{Bernardi:1987ek} (using the reinterpretation from~\cite{Ruchayskiy:2011aa}),
$\sqrt{s} = 13$ TeV CMS~\cite{Sirunyan:2018mtv},
LHCb (using the reinterpretation from Ref.~\cite{Antusch:2017hhu} of the displaced vertex search Ref.~\cite{Aaij:2016xmb} for masses $m_N > 4.5$ GeV),
BEBC~\cite{CooperSarkar:1985nh} and NA3~\cite{Badier:1986xz}.\footnote{Note that the interpretation of several past experiments is a subject of controversy, cf. e.g. refs.~\cite{Ruchayskiy:2011aa,Shuve:2016muy} for a discussion.}
MATHUSLA is then projected to significantly surpass the present sensitivity to sterile neutrino masses in the
few-GeV range, where the sterile neutrino is long-lived at sufficiently
large mixing angles to produce an appreciable number of sterile neutrinos $N$ at the LHC. For the minimal case $n = 2$, as in the $\nu$MSM, a sizable part of the parameter space with $m_N < 3$ GeV and an active-sterile mixing
between Eq.~(\ref{limit-on-mixing-from-BAU}) and Eq.~(\ref{limit-on-mixing-from-seesaw}), for which leptogenesis is possible, can be accessed with
MATHUSLA.

For the case $\alpha = e$, we note that current constraints from the absence of neutrinoless double beta ($0\nu\beta\beta$) decay
can place stringent limits in the ($m_N,\, |U_{e N}|^2$), as shown e.g.~in Fig.~(1)-(3) from~\cite{Helo:2013esa}.
However, for $0\nu\beta\beta$ decay one needs to sum over all (virtual) mass eigenstates and in the presence of
non-zero Majorana phases between different contributions this can lead to
cancellations in the sum. This significantly weakens the $0\nu\beta\beta$ decay sensitivity compared to
what can be achieved with MATHUSLA. This result holds, for instance, in the case of an approximate $B-L$ symmetry~\cite{Shaposhnikov:2006nn, Kersten:2007vk}, which both leads to larger mixing angles (increasing the MATHUSLA sensitivity) and suppressed contributions to $0\nu\beta\beta$ decay.

We also stress that both CMS and ATLAS detectors can perform searches for long-lived sterile neutrinos using displaced vertices.
The main detectors will obviously have superior sensitivity for short RH neutrino lifetimes, but in the long-lifetime regime, the acceptance for LLP decays at MATHUSLA is about the same as for ATLAS or CMS. As discussed in Section~\ref{s.LHCLLPcomparison}, this means that MATHUSLA will have superior sensitivity if the main detector LLP search suffers from any bottlenecks due to triggering, cut efficiencies, requirements on the LLP decay, or backgrounds. 
For $m_N \lesssim 5 \gev$ regime, many of these in some way restrict the sensitivity of main detector searches, since the low mass means that triggering and reconstruction of the DV will likely suffer from some inefficiencies and backgrounds, see Section~\ref{s.LHCLLPcomparison}.\footnote{However, proposals exist for search strategies to minimize  backgrounds for
models with light sterile neutrinos \cite{Izaguirre:2015pga}.}
The true performance of ATLAS/CMS searches will depend critically on details of the HL-LHC detector upgrades, but it is expected that MATHUSLA will have significantly better sensitivity to these light RHNs than the main detectors.

\begin{table}
\begin{center}
\begin{tabular}{cccc}
\hline
	&	$z$ [m] & $y$ [m] & $x$ [m] \\
\hline
``standard'' & [100,300] & [100, 120] & [-100, 100] \\
\hline\hline
	&	$z$ [m] & $r$ [m] & $\phi$ [m] \\
\hline
``forward'' & [20,40] & [5,30] & [0, $2 \pi$] \\
\hline
\end{tabular}
\end{center}
\caption{Possible detector geometries for MATHUSLA at FCC-hh. The origin of the coordinate system is the
IP, with $(z,y,x)=(0,0,0)$, with the $z$ axis pointing along the direction of the beam, and $y$ in the vertical and $x$ in the horizontal direction.
The ``standard'' geometry is the benchmark shown in Fig.~\ref{f.mathuslalayout} and assumed throughout this paper for HL-LHC.
The ``forward'' detector variant is assumed to be symmetric in the angle $\phi$ (which rotates in the $x$-$y$ plane) and with the fiducial detector
volume starting outside of an inner circle with radius 5 m (to account for the beam pipe).}
\label{tab:MATHUSLA-variants}
\end{table}

\begin{figure}
\hspace*{-5mm}
\begin{tabular}{m{0.5\textwidth}m{0.5\textwidth}}
\includegraphics[width=0.498\textwidth]{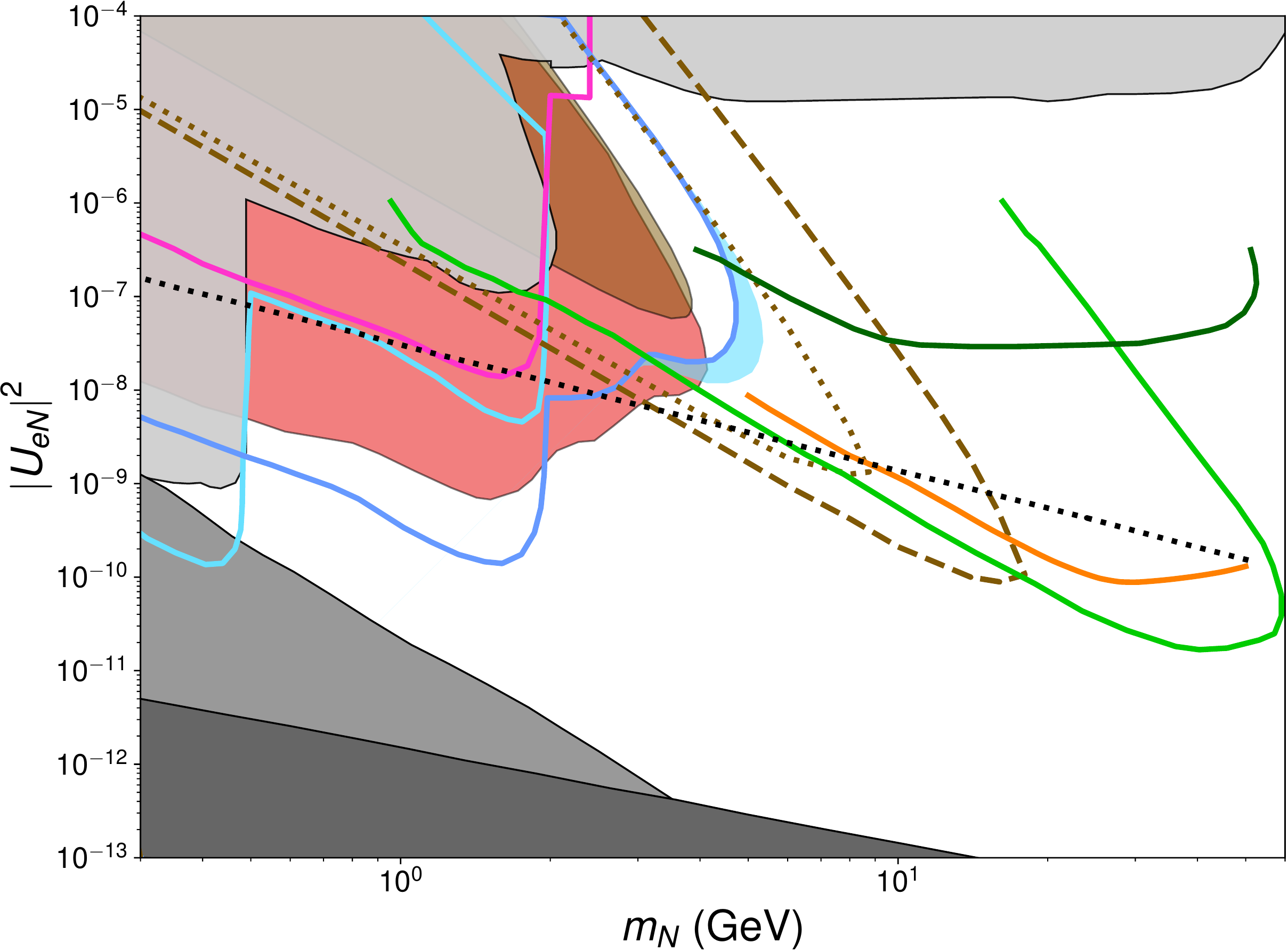}
&
\includegraphics[width=0.498\textwidth]{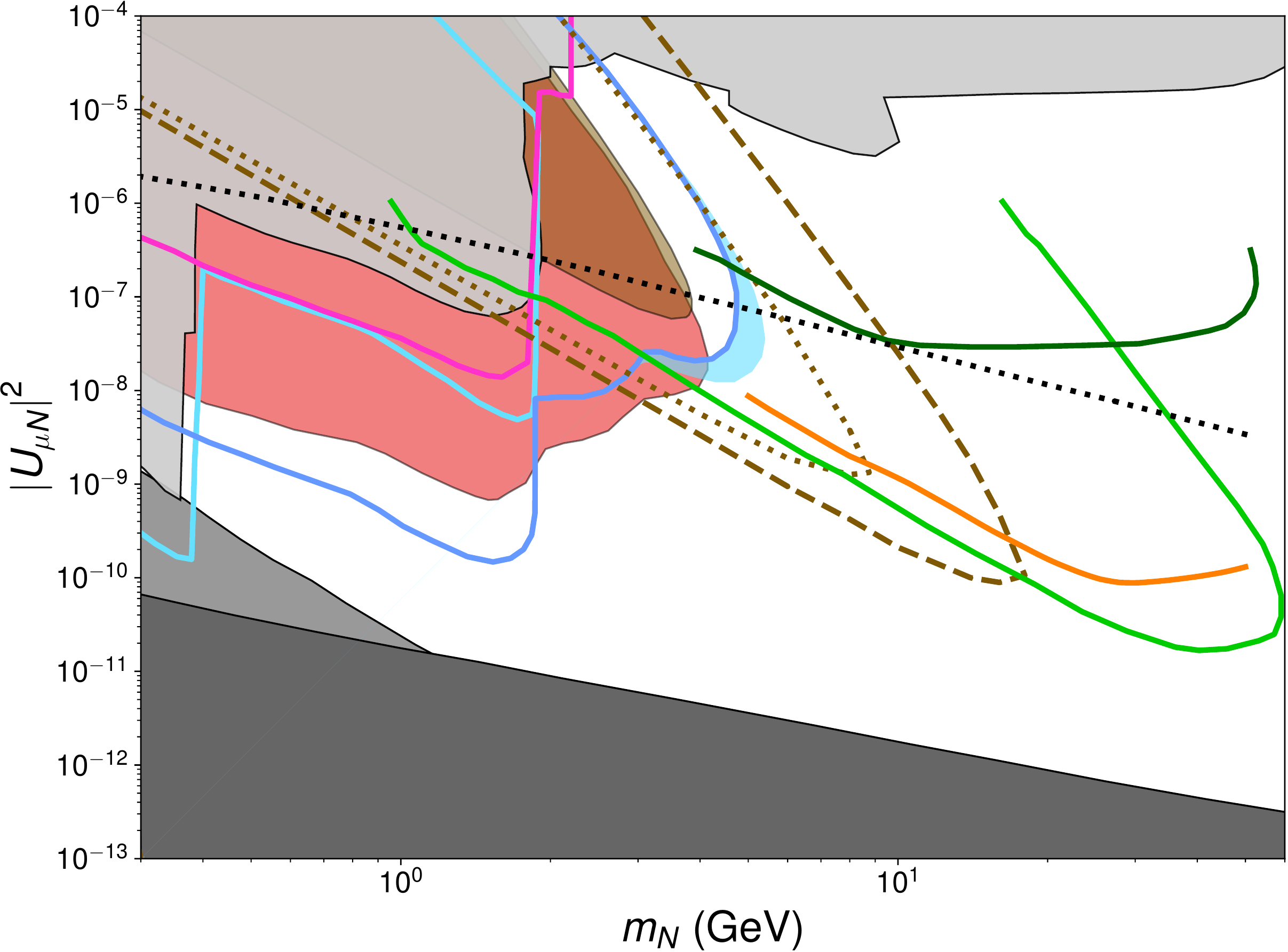}
\\
\includegraphics[width=0.498\textwidth]{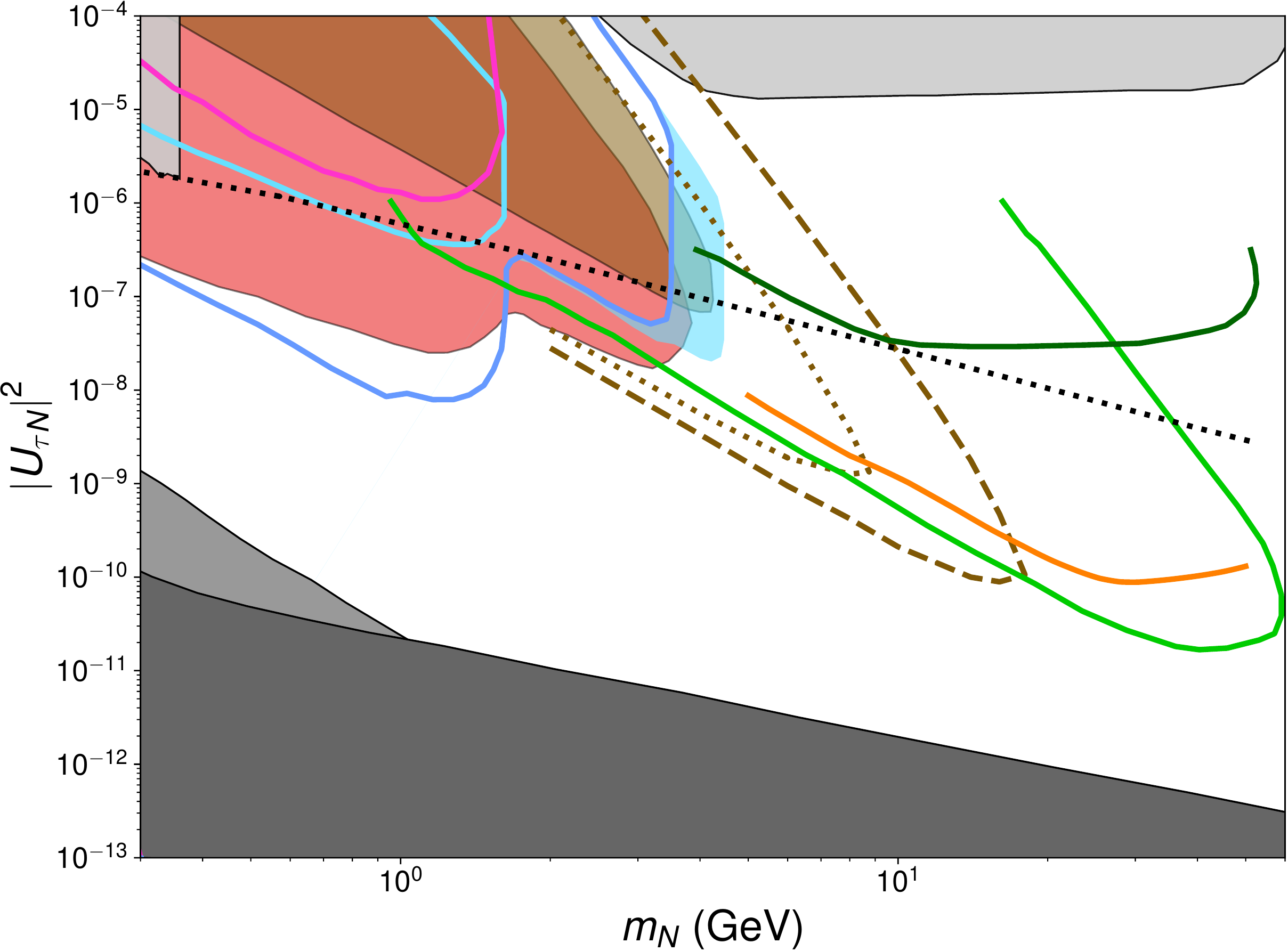}
&
\begin{tabular}{m{0.02\textwidth}m{0.2\textwidth}m{0.23\textwidth}}
&
\includegraphics[width=0.2\textwidth]{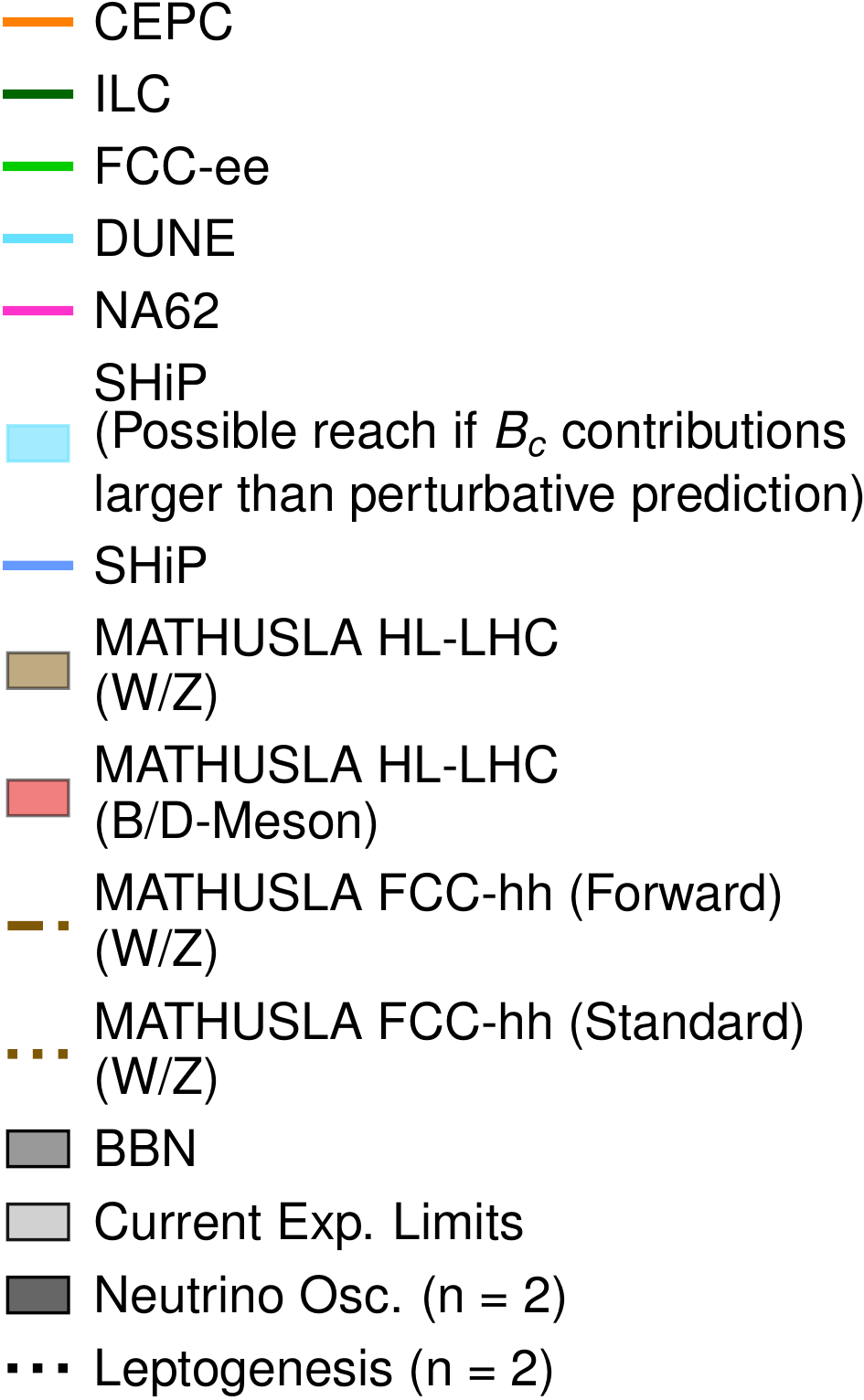}
&
\end{tabular}
\end{tabular}
\caption{Projected sensitivity (4 events) in the ($m_N,\, |U_{\alpha N}|^2$) plane to sterile neutrinos, $N$, produced
in $W/Z$ decays at FCC-hh for MATHUSLA ``standard'' benchmark surface version (dotted brown) and ``forward'' version (dashed brown),
cf.\ Table~\ref{tab:MATHUSLA-variants} and Fig.~\ref{f.mathuslalayout}, for $\alpha = e$ (top-left), $\alpha = \mu$ (top-right) and $\alpha = \tau$ 
(bottom-left, only shown for $m_N > 2$ GeV, see text for details). 
Also shown are the MATHUSLA sensitivities for HL-LHC from Fig.~\ref{fig:Sterile2},
as well as the projected future sensitivity of various facilities:
SHiP~\cite{Beacham:2019nyx},
FCC-ee~\cite{Blondel:2014bra}, CEPC~\cite{Antusch:2016vyf,Antusch:2017pkq},
ILC~\cite{Antusch:2016vyf,Antusch:2017pkq}, NA62~\cite{Drewes:2018gkc} (see also~\cite{Talk:Lanfranchi}), and DUNE~\cite{Krasnov:2019kdc}.
For the projected sensitivity of FASER, see~\cite{Kling:2018wct}, and for comparisons to other proposed LLP detectors see~\cite{Beacham:2019nyx}.
The light blue shaded region indicates the uncertainty in the SHiP reach due to $B_c$ production, with $\sigma(B_c)/\sigma(B)$ set to the LHC value at its outer boundary.
The present limits on ($m_N,\, |U_{\alpha N}|^2$) from Fig.~\ref{fig:Sterile2} are shown as a light-grey region.
The region excluded from primordial nucleosynthesis (BBN) is shown in medium-grey.
The upper limit on $|U_{\alpha N}|^2$ from viable leptogenesis for the minimal case $n= 2$ (assuming normal neutrino mass hierarchy, see Ref.~\cite{Drewes:2016jae}) and
the lower exclusion on $|U_{\alpha N}|^2$ from the active neutrino oscillation data for the minimal case $n= 2$
(for normal neutrino mass hierarchy, see Ref.~\cite{Drewes:2016jae}) are respectively shown as a black dotted line and a dark-grey region. For $n = 3$, the leptogenesis region extends up to the present DELPHI bounds~\cite{lucente_michele_2018_1289773, Canetti:2014dka}.
}
\label{fig:sensitivity-FCC}
\end{figure}

Regarding future $pp$ colliders, the Future Circular Collider (FCC) in its hadron-hadron
mode (FCC-hh) \cite{Golling:2016gvc,Mangano:2016jyj,Contino:2016spe} (or, equivalently, the SppC \cite{Tang:2015qga}) would be
excellent facilities to search for sterile neutrinos with very long lifetimes. We investigate the potential sensitivity of two variants
of MATHUSLA at FCC-hh, namely the ``standard'' surface version used as a benchmark in this document, see Fig.~\ref{f.mathuslalayout}, and an alternative ``forward'' version in the shape of a cylindrical ring aligned with the beamline,
as defined in~\cite{Chou:2016lxi}, with respective detector geometries given in Table~\ref{tab:MATHUSLA-variants}.
We consider sterile neutrino production through charged and neutral Drell-Yan processes. The cross-sections are evaluated for $\sqrt{s} = 100$ TeV  at Leading Order with WHIZARD~\cite{Kilian:2007gr,Moretti:2001zz}
and Madgraph5\_aMC@NLO~\cite{Alwall:2014hca} using
the parton distribution function CTEQ6L (neglecting theoretical uncertainties and uncertainties on the input parameters), and
we perform a similar analysis to the one carried above for the MATHUSLA sensitivity to sterile neutrinos from $W,\,Z$ decays,
considering in this case a total integrated luminosity of 20 ab$^{-1}$, as suggested in Ref.~\cite{Hinchliffe:2015qma}.

In Fig.~\ref{fig:sensitivity-FCC} we show the sensitivity (4 events) in the ($m_N,\, |U_{\alpha N}|^2$) plane
for the two MATHUSLA variants at FCC-hh, i.e.~the ``standard'' surface version (dotted) and the
alternative ``forward'' version (dashed), for $\alpha = e$ (top-left), $\alpha = \mu$ (top-right) and $\alpha = \tau$ 
(bottom-left).\footnote{For $\alpha = \tau$ we choose to only show the region $m_N > 2$ GeV. For lighter masses we expect a slight departure from the above  
sensitivities (in particular for $m_N < m_{\tau}$, due to the contribution from Drell-Yan tau lepton production).}
We also show the sensitivities for MATHUSLA at HL-LHC from Fig.~\ref{fig:Sterile2}, as
well as the expected sensitivities from other proposed facilities. 
In the relatively near future,  experiments like 
NA62~\cite{Talk:Lanfranchi,Drewes:2018gkc}
and
SHiP~\cite{Alekhin:2015byh,Anelli:2015pba, SHiP:New} could explore new regions of parameter space.
Note that the reach projections we show for SHiP must be regarded as preliminary. The solid blue line includes secondary $B$ production in the fixed target. 
On the other hand, the contribution from $B_c$ production is not yet understood and needs further study. The perturbative prediction for $\sigma(B_c)/\sigma(B)$ at SHiP is roughly two orders of magnitude below the measured value at the LHC~\cite{Baranov:1997wy, Kolodziej:1997um}, but given the unknown non-perturbative effects, it is in principle possible that this prediction is too small by up to two orders of magnitude. Therefore, the blue shading indicates the uncertainty in the SHiP reach due to $B_c$ production, where $\sigma(B_c)/\sigma(B)$ is set to the measured LHC value as an absolute upper limit on the outer boundaries of the shaded region. 
A LBNE/DUNE-like facility~\cite{Adams:2013qkq} could have the best sensitivity for very small mixing angles at sub-GeV RH neutrino masses, but detailed estimates for DUNE's updated detector design are not yet available.
On time scales relevant for the FCC-hh, other future colliders like 
FCC-ee~\cite{Blondel:2014bra}, 
CEPC~\cite{Antusch:2016vyf,Antusch:2017pkq} and
ILC~\cite{Antusch:2016vyf,Antusch:2017pkq} would greatly extend sensitivity.
The envelope of the excluded region from current experiments (from Fig.~\ref{fig:Sterile2}) is shown in grey,
together with the constraint from the generation of
light active neutrino masses via the see-saw mechanism (for normal neutrino mass hierarchy, see \emph{e.g.,}~Ref.~\cite{Drewes:2016jae}) and
from the viability of leptogenesis (also for normal neutrino mass hierarchy)~\cite{Drewes:2016jae} for the minimal scenario $n = 2$. We 
also require the sterile neutrinos to decay  before primordial nucleosynthesis ($\tau_{N} \leq 1$ s). 

Clearly, a MATHUSLA-like detector at a 
future 100 TeV collider would probe previously unexplored regions of RHN parameter space. 
Furthermore, MATHUSLA at the HL-LHC and SHiP explore similar and complementary regions of parameter space.

\subsection{The $B-L$ Gauge Portal}\label{sec:bminusl}

The active neutrino masses can also have weak-scale origins if the Majorana masses of the sterile neutrinos arise from a local $B-L$ symmetry that is broken at the weak scale, implementing the type-I seesaw at low energies~\cite{Minkowski:1977sc, Mohapatra:1979ia, Yanagida:1979as, GellMann:1980vs, Glashow:1979nm}. We consider two scenarios:~one with new vector bosons at masses well below the weak scale, with sensitivity to long-lived right-handed neutrino decays; and a UV-complete model of TeV-scale $B-L$ breaking with long-lived exotic Higgs scalar states.

\subsubsection[Low-Mass $Z'$]{Low-Mass $Z'$\footnote{Brian Batell}}\label{sec:bminusl_lowmass}

A simple and well-motivated extension of the SM is a model based on a local $U(1)_{B-L}$ symmetry. Neutrino masses naturally emerge in this model once the $U(1)_{B-L}$ symmetry is spontaneously broken, resulting in a type-I seesaw mechanism. In particular, unlike in the SM, three right-handed neutrinos are required in the $B-L$ model to cancel gauge anomalies. We assume that the $B-L$ gauge symmetry does not contribute to electric charge so that its coupling can be chosen arbitrarily small.
As we demonstrate below, this gives rise to a particularly attractive discovery scenario for MATHUSLA. The RHN decays via the same small mixing angle as in the minimal model discussed in Section~\ref{sec:minimalRHN}, but acquires a additional production mode through Drell-Yan like processes involving the on-shell $Z'$ gauge boson of the $U(1)_{B-L}$ broken gauge symmetry. MATHUSLA and the main detectors will then cover different but equally motivated regions of the scenarios parameter space.

In this context, it is perhaps natural to expect the $N$ mass to be correlated with the $B-L$ gauge boson mass since they are both governed by  symmetry breaking in the $B-L$ sector.  In this case, the $B-L$ gauge interaction opens up new production channels for the RHNs and potentially allows accelerator experiments to probe the seesaw mechanism in the laboratory. Here we explore the sensitivity of MATHUSLA to a particular phenomenologically viable benchmark scenario in this model. We focus on the following simplified approach following Ref.~\cite{Batell:2016zod}. The effective interaction Lagrangian after electroweak and $B-L$ symmetry breaking is taken to be
\begin{eqnarray}
{\cal L} & = &
g'V_\mu \left( \sum_{\rm SM} Q_{B-L}\bar\psi \gamma^\mu \psi +\bar N \gamma^\mu \mathcal{P}_{\rm L} N  \right)
+ U_{\mu N} ~ \frac{g_W}{\sqrt{2}}\left(  \bar\mu_L \gamma^\mu W_\mu^- \mathcal{P}_{\rm L} N    + \mathrm{h.c.} \right) + \dots~.~~~~~~~
\end{eqnarray}
We have included the sterile neutrino $N$, the $B-L$ gauge boson $V_\mu$, along with the relevant SM fields.  Under $B-L$, the SM lepton fields have charges $-1$, SM quark fields have charges $+1/3$, and and the $N$ fields have charge $+1$.

\begin{figure}[tb]
\centering
\includegraphics[width=0.55 \textwidth ]{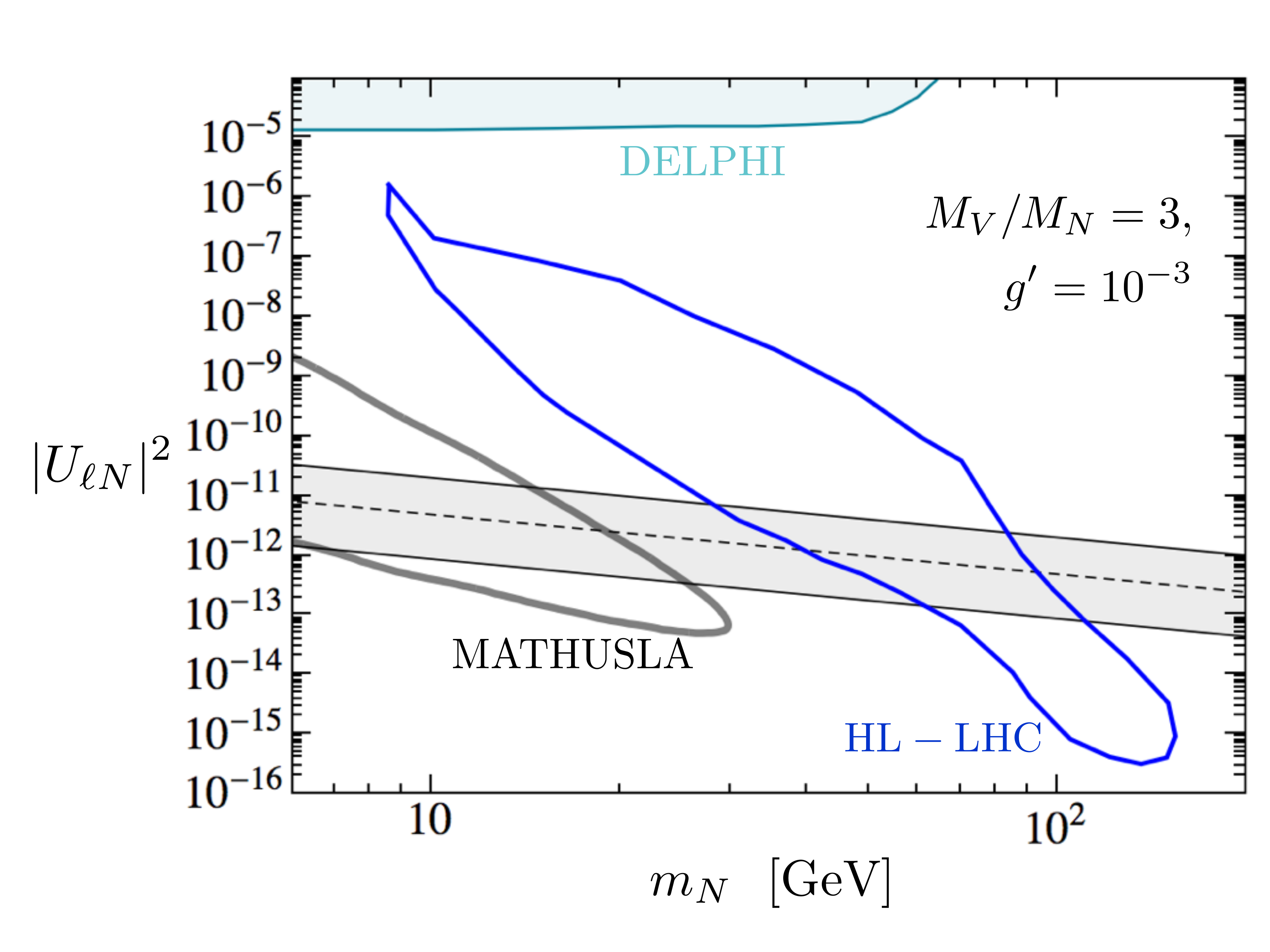}
\caption{
Dark gray contour corresponds to 50 events in MATHUSLA from sterile neutrinos in the gauged $B-L$ model (gray) for the case $m_V/m_N = 3$, $g' = 10^{-3}$. Also shown are limits from DELPHI~\cite{Abreu:1996pa} (teal) and a projection from a proposed displaced vertex search at ATLAS and/or CMS during the high luminosity run (blue)~\cite{Batell:2016zod}. The grey shaded region indicates the parameters favored in a minimal type-I see-saw to give light neutrino masses ranging from $\sqrt{\Delta m_{\rm sol}^2}$ to the Planck upper limit \cite{Ade:2015xua}. Larger couplings are also allowed in $B-L$ models.
This estimate assumes the $200m \times 200m \times 20m$ benchmark geometry of Fig.~\ref{f.mathuslalayout}.
}
\label{fig:B-L}
\end{figure}

For simplicity, we  consider only one sterile neutrino which mixes exclusively with the muon flavoured SM neutrino $\nu_\mu$ (similar results hold for mixing with $\nu_e$). This Lagrangian has four unknown parameters:~$M_V$, $g'$, $M_N$, and $U_{\mu N}$. Taking the simplest seesaw motivated parameter choices corresponding the scale $\sqrt{|\Delta(m_\nu^2)_{\rm atm}|} \sim 0.05$ eV suggests a mixing angle
\begin{equation}\label{eq:naiveseesaw}
U_{\mu N}^2 \approx \frac{m_\nu}{m_N} \sim 5 \times 10^{-11} \times \left(\frac{1 \,\rm GeV}{m_N}\right).
\end{equation}
Such a small mixing angle is difficult to directly  probe if the only production channels for $N$ occur through the weak interactions via mixing. However, in the $B-L$ gauge extension there are additional channels present.  Here we consider $pp\rightarrow V\rightarrow NN$ at the LHC. The production rate can be dramatically enhanced over the usual weak-interaction production process. Once produced, the RHN will decay via mixing through the weak interactions. If the RHN is lighter the $W$ boson, it will generically be long-lived. The mixing angle can, of course, also be larger than suggested by Eq.~\ref{eq:naiveseesaw} due to cancellations in the active neutrino mass matrix, Eq.~\ref{active-masses} (this is particularly true in $B-L$ models or other models with lepton flavour symmetries), in which case production modes from neutrino mixing can potentially be competitive.

In Fig.~\ref{fig:B-L} we show the sensitivity of MATHUSLA in the $m_N-|U_{\mu N}|^2$ plane. For concreteness, we  choose the parameters $m_V  = 3 m_N$ and $g' = 10^{-3}$, which is below the sensitivity of current direct dilepton searches for $V$ \cite{Hoenig:2014dsa}. Since the only role of $g'$ is in the production rate of $N$, the sensitivity to $U_{\mu N}^2$ scales inversely to ${g'}^2$. Thus, the 50 event contour in Fig.~\ref{fig:B-L} is equivalent to a 5 event contour with $g'\sim3\times10^{-4}$. This exceeds the sensitivity of optimistic projections for the high-luminosity LHC, which could discover $Z'$ in Drell-Yan production for $g'\gtrsim 5-10\times10^{-4}$ \cite{Hoenig:2014dsa,Batell:2016zod}. It also complements the parameter space for displaced vertex searches at the LHC \cite{Batell:2016zod}.

We observe that MATHUSLA has the potential to probe the seesaw-motivated parameter space of theories with a local $B-L$ symmetry where both vector and RH neutrino masses are at or below the weak scale. We also remark that our findings apply quite generally to new light gauge bosons that couple to $N$, and in some of these cases the typical dilepton constraints for $B-L$ models are significantly relaxed, giving MATHUSLA sensitivity to an otherwise uncovered parameter space. This parameter space could be otherwise challenging to probe at the LHC due to low reconstruction efficiencies for low-mass LLPs and possible sources of background in high-luminosity running (see Section~\ref{s.LHCLLPcomparison}).

\subsubsection[TeV-Scale $B-L$ Symmetry Breaking and Long-Lived Scalars]{TeV-Scale $B-L$ Symmetry Breaking and Long-Lived Scalars\footnote{P.~S.~Bhupal Dev, Rabindra Mohapatra, Yongchao Zhang}}
\label{s.BLnuRLLPscalars}

In this section, we examine whether a Higgs boson that breaks the $B-L$ symmetry  at the TeV scale can be accessible at MATHUSLA. We specifically consider the low-mass regime for this scalar field, which can arise if the $B-L$ symmetry is radiatively broken.  Any evidence for such low mass scalars can play a crucial role in the elucidating the seesaw mechanism. The mass and couplings of this new Higgs field are, to a large extent, \emph{a priori} unrestricted and we show that certain parameter ranges of this boson can be probed at the MATHUSLA detector.

Both the $U(1)$ and left-right symmetric completions of $B-L$ theories can have a light ($\sim$ GeV-scale) neutral scalar field which will be long-lived. Depending on the details of the theories, the MATHUSLA detector may provide an appropriate venue for searching for these displaced vertices. In this section, we focus on simple $B-L$ models based on $SU(2)_L\times U(1)_{I_{3R}}\times U(1)_{B-L}$ local symmetry, and will discuss TeV-scale left-right models in Section \ref{sec:LLPscalar_LR} of which this is a subgroup. In this case the $B-L$ symmetry therefore clearly contributes to electric charge. 

The SM fermions are charged under the gauge group $SU(2)_L\times U(1)_{I_{3R}}\times U(1)_{B-L}$ (with  the gauge couplings $g_L$, $g_R$, and $g_{BL}$, respectively) as
\begin{align}
& Q=(u_L, \: d_L)^{\sf T}: \left({\bf 2}, 0, \frac13\right);  \quad
L=(\nu, \:  e_L)^{\sf T}: \left({\bf 2}, 0, -1 \right); \nonumber \\
& u_R: \left({\bf 1}, \frac12, \frac13 \right); \quad
d_R: \left({\bf 1}, -\frac12, \frac13 \right); \quad
e_R: \left({\bf 1},-\frac12, -1 \right). \nonumber
\end{align}
Anomaly freedom requires that this model have three right-handed neutrinos (RHNs) $N_i$ ($i$=1,2,3) with gauge quantum numbers ({\bf 1}, 1/2, $-1$). The minimal Higgs fields to break the symmetry to the SM gauge group are $H({\bf 2},-1/2, 0)$ and $\Delta({\bf 1},-1,2)$ with the following Yukawa couplings:
\begin{eqnarray}
{\cal L}_Y \ = \ h_u\overline{Q}Hu_R+h_d\overline{Q}\widetilde{H}d_R+h_e\overline{L}\widetilde{H}e_R
+h_\nu\overline{L}{H}N+f\overline{N}^c\Delta N+ {\rm H.c.} \, .
\label{eq:1}
\end{eqnarray}
Note that $\langle\Delta^0\rangle=v_R$ breaks the gauge symmetry down to the SM gauge group, which in turn is  further broken by  $\langle H^0\rangle=v_{\rm EW}$ to $U(1)_{\rm em}$. From the Yukawa interactions in Eq.~\ref{eq:1} it is clear that after symmetry breaking this leads to the type-I seesaw formula for neutrino masses. In this model, $H_3 = {\rm Re}(\Delta^0)$ is the light scalar candidate, though in principle it is allowed to have mass in a wide range starting from below GeV-scale to $v_R$~\cite{Dev:2016dja, Maiezza:2016ybz, Dev:2016vle, Dev:2017dui}.
As argued in Ref.~\cite{Dev:2016vle, Dev:2017dui}, a small mass $m_{H_3}$ can be stable under  radiative corrections in the presence of direct couplings of $H_3$ to both the bosonic and fermionic particles.

We consider $H_3$ masses $\geq100$ MeV because lighter masses can be constrained by Big Bang Nucleosynthesis (BBN) (assuming $M_{H_3}\leq M_N$) and supernova energy considerations~\cite{Dev:2017dui}. 
For $M_{H_3}$ in the GeV range, the $H_3$ boson could be sufficiently long-lived and give rise to displaced vertex signatures accessible to the MATHUSLA detector.

The decay of the neutral scalar $H_3$ is dominantly governed by its mixing with the SM Higgs, parameterized by the angle $\sin\theta$. The tree level couplings of $H_3$ to the SM fermions are proportional to the SM Yukawa couplings, rescaled by the mixing angle $\sin\theta$, all of which are flavour conserving.  If $m_{H_3} \lesssim {\rm GeV}$, it decays predominantly into the SM fermions at tree level, and into $\gamma\gamma$ and $gg$ at one-loop level. 
The branching fractions do not depend on the mixing angles but only on $H_3$ mass, as all the couplings are universally proportional to the mixing angle. The $H_3$ lifetime can be long, when the couplings of the SM Higgs to sub-GeV particles are small.

Flavour-changing couplings, such as $H_3 \bar{s} b$, can arise at one-loop level through mixing with the SM Higgs, which leads to the flavour-changing rare decays of the $K$ and $B$ mesons mediated by the light scalar such as $B \to K H_3 \to K \mu^+ \mu^-$ (see discussions of the flavour limits in the SM+S model in Section \ref{sec:singlets}). However, the flavour limits in the $U(1)_{B-L}$ model are much weaker than those in the LR model, as in the latter case all these flavour-changing couplings occur at the tree level. The flavour limits in the $U(1)_{B-L}$ model could yet reach the level of $10^{-4}$ or below in  future high-intensity experiments such as SHiP and DUNE; however, the detailed reach  depends sensitively on the $H_3$ mass and the flavour decay modes. For details, see Ref.~\cite{Dev:2017dui}.

In the $U(1)_{B-L}$ model, the light scalar $H_3$ could be produced either from mixing with the SM Higgs (scalar portal) or through the gauge interaction with the heavy $Z_R$ boson (gauge portal).
The scalar portal production scenario is equivalent to the SM+S simplified model discussed in Section~\ref{sec:singlets}. As shown in Fig.~\ref{fig:HBRScalar}, MATHUSLA can cover significant part of that scenario's parameter space that are inaccessible to the main detectors and highly complementary to the reach of proposed experiments like SHiP.

The light scalar $H_3$ could also be produced through the gauge portal, i.e. via interactions with the $Z_R$ gauge boson. This provides another channel to probe the theory, where the light scalar $H_3$ could be produced in association with the heavy $Z_R$ boson by analogy with the SM Higgs-strahlung process. The $Z_R$ which  further into the SM quarks and charged leptons (for simplicity we have neglected here the heavy and light neutrino decay modes), 
i.e.
\begin{eqnarray}
pp \to Z_R^{\ast} \to  H_3 Z_R \,, \quad
Z_R \to q\bar{q},\, \ell^+ \ell^- \,.
\end{eqnarray}
$H_3$ could also be produced by the vector-boson fusion (VBF) of two heavy $Z_R$ bosons, i.e. $pp \to Z_R^\ast Z_R^\ast jj \to H_3 jj$
(with $j = u,\,d,\,s,\,c$), which is subleading to the associated production mode with an on-shell $Z_R \to jj$; thus, we focus on the associated production mode.

With the heavy $Z_R$ taking away most of the energy in the final state, the light scalar $H_3$ tends to be  soft, with a transverse momentum $\lesssim 100$ GeV for most of the events. Therefore only a small portion could arrive at MATHUSLA, similar to the scalar portal. The high-$p_T$ jets/leptons (typically $\gtrsim$ TeV) allow events to pass the trigger in ATLAS and CMS. The combination of the high-$p_T$ jets/leptons and the LLP which should also mean that backgrounds for this search will be low, but the low mass and high boost of the LLP might still mean that reconstruction is difficult (and there may still be some backgrounds;  for now, we assume LHC has zero background, and for a fuller discussion, see Section 5.1). We take possible inefficencies into account by showing LHC curves with representative LLP reconstruction efficiency of 1 and 0.1. In Fig.~\ref{fig:LLP}, we show the parameters giving a rate of at least four signal events after requiring that the light scalar decays in the tracker with approximate decay length of $1\,{\rm cm} \lesssim \gamma c\tau_0 \lesssim 1\,{\rm m}$, where the boost factor $\gamma= E_{H_3} / m_{H_3}$ has been taken into consideration. The ultimate LHC sensitivity likely lies somewhere between these curves.

A major limiting factor on the sensitivity to associated production of $H_3$ is due to dilepton limits on $Z_R$~\cite{ATLAS:2016cyf,CMS:2016abv, Patra:2015bga, Lindner:2016lpp}. Considering only benchmark points that are not yet excluded by direct searches for $Z_R$, an optimistic benchmark scenario is $g_R/g_L = 0.835$, for which the $Z_R$ mass limit is 3.64 TeV. When the $g_R$ coupling becomes smaller, the gauge coupling $g_{BL} = g_Y g_R / \sqrt{g_R^2 - g_Y^2}$ becomes larger which would enhance the production cross-section of $Z_R$ at the LHC and makes the dilepton mass limits on $Z_R$ more stringent.\footnote{With the dilepton limits on $Z_R$ becoming stronger at the LHC, it is very likely that the $Z_R$ boson is so heavy that we could not have 4 events at MATHUSLA, even with the ultimate luminosity of 3000 fb$^{-1}$.} In this optimistic benchmark scenario, the cross-section in the gauge portal is $\sigma (pp \to H_3 JJ) = 0.97 \, {\rm fb}$ after applying a $k$-factor of 1.2 ($J$ runs over all the SM quark and charged leptons). This rate does not depend on the mixing angle $\sin\theta$ in the scalar sector. We apply simple cuts $p_T (J) > 25$ GeV and $\Delta \phi (JJ) > 0.4$ on all the quark and charged leptons in {\tt MadGraph5\_aMC@NLO}~\cite{Alwall:2014hca}.
With the small production cross-section, suppressed by the large $Z_R$ mass, only a narrow region of $m_{H_3} - \sin\theta$ could be probed by the MATHUSLA detector, as shown in Fig.~\ref{fig:LLP}.
With much more signal events expected at ATLAS/CMS, the displaced vertex searches is largely complementary to the ULLP searches at MATHUSLA. In other $U(1)$ models, \emph{e.g.,} those motivated from Grand Unified Theories (GUTs)~\cite{Langacker:2008yv}, the LHC constraints on $Z'$ mass might be somewhat weaker, and the production cross-section $\sigma (pp \to Z^\prime H_3)$ gets larger; in such a scenario, a broader region of the scalar mass $m_{H_3}$ and mixing angle $\sin\theta$ could be probed at the HL-LHC. In addition, it might also be more promising to test the light scalar at the future 100 TeV collider and the dedicated forward detector, in searches of the ULLP events~\cite{Dev:2016dja, Contino:2016spe}. 

In summary, long-lived scalars can arise as a result of $B-L$ symmetry breaking in UV completions of the minimal sterile RH neutrino scenario. For $m_{H_3} \lesssim 5 \gev$, the best small-mixing-angle sensitivity would likely come from MATHUSLA searches for LLPs produced in meson decays, see Fig.~\ref{fig:HBRScalar}. For larger scalar masses, the gauge portal likely provides the best sensitivity, and while main detector searches have excellent sensitivity in this regime, MATHUSLA will likely be able to expand coverage to somewhat lower mixing angles.

\begin{figure}[t]
  \centering
  \includegraphics[width=0.55\textwidth]{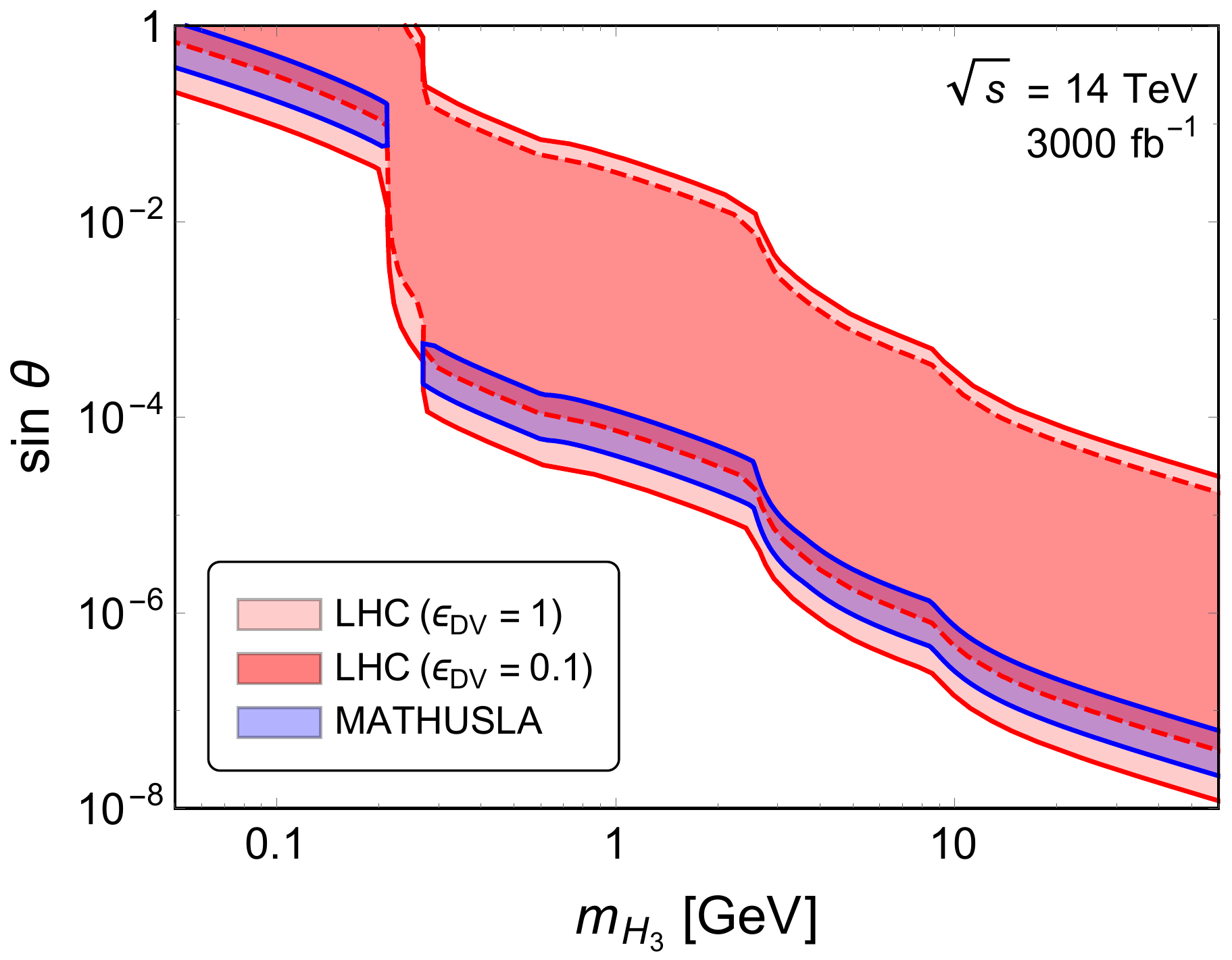}
  \caption{LLP search sensitivities at LHC and MATHUSLA in the $U(1)_{B-L}$ model, with $\sqrt{s} = 14$ TeV and an integrated luminosity of 3000 fb$^{-1}$, in the gauge portals by coupling to the $Z_R$ boson with a benchmark gauge coupling of $g_R = 0.835 g_L$. (See also Section~\ref{sec:singlets} and Fig.~\ref{fig:HBRScalar} for constraints from production through the scalar portal, as well as other limits on the mixing angle. For $m_{H_3} \lesssim 5 \gev$, those MATHUSLA sensitivity is better than the gauge portal sensitivity shown here.) For the LHC reaches we assume a signal efficiency factor of 1 or 0.1, with at least 4 signal events for both LHC and MATHUSLA. Due to the small mass, the lower efficiency is likely more realistic, but the assumption of no background at the main detector is likely justified due to the hard dilepton and dijet pair produced from decay of the on-shell $Z_R$. 
Note that MATHUSLA searches for the production of low-mass $H_3$ with $m_{H_3} \lesssim \gev$ from $Z'$ decay may suffer from some backgrounds or lower reconstruction efficiency depending on the final detector design, see Section~\ref{s.energythresholds}.
This estimate assumes the $200m \times 200m \times 20m$ benchmark geometry of Fig.~\ref{f.mathuslalayout}.
 }
  \label{fig:LLP}
\end{figure}

\subsection{Left-Right Symmetric Model}
\label{sec:LRSYMM}

\subsubsection[Long-Lived Right-Handed Neutrinos in the Left-Right Model]{Long-Lived Right-Handed Neutrinos in the Left-Right Model\footnote{P.~S.~Bhupal Dev, Juan Carlos Helo, Martin Hirsch, Rabindra Mohapatra, Yongchao Zhang}}
\label{sec:LRSYMMnuR}

The standard left-right symmetric model \cite{Pati:1974yy,Mohapatra:1974gc,Senjanovic:1975rk,Senjanovic:1978ev} has the gauge group
$SU(2)_{L}\times SU(2)_{R} \times U(1)_{B-L}$ and provides an ideal setting for a low scale seesaw model for neutrino masses.  Gauge couplings are
denoted respectively as $g_{L}$, $g_{R}$ and $g_{BL}$.
The gauge charges of quarks and leptons in the LR gauge group
are, respectively,
\begin{eqnarray}
&& Q_L=(u_L, \: d_L)^T: ({\bf 2}, {\bf 1}, 1/3);  \quad L=(\nu, \:  e_L)^T: ({\bf 2}, {\bf 1}, -1) \,; \quad \nonumber  \\
&& Q_R=(u_R, \: d_R)^T: ({\bf 1}, {\bf 2}, 1/3); \quad R=(N, \:  e_R)^T: ({\bf 1},{\bf 2}, -1) \,.
\end{eqnarray}
In this setup, RHNs $N$ are automatically included in the theory.
At a scale $v_R$, the LR symmetry is broken to the SM
gauge group.
The minimal Higgs sector includes the fields $\Phi({\bf 2}, {\bf 2}, 0)$ and $\Delta_R({\bf 1}, {\bf 3}, 2)$. The Yukawa couplings are given by
\begin{eqnarray}
{\cal L}_Y \ = \ h_u\overline{Q}_L\phi Q_R+h_d\overline{Q}\widetilde{\phi}Q_R+h_e\overline{L}\widetilde{H}R+h_\nu\overline{L}{\phi}R+f\overline{R}^c\Delta R+ {\rm H.c.} \, .
\end{eqnarray}
After  symmetry breaking we get the seesaw formula for neutrino masses; however, we emphasize that  the $B-L$ group in the left-right model is very different from that studied above.

The charged-current and neutral-current interactions relevant for our analysis are
\begin{eqnarray}\label{CC-NC-LR}
{\cal L} &=& \frac{g_{R}}{\sqrt{2}}
\left(\bar{d}  \gamma^{\mu} P_R u
+  V^{R}_{l N}\cdot \bar l \gamma^{\mu} P_R N\  \right) W^-_{R \mu} + {\rm H.c.}\,,
\end{eqnarray}
where $V^{R}$ is the neutrino mixing matrix in the right-handed sector. The
gauge $W_{L,R}$-boson states can be rewritten in terms of the mass
eigenstates as:
\begin{eqnarray}\label{W-MES}
W^{-}_{L} &=& \cos\zeta\cdot  W^{-}_{1} - \sin\zeta\cdot W^{-}_{2},\\
\nonumber
W^{-}_{R} &=& \sin\zeta\cdot W^{-}_{1} + \cos\zeta\cdot W^{-}_{2},
\end{eqnarray}
whith the mixing angle given by
\begin{eqnarray}\label{WL-WR-mix}
\tan 2\zeta &=& \frac{2 g_{L}g_{R} M^{2}_{W_{L}} \cdot
\sin 2\beta}{g_{R}^{2} M_{W_{L}}^{2} + g_{L}^{2}(M_{W_{R}}^{2} - M_{W_{L}}^{2})} \\
\nonumber
&\approx&  2\  \frac{g_{R}}{g_{L}} \frac{M^{2}_{W_{L}}}{M^{2}_{W_{R}}} \sin 2\beta .
\end{eqnarray}
Here $\tan\beta = \kappa^{\prime}/\kappa$ is the ratio of the two
vev's of the bidoublet Higgs $\Phi$.

The RHNs are typically produced in the on-shell decays of $W_R$. The RHN subsequently decays back into an off-shell $W_R^{\ast}$, which decays exclusively into the light SM quarks with almost a BR of 100\%, \emph{i.e.,} $N \to \ell W_R^\ast \to \ell jj$ where $j$ are the jets from the quarks $u,\,d,\,s,\,c$. In the mass range under consideration, the widths of the three heavy neutrinos of the LR models are approximately
\cite{Helo:2013esa}:
\begin{eqnarray}\label{Dec-Rate-RR}
\Gamma_{N}
\approx \frac{3  G^{2}_{F}}{32 \pi^{3}} m_{N}^{5}
\left(\frac{M_{W_{L}}}{M_{W_{R}}}  \frac{g_{R}}{g_{L}}\right)^{4}
\left[1 + \sin^{2} 2\beta \right] \sum_{l} |V^{R}_{l N}|^{2}, \ \
\end{eqnarray}
where we neglected the masses of all the final state particles.
In our numerical study we will consider a simplified case with
only one heavy neutrino in the relevant mass range.
If the RHN mass is order 
GeV
then its {\it proper} lifetime for a $W_R$ mass of a few TeV would be at the 
100 m
level:
\begin{eqnarray}
\label{eqn:lifetime}
\tau^0_N \ \simeq \ \left( 290 \ {\rm m} \right)
\left( \frac{m_N}{2\, {\rm GeV}} \right)^{-5}
\left( \frac{M_{W_R}}{3\, {\rm TeV}} \right)^{4}
\left( \frac{g_R}{g_L} \right)^{-4} \,
 \,.
\end{eqnarray}
If the $N$ mass is even lighter, then the RHN can be produced in meson decays such as $D_s \to \ell N$, with the subsequent decay $N \to \ell \pi$~\cite{Castillo-Felisola:2015bha}. Both the production of $N$ from mesons as well as their decays into lighter states are mediated by the $W_R$ gauge interaction. The masses $M_{W_R}$, $m_N$ and the gauge coupling $g_R$ can thus be probed at dedicated beam-dump experiments such as SHiP~\cite{Helo:2015ffa, Alekhin:2015byh, Castillo-Felisola:2015bha}, as shown in Fig.~\ref{fig:RHN}, in addition to high-energy colliders.

\begin{figure}[!t]
  \centering
  \includegraphics[width=0.48\textwidth]{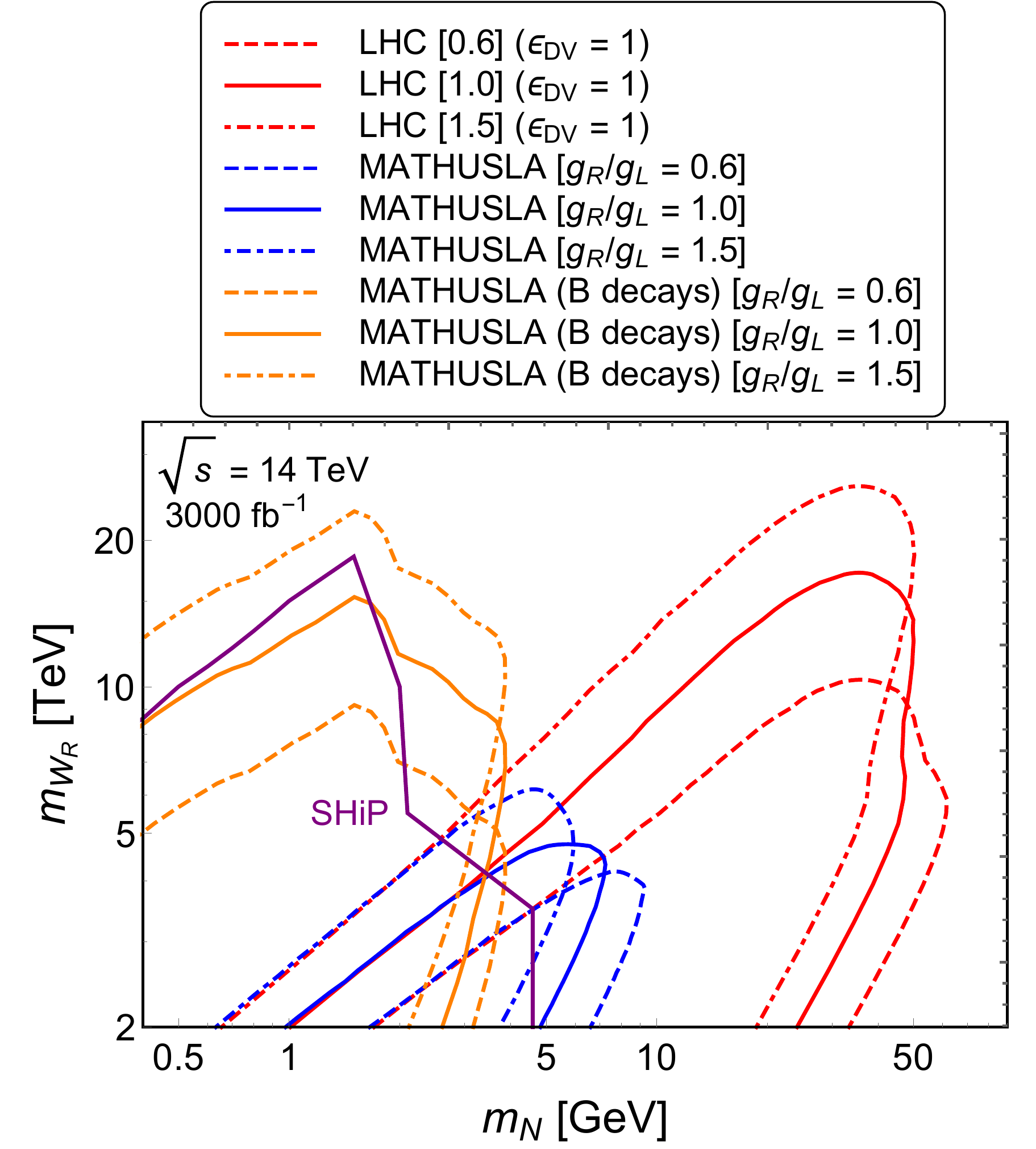}
  \includegraphics[width=0.48\textwidth]{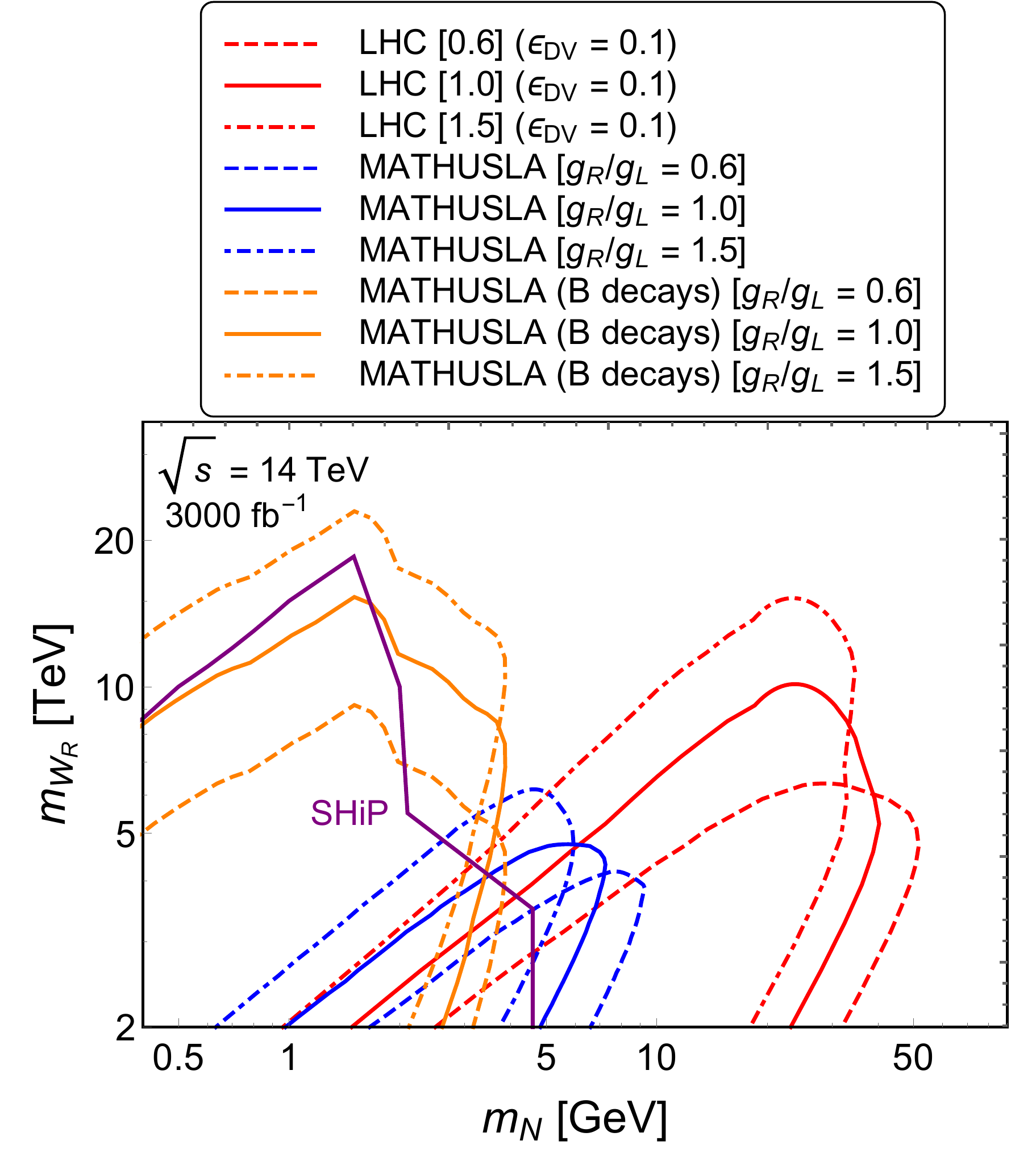}
  \caption{Light RHN sensitivity in the minimal LR model from the (U)LLP searches at the $\sqrt s=14$ TeV LHC (red) and MATHUSLA (blue) with the RHN produced from (on-shell) $W_R$ decay. We also show the MATHUSLA prospects (orange, assuming the $200m \times 200m \times 20m$ benchmark geometry of Fig.~\ref{f.mathuslalayout}) from the decays of $B$ mesons via $B \to \ell N$, for three different values of $g_R/g_L = 0.6$, 1 and 1.5.
  To take account possible signal inefficiencies in DV reconstruction to reject backgrounds, we show curves for an efficiency factor of $\epsilon_{\rm DV} = 1$ (left) and 0.1 (right) for the LHC reaches. Below these curves we can have at least 4 signal events for both LHC and MATHUSLA. The purple lines indicate the DV prospect at SHiP for $g_R = g_L$~\cite{Helo:2015ffa, Alekhin:2015byh, Castillo-Felisola:2015bha}.}
  \label{fig:RHN}
\end{figure}

In the minimal LR group, where the $SU(2)_R$ gauge symmetry is broken by a RH triplet scalar $\Delta_R$, we have $M_{Z_R}>M_{W_R}$. Thus the dominant production of RHNs at the LHC is through the $s$-channel (on-shell) $W_R$: $pp \to W_R^{(\ast)} \to \ell N$, followed by the three-body decay $N\to W_R^*\ell \to \ell jj$~\cite{Keung:1983uu, Deppisch:2015qwa}.  
With $M_{W_R} \gtrsim 3$ TeV$(g_R/g_L)^4$, as required to satisfy the direct LHC constraints~\cite{Khachatryan:2014dka, Aad:2015xaa}, as well as the low-energy flavour-changing neutral current (FCNC) constraints~\cite{Zhang:2007da, Bertolini:2014sua}, the production cross-section could reach few tens of fb, depending on the $W_R$ mass as well as the gauge coupling $g_R$.
The sensitivity contours for MATHUSLA are shown in Fig.~\ref{fig:RHN}, assuming at least 4 signal events, for three different values of the gauge coupling $g_R/g_L = 0.6$, 1 and 1.5.
For concreteness, we have assumed only the electron flavour $\ell = e$ without RH leptonic mixing. Though the effective cross-section is small, due to the small effective solid angle, MATHUSLA is sensitive to light RHN with mass as low as $\sim 1$ GeV. For the purpose of illustration, we also show the proper lifetime of RHN for $g_R = g_L$, estimated from Eq.~(\ref{eqn:lifetime}); for the values of $g_R \neq g_L$ the lifetime should be rescaled via $(g_R / g_L)^{-4}$ accordingly.

If kinematically allowed, the light RHN could also be produced at the LHC from the decays of $D$ and $B$ mesons, as in the simplest seesaw mechanism in Section~\ref{sec:minimalRHN}. In the minimal LR model, the RHN decay and production are mediated by a heavy $W_R$ boson; by comparing the decay width in Eq.~(\ref{eqn:RHNwidth}) and (\ref{Dec-Rate-RR}), the MATHUSLA prospects on the effective heavy-light neutrino mixing angles in Section~\ref{sec:minimalRHN} can be cast onto the $W_R$ mass in the LR model, depending on the $g_R$ coupling. To be specific, three benchmark values of $g_R/g_L = 0.6$, 1 and 1.5 are shown in Fig.~\ref{fig:RHN}, with the RHN from $B$ meson decays (ornage lines).  It is apparent that the meson decay prospects are largely complementary to those from the (on-shell) $W_R$ decay in Fig.~\ref{fig:RHN}.

For the sake of comparison we  also estimate the sensitivity to RHN LLPs at the main ATLAS or CMS detectors. We require that the RHN decay inside the tracker with a decay length 1 cm $\lesssim \gamma c \tau_0 \lesssim$ 1 m, as shown in Fig.~\ref{fig:RHN}. To account for possible inefficiencies in the reconstruction of the displaced vertex, possible benchmark values for the DV efficiency are set to $\epsilon_{\rm DV} = 1$ and 0.1 in the left and right panels of Fig.~\ref{fig:RHN}.
Note  in Fig.~\ref{fig:RHN}  that, even when the heavy $W_R$ is off-shell, \emph{i.e.,} $M_{W_R} \gtrsim 6$ TeV, the light RHN could yet be produced abundantly. In fact, regardless of whether the $W_R$ boson is on-shell or off-shell, the RHN tends to have a huge boost factor of $\gamma \simeq m_{W_R^{(\ast)}} / 2 m_N \sim 10^{3}$. The decay products of $N$ are consequently highly collimated, and the reconstruction of the LLP events would be rather challenging. In the optimistic case, depending on $g_R$, the general-purpose detectors at LHC could probe the {\it proper} lifetime $\tau_N^0$ from $\sim 10$ m to below $0.01$ cm, and the RH sector can be probed up to $M_{W_R} \simeq 20$ TeV for a large $g_R /g_L = 1.5$, which is largely complementary to the ULLP searches at MATHUSLA. With a more pessimistic $\epsilon_{\rm DV} = 0.1$, the probable regions shrink significantly. The reach of ATLAS and CMS may be even worse when realistic efficiencies of reconstructing boosted LLPs and their associated backgrounds are taken into account, see discussion in Section~\ref{s.LHCLLPcomparison}. 
However, even with these optimistic main detector projections, it is clear that MATHUSLA can provide the best sensitivity in the low-mass regime $m_N \lesssim 5 \gev$ which are very difficult to probe at ATLAS or CMS. Even at higher masses, MATHUSLA would likely extend the coverage provided by the main detectors.

\subsubsection[Long-Lived Scalars in the Left-Right Model]{Long-Lived Scalars in the Left-Right Model\footnote{P.S.~Bhupal Dev, Rabindra Mohapatra, Yongchao Zhang}}\label{sec:LLPscalar_LR}

As in the $U(1)_{B-L}$ model, the symmetry-breaking sector of the left-right model can yield interesting dynamics that can be probed at the LHC and detectors such as MATHUSLA. Due to the expanded gauge structure $SU(2)_L \times SU(2)_R \times U(1)_{B-L}$,
this model has a bidoublet $\Phi$ and a RH Higgs triplet $\Delta_R$ in the scalar sector that breaks the gauge symmetry and is responsible for implementing the seesaw mechanism:
\begin{eqnarray}
\Phi \ = \ \left(\begin{array}{cc}\phi^0_1 & \phi^+_2\\\phi^-_1 & \phi^0_2\end{array}\right) : ({\bf 1}, {\bf 2}, {\bf 2}, 0)\, , \qquad
\Delta_R \ = \ \left(\begin{array}{cc}\Delta^+_R/\sqrt{2} & \Delta^{++}_R\\\Delta^0_R & -\Delta^+_R/\sqrt{2}\end{array}\right) : ({\bf 1}, {\bf 1}, {\bf 3}, 2) \, .
\label{eq:scalar}
\end{eqnarray}
There are three physical neutral scalars in this model:~the SM Higgs $h$, a new heavy Higgs field $H^0_1$, and the remnant of the $SU(2)_R$-breaking scalar $H_3$ (see Ref.~\cite{Dev:2016dja} for nomenclature of these scalars):
\begin{eqnarray}
h \simeq {\rm Re}\, \phi_1^0 \,, \quad
H_1 \simeq {\rm Re}\, \phi_2^0 \,, \quad
H_3 \cong {\rm Re}\, \Delta_R^0 \,, \quad
\end{eqnarray}
in the limit of $\langle \phi_1^0 \rangle \gg \langle \phi_2^0 \rangle$. There is almost no absolute {\it lower} mass limit on $H_3$ (except those from the cosmological and astrophysical observations such as BBN and supernovae which requires that $m_{H_3} \gtrsim 100$ MeV), which renders it to be the only LLP candidate in the scalar sector of minimal LR model~\cite{Dev:2016dja, Dev:2016vle, Dev:2017dui}.
The role of $H_3$ is analogous to the light scalar in the $U(1)_{B-L}$ case but their properties are very different as we now show.

In analogy with the $B-L$ model, the smallness of $m_{H_3}$ in the LR model is also stable against  loop corrections due to  heavy particles in the model; thus the $H_3$ field has a wide range of viable masses, and at low masses it could be sufficiently long-lived to give displaced signatures at MATHUSLA. The mixings of $H_3$ to the SM Higgs $h$ and the heavy scalar $H_1$ are governed by two free parameters that represent the mixing between the ${\rm Tr} (\Delta^\dagger \Delta)$ term with the  ${\rm Tr} (\Phi^\dag \Phi)$ term in the scalar potential.
However, as a result of the tree-level FCNC couplings of $H_1$, it turns out that mixing of $H_3$ with $h$ and $H_1$ are highly suppressed, $\lesssim 10^{-4}$ and $\lesssim 10^{-5}$, in the low mass range $M_{H_3}\leq 5-20$ GeV~\cite{Dev:2016vle, Dev:2017dui}.

The crucial difference from the $U(1)$ case is the presence of the right-handed $W_R$ boson (along with additional charged Higgs bosons $H_1^\pm \simeq \phi_2^\pm$ and $H_2^{\pm\pm} \cong \Delta_R^{\pm\pm}$) at the TeV scale. Due to its suppressed coupling to quarks and leptons, 
the dominant decay mode of $H_3$  is to two  photons  from a $W_R$ loop; this is analogous to the case of the $W^\pm$ loop for the SM Higgs decay $h \to \gamma\gamma$ (with subleading contributions from the heavy charged scalars). When the $H_3$ is boosted and long-lived, this gives rise to two, collimated, displaced photons.
As the mixing of $H_3$ to $h$ and $H_1$ is tightly constrained by low-energy FCNC limits from $B$ and $K$ meson decays and oscillations, the long lifetime of $H_3$ in this case is guaranteed. 

In the LR model, the light scalar $H_3$ decays almost 100\% into two photons via the $W_R$ and charged scalar loops~\cite{Dev:2016vle, Dev:2017dui}:
\begin{eqnarray}
\label{eqn:H3diphoton}
\Gamma (H_3 \to \gamma\gamma) & \ = \ &
\frac{\alpha^2 m_{H_3}^3}{18 \pi^3 v_R^2} \ .
\end{eqnarray}
with the factors in the parentheses from the loop functions for the vector bosons and scalars in the limit of $m_{H_3} / m_{W_R} \to 0$. In fact, the decay length of $H_3$ in the LR model is determined solely by the RH scale $v_R$, as well as the scalar mass $m_{H_3}$, when the scalar mixing angles are small. Therefore, in the presence of the extended gauge symmetry, the displaced (collimated) photon signal is rather unique. The effects at LHC and MATHUSLA might provide distinctive evidence of the parity-symmetric LR theories and neutrino masses beyond the SM via these clean displaced photon events. A cautionary note is that MATHUSLA may or may not be able to detect photons depending on the ultimate detector design (see Section~\ref{s.mathuslabasic} and Ref.~\cite{Curtin:2017izq}).

\begin{figure}[!t]
  \centering
  \includegraphics[width=0.48\textwidth]{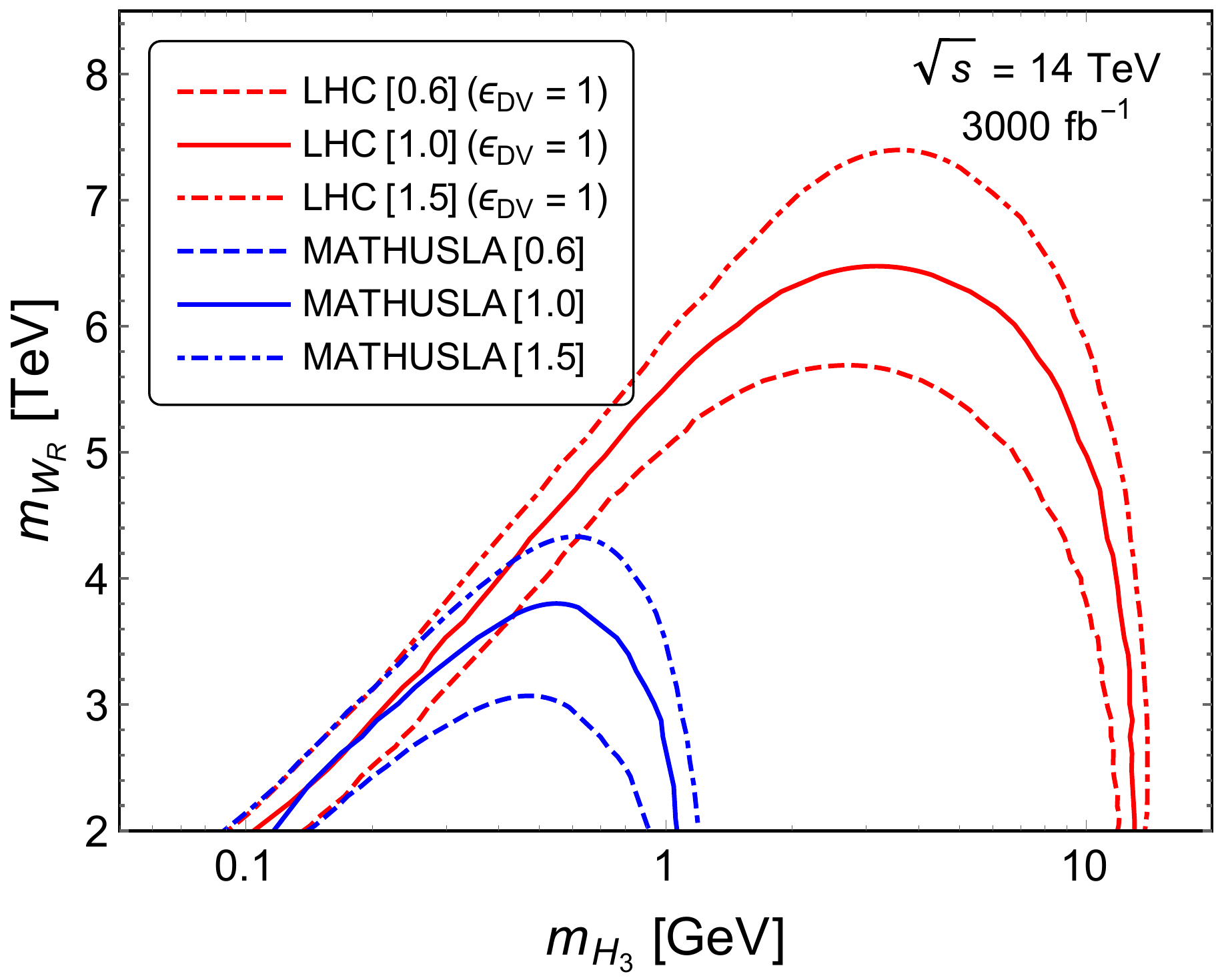}
  \includegraphics[width=0.48\textwidth]{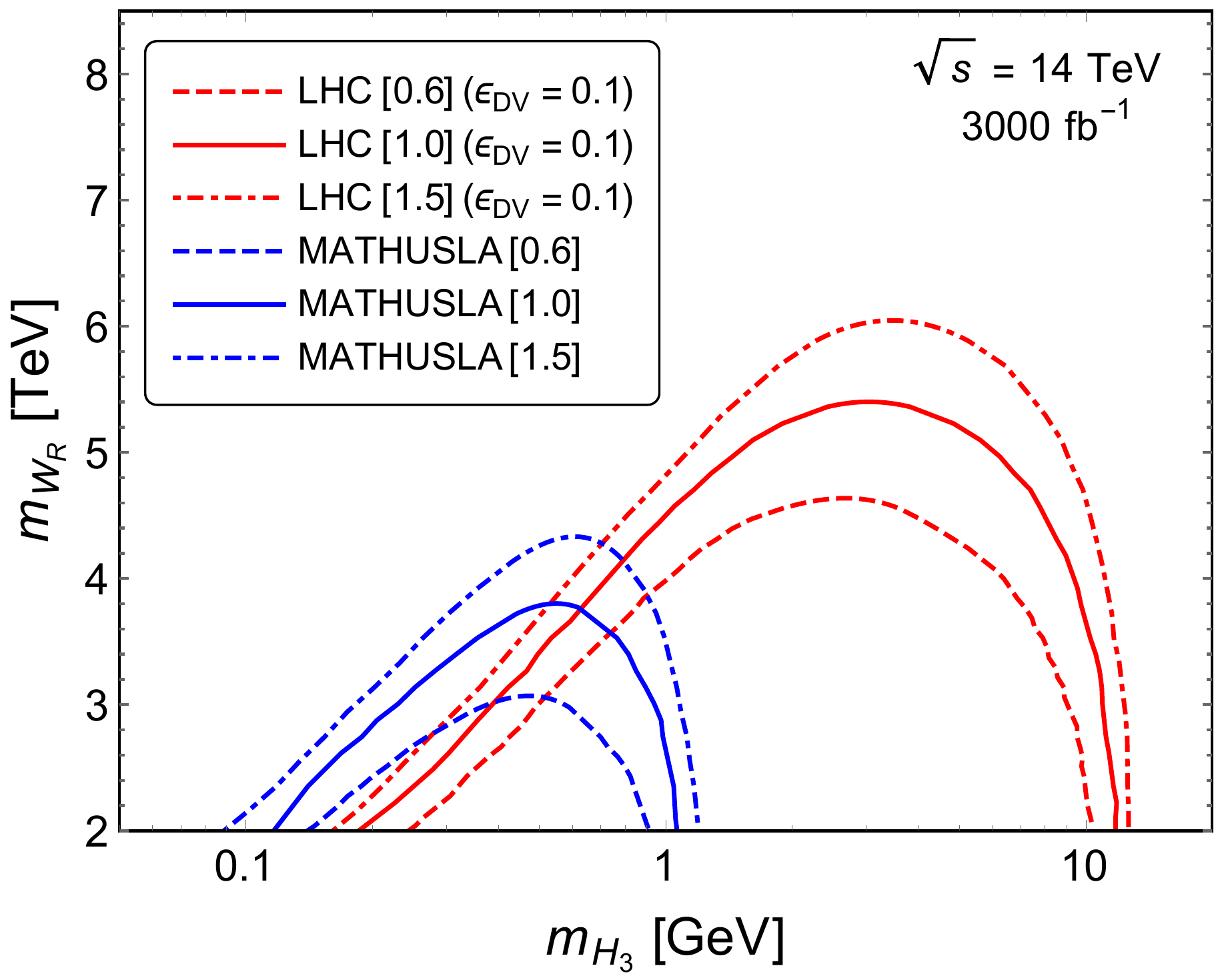}
  \caption{(U)LLP sensitivities at the $\sqrt s=14$ TeV LHC (red) and MATHUSLA (blue) for the lights scalar $H_3$ in the TeV LR model for three different values of $g_R/g_L=0.6, 1$ and 1.5. The grey contours show the decay lengths of $H_3$ in the laboratory frame with $g_R=g_L$; for $g_R \neq g_L$, the lifetime has to be rescaled by the factor of $(g_R/g_L)^{-2}$. To take into account the SM background, we assume an efficiency factor of $\epsilon_{\rm DV} = 1$ (left) and 0.1 (right) for the LHC reaches. Below these curves we can have at least 4 signal events for both LHC and MATHUSLA.
  Note that MATHUSLA searches for the production of low-mass $H_3$ with $m_{H_3} \lesssim \gev$ from $Z'$ decay may suffer from some backgrounds or lower reconstruction efficiency depending on the final detector design, see Section~\ref{s.energythresholds}.
  This estimate assumes the $200m \times 200m \times 20m$ benchmark geometry of Fig.~\ref{f.mathuslalayout}.
  }
  \label{fig:LLP2}
\end{figure}

Turning to the production of $H_3$ in the minimal LR model, it can be produced from its coupling to the heavy RH gauge bosons $W_R$ and $Z_R$, as well as through its coupling to the SM Higgs~\cite{Dev:2016dja, Dev:2016vle, Dev:2017dui}. As the mixing of $H_3$ to the SM Higgs is severely constrained by the flavour data, we focus here only on the gauge portal production, which is dominated by the associated production of $H_3$ with a heavy $W_R$ boson. The $W_R$ subsequently decaying predominantly into the SM quarks ($J = u,\, d,\, s,\, c,\, b,\, t$):
\begin{eqnarray}
pp \ \to \ W_R^\ast \ \to \ W_R H_3 \,, \quad
W_R \to JJ \,.
\end{eqnarray}
Here for simplicity we have assumed that the decay mode to on-shell heavy RHNs,  \emph{i.e.,} $W_R \to \ell N$, is kinematically forbidden. One should note that the $H_3 jj$ processes (with $j = u,\, d,\, s,\, c$) also receive (small) contributions from the heavy vector boson fusion (VBF) $pp \ \to\  W_R^{\ast} W_R^\ast jj \ \to \ H_3 jj$, which is however suppressed by the three-body phase space and the off-shell $W_R$ propagator. At the LHC, limited by the total center-of-mass energy, the associated production with the $Z_R$ boson is always highly suppressed, as it is heavier than the $W_R$ boson in the minimal LR scenario.
When $m_{H_3} \lesssim 10$ GeV, the production rate is almost constant for a given $v_R$, and is sensitive only to the gauge coupling $g_R$ (for the phenomenologically favored TeV range $v_R$, the production cross-section is at the fb level). For smaller $g_R < g_L$, the $W_R$ boson is lighter and the production of $H_3$ can be significantly enhanced. As in the case of $U(1)$ model above, the associated  jets from $W_R$ decay tend to have a large $p_T$ (typically $\gtrsim$ 1 TeV), and can be easily used for triggering of the LLP events.

In Fig.~\ref{fig:LLP2}, we give the range of $H_3$ masses and $W_R$ masses that can be probed in the displaced diphoton channel at the LHC (red), as well as at MATHUSLA (blue) for three different values of $g_R/g_L = 0.6$, 1 and 1.5, with $\sqrt{s} = 14$ TeV and an integrated luminosity of 3000 fb$^{-1}$, assuming at least 4 signal events.
For the LHC case, we assume the scalar $H_3$ decays inside the tracker with a decay length of $1\, {\rm cm} \lesssim \gamma c\tau_0 \lesssim 1\, {\rm m}$. Note that with the TeV-scale $W_R$ boson taking away most of the energy in the final state, reconstruction of the low-mass and highly boosted $H_3$ would be rather challenging. In addition, the displaced photons are more difficult than displaced charged particles, due to absence of tracking information. Therefore an efficiency factor of $\epsilon_{\rm DV} = 0.1$ might be closer to be realistic in Fig.~\ref{fig:LLP2}.
The (U)LLP searches at LHC and MATHUSLA are largely complementary to each other, as in the case of $U(1)$ model, and could probe a $W_R$ boson up to 6 TeV or so for $g_R = g_L$, which is complementary to the direct searches of $W_R$ at the LHC in the same-sign dilepton plus jets events.
As noted earlier, the virtually background-free environment at MATHUSLA might have superior sensitivity relative to the larger acceptance of the LHC for lighter $H_3$ masses.

\subsection[Neutrinos from the Higgs Portal]{Neutrinos from the Higgs Portal\footnote{Elena Accomando, Luigi Delle Rose, Stefano Moretti, Emmanuel Olaiya, Claire H.~Shepherd-Themistocleous}}\label{sec:Higgportalneut}

In addition to supersymmetry, abelian extensions of the SM at the TeV scale represent an intriguing possibility among all the BSM scenarios. Indeed, the remarkable convergence of the gauge couplings, although only approximate, predicted by their renormalisation group evolution at around $10^{15}$ GeV, strongly hints in favour of GUTs. One of the main features of such theories is the appearance of an extra $U(1)'$ gauge group with the associated gauge boson $Z'$ within reach of LHC energies; see, for instance, Ref.~\cite{Langacker:1980js,Hewett:1988xc} for a review. 

The discovery of a new massive vector boson at the TeV scale would have further  interesting implications. Indeed, the breaking of the extra abelian gauge symmetry would require the existence of an enlarged scalar sector with a heavy scalar field mixing with the SM Higgs doublet and giving mass to the $Z'$. Moreover, anomaly cancellation naturally predicts exotic fermionic states. These could be SM-singlet right-handed neutrinos, which give mass to the SM neutrinos through a seesaw mechanism. In the simplest realisation of a type-I seesaw scenario with one singlet fermion for each flavour generation (other seesaw realisations can be envisaged as well), heavy neutrinos typically have an extremely small coupling to the SM gauge bosons induced by their small mixing with the light, active neutrinos.
Therefore, the decay width of a heavy-neutrino could be small and its lifetime particularly large over a substantial region of the available parameter space, making the $U(1)'$ extension one of the best BSM scenarios predicting LLPs. (See e.g. Section~\ref{sec:bminusl}).

The heavy neutrino, $N$, has a decay length determined by its mass and by the neutrino mixing matrix, as outlined in Sec.~\ref{sec:minimalRHN}. In the mass range $1 \, \textrm{GeV} < m_{N} < 100 \, \textrm{GeV}$, the $N$ proper decay length spans from $10^9$ m to few cm \cite{Accomando:2016rpc,Accomando:2017qcs,Batell:2016zod} for parameters motivated by the minimal type-I seesaw, reaching the BBN limit \cite{Kawasaki:2004qu,Jedamzik:2006xz} for $m_{N}$ of few GeV; see Fig.~\ref{fig:decay.prod}~(a). 
In this mass range the heavy neutrinos decay through the processes $N \rightarrow l^\pm W^{\mp*}$ and $N \rightarrow \nu_l Z^*$ with off-shell gauge bosons, leading to four available modes, $qql$, $l^+l^-\nu_l$, $qq\nu_l$ and $\nu_l\nu_l\nu_l$ with BRs approximately given by 50\%, 24\%, 18\% and 8\%, respectively.
Long-lived heavy neutrinos are pair produced at the LHC via the s-channel exchange of the $Z'$, the 125 GeV Higgs $h$ and its heavy partner $H$.
The presence of these mechanisms is the main difference between $U(1)'$ extensions and the simple seesaw-extended SM. Indeed, in the minimal see-saw model, the production of $N$ is suppressed by the square of the mixing between left- and right-handed neutrino components, whereas in $U(1)'$ models there is a large region of the parameter space where production can be much larger due to $Z'$ and Higgs production of $N$. The decay modes of the heavy neutrinos remain the same in both scenarios.

The heavy scalar $H$ is responsible for the dynamical generation of the $N$ Majorana mass and, through the mixing $\alpha$ between $H$ and the SM Higgs in the scalar sector, acts as a portal providing an exotic heavy neutrino coupling to the SM Higgs.
As discussed in Sections~\ref{sec:exhdecays}, \ref{sec:darkphotons}, and \ref{sec:singlets}, such mixings are generic since it cannot be forbidden by any symmetry.
The $B-L$ scenario discussed in Sec.~\ref{sec:bminusl} represents an explicit realisation of a heavy scalar portal motivated by abelian extensions of the SM. The mixing angle, $\alpha$, scales all the interactions of the SM-like (heavy) Higgs to SM particles with $\cos \alpha$ ($\sin \alpha$). Interactions of scalars with other particles in the extended spectrum of the $U(1)'$ model, such as the $Z'$ and RH neutrinos,  are proportional to the complementary angle ($\cos\alpha$ for $H$, $\sin\alpha$ for $h$). In particular, this gives a $\cos^2 \alpha$ scaling of the cross-section of the standard $h$ production mechanisms, and  a $\sin^2 \alpha$ scaling of the partial width of the exotic $h$ decay into heavy neutrinos.

Notice also that extensions of the SM scalar sector affect the running of the quartic scalar couplings and help in the stabilisation of the vacuum \cite{Gonderinger:2009jp,Basso:2010jm,EliasMiro:2012ay,Coriano:2014mpa,Coriano:2015sea,Falkowski:2015iwa,Accomando:2016sge}. In particular, $\alpha \gtrsim 0.1$ may allow to achieve a stable and pertubartive vacuum up to the GUT scale over a wide range of heavy Higgs masses while complying with LHC Higgs searches, see for instance \cite{Falkowski:2015iwa,Accomando:2016sge}.

\begin{figure}
\centering
\begin{tabular}{ccc}
\includegraphics[scale=0.79]{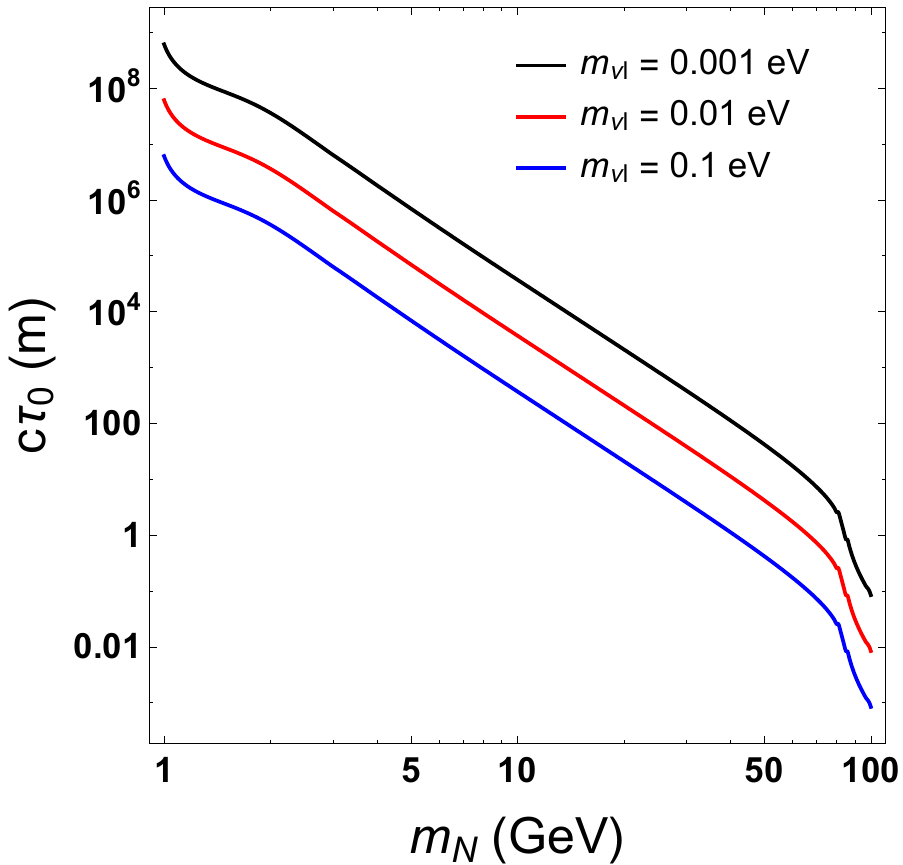}
&\quad&
\includegraphics[scale=0.56]{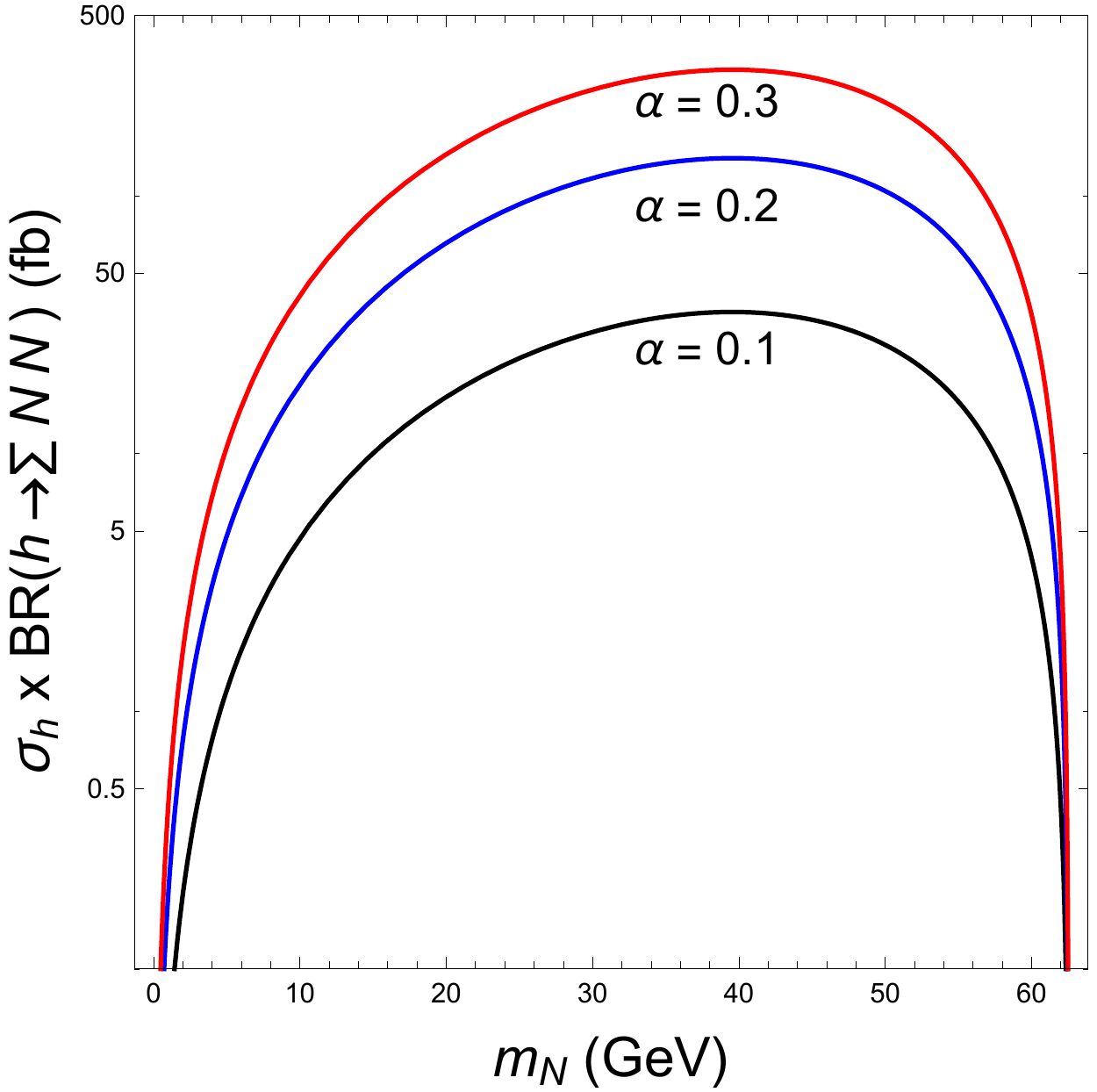}
\\
(a) & & (b)
\end{tabular}
\caption{(a) Proper decay length of the heavy neutrino as a function of its mass $m_{N}$ for three different values of the light neutrino mass. For illustrative purpose we have assumed a diagonal Dirac neutrino mass matrix. (b) $\sigma \times \textrm{BR}$ for the process $pp \rightarrow h \rightarrow \sum N N$ in the gluon channel at the 14 TeV LHC as a function of the heavy neutrino mass for three different values of the scalar mixing angle and $x = 4$ TeV. \label{fig:decay.prod}
This estimate assumes the $200m \times 200m \times 20m$ benchmark geometry of Fig.~\ref{f.mathuslalayout}.}
\end{figure}

\begin{figure}
\centering
\includegraphics[scale=0.79]{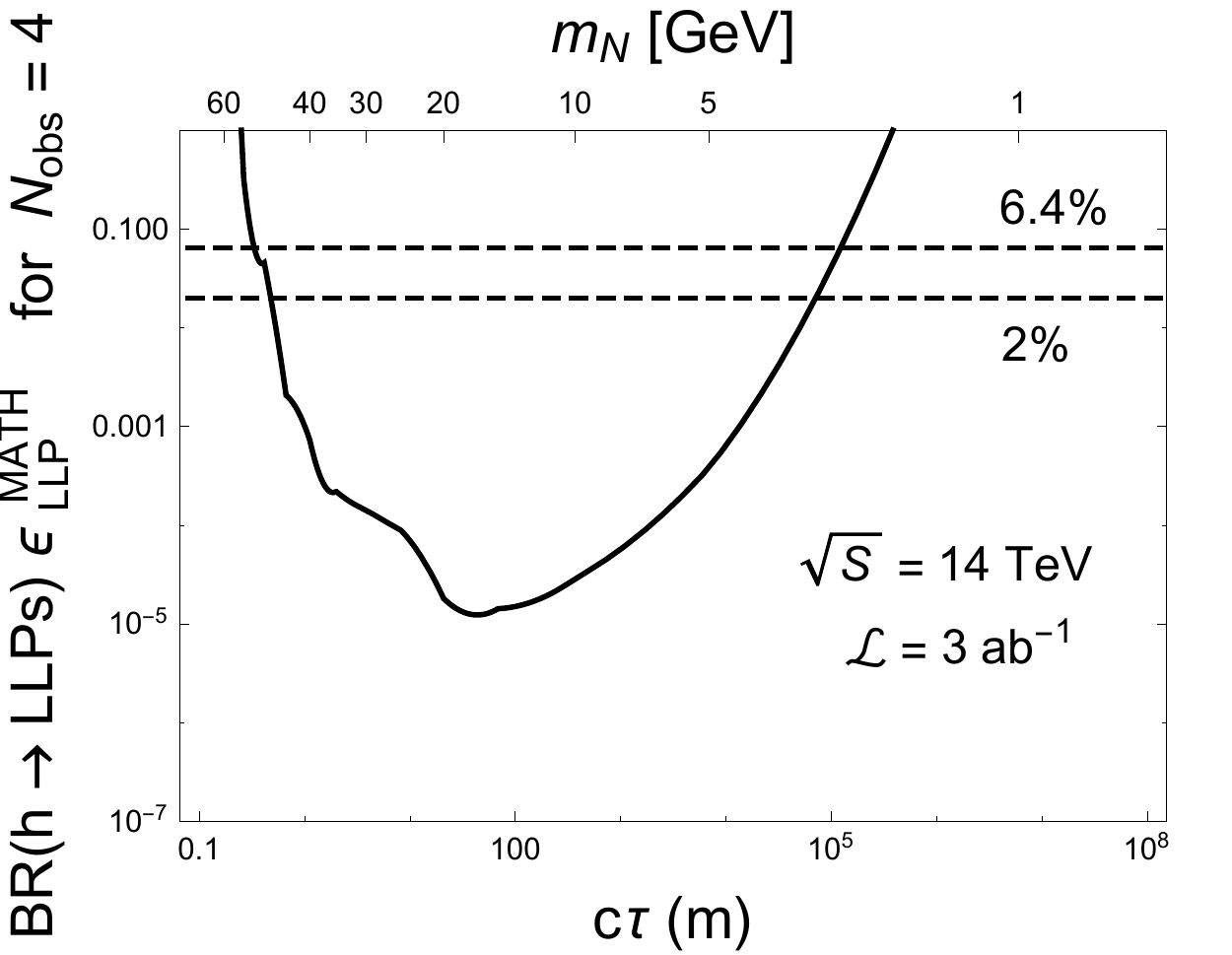}
\caption{Sensitivity estimate for the MATHUSLA detector at the HL-LHC (assuming the $200m \times 200m \times 20m$ benchmark geometry of Fig.~\ref{f.mathuslalayout}) for the process  $pp \rightarrow h \rightarrow \sum N N$. We have assumed a diagonal Dirac neutrino mass matrix and $m_{\nu_l} = 0.1$ eV to fix the RH neutrino lifetime.
The cross-section required to see 4 events has been normalised to the detection efficiency $\epsilon_\textrm{LLP}^\textrm{MATH}$ for a displaced vertex. 
The 2\% and 6.4\% lines are representative of possible $\mathrm{Br}(h \to \mathrm{invis})$ exclusions achievable by the HL-LHC, see Sections~\ref{s.METcomparison} and \ref{sec:exhdecays}.
\label{fig:MATHnuHsensitivity}}
\end{figure}

If $m_{N} < m_{h}/2$ and $\alpha \neq 0$, all three heavy neutrino production modes are kinematically accessible with the SM-like Higgs mediation being the dominant channel in a large region of the parameter space.
The corresponding partial decay width is
\begin{equation}
\label{eq:portal:partialwidth}
\Gamma(h \rightarrow \sum N N) = \sum_{i} \frac{m_{N_{i}}^2}{x^2} \sin^2 \alpha \frac{m_h}{16 \pi} \left( 1 - \frac{4m_{N_{i}}^2}{m_h^2} \right)^{3/2},
\end{equation}
where $i$ sums over the heavy neutrino families and $x$ is the vacuum expectation value of the extra scalar. This expression is common to every extension of the SM in which the Majorana mass of the heavy neutrinos is generated by the vev $x$ through spontaneous symmetry breaking of a SM-singlet scalar mixed with the SM $SU(2)$ Higgs doublet. In a $U(1)'$ scenario in which the $Z'$ mass $M_{Z'}$ arises by the same mechanism, the vev $x$ is not a free parameter but is fixed by $x = M_{Z'}/(z_S g')$, where $z_S$ and $g'$ are, respectively, the $U(1)'$ charge of the scalar singlet and the $U(1)'$ gauge coupling, and, therefore, it is constrained by the $Z'$ searches at the LHC in the di-lepton channel \cite{ATLAS:2016cyf,CMS:2016abv} (see \cite{Accomando:2016rpc} for a reinterpretation of the limits in terms of a generic $U(1)'$ charge assignement).

Let us assume that the $Z'$ mass arises from the same vev $x$. The BR$(h \rightarrow \sum N N)$ is constrained to be $\lesssim 1\%$, but it is compensated by a large Higgs production cross-section $\sigma_h = \cos^2 \alpha \, \sigma_{h_\textrm{SM}}$, Fig.~\ref{fig:decay.prod}(b), with $\sigma_{h_\textrm{SM}} = 54.67$ pb in the gluon-fusion channel at the 14 TeV LHC \cite{deFlorian:2016spz}.
The shape of the $N$ production cross-section in Fig.~\ref{fig:decay.prod}(b) is determined by the heavy neutrino mass. In particular its zeros are located at $m_{N} \simeq 0$ and $m_{N} = m_h/2$ where, respectively, the coupling of the Higgs to right-handed neutrinos is strongly suppressed or the process is kinematically closed.
 The dependence on $\alpha$ is mainly controlled by the scaling factors $\cos^2 \alpha$ and $\sin^2 \alpha$, affecting, respectively, the Higgs production cross-section and the partial decay width in Eq.~(\ref{eq:portal:partialwidth}). A mild residual dependence on $\alpha$, which is only seen for large values of $\alpha$, appears in the Higgs total decay width, $\Gamma_\textrm{tot} = \cos^2 \alpha \Gamma_\textrm{tot}^\textrm{SM} + \Gamma(h \rightarrow \sum N N)$, that normalises the BR$(h \rightarrow \sum N N)$.
 Moreover, if the vev $x$ is not constrained by the $Z'$ mass, the BR$(h \rightarrow \sum N N)$ is bounded from above only by the upper limit on the invisible Higgs decay \cite{Khachatryan:2016whc,ATLAS:2017zkz}.

For $m_{N} > m_{h}/2$, the $h$ channel is, instead, kinematically closed. For $M_{Z'}\gtrsim3$ TeV, as required by the recent di-lepton searches at the LHC \cite{ATLAS:2016cyf,CMS:2016abv}, the dominant $N$ production mode typically comes from decays of the heavy Higgs, $H$. However, if the heavy scalar sector is decoupled from the SM one, namely $\alpha \simeq 0$, both the $h \rightarrow N N$ decay and the heavy Higgs production from $pp$ collisions are suppressed and the heavy neutrino pair production via the $Z'$ remains the only available possibility over the entire range of the heavy neutrino masses. As an example, this provides, for $M_{Z'} = 4$ TeV and $m_{N} \ll M_{Z'}$, $\sigma_{Z'} \textrm{BR}(Z' \rightarrow \sum N N) \sim 0.6$ fb at the 14 TeV LHC. The production of $N$ from lower-mass $Z'$ with smaller gauge couplings $g'$ was discussed in Sec.~\ref{sec:bminusl_lowmass}.

Here we consider  long-lived heavy neutrino production from decay of the SM-like Higgs, and we present in Fig.~\ref{fig:MATHnuHsensitivity} an estimate of the sensitivity for the MATHUSLA detector at the HL-LHC.
In particular, we show the BR$(h \rightarrow \sum N N)$ required to observe 4 signal events as a function of the heavy neutrino proper decay length $c \tau_0$. The BR has been normalised to the displaced vertex detection efficiency in MATHUSLA, $\epsilon_\textrm{LLP}^\textrm{MATH}$.
The Standard Model Higgs production cross-section $\sigma_{h_\textrm{SM}}$ has been used. The $\cos^2 \alpha$ correction induced by the scalar mixing angle in the range $0 < \alpha \lesssim 0.3$ (larger values are disfavoured by the signal strength measurements of the Higgs) only reduces the Higgs cross-section by a factor less than $9\%$ and does not considerably affect the result in Fig.~\ref{fig:MATHnuHsensitivity}.
For the sake of simplicity we have also assumed a diagonal Dirac neutrino mass matrix and $m_{\nu_l} = 0.1$ eV.
Notice that, for a given choice of the light neutrino mass $m_{\nu_l}$, the mass of the heavy neutrino is completely determined by its decay length as shown in Fig.~\ref{fig:decay.prod}(a).

It is instructive to compare the MATHUSLA sensitivity to the Higgs portal $N$ production with the capabilities of the ATLAS and CMS detectors at the HL-LHC.
Since about 76\% of the heavy neutrino decay channels contain a lepton in the final state and the heavy neutrinos are pair produced in the exotic Higgs decay, a search for two LLPs in the trackers of the main detectors may be characterised by low background contamination. Therefore, exploiting dilepton triggers may result in a better sensitivity than MATHUSLA in the short lifetime regime, $c \tau \lesssim 10$ m. On the other hand, the same search in the long lifetime regime would provide a very poor sensitivity. In that case, the most direct comparison to MATHUSLA would be a search for a single right-handed neutrino.
In the most optimistic scenario one should focus on decays inside the ATLAS and CMS trackers and rely on the fully muonic decay mode, thus employing the dimuon trigger at Level 1.
Assuming the presence of a muon and a DV would be sufficient to eliminate all backgrounds at the main detectors, we can follow the procedure outlined in Section~\ref{s.LHCLLPcomparison} to compute the long-lifetime sensitivity gain $R_s$ of MATHUSLA compared to the main detector. 
Taking DV reconstruction in the main detector tracker to have roughly a quarter the geometric  acceptance and half the efficiency as MATHUSLA, along with the $\epsilon_\mathrm{cuts}^\mathrm{LHC} \sim 0.5$ requirement of a muon produced in RH neutrino decay,  we arrive at $R_s \sim 15$. 
Therefore, even though this signature would be background free in the main detectors, the sensitivity to the cross-section (and hence long lifetime) achieved by MATHUSLA would be at least an order of magnitude better. 
Furthermore, it is worth noting that the hypothesis of negligible background for displaced vertices in the tracker may be too optimistic. The very low sterile neutrino masses in the long lifetime regime may make their reconstruction and the background rejection very challenging.

The study presented above relies on the existence of a portal mediated by a SM singlet scalar to heavy neutrinos for the SM Higgs. A natural realisation of this scenario is given by an extension of the SM with a spontaneously broken abelian symmetry where both states, the heavy neutrino and the extra scalar, are naturally required by the gauge symmetry and the generation of the $Z'$ mass.
Nevertheless, the $U(1)'$ symmetry is not mandatory and other scenarios with a global symmetry can still provide exotic Higgs decay to long-lived particles.
A model-independent approach to such scenarios has been presented in \cite{Caputo:2017pit} where the contribution of heavy new physics degrees of freedom (such as the heavy scalar and the $Z'$ discussed above) has been parameterised in terms of a low-energy effective field theory (EFT) whose leading effects are encoded in dimension-5 operators. When the theory is augmented with SM singlet fermions with masses around or below the EW scale, new dimension-5 operators appear in the EFT \cite{Graesser:2007yj,delAguila:2008ir,Aparici:2009fh} besides the well-known Weinberg operator \cite{Weinberg:1979sa}. One of them, $(\lambda_{ij}/\Lambda) \overline{\nu_{R_i}^c} \nu_{R_j} \, \Phi^\dag \Phi$, contributes to the Majorana mass $M_M$ of the right-handed neutrinos and provides additional couplings to the SM Higgs which are not necessarily proportional to $M_M$. In particular, for $m_{N_i} < m_h/2$, the Higgs can decay to heavy neutrinos via the coupling $v/(2 \Lambda) h \, \overline{\nu_{R}^c} \lambda \nu_{R}$.
After the identification $v/(2 \Lambda) \lambda_{ij} \equiv M_{M_{ij}}/(2 x) \sin \alpha$, one can recover the Higgs partial decay width in Eq.~(\ref{eq:portal:partialwidth}) and easily reach similar conclusions to those shown here.

\subsection[Discrete Symmetries and FIMP Dark Matter]{Discrete Symmetries and FIMP Dark Matter\footnote{Alejandro Ibarra}}
\label{sec:inertdoubletportal}

In the models discussed so far, we have focused on the type-I seesaw model for neutrino masses. An alternative explanation for the smallness of neutrino masses results from new particles charged under a discrete or continuous symmetry (global or local), such that the dimension-5 Weinberg operator responsible for SM neutrino masses does not arise at tree level but instead at the $N$-loop level. In this class of models, the active neutrino mass matrix can be expressed as:
\begin{equation}
{\cal M}^\nu_{ij}\sim \left(\frac{1}{16\pi^2}\right)^N \frac{\alpha_{ij}}{\Lambda}\langle \Phi^0\rangle^2\;,
\end{equation}
where $\Lambda$ is the scale of the new physics and $\alpha_{ij}$ are effective couplings, accounting for the flavour structure of the neutrino mass matrix. Notably, the suppression by the loop factor of the radiatively generated neutrino masses lowers the mass scale of the new physics, $\Lambda$, thus opening the exciting possibility of producing in colliders the particles responsible for the neutrino mass generation. Furthermore, if the new symmetry is unbroken (or mildly broken) in the electroweak vacuum, the lightest particle of the sector responsible for neutrino masses is long-lived on cosmological time-scales and therefore constitutes a dark matter candidate. Collider experiments, in this case, may also produce dark matter particles, either directly or in cascade decays.

One of the simplest models of radiative neutrino mass generation incorporating a dark matter candidate is the so-called scotogenic model \cite{Ma:2006km}. In this model, the particle content of the Standard Model (SM) is extended with one additional scalar doublet $H_2\equiv (H^+,H_2^0)$ and at least two fermion singlets $N_j$ ($j=1,2,\ldots$). The model also postulates that the electroweak vacuum is invariant under a discrete $Z_2$ symmetry, under which all SM fields are even, whereas $N_j$ and $H_2$ are odd. The Lagrangian of the model reads
\begin{align}
\mathcal{L} =  \mathcal{L}_{\rm SM}\,+ \mathcal{L}_{H_2}\,+ \mathcal{L}_{N}\,+ \mathcal{L}_{\rm int}\;,
\end{align}
where $\mathcal{L}_{\rm SM}$ is the SM Lagrangian, $\mathcal{L}_{H_2}$ and $\mathcal{L}_{N}$ are, respectively, the terms in the Lagrangian involving only the fields $H_2$ and~$N_j$,
\begin{align}
\mathcal{L}_{H_2} & =  \left(D_\mu H_2\right)^\dagger(D^\mu H_2)-\,\mu_{2}^{2}\,(H_{2}^{\dagger}\,H_{2})\,-\,\lambda_{2}\,(H_{2}^{\dagger}\,H_{2})^{2}\;, \\
\mathcal{L}_{N} &= \frac{i}{2}\,\overline{N}_j\partial_\mu\gamma^\mu N_j-\frac{1}{2} M_{j}\,\overline{N^c_j} \,N_{j}+{\rm h.c.}\;,
\end{align}
and $\mathcal{L}_{\rm int}$ contains the interaction terms of the $Z_2$-odd fields with the Standard Model fields,
\begin{align}
\mathcal{L}_{\rm int}=&\,-\lambda_{3}\,(H_{1}^{\dagger}\,H_{1})\,(H_{2}^{\dagger}\,H_{2})
\,-\,\lambda_{4}\,(H_{1}^{\dagger}\,H_{2})\,(H_{2}^{\dagger}\,H_{1})\,-\, \frac{\lambda_{5}}{2}\,\left[(H_{1}^{\dagger}\,H_{2})^{2}\,+\,{\rm h.c.}\right] \nonumber \\
&+\Big[Y^{\nu}_{\alpha i}\,(\overline{\nu}_{\alpha L}\,H_{2}^{0}\,-\,\overline{\ell}_{\alpha L}\,H^{+})\,N_{i}	 \,+\,{\rm h.c.}\Big] \;.
\label{eq:L_int}
\end{align}
where $H_1$ is the SM Higgs doublet, $\ell$ are the  charged leptons and $\nu$ are the active neutrinos. The parameters of the scalar potential are chosen such that $\langle H_{1}\rangle =(0, v/\sqrt{2})$, with $v\simeq 246$ GeV, and  $\langle H_{2}\rangle =0$, hence the minimum of the potential breaks the electroweak symmetry while preserving the  $Z_2$ symmetry. A variant of the model has instead one $Z_2$-odd fermion singlet, and at least two $Z_2$-odd scalar doublets~\cite{Hehn:2012kz}, which naturally lead to a mild hierarchy between the solar and atmospheric mass scales.

The conservation of the $Z_2$ symmetry ensures that the lightest $Z_2$-odd particle is absolutely stable, which then constitutes a dark matter candidate if it is electrically neutral. The dark matter candidates of the model are the CP-even and CP-odd neutral scalars, $H^0$ and $A^0$, and the lightest singlet fermion $N_1$.  Here we focus on the latter candidate, concretely in the region of the parameter space where it constitutes a Feebly Interacting Massive Particle (FIMP), see also Section~\ref{sec:freezein}. If this is the case, heavier $Z_2$-odd particles can be very long-lived and decay inside MATHUSLA.

The signals of the scotogenic FIMP scenario crucially depend on the mass spectrum of the $Z_2$-odd sector. Of particular interest for MATHUSLA is the scenario where $M_1 < M_2 < M_3 < m_H$, where $m_H$ denotes the overall mass scale of the $Z_2$-odd scalar sector. The $Z_2$-odd scalars $H^0, \,A^0$ and $H^\pm$, can be produced at the LHC via neutral and charged current Drell-Yan  processes (as well as in gluon fusion with an off-shell Higgs in the s-channel~\cite{Hessler:2014ssa}) and subsequently decay into singlet fermions, mostly $N_2$ and $N_3$, due to the suppressed coupling of the FIMP to the $Z_2$-odd scalars. 
As shown in Fig.~\ref{f.benchmarkxsecs}, MATHUSLA could probe such EW production processes for LLPs for mass scales up to a TeV, depending on the lifetime. 
The singlet fermions $N_2$ and $N_3$ in turn decay producing visible particles in the final state with rates:\cite{Molinaro:2014lfa}
\begin{align}
\Gamma(N_{j} \to \ell^-_{\alpha} \ell^+_{\beta} N_1)
&\simeq  \frac{M_{j}^{5}}{6144\,\pi^{3}\,m_{H}^{4}}\left(\left|Y^{\nu}_{\beta 1} \right|^{2} \left|Y^{\nu}_{\alpha j} \right|^{2}+\left|Y^{\nu}_{\alpha 1} \right|^{2} \left|Y^{\nu}_{\beta j} \right|^{2}\right)\,,~~j=2,3,\label{N2decay}
\\
\Gamma(N_{3} \to \ell^-_{\alpha} \ell^+_{\beta} N_2)
&\simeq  \frac{M_{3}^{5}}{6144\,\pi^{3}\,m_{H}^{4}}\left(\left|Y^{\nu}_{\beta 2} \right|^{2} \left|Y^{\nu}_{\alpha 3} \right|^{2}+\left|Y^{\nu}_{\alpha 2} \right|^{2} \left|Y^{\nu}_{\beta 3} \right|^{2}\right)\,.\label{N3decay}
\end{align}
Here, the masses of the final fermions have been assumed to be negligibly small compared to the mass of the decaying fermion.

For FIMP dark matter, the requirement of reproducing the observed dark matter abundance fixes the size of the Yukawa coupling as a function of the FIMP mass and the charged scalar mass, thus giving proper decay-lengths for $N_2$ and $N_3$, given by~\cite{Hessler:2016kwm}
\begin{align}
&c\tau(N_2)\;\approx\; 2\times 10^{13}\,\text{m}\,\left(\frac{M_1}{\text{10 keV}}\right)\left(\frac{m_{H}}{\text{500 GeV}}\right)^3\left(\frac{\text{100 GeV}}{M_2}\right)^5\left(\frac{10^{-3}}{y_2}\right)^2\;,
\label{eq:N2toN1} \\
&c\tau(N_3)\;\approx\; 0.4\,\text{m}\left(\frac{\text{100 GeV}}{M_3}\right) \left(\frac{m_{H}}{M_3}\right)^4\left(\frac{10^{-3}}{y_2}\right)^2\left(\frac{10^{-3}}{y_3}\right)^2\;,
\label{eq:N3toN2}
\end{align}
where $y_k\equiv (\sum_\alpha |Y_{\alpha k}|^2)^{1/2}$.
$N_2$ is then stable even at distance scales of the Solar System. However, $N_3$ can be stable within the ATLAS detector and decay some distance away, possibly inside MATHUSLA, producing two charged leptons (in general with different flavour) and missing energy, carried away by $N_2$. 
The lifetime can easily be in the 100m or above range. 
MATHUSLA could then potentially supply the best sensitivity for production of such LLPs, in particular in the low-mass regime $m_{N_3} \lesssim 10 \gev$ where searches for displaced lepton-jets at the main detectors suffer from some backgrounds, see discussion in Section~\ref{s.LHCLLPcomparison}. 
Furthermore, MATHUSLA may offer the possibility of discriminating the three body final state $\ell_\alpha^- \ell_\beta^+ + \slashed E_T$ from the two body final state $\ell_\alpha^- \ell_\beta^+$, from the angular distribution of the charged leptons inside the detector~\cite{IMV}.

\subsection[Enhanced Residual Symmetry (ERS) Scenarios and Freeze-Out Leptogenesis]{Enhanced Residual Symmetry (ERS) Scenarios and Freeze-Out Leptogenesis\footnote{P.~S.~Bhupal Dev, Claudia Hagedorn and Emiliano Molinaro}}\label{BhupalClaudiaEmiliano}

The models considered in the previous sections treat neutrino masses and mixings as input parameters. 
In a complete theory, however, it is likely that symmetries play a crucial role in setting the observed pattern of masses, mixings and $CP$ phases.

The scenario explored in this section belongs to a class of models which postulate a particular flavor symmetry breaking pattern to derive the low-energy parameters of the type I see-saw.
It implements a type-I seesaw scenario with flavour symmetry $G_f$ and a CP symmetry~\cite{usinpreparation}
that strongly constrain lepton mixing angles, and both low- and high-energy CP phases~\cite{Feruglio:2012cw}.
 The three right-handed (RH) neutrinos $N_i$ have (almost) degenerate masses.
Their decays are responsible for the generation of the baryon asymmetry $\eta_B$ of the Universe via
resonant leptogenesis~\cite{Pilaftsis:1997jf, Pilaftsis:2003gt}.

This not only explains the measured values of the lepton mixing angles, but also makes predictions for leptonic CP violation in neutrino oscillations,  neutrinoless double beta decay, as well as connect low energy CP phases with those relevant for leptogenesis.

The dynamical mechanism of flavor symmetry breaking is not specified, but models which implement such mechanisms ~\cite{King:2013eh} often feature flavour symmetry breaking fields with aligned vacuum expectation values, resulting in a vacuum which respects an \emph{enhanced residual symmetry} (ERS) compared to the naive expectation. 
This ERS is slightly broken by higher-dimensional operators, leading to observables which depend on various powers of a small breaking parameter.

Points of ERS are motivated for phenomenological reasons, offering an explanation for the smallness of the reactor mixing angle $\theta_{13}$~\cite{Feruglio:2013hia} and enhancing the yield of leptogenesis. 
It also leads to \emph{some} of the RH neutrinos to have lifetimes much longer than the naive expectation of freeze-out leptogenesis. 

In this scenario, it leads to a very long-lived RH neutrino $N_3$ that could be detected at MATHUSLA~\cite{Chou:2016lxi}, while $N_{1,2}$
 can be searched for via either prompt or displaced vertex signals at the LHC~\cite{Deppisch:2013cya,Helo:2013esa}.

{\bf Framework}. A key feature of this scenario is the presence of a flavour $G_f$ and a CP symmetry. Both symmetries are
broken non-trivially to residual symmetries $G_\nu$ and $G_e$  in the neutrino and charged
lepton sector, respectively.
We choose in the following $G_f=\Delta (3 \, n^2)$~\cite{Luhn:2007uq} or $G_f=\Delta (6 \, n^2)$~\cite{Escobar:2008vc} ($n$ even, $3 \nmid n$, $4 \nmid n$).
These groups are all subgroups of $SU(3)$ and allow the three generations of leptons to be unified in one representation ${\bf 3}$ for $n \geq 2$, a hypothesis which is supported by the
fact that two of the three lepton mixing angles are large. CP is an additional symmetry of this scenario as part of $G_\nu$, since in this way low- and
high-energy CP phases, of Dirac as well as Majorana type, can be predicted.

 Left-handed (LH) lepton doublets $l_\alpha$ ($\alpha=e, \, \mu, \, \tau$)
transform in an irreducible faithful representation ${\bf 3}$, $N_i$
are in an irreducible real representation ${\bf 3^\prime}$,
whereas RH charged leptons $\alpha_R$ are assigned to ${\bf 1}$ of $G_f$. The latter are distinguished by an auxiliary symmetry
$Z_3^{\rm (aux)}$, under which $l_\alpha$ and $N_i$ are invariant. The CP symmetry is given
by the CP transformation $X (s) ({\bf r})$ in the representation ${\bf r}$ and depends on the integer parameter $s$,  $0 \leq s \leq n-1$,
(see Case 1 in Ref.~\cite{Hagedorn:2014wha}).
The forms of the neutrino Dirac mass matrix  $m_D$ and the charged lepton mass matrix $m_l$ are determined by the residual symmetries
$G_\nu=Z_2 \times CP$
and $G_e=Z_3^{\rm (D)}$ (the diagonal subgroup of $Z_3$ in $G_f$
and $Z_3^{\rm (aux)}$), respectively.
The Majorana mass matrix $M_M$ of RH neutrinos is invariant
under $G_f$ and CP.

The matrix $m_D$
can be written as~\cite{Hagedorn:2016lva}
\begin{equation}
m_D  = v\,\,  \Omega(s) ({\bf 3}) \, R_{13} (\vartheta_L) \, \left(
\begin{array}{ccc}
 y_1 & 0 & 0\\
 0 & y_2 & 0\\
 0 & 0 & y_3
 \end{array}
 \right) \, R_{13} (-\vartheta_R) \,  \Omega(s) ({\bf 3^\prime})^\dagger\,,
\end{equation}
where the unitary matrices $\Omega (s) ({\bf r})$
are determined by the CP transformation $X (s) ({\bf r})$
and $R_{13} (\vartheta)$ is a rotation in the (13)-plane through the angle $\vartheta$. Note that here we use the convention that
 $v \approx 174 \; \mbox{GeV}$ is the vacuum expectation value (VEV) of the Standard Model (SM) Higgs.
There are five real parameters:~$y_i$, $\vartheta_L$ and $\vartheta_R$, in $m_D$.
 The charged lepton mass matrix $m_l$ is diagonal with three undetermined entries corresponding to the charged lepton masses.
The Majorana mass matrix $M_M$ is of the form
\begin{equation}
\label{formLOMR}
M_M = M \, \left(
\begin{array}{ccc}
 1 & 0 & 0\\
 0 & 0 & 1\\
 0 & 1 & 0
 \end{array}
\right) \, ,
\end{equation}
and thus features three RH neutrinos with degenerate masses.
We consider the two cases of strong normal (NO) and inverted ordering (IO):
$(a)$ strong NO arises for $y_1=0$ so that $m_1$ vanishes, $m_2 = y_2^2 \, v^2/M$
and $m_3= y_3^2 \, |\cos 2 \, \vartheta_R| \, v^2/M$, while
$(b)$ strong IO arises for $y_3=0$ so that $m_3=0$, $m_1= y_1^2 \, |\cos 2 \, \vartheta_R| \, v^2/M$
and $m_2= y_2^2 \, v^2/M$.
The two non-vanishing Yukawa couplings are fitted to the solar and the atmospheric mass squared
differences $\Delta m_{\mathrm{sol}}^2$ and $\Delta m_{\mathrm{atm}}^2$ whose best fit values are taken from Ref.~\cite{Esteban:2016qun}. 

For particular values of $\vartheta_L$ and $\vartheta_R$, the residual symmetry $G_\nu=Z_2 \times CP$ can be enhanced.
 If $\vartheta_L=0, \pi$, the combination $m_D m_D^\dagger$ becomes invariant under a further $Z_2$ subgroup of $G_f$.
Similarly, for the choices $\vartheta_R=0, \pi/2, \pi, 3 \pi/2$ the combination $m_D^\dagger m_D$ preserves a symmetry larger than $G_\nu$.
 This symmetry is also larger than the one of $m_D m_D^\dagger$ for $\vartheta_L=0, \pi$, since RH neutrinos transform as the real representation ${\bf 3^\prime}$ of $G_f$
that is unfaithful for $n > 2$.

These points of ERS are of particular relevance for phenomenology, since $\vartheta_L$ deviating
from $\vartheta_{L,0} =$ $0$ or $\pi$ leads to a non-zero value of the reactor mixing angle $\theta_{13}$. $\vartheta_R$ close to $\vartheta_{R,0} =$ $0$, $\pi/2$, $\pi$ or $3\pi/2$ makes it possible for
the RH neutrino $N_3$ to be long-lived enough for being detected with the MATHUSLA detector (see Eq.~\ref{GammaNi} and Fig.~\ref{fig:length}), while simultaneously maximizing the CP
asymmetries $\epsilon_{i \alpha}$ relevant for leptogenesis (see Eqs.~\ref{eps1NO} and \ref{eps1IO}). One can argue that the larger the ERS is, the smaller the deviation
from points of ERS  will be, i.e.~$\vartheta_R$ is expected to deviate from $\vartheta_{R,0}$ by
\be
\delta\vartheta_R=|\vartheta_R-\vartheta_{R,0}|\lesssim 0.01 ~~,
\label{eq:eqsplitR}
\ee
while $\vartheta_L$ can deviate from $\vartheta_{L,0}$ up to
\be
\delta\vartheta_L=|\vartheta_L-\vartheta_{L,0}|\sim 0.2 ~~.
\label{eq:eqsplitL}
\ee
These splittings will thus determine the resulting phenomenology.

In one type of explicit model~\cite{King:2013eh}, the flavour and CP symmetry are spontaneously broken to the residual symmetries $G_\nu$ and $G_e$
with the help of flavour symmetry breaking fields and a peculiar alignment of their VEVs, achieved with a potential with a particular form. Depending on the fields and the form of the potential, an ERS larger than $G_\nu$ and $G_e$
 can be preserved at leading order. Higher-dimensional operators then induce small
deviations from these points of ERS, thus explaining the particular sizes of $\vartheta_L$ and $\vartheta_R$.
 An example can be found in Ref.~\cite{Feruglio:2013hia}, where the correct size of $\vartheta_L$ and thus the reactor mixing angle $\theta_{13}$ is generated in this way.

Higher-dimensional operators connecting different sectors of the theory
are responsible for the eventual breaking of the residual symmetries $G_\nu$ and $G_e$ and thus
 affect the given form of $m_D$, $m_l$ and $M_M$. In particular, they are also the source of
corrections leading to a small splitting in the RH neutrino masses. This splitting is crucial for resonant leptogenesis. In the following,
 we focus on contributions to $M_M$ that possess the residual
symmetry $G_e$. These are proportional to $\kappa$, a positive power of the symmetry breaking parameter, measured in units of $M$.
  A small splitting of the RH neutrino masses therefore arises:
\begin{equation}
\label{eq:massesNi}
M_1= M\, ( 1 + 2 \, \kappa) \;\; \mbox{and} \;\; M_2=M_3= M \, (1-\kappa) \, .
\end{equation}

{\bf Lepton Mixing and Low-Energy CP Phases}.
In the limit of residual symmetries $G_\nu$ and $G_e$, we obtain that
 the lepton mixing angles can be accommodated well for $\vartheta_L \approx 0.18 \, (2.96)$~\cite{Hagedorn:2014wha}, i.e.
$\sin^2 \theta_{13}\approx 0.0219$,  $\sin^2 \theta_{12}\approx 0.341$ and $\sin^2 \theta_{23} \approx  0.605 \, (0.395)$.
Regarding the two physical CP phases in the cases of strong NO and IO, we find that the Dirac phase $\delta$ is trivial,
 whereas the Majorana phase $\alpha_2$ depends on the chosen CP transformation $X (s)$
 \begin{equation}
\label{case1sina}
\sin\alpha_2= (-1)^{k+r+s} \, \sin 6 \, \phi_s \;\;\; \mbox{and} \;\;\; \cos\alpha_2= (-1)^{k+r+s+1} \, \cos 6 \, \phi_s \;, \; \mbox{with} \; \phi_s=\frac{\pi \, s}{n} \,  ,
\end{equation}
where $k=0$ ($k=1$) for $\cos 2 \, \vartheta_R > 0$ ($ \cos 2 \, \vartheta_R < 0$)
and $r=0$ ($r=1$) for strong NO (IO).
 The value of the effective Majorana neutrino mass $m_{\beta\beta}$, accessible in neutrinoless double beta decay experiments,
 crucially depends on the choice of the CP symmetry and is in this scenario considerably restricted~\cite{Hagedorn:2016lva}.
 For $n=26$, $\vartheta_L \approx 0.18$ and strong NO with $k=1$,
we get $0.0019 \, \mathrm{eV} \lesssim m_{\beta\beta} \lesssim 0.0040 \, \mathrm{eV}$,
while for strong IO with $k=0$,
 we find $0.016 \, \mathrm{eV} \lesssim m_{\beta\beta} \lesssim 0.048 \, \mathrm{eV}$,
using the best fit values for $\Delta m_{\mathrm{sol}}^2$ and $\Delta m_{\mathrm{atm}}^2$~\cite{Esteban:2016qun}.
For strong IO, most of the admitted values of $m_{\beta\beta}$ can be tested with the proposed experiment LEGEND~\cite{Abgrall:2017syy}
and all of them can be explored with nEXO~\cite{Albert:2017hjq}, whereas it is challenging to test the values of $m_{\beta\beta}$ predicted for strong NO
 with current and future experiments.

{\bf High-Energy CP Phases and the Leptogenesis Connection}.
 Including the small mass splitting of the RH neutrinos,
 their out-of-equilibrium decays can generate the baryon asymmetry, $\eta_B$, via resonant leptogenesis~\cite{Pilaftsis:1997jf, Pilaftsis:2003gt}. The CP asymmetries $\epsilon_{i \alpha}$
due to the decay of $N_i$ and in the lepton flavour $\alpha$ read
\begin{equation} \label{epsilonia}
\epsilon_{i \alpha} \approx \frac{1}{v^4} \sum_{j \neq i}   \mathrm{Im} \left( \hat{m}^\ast_{D, \alpha i} \hat{m}_{D, \alpha j} \right) \, \mathrm{Re} \left( \left(\hat{m}_D^\dagger \hat{m}_D \right)_{i j} \right)  \, \mathcal{F}_{ij} \, ,
\end{equation}
with $\hat{m}_D$ being $m_D$ in the RH neutrino mass basis and
 $\mathcal{F}_{ij}$ related to the regulator that is proportional to the mass splitting of $N_i$~\cite{Dev:2014laa}.

We find the real part of $(\hat{m}_D^\dagger \hat{m}_D)_{i j}$ to be zero, if either $i=3$ or $j=3$. Hence, $\epsilon_{3 \alpha}=0$
for all $\alpha$ and $\epsilon_{i  \alpha}$ only has one contribution for $i=1,2$. The imaginary part of $\hat{m}^\ast_{D, \alpha 1} \hat{m}_{D, \alpha 2}$
is proportional to $\sin 3 \, \phi_s$ for even $s$ and to $-\cos 3 \, \phi_s$ for odd $s$, independent of the flavour $\alpha$. If $\alpha$ is summed over, $\epsilon_1$ and $\epsilon_2$ both vanish.
 For strong NO and even $s$, the CP asymmetries $\epsilon_{1 \alpha}$ read
\begin{equation}
\label{eps1NO}
\epsilon_{1 \alpha} \approx \frac{y_2 \, y_3}{9} \, (-2 \, y_2^2+y_3^2 \, (1- \cos 2 \, \vartheta_R)) \, \sin 3 \, \phi_s \, \sin\vartheta_R \, \sin\vartheta_{L, \alpha} \, \mathcal{F}_{12} \, ,
\end{equation}
and for strong IO, we find
\begin{equation}
\label{eps1IO}
\epsilon_{1 \alpha} \approx \frac{y_1 \, y_2}{9} \, (-2 \, y_2^2+y_1^2 \, (1+ \cos 2 \, \vartheta_R)) \, \sin 3 \, \phi_s \, \cos\vartheta_R \, \cos\vartheta_{L, \alpha} \, \mathcal{F}_{12} \, ,
\end{equation}
with $\vartheta_{L, \alpha} = \vartheta_L + \rho_\alpha \, 4 \pi/3$ and $\rho_e=0$, $\rho_\mu=1$, $\rho_\tau=-1$.
For strong NO (IO) $\epsilon_{i \alpha}$ becomes very small, if $\vartheta_R \approx 0 , \, \pi$ ($\vartheta_R\approx\pi/2,3\pi/2$). In addition, $\mathcal{F}_{ij}$
vanishes for $\cos 2 \, \vartheta_R=0$.
The CP asymmetries $\epsilon_{2 \alpha}$ are the negative of $\epsilon_{1 \alpha}$ with $\mathcal{F}_{12}$ being replaced by $\mathcal{F}_{21}$.
We note that different values of $s$ can lead to the same value of $\epsilon_{i \alpha}$. In particular, we find
\begin{equation}
\label{eps_srel}
\epsilon_{i\alpha} (s) = (-1)^s \, \epsilon_{i\alpha} (n-s) = \epsilon_{i\alpha} (n/2-s) = (-1)^{s+1}\, \epsilon_{i\alpha} (n/2+s) \;\; \mbox{for} \;\; s \leq n/2 \, .
\end{equation}
Eqs.~\ref{case1sina}, \ref{eps1NO} and \ref{eps1IO} show the close correlation between CP violation at low and high energies.

{\bf Decay Lengths of RH Neutrinos}.
The decay widths $\Gamma_i \approx M_i \, (\hat{m}_D^\dagger \, \hat{m}_D)_{ii}/(8 \, \pi \, v^2)$
of the RH neutrinos $N_i$ are
\small
\begin{eqnarray}
\label{GammaNi}
\Gamma_1&\approx& \frac{M}{24 \, \pi} \, \left( 2\, y_1^2 \, \cos^2 \vartheta_R + y_2^2 + 2\, y_3^2 \, \sin^2 \vartheta_R \right)  \, , \,\,
\Gamma_2 \approx  \frac{M}{24 \, \pi} \, \left( y_1^2 \, \cos^2\vartheta_R + 2 \, y_2^2 + y_3^2\, \sin^2 \vartheta_R  \right) \, , \;  \nonumber \\
\Gamma_3 & \approx & \frac{M}{8 \, \pi} \, \left( y_1^2 \, \sin^2 \vartheta_R + y_3^2 \, \cos^2 \vartheta_R \right) \, .
\end{eqnarray}
\normalsize
For $M$ in the few hundred GeV range,
we expect $y_i \sim 10^{-7}$ and thus mostly non-prompt decays at the LHC.
%
\begin{figure}[t!]
\centering
\includegraphics[width=0.5\textwidth]{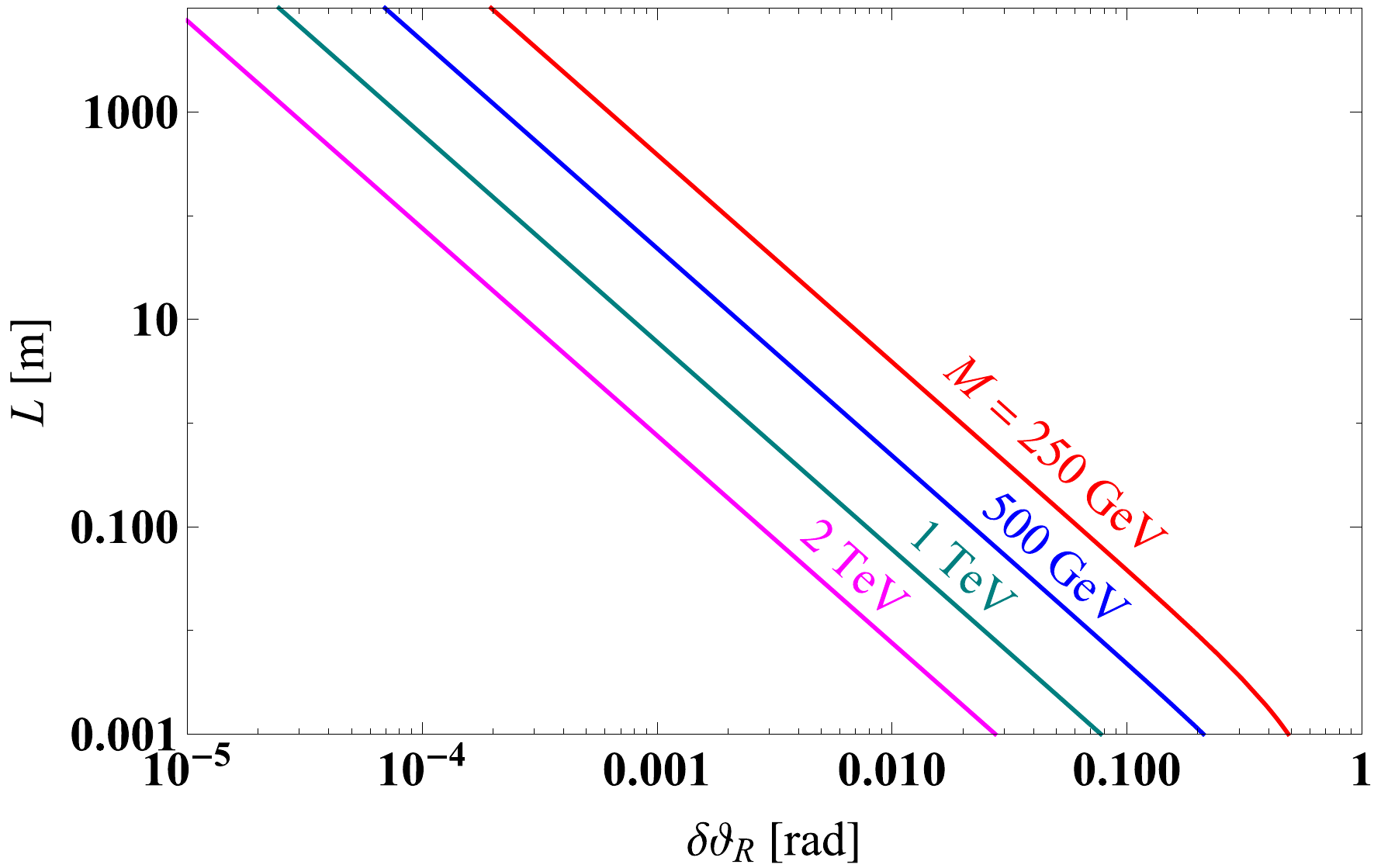}
\caption{{\small Decay length $L$ of $N_{3}$ (in meters) in the laboratory frame as a function of $\delta \vartheta_R$ (defined in Eq.~\ref{eq:eqsplitR}) for different values of $M$ assuming production at the LHC in the decay of a 4 TeV parent particle.}
}
\label{fig:length}
\end{figure}
%
If $\vartheta_R \approx \pi/2, \, 3 \pi/2$ (for strong NO) or $\vartheta_R \approx 0, \, \pi$ (for strong IO), i.e. $\vartheta_R$ close to points of ERS, $N_3$ can have a very long lifetime, since $\Gamma_3$ tends to zero.
 Thus, $N_3$ could be searched for with the MATHUSLA detector, if it is produced at the LHC with sufficient cross-section. This is shown in Fig.~\ref{fig:length} where we plot the decay length $L$ of $N_3$ in the
 laboratory frame as a function of the deviation of $\vartheta_R$ from points of ERS for different values of the mass of $N_3$.
 In doing so, we assume that $N_i$ are produced in the decay of a parent particle with mass $m_{parent} = 4 \tev$, corresponding to an average Lorentz boost factor of $\gamma=m_{parent}/(2M)$.
  For $M$ in the few hundred GeV to TeV range and $10^{-4} \lesssim \delta\vartheta_R \lesssim 10^{-2}$, $N_3$
would decay, on average, within the MATHUSLA detector.
 If $10^{-3} \lesssim \delta \vartheta_R \lesssim 10^{-1}$, $N_3$ would decay on average within the LHC detectors, along with $N_{1,2}$ decays, where the latter giving rise to either prompt or displaced vertex signals at the LHC, depending on the choice of $\vartheta_R$.

As these weak-scale right-handed neutrinos are not produced efficiently through the minimal interactions required for generating neutrino masses, to observe $N_i$ decays at colliders, an efficient production mechanism is required.  As in previous sections, there are a number of possibilities for additional UV  states to be produced at colliders, which subsequently decay to final states including $N_i$'s.
To mention just one example, one might consider theories related to the $B-L$ extensions discussed in the previous sections, where the RH neutrinos could be produced in the decay of a TeV-scale $Z'$. 
This could produce these weak-scale LLPs in sufficient numbers for detection. 
In some models this may allow the region of parameter space relevant for leptogenesis to be probed.

{\bf Summary.} We have presented a type-I seesaw scenario with a flavour and CP symmetry as well as three RH neutrinos with almost degenerate masses in the few hundred GeV to TeV
range. This class of models can be connected to the observed baryon asymmetry through resonant leptogenesis.  One of the RH neutrinos can be long-lived enough to be discovered at the MATHUSLA detector if the production cross-section is sufficent. The other two can be searched for at the LHC main detectors. This would allow both detectors working in conjunction to diagnose the mechanism of leptogenesis generating the baryon asymmetry of our universe.

\clearpage


\section{Theory Motivation for LLPs: Bottom-Up Considerations}
\label{sec:bottomup}

In this section we study a variety of well-defined, generic
possibilities for physics beyond the Standard Model, and demonstrate
that in many circumstances they naturally yield interesting LLP
signals for MATHUSLA.  The scenarios we study here represent plausible and
consistent possibilities for physics beyond the SM, which  can and
should be studied independently of particular `top-down' theoretical
motivations for specific forms of new physics.  We emphasize that this
section is highly complementary to the previous sections.  Two of the
main topics covered here are hidden valleys (Section~\ref{sec:hiddenvalleys}) and exotic Higgs decays (Section~\ref{sec:exhdecays} and \ref{sec:bbn}), 
and in both cases specific realizations of these general scenarios
have appeared in multiple contexts in the previous sections.  
We also discuss MATHUSLA's sensitivity to the minimal extensions of the
SM with (1) a Higgs portal-mixed singlet scalar (`SM$+$S', Section~\ref{sec:singlets}); (2) a massive
Abelian dark vector boson kinetically mixing with hypercharge
(`SM$+$V', Section~\ref{sec:darkphotons}); or (3) an axion-like particle (ALP, Section~\ref{sec:alp}).  
All of the above simple models are well-motivated from effective field theory considerations, and are, for example, realized in theories of Dark Matter (Section~\ref{sec:simps}) or Neutral Naturalness (Section~\ref{sec:neutralnaturalness}).
The results of this section thus apply to a broad range of well-motivated theories of physics
beyond the SM, and illuminate the essential features that make MATHUSLA a uniquely sensitive instrument for discovering SM-singlet LLPs in all of them.

\subsection[Hidden Valleys and High Multiplicity Scenarios]{Hidden Valleys and High Multiplicity Scenarios \footnote{Simon Knapen, Dean Robinson, Daniel Stolarski}}
\label{sec:hiddenvalleys}

\subsubsection{Motivation}
In this section, we consider hidden valley (HV) models~\cite{Strassler:2006im,Strassler:2006ri} that give rise to high multiplicity final states. HVs are a class of models where there are relatively light states whose only coupling to Standard Model (SM) states is via a heavy mediator. This setup naturally allows for long lifetimes, while maintaining a sizable production cross-section at the LHC. The hidden valley framework is very general, and appears in many well-motivated scenarios. 
It also arises naturally from the structure of gauge theories,  which makes disconnected  sectors a straightforward possibility. 
 For this reason, HV models are discussed throughout this report, including 
Section~\ref{sec:neutralnaturalness} on neutral naturalness and Section~\ref{sec:simps} on SIMPs/ELDERs.

In this section, we consider HVs with a confining gauge group so that showering and hardonization will lead to large particle multiplicity when the HV is accessed. Our only theory prior will be to assume a hadronization scale in the $\sim$ GeV range, as well as a much heavier mediator which can be accessed at the LHC. We will show how the phenomenology depends on the strength of the hidden sector coupling by considering two limiting cases.

\subsubsection{Hidden sector fragmentation and hadronization}

The phenomenology of confining hidden valleys most notably depends on the strength of the hidden sector gauge coupling. In particular, if the `t Hooft coupling is relatively small, the fragmentation process is dominated by soft and collinear branchings, as these  are typically enhanced by large logarithms. The result is a fairly collimated spray of particles with sizable hierarchies in their momentum distribution, where the average particle multiplicity scales as a powerlaw of the ratio of the hard scattering scale $Q$ and the hadronizations scale $\Lambda$. Specifically, one can show~\cite{Ellis:1991qj}
\begin{equation}
	\label{eqn:HVscaling}
\langle n(Q)\rangle \propto \left(\frac{Q}{\Lambda}\right)^{2\gamma_T(j)}\bigg|_{j = 1}\,.
\end{equation}
where $\gamma_T(j)<1/2$ the timelike (fragmentation) anomalous dimension. After hadronization, the dark hadrons decay to SM fields, but because the mediator is heavier, the decay length is often macroscopic and can be quite long. Therefore a jet of dark hadrons becomes an ordinary jet at long distances. The phenomenology of these `emerging jets' was studied~\cite{Schwaller:2015gea} using benchmarks motivated by~\cite{Bai:2013xga}. 

On the other hand, in the regime where `t Hooft coupling is large, the fragmentation process is very efficient regardless of the phase space configuration of the branchings. On average one therefore expects isotropic branchings with more or less equal energy sharing between the partons.  Moreover, if the `t~Hooft coupling remains large over a large energy window without triggering a mass gap, as can be the case in walking or quasi-conformal hidden sectors,  a high multiplicity of hidden sector hadrons are produced, with momenta of the order of the hadronization scale.  In this case the average multiplicity scales as 
\begin{equation}
	\label{eq:multiplscaling}
	\langle n(Q) \rangle \propto \frac{Q}{\Lambda}\,.
\end{equation}
Rather than a jet-like structure, the result of such efficient showering is an approximately spherically symmetric event, with an approximately democratic energy distribution \cite{Strassler:2008bv,Polchinski:2002jw,Hofman:2008ar,Lin:2007fa}. This topology is referred to as SUEP (Soft Unclustered Energy Patterns). The energy distributions of the final states can be regarded as almost thermal, parametrized by an effective Hagedorn temperature~\cite{Hagedorn:1965st}, $T \sim \Lambda$, a feature which is also borne out by AdS/CFT calculations \cite{Hatta:2008qx}. The case for which the hidden hadrons decay promptly poses significant trigger challenges at ATLAS and CMS, as was studied in \cite{Knapen:2016hky}.

The low-lying spectrum of hadrons can be very rich within an HV sector. Depending on the number of light and heavy HV quark flavors, as well as their HV charges, the spectrum may typically involve the analogues of glueballs, onium states, baryons, $\eta$' and lighter pions. In order to capture the leading features of a plausible hadronization model, we assume the low lying spectrum contains only a single flavor of a long-lived (pseudo)scalar $\phi$, with $m_{\phi} \sim \Lambda$. 
MATHUSLA's ability to reconstruct DVs for LLPs with masses below a GeV is likely to depend on the production mode and details of the detector design, see Section~\ref{s.energythresholds}. 
Anticipating that the low lying HV hadron $\phi$ will decay either to two or four SM states,  we therefore choose $\Lambda \simeq m_{\phi} = 1$~GeV for the hadronization scale and scalar hadron mass, in order to maximize the multiplicity of detectable final state particles. Assuming that detector efficiencies turn-on sharply at this detection threshold, this maximized multiplicity benchmark corresponds to the best-case scenario for detection.

The $\phi$ proper lifetime should fall between $\sim 10$ and $\sim 10^7$ meters in order to have the majority of $\phi$'s decay outside of ATLAS and CMS,
and avoid potential BBN constraints, respectively. Since $\Lambda \sim m_\phi = 1$~GeV, possible decay products include SM leptons, pions or photons. In the rest of this work we take the lifetime to be a free parameter.

For the MATHUSLA detector, the phenomenology between both limiting cases differs in the following ways:
\begin{itemize}
	\item For a similar confinement scale, the particle multiplicity for SUEP-like dynamics is higher than for jet-like dynamics. In both cases, the multiplicity is high enough such that two or more vertices can simultaneously occur in the detector, despite the relatively low geometric acceptance.
	\item For SUEP-like events, the multiplicity scales linearly with hard scattering scale, but the momentum distribution of the final state particles is determined only by the SUEP temperature. 
	For jet-like events, increasing the hard scattering scale increases the boost of the particles, while the multiplicity of the events scales as a sublinear power law, as encoded in the anomalous dimension Eq.~(\ref{eqn:HVscaling}).
\end{itemize}
In what follows, we discuss both cases separately.

\subsubsection{Emerging Jets}

Inspired by~\cite{Bai:2013xga,Schwaller:2015gea} we consider a QCD-like dark sector with $SU(N_d)$ confining gauge group and $N_f$ flavours. It has a confinement scale near the QCD scale. The mediator to the dark sector is a heavy scalar which is a bifundamental under both QCD and dark-QCD. The mediator, $X_d$, has a Yukawa coupling to quarks (q) and dark quarks (Q):
\begin{equation}
\kappa X_d \bar{Q} q + {\rm h.c}.
\label{eq:yuk}
\end{equation}
Here we have suppressed flavour indices in both the SM and dark sectors, but we assume that the coupling of the $X_d$ is dominantly to light flavours on the SM side. The production process for the mediator is shown in Fig.~\ref{fig:Feyn}. The mediator is pair produced via its QCD interaction, and the production cross-section is that of a scalar top from supersymmetry~\cite{Borschensky:2014cia} times $N_d$. In this work we take $N_d=3$, and we take two benchmark masses for $X_d$, $M = 1000,$ 1500 GeV. 

\begin{figure}[t]
\begin{center}
\includegraphics[width=0.45\textwidth]{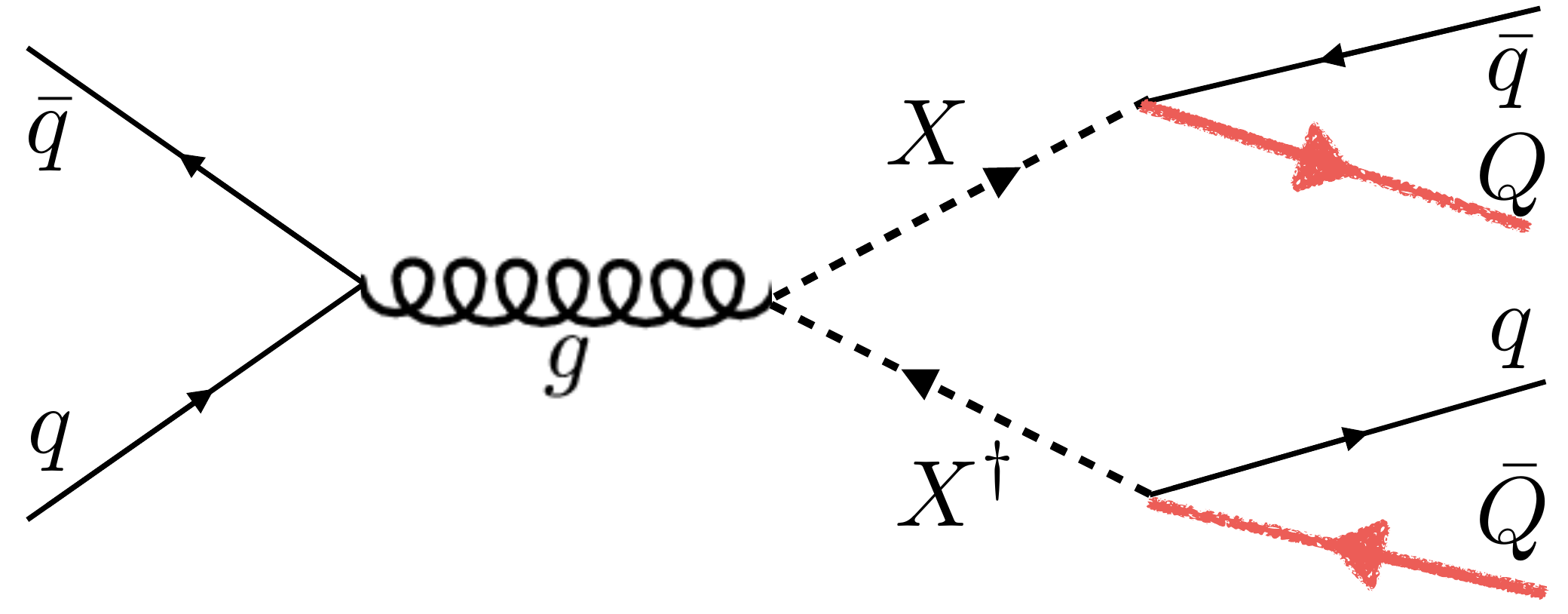}
\end{center}
\caption{Feynman diagram for one of the production processes for the emerging jet scenario at the LHC. The $X$ scalar is pair-produced, and each $X$ decays to an ordinary quark and a dark quark (Q, shown by the thick red line). This leads to events with two QCD jets and two emerging jets.  
\label{fig:Feyn}}
\end{figure}

Each mediator decays to one quark and one dark quark via the Yukawa coupling of Eq.~(\ref{eq:yuk}). The quark and dark quark then shower and hadronize, so each event contains two ordinary jets from the quarks and two emerging jets from the dark quarks. As mentioned above, we take the simplifying assumption that the emerging jets contain only a single species of dark hadron whose mass is 1 GeV.  The dark pions then decay to Standard Model quarks via a virtual heavy mediator. The fact that the mediator is much heavier than the dark hadrons gives the dark pions a naturally long lifetime. One can calculate the width of the dark hadron assuming it is a pion-like state using dark chiral perturbation theory\footnote{This computation assumes that the SM final states can be treated as free quarks, so for dark hadrons with masses not too far above the QCD scale, there will be large corrections from strong QCD effects, but these will not change the qualitative picture.}~\cite{Bai:2013xga}:
\begin{equation}
\Gamma( \phi \rightarrow \bar{q}q) = \frac{\kappa^4 N_c f_{\phi}^2 m_q^2 m_{\phi}}{32 \pi M_{X_d}^4}
\end{equation}
where $f_\phi$ is the dark hadron decay constant and $m_q$ is the mass of the Standard Model quark that participates in the decay. From this equation, we can get the lifetime
\begin{equation}
c\tau_0 = \frac{c\hbar}{\Gamma} \approx 300 \,{\rm m} \times \left(\frac{0.3}{\kappa}\right)^4
\times\left(\frac{1\, {\rm GeV}}{f_{\phi}}\right)^2
\times\left(\frac{5\, {\rm MeV}}{m_{q}}\right)^2
\times\left(\frac{1\, {\rm GeV}}{m_{\phi}}\right)
\times\left(\frac{M_{X_d}}{1\, {\rm TeV}}\right)^4,
\end{equation}
and we see that for the parameters chosen here, the dark pions will be quite long-lived for ATLAS and CMS, but in right in the ballpark for MATHUSLA. The lifetime is very sensitive to the Yukawa coupling $\kappa$, which is unknown and can vary widely, so we see that very short and very long lifetimes are possible.

We simulate emerging jets events at the 13 TeV LHC using the hidden valley implementation~\cite{Carloni:2010tw,Carloni:2011kk} of Pythia 8~\cite{Sjostrand:2007gs} with the implementation of gauge coupling running from~\cite{Schwaller:2015gea}. The typical number of dark hadrons in an event depends strongly on the dark hadron mass, but only weakly on the mediator mass (see Eq.~(\ref{eqn:HVscaling})). For the benchmark used here of a 1 GeV dark hadron and a 1 TeV mediator, the typical number of dark hadrons is $\sim100$, while for a 1.5 TeV mediator it is $\sim 120$. 

From the simulation we can estimate the fraction of emerging jets events which will have one or two dark pions decaying in MATHUSLA. This is shown in the left panel of Fig.~\ref{fig:ejsensitivity}, with the solid lines corresponding to one dark pion decay, and the dot-dashed line being two dark pion decays. We have used the squark production cross-section with $N_d=3$ and 3,000 fb$^{-1}$ integrated luminosity. All the results are assuming perfect efficiency for detection of dark hadron decays MATHUSLA, $\epsilon_{\rm DV}= 1$. This of course is an unrealistic assumption, but all results can simply be scaled by the actual value of $\epsilon_{\rm DV}$.

\begin{figure}[t]
\includegraphics[width=0.45\textwidth]{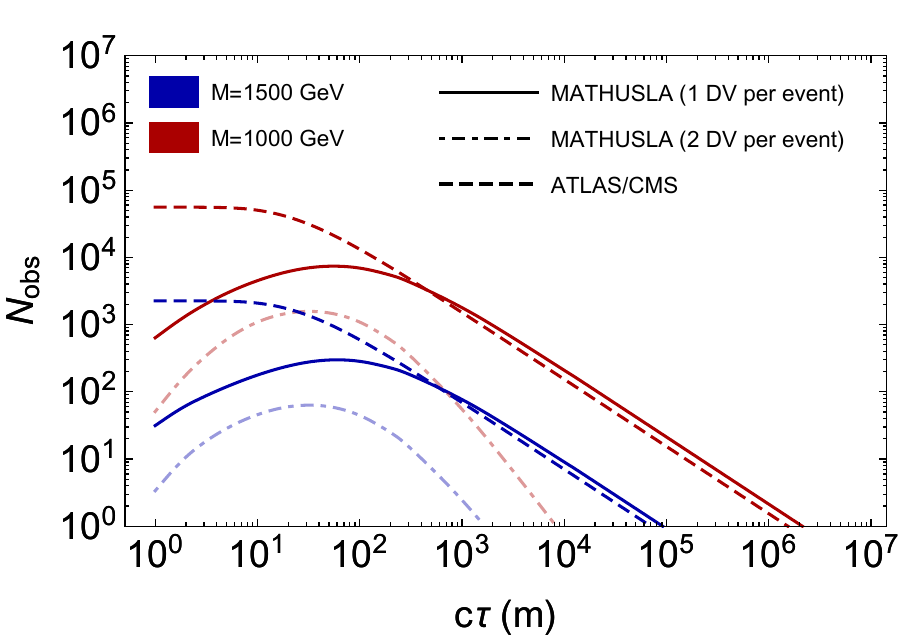}\hfill
\includegraphics[width=0.45\textwidth]{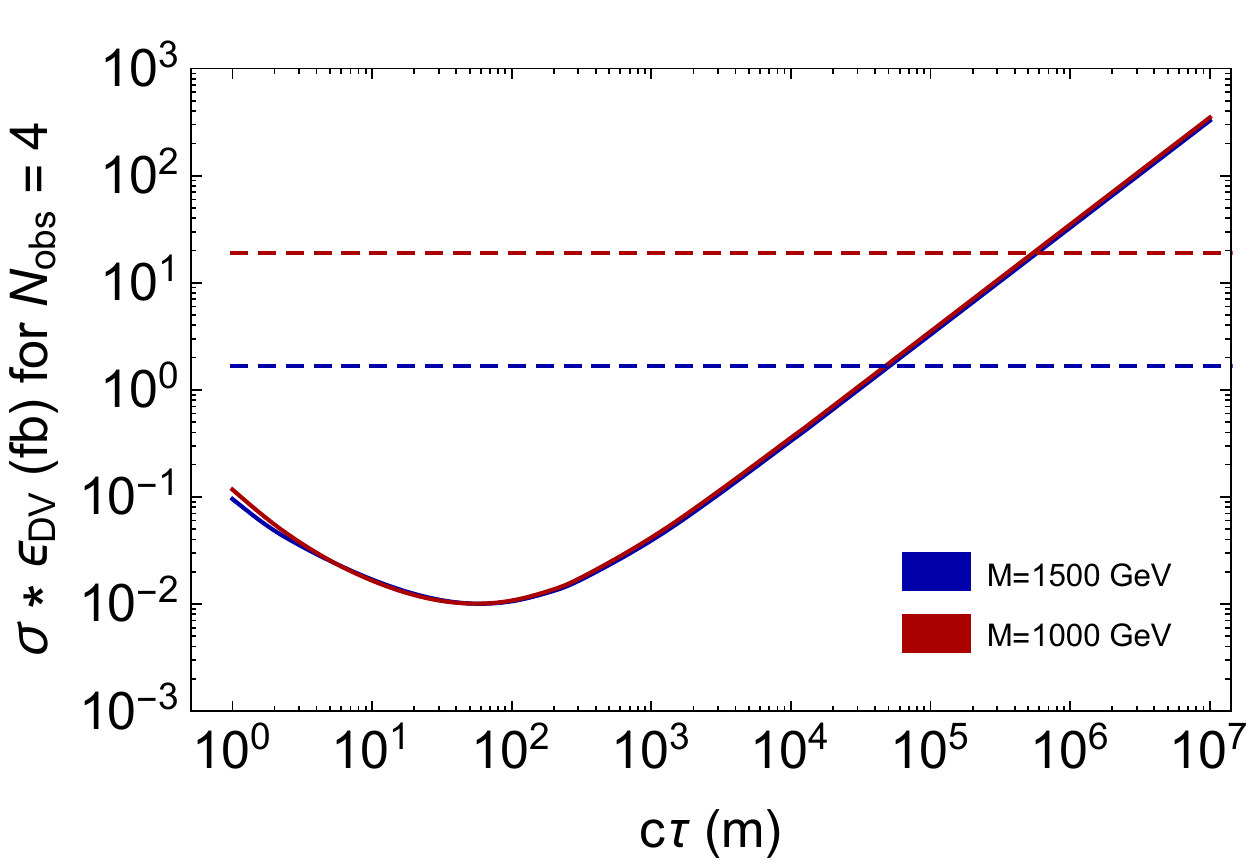}
\caption{Sensitivity plots for the emerging jets scenario with a mediator mass of 1 TeV (red) and 1.5 TeV (blue). 
 Left: Number of expected events in MATHUSLA with at least one (two) displaced vertices are shown in the solid dark (dashed light) lines for our two benchmark models. We also show the approximate number of expected decays in ATLAS or CMS tracker in the dashed lines. Right: MATHUSLA exclusion potential requiring four events with at least one displaced vertex (solid), compared to the projected jets plus missing energy limit in the long lifetime regime for the high luminosity LHC from~\cite{Cohen:2013xda} (dashed).  
 This estimate assumes the $200m \times 200m \times 20m$ benchmark geometry of Fig.~\ref{f.mathuslalayout}.
\label{fig:ejsensitivity}}
\end{figure}

We see that the optimum lifetime for MATHUSLA is $\sim$100 m, the distance it is from the interaction point, but that there are a few events in MATHUSLA for lifetimes as long as $10^6$ m. We can turn this around and ask what the limit that MATHUSLA can place on this model assuming it does not see any events. This is shown on the right panel of Fig.~\ref{fig:ejsensitivity}, and computed by asking what cross-section would give 4 events in the detector. We see that MATHUSLA can be sensitive to cross-sections as small as 0.01 fb with the full run of the LHC. This is significantly below the benchmark minimum signal cross-section Eqn.~\ref{e.MATHUSLAminxsec} due to the very high dark hadron multiplicity. We also note that in this plane our two benchmarks are quite similar, meaning that their difference in number of observed events is due almost entirely to the different cross-section. The two benchmarks are similar because the number of hadrons in an emerging jet scales as a very small power of the mediator mass as seen in Eq.~(\ref{eqn:HVscaling}), so the difference of 50\% in the mediator mass is barely visible on a log-log plot. This is to be contrasted with the SUEP scenario discussed in the next section where there is a much stronger dependance on the mediator mass. 

This scenario of physics beyond the SM could also be discovered or bounded with more traditional search strategies at ATLAS and CMS.  Because each event has two hard QCD jets, for sufficiently heavy mediators ($M_X \gtrsim 500$ GeV), these events become trivial trigger on. There are then two obvious ways to look for the models. The first is to search for displaced decays of the dark hadrons in ATLAS or CMS. For the light hadron masses considered here, decays in the calorimeter or muon system become difficult to resolve (see Section~\ref{s.LHCLLPcomparison} for more details). Therefore, we estimate the reach by requiring one event in tracker of ATLAS or CMS, and we very crudely model the tracker as a solid sphere of radius 1 meter. This estimate is shown as the dashed lines in the left panel of Fig.~\ref{fig:ejsensitivity}. We see that at short lifetimes, ATLAS and CMS see many more events, but for lifetimes longer than $\sim 100$ m, the MATHUSLA sensitivity becomes slightly better than that of current detectors. 
This ATLAS/CMS estimate assumes 100\% signal efficiency, but it is not clear how good the efficiency would actually be in the high pile-up environment of the LHC. 
In the short lifetime regime, many events will have multiple decays within the tracker, which can make the event appear more spectacular, but which could also degrade the tracking and detection efficiency. The estimate in Fig.~\ref{fig:ejsensitivity} also assumes zero background, but this is also probably not realistic, as there will be secondary interactions of hadrons in material, as well as decays of ordinary hadrons. We leave all these questions to future study, but here we note that the estimate for the ATLAS/CMS reach for searching for decays is a \textit{best case} estimate, and likely to be worse than what is given here.

For lifetimes $\gtrsim 10$ m, the majority of dark hadrons escape the detector and the emerging jets simply appear as missing energy in ATLAS and CMS. Therefore, the event topology for this model becomes jets and missing energy, an extremely well studied signature. The limits on this topology for the high luminosity LHC were studied in~\cite{Cohen:2013xda}\footnote{We thank Mike Hance for providing us with more detailed results.}, and we show them as the dashed lines in the right panel of Fig.~\ref{fig:ejsensitivity}. We see that for intermediate lifetimes MATHUSLA puts a significantly stronger limit than the jets plus missing energy search. Therefore, we see that MATHUSLA can extend the reach for this scenario significantly relative to jets plus missing energy, and is competitive or possibly superior to the search for displaced vertices within ATLAS and CMS depending on signal efficiencies.

\subsubsection{Soft Unclustered Energy Patterns Patterns (SUEP)}

Production of the SUEP event morphology can be encoded generally in an operator product of the form $M^{4- \Delta_{\text{vis}} - \Delta_{\text{HV}}}\mathcal{O}_{\text{vis}} \mathcal{O}_{\text{HV}}$, in which $\mathcal{O}_{\text{vis}}$ is an SM neutral operator consisting of either SM degrees of freedom or heavy exotic mediators. While SUEP production need not be associated with the production of an intermediate resonant state between the visible and HV sectors,  it is simpler and representative to focus on two cases, following the discussion in  \cite{Knapen:2016hky}:
\begin{itemize}
\item $\mathcal{O}_{\text{vis}} = S$, a heavy (pseudo)scalar field, that is singly produced by e.g. gluon fusion production, $S G_{\mu\nu} G^{\mu \nu}$ or $S G_{\mu\nu} \tilde{G}^{\mu \nu}$. For this scenario we consider two benchmarks, with scalar mass $M=750$ GeV and $M=400$ GeV.
\item  $\mathcal{O}_{\text{vis}} = h$, the SM-like Higgs scalar, produced by gluon fusion, vector boson fusion or associated production. The SUEP is then the result of an exotic Higgs decay into the strongly coupled hidden valley.
\end{itemize}
Other mediators are possible as well, and the phenomenology is largely independent of the choice of mediator. An important exception is the trigger efficiency by ATLAS and CMS, which greatly benefits from prompt, hard objects in the event, as discussed in the previous section. 
We assume a simplified fragmentation model, in which $\phi$ are produced spherically symmetrically, with a Maxwell-Boltzmann momentum distribution characterized by $T \sim \Lambda$. The multiplicity scales linearly with $\Lambda$, while for $\Lambda \sim T$ the typical boost remains roughly constant. This means that the sensitivity for different values of $\Lambda \sim T$ can be obtained by a simple rescaling.

\begin{figure}[t]
\includegraphics[width=0.45\textwidth]{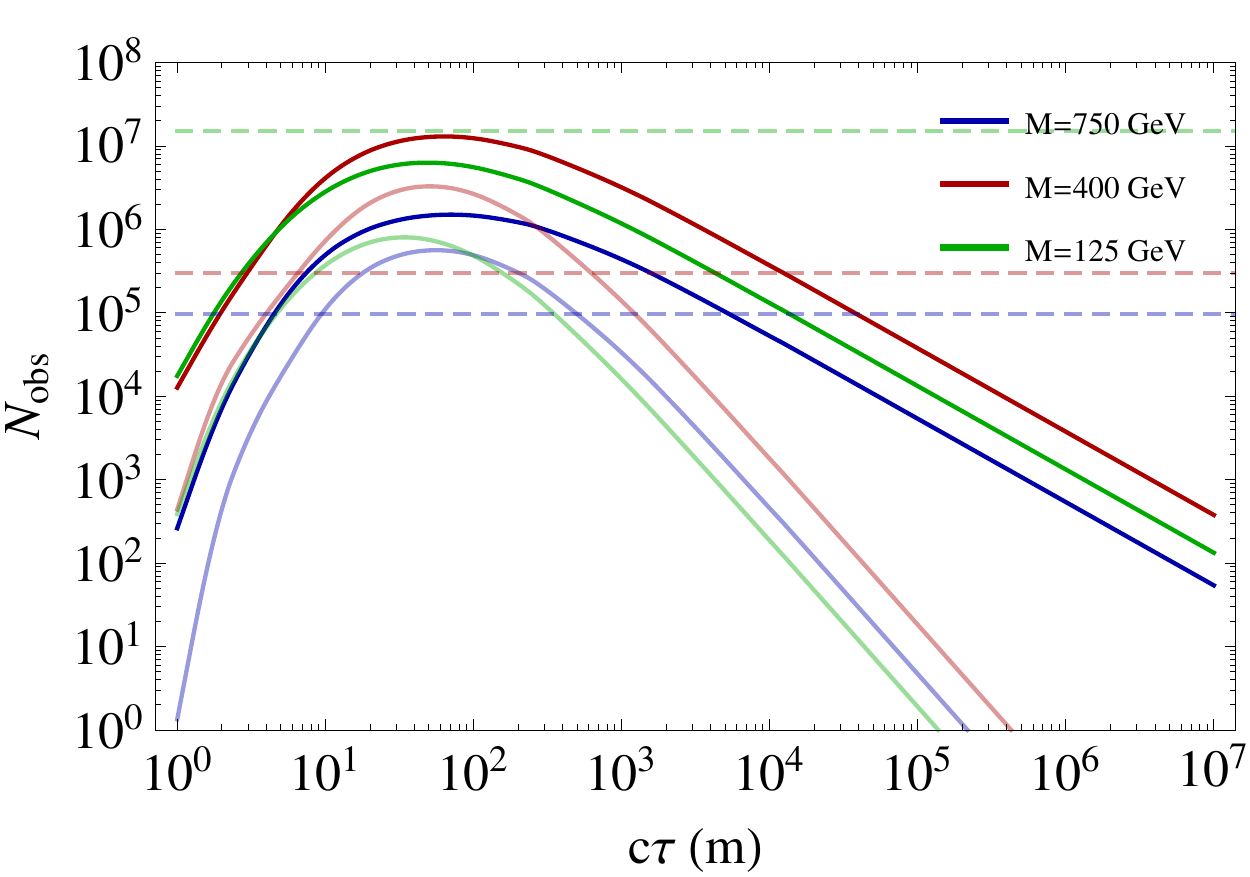}\hfill
\includegraphics[width=0.45\textwidth]{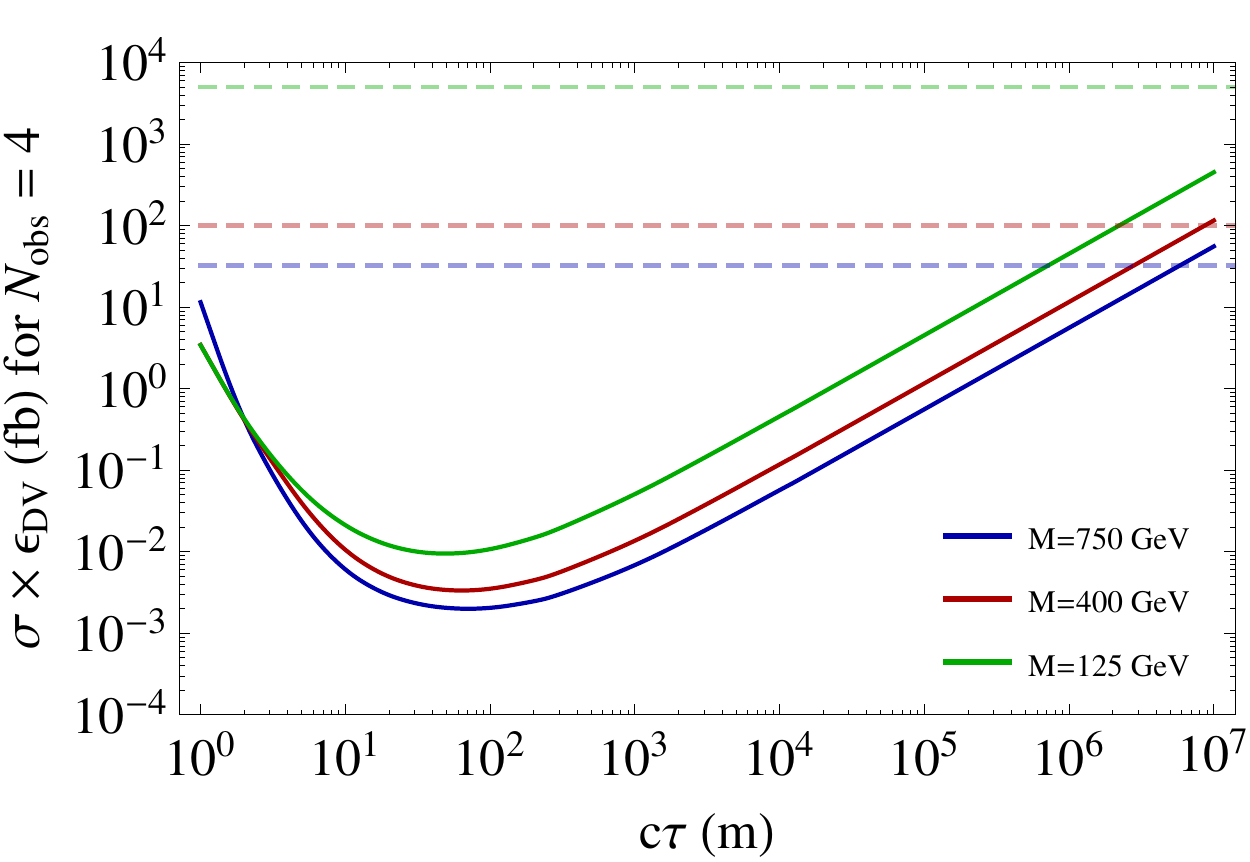}
\caption{Left: Number of expected events in MATHUSLA with at least one (two) vertices are shown in the dark (light) lines for our three benchmark models. For the $M=400$ GeV and $M=750$ GeV, we assumed a cross-section equal to that of a standard model Higgs boson with this mass \cite{Heinemeyer:2013tqa}. The 125 GeV curve assumes production in exotic Higgs decays with branching fraction 10\%. Also shown is the maximum number of events allowed by the expected mono-jet limits (dashed lines). Right: MATHUSLA exclusion potential (solid), compared to the projected mono-jet limits (dashed).  
This estimate assumes the $200m \times 200m \times 20m$ benchmark geometry of Fig.~\ref{f.mathuslalayout}.
\label{fig:bombsensitivity}}
\end{figure}

There are many possibilities for the decay of the long-lived HV particles $\phi$. 
Among many options for their decay, one might consider a kinetic mixing portal of the form $(\varepsilon/2)A'_{\mu\nu} B^{\mu\nu}$, with $B$ the hypercharge field strength, and $A'$ a hidden photon with mass $m_{A'} \ll m_Z$ that couples to $\phi$ via a chiral anomaly. 
Decays $\phi \to \ell \ell$ are not generated at tree level by this portal, but may occur at one-loop, assuming $m_{A'} > m_{\phi}$.  The tree-level double-Dalitz process $\phi \to 2A'^* \to 4\ell$ may also proceed. However, the lifetime for these processes typically far exceeds $1$\,s for $\varepsilon \lesssim 10^{-4}$ and  $m_{A'} \gtrsim 3$\, GeV, which corresponds to the projected reach of other experiments sensitive to hidden photons, such as Belle II. A simple alternative is to consider the regime $2m_{\ell,\pi} < m_{A'} < m_{\phi}/2$, such that the $\phi \to 2A'$ proceeds promptly via the chiral anomaly, followed by $A' \to \ell \ell$ or $\pi \pi$ and so on, with rate $\Gamma \sim m_{A'} \varepsilon^2 \alpha$.  The hidden photon parametric region $\varepsilon < 10^{-8}$ and  $1\,\text{MeV} < m_{A'} < 0.5$\,GeV is unconstrained~\cite{Hewett:2012ns}, and for example, for $\varepsilon \sim 10^{-8}$, this range of $A'$ masses produces lifetimes  $10^{-6}\,\text{s} \lesssim \tau \lesssim 10^{-4}$\,s and lifelengths $1\,\text{km} \lesssim \lambda \lesssim 10^3$\,km. In this scenario, rather than $\phi$, the long-lived $A'$ effectively generates an effectively weak SM-HV portal, such that  $\phi \to 2(A' \to \ell\ell)$ effectively  has a very long lifetime. Based on this simple example, we therefore pick $\phi \to 4\ell$ as our benchmark $\phi$ decay mode. 
For $m_{\phi} = 1 \gev$, the main detector signal could therefore be that of a displaced lepton-jet.
For such low LLP mass and assuming production in exotic Higgs decays (or another process without the guaranteed presence of conspicuous prompt objects to trigger on), this decay is actually the best-case scenario for a main detector LLP search. Even so, as discussed in  Section~\ref{s.LHCLLPcomparison}, there are likely to be significant challenges triggering on and reconstructing these DVs without backgrounds.
As such, MATHUSLA is likely to have significantly higher sensitivity than the main detectors in the long-lifetime limit. This advantage would be further compounded if the HV particles decayed dominantly into hadrons.

The resulting sensitivity for MATHUSLA as a function of the lifetime is shown in figure \ref{fig:bombsensitivity}, as compared to the 3000 $\text{fb}^{-1}$ projected jet+MET limits from ATLAS/CMS. For the $M=750$ GeV and $M=400$ GeV benchmarks, the jet+MET limits are adapted from \cite{Zhou:2013raa}; for the Higgs portal benchmark we assume a maximum invisible branching ratio of 10\% \cite{ATL-PHYS-PUB-2013-014}. We find that MATHUSLA would significantly extend the reach of ATLAS and CMS for all benchmarks over almost all of the lifetime reach. At its peak sensitivity, MATHUSLA's detection efficiency for this signal is essentially order one, which means that the reach becomes luminosity limited. The reason is the very high multiplicity of spherically distributed, displaced decaying particles in each event, which effectively compensates for the loss in geometric acceptance as compared to a hermetic detector. Interestingly, there is also a sizable part of parameter space where one can expect events with more than one displaced vertex, which would be a smoking gun for a strongly coupled hidden valley. It is worth noting that a priori ATLAS and CMS themselves will have a significant geometric acceptance for displaced vertices from SUEP events. However with $\Lambda \sim T\sim 1$ GeV the average energy for the decay products, would be an extremely challenging, if not impossible search, and we do not attempt to estimate its sensitivity here. For $\Lambda \sim T\gtrsim 10$ GeV searches for displaced vertices at ATLAS and CMS could however become competitive with MATHUSLA if the final states are predominantly leptons. Hadronic final states pose a greater challenge for ATLAS and CMS in terms of triggering and background rejection, likely requiring a higher $\Lambda \sim T$.  Finally, for $c\tau\lesssim 10$ m, it is likely to that ATLAS and CMS could constrain this scenario regardless the value of $\Lambda$, by searching for a specific pattern of hits in the inner detector \cite{bombdisplaced}.

\subsection[Exotic Higgs Decays]{Exotic Higgs Decays\footnote{David Curtin, Jessie Shelton}}
\label{sec:exhdecays}

\begin{figure}
\begin{center}
\includegraphics[width=12cm]{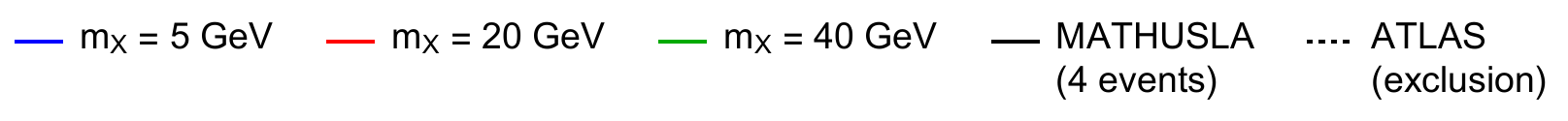}
\\
\includegraphics[width=12cm]{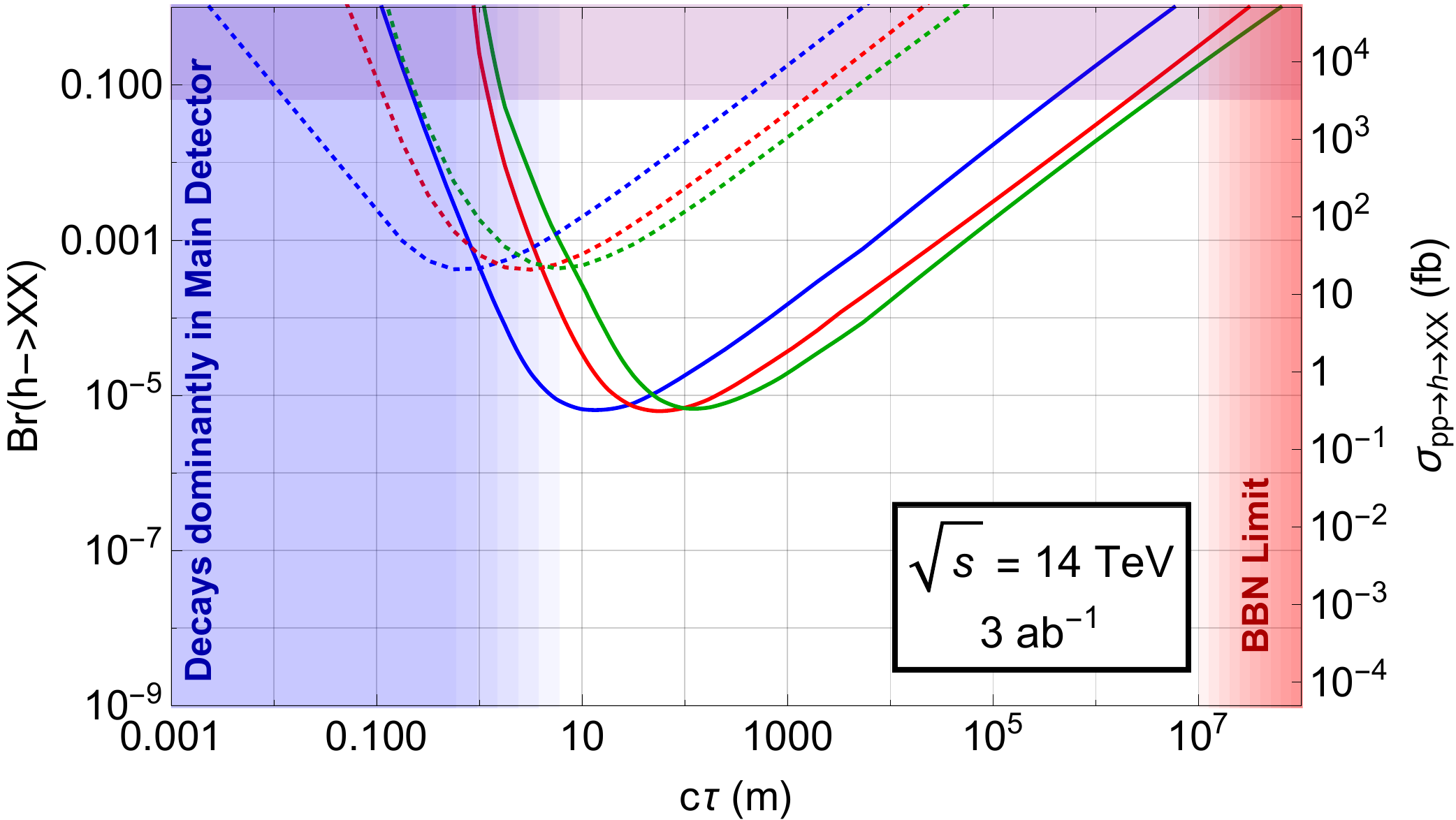}
\end{center}
\caption{
Solid lines show the MATHUSLA sensitivity to new particles $X$ pair-produced in exotic Higgs decays, as a function of $c\tau_X$ and assuming the $200 \,\mathrm{m} \times 200 \,\mathrm{m} \times 20 \,\mathrm{m}$ benchmark geometry of Fig.~\ref{f.mathuslalayout}. The purple shading at the top of the plot shows projected exclusions from CMS $Br(h \to \mathrm{invisible})$ searches \cite{Dawson:2013bba} (although some projections are an $\mathcal{O}(1)$ factor better, see Section~\ref{s.METcomparison}), which would also be sensitive to $X$s outside the blue shaded region. Dotted lines show projected ATLAS $Br(h \to XX)$ exclusions \cite{Coccaro:2016lnz}, which represent the best-case main detector reach projections for LLPs with very long lifetimes produced in exotic Higgs decays. Figure taken from Ref.~\cite{Chou:2016lxi}.
}
\label{f.exHlifetimeplot}
\end{figure}

One of the major discovery opportunities offered by the LHC is the search for new physics produced in exotic decays of the SM-like Higgs boson \cite{Curtin:2013fra}.  As for all newly discovered particles, a detailed characterization of the Higgs' decay modes is imperative.  However, the SM Higgs is especially sensitive to the potential existence of new light degrees of freedom. The Higgs portal operator, $|H|^2$, is one of the two leading operators that can couple to new SM-singlet degrees of freedom, making the Higgs a natural window onto low-mass dark states. 
The very fact that $|H|^2$ is a low-dimensional operator and a singlet under all known symmetries of the SM, which lies at the root of the hierarchy problem, is what generically enables the Higgs to couple to all new physics to some degree.
Discovery prospects are further enhanced thanks to the fortunate accident that all SM decay channels of the Higgs boson are suppressed, whether by phase space ($WW^*$, $Z Z^*$), loop factors ($gg$, $\gamma\gamma$, $Z\gamma$), or small Yukawa couplings ($b\bar b$, $c\bar c$, $\tau\bar\tau$, \ldots), resulting in an accidentally tiny SM Higgs width: $\Gamma (h\to \mathrm{SM}) = 4.10 $ MeV $\pm 0.73\%$ \cite{deFlorian:2016spz} for a $m_h = 125.09$ GeV Higgs boson \cite{Aad:2015zhl}.  Thus even small couplings of the Higgs to new light degrees of freedom can easily yield substantial exotic branching fractions.  
The 3 ab$^{-1}$ of data anticipated at HL-LHC will yield more than $10^8$ Higgs bosons.  This enormous data set could enable the discovery of exotic branching fractions as small as $\sim 10^{-6}$, provided that the signal can be both recorded and separated from background, which is frequently a stiff challenge at the LHC thanks to the low mass scales of Higgs events.  MATHUSLA naturally provides a nearly background-free environment, enabling it to take full advantage of the Higgs sample produced at the HL-LHC.  As shown in Fig.~\ref{f.exHdecaybounds}, MATHUSLA will be able to access Higgs branching fractions to LLPs down to the $10^{-5}$ level.

Exotic Higgs decays to LLPs appear throughout this document. In particular, they are one of the leading signals of many theories of neutral naturalness, extensively discussed in Sec.~\ref{sec:neutralnaturalness}.  Sec.~\ref{sec:hiddenvalleys} discusses Higgs decays to (other) hidden valleys, while in Sec.~\ref{sec:darkphotons} below we demonstrate that Higgs decays into dark photons offer a deep window into the parameter space of a kinetically mixed $U(1)$.  The minimal Higgs portal model SM$+$S  also has exotic Higgs decays as one of its leading signatures, and yields closely related signatures when the new scalar $S$ is  light enough to be produced in meson decays, as we discuss in Sec.~\ref{sec:singlets} below.  SM$+$S signatures arise naturally as signals of hidden sector dark matter or relaxion solutions of the hierarchy problem (Section~\ref{sec:relaxion}).
Finally, unification considerations motivate Higgs portal production of right-handed neutrino states (Section~\ref{sec:Higgportalneut}), giving access to much higher sterile neutrino masses than production in meson decays.
 In many of these examples, e.g. neutral naturalness, the produced LLPs decay back to the SM through Higgs portal couplings as well, predicting large LLP branching ratios to hadronic final states.  These low-mass hadronic final states can be challenging at the LHC main detectors, but offer excellent prospects for MATHUSLA.

Higgs decays to ultra-LLPs $X$ would dominantly appear as missing energy at the LHC when pair-produced in $h\to XX$.  Given $3 \,\mathrm{ab}^{-1}$ of data at 14 TeV, ATLAS and CMS have projected sensitivities to an invisible Higgs branching fraction of $\gtrsim\mathcal{O}(10^{-1})$ \cite{Dawson:2013bba}; MATHUSLA would be able not only to establish the finite lifetime of the $X$ particle but probe branching ratios four orders of magnitude smaller \cite{Chou:2016lxi}.  While proposed main detector searches for single LLPs that decay in the muon system can potentially approach the $Br(h\to XX)\gtrsim \mathcal{O}(10^{-3})$ level for $X$ proper lifetimes in the range $1 \,\mathrm{m}\lesssim c\tau_X\lesssim \,\mathrm{10 \,m}$  \cite{Coccaro:2016lnz}, Fig.~\ref{f.exHlifetimeplot} demonstrates that MATHUSLA will have orders of magnitude greater reach both in branching ratio and proper lifetime.

\begin{figure}
\begin{center}
\includegraphics[width=0.7 \textwidth]{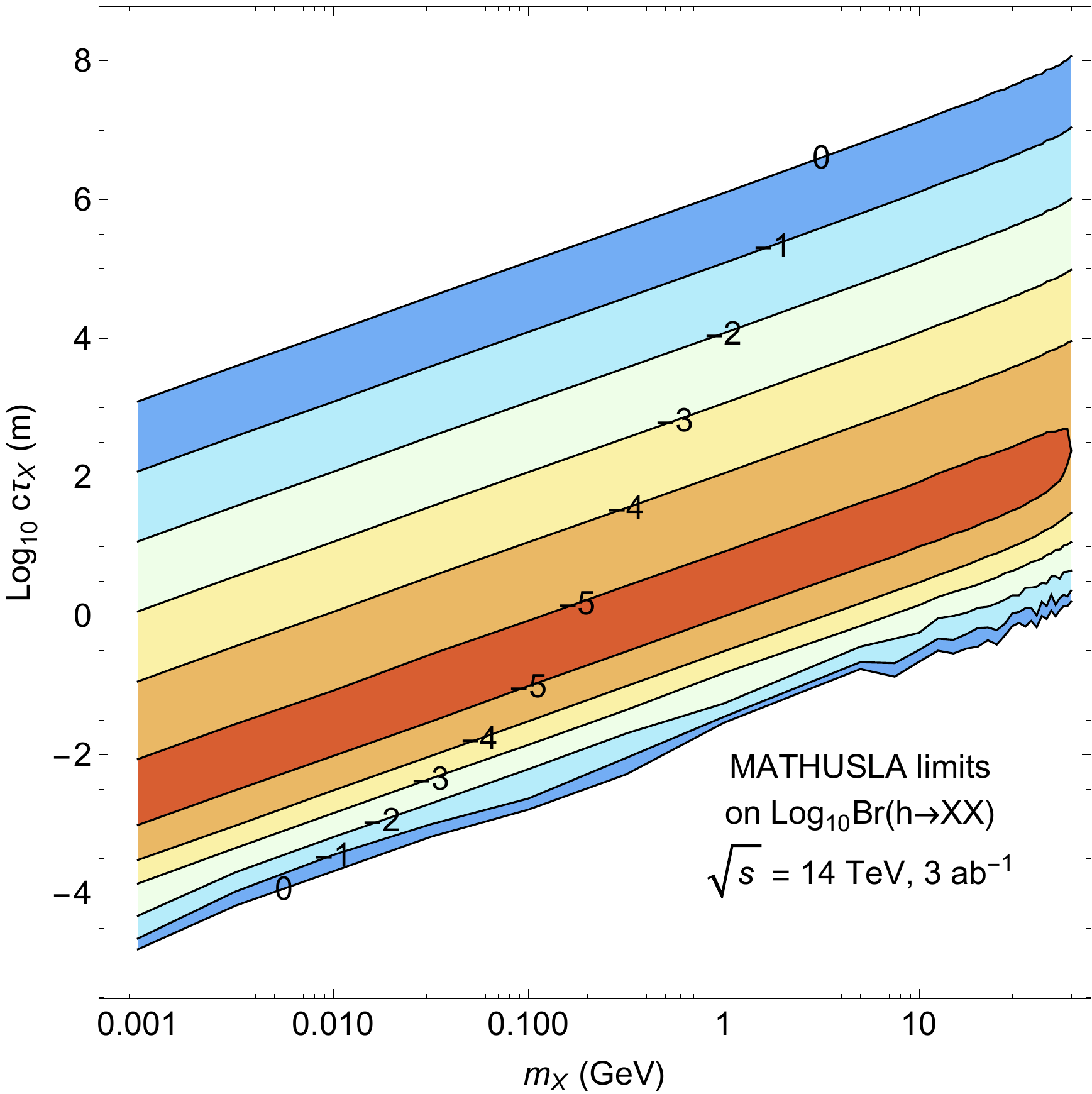}
\end{center}
\caption{
MATHUSLA sensitivity to new particles $X$ pair-produced in exotic Higgs decays. Contours indicate the value of $\log_{10} \mathrm{Br}(h\to XX)$ that could be excluded at $95\%$ CL, assuming SM Higgs production in gluon fusion with 3 $\mathrm{ab}^{-1}$ of data at $\sqrt{s}=14$ TeV.
Note that MATHUSLA searches for the production of low-mass LLPs with $m_{X} \lesssim \gev$ from exotic Higgs decays may suffer from some backgrounds or lower reconstruction efficiency depending on the final detector design, see Section~\ref{s.energythresholds}.
This estimate assumes the $200 \,\mathrm{m} \times 200 \,\mathrm{m} \times 20 \,\mathrm{m}$ benchmark geometry of Fig.~\ref{f.mathuslalayout}.
}
\label{f.exHdecaybounds}
\end{figure}

In Fig.~\ref{f.exHdecaybounds} we show a general-purpose estimate of MATHUSLA's sensitivity to a LLP $X$ pair-produced in exotic Higgs decays. Contours show the branching fraction $\mathrm{Br} (h\to X X)$ that can be tested at $95\%$ CL, assuming a SM Higgs production cross-section.  At fixed branching fraction and proper lifetime, lighter values for $m_X$ result in a higher boost in the lab frame, and thus a suppressed probability of decaying within the detector volume.  An exotic branching fraction of  10\% can be tested for $X$ lifetimes as long as $c\tau_X = 10^7$ m, or $\tau_X = 0.03$ s.   As we discuss in the next section, MATHUSLA's excellent reach allows it to approach the cosmic upper bound on possible $X$ lifetimes.
This also has important consequences for the interpretation of a $h \to \mathrm{invisible}$ signal, if one is found at the HL-LHC: a (say) 10\% invisible Higgs branching fraction could lead one to expect a signal at MATHUSLA if it is due to the production of LLPs with a lifetime below the BBN limit. Conversely, if no signal at MATHUSLA is found, this might add significant weight to the interpretation that the invisible Higgs decay did indeed produce cosmologically stable states, and hence candidates for DM.\footnote{There are important caveats in this argument, most importantly the fact that for lighter LLPs, the resulting boost means that proper lifetimes right at the BBN ceiling are no longer probed. Nevertheless, the absence of a MATHUSLA signal would strengthen the case for the DM interpretation.}

While for definiteness we show quantitative results for the simple decay $h\to XX$, when $X$ is part of a larger hidden sector, it may dominantly appear together with additional particles, $h\to X + \ldots$. 
Such multi-particle exotic Higgs decays can frequently be more challenging for the main detectors, as they distribute the relatively small Higgs energy among a large number of particles. For instance, ATLAS and CMS searches for $h\to \mathrm{invisible}$ rely on large MET for triggering as well as background rejection, and additional soft visible particles appearing in the exotic decay can substantially suppress sensitivity.  By constrast, MATHUSLA's sensitivity does not strongly depend on the number or type of other particles produced together with $X$, and in fact tends to have {\em better} prospects for detecting such multi-species decays, since the ULLP $X$ has a smaller boost and therefore a larger chance of decaying within the detector.  Moreover, when large final state multiplicities arise through showering, this naturally results in a larger multiplicity of LLPs, and thus a larger probability of a decay within MATHUSLA.

\subsection[The BBN Bound]{The BBN Bound\footnote{Anthony Fradette, Maxim Pospelov}}
\label{sec:bbn}

\subsubsection{Introduction}

The MATHUSLA proposal aims to search for long-lived exotic particles
(LLPs) decaying away from the production point at the
LHC~\cite{Chou:2016lxi,Curtin:2017izq}. These LLPs are well-motivated
theoretically (see {\em e.g.}
\cite{Graham:2012th,Craig:2015pha,Izaguirre:2015pga,Batell:2016zod})
and both ATLAS and CMS provide robust lower bounds on their lifetimes
from displaced vertices
searches~\cite{CMS:2014hka,Aad:2015rba,Aad:2015uaa}. A natural place
to look for an upper bound is cosmology, which together with the
current collider constraints would define a clear band of interest in
lifetimes for MATHUSLA. 
Ideally, for such an experimental search, the
LLPs would have different coupling constants moderating their
production and decays, with $\lambda_{\rm production} \gg \lambda_{\rm
  decay}$. Otherwise if one and the same $\lambda$ were responsible
for both the production and decay, large displacements would imply
very inefficient production rates. 
Here we consider exotic decays of the
Higgs boson to a pair of metastable states $S$.

Big bang nucleosynthesis (BBN), and its overall agreement with
observations~\cite{Cyburt:2015mya} (apart from the unclear status of
$^7$Li) can provide a limit on the lifetimes of such particles with
minimal assumptions. The thermal evolution to BBN temperatures
involves self-depletion via $SS\to {\rm SM}$ due to $\lambda_{\rm
  production} $, in an expected WIMP-type annihilation process, and
late-time decay of $S\to {\rm SM}$ where depending on lifetimes and
decay products the BBN outcome may get affected. These mechanisms are
well-understood in the BBN literature (see {\em e.g.}
\cite{Jedamzik:2009uy,Pospelov:2010hj} for reviews) and a small and
acceptable perturbation to the standard BBN (SBBN) outcome can be
turned into a $\tau_S$ limit.

In this work, fully presented in Ref.~\cite{Fradette:2017sdd}, we
analyze a fairly minimal model, where a new singlet scalar has
predominantly a quadratic coupling to the Higgs boson that regulates
both its production at colliders and its metastable cosmological
abundance. We find that for most of the analyzed parameter space with
$m_S < m_h/2$, the intermediate abundance of such particles is large
enough to affect the neutron-proton freeze out ratios at relevant
temperatures. This allows us to set fairly robust bounds on lifetimes
of such particles, which come out to be remarkably strong, and shorter
than $0.1$ seconds, with mild dependence on the mass scale of $S$. In
what follows we briefly review the model, its impact on the BBN; and
present a summary of our results with a short discussion.

\subsubsection{The minimal Higgs portal model}

We consider the simplest extension of the SM by a singlet scalar field $S$. 
At the renormalizable level, the Lagrangian of the singlet sector (including the SM) generically takes the form
\beq
\mathcal{L}_{H/S} = \mu^2 H^\dagger H - \lambda_H\left(H^\dagger H \right)^2  - V(S) - A S H^\dagger H - \lambda_S S^2 H^\dagger H+ \mbox{kin. terms}.
\eeq
The self-interaction potential $V(S) = \lambda_4 S^4 +\lambda_3 S^3 + \frac{m_{S0}^2}{2} S^2$ can be redefined in such a way that the linear term is absent. It is important that the $A,~\lambda_3 \to 0$ and $\langle S \rangle =0$ limit would correspond to the case of stable $S$ particles. To simplify the discussion without sacrificing much generality, we take $\lambda_{3,4} \to 0$ and assume $Av \ll m_{S0}^2, ~\lambda_S v^2$. 

The physical mass of $S$ receives a contribution from the electroweak symmetry breaking,  $m_S =\sqrt{m_{S0}^2 + \lambda_S v^2}$. The two scalars develop a mixing angle and renders the $S$ unstable via
\beq
{\cal L}_{\rm decay} =  S \times \theta \sum_{\rm SM} O_{h}, \qquad \qquad \theta = \frac{Av}{m_h^2-m_S^2}\left(1-\frac{\lambda_Sv^2}{m_S^2}\right). \label{Oh}
\eeq
where $O_h$ is the set of the standard Higgs interaction terms, with the Higgs field removed: {\em e.g.} $O_h = (m_f/v) \bar ff $
for an elementary SM fermion $f$. 

Both $\theta$ and $\lambda_S$ on their own are already subject to many
observational constraints~(see
Refs.~\cite{Alekhin:2015byh,Cline:2013gha,Athron:2017kgt} for recent
reviews). A generic feature is that $\lambda_S$ is bounded by a
maximal invisible Higgs branching ratio of 0.19 (at
$2\sigma$)~\cite{Belanger:2013xza}. With the well-predicted decay rate
of the SM Higgs into SM particles of $\Gamma_{SM} = 4.08$ MeV \cite{deFlorian:2016spz} and
\beq
\Gamma_{h\to SS} =\frac{\lambda_S^2 v^2}{8\pi m_h} \sqrt{1-\frac{4m_S^2}{m_h^2}}, \qquad \qquad
Br(h\to SS ) = \frac{\Gamma_S}{\Gamma_S + \Gamma_{SM}} \simeq 10^{-2} \,\left(\frac{\lambda_S}{0.0015}\right)^2,
\label{Gh}
\eeq
this translates into an upper bound on $\lambda_S \lesssim 0.007$ for $m_S \ll m_h$. If $S$ is to be stable ($\theta \to 0$), such small $\lambda_S$ would lead to an excessive abundance of $S$, which invalidates the $Z_2$ symmetric case, 
and forces us to include a non-zero decay term. From now on, we will consider $\theta \neq 0$, or in other words the case of unstable $S$ particles. Since our analysis is motivated by the 
LHC physics, we will use $Br(h\to SS )$ as an input parameter, and substitute $\lambda_S$ everywhere employing (\ref{Gh}).

\paragraph{Decay products}

Since $S$ interacts with the SM in the same fashion as the Higgs with
an additional $\theta$ mixing factor (\ref{Oh}), its decay properties
are similar to those of a light Higgs boson. For the derivations of
the actual limits on the lifetime of $S$, we need to know its mesonic
and nucleonic decay branching ratios, which are still poorly
understood and can vary by a few orders of magnitude near the di-kaon
threshold~\cite{Clarke:2013aya}. In particular, the metastable mesons,
such as $\pi^\pm$ and $K^\pm,\bar K^0, K^0$ are ``important'' decay
products, as they can participate in the charge-exchange reactions
with nucleons and shift the $n-p$ balance, hence affecting the whole
nucleosynthetic chain. We will show that two different decay models
obtain similar constraints on the lifetime, thus minimizing the
uncertainty in the decay rate.

The leptonic decay channels are straightforward, with the decay rate given by
\beq
\Gamma_{S\to l \bar{l}} = \frac{\theta^2}{8\pi} \frac{m_l^2}{v^2} m_S \left(1-\frac{4m_l^2}{m_S^2}\right)^{3/2}. \label{eq:Gamff}
\eeq
If the decaying product is a pair of heavy quarks, there are $\mathcal{O}(1)$ corrections coming from the 1-loop QCD vertex correction~\cite{Cline:2013gha}. For $m_S > 2.5\GeV$, we use the higher order perturbative results from the \textsc{HDecay} code~\cite{Djouadi:1997yw} which includes all necessary corrections. 

In the mass range where the perturbative QCD calculations are no longer valid, we base our baseline calculations on Ref.~\cite{Bezrukov:2009yw}. The scalar-pion interaction can be extracted from the low-energy expansion of the trace of the QCD energy-momentum tensor~(see for {\em e.g.}.~\cite{Voloshin:1985tc,Leutwyler:1989xj}) yielding the effective decay rate~\cite{Bezrukov:2009yw}
\beq
\Gamma_{S\to \pi^+ \pi^-} = 2\Gamma_{S\to \pi^0 \pi^0} = \frac{\theta^2}{16\pi}\frac{m_S^3}{v^2} \left(\frac{2}{9} + \frac{11}{9} \frac{m_\pi^2}{m_S^2} \right)^2
\sqrt{1- \frac{4m_\pi^2}{m_S^2}}. \label{eq:LowEPi}
\eeq
Final-state resonances however spoil this expression far from the threshold. To include kaons, we use an interpolation from Ref.~\cite{Donoghue:1990xh}, matching low-energy theorems to the dispersion results from the $\pi \pi$ phase-shift analysis above 600 MeV~\cite{Hyams:1973zf}. The photon decay channel is added with the prescription detailed in Ref.~\cite{Spira:1997dg}. Finally, there is a gap for $1.4~\GeV < m_S < 2.5~\GeV$ where no analytical treatment is entirely trustworthy, we simply follow Ref.~\cite{Bezrukov:2009yw} and interpolate between the two regimes, under the assumption that there is no order of magnitude deviation in this mass range. The branching ratios and the lifetime for $\theta = 10^{-6}$ are displayed in Fig.~\ref{fig:S_Br_tau}.

\begin{figure}[t]
\centering
 \includegraphics[width= 0.45\columnwidth]{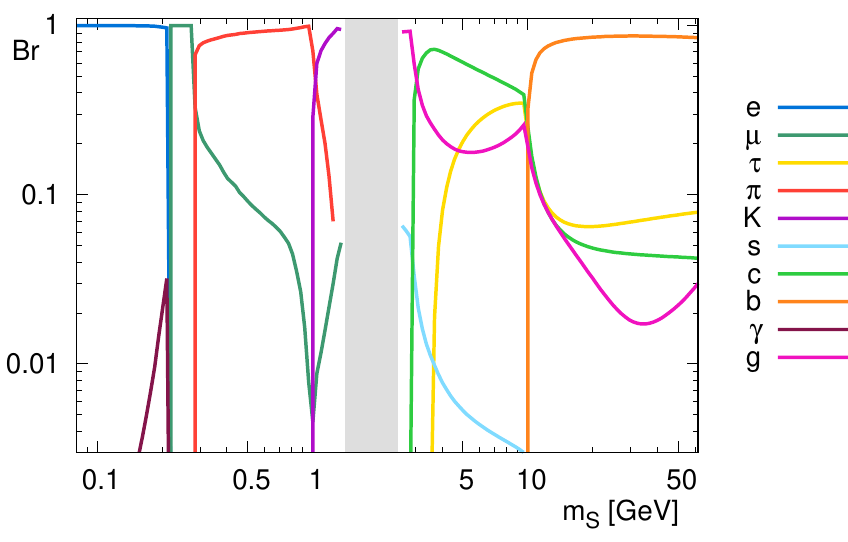} \hspace{0.5cm}
  \includegraphics[width= 0.4\columnwidth]{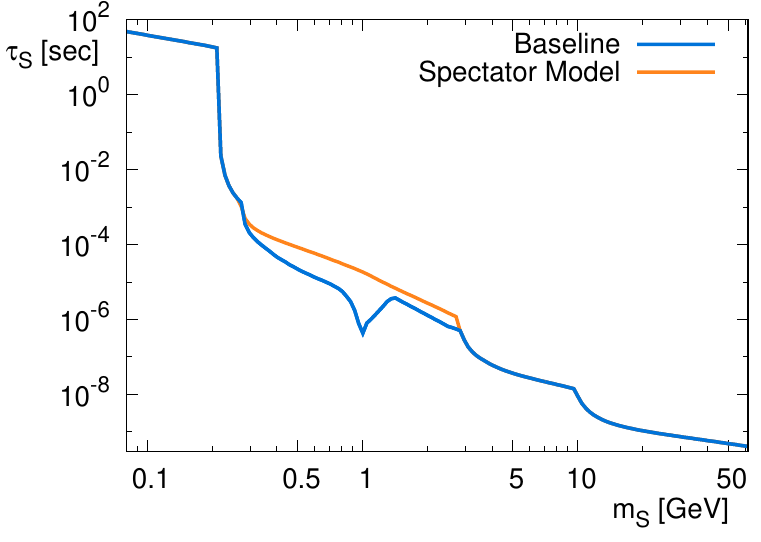}
\caption{\textit{Left}: Branching ratios of the scalar $S$ in our baseline decay model. See text for details. \textit{Right}: Scalar $S$ lifetime of our baseline model and the spectator model for the mixing angle $\theta = 10^{-6}$.} 
\label{fig:S_Br_tau}
\end{figure} 

As an alternative decay spectrum model, we also display the perturbative spectator approach~\cite{Gunion:1989we,McKeen:2008gd,Alekhin:2015byh}, where the relative decay width above the kaon threshold are given by
\beq
\Gamma_{\mu^+\mu^-} : \Gamma_{KK} : \Gamma_{\eta \eta} = 
m_\mu^2 \beta_{\mu}^3 : 3\frac{9}{13}m_s^2\beta_K^3 : 3\frac{4}{13}m_s^2\beta_\eta^3 , \label{eq:SpectBr}
\eeq
with $\beta_i = \sqrt{1-4m_i^2/m_S^2}\Theta(m_S - 2m_i)$, $\Theta$ being the step-function, and we adopt the running of $s$ quark mass following Ref.~\cite{Spira:1997dg}. The pion contribution is kept as in equation~(\ref{eq:LowEPi}) and then we use the \textsc{HDecay} output at the $c$-quark threshold and above to match our baseline model. 

For $m_S$ of several GeV and heavier, decays with final state nucleon-antinucleon pairs are possible. Even though the branching to such states are generally lower than 10\%, the effect on BBN can be quite significant, and therefore these are by far the most important channels for $\tau_S \gtrsim 1\;\sec$. On top of direct and for the most part subdominant contributions from $S\to \bar nn,...$, we need to take into account the (anti-)nucleon states that emerge from the hadronization of the quark decay products and heavy $B$-meson fragmentations. 

\paragraph{Cosmological metastable abundance}

Starting in thermal equilibrium with the SM, the $S$ population eventually freezes out to a metastable abundance, via the $s$-channel annihilation $SS \to h^* \to XX$, where on the 
receiving end are the pairs of the SM states $XX$ created by a Higgs-mediation process. The annihilation cross-section $\sigma v$ generically takes the form
\beq
\sigma v(s) = \frac{8 \lambda_S^2 v^2}{(s-m_h^2)^2 + m_h^2 \Gamma_{{\rm SM}+S}^2}\frac{\Gamma_{{\rm SM}}^{m_h\to\sqrt{s}}}{\sqrt{s}}, \qquad \qquad
\langle \sigma v \rangle = \frac{\int_{4m_S^2}^\infty ds\;  \sigma v (s)\;s \sqrt{s-4m_S^2}K_1\left(\frac{\sqrt{s}}{T}\right)}{16 T m_S^4 K_2^2\left(\frac{m_S}{T}\right)}.
\label{eq:sigmav}
\eeq
This formula recasts the rate in terms of a SM Higgs width
$\Gamma_{{\rm SM}}^{m_h\to\sqrt{s}}$ evaluated at a fictitious mass of
$\sqrt{s}$, thus encompassing both perturbative and non-perturbative
channels in the $h^*$ decay rate, which is the same as $\Gamma_S$ with
$\theta =1$. Since the nonrelativistic annihilation cross-section in
the minimal Higgs portal model ranges from $10^{-3}$ to $10^{-14}$ pb
for $m_S \sim 1\MeV - 60 $ GeV and \mbox{$Br(h\to SS) \sim 0.1 -
  0.001$}, the standard nonrelativistic WIMP freeze-out approximation
is not applicable and we numerically integrate the standard Boltzmann
equation to determine the metastable $S$ abundance. The results are
shown in Fig.~\ref{fig:YS_abundance}, normalized to the baryon number
density for a more intuitive interpretation of its impact on BBN in
the following section.  Qualitatively, it is clear that the relative
inefficiency of annihilation through the Higgs portal will leave
behind a fairly significant population of $S$ particles, which will
eventually lead to strong constraints on $\tau_S$.

For very light $m_S$, one can see that the freeze-out abundances are
large, and the relative spread between different input values of
$Br(h\to SS)$ gets smaller, as the annihilation cross-section becomes
very small. Such small cross-sections mean that freeze-out happens in
the semi-relativistic regime $x_{\rm f.o.} \sim \mathcal{O}(1)$ and
asymptote to the $Y_{\rm eq}$ relativistic plateau for small
$m_S$. The only difference at the lightest masses is from $Y_{\rm
  eq}^{rel} \propto 1/g_{*S}(T)$, where $g_{*S}$ is the number of effective degrees of freedom
appearing in the entropy density. Since
$g_{*S}$ is a monotonic function of temperature, weaker
annihilation cross-sections freeze out earlier, at a higher
temperature, thus yielding smaller abundances (as seen in the $m_S =
5\MeV$ curves in Fig.~\ref{fig:YS_abundance}). This is in contrast
with the standard freeze out in the non-relativistic regime, with
final abundances inversely proportional to the cross-section.  We note
in passing that the strong-interaction-related uncertainty
``propagates'' outside the $m_S \sim 2m_\pi - 2 m_c$ window. For
example, because of the relativistic freeze-out, for $m_S < 2
m_\pi$ the hadronic channels may turn out to be important.

\begin{figure}
\centering
 \includegraphics[width= 0.4\columnwidth]{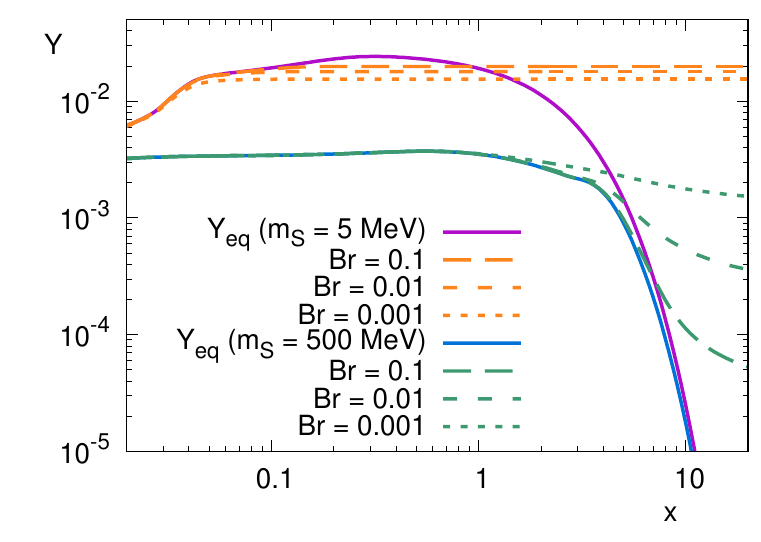} \hspace{1cm}
  \includegraphics[width= 0.4\columnwidth]{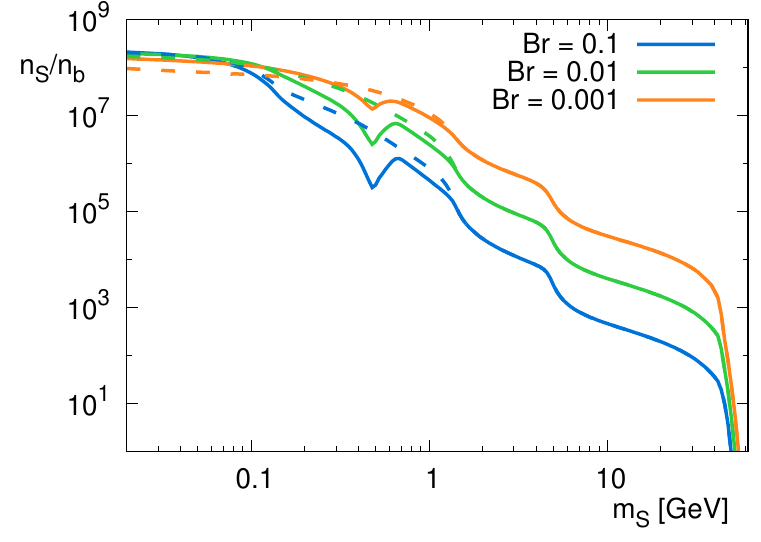}
 \caption{\textit{Left}: Temperature evolution ($x=m/T$) of the $Y_S$ intermediate abundance for $m_S = 5\MeV$ and 500 MeV for the three benchmark higgs branching ratios. \textit{Right}: Metastable abundance of $S$ prior to its decay normalized over the baryon density. Values shown for $Br(h\to SS) = 10^{-1}$, $10^{-2}$ and $10^{-3}$. The dashed lines correspond to the perturbative spectator model.} 
\label{fig:YS_abundance}
\end{figure} 

\subsubsection{Big Bang Nucleosynthesis}

The formation of light nuclei is one of the earliest probes of NP in
cosmology and is well-understood within SM physics. Modulo the $^7$Li
discrepency~\cite{Cyburt:2015mya}, the overall success of BBN in
predicting the more populous element abundances can be used to
constrain various types of NP~\cite{Pospelov:2010hj}.

The initial BBN stage is the neutron-proton ratio $n/p$ freeze
out. Maintained in equilibrium by electroweak interactions at high
temperatures, the neutron abundance follows $n/p \sim e^{-Q/T}$, where
$Q=m_n - m_p -m_e \simeq 1.293\MeV$, until the epoch when the weak
processes decouple around temperatures of 0.7 MeV. The outcome, $n/p
\simeq 1/6$, is quasi-stable, decreasing to $n/p \simeq 1/7$ at the
end of the ``deuterium bottleneck''. At $t_{\rm deut} \sim 200 $
seconds, $^4$He formation is very efficient and most neutrons end up
in the final $^4$He abundance (expressed in mass fraction from the
total baryon mass) $Y_p \simeq 2 (n/p)/\left( 1 + n/p\right) \simeq
0.25$.

For the problem at hand - the determination of the upper limit on the $S$ lifetime - few of the finer BBN details matter. The ample decaying abundances of $S$ particles ($n_S \sim 10^{2}-10^{9} \times n_b$) will flood the neutron-proton bath with SM particles prior to the bottleneck, inducing charge-exchange reactions that will modify the $n/p$ freeze out ratio. For each decay products $X$ that can modify $n/p$, we solve the neutron-proton Bolztmann equation
\beq
\frac{d X_n}{dT} = \frac{\Gamma_{n \nu_e \to p e^- }+\Gamma_{n e^+ \to p \bar{\nu}_e } }{T H(T)} \left(X_n -(1-X_n)e^{-Q/T} \right) +
\frac{\Gamma_n X_n}{T H(T)} + \left.\frac{d X_n}{dT}\right|_{X}, \label{eq:Boltzmann}
\eeq 
including the new charge-exchange term and require that $Y_p$ does not deviate from SBBN by more than 4\%, 
\beq
\Delta Y_p \equiv |Y_p -Y_p^{\rm SBBN}| < 0.01,
\eeq 
which is a rather generous allowance for the errors, considering the tight observational constraints on primordial helium abundance~\cite{Cyburt:2015mya}. 
Consequently, it will result in conservative limits of $\tau_S$. 
We show typical deviations from the SBBN case for pion, kaon, direct baryon~\footnote{We tuned the injected baryons after hadronization to $N_n = \kappa N_p$ and $N_{\bar{n}} = \kappa N_{\bar{p}}$, where a phenomenological hadronization parameter $\kappa$ is expected to be $ \simeq 0.5$ from simple quark counting of the main weak decay chain \cite{Fradette:2017sdd}. }
and neutrino (from muon decays) injections in figure~\ref{fig:Xn_Gamtau}, along with the maximal abundance (weighted by its $S$ branching ratio $\xi$) that satisfy the $\Delta Y_p < 0.01$ requirement. 
We refer the reader to the complete paper~\cite{Fradette:2017sdd} for additional detail, including direct $SS$ annihilations to charged pions and $N_{eff}$ deviations from energetic electron or muons.

\begin{figure}
\centering
 \includegraphics[width= 0.38\columnwidth]{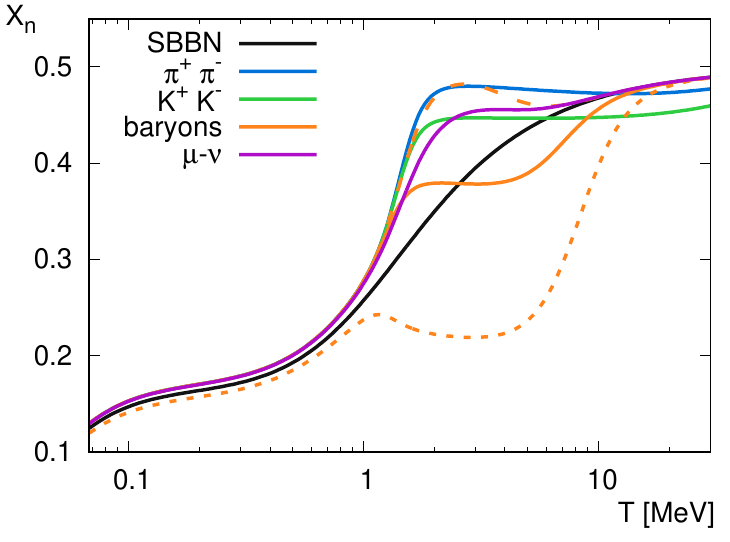} \hspace{1cm}
  \includegraphics[width= 0.42\columnwidth]{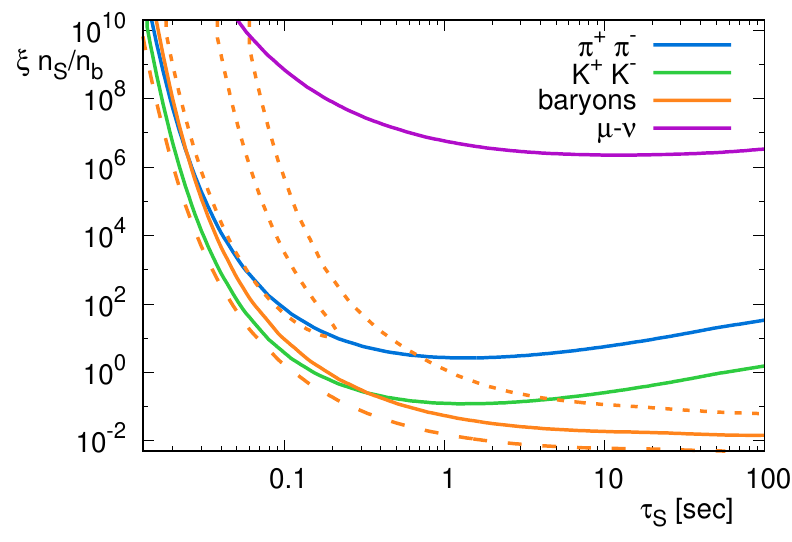}
  \caption{\textit{Left}: $X_n$ evolution for the SBBN and the
    injection of pions, kaons, baryons and muons (neutrinos) as
    described in the text for lifetimes of $0.05$ seconds with the
    initial $Y_S$ abundance tuned to yield $\Delta Y_p=0.01$ (maximum
    allowed shift of $Y_p$). The baryonic injection is taken at
    $\kappa = 0.5$ (full line), the lines for $\kappa = 1$ (dashed)
    and $\kappa = 0.2$ (dotted) are also displayed. \textit{Right}:
    Limit of injected pairs for each channel as a function of the $S$
    lifetime. The upper-right dotted line for $\kappa = 0.2$ is at
    $Y_p = 0.26$, the upper-left dotted island yields $Y_p = 0.24$. }
\label{fig:Xn_Gamtau}
\end{figure} 

\subsubsection{Results and Discussion}

Combining the constraints on each energy injection mode, the surviving
parameter space of the minimal Higgs model is shown in
Fig.~\ref{fig:S_paramspace} as a function of the scalar mass $m_S$ and
lifetime $\tau_s$.
The assumptions considered in each mass range, labelled from
A to G are described in Ref.~\cite{Fradette:2017sdd}. 

\begin{figure}
\centering
 \includegraphics[width= 0.8\columnwidth]{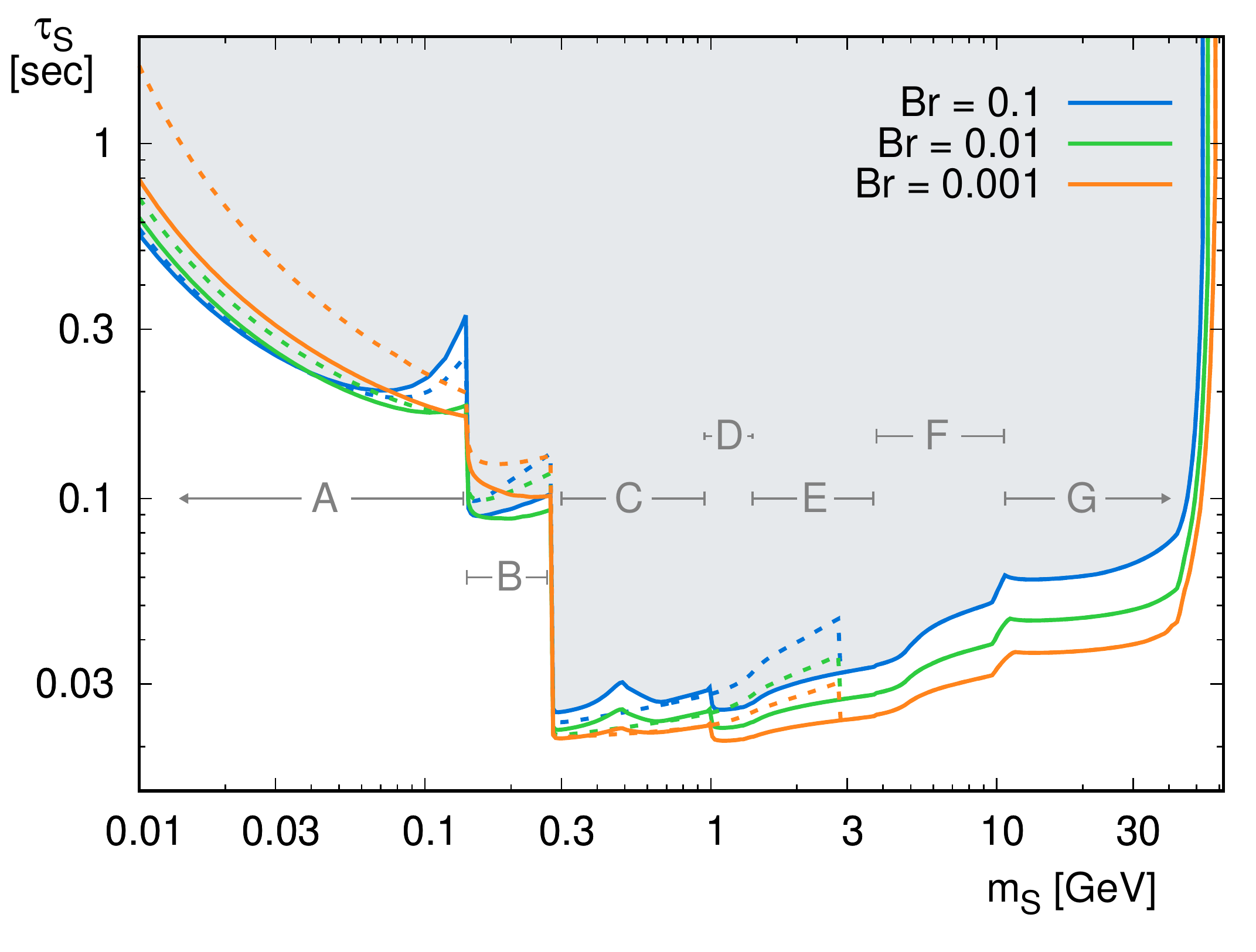} 
  \caption{ Lifetime constraint as a function of the $S$
    mass for three $h\to SS$ branching ratios. The lettered regions
    represent different assumptions or physics and are described in
    the text. The dotted lines correspond to the perturbative
    spectator model. 
}
\label{fig:S_paramspace}
\end{figure} 

We find that throughout almost the whole mass range considered in this
work, $2m_\mu < m_S < m_h/2$, the constraints on the lifetime of $S$
particles are stronger than $0.1$ seconds, with only a mild dependence on
$Br(h\to SS)$ as the
amount of decaying $S$ is much larger than the baryon abundance in all
cases. From the standpoint of LHC physics, the most notable portion of
parameter space is at relatively large masses, where $m_S$ is not far
below $m_h/2$.   In that case, the proper decay length has to be on
the order of or smaller than $2\times 10^7$ meters, and comparing with
Fig.~\ref{f.exHdecaybounds}, this places the upper bound on $X$ lifetimes in the same overall range that can be probed by MATHUSLA. 

The above considerations can be generalized to other models of the
Higgs-portal-coupled particles or even different types of
interactions, via $Z$, $Z'$ etc. For example, consider a fermion
$\chi$, coupled to the Higgs via $H^\dagger H (\bar\chi \chi)$ or
$H^\dagger H(\bar \chi i \gamma_5 \chi)$ dimension-five operators, and
having a small decay term such as {\em e.g.} neutrino portal $LH
\chi$. The main analysis of our work can be recast for that model,
especially in the part that connects Higgs decays with a metastable
abundance of $\chi$. Evidently, for the same input values of $Br(h\to
\chi\bar\chi )$ and $ Br(h\to SS) $, one will end up with $Y_\chi \sim
Y_S$. The only change will be in the yields of mesons and baryons in
the decays of $\chi$ compared to $S$.  However, as the yield of pions and kaons in $\chi$ decays is already known to be
substantial for $m_\chi > 250$ MeV~\cite{Alekhin:2015byh}, we expect that for the most part the constraints we have
derived for $\tau_S$ will translate to similar limits on
$\tau_\chi$. On the other hand, these constraints can be evaded if
there are additional degrees of freedom to deplete the energy outside
the BBN lifetime window, at the expense of additional complication of
the model.

\subsection[SM + S: Singlet Extensions]{SM + S: Singlet Extensions\footnote{Jared Evans}}
\label{sec:singlets}

Hidden sectors populated with new particles that are only very weakly coupled to the standard model are well-motivated.  A simple extension that includes new weakly-coupled  scalars has been used to explain a wide variety of outstanding deficiencies in the standard model, such as dark matter \cite{Tulin:2013teo,Martin:2014sxa,Krnjaic:2015mbs}, the  $(g-2)_\mu$ anomaly \cite{Zhou:2001ew,Pospelov:2008zw},  inflation \cite{Bezrukov:2009yw}, naturalness \cite{Chacko:2005pe,Burdman:2006tz},  neutrino masses \cite{Dev:2016vle}, and the proton radius puzzle \cite{TuckerSmith:2010ra,Barger:2010aj,Batell:2011qq}.  A new scalar, $S$, can be coupled to the standard model via the renormalizable Higgs portal interaction, $\epsilon \left|S\right|^2 H^\dagger H$.   A minimal simplified model can be constructed with the scalar Lagrangian \cite{Evans:2017lvd},
\beq
\mathcal L_{scalar} = \mathcal L_{kin} - \frac 12 \epsilon S^2 H^\dagger H +\frac 12 \mu_S S^2- \frac {\lambda_s}{4!} S^4 +\mu_H^2 H^\dagger H - \lambda_H \left( H^\dagger H\right)^2
\label{eq:scalarLag}
\eeq
 for real scalar, $S$, with an imposed discrete symmetry $S\to-S$ to prevent all terms cubic and linear in $S$.  Adding these terms complicates the sector, but does not qualitatively alter the physics from the story presented here.   If both $S$ and $H$ have nonzero vacuum expectation values, $S=s +v_s$ and $H = (h+v_h)/\sqrt2$, the two scalar states will mix.  For portal coupling $\epsilon \ll 1$, $v_h$ and one of the mass eigenvalues can be identified as the usual Higgs vev, $v_h=246$ GeV, and observed Higgs mass, $m_h=125$ GeV, while the remaining three parameters in (\ref{eq:scalarLag}) can be identified with the mass of the scalar, $m_s$, a mixing angle between the two sectors, $\sin \theta= \frac{\epsilon v_h v_s}{m_h^2 -m_s^2} + \mathcal O (\epsilon^3)$, and the coupling of the Higgs with two hidden sector scalars, 
 \beq
\mathcal L \ni \frac\kappa2 hs^2 = \frac12 \sqrt{\frac{\lambda_s}{3}}\sin\theta \left(\frac{m_h^2 +2 m_s^2}{m_s}\right) hs^2.
 \eeq

The hidden sector scalar couples to standard model fermions and vector bosons as a standard model Higgs, but with strength reduced by a factor of $\sin\theta$.   The scalar width is thus reduced by $\sin^2\theta$ from a standard model-like Higgs of the same mass, i.e., $\Gamma_s = \sin^2\theta \Gamma_{h,SM}(m_s)$, which, for sufficiently small mixings, results in a particle that is long-lived on collider timescales.  The width of a hadronically decaying light scalar has a high degree of uncertainty for scalar masses between $2m_\pi$ and $\sim 4 \gev$ (see refs.~\cite{Gunion:1989we,Clarke:2013aya} for more details). 
It is common in the literature to use a perturbative spectator model in this region, see e.g.~\cite{Gunion:1989we,McKeen:2008gd,Alekhin:2015byh,Fradette:2017sdd}, but there are some reasons to suspect that this approach may be underestimating the scalar's partial width into hadrons.    
As the scalar mass is increased, it crosses through several dozen hadron mass thresholds that open up more and more accessible decay channels, and, as the Higgs-mixed scalar couples to mass and $\Lambda_{QCD} > m_s$, these channels may provide large corrections.   Across this region there will be many scalar meson resonances for the state to mix with (including the observed $f_0(1370)$, $f_0(1500)$, $f_0(1710)$, and the $f_0(980)$, the latter of which is responsible for the extrema near 1 GeV in ref.~\cite{Donoghue:1990xh}, and shown in figure \ref{fig:scalarBRJAE}), and even rather broad states that have never been observed could appreciably amplify the hadronic decay width \cite{Raby:1988qf} relative to the predictions of the perturbative spectator model.   Motivated by these uncertainties, in this region we use the branching fractions in ref.~\cite{Evans:2017lvd}, shown in figure \ref{fig:scalarBRJAE}.\footnote{The larger hadronic partial width and smaller muon branching ratio in this interpretation combine to enlarge the gap in coverage between LHCb (from above) and long lifetime experiments like MATHUSLA, SHiP \cite{Anelli:2015pba}, CODEX-b \cite{Gligorov:2017nwh}, or FASER \cite{Feng:2017uoz} (from below), which makes this choice conservative with regards to ensuring complete coverage in the gap between the different experimental approaches.}

  \begin{figure}[!t]
\begin{center}
\includegraphics[scale=0.6]{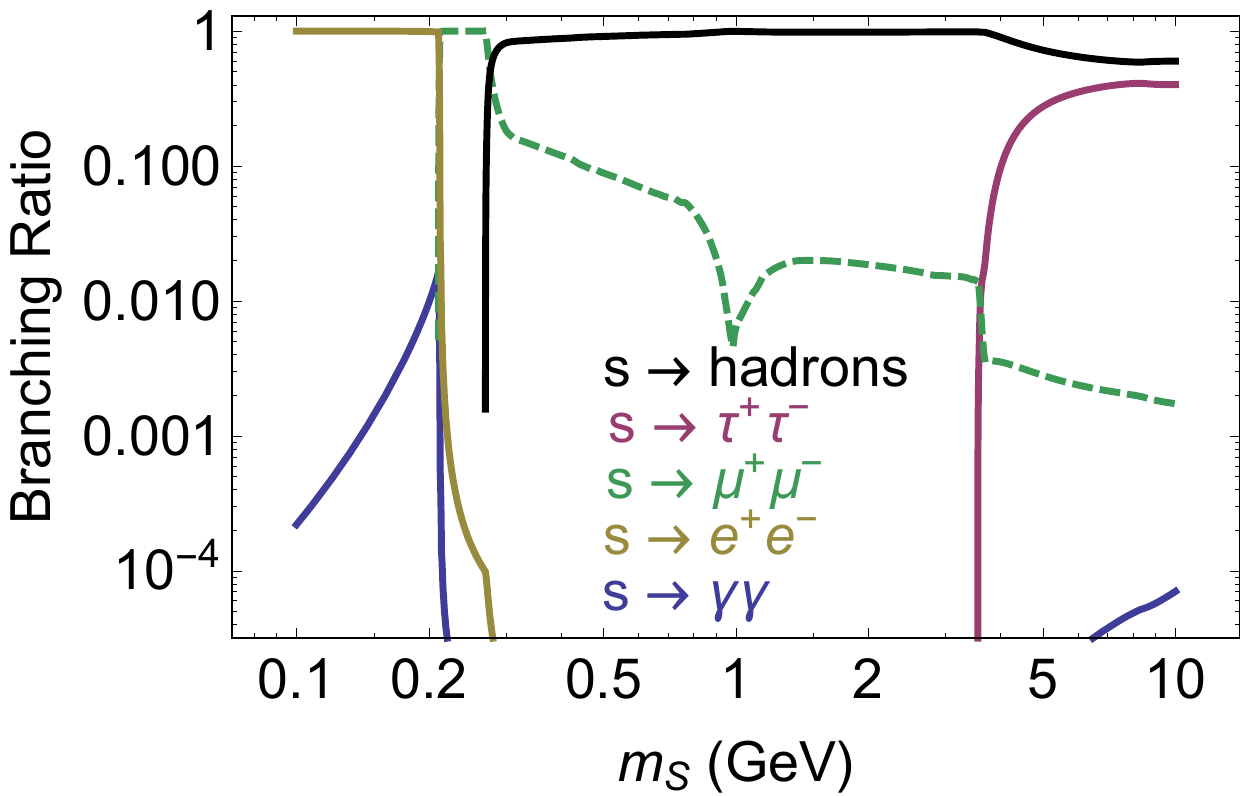}
\end{center}
\caption{Branching ratios assumed in the SM+S model for the additional scalar in the light hadron region. 
For masses below $\sim 1.4$ GeV, the calculation of ref.~\cite{Donoghue:1990xh} is used. We implement an extrapolation in the region from $1.4\,\mathrm{GeV}$ to $2m_D$ that yields a larger partial width into hadrons than predicted in the perturbative spectator model \cite{Gunion:1989we}.}
\label{fig:scalarBRJAE}
\end{figure}

At the LHC, the long-lived scalars can be produced in exotic Higgs decays. 
 Allowing for $m_s$ and $\sin\theta$ to assume any value still places a restriction on how large $\kappa$ can be when one mandates perturbativity of $\lambda_s$ ($\lambda_s <16\pi^2$).  The maximum allowed $h\to ss$ branching ratio is then,
 \beq
 \mbox{BR}(h\to ss)_{MAX} = \frac{\pi \sin^2\theta m_h^3}{3m_s^2\Gamma_{h,tot}} \left( 1 +2 \,\frac{m_s^2}{m_h^2}  \right)^2\sqrt{1 -4\,\frac{m_s^2}{m_h^2}}
 \label{eq:scalarBR}
 \eeq
 where $\Gamma_{h,tot}\approx 4.15$ MeV.   While it is possible that the addition of linear and cubic interactions could relax this constraint slightly, it cannot be modified parametrically without introducing additional particles, for instance, a two-site model with $h\to s_1s_1$ followed by $s_1\to s_2s_2$.

 \begin{figure}[t]
\begin{center}
\includegraphics[scale=1.]{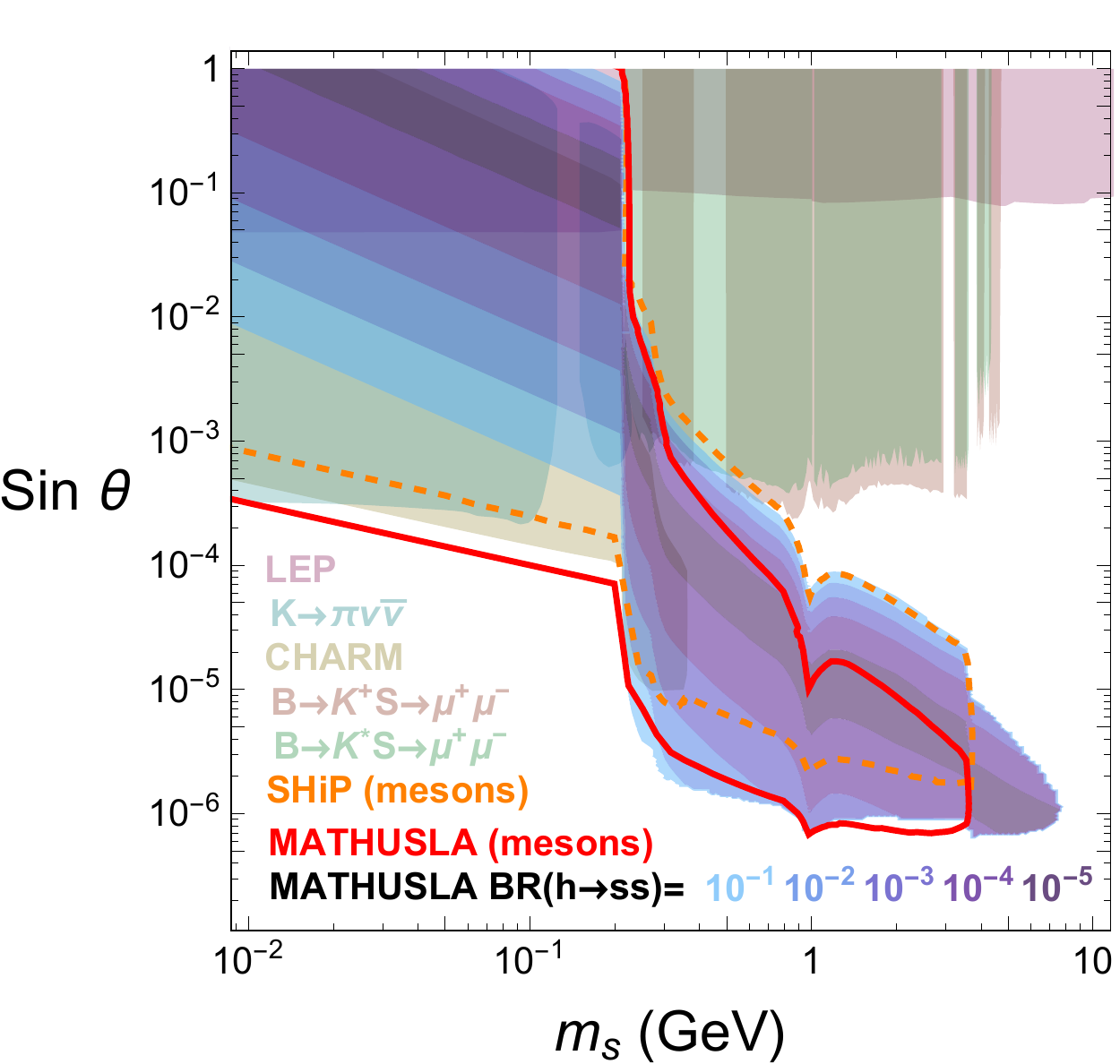}
\end{center}
\caption{The projected sensitivity of MATHUSLA to scalar LLPs in the minimal SM+S extension after 3 ab$^{-1}$ of 14 TeV LHC assuming 4 events, and assuming the $200m \times 200m \times 20m$ benchmark geometry of Fig.~\ref{f.mathuslalayout}.  The red contour is the sensitivity to $B$-meson decays (Kaons would provide percent level corrections). The different blue-purple contours illustrate the minimum BR$(h\to ss)$ value to which MATHUSLA would be sensitive.  The feature near 1 GeV is due to the peak in the partial width to hadronic states that appears in the Donoghue, Gasser, Leutwyler \cite{Donoghue:1990xh} modeling used in this region (between $1.4-4$ GeV, the interpolation from \cite{Evans:2017lvd} is used).  With decreasing (increasing) $\sin\theta$ or $m_s$ the lifetime grows (shrinks).  The overall shape of the sensitivity at higher masses is heavily sculpted by the maximum allowed value of BR($h\to ss$) consistent with perturbative couplings in the theory.  The projected constraint contour for the SHiP experiment \cite{Lanfranchi:2243034,Anelli:2015pba,Alekhin:2015byh} is shown by the dashed orange contour.  Current constraints are described in the text.}  
\label{fig:HBRScalar}
\end{figure}

 Additionally, this scalar could be emitted in rare meson decays.  For the purpose of MATHUSLA, $B$-mesons are the most relevant production mechanism, and the inclusive branching fraction can be expressed as \cite{Grinstein:1988yu, Gligorov:2017nwh}
\beq
\mbox{BR}(B\to sX_s) = \frac{27\sqrt 2 G_Fm_t^4}{64\pi^2 \Phi m_b^2}\! \left|\frac{V_{ts}^* V_{tb}}{V_{cs}}\right|^{2}\!\! \left(1-\frac{m_s^2}{m_b^2}\right)^{2}\!\!\! \sin^2\theta \mbox{ BR}_{B\to X_c e \nu_e}
 \approx 6.2  \left(1-\frac{m_s^2}{m_B^2}\right)^2 \!\!\! \sin^2\theta.
\label{eq:firstBR}
\eeq
where $\Phi\approx 0.5$ \cite{Lenz:2014jha} is a phase space factor for the semi-leptonic decay.

Following \cite{Evans:2017lvd}, we can estimate the sensitivity of MATHUSLA to $s$ particles produced in either meson decays or exotic Higgs decays.  Events are generated for  $h\to s s$ and $b\bar b$ production in Pythia 8 \cite{Sjostrand:2014zea} at 14 TeV.    From the kinematic distributions, we can compute the probability that $s$ decays within the MATHUSLA detector volume.   Joining these to either the 14 TeV Higgs production cross-section of 62.6 pb \cite{deFlorian:2016spz,Anastasiou:2016cez} or an assumed $b\bar b$ production cross-section of 0.5 mb \cite{Aaij:2016avz} and a projected luminosity of 3 ab$^{-1}$ to determine the number of events that would decay in MATHUSLA.   We project to 95\% CL constraints assuming zero background and a 75\% detection efficiency (4 scalar LLP decays observed). 
 In the case of meson decays, most of the scalars have energies above 2 GeV \cite{Evans:2017lvd}, but for masses below $\sim 10 \mev$, their detection efficiency may be significantly degraded and depends on details of the detector design, see Section~\ref{s.energythresholds}.

 The projected sensitivity for both meson decays and exotic Higgs decays are shown in Fig.~\ref{fig:HBRScalar} in the plane of $m_s$ vs mixing angle $\sin\theta$.  While the meson decay constraints are robustly determined by the position in this parameter space, the exotic Higgs decay constraints depends on an additional parameter, the $h\to ss$ branching ratio, which has no lower bound.  For arbitrarily small $\lambda_s$ (i.e., large $v_s/m_s$), all sensitivity to this channel can vanish.  We show with the blue-purple contours in  Fig.~\ref{fig:HBRScalar}   constraints that will arise for different choices of the  branching ratio.   We additionally require that the maximum allowed branching ratio (\ref{eq:scalarBR}) is consistent with the resulting limit, which sculpts the shape of the contours at high mass.  Also shown (computed with the same assumptions for scalar decay widths) are the projected constraint contour for the SHiP experiment \cite{Lanfranchi:2243034,Anelli:2015pba,Alekhin:2015byh} in dashed orange, and current constraints on the parameter space from LEP Higgs searches (light red) \cite{Acciarri:1996um,Buskulic:1993gi}, $K^\pm \to \pi^\pm +\mbox{invisible}$ at E949 \& E787 (light blue) \cite{Artamonov:2009sz}, the CHARM beam dump (gold) \cite{Bergsma:1985qz}, and rare $B$ decays at LHCb (light green and brown) \cite{Aaij:2015tna,Aaij:2016qsm}.  Not shown are other proposals to find light, Higgs-mixed scalars at CODEX-b \cite{Gligorov:2017nwh} and FASER \cite{Feng:2017uoz}.
 
 In summary, MATHUSLA would allow us to peer deeply into the SM+S parameter space, both via exotic Higgs decays and via meson decays. In the former case, the sensitivity is orders of magnitude better than main detector LLP searches, as discussed in Section~\ref{sec:exhdecays}. The latter are even more challenging at the main detectors, and is the target of proposed experiments like SHiP. MATHUSLA would be able to extend the reach of these experiments to significantly smaller mixing angles.

\newcommand{\SUWeak}{\mathrm{SU}(2)_{\rm W}}
\newcommand{\Lag}{\mathcal{L}}
\newcommand{\Lagtree}{\mathcal{L}_{\rm tree}}
\newcommand{\mHpm}{m_{H^\pm}}
\newcommand{\Ap}{A^\prime}
\newcommand{\hd}{h_{\rm D}}
\newcommand{\Hd}{H_{\rm D}}
\newcommand{\ed}{e_{\rm D}}
\newcommand{\ad}{\alpha_{\rm D}}
\newcommand{\vd}{v_{\rm D}}
\newcommand{\lamd}{\lambda_{\rm D}}
\newcommand{\mZ}{m_Z}
\newcommand{\mHd}{m_{h_{\rm D}}}
\newcommand{\mAp}{m_{A^\prime}}
\newcommand{\fb}{\mathrm{fb}}
\newcommand{\UD}{U(1)_{\rm D}}
\newcommand{\thh}{\theta_h}
\def\babar{\mbox{\slshape B\kern-0.1em{\smaller A}\kern-0.1em
    B\kern-0.1em{\smaller A\kern-0.2em R}}}

\subsection[SM + V: Dark Photons]{SM + V: Dark Photons\footnote{Nikita Blinov, Jae Hyeok Chang, David Curtin, Rouven Essig, Brian Shuve}}
\label{sec:darkphotons}

Dark sectors can contain mediator particles that allow for interactions with SM particles through portals, see e.g.~\cite{Jaeckel:2010ni,Hewett:2012ns,Essig:2013lka,Alexander:2016aln,Battaglieri:2017aum} for recent reviews.  
If the dark sector contains a dark abelian gauge group $U(1)_D$, this may give rise to what is known as the ``vector portal'', a renormalizable kinetic mixing between the \emph{dark photon} and the SM hypercharge gauge boson:~\cite{Holdom:1985ag,Galison:1983pa},  
\begin{equation} \label{kmix} 
{\cal L} \supset -\frac{\epsilon}{2\cos\theta_W} F'_{\mu\nu} F^{\mu\nu}_Y, 
\end{equation}
Here $\epsilon$ is the kinetic-mixing parameter, $\theta_W$ is the Weinberg mixing angle, 
$F'_{\mu\nu}=\partial_{\mu}A'_{\nu}-\partial_{\nu}A'_{\mu}$ 
is the $U(1)_D$ field strength, and similarly $F^{\mu\nu}_Y$ denotes the SM hypercharge $U(1)_Y$ field strength. 
This mixing allows $A'$s to be produced in charged particle interactions.
The value of $\epsilon$ is arbitrary, but a value of $\epsilon^2 \sim 10^{-8}-10^{-4}$ is natural if generated by quantum effects of heavier particles charged under $U(1)_D$ and $U(1)_Y$. Since the operator is renormalizable, new physics effects at any scale can generate detectable kinetic mixings.
If the SM forces unify in a Grand Unified Theory, 
then $\epsilon^2 \sim 10^{-12}-10^{-6}$ is natural~\cite{ArkaniHamed:2008qp,Baumgart:2009tn,Essig:2009nc}.

\begin{figure}[t]
\begin{center}
\hspace*{-1cm}
\begin{tabular}{cc}
\includegraphics[width=0.5\textwidth]{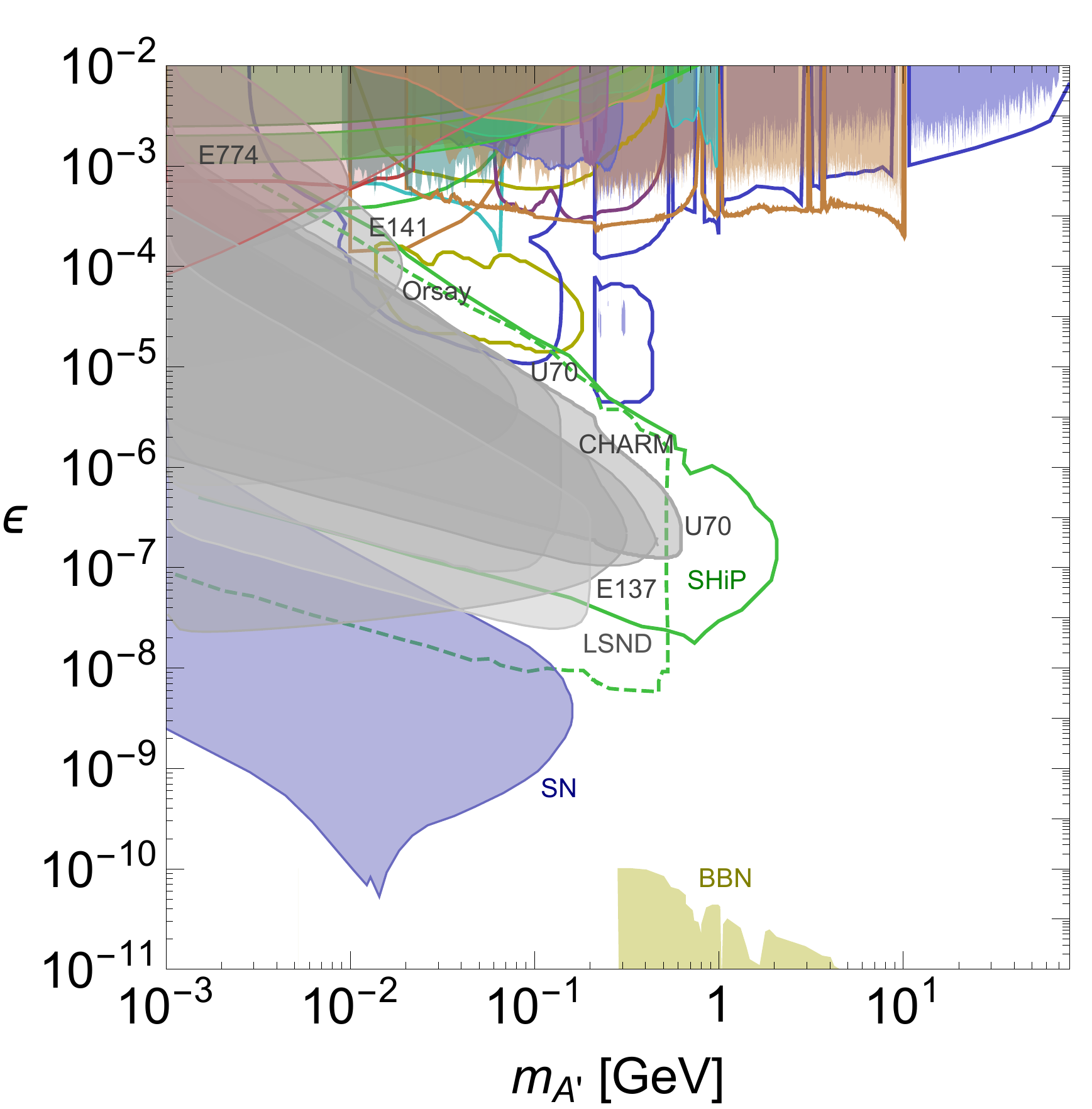}
&
\includegraphics[width=0.5\textwidth]{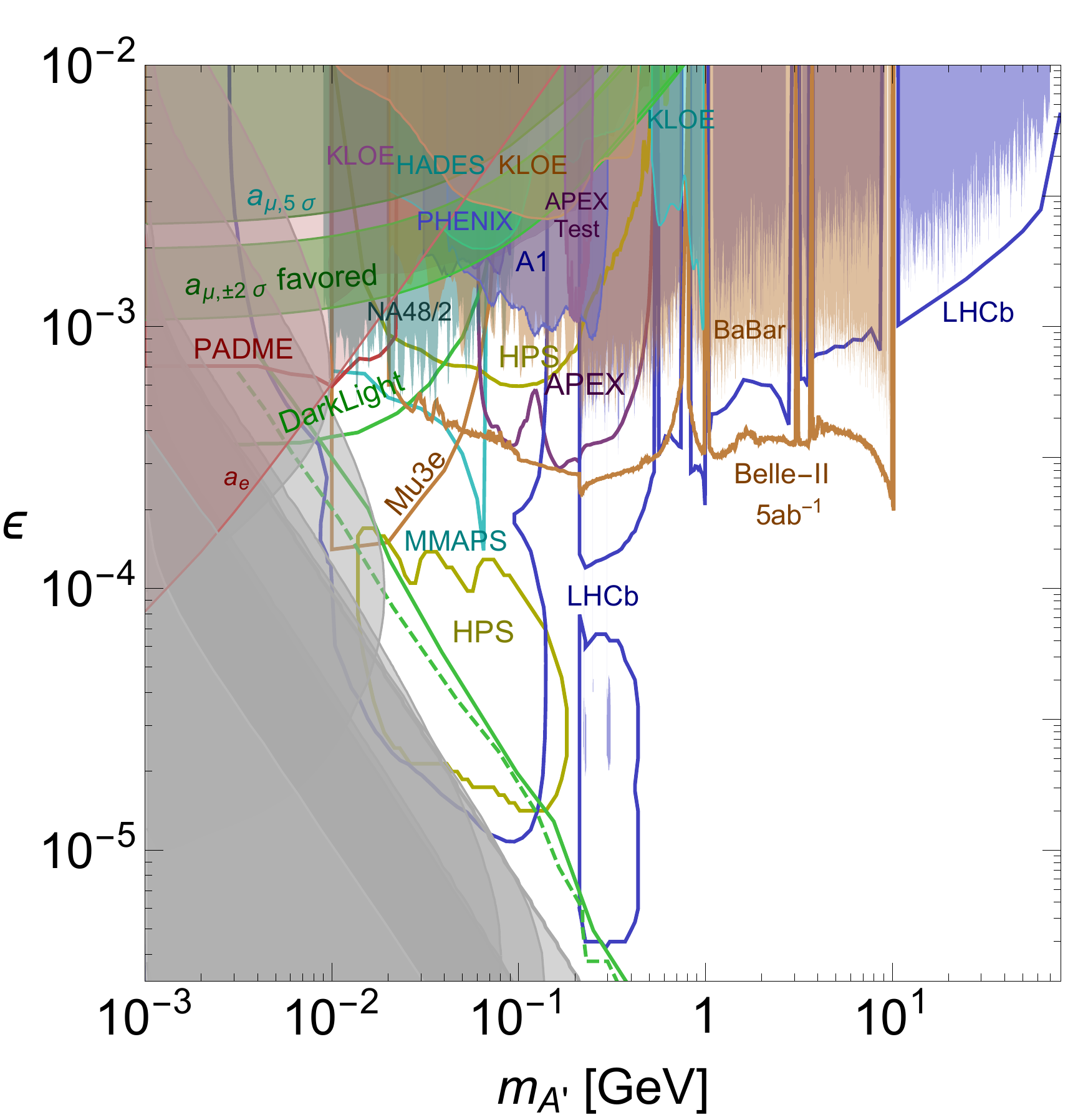}
\end{tabular}
\caption{
Existing 95\% confidence level limits (gray shaded regions) on dark photons ($A'$) and proposed experimental searches. An $A'$ inside the green band can explain the discrepancy between the measured and 
calculated value of the muon anomalous magnetic moment. 
For a figure showing all proposed searches and references, see e.g.~\cite{Alexander:2016aln}. }
\label{fig:A'}
\end{center}
\end{figure}

The dark photon can be massive if $U(1)_D$ is broken, the most obvious mechanism for this breaking being a dark Higgs mechanism at a scale close to the dark photon mass.
This is called the Hidden Abelian Higgs Model, see e.g.~\cite{Curtin:2014cca} for a full description.
The massive dark photon can then decay into SM particles through the small kinetic mixing, making it a possible LLP. 
The dark Higgs $h_D$ is expected to have some degree of mixing with the SM Higgs via the Higgs portal 
\begin{equation}
\mathcal{L} \supset \kappa |h_D|^2 |H|^2 \ ,
\end{equation}
since such a term cannot be forbidden by symmetries. This mixing would provide another production mode for dark photons in exotic Higgs decays. 

Fig.~\ref{fig:A'} shows constraints on an $A'$ with mass between the MeV and the weak scale, assuming $A'$ decays only to SM particles~\cite{Riordan:1987aw,Bjorken:1988as,Bross:1989mp,Batell:2009yf,Bjorken:2009mm,Blumlein:2011mv,Andreas:2012mt,Pospelov:2008zw,Reece:2009un,Aubert:2009cp,Archilli:2011zc,Abrahamyan:2011gv,Merkel:2011ze,Merkel:2014avp,Babusci:2012cr,Babusci:2014sta,Lees:2014xha,Batley:2015lha,::2016lwm,Babusci:2015zda,Chang:2016ntp}.  
The $A'$ mass is arbitrary, but this range arises naturally in several models~\cite{Cheung:2009qd,ArkaniHamed:2008qp,Fayet:2007ua,Morrissey:2009ur}. 
An $A'$ can also explain the discrepancy between the measured and calculated value of the muon anomalous magnetic moment~\cite{Gninenko:2001hx,Pospelov:2008zw} (Fig.~\ref{fig:A'}, green band), although non-SM decays are needed.  

\subsubsection{\bf $A'$ as a LLP}

\begin{figure}
\begin{center}
\includegraphics[width=0.6\textwidth]{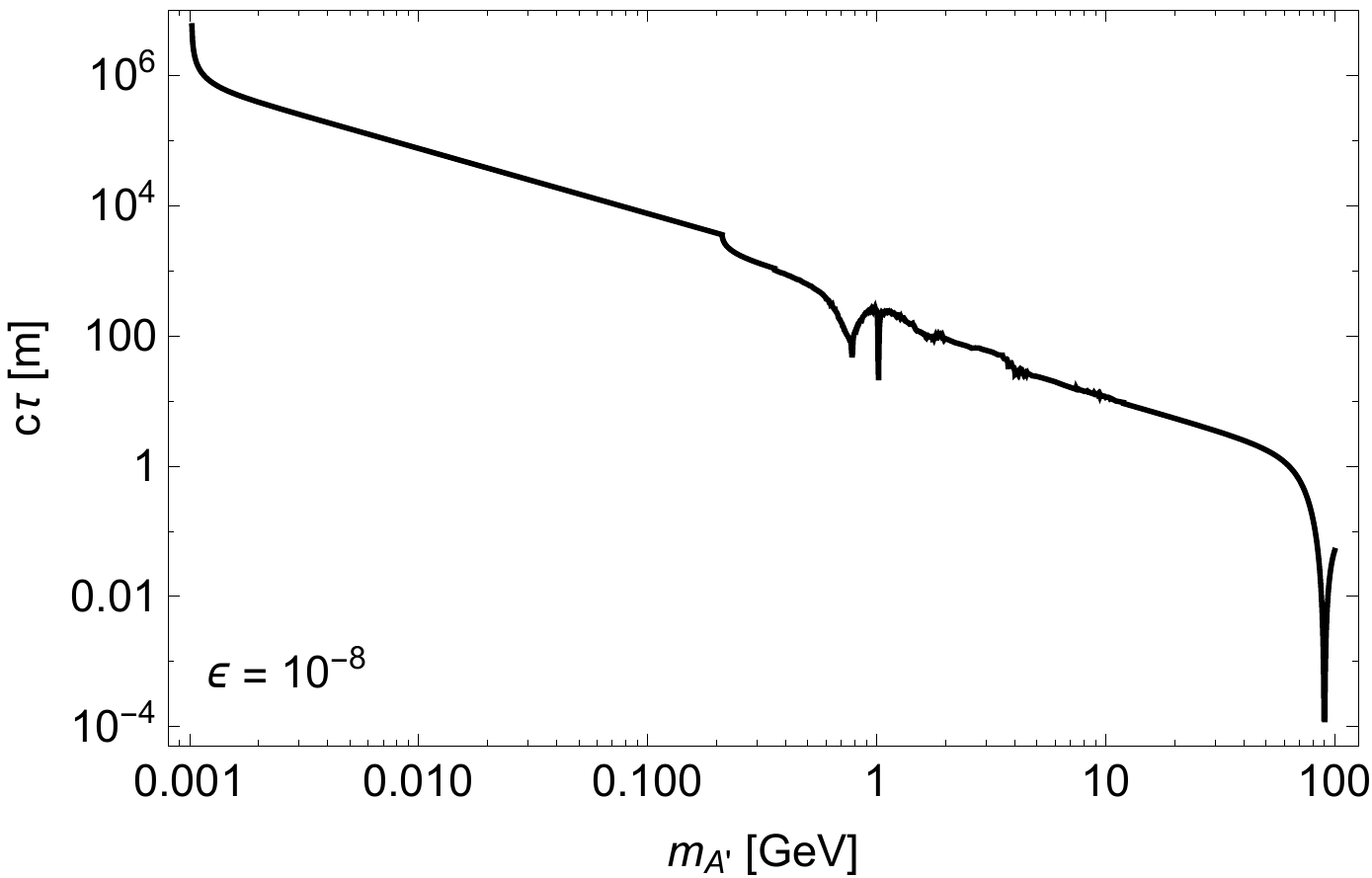}
\end{center}
\caption{
Dark photon lifetime for $\epsilon = 10^{-8}$ when $A'$ can only decay to SM particles. $c\tau$ scales as $\epsilon^{-2}$.
}
\label{f.darkphotonlifetime}
\end{figure}

\begin{figure}
\begin{center}
\includegraphics[width=0.8\textwidth]{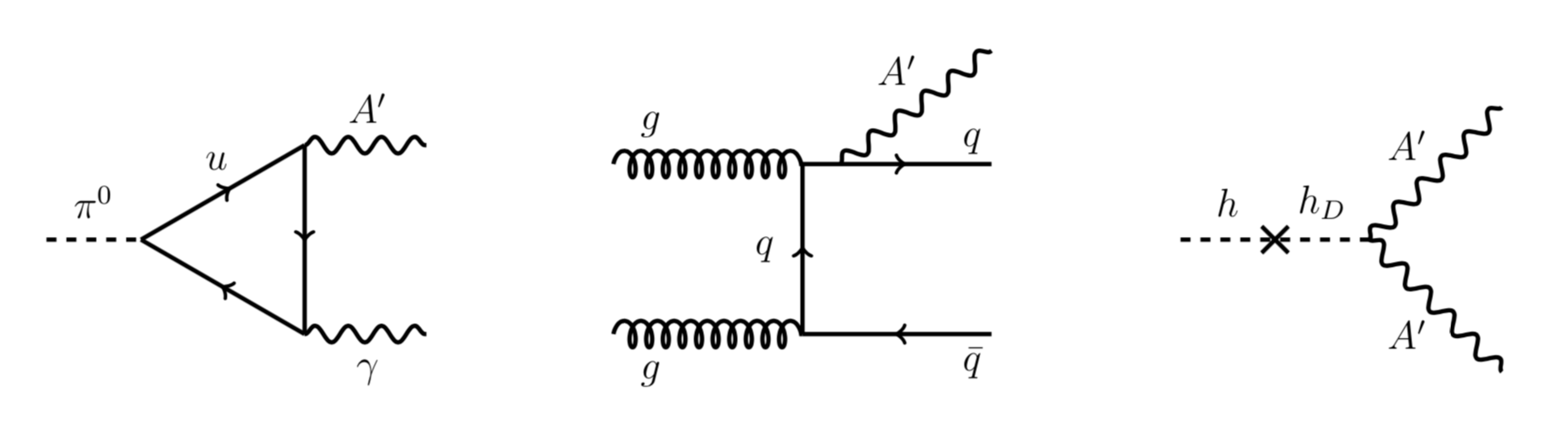}
\caption{Examples of the Feynman diagrams for dark photon production at the LHC. Left: pion decay to the dark photon and a SM photon. Middle: bremsstrahlung of the dark photon during gluon-gluon collision. Right: exotic Higgs decay to two dark photons via mixing with the dark Higgs.}
\label{fig:darkphotonfd}
\end{center}
\end{figure}

We first consider the case that the dark photon is the lightest (or only) dark-sector particle. 
In this case, once produced, it can decay only to three photons (for $m_{A'}<2m_e$) or 
two charged SM particles (for $m_{A'}>2m_e$). 
However, the decay length to three photons is much larger than 200~m even for $\epsilon \sim 1$. Since the reconstruction of such a light LLP that decays only to photons would anyway be very challenging at MATHUSLA, we only consider the regime $m_{A'}>2m_e$.
In this case, the decay width to electrons is given by
\begin{equation}
\Gamma(A' \rightarrow e^- e^+) = \frac{1}{3}\epsilon^2 \alpha \left(1+\frac{2m_e^2}{m_{A'}^2}\right) \sqrt{m_{A'}^2-4m_e^2}
\,\Theta(m_{A'}-2m_e)\,.
\end{equation} 
with similar expressions for the other SM fermions at tree-level. Below the $\bar b b$ threshold, threshold effects and hadronic corrections cannot be neglected, but they can be accounted for using $e^+ e^- \to \mathrm{hadrons}$ experimental data, see e.g. \cite{Curtin:2014cca} for details. 
For small $\epsilon$, the dark photon will be long-lived, see Fig.~\ref{f.darkphotonlifetime}.

\begin{figure}[t]
\begin{center}
\includegraphics[width=0.6\textwidth]{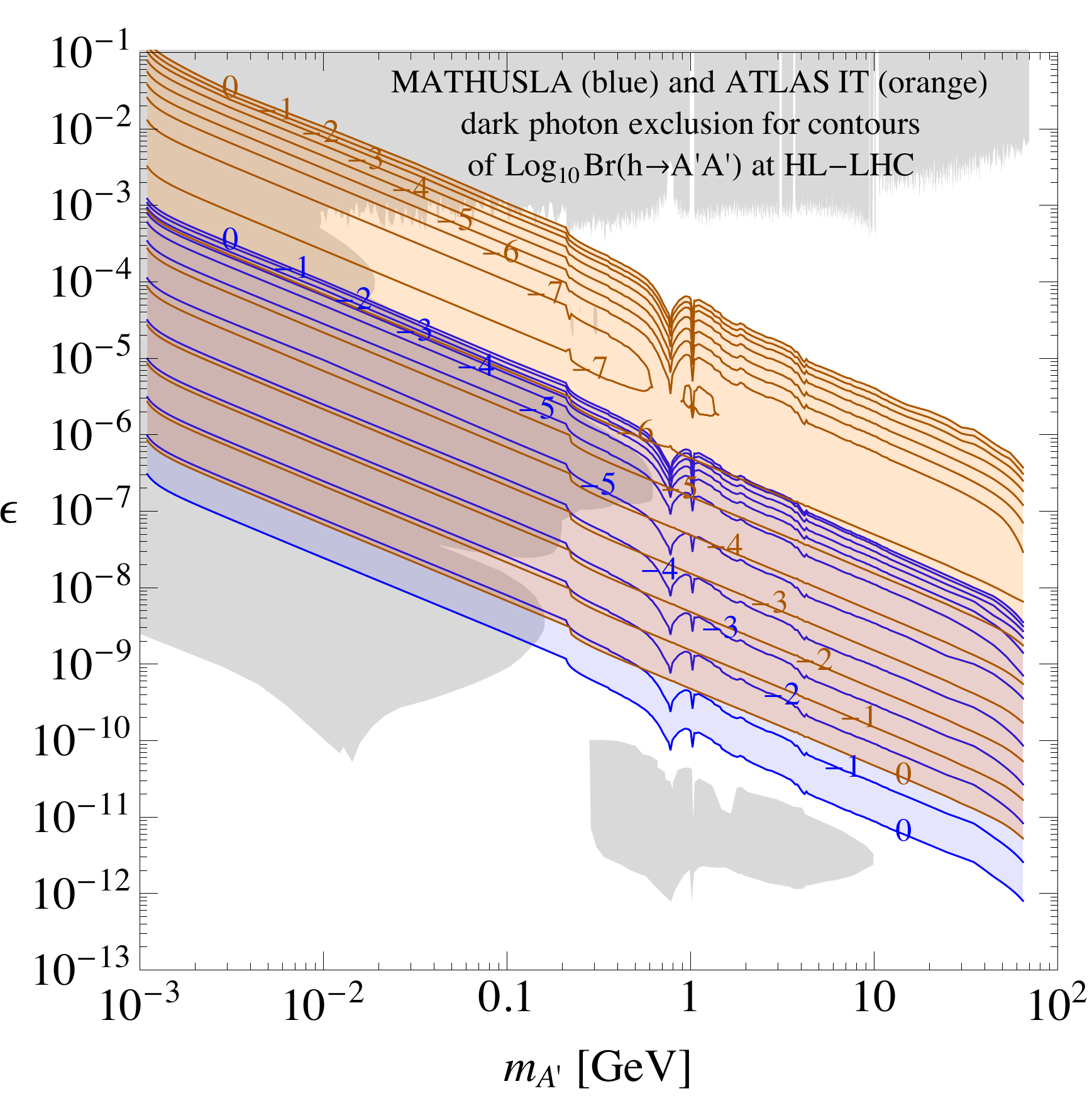}
\caption{
Regions of dark photon parameter space that could be probed by an LLP search at MATHUSLA (blue) or the ATLAS inner tracker (orange) assuming dark photon production in exotic Higgs decays with the indicated $\mathrm{Br}(h \to A' A')$. 
Note that MATHUSLA searches for the production of low-mass dark photons with $m_{A'} \lesssim \gev$ from exotic Higgs decays may suffer from some backgrounds or lower reconstruction efficiency depending on the final detector design, see Section~\ref{s.energythresholds}.
This estimate assumes the $200m \times 200m \times 20m$ benchmark geometry of Fig.~\ref{f.mathuslalayout}.
}
\label{fig:higgsdecay}
\end{center}
\end{figure}

There are several production channels for dark photons at the LHC. Two important processes that assume only the presence of kinetic mixing are meson decays and bremsstrahlung processes, see Fig.~\ref{fig:darkphotonfd}.   
Assuming only kinetic mixing, the abundant meson production rates at the LHC in QCD jets mean their decay is the dominant production mechanism for dark photons, as long as the dark photon mass is below the meson mass.  
The most important meson decays are pion and eta decays, while other mesons 
contribute only a negligible amount to the total production.   
At higher masses, bremsstrahlung processes are the dominant source of directly produced dark photons. However, since both of these processes have dark photon production rates which scale as $\epsilon^2$, they lead to very small signal rates in the long-lifetime regime that is accessible by MATHUSLA. As a result, the only regions of $(m_{A'}, \epsilon)$ parameter space that lead to a MATHUSLA signal assuming dark photon production via kinetic mixing are already excluded from past beam-dump experiments~\cite{Alexander:2016aln}.

Dark photons can also be produced in exotic Higgs decays, shown on the right in  Fig.~\ref{fig:darkphotonfd}. This production rate depends on the mixing between the dark and the visible Higgs, leading to possible branching ratios as large as $\sim 10\%$. Since this production rate does not depend on $\epsilon$, it allows for very small kinetic mixings to be probed if the search is sensitive to long lifetimes.

Similar to the LLP searches studied in~\cite{Curtin:2014cca}, we can show the regions of $(m_{A'}, \epsilon)$ parameter space that can be probed assuming a certain exotic Higgs branching ratio $\mathrm{Br}(h \to A' A')$ in Fig~\ref{fig:higgsdecay}.
The MATHUSLA sensitivity, corresponding to 4 decays in the detector, is shown as the blue contours. For comparison, we also show the sensitivity of an ATLAS search for a single LLP decay in the inner tracker (see definitions in Section~\ref{s.LHCLLPsignal}), where the dark photon LLP is required to decay leptonically for triggering and background rejection purposes. This main detector search is also assumed to be background-free, but this is likely too optimistic, especially for dark photon masses below $\sim$ 10 GeV, see discussion in Section~\ref{s.LHCLLPcomparison}. 
Even with these generous assumptions for the ATLAS search, MATHUSLA is able to probe about an order of magnitude smaller kinetic mixings down to $\epsilon \sim 10^{-12}$ , representing the greatest sensitivity to small mixing possible at any experiment in that mass range.

\subsubsection{\bf $A'$ decaying to LLP's}

\begin{figure}[t]
\begin{center}
\includegraphics[width=0.5\textwidth]{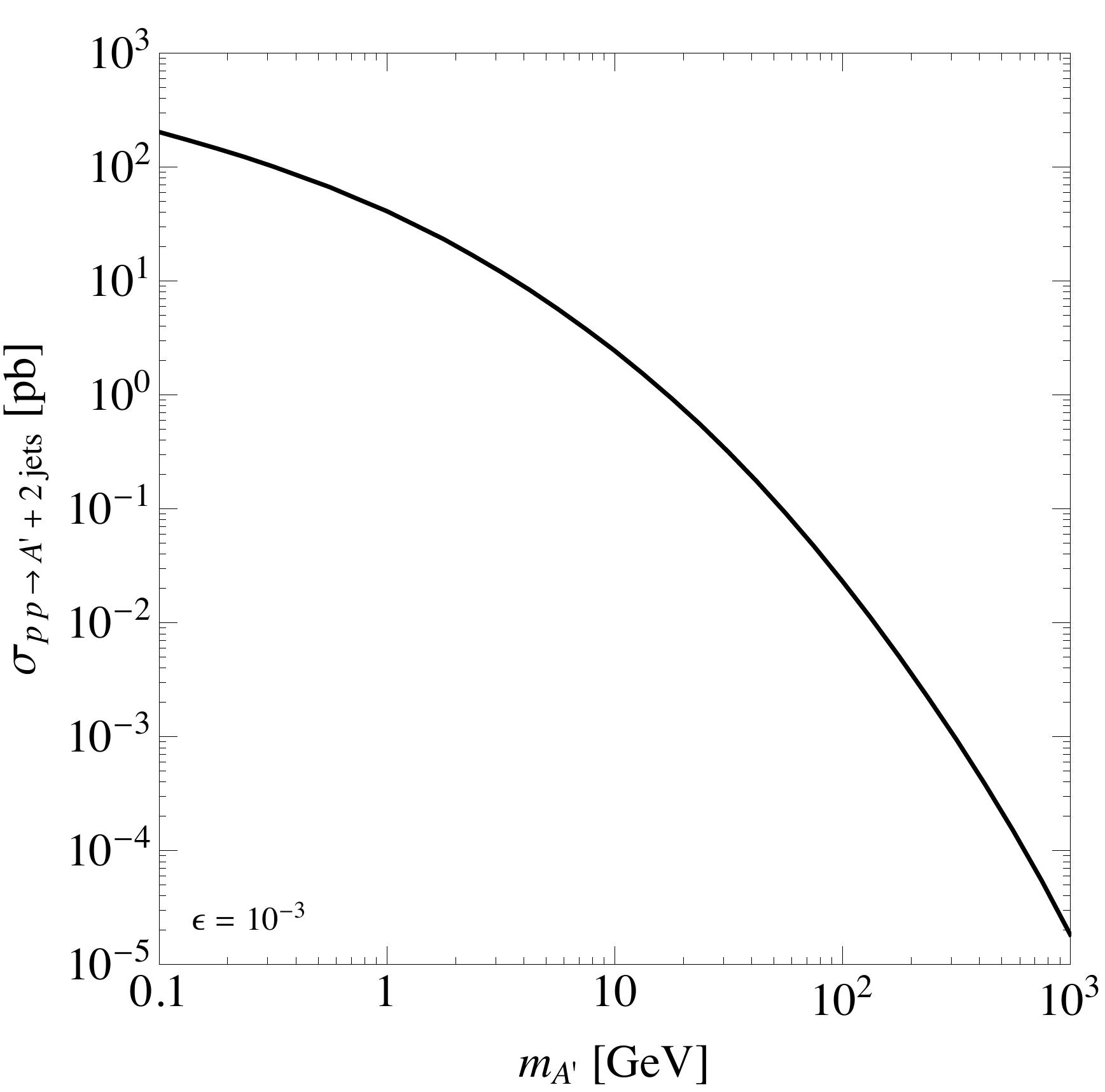}
\caption{Cross-sections for the bremsstrahlung of dark photon at LHC in terms of dark photon mass. Here, $\epsilon = 10^{-3}$ is chosen, and the cross-section is proportional to $\epsilon^2$.}
\label{fig:pptoapjjxsec}
\end{center}
\end{figure}

\begin{figure}[t]
\begin{center}
\includegraphics[width=0.5\textwidth]{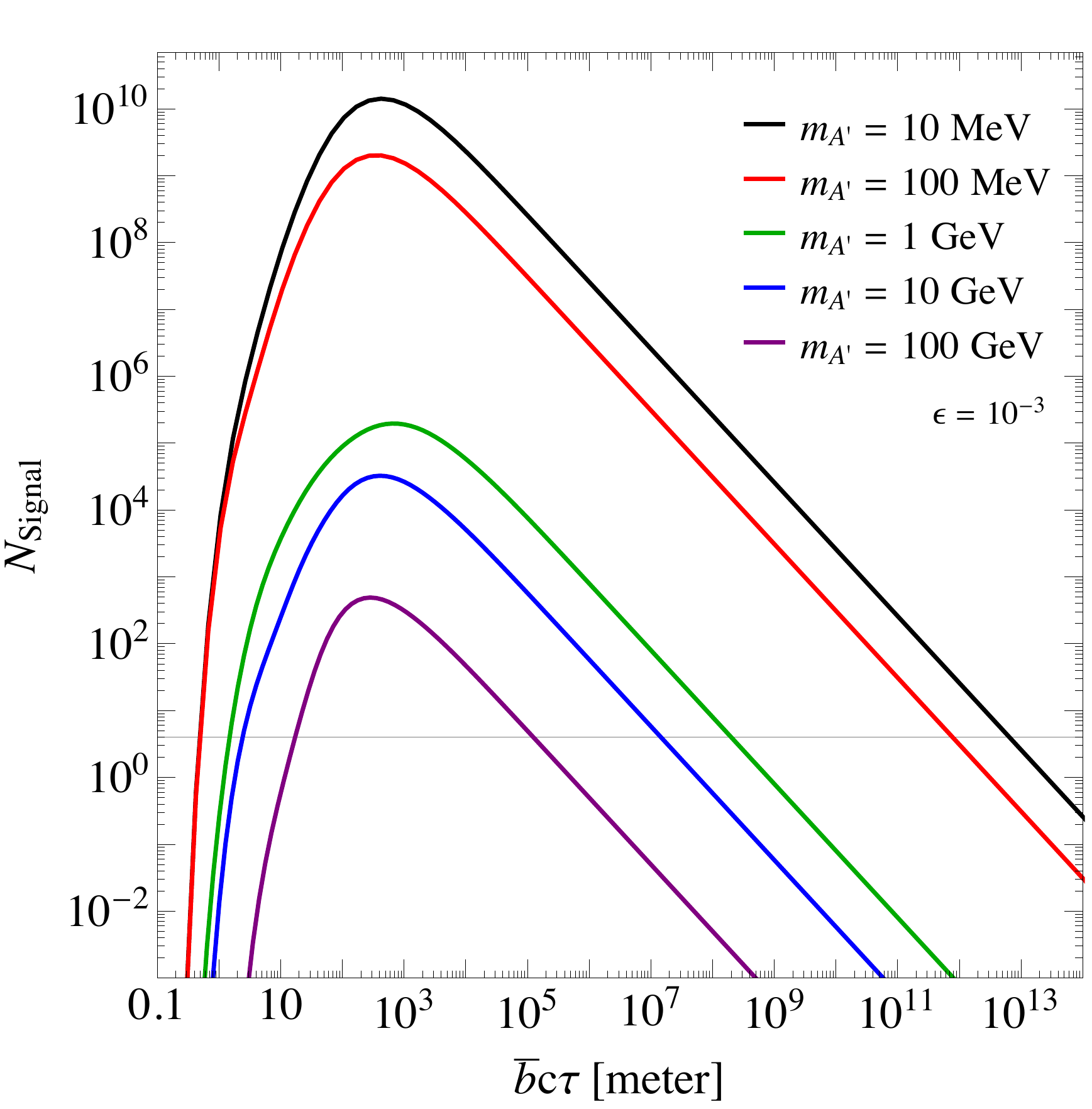}
\caption{Projections at MATHUSLA for the case that the dark photon decays to two LLP's. The number of signal is presented in terms of $\bar b c \tau$ of the LLP for different dark photon masses. Here, $\epsilon=10^{-3}$ is used, and $N_{\textup{Signal}}$ is proportional to $\epsilon^2$. For $m_{A'}=10 \textup{MeV}$ and $m_{A'}=100 \textup{MeV}$, the dominant dark photon production process is meson decay, and bremsstrahlung of dark photon for higher masses. The grid line is for $N_{\textup{Signal}}=4$. $\bar b$'s from the numerical simulations are $\bar b = 1.7 m_\pi/m_\textup{LLP}$ for $m_{A'}=10\textup{MeV}$ and $m_{A'}=100\textup{MeV}$, $\bar b = 2.0 m_{A'}/m_\textup{LLP}$ for $m_{A'}=1\textup{GeV}$, $\bar b = 0.80 m_{A'}/m_\textup{LLP}$ for $m_{A'}=10\textup{GeV}$, and $\bar b = 0.45 m_{A'}/m_\textup{LLP}$ for $m_{A'}=100\textup{GeV}$.
This estimate assumes the $200m \times 200m \times 20m$ benchmark geometry of Fig.~\ref{f.mathuslalayout}.
}
\label{fig:pptoapjjdecay}
\end{center}
\end{figure}



We next consider the highly generic possibility that there are additional dark-sector particles charged under $U(1)_D$ to which the dark photon can decay. 
For example, the $U(1)_D$ could be part of a confining hidden valley~\cite{Strassler:2006im,Essig:2009nc} that gives rise to bound states of the hidden QCD-like interaction in the IR. If these hidden hadrons have mass scale $m_D$ below the dark photon mass, the width for their decay to SM fermions via an off-shell $A'$ is roughly
\begin{equation}
\label{e.darkphotonHVdecaywidth}
\Gamma \sim \frac{\alpha_D \alpha \epsilon^2}{18 \pi} \frac{m_D^5}{m_{A'}^4} \ .
\end{equation}
It is then possible for $\epsilon$ to be relatively large, leading to sizable dark photon production rates through the kinetic mixing operator, while the dark hadron decay length could easily be much larger than the size of the main detectors for modest hierarchies of $m_D/m_{A'}$.
Just as was the case for dark photon LLP production via exotic Higgs decays, LLP lifetime is now largely decoupled from the LLP production rate (now via dark photon decays). 
This is  a prime signal to search for at MATHUSLA, especially (but not only) in the regime where the hidden hadrons have mass below $\sim 10 \gev$, leading both to long lifetimes and making the background-free reconstruction of the associated displaced vertices at the main detectors more difficult. 

To understand the number of LLPs that might be produced in dark photon decay, we first need the \emph{total} dark photon production rate at the LHC as a function of $\epsilon$ and $m_{A'}$. (Here we assume only production processes that rely on kinetic mixing.)
For $m_{A'} \lesssim m_\eta$, meson decay is the dominant dark photon production mode. At higher masses, bremsstrahlung production $p ~ p \rightarrow A' + 2 ~ \textup{jets}$ is the most important process (we checked that $p ~ p \rightarrow A' + 1 ~ \textup{jet}$ is smaller). 

To estimate dark photon production in meson decays, we use the event generator SIBYLL2.3 \cite{Ahn:2009wx,Riehn:2015oba} to compute the total number of pions and etas produced at the HL-LHC: 
\begin{equation}
N_\pi=1.4\times10^{19} \ \ \ \ \ \mathrm{and} \ \ \ \ \ \ N_\eta=3.0\times10^{17}.
\end{equation}
The branching fraction to dark photons are
\begin{eqnarray}
\textup{Br}(\pi[\eta] \rightarrow A' \gamma) &=& 2 \epsilon^2 \textup{Br}(\pi[\eta] \rightarrow 2\gamma) \left(1-\frac{m_{A'}^2}{m_{\pi[\eta]}^2}\right)^3 \Theta(m_{\pi[\eta]}-m_{A'})\\
\end{eqnarray}
The dark photon production rate in bremsstrahlung processes is computed in MadGraph~5~\cite{Alwall:2011uj} with a minimum jet $p_T$ of 10~GeV.\footnote{A more sophisticated matched calculation would give a somewhat higher dark photon yield, but our conservative estimate is sufficient to demonstrate the importance of dark photons as a potential LLP source.}
It also scales with $\epsilon^2$ and is shown as a function of dark photon mass in  Fig.~\ref{fig:pptoapjjxsec}.

We now assume that the $A'$ decays directly to two long-lived LLPs with mass $m_{\rm LLP}$ and lifetime $c\tau$.  
Given the possible high multiplicity of the produced dark mesons after hidden sector hadronization, this is a simplistic assumption that likely underestimate the LLP  signal. Nevertheless, it is instructive to show what range of hidden sector lifetimes could be probed at MATHUSLA for different kinetic mixings. 

We use simulations (as described above) to calculate the average boost of the LLP, and confirm that Eqn.~(\ref{e.NobsMATHUSLA}) gives a very good analytical estimate of the number of observed decays in MATHUSLA in the long-lifetime regime if we use $\epsilon_{\textup{geometric}} \sim 0.02$. 
Fig.~\ref{fig:pptoapjjdecay} then shows the resulting number of LLPs that decay in the MATHUSLA detector for various $m_{A'}$.  
For $m_{A'} = 10~\textup{MeV}$ and $m_{A'} = 100~\textup{MeV}$, production from meson decay dominates, so we use the results from SIBYLL2.3 with assumptions that the LLPs are co-linear to the decaying mesons and carry a quarter of the decaying meson's momentum. For the other values of $m_{A'}$ shown in the figure, bremsstrahlung is the only relevant production process, and we use MadGraph 5 to simulate $p ~ p \rightarrow 2 ~ \textup{jets} + (A' \rightarrow X ~ \bar X)$, where $X$ is the LLP charged under $U(1)_D$, assumed to have mass $m_X \ll m_{A'}$.
As expected, the number of events has a peak near $\bar b c \tau \sim 200~\textup{m}$. 
In this model scenario, MATHUSLA can probe the parameter space between the intersections of the thick lines and the grid line, where the 
grid line indicates 4 signal events.

\subsubsection{Long-lived dark Higgs production in exotic $Z$ decays}

An important test of the Hidden Abelian Higgs Model is the verification of the origin of symmetry breaking in the hidden sector. If a dark Higgs boson, $h_{\rm D}$, is responsible for giving mass to a hidden photon, $A'$, then the dark Higgs can be produced in association with a dark photon:~this is the dark Higgs-strahlung process \cite{Batell:2009jf,Blinov:2017dtk}. Depending on the magnitude of the hidden-sector gauge coupling, $\alpha_{\rm D}$, the dark Higgs-strahlung process can give the best sensitivity to hidden sector parameters. There exist searches for dark Higgs-strahlung at $B$-factories via the process $e^+e^-\rightarrow {A'}^*\rightarrow A'h_{\rm D}$, which set the best limits on the hidden sector for large values of $\alpha_{\rm D}\gtrsim10^{-3}-10^{-2}$ \cite{Lees:2012ra,TheBelle:2015mwa}. However, there are currently no searches at the LHC that are sensitive to this process.

  \begin{figure}
    \centering
    \includegraphics[width=0.35\textwidth]{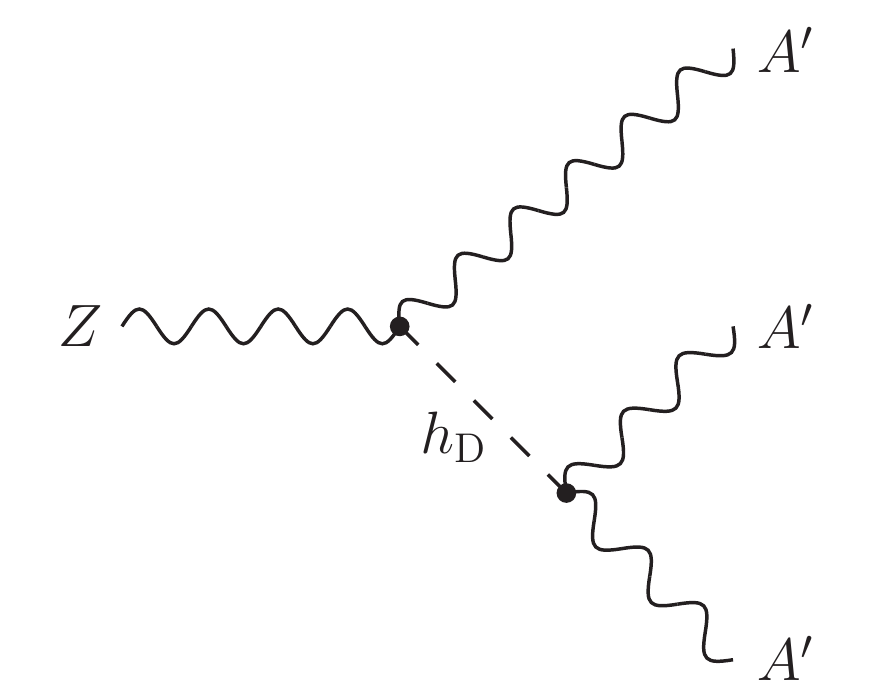}
    \caption{Feynman diagram illustrating dark Higgs ($\hd$) and dark photon ($A'$) production in rare $Z$ boson decays. 
      The dark photons in $\hd$ decay can be on- or off-shell.
    \label{fig:feynman_Zdecay}}
  \end{figure}

  \begin{figure}
    \centering
    \includegraphics[width=0.7\textwidth]{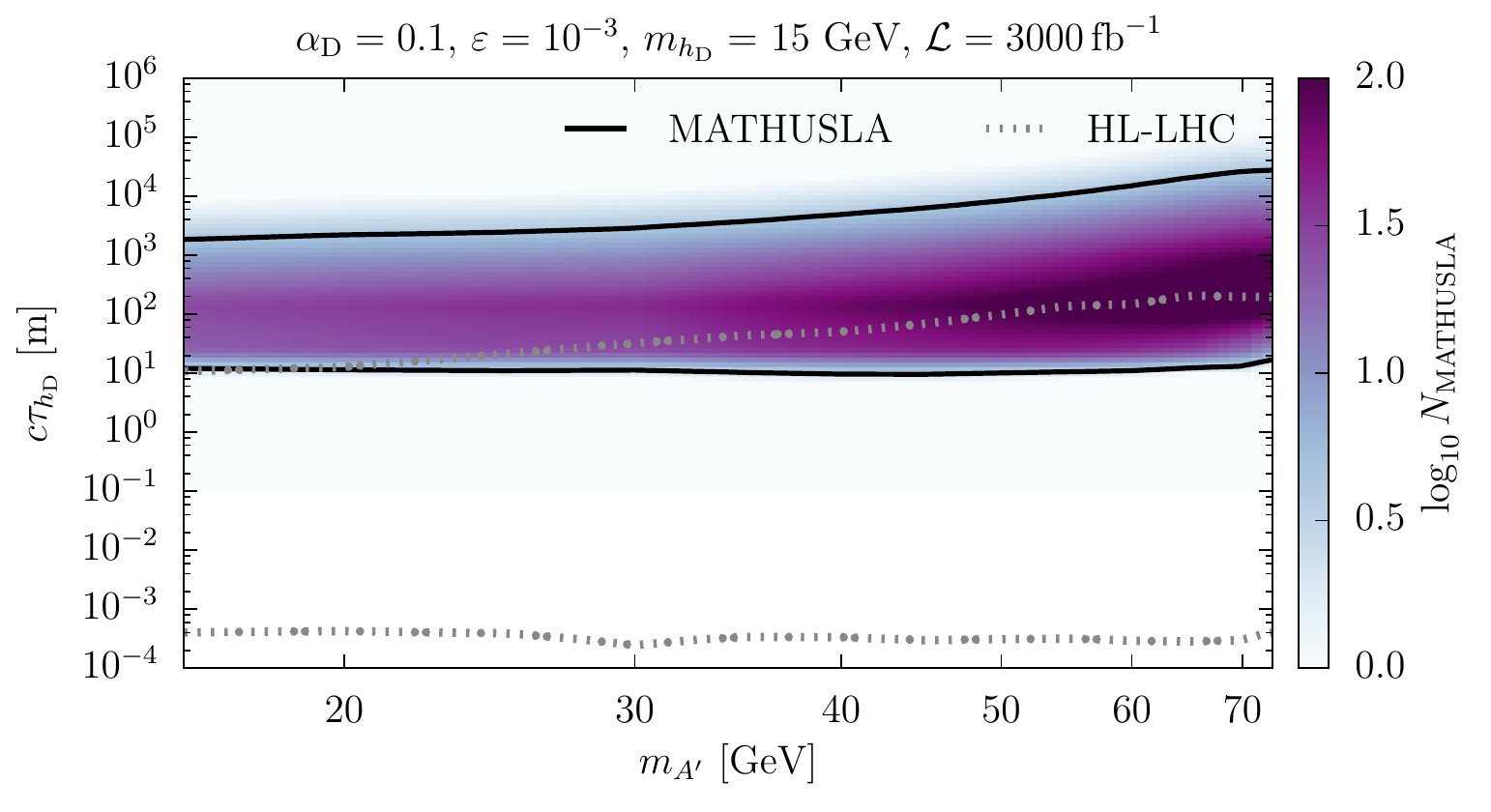}
    \caption{Sensitivity of MATHUSLA to the dark Higgs-strahlung process is shown in bold solid lines for the case where $\ad=0.1$, $\varepsilon=10^{-3}$, and $\mHd=15$ GeV with varying $\mAp$ mass. The colormap shows the approximate event yield and demonstrates that MATHUSLA has sensitivity to proper lifetimes in the $10-10^4$ m range. By comparison, the sensitivity of ATLAS and CMS is shown by the region within the gray dotted lines; their reach is optimal for lower lifetimes.
    This estimate assumes the $200m \times 200m \times 20m$ benchmark geometry of Fig.~\ref{f.mathuslalayout}.
    \label{fig:displaced_sensitivity_mHd_15_L3000_mA_vs_ctau_mathusla}}
  \end{figure}
  
  \begin{figure}
    \centering
    \includegraphics[width=0.7\textwidth]{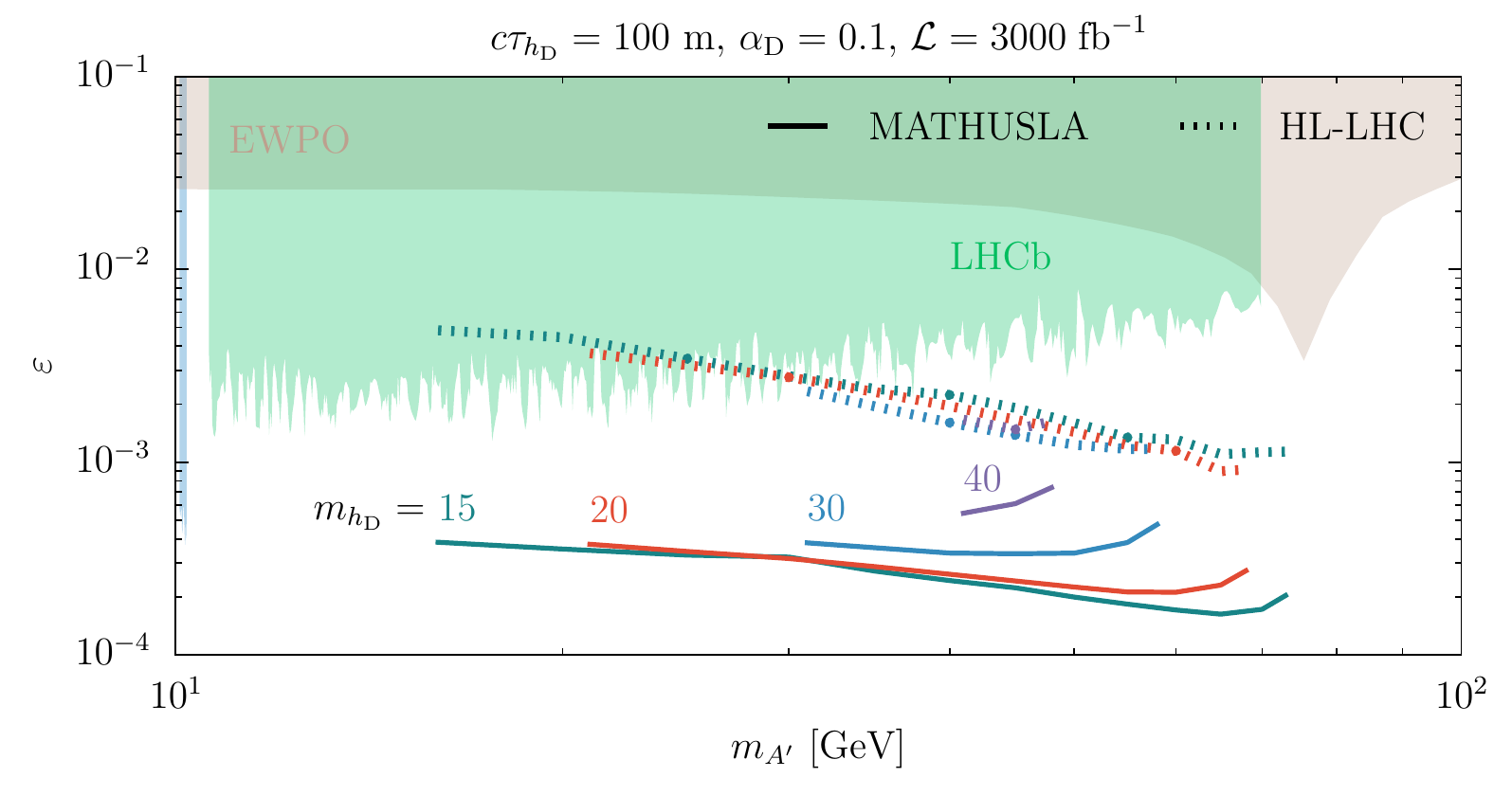}
    \caption{Sensitivity of MATHUSLA to the dark photon parameter space for the particular case of $c\tau=100$ m, $\ad=0.1$. MATHUSLA sensitivity is shown with solid lines while ATLAS/CMS sensitivity is shown with dashed lines. The $\hd$ masses are given in GeV. For comparison, constraints from LHCb \cite{Aaij:2017rft}, electroweak precision observables (EWPO) \cite{Hook:2010tw}, and at the far left, BABAR \cite{Lees:2014xha}.
    This estimate assumes the $200m \times 200m \times 20m$ benchmark geometry of Fig.~\ref{f.mathuslalayout}.
    \label{fig:displaced_sensitivity_L3000_mA_vs_eps_ct_100m_mathusla}}
  \end{figure}

  When kinematically accessible, the dominant Higgs-strahlung process at the LHC is in the rare $Z$ boson decay, $Z\rightarrow A' h_{\rm D}$, shown in Fig.~\ref{fig:feynman_Zdecay}. The dark Higgs subsequently decays into on- or off-shell $A'$, so that the final signature is $Z\rightarrow A'{A'}^{(*)}{A'}^{(*)}$, leading to final states with multiple resonances and high multiplicities of soft leptons. This scenario has been proposed and recently studied in detail in Ref.~\cite{Blinov:2017dtk}. In particular, this rare $Z$ decay mode is promising for moderately large $\alpha_{\rm D}\gtrsim0.05$, and small SM-dark Higgs mixing ($\kappa^2\ll \alpha_{\rm D}$). The branching fraction of $Z\rightarrow A' h_{\rm D}$ is
  \be
  \mathrm{Br}(Z\rightarrow \Ap \hd) = 
  \frac{2\ad \varepsilon^2 \tan^2\theta_{\rm W} \,\mAp^2 \mZ}{3(\mZ^2-m_{A'}^2)^2}
  \left[1+\frac{(\mZ^2+\mAp^2-\mHd^2)^2}{8\mZ^2 \mAp^2}\right]
  \beta\left(\frac{\mAp}{\mZ},\frac{\mHd}{\mZ}\right),\label{eq:z_branching}
  \ee
  where $\beta(x,y) = [(1-(x-y)^2)(1-(x+y)^2)]^{1/2}$ and we assumed that $\varepsilon\ll1$ and that the $h-h_{\rm D}$ mixing angle is negligibly small.

Most relevant for MATHUSLA is the parameter regime where the dark Higgs, $h_{\rm D}$, is long-lived. This occurs when $m_{h_{\rm D}}< m_{A'}$, in which case the dark Higgs cannot decay to on-shell dark photons. In this case, the $h_{\rm D}$ typically decays radiatively via off-shell dark photons into two SM fermions $f\bar{f}$,
\begin{eqnarray}
\Gamma(\hd\rightarrow f \bar{f}) & \sim & \frac{\alpha^2 Q_f^4 \ad \varepsilon^4}{32\pi^2}\left(\frac{m_f}{\mAp}\right)^2 \mHd\label{eq:width-displaced-hypercharge}\\
&\sim& \left(\frac{1}{100\,\,\mathrm{m}}\right)\left(\frac{\ad}{0.1}\right)\left(\frac{\varepsilon}{3\times10^{-3}}\right)^4\left(\frac{15\,\,\mathrm{GeV}}{\mAp}\right).\nonumber
\end{eqnarray}
Additionally, radiative corrections induce a mixing between the SM and dark Higgs even in the absence of a tree-level coupling. The result is UV-sensitive, and therefore the lifetime can take on a wide range of values from the mm-scale to $\gtrsim100$ m depending on the couplings. Because of this dependence on otherwise unobservable UV model parameters, we adopt an approach where we take the dark Higgs lifetime, $c\tau_{h_{\rm D}}$, to be a free parameter of the theory, while the production of the dark Higgs is governed by the kinetic mixing, $\varepsilon$, and the dark gauge coupling, $\alpha_{\rm D}$.

In Figs.~\ref{fig:displaced_sensitivity_mHd_15_L3000_mA_vs_ctau_mathusla} and \ref{fig:displaced_sensitivity_L3000_mA_vs_eps_ct_100m_mathusla}, we project the sensitivity of the proposed MATHUSLA experiment to the dark Higgs-strahlung production of $\hd$, requiring four events in MATHUSLA with $\sqrt{s}=14$ TeV and $\mathcal{L}=3000\,\,\mathrm{fb}^{-1}$ of integrated luminosity. In Fig.~\ref{fig:displaced_sensitivity_mHd_15_L3000_mA_vs_ctau_mathusla}, we show the sensitivity to $\mAp$ and $c\tau$ for fixed $\mHd$ and dark photon couplings, while in Fig.~\ref{fig:displaced_sensitivity_L3000_mA_vs_eps_ct_100m_mathusla}, we show the sensitivity to $\mAp$, $\mHd$ and $\varepsilon$ for fixed $c\tau$ and $\alpha_{\rm D}$. For comparison, we also show the sensitivity of the main ATLAS or CMS detectors to the long-lived $\hd$ scenario, using the selection criteria from Ref.~\cite{Blinov:2017dtk}. For the ATLAS/CMS sensitivity, we require that the $A'$ produced directly in the $Z\rightarrow A'\hd$ process decays promptly to leptons, and that these leptons pass standard dilepton triggers of two opposite-sign, same-flavour (OSSF) muons with $p_{\rm T}>17,\, 8$ GeV, or two OSSF electrons with $p_{\rm T}>23,\,12$ GeV, respectively \cite{Sirunyan:2017lae}. We demand the impact parameters of tracks coming from the $\hd$ decay to satisfy $1\,\,\mathrm{mm}<|d_0|<200\,\,\mathrm{mm}$ and that the physical decay occur within 200 mm of the primary vertex. We additionally assign a 50\% reconstruction penalty for the $\hd$ displaced vertex; backgrounds are expected to be negligible due to the resonant reconstruction of the $\Ap$ mass in the two prompt leptons. It is evident that, for long-lived $\hd$ with $c\tau\sim100$ m, MATHUSLA substantially outperforms ATLAS or CMS; the main LHC detectors have excellent, complementary coverage for lower lifetimes. Thus, MATHUSLA has unique sensitivity to parameters that are well-motivated by the Hidden Abelian Higgs Model and are otherwise unconstrained by existing experiments.

\subsection[Axion-Like Particles]{Axion-Like Particles\footnote{Martin Bauer, Matthias Neubert, Anson Hook, Andrea Thamm}}
\label{sec:alp}

Axion like particles (ALPs) is a collective name for pseudo Nambu-Goldstone bosons with unspecified derivative couplings to Standard Model (SM) particles. The name is inspired by the axion which is the pseudo Nambu-Goldstone boson of the Peccei-Quinn symmetry \cite{Peccei:1977hh,Peccei:1977ur,Weinberg:1977ma,Wilczek:1977pj} introduced to solve the strong CP-problem, but an ALP appears in any theory with a spontaneously broken global symmetry~\cite{Merlo:2017sun, Arias-Aragon:2017eww, Gavela:2018paw, Linster:2018avp, Ahn:2018cau, Ema:2018abj, Bjorkeroth:2017tsz, Brivio:2017sdm, Choi:2017gpf, DiLuzio:2017tjx, Calibbi:2016hwq, Ema:2016ops, Ahn:2014gva}. For some large breaking scale $f$, the ALP can be the harbinger of an ultraviolet sector of physics with masses $M_\text{UV}\propto f$ that is otherwise inaccessible by current and future collider experiments.  Since ALP couplings instead scale as $1/f$, they can be long-lived if the New Physics is heavy, making them prime candidates for experiments probing displaced vertices.\footnote{
This section focuses on ALP production in high-energy processes exclusive to the LHC. Low-mass axions can also be directly produced via their minimal gluon, photon or fermion couplings.  After this whitepaper first appeared as a preprint, the sensitivity of MATHUSLA to such minimally coupled ALP's was computed and compared to the reach of SHiP and other proposed experiments in~\cite{Beacham:2019nyx}. MATHUSLA is highly competitive to minimal light axions with gluon and fermion couplings.
}
Measuring the ALP couplings to SM particles can therefore reveal non-trivial information about a whole New Physics sector. In addition, ALPs can be non-thermal candidates for Dark Matter \cite{Preskill:1982cy}. 
In order for the decays of the ALP Dark Matter not to disturb cosmology, the ALP has to decay before Big Bang Nucleosynthesis \cite{Arias:2012az} (see also Section~\ref{sec:bbn}). This means that the lifetime of the ALP must be  $c\tau_a \lesssim 10^8$ m, providing additional motivation for displaced vertex searches.\footnote{Due to their light masses, ALPs are generically displaced from their minimum during inflation.  After reheating, their energy density behaves like dark energy until Hubble is of order their mass.  Afterwards, they dilute away as normal matter.  Because their energy density does not dilute away like matter until very late, they generically overclose the universe unless they decay or
$
m_a f_a^4 \lesssim (10^7 \, \text{GeV} )^5
$
where we have made the optimistic assumption that the axions start oscillating as soon as they can.  This assumption is not true for some ALPs, e.g. the QCD axion, where the mass term is not present at early times. See also discussions in Sections~\ref{sec:axino-ewino} and~\ref{sec:freezein}}

Up to operators of dimension five, the couplings between the ALP and SM particles are given by the operators
\begin{align}
\label{eq.axionL5}
\mathcal{L}_5= c_{G}\frac{g_s^2}{16\pi^2}\frac{a}{f}\,G_{\mu\nu}^A\tilde G^{\mu\nu,\,A}+c_{W}\frac{g^2}{16\pi^2}\frac{a}{f}\,W_{\mu\nu}^A\tilde W^{\mu\nu,\,A}+c_{B}\frac{g^{\prime\,2}}{16\pi^2}\frac{a}{f}\,B_{\mu\nu}\tilde B^{\mu\nu}+\frac{\partial_\mu a}{f}\sum_i \frac{c_{i}}{2}\,\bar \psi_i \gamma_\mu\gamma_5 \psi_i\,,
\end{align}
where $c_{\gamma\gamma}=c_{W}+c_{B}$ and $c_{\gamma Z}=\cos^2\theta_w\,c_W-\sin^2\theta_W\,c_B$ and $c_{ZZ}= \cos^4\theta_w\,c_W+\sin^4\theta_W\,c_B$ are the relevant Wilson coefficients in the electroweak broken phase, and the couplings to fermions $c_i$ are assumed to be flavour universal. 
Here, $f$ sets the scale of the UV completion and is related to the ALP decay constant by $f=-2c_{G}f_a$. Operators that introduce couplings between the ALP and the Higgs boson $H$ only arise at dimension six and higher,
\begin{align}
\mathcal{L}_{>5}=\frac{c_{ah}}{f^2}\left(\partial_\mu a\right)\left(\partial^\mu a\right) H^\dagger H+\frac{c_{Zh}}{f^3}(\partial^\mu a)\left(H^\dagger iD_\mu  H +\text{h.c.}\right) H^\dagger H\,+\ldots\,,
\end{align}
where the Higgs portal allows for $h \to aa$ decays, whereas the coupling to the Higgs current introduces the decay $h \to Z a$. A possible dimension five operator coupling the ALP to the Higgs current is redundant unless it is induced by integrating out new massive particles that obtain most of their mass from the electroweak scale \cite{Brivio:2017ije, Bauer:2016ydr, Bauer:2016zfj, Bauer:2017nlg, Bauer:2017ris}. An ALP mass can be generated through some external breaking of the corresponding symmetry, or can be dynamically introduced through its couplings to the QCD condensate. In the latter case, the ALP mass is directly related to the decay constant $m_a \propto f_\pi m_\pi/f_a$ with $f_\pi $ and $m_\pi$ the pion decay constant and the pion mass, respectively. In the more general case there is no such relation and $m_a$ and $f$ are independent parameters. \\

\begin{figure}
\includegraphics[width=\textwidth]{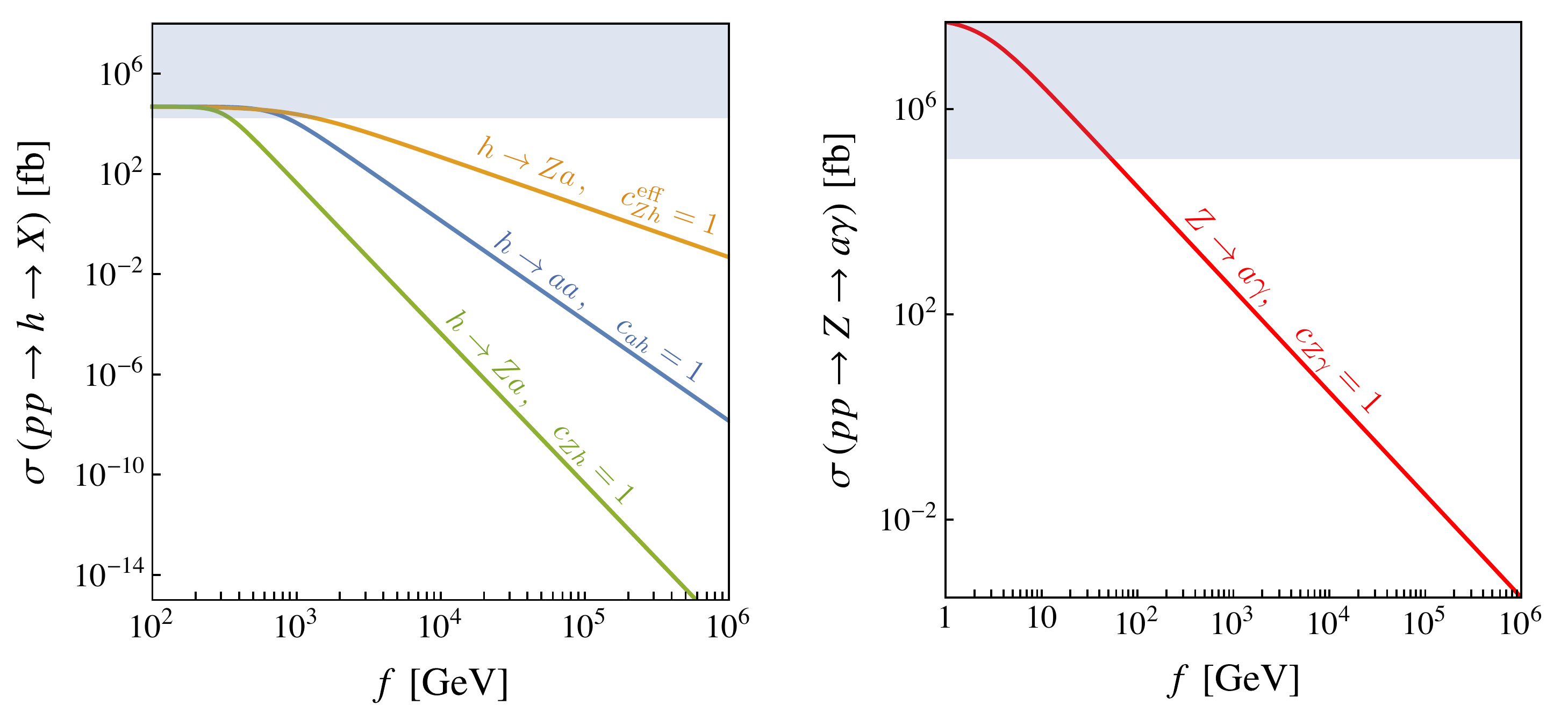}\caption{\label{fig:ALPpsec} Production cross-section of ALPs in the decays of heavy SM particles. }
\end{figure}

ALPs at the LHC suffer from a small production cross-section $\sigma (pp \to a) $ - if $f$ is large - or decay promptly - if $f$ is small. Beyond resonant production, ALPs can be produced in decays of heavy SM particles. In this case, larger scales $f$ correspond to smaller branching ratios and delayed ALP decays. In the following, we will discuss the reach of the MATHUSLA detector for ALPs produced in the decays $Z\to a \gamma$, $h \to a Z$ and $h \to aa$. In Fig.\ref{fig:ALPpsec}, we show the corresponding production cross-sections in dependence of the breaking scale $f$ for a mass $m_a=0$, using the relevant branching ratios 
\begin{align}
\Gamma(h \to Za)&=\frac{m_h^3}{16\pi\,f^2}|c_{Zh}^\text{eff}|^2\lambda^{3/2}\Big(\frac{m_Z^2}{m_h^2},\frac{m_a^2}{m_h^2}\Big)\,,\\
\Gamma(h \to aa)&=\frac{m_h^3\,v^2}{32\pi\,f^4}|c_{ah}|^2\left(1-\frac{2m_a^2}{m_h^2}\right)^2\sqrt{1-\frac{4m_a^2}{m_h^2}}\,,\\
\Gamma(Z\to a \gamma)&=\frac{\alpha\,\alpha(m_Z)m_Z^3}{96\pi^3\sin^2\theta_W\cos^2\theta_W\,f^2}|c_{Z\gamma}|^2\left(1-\frac{m_a^2}{m_Z^2}\right)^3\,,
\end{align}
where we define $c_{Zh}^\text{eff}=c_{Zh}^5+2c_{Zh}\,v^2/f^2$ in order to take into account possible contributions from chiral new physics (that arise for example by integrating out the top quark). The cross-sections clearly show the different scaling for the dimension five, six, and seven operators. The shaded region is excluded by Higgs coupling measurements constraining general beyond the SM decays of the Higgs $\text{Br}(h\to\text{BSM})<0.34$  \cite{Khachatryan:2016vau} and the error on the measurement of the total $Z$ width, which corresponds to $\text{Br}(Z\to \text{BSM})<0.0018$ \cite{ALEPH:2005ab}.\\


%
\begin{figure}[t]\centering
\includegraphics[width=.9\textwidth]{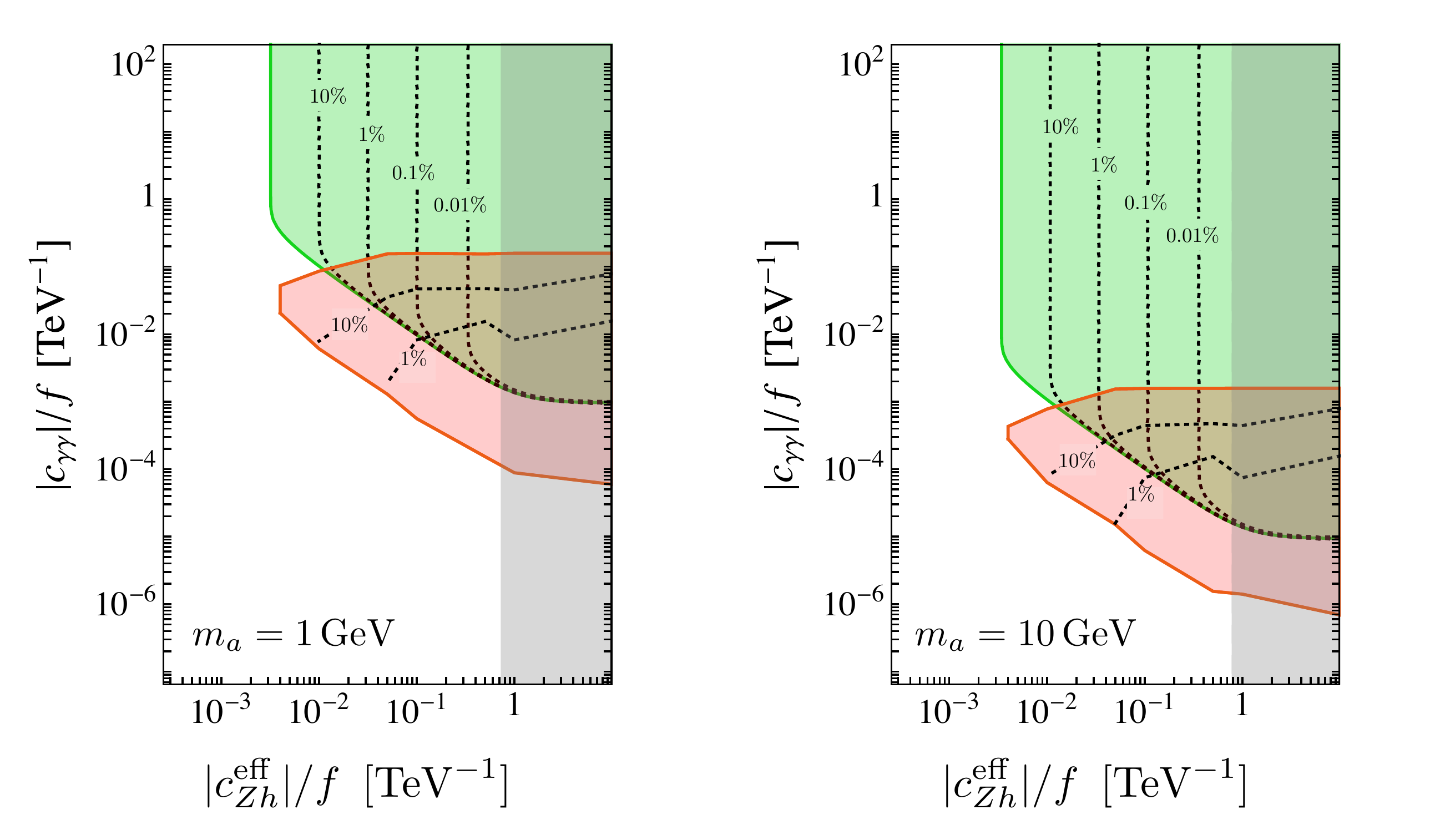}
\caption{\label{fig:hZa} Projected reach in searches for $h \to Za \to \ell^+\ell^-+2\gamma $ decays with ATLAS/CMS (green) and MATHUSLA (red) with $\sqrt{s}=14$ TeV center-of-mass energy and $3000$ fb$^{-1}$ integrated luminosity. The parameter region with the solid contours correspond to a branching ratio of $\text{Br}(a\to \gamma\gamma)=1$, and the contours showing the reach for smaller branching ratios are dashed.  This estimate assumes the $200m \times 200m \times 20m$ benchmark geometry of Fig.~\ref{f.mathuslalayout}.
}
\end{figure}
%

%
\begin{figure}[t]\centering
\includegraphics[width=.9\textwidth]{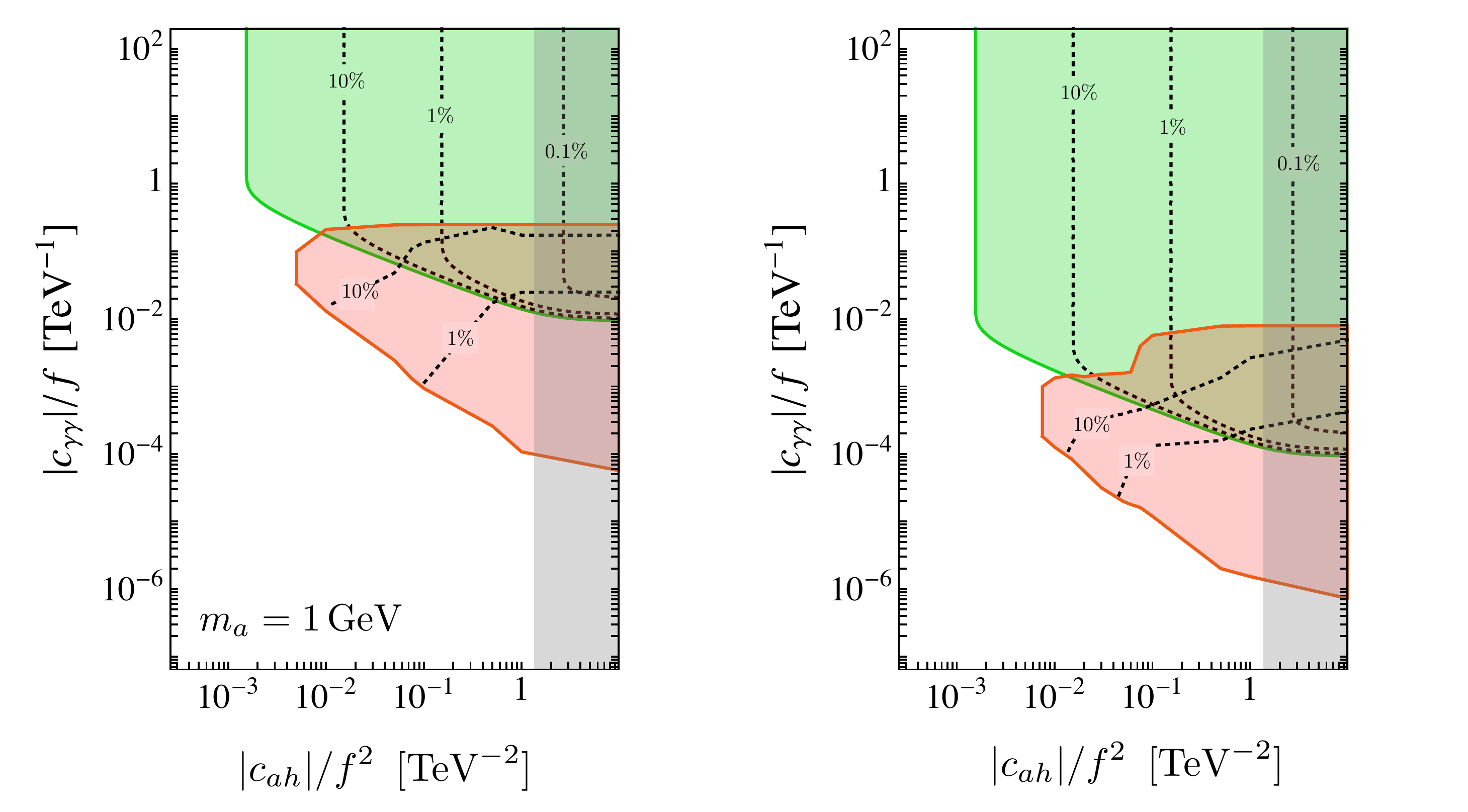}\caption{\label{fig:haa} Projected reach in searches for $h \to aa \to 4\gamma$ decays with ATLAS/CMS (green) and MATHUSLA (red) with $\sqrt{s}=14$ TeV center-of-mass energy and $3000$ fb$^{-1}$ integrated luminosity. The parameter region with the solid contours correspond to a branching ratio of $\text{Br}(a\to \gamma\gamma)=1$, and the contours showing the reach for smaller branching ratios are dashed.
This estimate assumes the $200m \times 200m \times 20m$ benchmark geometry of Fig.~\ref{f.mathuslalayout}.}
\end{figure}

%
In the following discussion, in order to evaluate the reach of ATLAS, CMS and the MATHUSLA detector, we consider ALP decays into photons, leptons and gluons as exemplary final states, but other final states are equally interesting, if ALPs are heavy enough. 
Depending on their mass, ALPs from Higgs or $Z$ decays can be highly boosted with the usual relativistic factor
\begin{align}
\gamma_a=\begin{cases}\displaystyle{ \frac{m_h^2-m_Z^2+m_a^2}{2m_am_h}}\,,& \text{for}\,\, h \to Z a\,,\\[14pt]
\displaystyle \frac{m_h}{2m_a}\,,&\text{for}\,\, h \to aa\,,\\[14pt]
\displaystyle\frac{m_a^2+m_Z^2}{2m_Zm_a}\,,&\text{for}\,\, Z \to a\gamma\,.\\
\end{cases}
\end{align}
For searches with ATLAS or CMS, we demand that all final state particles are detected in order to reconstruct the decaying SM particle and that the decays into photons occur before the electromagnetic calorimeter, $R =1.5$ m, and the decays into leptons before the inner tracker $R=2$ cm. For example, for $h \to Za \to \ell^+\ell^- \gamma\gamma$ decays, we ask for the $Z$ to be reconstructed in dileptons \emph{and} the ALP to decay inside the detector. We therefore define the effective branching ratios
\begin{align}
\text{Br}(h\to Za\to \ell^+\ell^-+\gamma\gamma)\big\vert_\text{eff}&=\text{Br}(h\to Za)\,\text{Br}(a\to \gamma\gamma)f_\text{dec}^a\,\text{Br}(Z\to \ell^+\ell^-)\,,\label{eq:LHChZa}\\
\text{Br}(h\to aa\to \gamma\gamma+\gamma\gamma)\big\vert_\text{eff}&=\text{Br}(h\to aa)\,\text{Br}(a\to \gamma\gamma)^2f_\text{dec}^{aa}\,,\label{eq:LHChaa}\\
\text{Br}(Z\to a\gamma\to \gamma\gamma \gamma)\big\vert_\text{eff}&=\text{Br}(Z\to a\gamma)\,\text{Br}(a\to\gamma\gamma)f_\text{dec}^a\,,\label{eq:LHCZag}
\end{align}
for the different processes considered.
$f_\text{dec}^a = f_\text{dec}^a(\gamma_a)$ is the fraction of axions decaying in the main detector, which is approximated as an infinitely long cylinder with the above mentioned inner and outer radii. 
 Analogous expressions hold for the ALP decaying into leptons and gluons. We further do not distinguish displaced from prompt decays and derive the reach for a number of $100$ signal events, which is typically needed to suppress backgrounds in searches for New Physics with prompt decays of $h$ and $Z$ bosons \cite{Khachatryan:2016vau, Khachatryan:2016are, Aad:2015bua}. For MATHUSLA, it is impossible to detect both final state particles in $h\to Za$ and $Z\to a \gamma$ decays and highly unlikely to see both ALPs from $h\to aa$ decays in the decay volume. However, because of the much lower background, single ALPs can be detected irrespective of their origin. 
 %
%
\begin{figure}[t]\centering
\includegraphics[width=.9\textwidth]{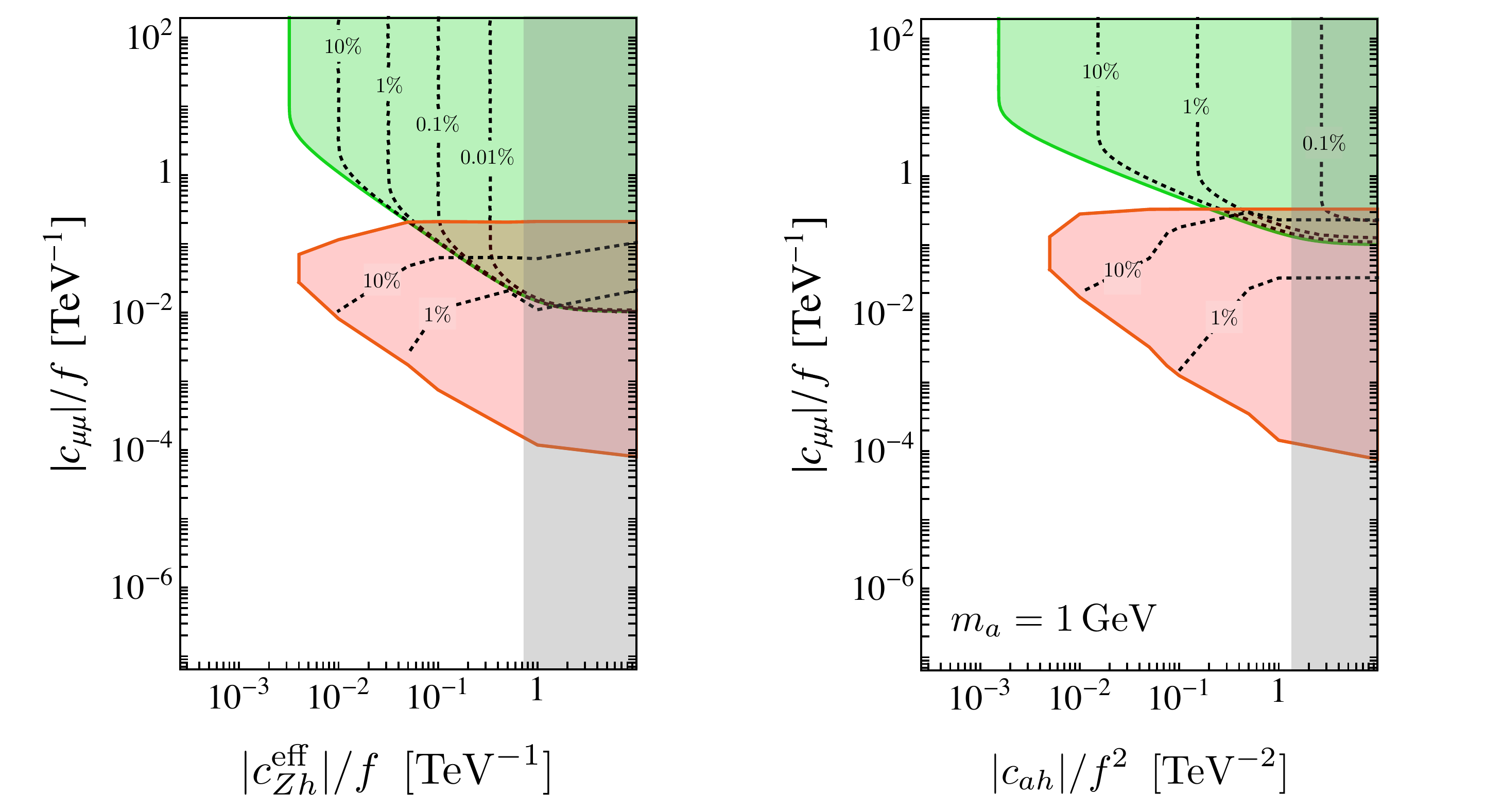}
\caption{\label{fig:muons} Projected reach in searches for $h \to Za \to \ell^+\ell^-+\mu^+\mu^- $ (left) and $h \to aa \to \mu^+\mu^-+\mu^+\mu^- $ (right) decays with ATLAS/CMS (green) and MATHUSLA (red) with $\sqrt{s}=14$ TeV center-of-mass energy and $3000$ fb$^{-1}$ integrated luminosity. The parameter region with the solid contours correspond to a branching ratio of $\text{Br}(a\to \mu^+\mu^-)=1$, and the contours showing the reach for smaller branching ratios are dashed.  This estimate assumes the $200m \times 200m \times 20m$ benchmark geometry of Fig.~\ref{f.mathuslalayout}.}
\end{figure}
%
%
We ask for at least four ALP decays within the MATHUSLA volume to derive the reach of the detector, so that the corresponding effective branching ratios for ALP decays in MATHUSLA read 
\begin{align}
\text{Br}(h\to Za\to Z + \gamma\gamma)\big\vert_\text{eff}^\text{M}&=\text{Br}(h\to Za)\,\text{Br}(a\to \gamma\gamma)f_\text{M}^a\,\label{eq:MATHhZa}\,,\\
\text{Br}(h\to aa\to a + \gamma\gamma)\big\vert_\text{eff}^\text{M}&=2\text{Br}(h\to aa)\,\text{Br}(a\to \gamma\gamma)f_\text{M}^a\,\label{eq:MATHhaa}\,,\\
\text{Br}(Z\to a \gamma\to \gamma + \gamma \gamma)\big\vert_\text{eff}^\text{M}&=\text{Br}(Z\to a\gamma)\,\text{Br}(a\to \gamma\gamma)f_\text{M}^a\, \label{eq:MATHZag}.
\end{align}
Note that states to the left of the ``$+$'' on the LHS are visible not in MATHUSLA but in the main detector. $f_\text{M}^a = f_\text{M}^a(\gamma_a)$ is the fraction of LLPs which decay in the MATHUSLA decay volume.
Again, the expressions for ALP decays into leptons and gluons are analogous to Eqns.~(\ref{eq:MATHhZa})-(\ref{eq:MATHZag}). 
In order to fully capture the geometric acceptance of the MATHUSLA detector, we use \texttt{MadGraph5} to simulate the signal events at parton level and the code provided by the MATHUSLA working group to compute the acceptance.


%
\begin{figure}[t!] \centering
\includegraphics[width=.9\textwidth]{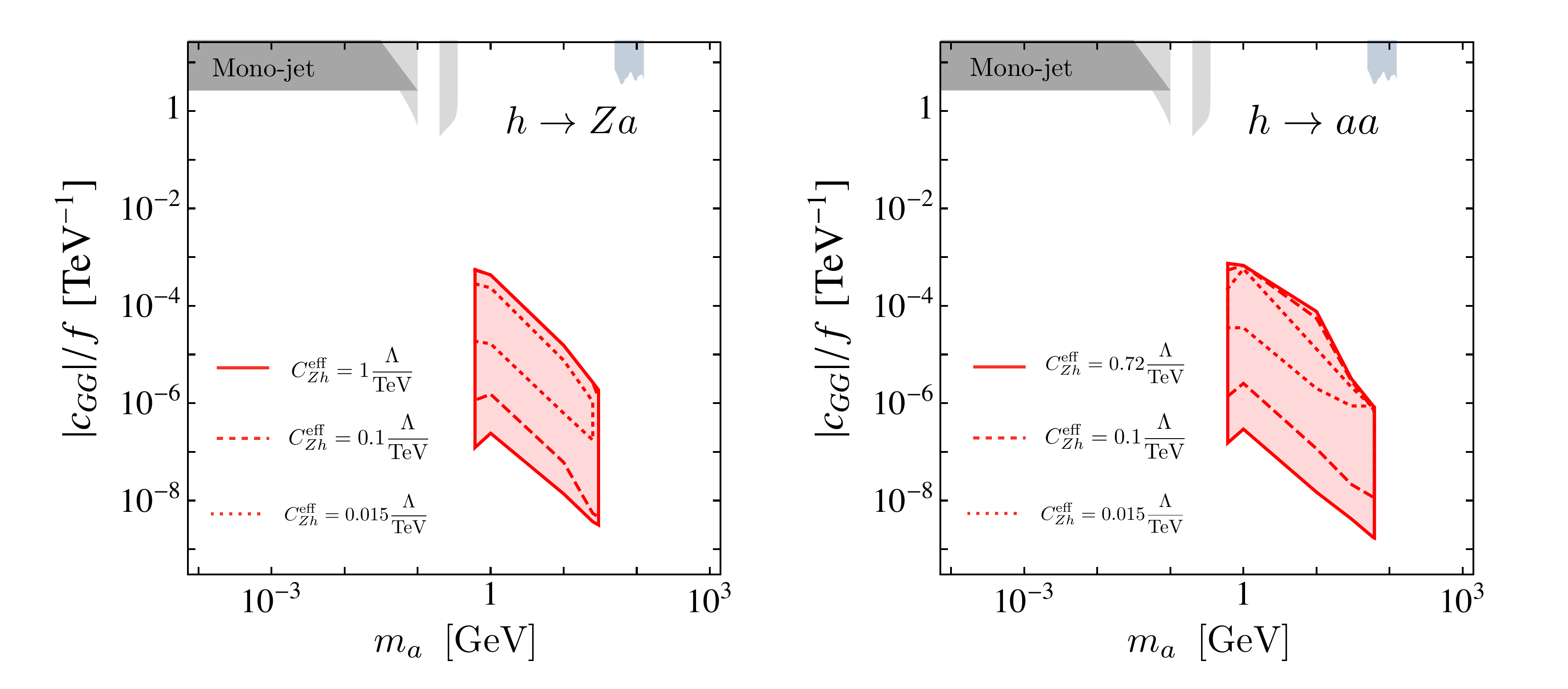}
\caption{\label{fig:jets} 
Projected exclusion contours for searches for $pp\to h \to Z a $  (left) and $pp\to h \to aa $ (right) with the subsequent ALP decay $a \to gg$ and $\text{Br}(a\to gg)=1$ within the MATHUSLA detector at the HL-LHC.  This is compared to projected HL-LHC monojet bounds~\cite{Mimasu:2014nea}. (See also~\cite{Aloni:2018vki}.)
This estimate assumes the $200m \times 200m \times 20m$ benchmark geometry of Fig.~\ref{f.mathuslalayout}. }
\end{figure}
%
%

We illustrate the reach for the ATLAS or CMS detector for discovering ALPs decaying into photons from $h \to aZ$ and $h \to a a$ decays in Fig~\ref{fig:hZa} and Fig.~\ref{fig:haa}, respectively. For the green parameter space with solid contours, ATLAS or CMS would see 100 events with a luminosity of $\mathcal{L}=3000$ fb$^{-1}$ and a branching ratio of $\text{Br}(a\to \gamma\gamma)=1$. For smaller branching ratios, larger couplings $|c^\text{eff}_{hZ}|$ and $|c_{ah}|$ are required to obtain the same number of events. 
Dashed lines show the lower limit for $\text{Br}(a\to \gamma\gamma)=0.1$, $\text{Br}(a\to \gamma\gamma)=0.01$ and $\text{Br}(a\to \gamma\gamma)=0.001$.\footnote{A smaller branching ratio for the given coefficients implies a larger total LLP width and hence shorter lifetime. In the long lifetime limit, the increased signal rate due to shorter lifetime offsets the lower branching ratio, making the lower boundaries of sensitivity independent of the branching ratio.} The red region with solid contours shows the parameter space for which four ALP decays are expected within the MATHUSLA detector for $\text{Br}(a\to \gamma\gamma)=1$ and $\mathcal{L}=3000$ fb$^{-1}$.  For $\text{Br}(a\to \gamma\gamma)=0.1$, $\text{Br}(a\to \gamma\gamma)=0.01$ and  $\text{Br}(a\to \gamma\gamma)=0.001$, MATHUSLA therefore looses sensitivity for  larger values of $|c_{\gamma\gamma}|/f$. In the case of $h \to a a$ decays, MATHUSLA will be able to probe smaller branching ratios than ATLAS or CMS. This underlines the complementarity between searches for prompt decays with ATLAS and CMS and searches for displaced ALP decays with MATHUSLA. 
In the event that MATHUSLA finds an LLP signal, event-by-event information on the LLP boost (and to some extent the final state, depending on the final detector capabilities) can be correlated with information prompt displaced object information from the main detector to elucidate the LLP production mode and eventually identify the LLP as an ALP. 

In Fig.~\ref{fig:muons}, we further show the reach for ALP decays into muons. Since at least approximate lepton flavour universality is expected in the couplings of the ALP, the muon decay mode is particularly well-motivated for $2m_\mu < m_a < 2m_\tau$. 


Finally, we present the reach of MATHUSLA for ALPs produced in Higgs decays with subsequent ALP decays into jets in Fig.~\ref{fig:jets}. 
We show the parameter space for which at least four $a\to jj$ events are expected within the MATHUSLA volume in the $m_a-c_{GG}$ plane in Fig.~\ref{fig:jets} for different values of $c_{Zh}^\text{eff}$ (left) and $c_{ah}$ (right). The expected minimal mass resolution of the MATHUSLA detector for ALPs in Higgs decays is of the order of $m_a\approx 100\,$MeV, assuming a spatial resolution of $1\,$cm. In Figure~\ref{fig:jets} we chose the lowest ALP mass to be $m_a= 600$ MeV. Note that for ALP masses below $m_a=1\,$GeV the ALP-gluon coupling $c_\text{GG}$ induces a sizable photon coupling through ALP-meson mixing, leading to additional constraints.
 In contrast to ALP decays into photons and  leptons, we refrain from showing projections for LHC LLP searches for $a\to jj$ decays,  given the large backgrounds for the Higgs decays $h\to Z a\to \ell^+\ell^- jj$ and $h \to aa \to 4j$. It has been shown in~\cite{Chou:2016lxi} that MATHUSLA has 1000$\times$ better sensitivity to LLP production cross-sections than ATLAS or CMS for this channel.
If decay to jets is the dominant axion decay mode, MATHUSLA provides by far the strongest sensitivity for the $c_G$ coefficient. (Larger $c_G$ values may be probed at the main detectors.)

%

%
%
%
\begin{figure}\hspace{2.5cm}
\includegraphics[width=.8\textwidth]{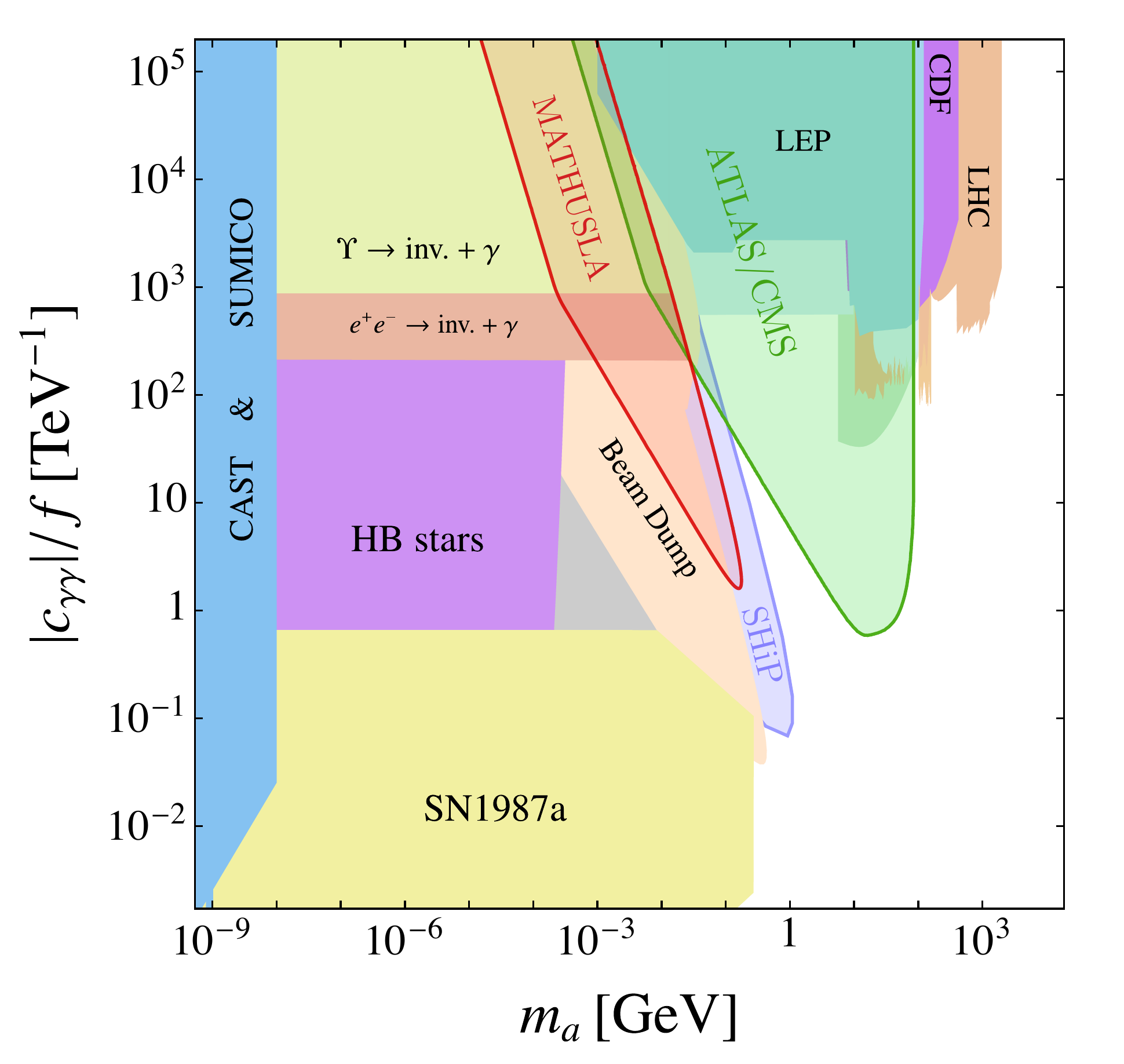}
\caption{\label{fig:ZAG} Projected reach in searches for $Z \to a\gamma \to 3\gamma$ decays with ATLAS/CMS (green) and MATHUSLA (red, assuming the $200m \times 200m \times 20m$ benchmark geometry of Fig.~\ref{f.mathuslalayout}) with $\sqrt{s}=14$ TeV center-of-mass energy and $3000$ fb$^{-1}$ integrated luminosity.  
%
}
\end{figure}
%

For $Z\to a \gamma$ decays with subsequent ALP decays into photons, the relevant Wilson coefficients $c_{\gamma\gamma} $ and $c_{Z\gamma}$ are linear combinations of $c_{W}$ and $c_{B}$. A scenario in which these coefficients are completely independent therefore appears to be fine-tuned.\footnote{Integrating out a single electroweak multiplet for example always generate $c_{W}$ and $c_B$ with the same sign, resulting in $|c_{Z\gamma}|\lesssim \cos^2\theta_w |c_{\gamma\gamma}|$.} We therefore show the reach of ATLAS or CMS and MATHUSLA in the $m_a - |c_{\gamma\gamma}|/f$ plane under the assumption that $c_W=0$ and $c_{Z\gamma}=-\sin^2{\theta_w}\, c_{\gamma\gamma}$, but impose a hard cut on $c_{Z\gamma}$ in the parameter space for which the constraint from the total $Z$ width, $\text{Br}(Z\to \text{BSM})<0.0018$, is violated \cite{Jaeckel:2015jla}. The corresponding exclusion region is shown in Fig.~\ref{fig:ZAG} together with various other constraints which only depend on $c_{\gamma\gamma}$. The kink in the exclusion region at $|c_{\gamma\gamma}|/f \approx 1$\, GeV and the resulting gap between the ATLAS/CMS and MATHUSLA reach occur because the values of $c_{Z\gamma}$ corresponding to smaller values of $c_{\gamma\gamma}$ would not yield enough events for small ALP masses. The reach of the MATHUSLA experiment overlaps with the existing limits from the E137 and E141 beam dump experiments (shaded light brown in Fig.~\ref{fig:ZAG}) \cite{Riordan:1987aw,Bjorken:1988as} 
and competes with the projected limits from the future SHiP experiment (shaded blue in Fig. 70) \cite{Alekhin:2015byh} and the future FASER experiment \cite{Feng:2018pew} (shaded yellow in Fig.~\ref{fig:ZAG}), though these have higher sensitivity.
It is also worth noting that the limits and sensitivity projections in Fig.~\ref{fig:ZAG} assume that all other coefficients in Eqn.~(\ref{eq.axionL5}) are zero. If any other coefficients are present, MATHUSLA could be sensitive to values of these coefficients (see Figs.~\ref{fig:hZa} - \ref{fig:jets})  that are orders of magnitude smaller than the $c_{\gamma \gamma}/f$ range shown in Fig.~\ref{fig:ZAG}.

In conclusion, Axion-like particles are a very general and well-motivated BSM scenario that can be probed by MATHUSLA at lifetimes much larger (and couplings much smaller) than possible with the LHC main detectors alone.

\clearpage 
\section{Executive Summary}
\label{sec:signatures}

This document has two main aims: to demonstrate that (1)  neutral LLPs are broadly and fundamentally motivated in BSM theories, and (2)  the construction of the MATHUSLA detector is necessary to fully leverage the LHC's vast discovery potential for new physics. 
In this section we summarize how the results presented here fulfill these objectives.

We have discussed the most bottom-up motivations for LLPs in the Introduction. Many particles in the SM have much longer lifetime than a na\"ive expectation from dimensional analysis might suggest, in many cases out to macroscopic distances.  A variety of mechanisms can suppress the decay width of an unstable particle: small couplings, heavy mediators, approximate symmetries, and/or phase space suppressions. 
These exact mechanisms can give rise to neutral LLPs in any BSM theory, including simple extensions of the SM such as hidden valleys (Section~\ref{sec:hiddenvalleys}), which are a generic consequence of the structure of gauge theories, and minimal benchmark models like dark scalars (Section~\ref{sec:singlets}), dark photons (Section~\ref{sec:darkphotons}), and Axion-like particles (Section~\ref{sec:alp}). 
These hypothesized new physics sectors are separated from the SM fields not (necessarily) by a mass hierarchy, but by an absence of large couplings between the hidden and visible sectors.
The very nature of a (possibly confining) hidden sector, only connected to the SM via a tiny portal at low energies, makes neutral LLPs an obvious signal to search for. 
Regardless of the details of the new physics, exotic Higgs decays (Section~\ref{sec:exhdecays}) are one of the most  motivated production modes for light new states including LLPs, due to the small SM Higgs width, its large LHC production rate, and the lack of symmetry protection for the $|H|^2$ operator allowing it to couple to any new physics.
Furthermore, MATHUSLA would be sensitive to LLPs produced in exotic Higgs decays out to lifetimes near the upper limit from BBN (Section~\ref{sec:bbn}), provided they are not too boosted.
Searches for these simplified IR scenarios are motivated in their own right to agnostically cover as much possible new physics theory- and parameter-space as we experimentally can, especially in light of recent LHC null results.

The bottom-up \emph{plausibility} of LLP signatures is therefore well-established.
However, one of the most important conclusions we can draw from the results presented in this document 
is that LLPs are not just plausible, but \emph{strongly fundamentally motivated for a broad variety of top-down reasons}.
They are ubiquitous in BSM scenarios that address longstanding mysteries like the naturalness of the weak scale (Section~\ref{sec:naturalness}),  Dark Matter (Section~\ref{sec:dmtheory}), Baryogenesis (Section~\ref{sec:baryogenesis}) and Neutrino Masses (Section~\ref{sec:neutrinos}).
Furthermore, they are often an  intrinsic part of the theory mechanism which addresses the fundamental mystery in the first place.

\begin{figure}
\begin{center}
\hspace*{-0.7cm}
\includegraphics[width=17cm]{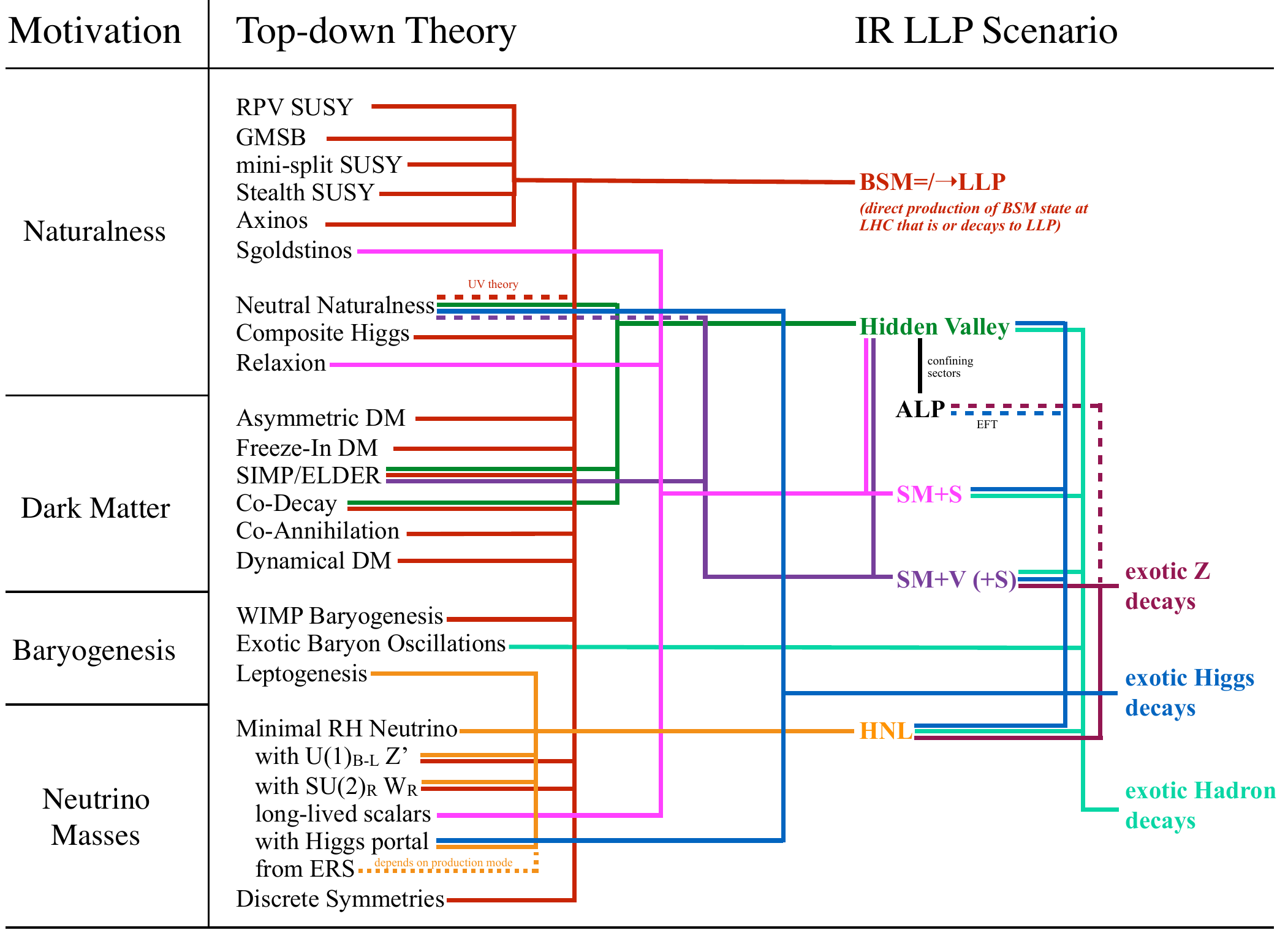}
\end{center}
\caption{
Qualitative overview of the top-down theory motivations for neutral LLPs discussed in this document, with colored lines (from left to right) indicating which IR LLP scenario they  motivate at the LHC.
Some of the IR LLP scenarios or simplified models in turn motivate specific signatures like exotic Higgs decays. 
We stress that these top-down theories are not the only motivations for the IR scenarios or simplified models shown here: hidden valleys, exotic Higgs decays, etc., are also motivated in their own right on generic, bottom-up grounds.
}
\label{f.mastersummary}
\end{figure}

One
way to map out these top-down motivations is sketched in Fig.~\ref{f.mastersummary}. 
This figure qualitatively illustrates which theories and frameworks discussed in the preceeding sections give rise to which ``IR LLP Scenarios", broadly defined to include general classes of bottom-up theories like hidden valleys, simplified models like SM+S or SM+V, and specific LLP signatures like exotic Higgs decays. 
A very common example of an neutral LLP signature is simply the direct production at the LHC of a BSM state with sufficiently sizable couplings to the SM, which either is an LLP itself, or decays promptly to an LLP (``BSM=/$\to$LLP'').
The theories examined in this document are hardly exhaustive, but the ubiquity of LLPs in top-down motivated BSM theories is evident.

One might wonder why such a coarse-grained classification of signatures is even helpful. After all, the ``BSM=/$\to$LLP'' scenario includes a wide variety of different LLP species with different production and decay modes.\footnote{This is to be contrasted with simplified models for LLP searches at the main detectors developed by the LHC-LLP Community working groups~\cite{LHCLLPwhitepaper}, which have to parameterize the large variety of displaced and associated prompt signals in considerable kinematic detail, in order to facilitate the development of concrete search and background rejection strategies.} 
Essentially, this is because almost any theory with LLPs can give rise to very long lifetimes, either because the long-lifetime regime is specifically motivated, or because the lifetime is practically a free parameter.\footnote{There are a few exceptions which prove the rule, e.g. pure higgsinos with a tiny mass splitting from electroweak symmetry breaking \cite{Schwaller:2013baa,
Low:2014cba,
Cirelli:2014dsa,
Mahbubani:2017gjh,
Fukuda:2017jmk}, with lifetimes below a cm. We did not study these examples here, but they of course add motivation for LLP searches at the LHC main detectors, and slight modifications of the model can yield longer lifetimes.}
As we review below, in this long-lifetime regime the discussion of LLP signatures at MATHUSLA, 
and comparing sensitivity to the LHC main detectors, becomes quite simple, and leads to the conclusion that MATHUSLA has highly general and robust advantages when searching for LLPs.

The basic MATHUSLA detector concept is described in Section~\ref{sec:mathuslasummary}. 
The benchmark design is an empty box on the surface with trackers in the roof and active vetoes surrounding the $200m \times 200m \times 20m$ air-filled detector volume. Neutral LLP decays into two or more charged particles are reconstructed as displaced vertices with stringent geometric and timing requirements.
MATHUSLA's position on the surface provides shielding from the deluge of SM particles produced at the collision point. 
The high-energy displaced signature of LLP decays is therefore even more distinctive in MATHUSLA than inside the LHC main detectors. 
The most important remaining backgrounds on the surface are cosmic rays, high-energy muons from the LHC, and neutrino scatterings. All of these can be rejected with extremely high fidelity, using simple requirements on the charged particle direction of travel as well as more elaborate geometrical and timing cuts. As a result, MATHUSLA can search for LLPs in effectively the background-free regime.

\begin{figure}
\begin{center}
\includegraphics[width=11cm]{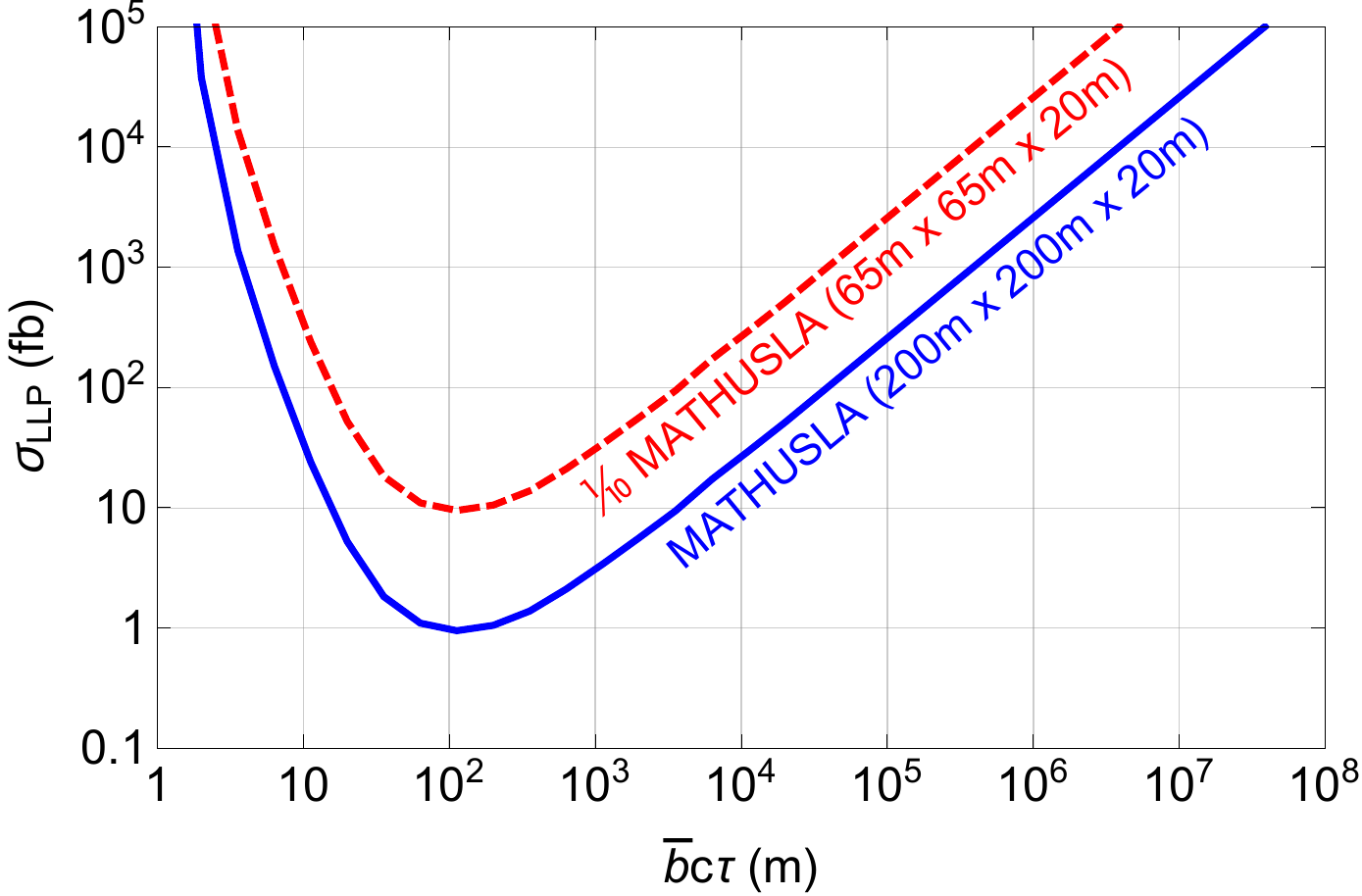}
\end{center}
\caption{
(Identical to Fig.~\ref{f.MATHUSLAsensitivitycartoon}.)
Schematic order-of-magnitude sensitivity of MATHUSLA, assuming $\mathcal{O}(1)$ produced LLPs per production event at the HL-LHC. $\bar b$ is the mean boost of the produced LLPs. The shape of the exclusion/discovery region at short lifetimes depends on the detailed boost distribution, but for long lifetimes $\bar b c \tau \gg 200 m$ depends only on the mean boost and is very model-independent up to an $\mathcal{O}(1)$ factor. Note that LLPs near the BBN lifetime limit of $c \tau \sim 10^7$m can be probed if they are produced with cross-sections in the pb range at the HL-LHC.
To emphasize the scalability of the MATHUSLA design, we also show the reach achievable with a version of MATHUSLA with only $1/10$ the detector volume of the $200m \times 200 m \times 20 m$ benchmark geometry. 
}
\label{f.MATHUSLAsensitivitycartoon2}
\end{figure}

In Section~\ref{sec:LLPSmathuslahllhc}, we take a model-independent approach to assess the sensitivity of MATHUSLA to neutral LLP production rates, and compare its sensitivity to main detector LLP searches.
In the long-lifetime regime $b c \tau \gg 100$m, MATHUSLA has comparable acceptance for LLP decays as ATLAS or CMS, with only very modest dependence on the production mode. However, unlike the underground detectors, which have to contend with a variety of backgrounds when searching for neutral LLPs, MATHUSLA can operate without backgrounds.
This allows for the detection of neutral LLPs with lifetimes near the BBN limit of $c \tau \sim 10^7 m$ (for order one boosts) if they are produced with $\sim$pb cross-section. Decay lengths of $\sim 100m$ can be detected for $\sim$fb cross-section at the LHC. This model-independent sensitivity is shown in Fig.~\ref{f.MATHUSLAsensitivitycartoon}, which we reproduce in this Section as Fig.~\ref{f.MATHUSLAsensitivitycartoon2} for convenience.
For the purposes of neutral LLP discovery, MATHUSLA can therefore be thought of as a version of the main detectors that \emph{sacrifices} sensitivity to shorter decay lengths in order to \emph{gain} the ability to search for LLPs without backgrounds or trigger limitations.

With the motivation for neutral LLP searches established, we must therefore ask: (1) how important is MATHUSLA's advantage of zero background and no trigger issues compared to the main detectors, and (2) how motivated is the long-lifetime regime (so that there is any signal at MATHUSLA)?

The general discussion of the first issue is provided in Section~\ref{s.MATHUSLAvsMAINDETECTORS} and can be summarized with a few simple qualitative conclusions.
Missing energy triggers at ATLAS/CMS are generally inefficient for neutral LLP searches unless the production rates are sizable and sufficiently energetic prompt objects are also present. 
Since MATHUSLA has a similar acceptance for LLP decays as the main detectors, but operates essentially free from backgrounds or trigger limitations, it will have superior sensitivity for any neutral LLP signal where either backgrounds, cut efficiency (including requirements on LLP decay and production mode) or triggers impede the main detector search. This includes:
\begin{itemize}
\item LLPs that decay with less than a few hundred GeV of visible hadronic energy. For example, MATHUSLA has 3 orders of magnitude better cross-section sensitivity to LLP that are produced in exotic Higgs decays and decay  via the Higgs portal.
\item LLPs that have subdominant leptonic branching fractions, for masses below a few hundred GeV.
\item LLPs lighter than $\sim 10 \gev$ that decay to lepton jets, where MATHUSLA may increase reach by 1-2 orders of magnitude in cross-section.
\end{itemize}
Another interesting scenario with potentially large sensitivity gains are LLPs that decay to photons, if MATHUSLA is configured for their detection and can search with low backgrounds.
Conversely, if an LLP is always produced in association with a hard lepton or decays into jets with more than several 100s of GeV of energy, the event can pass L1 triggers and the main detector LLP search has very few backgrounds, resulting in likely similar sensitivity to MATHUSLA in the long-lifetime regime. 
These simple arguments illustrate why MATHUSLA has far superior sensitivity to the main detectors for large classes of important  neutral LLP signals.

We now turn to the  motivation for LLPs in the long-lifetime regime.
Most of the theories discussed in this document feature LLP signals for which MATHUSLA could be our only discovery opportunity in large parts of parameter space. 
To more explicitly demonstrate the role of MATHUSLA in probing fundamentally motivated BSM theories, we provide Tables~\ref{t.summarytheorytable1} and \ref{t.summarytheorytable2}.

Table~\ref{t.summarytheorytable1} summarizes those BSM scenarios where the discoverable LLP is a strongly motivated intrinsic part of the theory mechanism. 
We attempt to summarize the main role that LLPs play in each theory, the motivations for the long-lifetime regime if any, and the role MATHUSLA would play in their discovery. 
Examples include Neutral Naturalness (Section~\ref{sec:neutralnaturalness}), where the very symmetry protection which stabilizes the weak scale gives rise to a hidden valley containing LLPs accessible via the Higgs portal; WIMP Baryogenesis (Section~\ref{sec:wimpbg}), where the LLP decay at long lifetimes is the very mechanism which generates the baryon asymmetry of the universe; and FIMP DM (Section~\ref{sec:freezein}), 
which can be produced both in the early universe and at the HL-LHC in the decay of a parent LLP with sizable SM couplings. 
In all these case, not only would MATHUSLA have the greatest sensitivity to discover many classes of these BSM scenarios, there are often arguments why the long lifetime regime might even be theoretically preferred. This makes the search for very long-lived particles even more urgent.

For other theories, summarized in Tables~\ref{t.summarytheorytable2}, LLPs are no less motivated, but their existence and/or long lifetime is simply one part of a much larger space of possible signals, depending on the specific model details and parameters. 
This includes the ubiquitous hidden valley idea, the general new physics discovery channel of exotic Higgs decays, as well as broad classes of dark matter models.

\begin{table}
\begin{center}
\vspace*{-1.2cm}
\hspace*{-1.4cm}
\begin{tabular}{|m{2.3cm}|m{5cm}|m{2.7cm}|m{5cm}|m{0.5cm}|m{0.5cm}|}
\hline
BSM Scenario & Role of LLPs & Typical $c\tau$ & Role of MATHUSLA & Sec. & Fig. 
\\
\hline 
\hline
Neutral \hspace{8mm} Naturalness 
& 
\small
Discrete symmetry stabilizing Higgs mass $\to$ hidden valley with Higgs portal. Cosmology $\to$ hidden valley particles are LLPs.
& 
\small
Any, but $\mathbb{Z}_2$ arguments favor lower $\hat \Lambda_{QCD}$ and hence long lifetimes.
&
\small
Smoking gun signal are mirror glueball LLPs. For long lifetimes, they can only be discovered at MATHUSLA. 
&
\small \ref{sec:neutralnaturalness}
&
\small \ref{f.NNLLPbounds}, \ref{f.MTHRHneutrinoMATHUSLAreach}
\\
\hline
WIMP \hspace{6mm} \mbox{Baryogenesis}
& 
\small
Out-of-equilibrium decay of WIMP-like LLP produces baryon asymmetry.
& 
\small
$\gtrsim$ cm for weak-scale LLP masses.
&
\small
Decays to baryons $\to$ MATHUSLA likely much greater sensitivity than main detectors. MCFODO
&
\small \ref{sec:wimpbg}
&
\small \ref{fig:WIMPBGexclusion}
\\%
\hline
FIMP DM  
& 
\small
Freeze-in via decay requires LLPs with  SM couplings.
& 
\small
Fixed by masses \& cosmology. Long lifetimes generic.
&
\small
Model-dependent, but in long-lifetime regime MCFODO.
&
\small \ref{sec:freezein}
&
\small \ref{fig:MasterPlots}, \ref{FI_Higgs_m1m2}, \ref{fig:axinos}, 
\\%
\hline
Co-decaying DM
& 
\small
Out-of-equilibrium decay of hidden sector LLP determines DM abundance. Also, small portal  $\to$ visible sector LLPs. 
& 
\small
For weak scale LLP masses, most of parameter space is long lifetimes.
&
\small
Depending on model details (production \& decay mode), MCFODO. 
&
\small \hspace*{-1mm}\ref{s.codecayingDM}
&
\small \ref{fig:codecaylimit}
\\%
\hline
\mbox{Co-annihilating} DM
& 
\small
DM relic abundance relies on small mass splitting with another state $\to$ other state is LLP.
& 
\small
Any, long lifetimes generic.
&
\small
Depends on model details, but e.g. for Higgs Portal implementations, MCFODO. &
\small
\ref{sec:coannihilation}
&

\\%
\hline
SUSY: Axinos
& 
\small
High PQ-breaking scale $V_{PQ}$ suppresses axion/axino couplings, making LOSP an LLP
& 
\small
Any, long lifetimes generic.
&
\small
For high $V_{PQ}$, MCFODO.
&
\small \hspace*{-2mm}
\ref{sec:axino-ewino}
&
\small \ref{fig:axinos}
\\%
\hline
SUSY: GMSB
& 
\small
Low SUSY breaking scale $F$ (motivated by flavor problem) leads to light gravitino and small couplings to LOSP, which can hence be LLP.
& 
\small
Any, long lifetimes generic.
&
\small
MCFODO, depending on spectrum and lifetime.
&
\small \hspace*{-2mm}
\ref{sec:gaugemediation}
&
\small \ref{fig:GMSBhiggsinoreach}
\\%
\hline
SUSY: RPV
& 
\small
small RPV couplings (motivated by avoiding flavor violation, proton decay, baryon washout) $\to$ LOSP can be LLP
& 
\small
Any, long lifetimes generic.
&
\small
MCFODO, especially for EW-charged LLPs or squeezed spectra.
&
\small \hspace*{-2mm}
\ref{sec:rpv}
&
\small \ref{fig:RPVplots}
\\%
\hline
SUSY: \hspace{6mm} Sgoldstinos
& 
\small
SUSY breaking scale $F$ suppresses sgoldstino coupling to supercurrents $\to$ can be LLP.
& 
\small
Any. Long lifetimes $\to$ smallest production, hardest to probe.
&
\small
Similar to SM+S. For masses $\lesssim 5 \gev$, MATHUSLA and/or SHiP may be only/first discovery opportunity. 
&
\small \hspace*{-2mm}
\ref{sec:sgoldstinos}
&
\\%
\hline
Exotic Baryon Oscillations
& 
\small
Exotic Baryon is LLP and induces oscillations that generate baryon number.
& 
\small
$\gtrsim 100$m
&
\small
Heavy baryon decays produce LLP. MATHUSLA and/or SHiP may be only/first discovery opportunity. 
&
\small \hspace*{-2mm}
\ref{sec:exoticbaryonoscillations}
&
\\%
\hline
minimal RH \phantom{bl} \mbox{neutrino model}
& 
\small
Type-1 see-saw $\to$ tiny mixing between $\nu_L$ and $\nu_R$ $\to$ \phantom{bla} $\nu_R$ LLPs
& 
\small
Any, long lifetimes favor lower $m_N$
&
\small
In long-lifetime/low-mass regime, MATHUSLA and/or SHiP may be only/first discovery opportunity.
&
\small \ref{sec:minimalRHN}
&
\small \ref{fig:Sterile2}, \ref{fig:sensitivity-FCC}
\\%
\hline
$\hookrightarrow$ with \phantom{blabl} $U(1)_{B-L} \ Z'$
& 
\small
Weakly gauged $B-L$ breaking generates $M_N$, additional $\nu_R$ production mode from $Z'$.
& 
\small
$m_{N} \sim1$-$10 \gev$ suggests long lifetime regime.
&
\small
For sub-weak-scale $m_{N}$, MCFODO.
&
\small \hspace*{-2mm} \ref{sec:bminusl_lowmass}
&
\small \ref{fig:B-L}
\\%
\hline
$\hookrightarrow$ with \phantom{blabl} $SU(2)_L \ W_R$
& 
\small
$\nu_R$  part of gauged $SU(2)_R$, breaking generates $M_N$. Additional $\nu_R$ production mode from $W_R^\pm$.
& 
\small
Any, long lifetimes favor lower $m_{N}$.
&
\small
For $m_{W_R} \sim 10 \tev$: main detector probes weak-scale $m_{N}$. MATHUSLA/SHiP only discovery opportunity for $m_{N} \lesssim 5 \gev$. 
&
\small \hspace*{-2mm} \ref{sec:LRSYMMnuR}
&
\small \ref{fig:RHN}
\\%
\hline
$\hookrightarrow$  with \phantom{blabl} Higgs Portal
& 
\small
GUT motivates extra broken $U(1)$ gauge groups, extended scalar sectors mix with Higgs $\to$ produce $\nu_R$ in Higgs and other scalar decays.
& 
\small
Any, long lifetimes favor lower $m_{N}$.
&
\small
MCFODO, improves Br reach of main detectors by at least order of magnitude.
&
\small \ref{sec:Higgportalneut}
&
\small \ref{fig:MATHnuHsensitivity}
\\%
\hline
\mbox{$m_{\nu}$ via discrete} symmetries
&
\small
Discrete sym. generates $m_{\nu}$ and stabilizes FIMP DM. 
& 
\small
See FIMP DM.
&
\small
LLPs with EW charge $\to$ MCFODO, especially for $m \lesssim 10 \gev$
&
\small \ref{sec:inertdoubletportal}
&
\\
\hline
\end{tabular}
\end{center}
\vspace{-3mm}
\caption{
BSM scenarios discussed in this document where neutral LLP signals at MATHUSLA are a strongly motivated intrinsic part of the theory mechanism, \emph{and} MATHUSLA Could be First or Only Discovery Opportunity (MCFODO). When discussing lifetimes, ``any'' means up to the BBN limit, ``long'' means the MATHUSLA regime. LOSP = lightest observable-sector supersymmetric particle. }
\label{t.summarytheorytable1}
\end{table}

\begin{table}
\begin{center}
\vspace*{-1.2cm}
\hspace*{-1.4cm}
\begin{tabular}{|m{2.3cm}|m{5cm}|m{2.7cm}|m{5cm}|m{0.5cm}|m{0.5cm}|}
\hline
BSM Scenario & Role of LLPs & Typical $c\tau$ & Role of MATHUSLA (long $c \tau$) & Sec. & Fig. 
\\
\hline 
\hline
\small
Hidden Valleys (HV)
& 
\small
Small portal to visible sector and possibly hidden sector confinement $\to$ meta-stable states.
& 
\small
Any.
&
\small
MCFODO, especially if LLPs are significantly below the weak scale or decay hadronically. 
&
\small \ref{sec:hiddenvalleys}
&
\small \ref{fig:ejsensitivity}, \ref{fig:bombsensitivity}
\\
\hline
\small
SM+S
& 
\small
Small mixing $\to$ scalar LLP, produce in exotic Higgs decays for $m_S < m_H/2$. Large mixing $\to$ $S$ could decay to HV LLPs. 
& 
\small
Any.
&
\small
MCFODO. Complementarity with SHiP.
&
\small \ref{sec:singlets}
&
\small \ref{fig:HBRScalar}
\\
\hline
\small
SM+V
& 
\small
Dark photon/dark Higgs LLP could be produced in exotic Higgs/$Z$ decays. Dark photon with non-tiny kinetic mixing could be copiously produced at LHC and decay to HV LLPs.
& 
\small
Any.
&
\small
MCFODO. 
Significantly extends main detector long-lifetime reach for dark photons and dark Higgs produced in exotic $H$ and $Z$ decays. For LLPs produced in dark photon decays, see HV. 
&
\small \ref{sec:darkphotons}
&
\small
\ref{fig:higgsdecay}, \ref{fig:pptoapjjdecay}, \ref{fig:displaced_sensitivity_mHd_15_L3000_mA_vs_ctau_mathusla}, \ref{fig:displaced_sensitivity_L3000_mA_vs_eps_ct_100m_mathusla}
\\

\hline
\small
Exotic \phantom{blabla} Higgs decays
& 
\small
Higgs coupling to new states, like HV or other LLPs, is highly generic and leads to large production rates at LHC.
& 
\small
Any.
&
\small
MCFODO for Br $\lesssim 0.1 - 0.01$.
Higgs portal motivates hadronic LLP decays, for which MATHUSLA has $10^3$ better Br reach than main detectors. MATHUSLA also has significantly better sensitivity for LLP masses $\lesssim 10 \gev$ even if they decay leptonically, or for LLPs with subdominant leptonic decays.
&
\small \ref{sec:exhdecays}
&
\small \ref{f.exHlifetimeplot}, \ref{f.exHdecaybounds}
\\

\hline
\small
\mbox{Asymmetric DM}
& 
\small
Relating DM to baryon abundance requires operator connecting DM number and Baryon/Lepton number $\to$ higher dimensional operator $\to$ LLPs
& 
\small
Any, depending on kind and scale of physics generating the operator.
&
\small
MCFODO (highly dependent on production and decay mode).
&
\small \ref{sec:adm}
&
\\

\hline
\small
Dynamical DM
& 
\small
Dark sector includes spectrum of states with varying life-time up to hyperstable DM states. 
& 
\small
Any, DDM ensemble contains short to hyperstable $c\tau$.
&
\small
MCFODO (highly dependent on production and decay mode).
&
\small \ref{sec:dynamicaldm}
&
\small
 \ref{fig:ddmfig1}, \ref{fig:ddmfig2}
\\

\hline
\small
SIMP/ELDER DM
& 
\small
Strong dynamics of HV generate DM abundance. HV $\to$ LLPs.
& 
\small
Any.
&
\small
See HV.
&
\small \hspace*{-2mm} \ref{s.SIMP}, \hspace*{-1.3mm}\ref{s.ELDER} 
&
\\

\hline
\small
Relaxion
& 
\small
Relaxion or other new scalars in theory generically mix with Higgs $\to$ SM+S. 
& 
\small
Any.
&
\small
See SM+S.
&
\small \ref{sec:relaxion}
&
\\

\hline
\small
Axion-like \phantom{bla} particles
& 
\small
ALP couplings to $h$ and $Z$ are generic in EFT framework. $1/f$ suppression makes ALP an LLP.
& 
\small
Any.
&
\small
MCFODO for low-scale $f$.
&
\small \ref{sec:alp}
&
\small \ref{fig:hZa}, \ref{fig:haa}, \ref{fig:muons}, \ref{fig:jets}, \ref{fig:ZAG}
\\

\hline
\small
Leptogenesis
& 
\small
Motivates minimal RH neutrino model and other neutrino extensions, which generically feature LLPs. 
& 
\small
Freeze-out LG favors weak-scale $m_N$ but not so for other scenarios. Lower $m_N$ favor long lifetimes.
&
\small
Generally very difficult to probe, especially at high leptogenesis scale. 
In long-lifetime/low-mass regime, MATHUSLA and/or SHiP may be only/first discovery opportunity.
&
\small \ref{sec:bgleptogenesis}
&
\\

\hline
\small
Scalars in neutrino extensions
& 
\small
Gauge extensions in neutrino models give rise to new scalars that can mix with Higgs $\to$ SM+S. Provides additional $S$ production modes via heavy gauge boson decay.
& 
\small
Any.
&
\small
See SM+S, with some additional production modes (new heavy gauge bosons).
&
\small \hspace*{-2mm} \ref{s.BLnuRLLPscalars}, \hspace*{-1.5mm} \ref{sec:LLPscalar_LR}
&
\\
\hline
\end{tabular}
\end{center}
\vspace{-3mm}
\caption{
BSM scenarios discussed in this document where neutral LLP signals at MATHUSLA are a strongly motivated generic possibility (often as part of a broad parameter or theory space), \emph{and} MATHUSLA Could be First or Only Discovery Opportunity (MCFODO).
When discussing lifetimes, ``any'' means up to the BBN limit, ``long'' means the MATHUSLA regime. HV = hidden valley.
Since lifetimes are mostly arbitrary here, we focus on the long lifetime regime when discussing the role of MATHUSLA.
}
\label{t.summarytheorytable2}
\end{table}

MATHUSLA is also important for investigating other possible theories of new physics. 
Split versions of composite Higgs (Section~\ref{sec:compositehiggs}) and supersymmetry (Section~\ref{sec:minisplit}) generically give rise to long-lived colored particles which may be discovered at the main detectors, but MATHUSLA would provide an additional discovery channel as well as important information about the behavior of such R-hadrons in matter. 
Extended versions of these theories, which avoid some of the constraints suffered by the minimal models, can also give rise to neutral LLPs for which MATHUSLA is the prime discovery tool. 
In many cases, new physics might be discovered at the main detectors but be mis-diagnosed. This is generally true if a MET search discovers what is actually a very long-lived particle, but can also be true for resonance searches where a discovery of new physics obscures the existence of a hidden sector containing light LLPs that are important for diagnosing the complete theory, such as might be the case in versions of Stealth SUSY (Section~\ref{sec:stealthsusy}). 
It is also possible for details of the UV theory to generate MATHUSLA signals in scenarios where we do not naively expect them from low-energy considerations, such as for neutrino models with Enhanced Residual Symmetry (ERS, Section~\ref{BhupalClaudiaEmiliano}). 
Finally, MATHUSLA has impressive capabilities as a cosmic ray telescope. This is briefly discussed in Section~\ref{s.mathuslaCR} and will be the subject of its own dedicated study. MATHUSLA's measurements could address many outstanding puzzles in cosmic-ray and astro-particle physics, and represent a guaranteed physics return on the investment of constructing the detector.

We close by pointing out that MATHUSLA is not only a very strongly motivated and relatively affordable way of extending the capabilities of the LHC, the concept is also exceedingly flexible, general and scalable.
Future proton colliders, like the 100 TeV FCC-hh~\cite{Golling:2016gvc,Mangano:2016jyj,Contino:2016spe} or SPPC~\cite{Tang:2015qga} should include as part of their design an underground, shielded, dedicated displaced vertex detector to maximize their discovery potential for new physics. 
At the HL-LHC, MATHULSA can be constructed incrementally in a modular fashion, and even a much smaller initial version than the 200m $\times$ 200m $\times$ 20m benchmark assumed in this document could quickly supply the world's best sensitivity to many LLP physics scenarios, with the possibility of discovery within a few years.
All of this makes MATHUSLA a uniquely exciting opportunity for the upcoming HL-LHC upgrade that would continue to yield physics dividends into the HE-LHC era and beyond.

\clearpage 

\newpage
\section*{Acknowledgements}

The editors would like to thank David Morrissey for valuable feedback on a draft version of this manuscript. 
The work of D.~Curtin was supported in part by the National Science Foundation under grant PHY-1620074, and by the Maryland Center for Fundamental Physics. The work of M.~Drewes was supported by the Collaborative Research Center SFB1258 of the Deutsche Forschungsgemeinschaft and by the DFG cluster of excellence "Origin and Structure of the Universe" (\texttt{universe-cluster.de}).  The work of P.~Meade was supported in part by the National Science Foundation under grant NSF-PHY-1620628 R. N. Mohapatra, R. Sundrum and Y. Tsai acknowledge the support of NSF grant No.PHY1620074. The work of J. Shelton was supported in part by DOE grant DE-SC0017840.      The work of B. Batell is supported in part by the US Department of Energy under grant DE-SC0015634.  The work of N. Blinov is supported by the U.S. Department of Energy under Contract No.DE-AC02-76SF00515 The work of A. Fern\'andez T\'ellez, K. S. Caballero-Mora, M. Rodriguez Cahuantzi and J. C. Arteaga Velazquez is supported in part by CONACyT-M\'exico, Grants CB  243290 The work of Jae Hyeok Chang is supported by DoE Grant DE-SC0017938.   The work of T.~Cohen was supported in part by the U.S. Department of Energy under grant number DE-SC0018191 and DE-SC0011640. The work of P.~Cox is supported by the World Premier International Research Center Initiative (WPI), MEXT, Japan. The work of N.~Craig is supported in part by the US Department of Energy under the grant DE-SC0014129.   Y.Cui is supported in part by the US Department of Energy Grant DE-SC0008541.    The research activities of K..~R.~Dienes are supported in part by the U.S. Department of Energy under Grant DE-FG02-13ER41976 (DE-SC0009913) and by the U.S. National Science Foundation through its employee IR/D program.   The opinions and conclusions expressed herein are those of the authors, and do not represent any funding agencies. J.~Dror is supported in part by the DOE under contract DE-AC02-05CH11231. R.~Essig acknowledges support from DoE Grant DE-SC0017938. The work of J.~A.~Evans was supported in part by NSF CAREER grant PHY-1654502 and by DOE grant DE-SC0015655.  The work of A. Fern\'andez T\'ellez, K. S. Caballero Mora, M. Rodriguez Cahuantzi and J. C. Arteaga Velazquez is supported in part by CONACyT-M\'exico, Grants 243290     The work of E.~Fuchs was supported in part by the Minerva Foundation. The work of T.~Gherghetta was supported in part by the U.S. Department of Energy under Grant DE-SC0011842.   The CP3-Origins centre (C.~Hagedorn) is partially funded by the Danish National Research Foundation, grant number DNRF90.   J.~C.~Helo is supported by Chile grants Fondecyt No. 1161463,Conicyt PIA/ACT 1406 and Basal FB0821. M.~Hirsch is supported by the Spanish MICINN grants FPA2017-85216-P, SEV-2014-0398 and PROMETEOII/2014/084 (Generalitat Valenciana) Y.~Hochberg is supported by the Israel Science Foundation (grant No. 1112/17), by the Binational Science Foundation (grant No. 2016155), by the I-CORE Program of the Planning Budgeting Committee (grant No. 1937/12), by the German Israel Foundation (grant No. I-2487-303.7/2017), and by the Azrieli Foundation. A.~Hook. is supported by the NSF under grant NSF-PHY-1620074 and the Maryland Center for Fundamental Physics A.~Ibarra is partially supported by the DFG cluster of excellence EXC 153 ``Origin and Structure of the Universe'' and by the Collaborative Research Center SFB1258 S.~Ipek. is supported by the University of California President's Postdoctoral Fellowship Program and parttialy supported by NSF Grant No.~PHY-1620638. S.~Joon is supported by Korea NRF-2017R1D1A1B03030820 and NRF-2015R1A4A1042542. Simon Knapen is supported by the DoE under contract 5DE-AC02-05CH11231, and the National Science Foundation (NSF) under grants No.PHY-1316783  and  No. PHY-1002399. The work of E.~Kuflik is supported by the I-CORE Program of the Planning Budgeting Committee of the Israel Science Foundation (grant No. 1937/12), by the Binational Science Foundation (grant No. 2016153), and the Israel Science Foundation (grant No. 1111/17). This manuscript has been authored by Fermi Research Alliance, LLC under Contract No. DE-AC02-07CH11359 with the U.S. Department of Energy, Office of Science, Office of High Energy Physics.  H. J. Lubatti and C.~Alpigiani are supported in part by National Science Foundation grantPHY-1509257. The research of D.~McKeen is supported in part by the National Research Council of Canada.  The work of S. Moretti is supported in part through the NExT Institute, the STFC CG ST/L000296/1 and the H2020-MSCA-RISE-2014 grant n. 645722 (NonMinimalHiggs). The work of N. Nagata was supported in part by the Grant-in-Aid for Scientific Research No. 17K14270  The work of J.~M.~No was partially supported by the European Research Council under the European Unions Horizon 2020 program (ERC Grant Agreement no.648680 DARKHORIZONS), and by the Programa Atraccion de Talento de la Comunidad de Madrid under grant n. 2017-T1/TIC-5202.  The work of G.~Perez is supported by grants from the BSF, ERC, ISF, Minerva, and the Weizmann-UK Making Connections Program.  Research at Perimeter Institute is supported by the Government of Canada through the Department of Innovation, Science and Economic Development and by the Province of Ontario through the Ministry of Research and Innovation. The work of M.~Reece was supported in part by the DOE Grant de-sc0013607. The work of D.~J.~Robinson was supported in part by NSF grant PHY-1720252. The work of A. Fern\'andez T\'ellez, K. S. Caballero Mora, M. Rodriguez-Cahuantzi and J. C. Arteaga Velazquez is supported in part by CONACyT-M\'exico, Grants 243290. M. Rod\'iguez Cahuantzi also thankfully acknowledge the computer resources, technical expertise and support provided by the Laboratorio Nacional de Superc\'omputo del Sureste de M\'exico and VIEP-BUAP.     Daniel Stolarski is supported in part by the Natural Sciences and Engineering Research Council of Canada (NSERC). The work was supported by the Instituto de Investigaciones F\'isicas (IIF) and Universidad Mayor de San Andr\'es (UMSA) in collaboration with Benem\'erita Universidad Aut\'onoma de Puebla (BUAP), Puebla, M\'exico.   The research activities of B. Thomas are supported in part by NSF grant PHY-1720430.   B.~Tweedie was supported by DoE grant No. DE-FG02-95ER40896 and by PITT PACC.   This research is supported by the Cluster of Excellence PrecisionPhysics, Fundamental Interactions and Structure of Matter (PRISMA-EXC1098). The work of B.~Zaldivar was supported in part by the "Investissements d' avenir" program of the French ANR, Labex "ENIGMASS", and by the Programa Atraccion de Talento de la Comunidad de Madrid under grant n. 2017-T2/TIC-5455. The work of Yongchao Zhang was partially supported by the IISN and Belgian Science Policy (IAP VII/37).   

\clearpage 

\bibliographystyle{report}
\bibliography{report}

\end{document}